\documentstyle[epsf]{elsart}

\newcommand{\hm}{h^{-1}Mpc}
\newcommand{\be}{\begin{equation}}
\newcommand{\bef}{\begin{figure}}
\newcommand{\eef}{\end{figure}}
\newcommand{\hmp}{h^{-1}Mpc}
\newcommand{\ee}{\end{equation}}

\def\spose#1{\hbox to 0pt{#1\hss}}
\def\ltapprox{\mathrel{\spose{\lower 
3pt\hbox{$\mathchar"218$}}
 \raise 2.0pt\hbox{$\mathchar"13C$}}}
\def\gtapprox{\mathrel{\spose{\lower 
3pt\hbox{$\mathchar"218$}}
 \raise 2.0pt\hbox{$\mathchar"13E$}}}
\def\inapprox{\mathrel{\spose{\lower 
3pt\hbox{$\mathchar"218$}}
 \raise 2.0pt\hbox{$\mathchar"232$}}}

\begin{document}
 \begin{frontmatter}
\title{Scale-invariance of galaxy clustering }

\author[Roma,Infm]{F. Sylos Labini},
 \author[Roma,Infm]{M. Montuori} and
\author[Roma,Infm]{L. Pietronero}
\address[Roma]{Dipartimento di Fisica, Universit\`a di Roma
``La Sapienza'' P.le A. Moro 2, I-00185 Roma, Italy.} 
\address[Infm]{Istituto Nazionale Fisica della Materia, 
Sezione di Roma 1}

\tableofcontents

\begin{abstract}

Some years ago we proposed a new approach to the analysis of galaxy 
and cluster correlations based on the {\it concepts and methods of 
modern statistical Physics}. This led to the surprising result that galaxy 
correlations are fractal and not homogeneous up to the limits of the 
available  catalogs.  The usual
statistical methods, which are based on the assumption of homogeneity, 
 are therefore inconsistent for all the length scales probed so far, and 
a new, more general, conceptual framework is necessary to identify
the real physical properties of these structures.
In the last few years the 3-d  catalogs
have been significatively 
improved and we have extended our methods  to the analysis of 
number counts and angular  catalogs. This has led to a complete 
analysis of all the available data that we  present     
in this review. In particular we discuss the properties of the 
following  catalogs: CfA, Perseus-Pisces, SSRS, IRAS, LEDA, APM-Stromlo,
Las Campanas and ESP for galaxies and Abell and ACO for galaxy 
clusters. The 
result is that galaxy structures are highly irregular and self-similar: 
all the available data are consistent with each other and 
show fractal correlations (with dimension $D \simeq 2$) 
up to the deepest scales probed so far
($1000 \hmp$) and even more as indicated from the new interpretation 
of the number counts. The evidence 
for scale-invariance of galaxy clustering 
is very strong up to $150 \hmp$ due to the 
statistical robustness of the data but becomes 
progressively weaker (statistically) at larger distances due
to the limited data. In addition the 
luminosity distribution is  correlated with the space distribution in a 
specific way. These facts lead to fascinating conceptual implications 
about our knowledge of the universe and to a new scenario for the 
theoretical challenge in this field.
\end{abstract}
\end{frontmatter}
\centerline{PACS: 02.50.+s; 98.60.-a; 98.60.Eg}
\vspace{2cm}
\section{Introduction}
\label{intro}

The four most important experimental  evidences of modern 
Cosmology are:
- {\it The space distribution of galaxies and clusters}: the recent 
availability 
of several three dimensional samples of galaxies and clusters permits 
the direct characterization of their correlation properties. - {\it The cosmic 
microwave background radiation} (CMBR), that shows an extraordinary 
isotropy and an almost perfect black body spectrum.
- {\it The linearity of the redshift-distance relation}, usually known as the 
Hubble law. This law has been established by measuring 
independently the redshift and the distance of galaxies. The relation 
with velocity and space expansion is not tested directly and comes 
from the theoretical interpretation.
- {\it The abundance of light elements in the universe} that presents the most 
stringent evidence of generating and destroying atomic nuclei.

Each of these four points provides an independent experimental fact. 
The objective of a theory should be to provide a coherent explanation 
of all these facts together and to explain their interconnections.
Our work refers mainly to the first point, {\it the space distribution of 
galaxies and clusters} (Fig.\ref{fig1}) 
\bef 
%\vspace{}
\epsfxsize 16cm
\centerline{\epsfbox{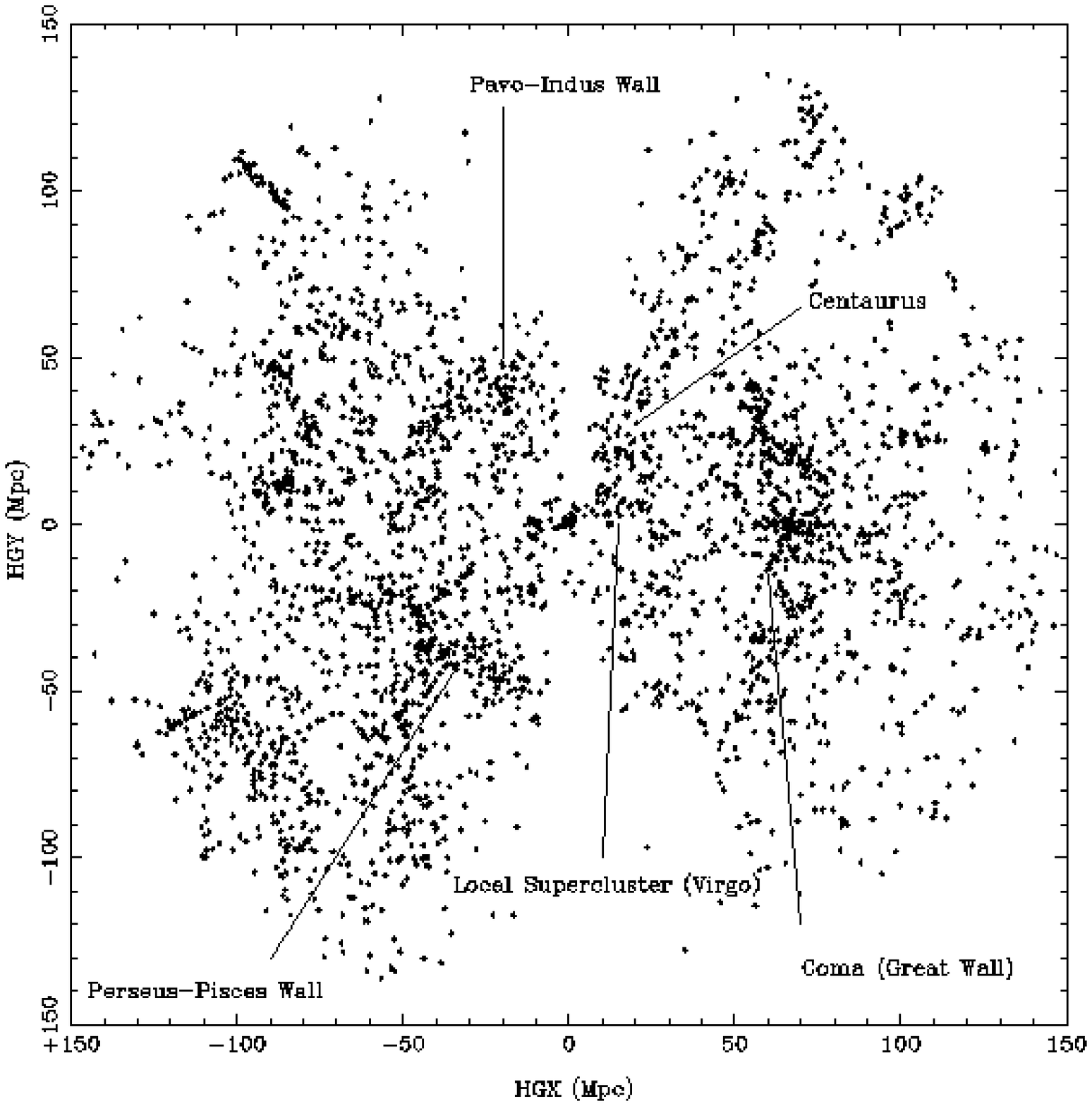}}
 \caption{\label{fig1}Face-on view of the hypergalactic plane, 
slice of $150\hmp$  on the Z axis from the LEDA database.
Each point corresponds to a galaxy, ours being located at the origins
of the coordinates. Our
   position is in the border with respect to
the Local Supercluster centered on Virgo cluster.
Artifacts of 
observations are clearly seen: zones without galaxy radial 
velocity because of the Milky Way extinction 
(vertical trace) and radial elongation of 
galaxy clusters (particularly in periphery).
The Local Supercluster seems to be surrounded by an empty region,
with a belt made of the neighbor superclusters:
Centaurus, Pavo-Indus, Perseus-Pisces and the Great Wall.  
Voids are filled up due to  a projection effect.
(Courtesy of H. Di Nella).}
\eef
which, however, is closely related to the 
interpretation of all the other points. In 
particular we claim that 
the usual methods of analysis are intrinsically inconsistent with
respect to the properties of the available samples.
The correct statistical analysis of the 
experimental data, performed with the methods of modern Statistical 
Physics, shows that the distribution of galaxies is fractal up to the 
deepest observed scales \cite{pie87,cp92}.
This result has caused a strong opposition from various authors in the 
field because it is in contrast with the usual assumption of large scale
 homogeneity which is at the basis of most theories. Actually 
homogeneity represents much more than a working hypothesis for 
theory, it is often considered as a paradigm or principle and for some 
authors it is conceptually absurd even to question it \cite{pee93}.
For other authors instead homogeneity is just the simplest working 
hypothesis and the idea that nature might actually be more complex is 
considered as extremely interesting \cite{wei72}. 
The two points of 
view are not so different after all because, if something considered 
absurd becomes real, then it must be really exciting. Given this 
situation, it may be interesting to analyze why this question develops 
such strong feelings. This  helps us to distinguish opinions from 
bare facts and to place the discussion in the appropriate perspective.
\bef 
%\vspace{}
\epsfxsize 18cm
\centerline{\epsfbox{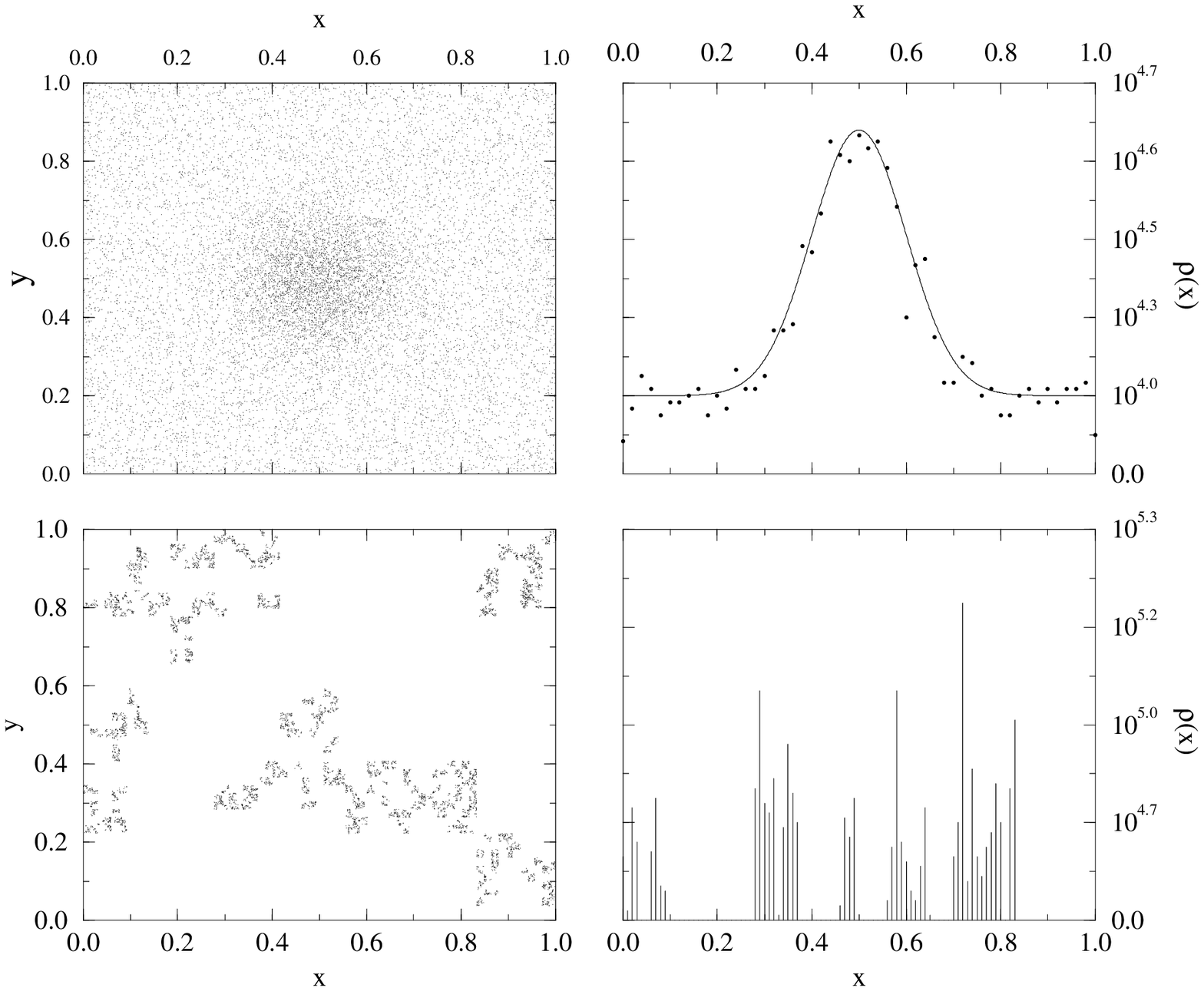}}
\caption{\label{fig2} 
Example of analytical and non-analytic structures. {\it Top panels:}
(Left)  A cluster in a homogeneous distribution. (Right)
Density profile. In this case the fluctuation
 corresponds to an enhancement
of a factor 3 with respect to the average density.
{\it Bottom panels:} (Left) Fractal distribution 
in the two dimensional Euclidean space. (Right) Density profile. In this 
case the fluctuations are non-analytical and there is no 
  reference value, i.e. the average density. The {\it average density
scales} as a power law from any occupied point of the structure.
}
\eef

The consensus on the homogeneity 
has never been quite broad. Early works of  Kant and Lambert
suggested a hierarchy of stars forming clusters 
forming galaxies which conform 
larger structures and so on.
Fournier d'Albe and Charlier \cite{char1,char2}
discussed a hierarchy where the mass 
within distance $d$ varies as $M \sim d$.  
De Vaucouleurs \cite{devac1,devac2} 
studied the possibility  that there is a 
universal density-radius power law as a basic factor in 
Cosmology, reflecting
a hierarchic distribution.  The hierarchical distribution 
proposed by De Vaucouleurs can be naturally developed within the 
framework of fractal geometry. From the theoretical point of view 
we refer to \cite{bslmp94} for a 
summary of different theoretical approaches
to this problem (see also the last section).
Interesting discussions about 
the case of hierarchical distributions of large scale structures 
can be found in  \cite{klein} and  in \cite{lerner}.

Most of theoretical Physics is based on analytical functions and 
differential equations. This implies that structures should be 
essentially smooth and irregularities are treated as single fluctuations 
or isolated singularities. The study of critical phenomena and the 
development of the Renormalization Group (RG) theory in the 
seventies was a major breakthrough 
\cite{wil81,amit86}.
In that field one  observes and describes
phenomena in which {\it intrinsic 
self-similar irregularities develop at all scales} and fluctuations cannot 
be described in terms of analytical functions. The theoretical methods 
to describe this situation cannot be based on ordinary differential 
equations because self-similarity implies the absence of analyticity 
and the usual mathematical Physics becomes not useful. In some 
sense the RG corresponds to the search of a space in which the 
problem becomes again analytical. This is the space of scale 
transformations but not the real space in which fluctuations are 
extremely singular. This peculiar situation seemed to be
characteristic of critical points, corresponding to the competition 
between order and disorder. However, in the past years, the 
development of Fractal Geometry \cite{man82},
has allowed us to 
realize that a large variety of structures in nature are intrinsically 
irregular and self-similar (Fig.\ref{fig2}). 

Mathematically these 
structures are described as singular in every point.  This property can be now 
characterized in a quantitative mathematical way by using the concept of 
fractal 
dimension and other   concepts developed in this field. However, given these subtle 
properties, it is clear that making a theory for the physical origin of 
these structures is   a rather challenging task. This is 
actually the objective of the present activity in the field 
\cite{epv95}.
The main difference between the popular fractals like coastlines, 
mountains, trees, clouds, lightnings etc., and the self-similarity of 
critical phenomena, is that criticality at phase transitions occurs only 
with an extremely accurate fine tuning of the critical parameters 
involved. In the more familiar structures observed in nature, instead, 
the fractal properties are self-organized, they develop spontaneously 
out of some dynamical process. It is probably in view of this important 
difference that the two fields of critical phenomena and Fractal 
Geometry have proceeded somewhat independently.

The fact that we are traditionally accustomed to think in terms of 
analytical structures has a crucial effect on the type of questions we 
ask and on the methods we use to answer them. If one has never been 
exposed to the subtlety of non-analytic structures, it is natural that 
analyticity is not even questioned. It is only after the above 
developments that we can realize that the property of analyticity 
can be tested experimentally and that it may, or may not, be present  
in a given physical system.

We can now appreciate how this discussion is directly relevant to 
Cosmology by considering the question of the {\it Cosmological Principle }
(hereafter CP). It is quite reasonable to assume that the Earth is not at 
a privileged position
in  the universe and to consider this as a principle, the CP. The 
usual   implication of this principle is that the universe 
must be homogeneous. This reasoning implies the hidden assumption 
of analyticity that often is not even mentioned. In fact the above 
reasonable requirement only leads to {\it local isotropy}.
 For an analytical 
structure this also implies homogeneity \cite{wei72}. However, if 
the structure is 
not analytical, the above argument  does not hold. For example, a 
fractal structure is locally isotropic but not homogeneous.
This means 
that a fractal structure satisfies the CP in the sense that all the points 
are essentially equivalent (no center or special points), but this does 
not imply that these points are distributed uniformly \cite{sl94}.

This important distinction between isotropy and homogeneity has 
other important 
consequences. For example, it clarifies that drawing 
conclusions about 3-d galaxy correlations from the angular 
distributions alone can be rather misleading. In addition, from this 
new perspective, the isotropy of the CMBR may appear less 
problematic in relation with  the highly irregular three dimensional 
distribution of matter and this may lead to theoretical approaches of 
novel type for this problem.
In the present work, however, we limit our discussion to 
the way to analyze the data provided
 by the galaxy  catalogs,
from the broader perspective in which analyticity and homogeneity 
are not assumed a priori, but they are explicitly tested. The main 
result is that the data of different galaxy catalogs 
become actually consistent with each other and 
coherently point to the same conclusion of fractal correlations up to 
the present observational limits.

\subsection{ Statistical Methods and Correlation Properties}
\label{introstat}

{\it - Usual arguments:} 
Before the extensive redshift measurements of the 80s, the 
information about   galaxy distributions was only given  
in terms of the two 
angular coordinates. These angular distributions appear rather 
smooth at relatively large angular scale, like for example the lower 
part of Fig.\ref{fig3}.
\bef 
%\vspace{}
\epsfxsize 14cm
\centerline{\epsfbox{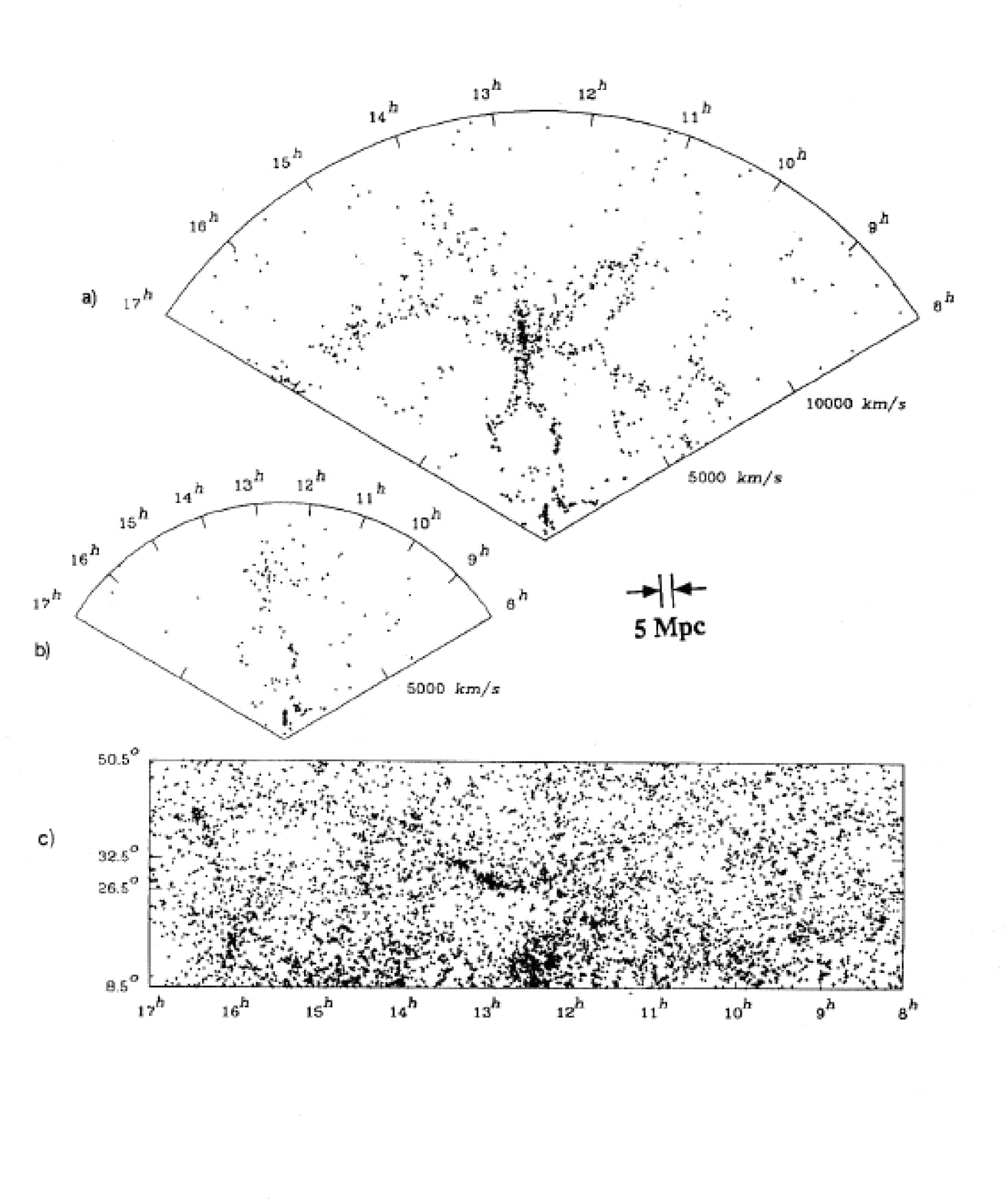}}
\caption{\label{fig3}
A slice of  the three dimensional galaxy distribution (old CfA1 catalog 
{\it (a)}
and new CfA catalog {\it (b)})
compared with the corresponding {\it (c)}
angular distribution (the portion between 
$26.5^{\circ}$ and $32.5^{\circ}$) - from the de Lapparent \etal 1988.
Note that the angular distribution appears relatively 
homogeneous while the real three distribution in space is much more irregular. 
In particular this picture points out the so-called Great Wall 
which extends over the entire sample (at least $170 \hmp$).
We also show the size of the galaxy "correlation-length" ($r_0 = 5 \hmp$)
derived from the standard analysis. The more 
general analysis we discuss  here implies instead that 
an eventual correlation length should be larger than any observable
structure, i.e. $\gg 170 \hmp$ and that the present data show well defined
fractal properties up to the sample limits.
}
\eef
Assuming that this smoothness corresponds to a real homogeneity
in 3-d space and estimating the characteristic depth of the angular 
 catalog from the magnitudes,  "a characteristic length"   
$r_0 = 5 \hmp$ has been measured \cite{pee93}. 
The idea was that beyond such a distance the 3-d galaxy 
distribution would become about homogeneous and it could be well 
approximated by a constant galaxy density. This value, apart from the 
eventual dark matter, is the one to
 use into Einstein equations 
to derive the Friedmann metric and the other usual concepts. 

Later on, the 
measurements of the galactic redshifts, plus the Hubble law, provided 
also the absolute distances and could identify the position of galaxies 
in space. However, the 3-d galaxy distributions turned out 
to be much 
more irregular with respect to their angular projections and unveiled 
{\it large structures and large voids},
 as shown in the upper part of Fig.\ref{fig3}. 
At first these irregular structures appeared to be in contradiction 
with the picture derived from the angular  catalogs and, as we discuss 
in what follows, 
they really are. However, in 1983, a
correlation analysis of the 3-d distribution 
CfA1 catalog \cite{huc83}
was performed by Davis \& Peebles \cite{dp83} 
and the 
result was again that the correlation length was $r_0 = 5 \hmp$ as for 
the angular  catalogs. This seemed to resolve the puzzle, because it 
was interpreted as if a relatively small correlation length can be 
consistent with the observation of large structures. This value for
 $r_0$ 
has not been seriously questioned, even after the observation of huge 
structures, like the Galaxy Great Wall, that extend up to $\sim 170 \hmp$ or more.

The usual correlation analysis is performed by estimating at which 
distance ($r_0$) 
the density fluctuations are comparable to the average density in the 
sample ($\xi(r)=  \langle n(0)n(r) \rangle/ \langle n \rangle^{2} -1$; 
$\xi(r_0) \equiv 1$). 
Now it's commonly accepted 
that there are fractal correlations at least at 
small scales \cite{dav97} (see also \cite{pmsl97}). 
The important physical question is 
therefore to identify the 
distance $\lambda_0$ 
at which, possibly, the fractal distribution has a 
crossover into a homogeneous one. The first question is whether 
there is, or not,  such a crossover. This is the
 {\it real correlation length},
beyond which the distribution can be approximated by an average 
density. The problem is therefore to understand the relation between $r_0$
 and $\lambda_0$: are they the same, or closely related, or do 
they correspond to different properties? 
This is, actually, a subtle point.
In fact, if the galaxy distribution becomes really 
homogeneous at a scale $\lambda_0$   within the sample in question, 
then the value of $r_0$
 is related to the real correlation properties of the distribution and one 
has $r_0 \simeq \lambda_0/2$. 
If, on the other hand, the fractal correlations extend up to the sample 
limits, then the resulting value of $r_0$ has nothing to do with the real 
properties of galaxy distribution, but it is fixed just by the size of the 
sample \cite{cp92}. 

{\it New Perspective:} Given this situation 
of ambiguity in relation with the real 
meaning of $r_0$, it is clear that the usual study of 
correlation  in terms of 
the function $\xi (r)$ is not an
appropriate method to clarify these 
basic questions.
The essential problem is that, by using the function $\xi(r)$, one 
defines the amplitude of the density fluctuations by normalizing them 
to the average density of the sample in question. This implies that the 
observed density should be the real one and it should not depend on 
the given sample or on its size, apart from Poisson fluctuations. 
However, if the distribution shows long range (fractal) correlations, 
this 
approach becomes meaningless. For example, if one studies a fractal 
distribution using the concept $\xi(r)$ as defined previously,
a characteristic length $r_0$
 can be identified, but this is clearly an artifact because the structure is 
characterized exactly by the absence of any defined  length \cite{cp92}.
It is important to stress that the so-called 
correlation length $r_0$
is just one of the statistical 
quantities used by  the standard approach, which are not suitable to describe 
irregular (scale-invariant) distributions.

The appropriate analysis of   
correlations should, therefore, be 
performed using methods that can check homogeneity or fractal 
properties without assuming a priori either one.
The simplest method to do this is to look directly at the conditional 
density  $\Gamma(r) \sim  \langle n(0)n(r) \rangle$,
without normalizing it 
to the average. There are several other methods that we 
discuss in what follows.
This is not all however,
because one has also to be 
careful not to make hidden assumptions of homogeneity in the specific 
procedure to evaluate these correlations. For example, if a galaxy is 
close to the boundary of the sample, it is possible that the sphere of 
radius $r$ around it, {where the conditional density is computed}, 
 may lie in part outside the sample boundary. In this 
case the usual procedure is to use weighting schemes of various types 
to include also these points in the statistics. In this way, one implicitly 
assumes that the fraction of sphere contained in the sample is 
sufficient to estimate the properties of the full sphere. This implies 
that the properties of a small volume are assumed to be
the same as for a larger 
volume (the full sphere). This is a hidden assumption of homogeneity 
that should be   avoided by including only the properties of those 
points for which a surrounding sphere of radius $ r$ is fully included in 
the sample. These procedures are fully standard in modern Statistical 
Mechanics and a detailed description can be found in \cite{cp92,pmsl97} 
(Sec.2. and Sec.3). %CP 1992.
This means that the statistical validity of a sample is limited to the 
radius of the largest sphere that can be contained in the sample. We 
call this distance $R_{eff}$ 
and it should not be confused with the sample 
depth $R_s$, which can be in general much larger, 
depending on the survey geometry.

In 1988, 
we  reanalyzed the CfA1  catalog \cite{cps88}.
The result was that the  catalog has   statistical validity up to 
$R_{eff} = 20 \hmp$  and, up to this length, it shows well defined fractal 
correlations with a value of the fractal dimension $D \approx  1.5$. This shows 
therefore that the "correlation length" $r_0=5\hmp$
 derived by \cite{dp83}, 
was a spurious result due to an 
inappropriate method of analysis and it has nothing to do with the real 
correlation properties of the system. A similar new analysis of the Abell 
cluster  catalog also showed 
fractal properties up to $R_{eff} = 80\hmp$,
 so  that also the cluster "correlation length" 
$r_0^c = 25\hmp$ \cite{bs83}
% (Bachall and Soneira, 1983) 
should be considered as spurious. One consequence of 
these results was that the so called galaxy-cluster mismatch could be 
automatically eliminated by the appropriate analysis. Also other properties 
like $\delta N/N$, directly related to $r_0$, 
suffer from the same consistency problems because the lack of a 
reference value \cite{bslmp94}. 

This situation 
led to a rather controversial debate in the field \cite{dav97,pmsl97}. In the meantime 
many more data became available and we decided to perform a 
complete analysis of all the data for galaxies and clusters. In what follows
we report our main results.

\subsection{Organization of the paper}
\label{introorg}

{\it Fractal
Geometry}
provides a quantitative mathematical framework 
for the analysis and the 
characterization of irregular non analytic structures, as 
well as regular and homogeneous ones. We introduce in Sec.\ref{statmec}
the main 
properties and concepts of fractal geometry which we use in the 
analyses presented in this work.
 In particular, we stress the conceptual consequence 
of the lack of any reference value, like 
the average density, in the case of self-similar structure 
and the consequent shift of the theoretical investigation
from {\it "amplitudes"} towards {\it "exponents"}. The properties of the 
(average) conditional density are discussed. Such a quantity is 
the most important statistical tool that we 
use for the characterization of the correlation properties of
galaxies and clusters.

The correlation analysis for various galaxy and 
cluster redshift surveys is presented in Sec.\ref{corran}, together 
with an broad 
and detailed discussion of the basic techniques that we employ. Moreover, 
we present several tests on the treatment of the boundary 
conditions, and
on the stability of the correlation analysis versus eventual 
systematic 
and random 
errors that can affect the real data. Finally,
 we  clarify the luminosity
segregation effect, as well as the galaxy cluster mismatch, showing that
these concepts arise only from an inconsistent data analysis.

The determination of the power spectrum is discussed 
in Sec.\ref{powerspect}. We point out the conceptual
difficulties of the standard analysis and we introduce a more general
determination of the power spectrum, that can be useful for the 
characterization of the properties of self-similar irregular
 systems, as well as regular ones.

How many galaxies should contain a galaxy sample, in order to be
statistically meaningful ? The discussion of this important question
 allows us, in Sec.\ref{validity}, to clarify the concept of {\it fair sample} 
and to derive 
a quantitative criteria to define the statistical validity of samples.

In Sec.\ref{radial}
 we introduce and discuss     another determination of the galaxy
space density: the radial density. This quantity is determined from a single 
point,
and allows one to reach very large distances
 especially in the case of narrow and
 deep surveys.
The price to pay, however, is that this quantity is not
an average one. Hence, it is subject to finite size and intrinsic 
fluctuations,
that 
must be study in great detail, in order to interpret 
correctly the experimental data. 
In particular the nature of intrinsic fluctuations
which are inherent to fractal sets, is
qualitatively different from the poissonian one.
In this section,
we summarize of the different determinations
of the space density, i.e. by the conditional density 
and the radial density.
The results of the correlation analysis 
in  the various available redshift surveys are shown to be 
compatible with each other.
 In such a
 way we 
 may present the full correlation analysis in the range $0.5 \div 1000 \hmp$.
The result is that 
galaxy properties are compatible in the different catalogs, and
there are fractal correlations with dimension $D = 2.0 \pm 0.2$ up
to the deepest scale observed for visible matter.

At the light of the interpretation of the radial density behavior,
 we   discuss in Sec. \ref{counts} the analysis of the number counts 
as a function of the apparent flux (or magnitude). 
The counts of various 
astrophysical objects are usually characterized
by  a spurious regime at 
bright fluxes, that seems to be nearly Euclidean.
At faint  apparent fluxes (or magnitudes) 
the number counts deviates from the Euclidean 
behavior, having a well defined exponent.
The crucial point in this case is that
the counts are determined by the Earth, 
 without performing
an average over different observers. This leads to finite
size spurious fluctuations, as well as intrinsic ones which are 
not smoothed out by averaging,
which dramatically affect the behavior at small scale, i.e. at
the bright end of the number counts. On the other hand at faint 
fluxes, as the space volume involved is large enough, it is possible 
to obtain the genuine scaling behavior {\it without performing averages}.
Our new interpretation shows that
 the counts of different astrophysical objects, 
such as galaxies
in the different 
spectral bands, X-ray sources, Radio-Galaxies, Quasars
are all {\it compatible}
 with a scale-invariant distribution with dimension  
$  D \approx 2$  up
to the faint end of
 the counts, that is to say the deepest scales ever investigated for visible 
matter (i.e., $m \sim 29$ in the $B$ band for optical galaxies). 
Finally we clarify the problems of angular correlations,
namely the   uniformity and isotropic nature of the angular projection
of fractal structures, and the scaling of the amplitude of the angular 
correlation function.

In Sec.\ref{lumspace}
 we consider, by  a quantitative analysis, an important observational
fact: galaxy positions and luminosities are strongly correlated.
This fact has lead to various morphological interpretation of
the distribution of galaxies with different luminosity. In this section,
we present the multifractal analysis of galaxy distribution. Such
an analysis allows us to consider the correlation between the 
space and luminosity distributions within 
a quantitative mathematical framework, 
and to unify  these two distributions. 
Moreover, we clarify the 
segregation of luminous galaxies in the core of clusters and fainter ones 
in the field, in terms of multifractals. 
The multifractality of matter distribution should therefore claim a 
central stage in 
theoretical investigations. 

Finally in Sec.\ref{conclusion}
we present our main conclusions.  We stress 
the paradoxical situation due to the coexistence of 
the fractal distribution of visible matter and the strictly linear Hubble
law at the same scales. In fact, in the standard scenario of the 
Friedmann models, the Hubble law is a 
consequence of the assumption of
homogeneity of matter distribution. 
We examine various possible solutions of such a problem
with particular emphasis on the role of the so-called {\it dark matter}. 
Moreover we discuss the theoretical implications 
and change of perspective implied by our results.
At the end of this section we briefly present our predictions for the 
forthcoming redshift surveys like CfA2, 2dF, and SLOAN.

In the Appendix we report all the details of the catalogs 
analyzed: CfA1, SSRS1, Perseus-Pisces, IRAS $2 Jy$, 
IRAS $1.2 Jy$,  Las Campanas Redshift Survey, the ESO Slice Project, 
and the Stromlo-APM redshift
 survey for galaxies (we have considered and discussed 
 also the SSRS2 and CfA2 redshift surveys, which are not yet
published) and Abell and ACO 
for galaxy clusters.

\section{Statistical methods and correlation properties for galaxy
 distributions}
\label{statmec}

We have discussed 
in a series of papers 
\cite{pie87,cps88,cp92,bslmp94,slgmp96,dmpps96} 
the conceptual problems of the standard correlation analysis
and we have introduced the correct correlation analysis that 
should be applied for the characterization of the statistical 
properties of irregular as well as regular distributions.
Here we briefly review the main results and 
we introduce the basic concepts of fractal geometry.

\subsection{Usual analysis, conceptual problems and correlation lengths}
\label{usualan}
 
In the following  description the galaxy is 
treated as a point (while in Sec.\ref{lumspace} 
we  consider whole mass distribution). 
In this case the statistical analysis is  
performed only on the number density, neglecting 
the masses of galaxies,
\be
\label{e21}
n(\vec{r}) = \sum_{i=1}^N \delta(\vec{r}-\vec{r}_i)  \, .
\ee
The standard statistical analysis considers only the number density 
 (Eq.\ref{e21}) and it consists in the computation of the so-called
$\xi(r)$ correlation function defined as \cite{pee80} 
\be
\label{e22}
\xi(r) = \frac{\langle n(\vec{r}_*) n(\vec{r}_*+\vec{r}) \rangle}{\langle
n \rangle  ^2} - 1
\ee
where the average is defined as 
\be
\label{e23}
\langle \cdot \cdot \cdot \rangle   = \frac{1}{V}
\int_V (\cdot \cdot \cdot) d\vec{r}_*
\ee
and 
\be
\label{e24}
\langle n \rangle   = \frac{N}{V}
\ee
is the average density of galaxies in the sample
($V$ is the sample volume and $N$ is the number of galaxies
contained in that volume). In the computation of 
$\xi(r)$ (Eq.\ref{e22}) it is performed the angular average over all
the possible directions, and only the radial dependence is considered.
From the very definition of $\xi(r)$ it follows that \cite{cp92}
\be
\label{e25}
\int_V \xi(r) dr =0 \; .
\ee
This means that if $\xi(r)$ is 
positive for some range of values of $r$, then there must
be other ranges in which it is negative, in view
of its definition as a measure of fluctuations from the average. 
$\xi(r)$  is an appropriate function to
characterize correlations for systems in which the
 average density 
(Eq.\ref{e24})
 is a well defined intrinsic property, as for example in liquids.
In this case the average density $\langle n \rangle$ 
is a well defined quantity beyond a certain scale (say several
times the mean interparticle separation), 
 and the definition of $\xi(r)$
is meaningful. 
In fact, if the system shows correlations then
$\langle n(\vec{r}_*)
 n(\vec{r}_* + \vec{r}) \rangle  \gg \langle n \rangle^2$  
so that $\xi(r) \neq 0$, while if there are no correlations 
$\xi(r)=0$. The distance $r_0$ 
defined as $\xi(r_0) = 1$, separates 
a correlated regime characterized by large fluctuations,
from a regime of negligible fluctuations.

In the  case of irregular systems, characterized by large structures and voids,
 the average density  is  not a well defined property, and the 
$\xi(r)$ analysis   gives rise to spurious results. In other words 
the homogeneity 
assumption used in the $\xi(r)$ analysis must be seriously questioned.

\subsubsection{Three dimensional galaxy and cluster distributions}
\label{threedim}
As we have already
mentioned in the previous section, the usual statistical analysis 
of galaxy correlation is performed by measuring the $\xi(r)$ function
in three dimensional samples. In the past twenty years, 
this kind of analysis has been performed for galaxy and cluster
distributions (Sec.\ref{corran}; 
for a detailed discussion;  we refer to  \cite{pee93,dav97,sw95,pee80} 
for a review of the state of the art in the field) and the 
results are quite similar.
At small distances, the standard two point correlation function
can be characterized by a power law behavior 
\be
\label{e26}
\xi_{GG}(r) \sim A_G r^{-\gamma_{GG}}
\ee
where $\gamma_{GG} \approx 1.7$. The so-called 
correlation length $r_0$ is defined as
$\xi(r) \equiv 1$, so that we have   \cite{pee80,dav97}
\be
\label{e27}
r_0^{G} = A_G^{\frac{1}{\gamma_{GG}}} \approx 5 \hmp \; .
\ee
At twice this distance 
the $\xi(r)$ function deviates from a power law behavior 
and becomes negative or zero (Fig.\ref{fig4}).
\bef 
%\vspace{}
\epsfxsize 8cm
\centerline{\epsfbox{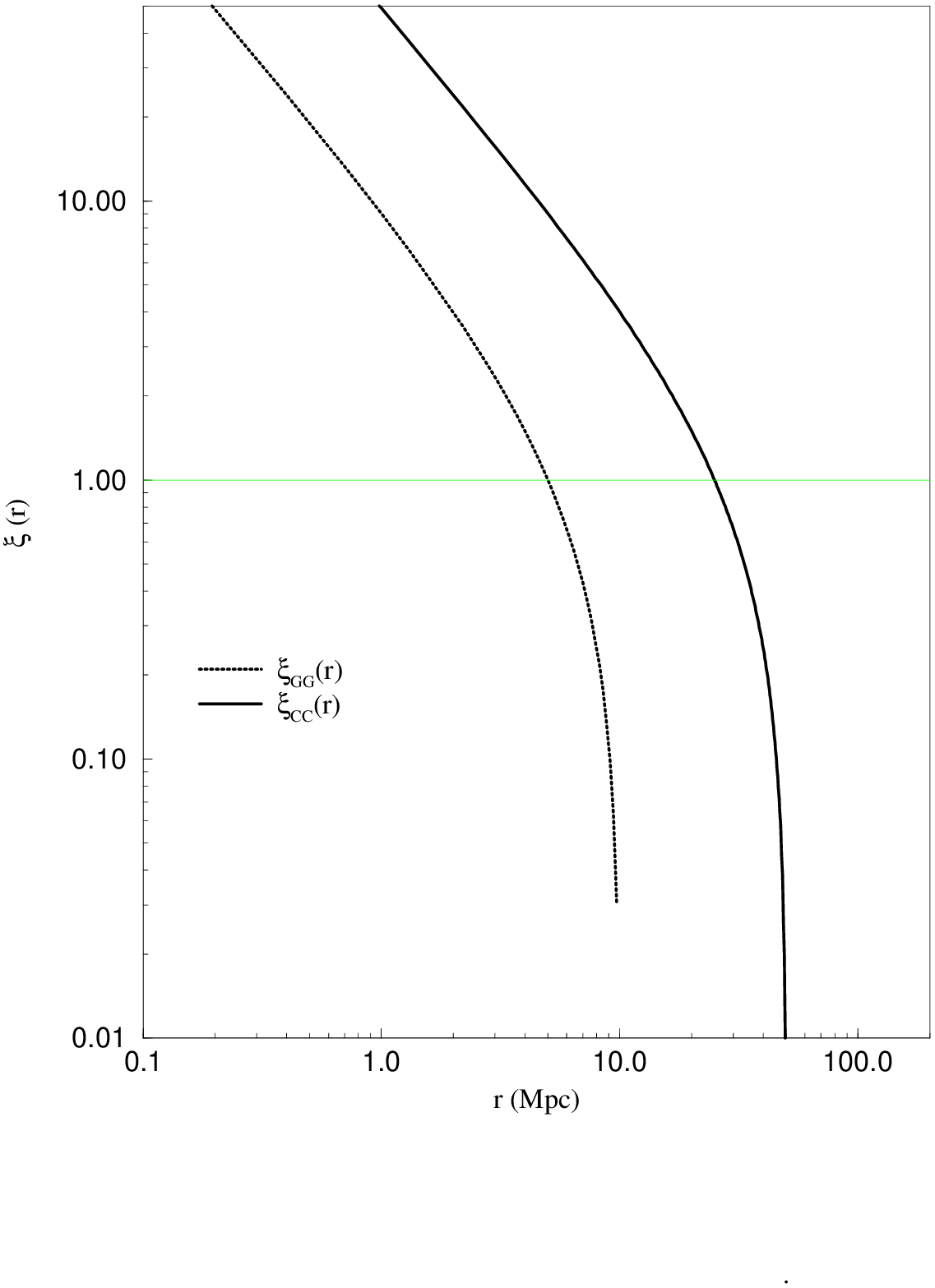}}
\caption{\label{fig4}  Standard analysis of galaxy $\xi_{GG}$
and cluster $\xi_{CC}$ distributions. The two functions show power
law behavior with about the same exponent, but the one corresponding 
to cluster has a larger amplitude.}
\eef
In the case 
of galaxy clusters, the volumes explored are larger than those of 
galaxy catalogs. However the behavior of 
the standard correlation function is found to be 
quite similar to that
of galaxies, and, at small scales, it has been measured \cite{bs83}
that
\be
\label{e28}
\xi_{CC}(r) \sim A_C r^{-\gamma_{CC}}
\ee
where  $\gamma_{CC} \approx 1.7$ as in the case of galaxies. 
The amplitude $A_C$ is larger than $A_G$, so that the cluster 
correlations length is
\be
\label{e29}
r_0^{C} = A_C^{\frac{1}{\gamma_{C}}} \approx 25 \hmp \; . 
\ee
The difference between galaxy and cluster  correlation length is known
as the {\it "galaxy - cluster mismatch"}. 
Various interpretations in literature
have been proposed in order to explain such a result. We
show later 
that the self-similar behavior of galaxy and cluster distributions 
naturally resolve this apparent mismatch (Sec.\ref{gammaclu}).

\subsubsection{Angular distributions}
\label{angdist}
The problem of the angular analysis consists
of  the reconstruction of the three dimensional properties
of galaxy distribution 
from the knowledge of 
the angular coordinates and the 
apparent magnitude.
This is a very complex problem (see also 
Sec.\ref{counts} and Sec.\ref{lumspace}), 
and the standard analysis does not consider
 some conceptual
and fundamental
  difficulties, 
which make very 
hard the deduction of the
  three dimensional properties from the angular information.

The standard method used to analyze angular catalogs,
is based on the assumption that galaxies are correlated only at small
distances. In such a way  the effect of the large spatial inhomogeneities
is not considered at all. Under this  assumption, that is
not
supported by any experimental evidence, it is possible to derive the
Limber equation \cite{lim70,lim71}. In practice,
the angular analysis is performed by computing the two 
point correlation function
\be
\label{e210}
\omega(\theta) = \frac{\langle n(\theta_0)n(\theta_0+\theta) \rangle}
{\langle n \rangle} -1
\ee
that is the analogue of Eq.\ref{e22} for the angular coordinates.
The results of such an analysis are quite similar to the 
three dimensional ones. In particular, it has been obtained that, 
in the limit of small angles,
\be
\label{e211}
\omega(\theta) = \theta^{-\gamma+1}
\ee
with $\gamma \approx 1.7$. It is possible to show \cite{pee80} that,
in the Limber approximation (Eq.\ref{e211}), the angular correlation
function 
 corresponds to
$\xi(r) \sim r^{-\gamma}$ for its three dimensional counterpart
(in the case $\gamma > 1$).
The determination of the correlation length is more complex, 
and it requires a comparison of catalogs with different depth, i.e. with 
different apparent magnitude limit. 
Using the Limber equation and the assumption that the luminosity distribution
is independent on the spatial one, it is possible to
derive the following relation
\be
\label{e212}
\omega'(\theta'_{12} = (R_s/R_s')\theta_{12}) = 
(R_s/R_s')\omega(\theta_{12}) \; .
\ee
Such an equation links the depth of a certain catalog $R_s$ to that 
of another catalog $R_s'$. In Fig.\ref{fig5}
\bef %\vspace{}
\epsfxsize 12cm
\centerline{\epsfbox{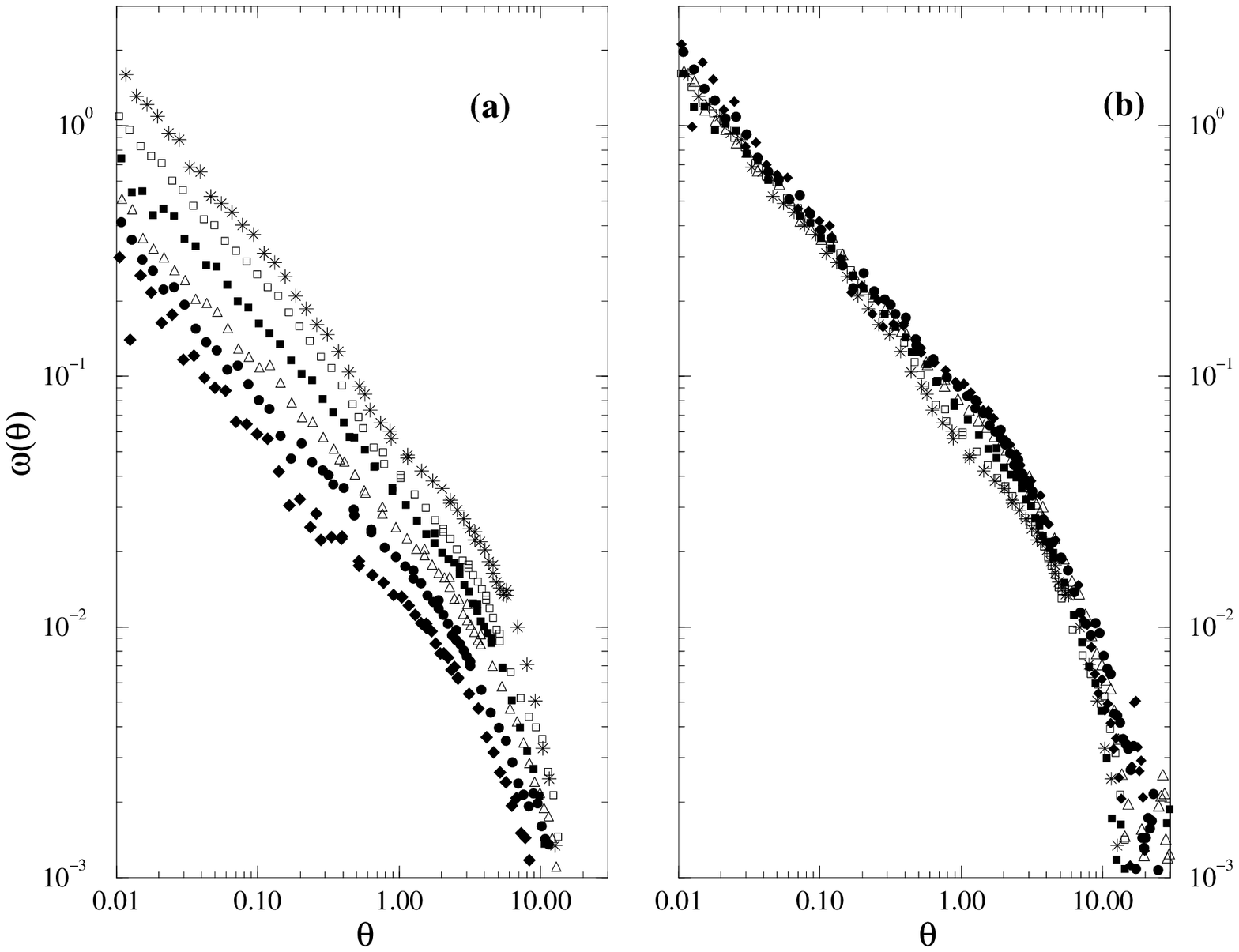}}
\caption{\label{fig5}  {\it (a)} Scaling test of the two point angular correlation
function with sample depth (different apparent magnitude limit) 
in the APM angular catalog (from Maddox \etal, 1990). 
In {\it (b)} it is shown the rescaling of the angular correlation function, 
by applying Eq.12, that holds for a homogeneous distribution. 
We refer to Sec.7
for a detailed discussion of such a test.}
\eef
it is shown the dependence of the amplitude of $\omega(\theta)$ with 
sample depth. The fact that Eq.\ref{e212}  is found to be satisfied in 
real catalogs, 
has been interpreted as an evidence of homogenization \cite{pee93}.
The corresponding correlation length is quite similar to that found
in the three dimensional catalogs, i.e. $r_0 \approx 5 \hmp$.
However this kind of analysis suffers of the same problems of $\xi(r)$, 
and it is based on untested assumptions. In particular the main 
point is that it does not take into account the effects of 
 spatial inhomogeneities.

\subsection{Large structures, 
long range correlations and fractal properties}
\label{largestr}
Fractals are simple but subtle.
In this section we provide a brief
description of their essential properties.
This description is intended to illustrate
 the consequences of the properties of
self-similarity so that, if this property is
actually present in the experimental data, we 
are able to detect it correctly; on the contrary,
if the data were not consistent  with the fractal
properties, we have to know well the properties
of fractals in order to eventually conclude that
observations are actually in contrast with them.

A basic element of fractal structures
is that if one magnifies a small portion
of them, this reveals a complexity comparable
to that of the entire structure. This is  geometric
 self-similarity and it has deep implications about
 the non-analyticity of these structures. In fact,
analyticity or regularity  implies that at
some small scale the profile becomes smooth
 and one can define a unique tangent. Clearly
this is impossible in a self-similar structure
 because at any small scale a new structure
appears and the structure is never smooth.
 Self-similar structures are therefore intrinsically
 irregular at all scales and this is why many familiar
phenomena have remained at the margins of scientific
investigation.  The usual mathematical concepts in
 Physics are mostly based on analytical functions
and, in this perspective, irregularities are seen as
imperfections. Fractal geometry changes completely
this perspective by focusing exactly on these intrinsic
irregularities and it allows us to characterize them in a
 quantitative mathematical way.

\subsubsection{The "Mass-length" relation}
\label{masselength}

A fractal is a  system in which more
and more structures appear at smaller and
smaller scales and the structures at small
scales are similar to the one at large scales.
In Fig.\ref{fig6}, we show an elementary (deterministic)
fractal distribution of points in space whose
construction is trivial. Starting from a
 point occupied by an object, we count how
many objects are present within a volume
characterized by a certain length scale, in
 order to establish a generalized "mass-length"
relation from which one can define the fractal
dimension.
\bef %\vspace{}
\epsfxsize 10cm
\epsfysize 10cm
\centerline{\epsfbox{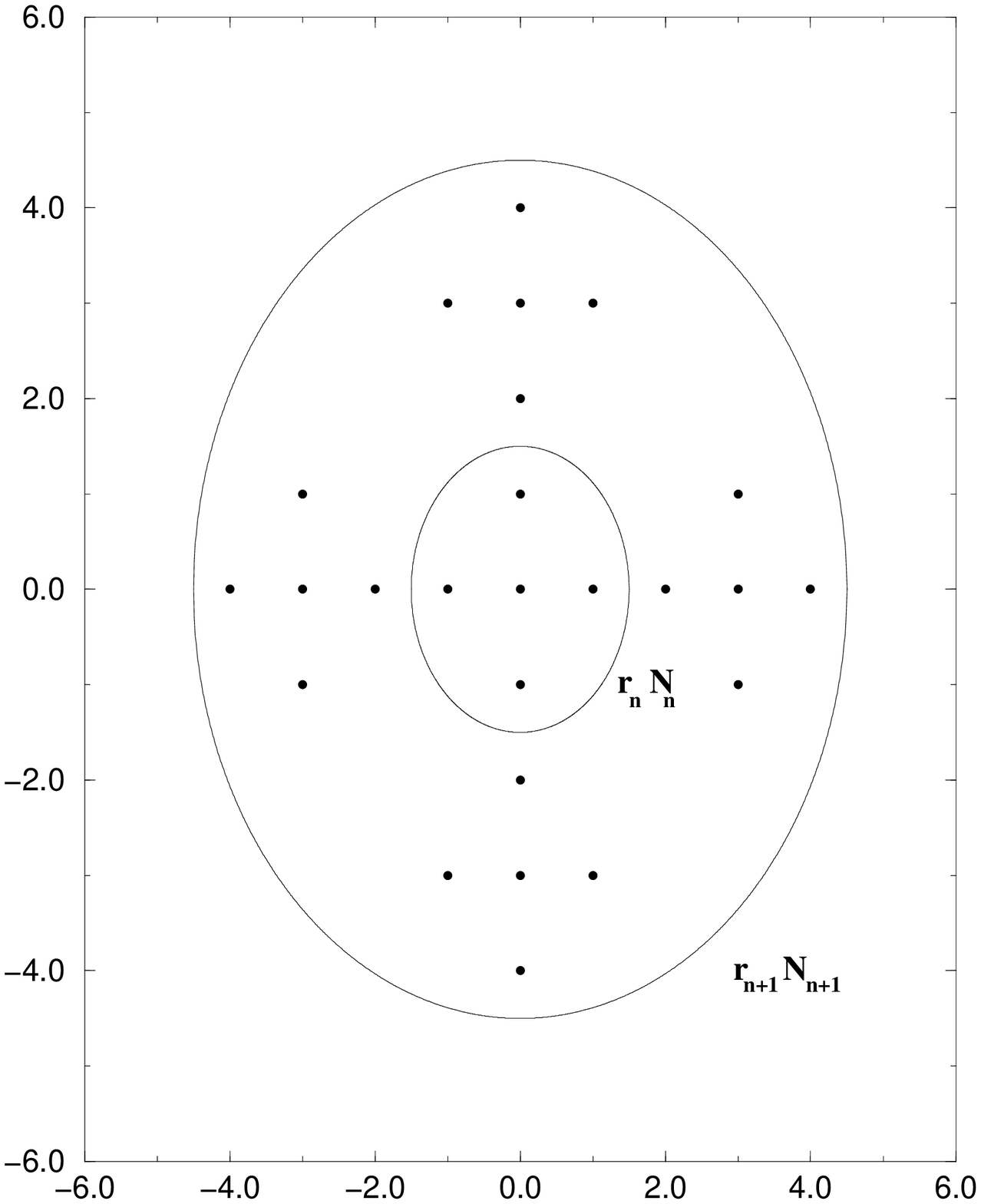}}
\caption{\label{fig6}  A simple example of a deterministic fractal
in the two dimensional Euclidean space.
The same structure repeats at different scales in a self-similar way.}
\end {figure}
Suppose that, in the structure of
Fig.\ref{fig6},  we can find $N_0$ objects in a volume of
 size $\:r_{0}$. If we consider a larger volume
of size $\:r_{1} = k \cdot r_{0}$, we   find
$\:N_{1} ={\tilde  k} \cdot N_{0}$ objects.
In a self-similar structure, the parameters $\:k$
 and $\:{\tilde k}$ are the same also for
other changes of scale. So, in general, in a
structure of size $\:r_{n} = k^{n}\cdot r_{0}$,
we   have $\:N_{n} ={\tilde  k^{n}} \cdot N_{0}$
objects. We can then write a (average) relation
 between $\:N$ ("mass") and $\:r$ ("length") of type
\be
\label{e213}
N(r) = B\cdot r^{D}
\ee
where $D$ is the fractal dimension
\be
\label{e214}
D = \frac{log{\tilde k}}{log k}
\ee
and depends on the rescaling factors $\:k$
and $\:{\tilde k}$. The prefactor $\:B$
is instead related to the lower cut-offs $\:
N_{0}$ and $\:r_{0}$ of the structure
\be
\label{e215}
B =\frac{ N_{0}}{r_{0}^{D}} \; .
\ee
It should be noted that Eq.\ref{e213} corresponds
to a smooth convolution of a strongly
fluctuating function as evident
in Fig.\ref{fig6}. Therefore a fractal
structure is always connected with large
 fluctuations and clustering at all 
scales.

From Eq.\ref{e213} we can readily compute the
average density $\:<n>$ for a sample of
 radius $\:R_{s}$ which contains a portion
of the fractal structure. The sample volume
is assumed to be a sphere 
($\:V(R_{s}) = (4/3)\pi R_{s}^{3}$) and therefore
\be
\label{e216}
<n> =\frac{ N(R_{s})}{V(R_{s})} = 
\frac{3}{4\pi } B R_{s}^{-(3-D)} \; . 
\ee
From Eq.\ref{e216} it follows that the average density
is not a meaningful concept in a fractal
because it depends explicitly on the sample
size $\:R_{s}$. Moreover for
$\:R_{s} \rightarrow \infty$ the average density
$\:<n> \rightarrow  0 $: this implies that 
a fractal structure is asymptotically
dominated by voids. Therefore   the
average density $\:<n>$ is not a well defined
quantity: the
amplitude of this function essentially
refers to the unit of measures given by the
lower cut-offs but it has no particular
 physical meaning.
We can also define the conditional density from any point
occupied as
\be
\label{e217}
\Gamma (r)= S^{-1}\frac{ dN(r)}{dr} = \frac{D}{4\pi } B r ^{-(3-D)}
\ee
where $\:S(r)$ is the area of  a spherical shell of radius $\:r$.
The {\it conditional average density},
as given by Eq.\ref{e217}, is well defined in terms of
its exponent, the fractal dimension.
Usually the exponent that defines the decay
 of the conditional density $\:(3-D)$ is called
the codimension and it corresponds to the
exponent $\:\gamma$ of the galaxy distribution.
In Fig.\ref{fig7}(b) we show a stochastic fractal
(generated by  the random-$\:\beta$-model algorithm
\cite{ben84}
in the  two dimensional Euclidean space - see Sec.\ref{angred})
constructed with a probabilistic algorithm
with a well defined fractal dimension $\:D = 1.2$.
\bef
%\vspace{}  
\epsfxsize 10cm
\centerline{\epsfbox{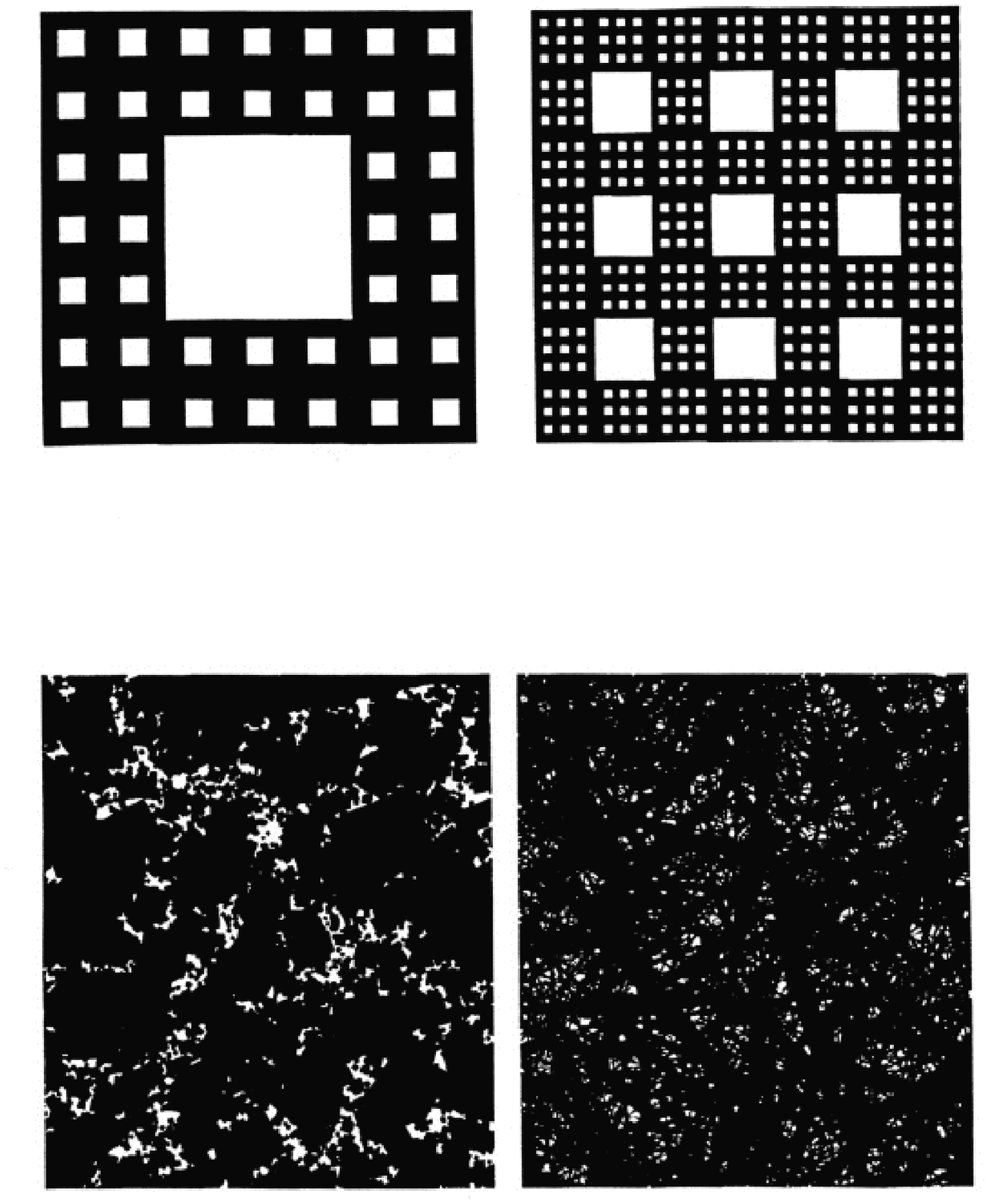}}
\caption{\label{fig7} {\it (a)}
 Two deterministic fractals of identical fractal
dimension and sharply distinct form. {\it (b)}
 Two random fractals of identical 
and different lacunarity. These structures are obtained by the random 
$\beta$ model algorithm, with different shaped voids (lacunarity) 
(Courtesy of B. Mandelbrot).}
\eef

\subsubsection{ Power laws, self-similarity  and non-analyticity }
\label{powerlaws}

From Fig.\ref{fig6} the geometrical self-similarity
is evident in the construction, while in Fig.\ref{fig7}, 
only a detailed analysis can show the self-similarity
of the structure.  From a
mathematical point of view self-similarity
implies that a rescaling of the length
by a factor $\:b$
\be
\label {e218}
r \rightarrow r ' = br
\ee
leaves the correlation function unchanged
apart from a rescaling that depends on
$\:b$ but not on the variable $\:r$.
 This leads to the functional relation
\be
\label {e219}
\Gamma (r') = \Gamma (b\cdot r) =
A(b)\cdot \Gamma (r)
\ee
which is clearly satisfied by a power law
with any exponent ($A(b)$ is a prefactor depending on $b$ only).
In fact, for
\be
\label {e220}
\Gamma (r) = \Gamma_{0} r ^{\alpha}
\ee
we have
\be
\label {e221}
\Gamma (r') = \Gamma_{0}(br)^{\alpha}
= (b)^{\alpha}\Gamma (r) \; .
\ee
Note that Eq.\ref{e219}  does not hold, for example,
for an exponential function
\be
\label {e222}
\Gamma ( r ) = \Gamma _{0} e ^{-r/r_{0}} \; .
\ee
This reflects the fact that power laws do
not possess a characteristic  length,
while for the exponential decay,
$\:r_{0}$ is a characteristic
length. Note that the characteristic
length has nothing to do with the
prefactor of the exponential
and it is not defined by the condition
$\:\Gamma ( r_{0} ) = 1$, but from the
intrinsic behavior of the function.
This brings us to a common
misconception that sometimes
occurs in the discussion of
galaxy correlations. Even for a
perfect power 
law as Eq.\ref{e220},
one might use the condition
$\:\Gamma(r_{0})=1$ to derive
a "characteristic length":
\be
\label {e223}
r_{0}=\Gamma_{0}^{-1/\alpha} \; .
\ee
This however is completely meaningless
because the power law refers to a fractal
structure constructed as self-similar
and, therefore, without a characteristic
length.
In Eq.\ref{e223}, the value of $\:r_{0}$
is just related to the power law amplitude  
 that, as we have already pointed out,
has no physical meaning.
The point is that the value
1 used in the 
relation $\:\Gamma(r_{0})=1$,
is not particular in any way, so one
may have used, as well, the
condition $\:\Gamma(r_{0})=10^{10}$
or $\:\Gamma(r_{0})=10^{-10}$
to obtain other lengths. This
is the subtle point of self-similarity;
{\it there is no reference value (like
the average density) with respect
to which one can define what is big or
small.}

We have discussed  how self-similarity
implies power laws: we consider now the inverse problem, namely, 
whether a power law implies self-similarity. This point
allows us to stress the non-analytic nature 
of fractal structures.
This question can be examined with a simple example.
We consider a density
 $\rho(r)$ 
that behaves, in three dimensions, as
\be
\label{e224}
\rho(r) = \frac{1}{r}.
\ee
One can argue that the mass-length relation  is given by
\be
\label{e225}
N(< r) \approx \int_0^R \rho(r) r^2 dr \approx R^2
\ee
and therefore interpret  the exponent $2$
 as the fractal dimension of this distribution.
This not correct, because Eq.\ref{e225} 
holds only for one specific origin ($r=0$),
while for a fractal it should hold for {\it any origin}.
For any other origin one obtains
\be
\label{e226}
N(< r)  \approx \int_0^R \rho(\vec{r}_0 + \vec{r}) d\vec{r}.
\ee
For  $|\vec{r}_0| >   R$ we can approximate the density as 
 (Fig.\ref{fig2} upper part)
\be
\label{e227}
\rho(\vec{r}_0 + \vec{r}) \approx \rho(\vec{r}_0) = \rho_0
\ee
and therefore 
\be
\label{e321}
N(< r) \approx \int_0^R \rho_0 r^2 dr \approx \rho_0 R^3.
\ee
This gives the standard dimension of the embedding space 
that shows we are dealing with a smooth (except for the 
point $\vec{r}=0$), nonfractal distribution. For homogeneous fractals,
one  should find the same nontrivial exponent
 $D < 3$, no matter which lower integration limit is considered.
The power law is non-analytic at the origin but this 
actually refers to each occupied point of the system. 
Thus the system is globally non-analytic because each point corresponds
to a singularity.

\subsubsection{Lacunarity and voids distribution}
\label{lacunarity}

So far we have quantified fractal structures by their dimension.
That this is not a sufficient characterization is illustrated in 
Fig.\ref{fig7}.
We illustrate the construction of two Cantor sets (one deterministic 
and one  stochastic)
with the same 
fractal dimension $D$ but with different {\it morphological properties}.
In order to distinguish such sets,  Mandelbrot \cite{man82} has 
introduced the concept of {\it lacunarity}  $F$ as 
\be
\label{e322}
Nr(\lambda >\Lambda) = F \Lambda^{-D}
\ee
where $ Nr(\lambda >    \Lambda)$ 
is the number of voids with a size 
 $ \lambda  > \Lambda$. 
The scaling behavior of  $ Nr(\lambda)$  is the same for both Cantor sets.
However the lacunarity $F$, i.e. the prefactor of the distribution, takes 
different values for the two Cantor sets.

In order to define lacunarity for random fractals 
we need a probabilistic form
of Eq.\ref{e322}. This can be done by introducing $P(\lambda)$,
which is the conditional probability that, given a box of size
 $\epsilon$
containing points of the set, 
this box is neighbored by a void of size $\lambda > \Lambda$.
Lacunarity is defined as the prefactor of the void distribution
\be
\label{e323}
P(\lambda >   \Lambda) = F \Lambda^{-D} \; .
\ee
It is easy to show \cite{sie89} that in the case 
of deterministic fractals this
definition gives the same value of the lacunarity defined in 
Eq.\ref{e322}. 
Lacunarity plays a very important role 
in the characterization of voids distribution in the available galaxy 
 catalogs \cite{man96} (Sec.\ref{radial} and Sec.\ref{counts}).

\subsubsection{Properties of orthogonal projection and  intersections}
\label{orthogonal}

We briefly present the properties of orthogonal projections and 
intersections of fractal structures. This discussion is 
useful in the
interpretation of angular and one dimensional (pencil beams)
catalogs (Sec.\ref{radial} and Sec.\ref{counts}).

Orthogonal
 projections preserves sizes of objects. If an object of fractal dimension
$D$, embedded in a  space of dimension $d=3$, is projected on a plane
(of dimension $d'=2$) it is possible to show that 
the projection has dimension $D'$ such that \cite{fal90,cp92}
\be
\label{epro1}
D'=D \; \; \mbox{if} \; \; D<d'=2 \; ; \; D'=d' \; \; \mbox{if} \; \; D>d'=2 \; .
\ee
This explains, for example, why clouds which have fractal dimension
$D \approx 2.5$, give rise to a compact shadow of dimension $D'=2$.
The angular projection represents a more complex problem 
due to the mix of very different 
length scales (Sec.\ref{counts} ). Nevertheless 
the theorem given by Eq.\ref{epro1} can be extended 
to the case of angular projections in the limit of small angles \cite{mon97b}. 

We discuss now a different but related problem: which are the properties
of the structure that comes out from the intersection of a fractal
with dimension $D$, embedded in the $d=3$ Euclidean space, with 
an object of dimension $D'$ ? The last can be for example a line
 ($D'=1$ - schematically a pencil beam survey), a plane ($D'=2$) or a
random distribution ($D'=3$). It is possible to show 
\cite{man82,cp92} 
that the law of codimension additivity gives for the dimension of the 
intersection
\be
\label{add1}
D_I=D+D'-d
\ee
If $D_{I} \le 0$, in 
the intersection it 
is not possible to recover any correlated signal \cite{cp92}.
Hence for example the intersection of a stochastic fractal with a
random distribution has the same dimension $D_I=D$ of the 
original structure. Such a property is useful in the discussion of 
surveys in which a random sampling has been applied 
(Sec.\ref{corran}).

\subsection{General properties of correlations}
\label{correlation}

In this section we discuss how to perform the correct correlation
analysis that can be applied to  
 an irregular distribution as well as to a regular one. 
We   start recalling  the 
concept of correlation. If the presence of an object at the point $r_1$ 
influences the probability of finding another object 
at $r_2$, 
these two points are correlated. Therefore there is a correlation
at  $r$ if, on average
\be
\label{e324}
G(r) = \langle n(0)n(r)\rangle   \ne \langle n\rangle  
\ee
where we average on all occupied points chosen as origin.
On the other hand, there is no correlation if
\be
\label{e325}
G(r) \approx \langle n\rangle  ^2.
\ee
The physically meaningful definition of  $\lambda_0$ 
is therefore the length scale which separates correlated regimes from
uncorrelated ones.

In practice, it  is useful 
to normalize the CF to the size  of the
sample   analyzed. Then we use, following \cite{cp92,gp84}
the average conditional density defined as
\be
\label{e326}
\Gamma(r) = \frac{<n(r)n(0)>}{<n>} = \frac{G(r)}{<n>}
\ee
where $\:<n>$ is the average density of the sample.  We stress
that this normalization does not introduce any bias even if the average
density is sample-depth dependent,
as in the case of fractal distributions,
because it represents
only an overall normalizing factor. 
In order to compare results from different catalogs
it is however more useful to use $\Gamma(r)$, in which
the size of a catalog only appears via the combination
$N^{-1}\sum_{i=1}^{N}$, so that a larger sample 
volume only enlarges the statistical sample over which averages are taken.
 $G(r)$ instead 
has an amplitude that is an explicit function of the sample's size
scale.
$\Gamma(r)$ (Eq.\ref{e326})
can be computed by the following expression
\be
\label{e327}
\Gamma(r) = \frac{1}{N} \sum_{i=1}^{N} \frac{1}{4 \pi r^2 \Delta r}
\int_{r}^{r+\Delta r} n(\vec{r}_i+\vec{r'})d\vec{r'} = 
\frac{BD}{4 \pi} r^{D-3}
\ee
where the last equality follows from Eq.\ref{e217}.
This function measures the average density at distance $\:\vec{r}$ from an
occupied point at $\vec{r_i}$
and it is called the {\it conditional density} \cite{cp92}.
If the distribution is fractal up to a certain distance $\lambda_0$,
and then it becomes homogeneous, we have 
that $\Gamma(r)$ has a power law decaying with
distance up to $\lambda_0$, and then it flattens 
towards a constant value.
Hence by studying the behavior of $\Gamma(r)$
it is possible to detect the eventual scale-invariant properties
of the sample. Instead the information given by the $\xi(r)$ is biased by the 
a priori (untested) assumption of homogeneity.

It is also very useful to use the {\it integrated conditional  density}
\be
\label{e328}
\Gamma^*(r) = \frac{3}{4 \pi r^3} \int_{0}^{r} 4 \pi r'^2 \Gamma(r') dr' =
\frac{3B}{4 \pi} r^{D-3} 
\ee
This function  produces an artificial smoothing of
rapidly varying fluctuations, but it correctly
reproduces global properties   \cite{cp92}.
                                        
For a fractal structure, $\Gamma(r)$ has a power law behavior
and the integrated conditional density is 
\be
\label{e329}
\Gamma^*(r)= \frac{3}{D} \Gamma(r).
\ee
For an homogeneous distribution ($D=3$) these two functions
are exactly the same and equal to the average density.

\subsubsection{ The $\xi(r)$ correlation function for a fractal }
\label{xitheo}

  Pietronero  and collaborators 
\cite{pie87,cps88,cp92} have clarified some crucial points of the
standard correlations analysis, and in particular they have discussed the 
physical meaning
of the so-called {\it "correlation length"}
  $\:r_{0}$ found with the standard
approach \cite{pee80,dp83} and defined by the relation:
\be
\label{e330}
\xi(r_{0})\equiv 1
\ee
where
\be
\label{e331}
\xi(r) = \frac{<n(\vec{r_{0}})n(\vec{r_{0}}+ \vec{r})>}{<n>^{2}}-1
\ee
is the two point correlation function used in the standard analysis.
The basic point in the present discussion,
is that the mean density, $<n>$,
used in the normalization of $\:\xi(r)$, is not a well defined quantity
in the case
of self-similar distribution and it is a direct function of the sample size.
Hence only in the case that 
homogeneity  has been reached well within the sample
limits the $\:\xi(r)$-analysis is meaningful, otherwise
the a priori assumption of homogeneity is incorrect and 
characteristic lengths, like $\:r_{0}$, became spurious.

For example from Eq.\ref{e216}  and Eq.\ref{e327}
the expression of the $\:\xi(r)$ in the case of
fractal distributions is \cite{cp92}:
\be
\label{e332}
\xi(r) = \frac{ 3-\gamma}{3} \left( \frac{r}{R_s} \right)^{-\gamma} -1
\ee
where $\:R_{s}$ is the depth of the spherical volume where one computes the
average density from Eq.\ref{e216}.
From Eq.\ref{e332} it follows that

i.) the so-called correlation
length $\:r_{0}$ (defined as $\:\xi(r_{0}) \equiv 1$)
is a linear function of the sample size $\:R_{s}$
\be
\label{e333}
r_{0} =\left(\frac{3-\gamma}{6}\right)^{\frac{1}{\gamma}}R_{s}
\ee
and hence it is a spurious quantity without  physical meaning but it is
simply related to the sample finite size (Fig.\ref{fig8}).
\bef 
%\vspace{}  
\epsfxsize 10cm
\centerline{\epsfbox{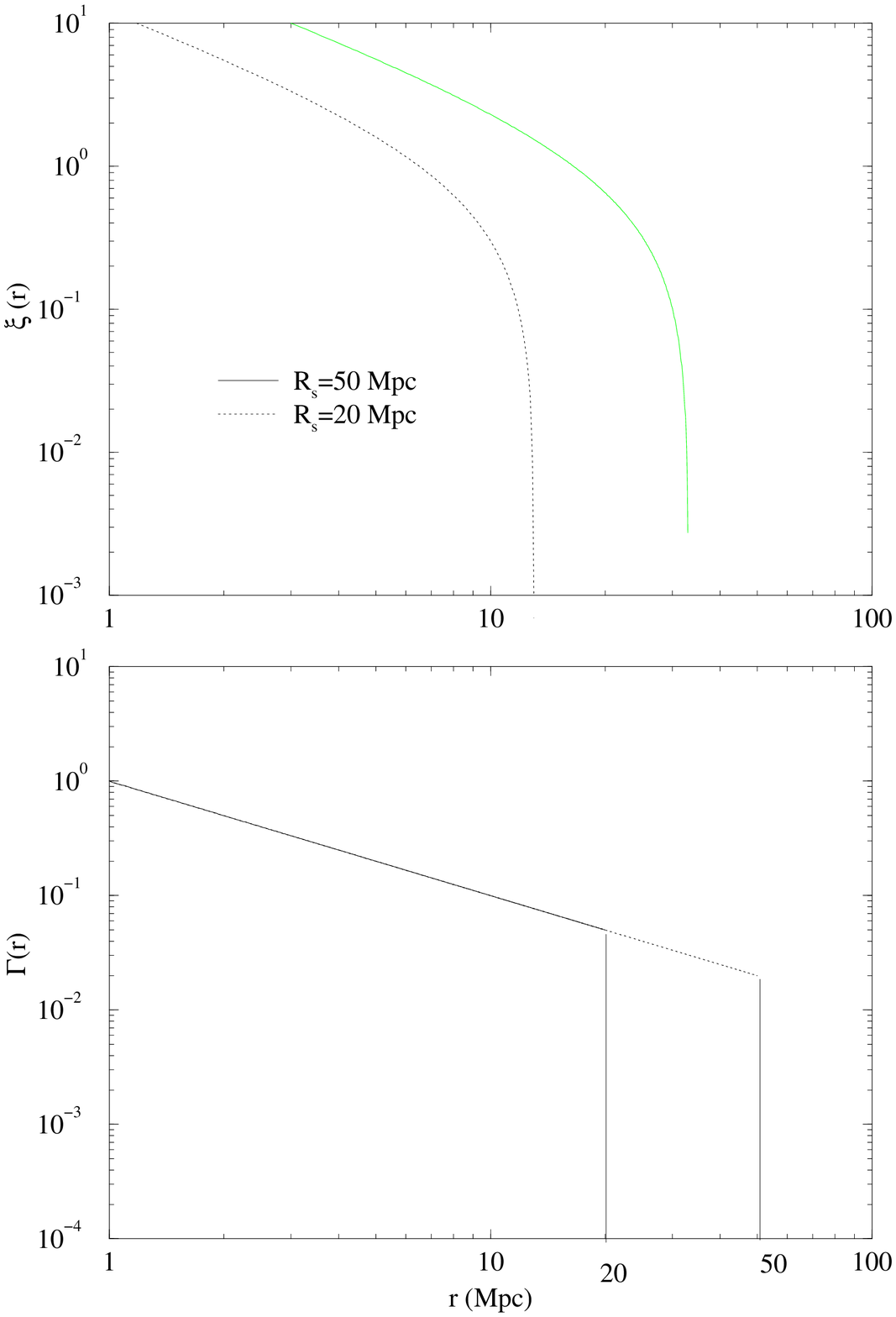}}
\caption{ \label{fig8} 
{\it Upper part:}  $\xi(r)$ computed for 
a self-similar structure contained in a sample of radius $R_s=20 Mpc$.
If one computes such a quantity in a sample of radius $R_s'=50 Mpc$, 
one can see
how inappropriate this function is
in view of its explicit dependence on the sample size.
{\it Bottom part:} It is shown the same exercise for $\Gamma(r)$.
The amplitude and the slope of the conditional density
perfectly matches the one obtained in the smaller sample.
}
\eef

ii.) $\:\xi(r)$ is power law only for
\be
\label{e334}
\left(\frac{3-\gamma}{3}\right) 
\left(\frac{r}{R_{s}}\right)^{-\gamma}  \gg 1
\ee
hence for $r \ll r_0$: for larger distances there is a clear deviation
from the power law behavior due to the definition of $\xi(r)$.
This deviation, however, is just due to the size of
 the observational sample and does not correspond to any real change
of the correlation properties. It is clear that if one 
estimates the  $\xi(r)$ 
exponent  at distances $r \ltapprox r_0$, one 
systematically obtains a higher value of the correlation exponent
 due to the break of $\xi(r)$ in the log-log plot. 
For example we can compute the $\xi(r)$ function for a fractal with dimension $D=2$
(i.e  $\gamma=1$) in two samples of different depths:
the first has $R_s=20 Mpc$ while the second has $R_s=100 Mpc$
(Fig.\ref{fig9}).
\bef %\vspace{} 
\epsfxsize 8cm
\centerline{\epsfbox{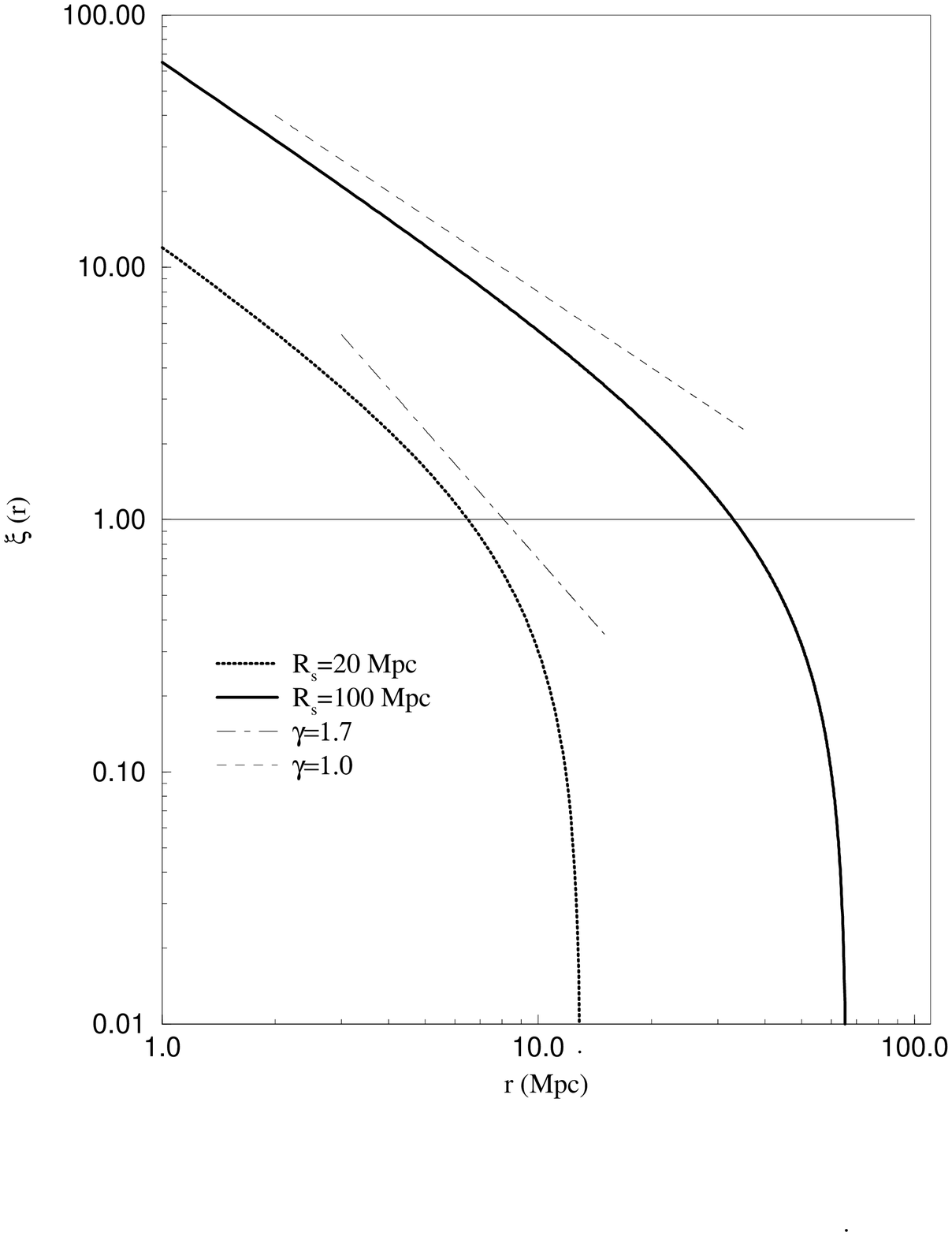}}
\caption{\label{fig9} The $\xi(r)$ function computed 
for a fractal with dimension $D=2$ ($\gamma=3-D=1$)
in two samples of different depth $R_s$. In the first case $R_s=20 Mpc$
one obtains an higher value for the correlation exponent $\gamma=1.7$
because one performs the fit in the region of length scales 
$r \sim r_0$. In the second case the power law behavior of $\xi(r)$
is more extended ($R_s=100 Mpc$)
and by evaluating the 
exponent in the region $r \ll r_0$ one obtains the correct value for
the correlation exponent $\gamma=1$.} 
\eef
In the first case one fits the 
$\xi(r)$ with a power law function 
in the region of length scales $r \sim r_0 $ and in such a way
one obtains a higher value for the correlation exponent, i.e.
$\gamma=1.7$. In the second case, as the power law behavior is 
more extended, one can measure the correlation exponent in
the range $r \ll  r_0$ and doing so, one obtains the 
correct value for $\gamma=1$. 
For the same reason one obtains 
a larger value of $\gamma$ by the angular correlation
$\omega(\theta)$ analysis. 

The analysis
 performed by $\xi(r)$ is therefore mathematically inconsistent, if
 a clear cut-off towards homogeneity has not been reached, because
 it gives an information that is not related to the real physical
 features of the distribution in the sample, but to the size of the
sample itself.

\subsection {Problems of treatment of 
boundary conditions and sample size effects}
\label{weight}

Given a certain spherical sample with solid angle $\Omega$ and depth $R_s$,
it is important to define which is 
 the maximum distance up to which it 
is possible to compute the correlation function ($\Gamma(r)$ or $\xi(r)$).
As discussed in \cite{cp92}, we   limit our analysis to an
effective  depth
$R_{eff}$ that is of the order of the radius of the maximum
sphere fully contained in the sample volume (Fig.\ref{fig10}).
\bef 
%\vspace{} 
\epsfxsize 8cm
\centerline{\epsfbox{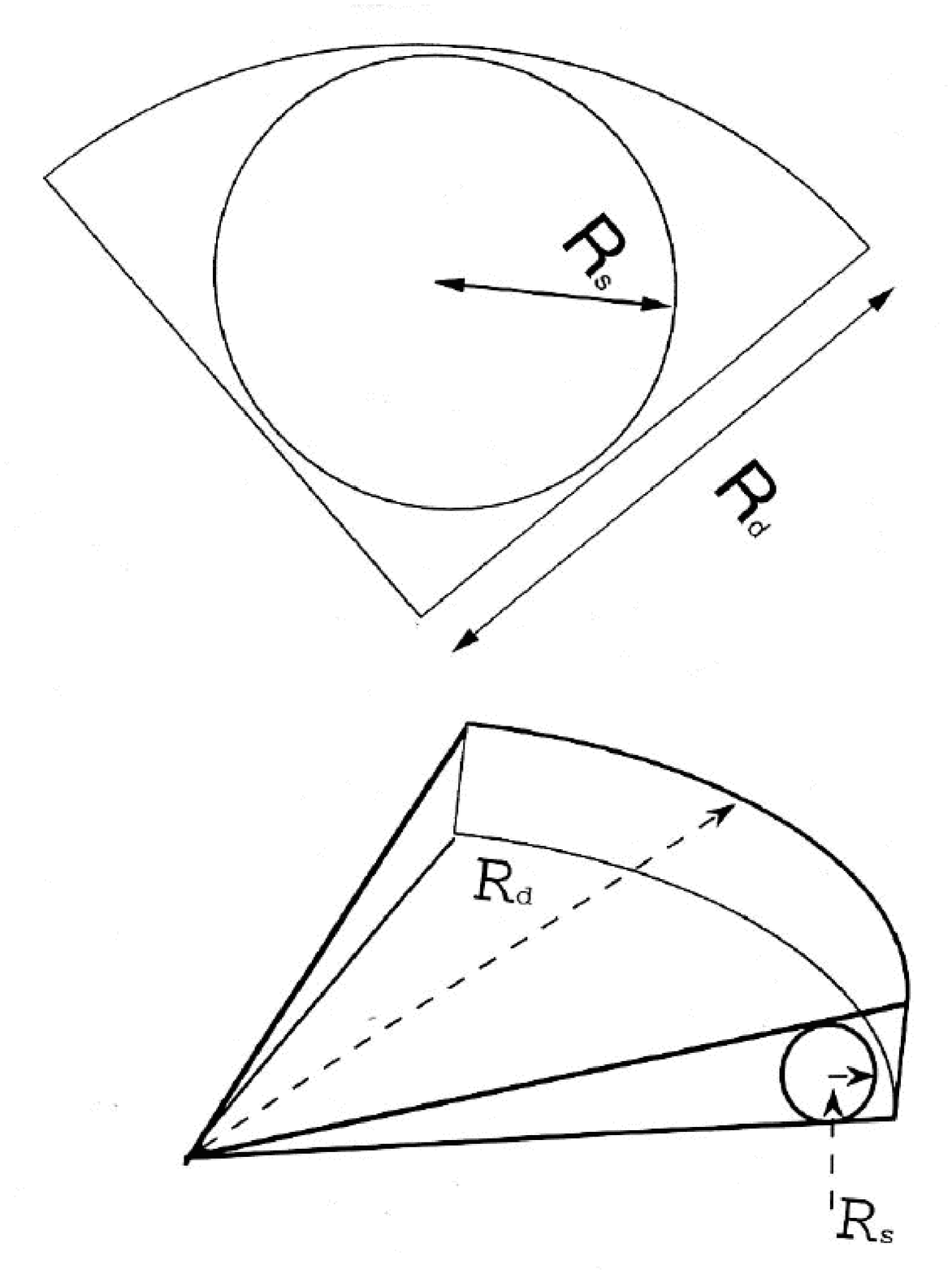}}
\caption{{\it Upper part:} A typical cone diagram for a wide angle
galaxy catalog 
(e.g. CfA, SSRS, Perseus-Pisces).
  The depth is $R_d$. The {\it effective depth}
 is $R_{eff}$ 
and it corresponds to the radius of the maximum
 sphere fully contained in the sample volume 
 ($R_{eff} \ltapprox R_s$).
{\it Bottom part:} A typical cone diagram for 
a narrow angle galaxy catalog (e.g. LCRS, ESP).   
In this case $R_{eff} << R_d$. 
\label{fig10}}
\eef
For example for a catalog with the limits in right ascension
($\alpha_1 \leq \alpha \leq \alpha_2$) and declination
($\delta_1 \leq \delta \leq \delta_2$) we have that 
\be
\label{ws1}
R_{eff} = \frac{R_s sin (\delta \theta /2)}{1+sin(\delta \theta/2)}
\ee
where $\delta \theta = min (\alpha_2 - \alpha_1, \delta_2-\delta_1)$.
In such a way  we do not consider in the statistics
the points for which a sphere of radius {\it r} is not
fully included within the sample boundaries.
Hence we do not make use of any weighting scheme
with the advantage of not making 
any assumption in the treatment of
the boundaries conditions.
For this reason we have a smaller number of points
and we stop our analysis  at a  smaller depth than that
of other authors.

 The reason why
$\Gamma(r)$ (or $\xi(r)$) cannot
be computed for $r > R_{eff}$
is essentially the following.
 When one evaluates the correlation
function
(or power spectrum) beyond $R_{eff}$,
then one  makes explicit assumptions on what
lies beyond the sample's boundary.  In fact, even in absence of
corrections for selection effects, one
is forced to consider incomplete shells
calculating $\Gamma(r)$ for $r>R_{eff}$,
thereby
implicitly assuming that what one  does not find  in the part of the
shell not included in the sample is equal to what is inside (or other
similar weighting schemes).
In other words, the standard calculation
introduces a spurious homogenization which we are trying to remove.

If one could reproduce via an analysis that uses weighting schemes, the 
correct properties of the distribution under analysis, it would be
not necessary to produce wide angle survey, and from a single pencil beam
deep survey it would be possible to study the entire matter
distribution up to very deep scales. It is evident that
this could not be the case.
By the way, we have done  a test on the homogenization effects
of weighting schemes on artificial distributions as well as on
real catalogs (Sec.\ref{testweight}), finding that the flattening of the 
conditional density
 is indeed introduced owing to  the weighting,
and does not correspond to any real feature in the galaxy distribution.

The conditional  density $\Gamma(r)$ (Eq.\ref{e327}) measures 
the density in a shell of thickness $\Delta r$ at distance $r$ from an 
occupied point, and then it is averaged over all the points of the sample. 
In practice, we have three possibilities
for $\Gamma(r)$: i) $D=3$: in this case this function is
simply a constant. ii) $0<D<3$ In this case the conditional density has a power law
decay  with exponent $-\gamma=D-3$. Finally iii) $D=0$: this is the limiting case 
in which there are no further points in the sample except the observer. In such 
a situation we have that $\Gamma(r)$ behaves as  $1/r^3$, i.e. as the 3-d volume.
Suppose now, for simplicity,  we have a spherical sample of volume $V$ in which there 
are $N$ points, and we want to measure the conditional density.
The {\it maximum depth}
is limited by the radius of the sample (as previously discussed), 
while the {\it minimum distance}
depends on the number of points contained in the volume. 
For a Poisson distribution the mean average distance between near neighbor 
is of the order $\ell \sim (V/N)^{\frac{1}{3}}$. Of course, such a 
relation does not express an useful quantity 
in the case of a fractal distribution, as well 
as the average 
density, while the meaningful 
measure is the {\it average minimum distance 
between neighbor galaxies} $\ell_{min}$, that is related 
to the lower cut-off of the distribution. 
If we measure the conditional density at distances $ r \ll \ell_{min}$, 
we are affected 
by a {\it finite size effect}. In fact, due the depletion of points at these distances 
we underestimate the real conditional  density finding an higher value 
for the correlation exponent (and hence a lower value for the fractal dimension). 
In the limiting case at  distances $ r 
<< \ell_{min}$, we can find almost no points and 
the slope is
$\gamma=-3$ ($D=0$). 
In general, when one  measures $\Gamma(r)$ at distances that correspond to 
a fraction of $\ell_{min}$, one finds systematically an higher value of the 
conditional density exponent. 
Such a trend  is completely spurious and due to the depletion of
points at such distances. 

For example in a real survey, in order to check this effect, 
one should measure 
$\Gamma(r)$ in samples with different values of 
$\ell_{min}$ (Sec.\ref{corran}). In general we find
that the sparser samples exhibit a change of slope, 
towards an higher value of  
the correlation exponent, 
 at small distances. For the samples 
 for which $\ell_{min}$ 
 is quite small, the change of slope at small distances 
is not found. In general for 
a typical sample of galaxies (Sec.\ref{corran} and the Appendix) 
$\ell_{min} \sim 1 \div 6 \hmp$, so that 
the behavior of $\Gamma(r)$ at distances of 
some Megaparsec is generally affected by this 
finite size effect. A way to reduce this 
effect is to chooses properly the thickness 
$\Delta r$ of the shell in which the conditional 
density is computed: this means that 
at small distances $\Delta r$ must be of the 
order of $\ell_{min}$
 and not 
smaller than this value. In general we have 
found that best way to optimize this 
estimate is to choose logarithm interval for 
$\Delta r$, as a function of the scale in which the 
conditional density is computed.

\subsection{Amplitude of fluctuations: linear and non-linear dynamics}
\label{amplitude}

Another argument often mentioned
in the discussion of large-scale
structures, is that it is true that
larger samples show larger structures
but their amplitudes are smaller
and the value of $\:\delta N / N$
tends to zero at the limits of the
sample \cite{pee93}; therefore one expects
that just going a bit further,
homogeneity may finally be
observed. Apart from the fact
that this expectation has been
systematically disproved,
the argument is conceptually
wrong for the same reasons
of the previous discussion.
In fact, we can consider a
portion of a fractal structure
of size $\:R_{s}$ and study
the behavior of  $\:\delta N / N$.
The average density $\:N$ is
just given by Eq.\ref{e216} while
the overdensity
$\:\delta N$, as a function
of the size $\:r$  ($\:r\leq R_{s}$) of a given
in structure is:
\be
\label {e3n1}
\delta N = \frac{N(r)}{V(r)} - <n>
= \frac {3}{4\pi} B (r^{-(3-D)}-R_{s}^{-(3 - D)}) \; .
\ee
We have therefore
\be
\label {e3n2}
\frac {\delta N}{N} =
\left(\frac{r}{R_{s}}\right)
^{-(3 - D)} - 1 \; .
\ee
Clearly for structures that
approach the size of the 
sample, the value of  $\:\delta N / N$
becomes very small and eventually
becomes zero at $\:r = R_{s}$
as shown in Fig.\ref {fig11}.
\bef 
%\vspace {8 cm}
\epsfxsize 8cm
\centerline{\epsfbox{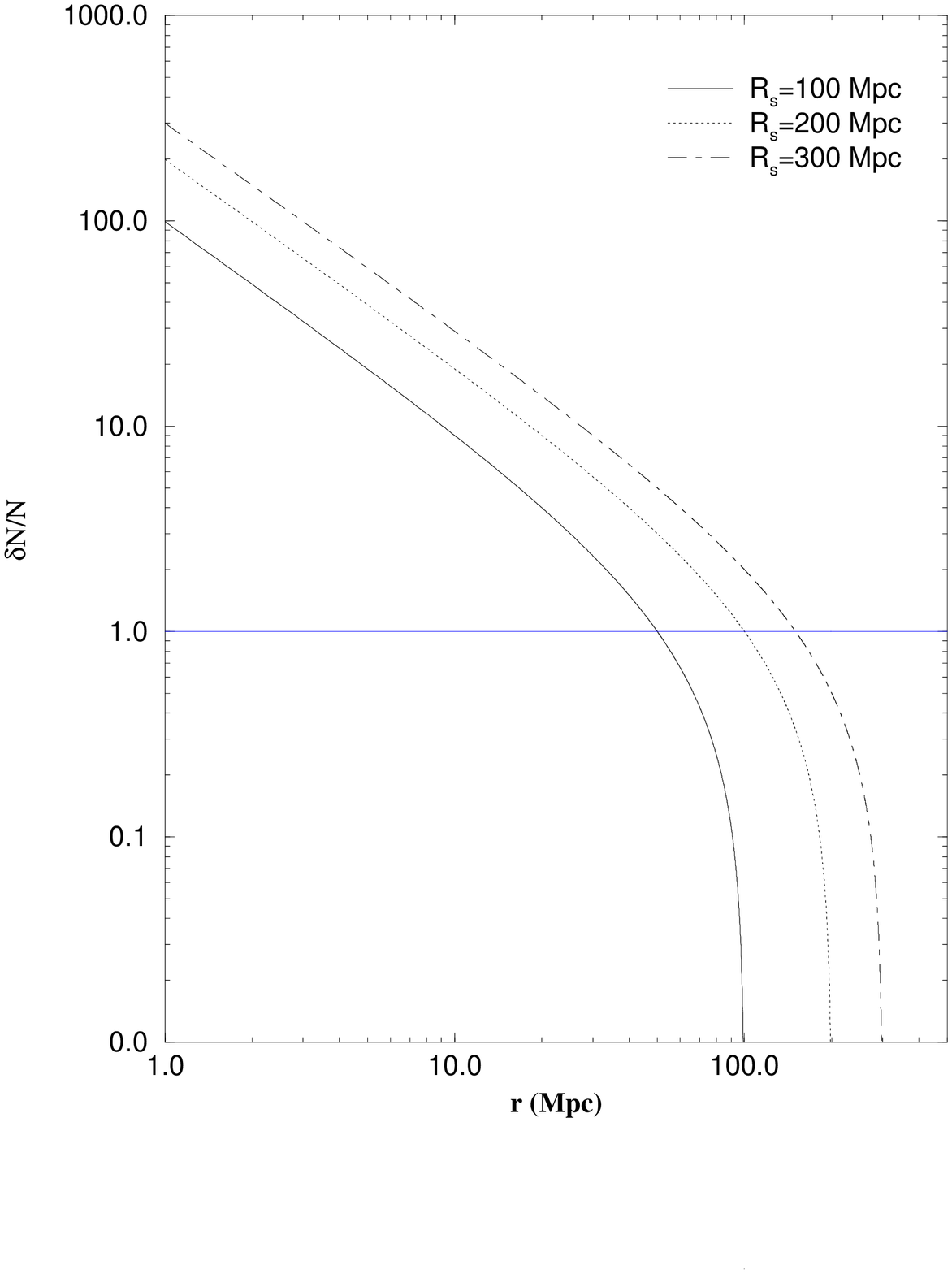}}
\caption{ Behavior of
$\:\delta N / N$ as a function of the size  $r$
in a portion of a fractal structure
for various depths of the sample: $\:R_{s}=100,200,300 Mpc$.
The average density is computed over the whole sample
of radius $\:R_{s}$.
The fact that $\:\delta N/N$ approaches
to zero does not mean that the
fluctuations are small and a homogeneous distribution
has been reached. The distance at which $\:\delta N/N =1$ scales
with sample depth and it has no physical meaning.
In the case of a fractal distribution the normalization
factor, i.e. the average density, is not an intrinsic quantity.
\label{fig11}}
\eef
This behavior, however, cannot be
interpreted as a tendency towards
homogeneity because again
the exercise refers to a self-similar
fractal by construction. Also
in this case, the problems
come from the fact that
one defines an "amplitude"
arbitrary, by normalizing
with the average density
that is not an intrinsic
quantity.
A clarification of this
point is very important
because the argument that
since  $\:\delta N / N$ becomes
smaller at large scale, there
is a clear evidence of
homogenization is still
quite popular \cite{pee93}
and it adds confusion to the
discussion.

The correct interpretation of
 $\:\delta N / N$ is also
fundamental for the development
of the appropriate theoretical
concepts. For example, a popular
point of view is to say that
 $\:\delta N / N$ is large $\:( \gg 1)$
for small structure and this
requires a  non linear
theory for the dynamics. On the other hand
 $\:\delta N / N$ becomes small
$\:(<1)$ for large structures,
which requires therefore a linear
theory.
The value of  $\:\delta N / N$
has therefore generated a conceptual
distinction between small structures
which   entails non linear dynamics
and large structures with small
amplitudes that corresponds instead
to a linear dynamics.
If one   applies the same
reasoning to a 
fractal structure we may conclude
that for a structure up to 
(from Eq.\ref {e3n2}):
\be
\label {e3n3}
r^{*} = 2^{-\left( \frac{1}{3-D} \right)} R_{s}
\ee
we have  $\:\delta N / N >1$ so that a non linear
theory is needed. On the other hand, for
large structures $\:(r>r^{*})$ we have
 $\:\delta N / N < 1$ which would correspond to a
linear dynamics. Since the fractal
structure, that we have used
to make this conceptual exercise,
has scale invariant properties
by construction, it follows that 
the distinction between
linear and non linear dynamics
is completely artificial
and wrong.
The point is again that the value
of $\:N$, we use to normalize
 fluctuations, is not
intrinsic, but it just
reflects the sample size 
that we consider ($R_{s}$).

If we have a sample with  depth $\:\tilde{R_{s}}$
larger  than the eventual scale of homogeneity $\:\lambda_{0}$,
then the average density is constant in the range
$\:\lambda_{0} < r < \tilde{R_{s}}$, apart from small amplitude
fluctuations. The distance at which $\:\delta N/N =1$ is
given by:
\be
\label {e3n4}
r^* = 2^{-\left( \frac{1}{3-D} \right)} \lambda_{0}
\ee
If, for example, $\:D=2$ and $\:\lambda_{0}=200 \hmp$ then
$\:r^{*} = 100 \hmp$.
Therefore a homogeneity scale of this order of magnitude is
incompatible with the standard normalization of
$\:\delta N/N =1$ at $\:8 h^{-1} Mpc$ (the same argument can be 
applied to statistical quantities like $\sigma^2$ or counts in cells,
which are usually used 
in the characterization of gaussian processes). 
The whole
discussion about large and small
amplitudes and the corresponding
non linear and linear dynamics,
has no meaning until an unambiguous
value of the average density has been defined. 
Only in this case  the concepts like large and small amplitudes
can take a physical meaning and be
independent on the size of the
 catalog.

The basic point of all this
discussion is that in a
self-similar structure one cannot
say that correlations are
"large" or "small", because
these words have no physical meaning due
to the lack of a characteristic
quantity with respect to which
one can normalize these properties.
The deep implication of this
fact is that one cannot
discuss a self-similar structure in terms
of amplitudes of correlation. The only
meaningful physical quantity is the
exponent that characterized the
power law behavior. Note that the "amplitude"
problem   is not
only present in the data analysis but
also in the theoretical models.
Meaningful amplitudes can only be
defined once one has unambiguous
evidence for homogeneity but this
is clearly not 
the case for galaxy and cluster distributions.

\subsubsection{Why a small 
correlation length {\it is not} compatible with large scale 
structures}
\label{smallcorr}

The distribution of galaxies
in space have been investigated very intensively
in the last years.
Several recent galaxy redshift
surveys such as CfA1 \cite{huc83}, 
CfA2 \cite{del88,dac94,par94}, 
SSRS1 \cite{dac88}, 
SSRS2 \cite{dac94},
Perseus Pisces \cite{hg88},
LCRS \cite{sch96}, IRAS \cite{str96}, 
pencil beams surveys \cite{bro90}
 and ESP \cite{vet94,zuc97}, 
have uncovered remarkable structures such as
filaments, sheets, superclusters
and voids (Fig.\ref{fig1} and Fig.\ref{fig3}). These galaxy  catalogs
probe scales from $\: \sim 100 \div 200 h^{-1} Mpc$
for the wide angle surveys, up to $\:\sim  1000 h^{-1} Mpc$
for the deeper pencil  beam surveys,
and show that
the Large-Scale Structures (LSS)
are the characteristic features of the visible matter distribution.
One of the most important issues
raised by these  catalogs is that the scale of {\em the
largest inhomogeneities}
are limited only by {\em the boundaries of
 the surveys} in which they are detected.
A new picture  emerges from these observations, 
 in which the
scale of homogeneity seems to shift to a very large value, not
still identified.

The usual correlation function analysis
performed by the $\xi(r)$ function, leads to the
identification of the "correlation
length" $r_0 \approx 5 h^{-1}Mpc$ \cite{dp83}.
This result appears
incompatible with the existence of LSS
of order of $50 \div 200 h^{-1}Mpc$.
In fact, according to this result, galaxy
distribution should become
smooth and regular at distances somewhat larger
 than $r_0$ without large fluctuations on larger scales.
The main problem of
the $\xi(r)$-analysis
 is 
the underlying {\it assumption}
of homogeneity.
The basic idea we address here, is to perform a
correlation analysis that does not require any a 
priori assumption
\cite{pie87,cps88}. This new correlation analysis 
reconciles the statistical studies
with the observed LSS.

In order to make  this discussion more quantitative  
we can consider a distribution that is
 fractal distribution up to a certain scale $\lambda_0$,
and beyond this length it becomes homogeneous.
This implies that if we locate a sphere 
of radius equal or larger than $\lambda_0$ randomly
in a three dimensional catalog 
(that for simplicity we suppose 
spherical  with
radius $R_{eff} \gg \lambda_0$), 
we should find that the number of points
inside this sphere is $N \pm \sqrt{N}$ everywhere in the catalog,
i.e. this number is constant apart the Poissonian fluctuations.
The average conditional density  becomes
\be
\label{3lss1}
\Gamma(r) = \frac{BD}{ 4 \pi} r^{D-3} \; , \; r < \lambda_0
\ee
and
\be
\label{3lss2}
\Gamma(r) = n_0 \; , \; r \geq \lambda_0\; .
\ee
The matching condition at $\lambda_0$ gives
\be
\label{3lss3}
n_0= \frac{B D}{4 \pi} \lambda_0^{D-3} \; . 
\ee
Therefore $\Gamma(r)$ has a power law decaying up to $\lambda_0$, 
followed by a constant behavior thereafter.
In such a case, and in the limit $\lambda_0 \ll R_{eff}$, it is easy 
to show \cite{cp92} that the usual correlation 
analysis performed by the $\xi(r)$ leads to the identification 
of a correlation length that is
\be
\label{3lss4}
r_0 = \lambda_0 \cdot 2^{1/(D-3)} \; .
\ee
In the case $D=2$ we simply obtain $\lambda_0 = 2 r_0$. This 
means that the value of $r_0$ {\it cannot} be much smaller (at most a factor 2)
than the largest structures observed in the sample, which are 
in this case of the order of $\lambda_0$. From this simple argument it follows 
that if we observe by eye structures and voids of order $ > 100 \hmp$
this cannot be compatible with a value of $r_0 \sim 5 \hmp$. 
However, we stress again that
the dimension of the largest structures is only limited by 
the boundaries of the surveys in which they are detected.

Peebles \cite{pee89} introduced the so-called {\it "egg-crate" model universe},
according to which galaxies are uniformly distributed on flat sheets, 
with the sheets placed at separation $L$ to
form a cubic lattice. He derived that in this case the correlation length
($\xi(r_0)\equiv 1$) is
\be
\label{ec1}
r_0 = \frac{L}{8} \; .
\ee
This example, tuned to show the maximum difference between $L$ and $r_0$,
is used to argue that 
 the existence of galaxy structures of an order of 
magnitude larger than the clustering length may not be contradictory.
However we  point out that:
i) the conditional density should be flat beyond the scale $L$,
and it should show 
a power law behavior at smaller distances (Fig.\ref{fig12}).
\bef %\vspace{} %\vspace{}
\epsfxsize 8cm
\centerline{\epsfbox{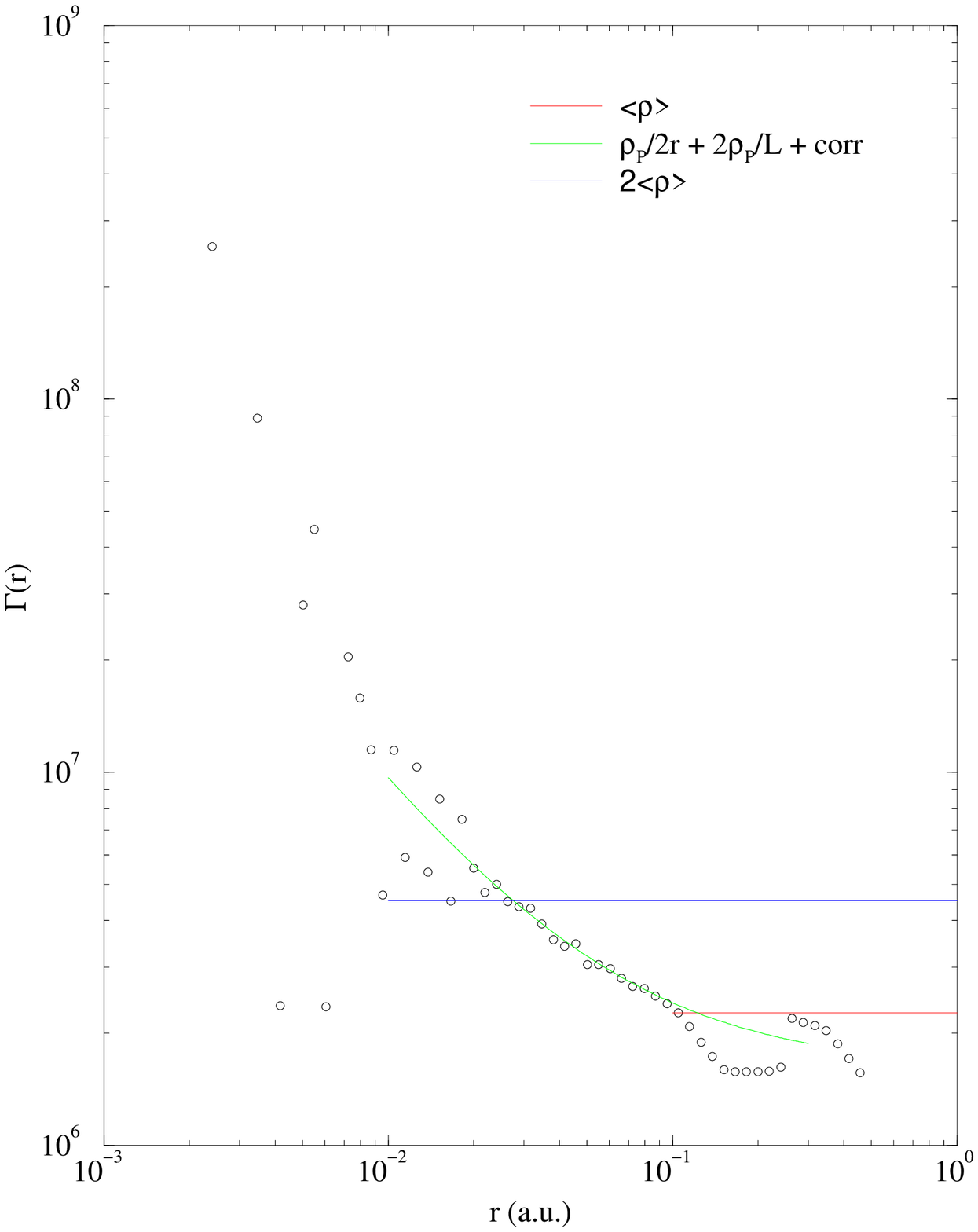}}
\caption{ \label{fig12}
The conditional density for the {\it egg-crate} model universe 
(in the horizontal axis distances are in arbitrary units (a.u.))
$\langle \rho \rangle$ is the average density.
This function shows a power law behavior up 
to $L=0.1$ and then it flattens towards a constant value. 
The dotted line shows the analytic prediction for such a model.
}
\eef
ii) The quantity $r_0$ in Eq.\ref{ec1} is a physically  real characteristic
length for the distribution, and it should be independent on the sample 
size.
iii) According to Eq.\ref{ec1} one should find structures and voids
of one order of magnitude larger than the correlation length
$r_0$, i.e. up to $L \sim 50 \hmp$, while this is not the case for
real galaxy structures. Of course, 
the difference between $r_0$ and $L$ is 
such a system is a consequence of a very particular distribution
(cubic lattice of uniform planes) that is different from the observed one.

In  real galaxy and cluster catalogs
we find that the conditional density does not show any tendency
towards homogenization in any of the available samples, and that $r_0$ is
indeed a linear function of
 the sample size.

%%%%%%%%%%%%%%%%%%%%%%%%%%%%%%%%%%%%%%%%%%%%%%%%%%%%%%%%%%%%%%%%%%%%%%%%%%%%

\section{Correlation analysis  for galaxy distributions}
\label{corran}

In this section we discuss the correlation properties of the 
galaxy distributions in terms of volume limited  catalogs \cite{dp83} 
arising from about 
100.000 redshift measurements that have been made to date.
To this end, we study the behavior of the {\it 
conditional (average) density}, that is the main 
statistical tool useful to characterize the properties 
of highly inhomogeneous systems as well as regular ones. In particular, 
this statistical quantity is the appropriate one to identify 
self-similar properties, if they are present in the 
distribution. 

A first important result is that the samples are 
{\it statistically rather good} 
and their properties are in agreement with each other.
 This gives a new perspective because, using the standard methods
of analysis, the properties of different samples appear contradictory with 
each other and, often, this is considered to be a problem of the data
(unfair samples) while, we show that this is due to the inappropriate 
methods of analysis. In addition, all the galaxy and cluster catalogs 
show well defined scale-invariant 
correlations up to their limits ($\sim 150 \hmp$)
and the 
fractal dimension is $D \approx 2$. We refer to Sec.\ref{radial} 
for a complete summary of all the available redshift samples.

\subsection{Detailed analysis of the available  catalogs}
\label{basicprop}

A three dimensional catalog contains for each galaxy the 
two angular coordinates $\alpha,\delta$, the redshift  $z$
and the apparent magnitude $m$. 
The analyses presented in this review 
are, in general, 
performed in galactic coordinates $(b,l)$, 
using, if needed, the limitation $|b| > 10^{\circ}$. 
In such a way we can exclude the 
region corresponding to the galactic  plane, 
where, because of dust absorption, it is problematic to 
measure redshifts.

We  introduce some basic definitions.
If $L$ is the absolute or intrinsic
luminosity of a galaxy at distance $r$,
this   appears with an apparent flux
\be
\label{e41}
f = \frac{L}{4 \pi r^2}  \; .
\ee
For historical reasons the apparent magnitude $m$
of an object with incoming flux $f$
is 
\be
\label{e42}
m = -2.5 \log_{10}f + constant,
\ee
while the absolute magnitude $M$
 is instead related to its intrinsic luminosity $L$ by
\be
\label{e43}
M=-2.5 \log_{10}L + constant'.
\ee
From Eq.\ref{e41} it follows that
the difference between the apparent and the absolute magnitudes
 of an
object at distance $r$ is (at relatively small distances,
neglecting relativistic effects)
\be
\label{e44}
m-M = 5 \log_{10}r + 25
\ee
where $r$ is expressed in Megaparsec ($1 Mpc= 3 \cdot 10^{24}
 \; cm$).

A catalog is usually obtained by measuring the redshifts 
of the all galaxies with apparent magnitude brighter
than a certain apparent magnitude limit $m_{lim}$,
 in a certain region of the sky defined by a solid angle $\Omega$.
An important selection effect exists, in that at every distance in the 
apparent magnitude limited survey, there is a definite limit in intrinsic 
luminosity
which is the absolute magnitude of the fainter galaxy
which can be seen at that distance. Hence at large distances, intrinsically
faint objects are not observed whereas at smaller distances they are observed.
In order to analyze the statistical properties of galaxy distribution,
a catalog which does not suffer for 
this selection effect
must be used. In general, it exists a very well known procedure to
obtain a sample that is not biased by this luminosity selection effect:
this is the so-called {\it  "volume limited"} (VL) sample.
A  VL  sample contains every galaxy in the volume
which is more luminous than a certain limit, so that in such a
 sample
there is no incompleteness for an observational
luminosity selection effect \cite{dp83,cp92}.
Such a sample is defined by a certain maximum distance $R_{VL}$
and the absolute magnitude limit $M_{VL}$ 
given by
\be
\label{e45}
M_{VL}=m_{lim}-5\log_{10}R_{VL} -25 -A(z)
\ee
where $A(z)$ takes into account various corrections (K-corrections, 
absorption, relativistic effects, etc.), and 
$m_{lim}$ is the survey apparent magnitude limit.
(In Sec.\ref{weight}, we have defined the effective depth $R_{eff}$ for 
the analysis of a sample.) Different VL samples, 
extracted from one catalog, have different $R_{eff}$, and 
deeper is a VL sample, larger its $R_{eff}$.
The (geometrical) limit of 
validity of the whole catalog is then 
the largest $R_{eff}$ 
that corresponds to the deepest VL sample 
with a robust statistics
(we refer to Sec.\ref{validity} 
for a discussion of the statistical fairness of samples).

The measured velocities of the galaxies have been
expressed in
the preferred frame of the Cosmic Microwave Background Radiation
(CMBR), i.e. the heliocentric velocities of
   the galaxies have been corrected
for the solar motion with respect to the CMBR, according
 with the formula
\be
\label{e46}
\vec{v} =\vec{v}_{m}+316 cos \theta \; \; km s^{-1}
\ee
where $\vec{v}$ is the corrected velocity,
$\vec{v}_{m}$ is the observed velocity and $\theta$
is the angle between the observed
velocity and the direction of the CMBR dipole anisotropy
($\alpha=169.5^{\circ}$ and $\delta=-7.5^{\circ}$).
From these corrected velocities, we have calculated
 the comoving distances
$r(z)$, with for example $q_0=0.5$, by using the Mattig's relation
\cite{bslmp94}
\be
\label{e47}
r(z)=6000\left(1-\frac{1}{\sqrt{(1+z)}}\right) \hmp \; .
\ee
In general, we have checked that the results of our analysis
depend very weakly on the particular value of $q_0$ adopted, 
except very deep surveys, and
we have also used the simple linear relation
\be
\label{e48}
r = cz/H_{0} \; .
\ee
In the nearby catalogs 
there is no any sensible change by using Eq.\ref{e48} 
instead of Eq.\ref{e47}.
Hereafter for the Hubble constant 
(unless it is not explicitly stressed), we use the value 
$H_{0}=100\cdot  h \cdot km sec^{-1} Mpc^{-1}$. 

All the analyses presented here have been 
performed in redshift space and we have not
adopted {\it any correction} to take into account the 
eventual effect of peculiar velocities (local distortion to the Hubble flow).
However we point out that peculiar velocities have an 
amplitude up to $\sim 500 \div 1000 km sec^{-1}$ 
and then their effect can be important only up
 to $ 5 \div 10 \hmp$, and not more. 

We briefly mention the characteristics 
of the {\it galaxy luminosity distribution}
that is useful in what follows 
analyses. The basic assumption 
we use to  compute all the following
 relations
is that:
\be
\label{e49}
\nu(L,\vec{r}) = \phi(L) \rho(\vec{r}) \; , 
\ee
i.e. that the number of galaxies for unit luminosity
and volume $\nu(L,\vec{r})$ can be expressed as
the product of the space density
$ \rho(\vec{r})$  and  the luminosity function $ \phi(L)$
($L$ is the intrinsic luminosity).
This is a crude approximation
in view of the multifractal properties of the distribution
(correlation between position and luminosity), and a detailed
discussion can be  found in  Sec.\ref{lumspace}.
However, for the purpose of the present discussion,
the approximation of Eq.\ref{e49} is rather good and the
explicit consideration of the multifractal properties
 have a minor effect on the properties we  
discuss \cite{slp96}.
                    
To each VL sample (limited by the 
absolute magnitude $M_{VL}$)
we can associate the luminosity factor
\be
\label{e410}
\Phi(M_{VL}) = \int_{-\infty}^{M_{VL}} \phi(M) dM  
\ee
that gives the fraction of galaxies for unit volume, 
present in the sample. Hereafter we adopt the following normalization
for the luminosity function
\be
\label{e411}
\Phi(\infty) = \int_{-\infty}^{M_{min}} \phi(M) dM  = 1 
\ee
where $M_{min} \approx -10 \div -12$ is the fainter galaxy present in 
the available samples. The luminosity factor of Eq.\ref{e410} is useful to
normalize the space density in different VL samples which 
have different $M_{VL}$ (Eq.\ref{e45}). The 
luminosity function measured in real catalogs has 
the so-called Schecther like shape (sec.\ref{lumspace})
\cite{sch76}
\be
\label{e412}
\phi(M)dM = A \cdot 10^{-0.4(\delta+1)M} e^{-10^{0.4(M^*-M)}}dM
\ee
where  $\delta \approx -1.1$ and $M^* \approx -19.5$ \cite{dac94}
\cite{vet94}, and the constant $A$ is given by the normalization
 condition of Eq.\ref{e411} (we refer to 
 Sec.\ref{lumspace} for a more detailed discussion
of this subject).

The scheme of the analyses presented in the next section
is the following.
We extract from the catalog several VL samples and we study: 
\begin{itemize}

\item The  average conditional density 
$\Gamma(r)$ and the integrated conditional  density $\Gamma^*(r)$. 
In such a way we  determine the fractal dimension and
detect, if present, the eventual crossover towards homogenization.
Moreover, we  present some tests to
check the statistical stability of the results versus
the possible incompleteness and systematic errors 
that could be present in the data.

\item Determination $\xi(r)$; in 
this way we can definitely establish
which kind of statistical information  such a function 
gives.

\item Determination of the dependence of 
the so-called "correlation length" $r_0$, defined as
 $\xi(r_0) \equiv 1$, on the sample depth $R_{eff}$.
We study also its 
eventual dependence on 
luminosity (i.e the so-called 
luminosity segregation phenomenon).

\item We refer to the Appendix for a detailed description of the VL 
samples of the various catalogs.
\end{itemize}

\subsubsection{CfA1}
\label{gammacfa1}

The CfA1 catalog  has been the first wide angle
redshift survey ($\Omega =1.83 \; sr$) available in the literature
\cite{huc83}.
Coleman, Pietronero 
and Sanders in 1988 \cite{cps88}
have analyzed 
 this catalog 
with the methods of modern statistical mechanics. 
They found in this sample galaxy
distribution 
shows power law (fractal)
correlations up to the sample limit of $20 h^{-1}Mpc$
without any tendency towards
homogenization.
In particular the main results are:

i) The CfA1  catalog is statistically a {\em fair sample}
up to $\: \sim 20 h^{-1}Mpc$:
a sample is statistically fair if it is possible
to extract from it an information that is statistically meaningful.
Whether it is homogeneous or not, it is a property that can
be tested and it is independent on sample  statistical fairness 
 (Sec.\ref{validity}).

ii) $\Gamma(r)$  shows a well defined power law
(fractal)
behavior up to the sample limit, $R_{eff} \sim 20 \hmp$, 
 (Fig.\ref{fig13})
without any tendency towards homogenization.
For CfA1 $\:\gamma = 3 -D \sim 1.1 \pm 0.2$ and 
the fractal dimension is $\:D =1.9 \pm 0.2$ 
(when measured by the $\Gamma^*(r)$).
The small discrepancy with the value reported by \cite{cp92}
(i.e. $D \sim 1.4 \pm 0.1$) is due basically to the fact that we have used 
$\Gamma^*(r)$ rather than $\Gamma(r)$ to estimate the 
correlation exponent. Moreover, we have used a logarithm 
variable value of
$\Delta r$ (the thickness of the shell in which the conditional density is computed):
in more dilute samples, like those of CfA1, 
such a procedure gives a more stable result for 
the correlation exponent 
(Sec.\ref{weight}).
\bef 
%\vspace{}  
\epsfxsize 8cm
\centerline{\epsfbox{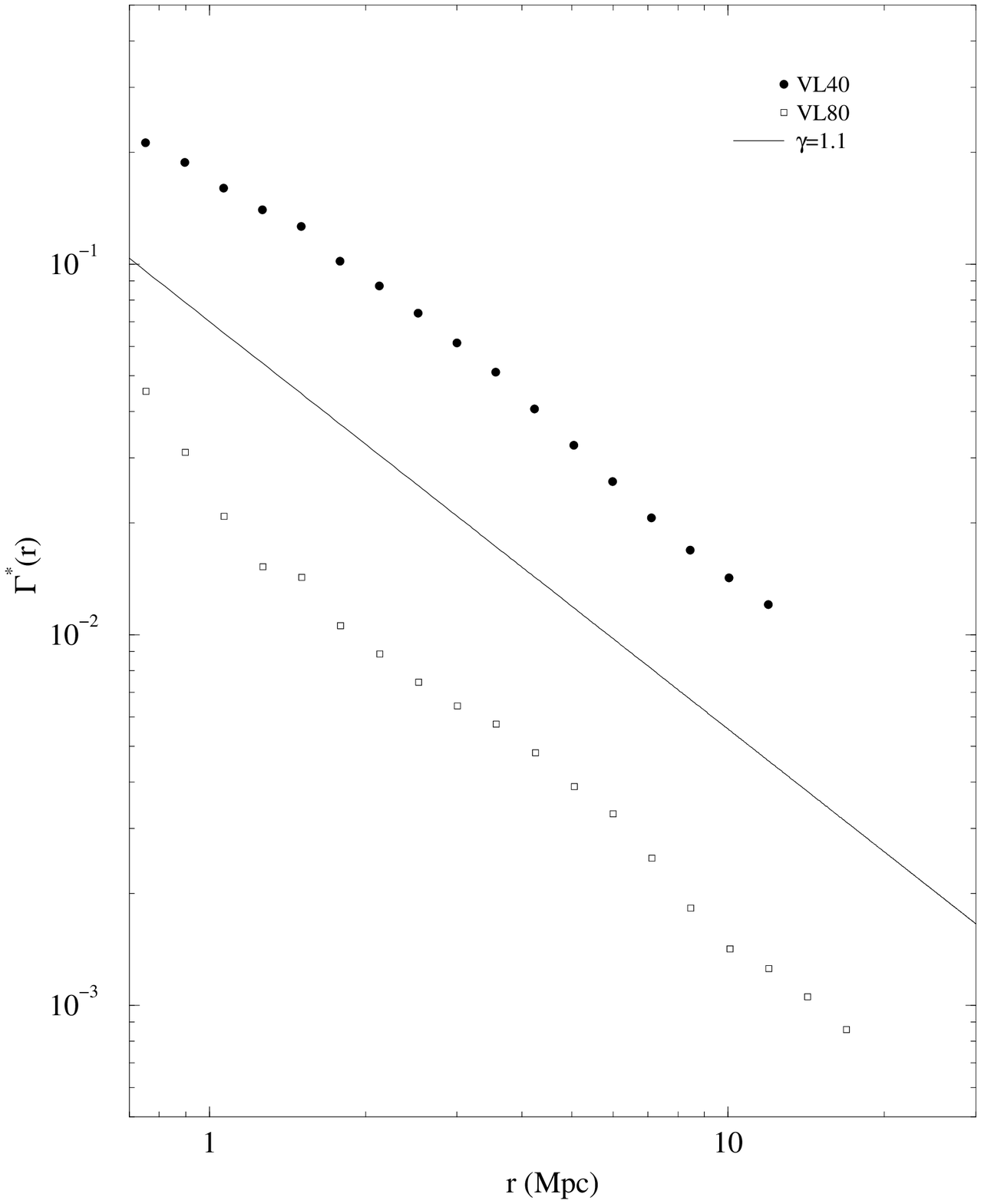}}
\caption{The conditional average density  $\Gamma^*(r)$ 
\label{fig13}
for the sample VL40 and VL80 of CfA1. The reference line has a slope 
$-\gamma=-1.1$.}
\eef

iii) 
The linear dependence of $\:r_{0}$ on 
the sample size $\:R_{eff}$
has been found (Fig.\ref{fig14})
\bef 
 %\vspace{}
\epsfxsize 8cm
\centerline{\epsfbox{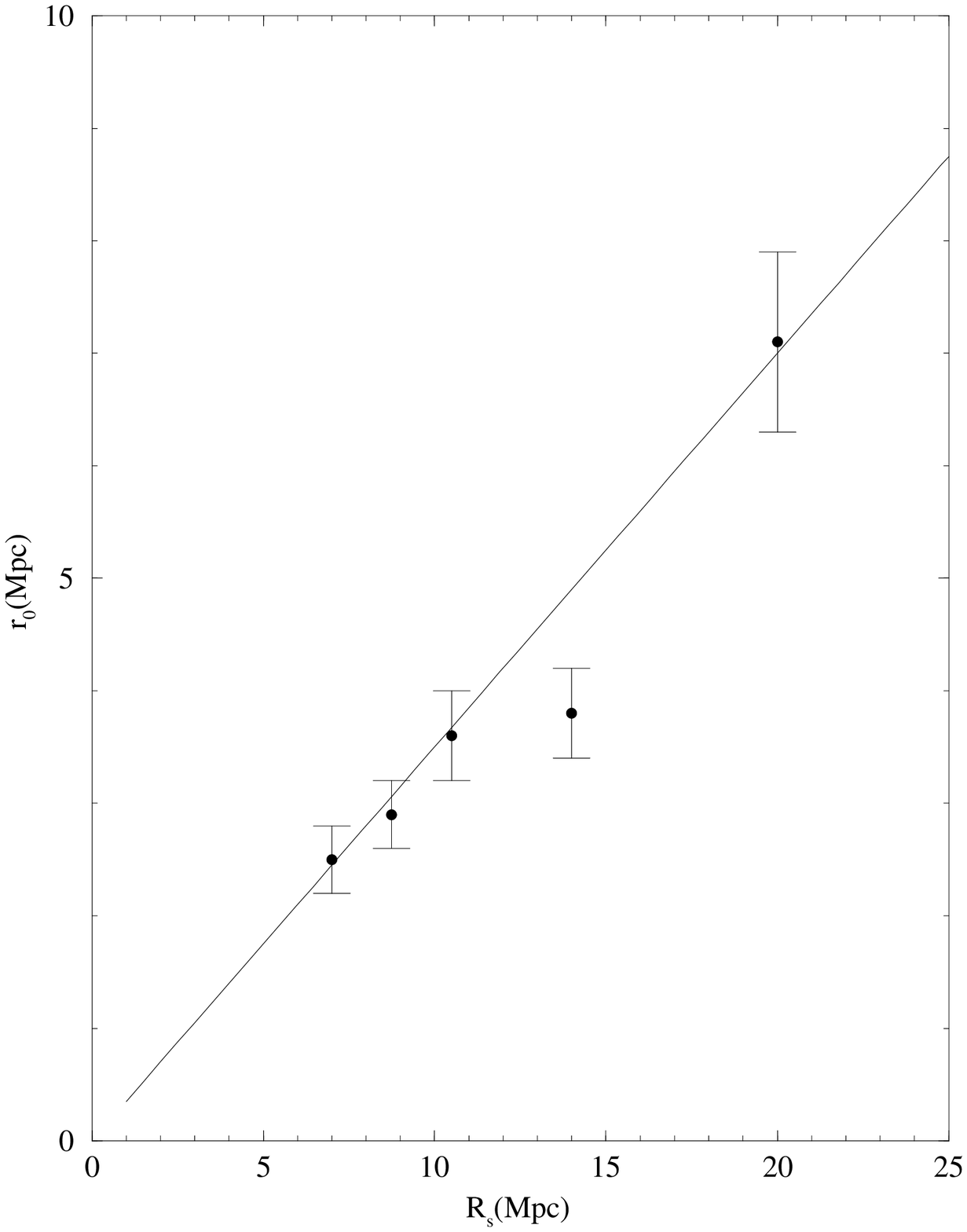}}
\caption{The so-called "correlation length"  $r_0$ 
as a function of the sample size $R_{eff}$. The reference line has a slope 
$1$, in agreement with the behavior for a fractal distribution.
\label{fig14}}
\eef 
in the whole CfA1 catalog \cite{cps88,cp92} 
 and it is naturally explained by the
the fractal nature of the galaxies distribution in this sample.
The extended CfA2 survey can proof (or not)
this relationship
over larger distances.

Davis \etal \cite{dav88} have 
reached a different conclusion.
In
agreement with our 
 results, they found
that $r_0$ increases with the sample depth. However the slope 
of linear regression between $\log r_{0}$ and $\log R_{eff}$ 
they find  is slightly smaller than 1. In fact,
their data are more consistent with an exponent of about 0.7-0.8
 instead of 
 0.5 as the authors claim. (This can be seen  in  their
 Table 1).   This minor discrepancy with the exponent 1 is probably due
 to their 
way of treating the boundary conditions (i.e. weighting schemes),
which, as we show  later, 
introduces spurious homogenization, and thus a systematic decrease
of the
$r_0$ scaling. However the CfA2 catalog, as  
  other deeper
redshift samples,   allows us to 
clarify also this controversial result.

\subsubsection{CfA2}
\label{gammacfa2}

The extended CfA survey represents
currently one of most complete catalog 
of visible matter distribution in the nearby Universe.
In this survey it is possible to study
the large scale structures distribution up to
$\approx 150 \hmp$. Since this survey is not yet published,
we   comment about the results of the analyses performed by 
various authors. Moreover we refer to Sec.\ref{powerspect} for a discussion on
the power spectrum results.

The old CfA survey (CfA1) is limited by an apparent magnitude
of $\:m_{b(0)} \le 14.5$ and contains about 1800 galaxies.
The extension of the CfA redshift survey
is up to $\:m_{b(0)} \le 15.5$ and includes $\:\sim 11000$ galaxies
($6478$ galaxies CfA North and  $\:4283$ galaxies
CfA south).
The main papers published about the CfA2 data analysis to
which we refer are \cite{del88,vog92,dac94,par94}.

The details on the subsamples
used in the CfA2 data analysis 
are the following (from \cite{par94}). 
The VL subsamples
correspond to depths $\:R_{s}=130 \hmp$ (CfA2-130),
101 (CfA2-101), 78.1 (CfA2-78), and 60 (CfA2-60),
respectively. The absolute magnitude limit
for CfA130 is $\:M=-20.3$: this sample has
the same absolute magnitude limit of the
VL sample CfA1-80
of CfA1
(CfA1-80 means that 
$R_{VL} = 80 h^{-1}Mpc$).

In \cite{par94}
 the correlation function $\xi(r)$
 (CF hereafter) has been estimated
from direct pair count distribution, normalizing these
counts to those for a
random distribution of points within the survey volume \cite{dp83}. 
In  Figure 10
of \cite{par94} (Fig.\ref{fig15})
\bef %\vspace{} %\vspace{}
\epsfxsize 8cm
\centerline{\epsfbox{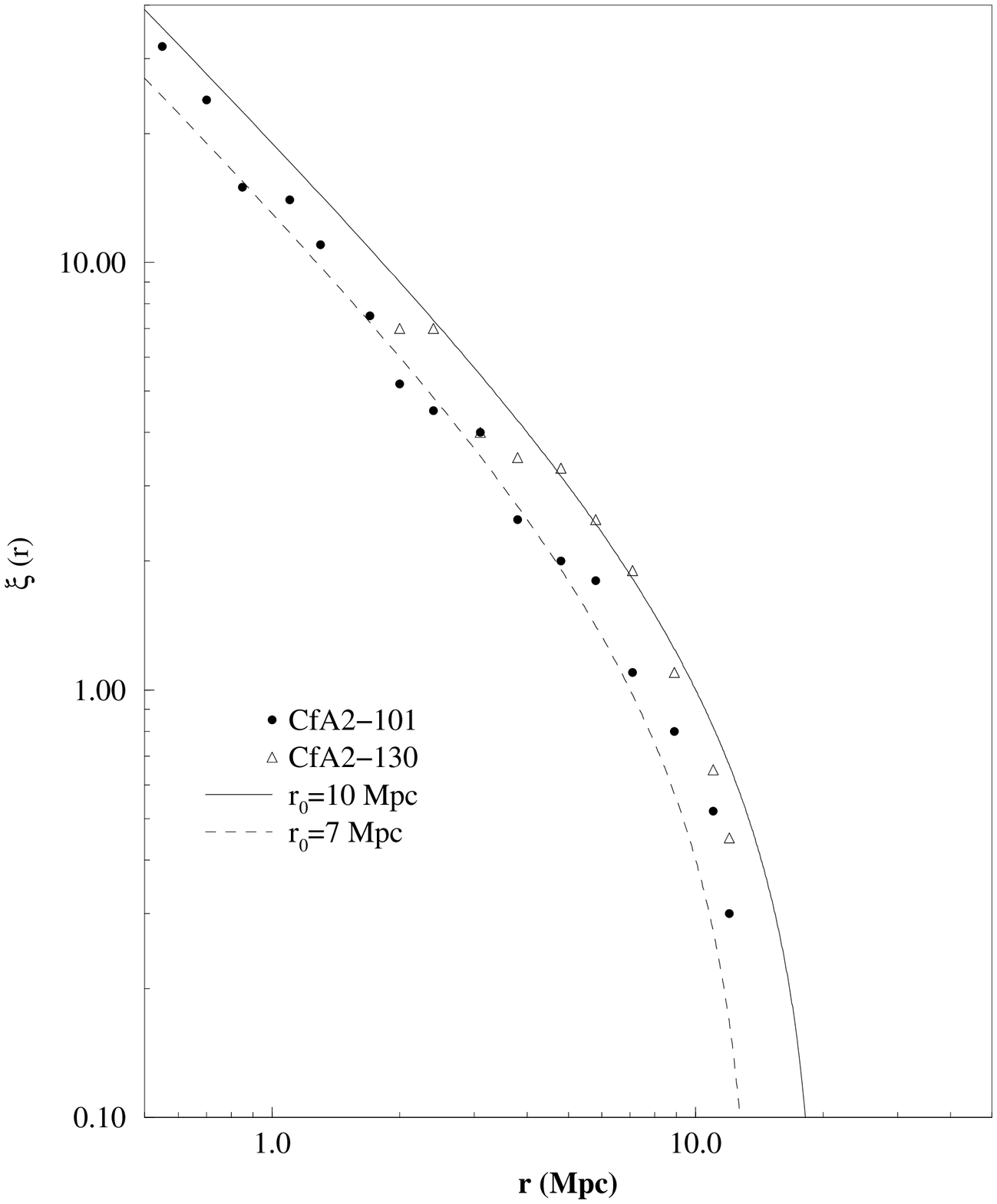}}
\caption{The correlation function $\xi(r)$ 
for two VL samples of CfA2: VL101 and VL130 (from Park \etal 1994).
The fitting curve are computed from the $\xi(r)$ function for a
fractal distribution. 
The correlation exponent, in both cases,
is $\gamma=1$, which corresponds to a fractal dimension $D=2$.
\label{fig15}}
\eef
it is shown the CF for two VL subsamples,
CfA101 and CfA130. 
The results of the CF-analysis are the following:
\smallskip

i) As shown in Fig.\ref{fig15}, 
the exponent of the CF $\:\xi(r)$ is $\:\gamma \sim 1.0$
so that the fractal dimension is $\: D \sim 2 = 3-\gamma$
in the range
between $\:2 h^{-1}Mpc$ and $\approx 10 h^{-1}Mpc$
in agreement with the results of the PS (Sec.\ref{powerspect}).
\smallskip

ii) The so-called correlation length $\:r_{0}$,
defined as $\:\xi(r_{0}) \equiv 1$, is $r_{0} \sim 7 \hmp$ 
for CfA2-101, and
$r_{0} \sim 10 h^{-1}Mpc$  for CfA2-130. 
The amplitude of the CF for
CfA2-130 is approximately $\:40\%$ higher than amplitude of the CF
for CfA2-101, in agreement with the larger amplitude
they find for the power spectrum 
for the same subsamples.
 The amplitudes 
of the CF in the two subsamples CfA2-130 and CfA2-101
are different and scale linearly with sample depth. This is the
same result found at lower distances:
the so-called correlation length $\:r_{0}$ is a linear
function of the  survey depth, if the
system has fractal properties, and it scales up to $\:10h^{-1}Mpc$
for the deeper sample (CfA2-130). 
Note that not  only the functional behavior, but also 
the values of $r_0$ is in agreement with the prediction
for a fractal with $D=2$: $r_0 \approx R_{eff}/3$. The value
of the effective depth  depends on
the solid angle; this 
is $\Omega \approx 1.1$ 
and $R_{eff} \approx 30 \hmp$ for CfA2-130, and 
$R_{eff} \approx 20 \hmp$ for CfA2-101.
Therefore the 
 linear dependence of $r_0$ on the sample depth $R_{eff}$ is
confirmed over the range of CfA1 and extended up the
deeper depth of CfA2.

The authors \cite{par94} comment that the amplitude of clustering may depend
on the luminosity of galaxies, because in the CfA130 subsample
the absolute magnitude of the galaxies is in average
higher than for the galaxies in CfA101.
We discuss this point later, but
 now we can go further by comparing the sample
of the new  catalog CfA2-130 with the
VL sample of the old  catalog
CfA1-80:
these samples have the same absolute magnitude limit. Hence 
 if 
these samples contain galaxies with the same distribution
of absolute magnitude but they have different depth, we can 
then test the following hypothesis: if the
amplitude of the CF  depends on the brightness of galaxies
one expects to find the same amplitude in both the subsamples
(CfA2-130 and CfA1-80) otherwise,
if the scaling of the amplitude of CF linearly depends 
on the sample depth, one expects to find a linear
proportion between amplitudes and depths. It is easy
to show that the second hypothesis is the case.

\subsubsection{Perseus-Pisces}
\label{gammapp}

We have studied the behavior of $\:\Gamma(r)$ and $\:\Gamma^*(r)$
in several  VL 
subsamples extracted from the
Perseus-Pisces\footnote{We warmly thank M. Haynes and R. Giovanelli 
for having given us the possibility of analyzing the Perseus-Pisces catalog.}
 (hereafter PP)
survey limited at the magnitude $m_b=15.5$  
(see Appendix) \cite{slmp96}. 
The total number of galaxies contained 
in such a sample is $3301$ and the solid angle is $\Omega = 0.9 \; sr$.
The effective depth (i.e. the radius of the maximum sphere fully contained in 
the sample) is $R_{eff} \approx 30 \hmp$.
The results are shown 
in Fig.\ref{fig16}.
\bef 
%\vspace{}  
\epsfxsize 8cm
\centerline{\epsfbox{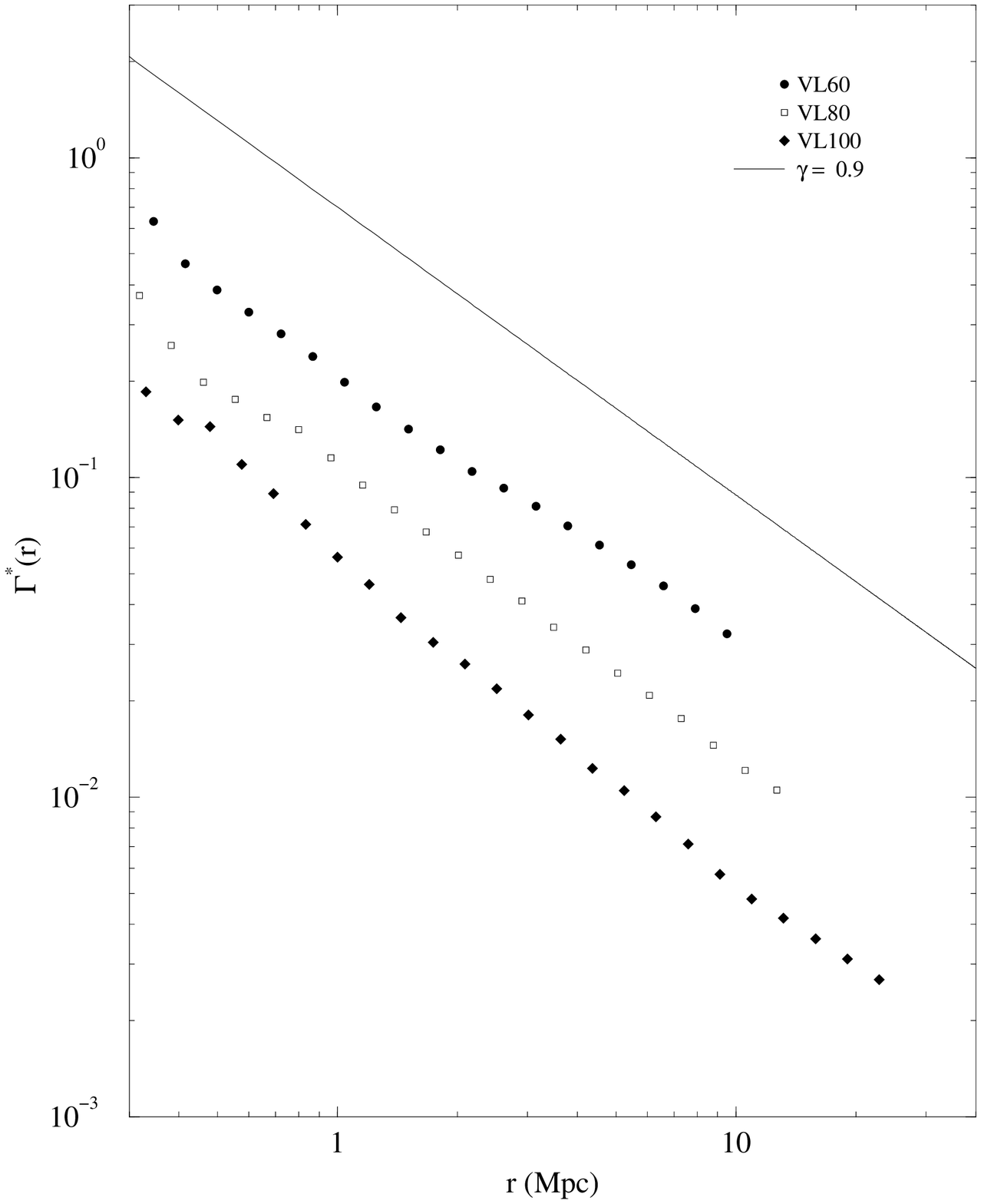}}
\caption{The average conditional density $\Gamma^*(r)$
plotted as a function of the length scale for various volume limited of
 Perseus-Pisces: VL60, VL80, VL100.
The difference in
amplitude is simply due to the different
luminosity selection in the various VL samples,
and can be renormalized by taking into account the different 
luminosity factor of these subsamples.
 A reference slope of $-\gamma=-0.9$ is indicated by the dashed line.
The fractal dimension of these samples turns out to be $D = 2.1$
up to $\sim 30\:h^{-1} Mpc$.
\label{fig16}}
\eef
A well defined power law behavior is detected up
to the sample limit ($R_{eff} \sim 30 \hmp$) 
without any tendency towards homogenization.
The codimension is, with very good accuracy,
$\gamma = 3-D \approx 0.9
$
so that $\:D \approx 2.1$.
Hence the PP redshift survey shows well defined fractal properties
up to the effective depth
$\:R_{eff} \approx 30 h^{-1} Mpc$.
It has consistent statistical
properties and hence
it is a {\it statistically fair} and not homogeneous sample \cite{slmp96}.

We can clarify here an important technical point in the 
computation of the conditional density. 
Guzzo \etal  \cite{gu92}, by 
measuring $\Gamma(r)$ in this catalog, have found
a  change of slope at $\sim 2 \hmp$ :
in particular they found that,  at these small scales,
$D=1.2$ rather than with $D \approx 2$. In order to understand why
this change of slope occurs, we recall 
that, in the  estimate of the conditional
density one computes the density in a shell of thickness $\Delta r$ at distance
$r$ from every occupied point, and then one performs the average.
When one computes this quantity in  real cases, one should define 
the appropriate
value for $\Delta r$. At small distances we are below the average minimum 
separation between nearest galaxies in the sample that is of the order
of some Megaparsec, depending on the sample considered.
For example for the various volume limited samples of Peruses-Pisces
this number is in the range $1 \div 4 \hmp$. At very small distances
one underestimates the number of galaxies in the shell of thickness
 $\Delta r$ for
a finite size effect that causes
 a steeper behavior for $\Gamma(r)$.
This finite size effect is the
real reason for the change of slope (Sec.\ref{weight}).

As we have already discussed in Sec.\ref{weight}
 (see also \cite{cp92}),
 we have limited our analysis to an
effective  depth
$R_{eff}$ that is of the order of the radius of the maximum
sphere fully contained in the sample volume.
In such a way we eliminate from the statistics
the points for which a sphere of radius {\it r} is not
fully included within the sample boundaries.
Doing so we do not 
make any assumption on the treatment of
the boundaries conditions.
Of course in
doing this, we have a smaller number of points
and we stop our analysis  at a  smaller depth than that
of other authors \cite{guz92},
 %(Guzzo et al., 1991),
with
the advantage, however, of not introducing any a priori
hypothesis.
                                                  
The different normalization in the various
VL samples  is due to the different
absolute magnitude limit ($M_{VL}$)
that define each VL subsample. To normalize the
density $\Gamma(r)$, we divide it for  a
luminosity factor given by Eq.\ref{e410} (we refer to 
Sec.\ref{radialpencil}. 
 for a detailed discussion
of such a normalization).

We have studied the $\:\xi(r)$ function
in the VL subsamples with different depth.  
We find that for $\:r \ltapprox r_0$
$\:\gamma \approx 1$.
The amplitude
of $\:\xi(r)$ is sample depth dependent:
in Fig.\ref{fig17}
\bef %\vspace{} %\vspace{}
\epsfxsize 8cm
\centerline{\epsfbox{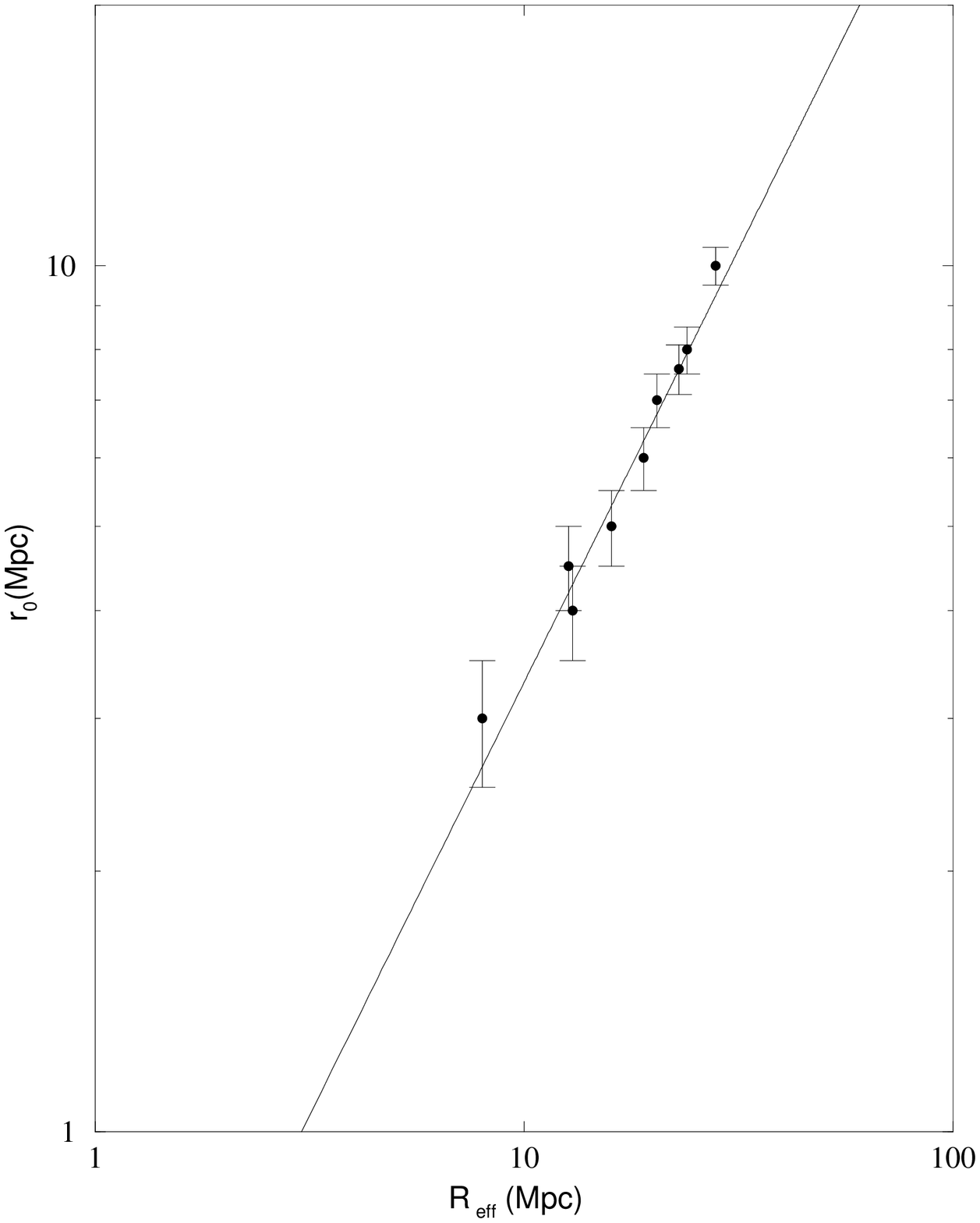}}
\caption{\label{fig17} The "Characteristic length scale" 
$r_0$ ($\xi(r_{0})\equiv 1$)
plotted as a function of the effective radius $R_{eff}$
for the samples of PP. If the catalog
is homogeneous we should find a constant value for $r_0$.
The fitting line has a slope of $(D/6)^{\frac{1}{3-D}}$ as 
predicted for the fractal case.
}
\eef
the behavior of $\:r_0$ is plotted as a
function of the sample depth $\:R_{eff}$
for the VL samples of PP.
The experimental data are very well fitted by 
the prediction for a fractal distribution.
This analysis is
in agreement with the one 
 of \cite{cp92} and with
the analysis done by the CF $\:\Gamma(r)$ discussed
previously. The so-called "correlation length" $\:r_0$
has therefore no physical meaning but
it only represents
a fraction of the sample size (we refer to 
Sec.\ref{lumsegr} 
for a test 
on luminosity segregation).

\subsubsection{SSRS1}
\label{gammassrs1}

This sample consists of 1773 galaxies covering an area
of $\Omega = 1.75 \; sr$ of the South galactic cap in the region
south of declination $-17^{\circ}.5$ and below galactic 
latitude $-30^{\circ}$ \cite{dac88,dac91}.  The sample
is complete down to a limiting galactic diameter given by $\log[D(0)]=0.1$,
where $D(0)$ is in arcminutes. 

The galaxy distribution exhibits 
prominent structures and large 
scale voids, delimited only 
by the extension of the survey: "....we can now recognize 
that the distribution of galaxies is very inhomogeneous ...." \cite{dac88}. 
A clear dependence of $r_0$ on the sample depth has been found also in 
this survey \cite{mau92}. However the authors \cite{mau92} attribute the 
scaling with depth of $r_0$ to the "luminosity segregation" phenomenon.

In order to investigate     this point we have extracted 
from the catalog various samples which are complete 
in absolute diameter (analogous a VL sample). We have considered the subsamples 
which have a depth of $40,60,80,100$ and $120 \hmp$ (see Appendix).
The radius of the maximum sphere fully contained in the  deepest subsample,
which is the limit of our statistical analysis, is $R_{eff} \sim 35 \hmp$.
We have computed the conditional average density and the conditional  density
for the various subsamples and we show the results 
in  Fig.\ref{fig18}.

\bef 
%\vspace{}
\epsfxsize 8cm
\centerline{\epsfbox{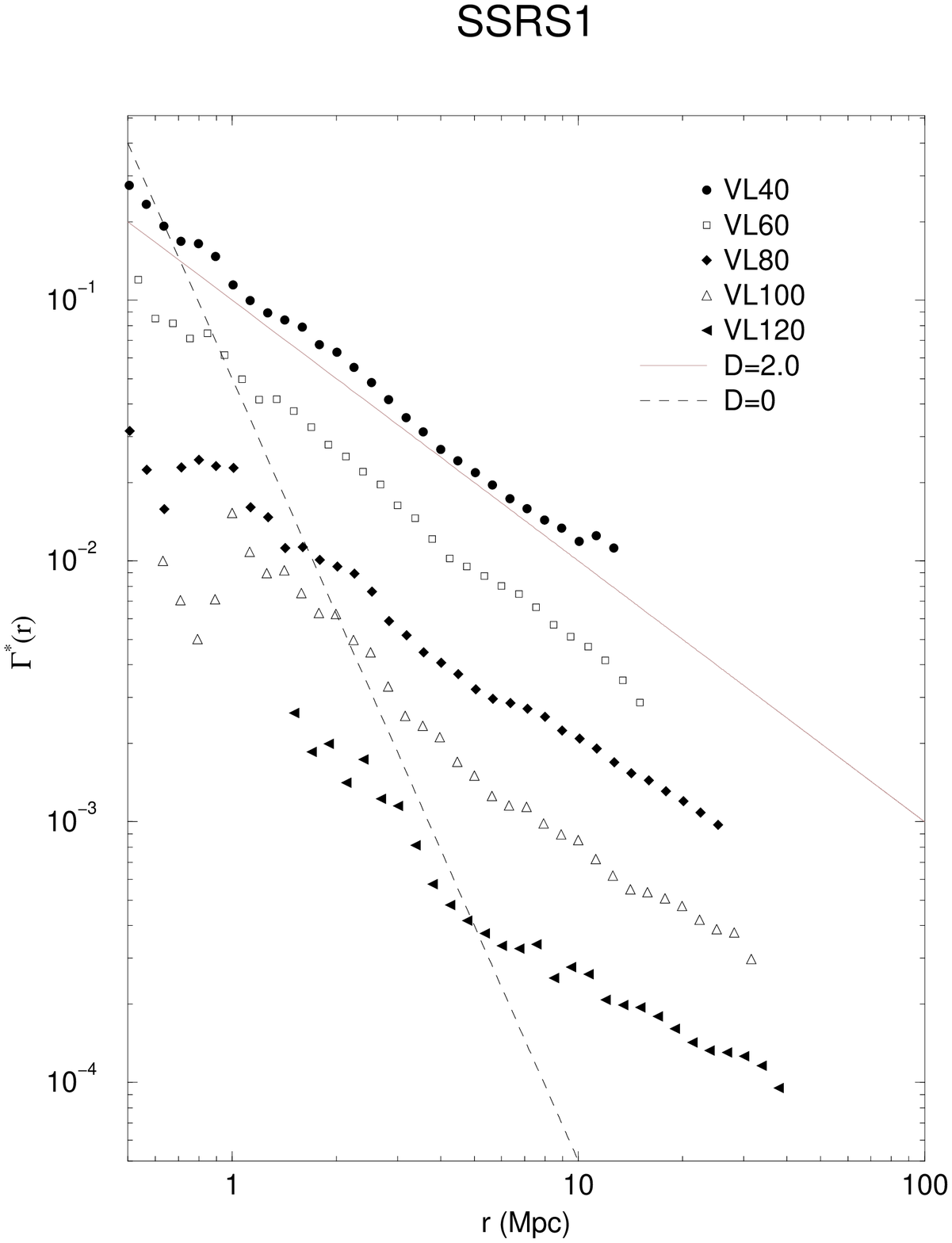}}
\caption{The average conditional density $\Gamma^*(r)$
plotted as a function of the length scale for various volume limited of
 SSRS1. A reference slope of $-\gamma=-1$ is indicated by the dashed line, that 
corresponds to a  fractal dimension of  $D\approx 2$
up to $35\:h^{-1} Mpc$. The dashed line has a slope $-3$ (D=0).
\label{fig18}}
\eef

We find that galaxy distribution in this sample is characterized by having long
range correlations up to $\sim 35 \hmp$, with fractal dimension 
$D = 2.0 \pm 0.1$. We find no evidence for a crossover towards homogenization
 in this catalog. The deepest sample VL120 has a very poor statistics,
and this is the reason why $\Gamma^*(r)$ is more noisy and has a
$1/r^3$ decay at small distances (Sec.\ref{angdist} 
and Sec.\ref{validity} for a detailed 
discussion of such an effect). This small scale behavior
 is due to the fact
that, in this sample,  in average one
 does not find any other galaxy for $r \ltapprox 7 \hmp$.
The conditional average density for the sample VL120 is   smoother and shows
a complete agreement with the all the other samples. 

We have then computed the so-called "correlation length"
by performing the standard correlation analysis by the $\xi(r)$ function.
As shown in Fig.\ref{fig19},
we have studied $\:\xi(r)$ in the VL 
samples with different depth.
We find that the amplitude
of $\:\xi(r)$ is sample depth dependent.
\bef %\vspace{}  
\epsfxsize 9cm
\centerline{\epsfbox{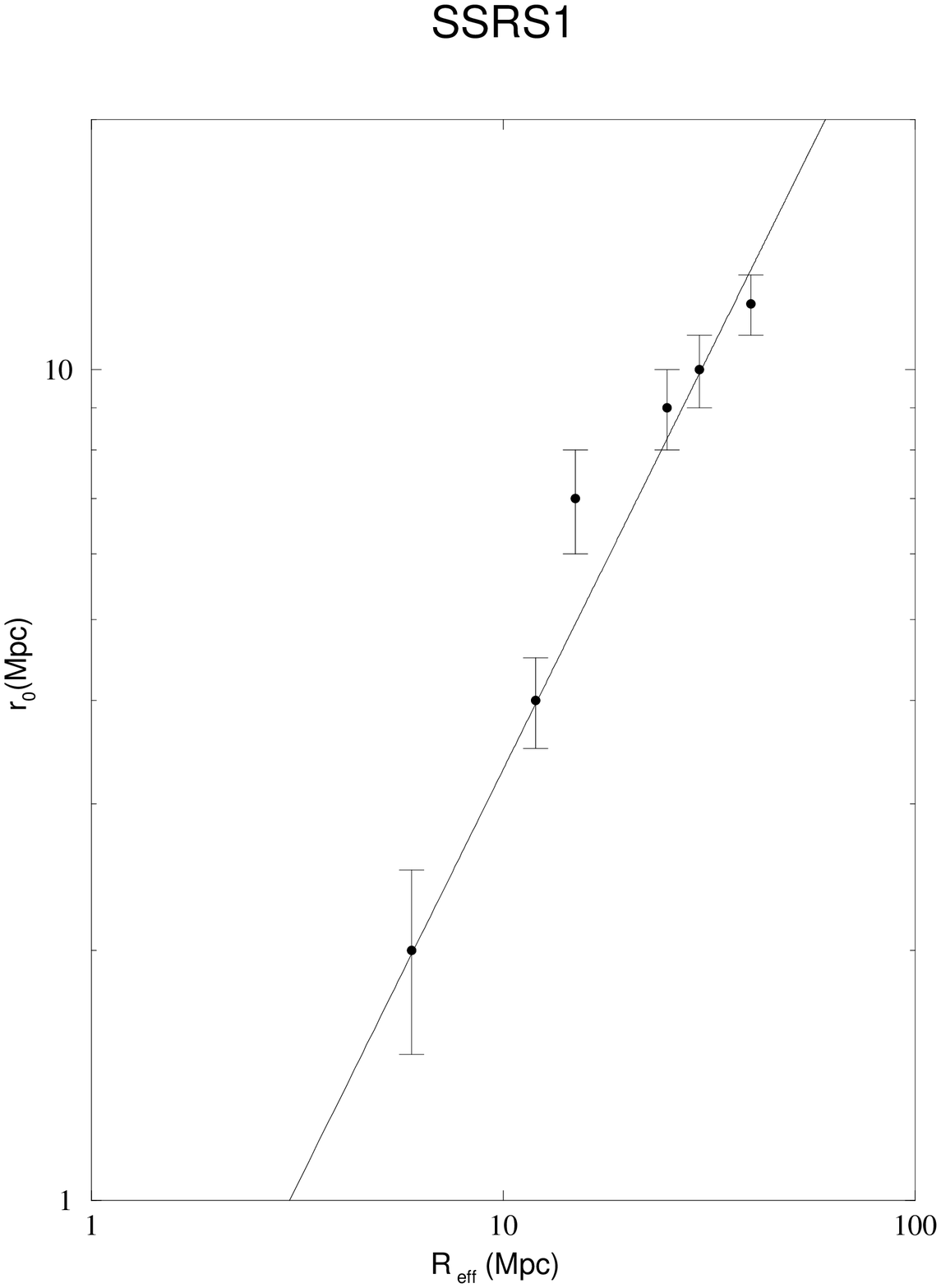}}
\caption{\label{fig19} The "Characteristic length scale"
$r_0$ ($\xi(r_{0})\equiv 1$)
plotted as a function of the effective radius $R_s$
for the samples of SSRS1. 
The fitting line has a slope of $(D/6)^{\frac{1}{3-D}}$ 
as predicted for the fractal case.
}
\eef
Even in this case 
the experimental data are very well fitted by
the prediction for a fractal distribution (Sec.\ref{correlation}).
Contrary to the results of 
Maurogordato, Schaeffer and da Costa \cite{mau92},
from the previous tests, we find no evidence for 
a luminosity dependence of $r_0$. Rather we find, in agreement with the
correlation analysis performed by the $\Gamma(r)$ function, that 
$r_0$ is a fraction of the sample size, as expected 
in the fractal case.

\subsubsection{SSRS2}
\label{gammassrs2}

The SSRS2 catalog is a magnitude limited survey that 
includes $\sim 3600$ 
galaxies and covers $\Omega = 1.13 \; sr$, complete
up to limiting 
magnitude $m_{B(0)} = 15.5$ \cite{dac94,ben96} 
(this catalog is  not yet published).
The characteristics of the VL samples 
are reported in the  Appendix. 
In the first paper about this survey \cite{dac94}, 
the authors concluded that "by examining 
the normalizing density fluctuations in the SSRS2 and CfA2 surveys
we suggest that the combined sample {\it is not large enough to be fair}:
there are large fluctuations (i.e. the South Wall) 
in shells at $100 \hmp$". It is clear that,
even in this case,
 the sample is declared to be "not fair" because it does not 
show a well defined average density, but on the contrary
it is dominated by fluctuations as large as the sample itself.

Benoist \etal \cite{ben96}
have measured the "correlation length" $r_0$
in several VL samples of different depth.
We show in Tab.\ref{tabssrs2}
\begin{table} \begin{center}
\begin{tabular}{|c|c|c|}
\hline
&      &                             \\
Sample       & $r_0 (\hmp)$    & $R_{eff} (\hmp)$    \\
&      &                             \\
 \hline
% \hline
  D48                & $3.5 \pm 0.2 $ & 12               \\
  D60                & $4.8 \pm 0.5 $ & 15            \\
  D74                & $5.5 \pm 0.4 $ & 18             \\
  D91                & $5.8 \pm 0.6 $ & 22             \\
  D112               & $5.5 \pm 0.6 $ & 27             \\
  D138               & $8.0 \pm 0.4 $ & 36             \\
  D168               & $15.8 \pm 2.9$ & 40              \\
&      &                             \\
\hline
\end{tabular}
\caption{Values of $r_0$ and $R_{eff}$ (which has been estimated by knowing 
the geometry of the sample) for the various VL samples
of the SSRS2 survey (from Benoist \etal 1996). 
There is a clear dependence 
of $r_0$ on the sample depth. \label{tabssrs2}}
\end{center} \end{table}
the values of $r_0$ and $R_{eff}$ for the various VL samples
of the SSRS2 survey. A clear dependence
of $r_0$ on sample size is shown. 
We  stress that
the authors \cite{ben96} fit the $\xi(r)$ with a power law behavior
in the wrong region of length scales, i.e. for $r \approx r_0$, and 
hence  they find  a higher value for the 
correlation exponent $\gamma \approx 1.5 \div 1.8$ (Sec.\ref{correlation}).
By fitting the $\xi(r)$ function with the expected shape for a fractal
distribution we find that there is a very good
agreement with the published data:
we show in Fig.\ref{fig20} 
\bef 
%\vspace{} 
\epsfxsize 8cm
\centerline{\epsfbox{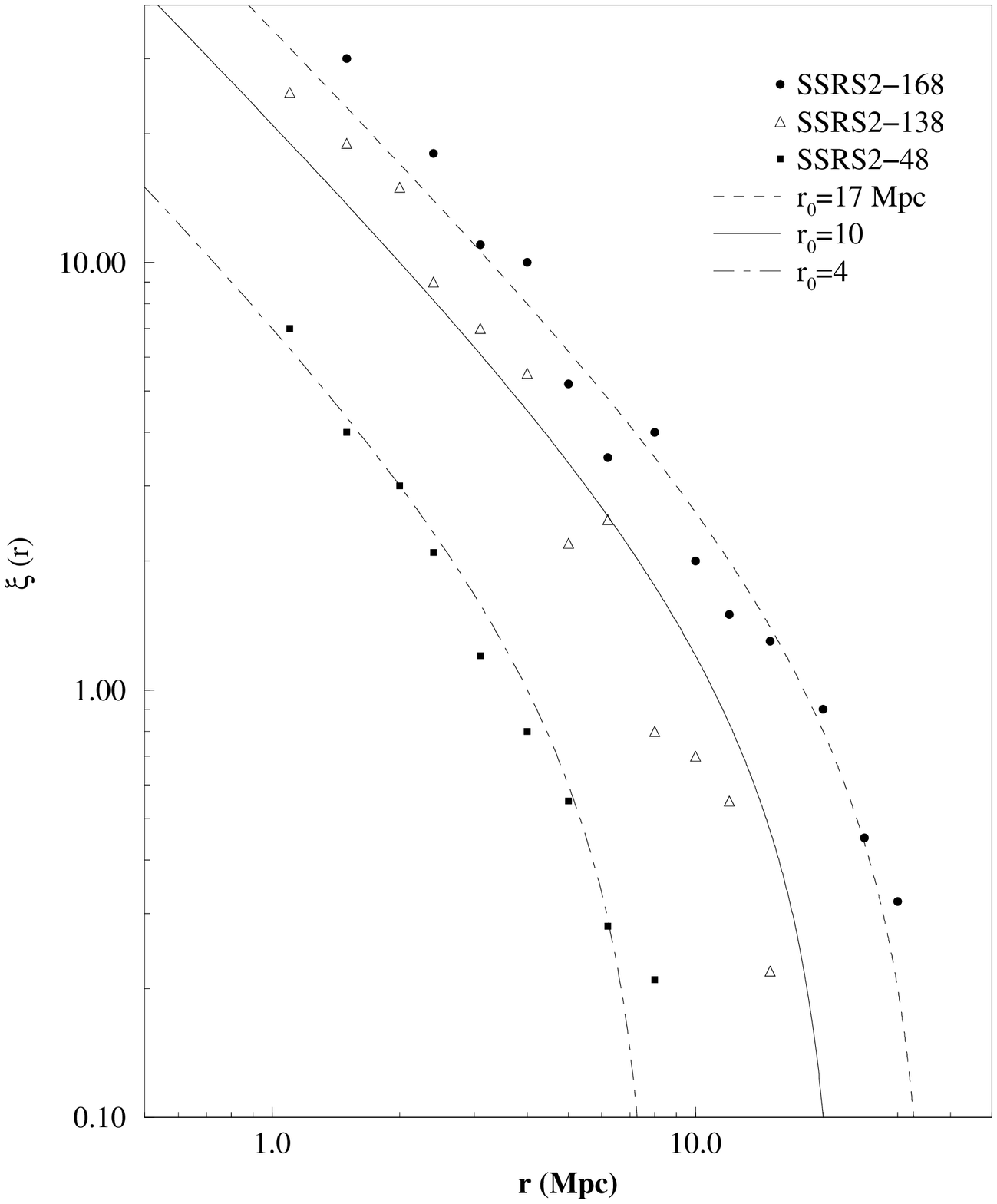}}
\caption{\label{fig20} 
The $\xi(r)$ function for
 SSRS2 in the VL samples D30, D138 and D168 (from Benoist \etal
1996). 
The fitting curve has been compute from the expression of $\xi(r)$
for a fractal distribution. In all the cases the slope is 
$\gamma=1$, i.e. the fractal dimension is $D=2$.
}
\eef
the $\xi(r)$ for several VL samples of SSRS2 together with 
a fitting curve that is $\xi(r) = [(3-\gamma)/3]
 (r/R_{eff})^{-\gamma} -1$, and
the value of the exponent is $-\gamma=-1$ in all the cases. 
We can then conclude that the fractal dimension in this case is
again $D \approx 2$.

Let us come to the discussion of the scaling of $r_0$
in the various VL samples. First of all we notice that
$r_0 \approx 16 \hmp$ is the highest value for 
the correlation length ever found in a redshift survey, after the 
results of LEDA \cite{dmpps96}. The
 authors \cite{ben96}
have done  several test in order to check if the dependence
of $r_0$ on sample size is due to 
a luminosity effect or to a depth effect.
In the first case it should be due to the so-called 
luminosity segregation phenomenon, while in the second
case it should be attributed to the intrinsic fractal
nature of galaxy distribution in this sample. 

First of all we   stress that in all the galaxy surveys,
 such as CfA1 and SSRS1 for example, there are galaxies with 
$M \leq -19.5$ because these are the most luminous ones. However,
in no one of these  catalogs 
a value of $r_0 \approx 16 \hmp$ has been 
ever measured. If luminosity segregation is the real
physical effect that produces the scaling of $r_0$,
 then such a value of $r_0$
should be present in all the other surveys, as it refers 
to the most luminous galaxies.

On the contrary, the authors \cite{ben96} 
conclude that the  variation  
  of the  $\xi(r)$ amplitude, observed in this catalog
has two contributions: sampling fluctuation for the faint galaxies, and
luminosity segregation for the brighter ones. 
However we can make the following comments
on these results:
i) The authors do not present {\it any quantitative argument} 
that explains the shift of $r_0$ with sample size (Fig.1 of \cite{ben96}), 
but  they just make several test on the deeper samples 
that seem to show a difference in the amplitude of $\xi(r)$
for brighter and fainter galaxies.
ii) The test only 
on luminosity segregation 
that gives a result that seems to be in agreement with such an effect
(but without any {\it quantitative discussion}), 
is performed only in the sample
D91 that contains 67 galaxies in the magnitude bin 
$-21.5/-20.5$ (Fig.2 of \cite{ben96}): they stress in fact
that "only in the large VL (D91) we found a significant effect". 
They do not show the results for the deeper samples,
nor they find a consistent argument that can explain why the 
amplitude of the correlation 
function reaches the value $r_0 \sim 16 \hmp$.
iii) In Fig.3 of \cite{ben96} it is reported the behavior of 
$\xi(r)$ for galaxies in the same interval of absolute magnitude
but in samples with different depth. 
About these results, we can comment 
that they have computed the $\xi(r)$ function by 
using the standard method of treatment of boundary conditions
 in samples with a small difference in the 
effective depth $R_{eff}$ (see Tab.\ref{tabssrs2} and their Fig.3).
Hence it is very difficult to make definitive conclusions from their
analyses.
iv) Moreover the authors \cite{ben96}
do not present the computation of $\Gamma(r)$
that is the clarifying test, in order to check 
the fractal versus homogeneous properties in this sample.

We came back on the problem of "luminosity segregation" later
by performing several specific tests. 
We  stress again that 
this effect has been invoked by several authors 
\cite{par94,ben96}
to explain the shift of the amplitude of $\xi(r)$ or of
the power spectrum in samples of different depth (and different 
average absolute magnitude), but in any case {\it it has never been 
presented any quantitative argument that can explain the 
behavior of the amplitude of $\xi(r)$ with absolute luminosity}.

Instead, in the case of SSRS2,
we cannot present here a detailed analysis because the 
survey is not yet published, but we can argue from the 
available results of the data analysis that 
galaxy distribution in SSRS2 is fractal with dimension $D \approx 2$ up
to the limiting depth of $R_{eff} \sim 50 \hmp$.
A test that may proof such a behavior is the measure of 
the conditional density that should have a power law behavior 
with exponent $-\gamma \approx -1$, i.e. $D \approx 2$, 
 up to $\sim 50 \hmp$.

\subsubsection{Stromlo-APM}
\label{gammasars}

The Stromlo-APM Redshift Survey (SARS) consists of 1797 with 
$b_J \leq 17.15$ selected randomly at a rate of 1 in 20 from APM scans
\cite{lov92,lov96}. The survey covers a solid angle of
 $\Omega = 1.3 sr$
in the south galactic hemisphere, delimited by
 $21^h \ltapprox \alpha \ltapprox 5^h$ and $-72.5^{\circ} \ltapprox \delta 
\ltapprox -17.5^{\circ}$ (see Appendix). 
An important selection effects exists: 
galaxies with apparent magnitude brighter than $b_j=14.5$ 
are not included in the sample because of photographic saturation (see 
Fig.\ref{fig21}).
\bef  
%\vspace{}
\epsfxsize 12cm
\centerline{\epsfbox{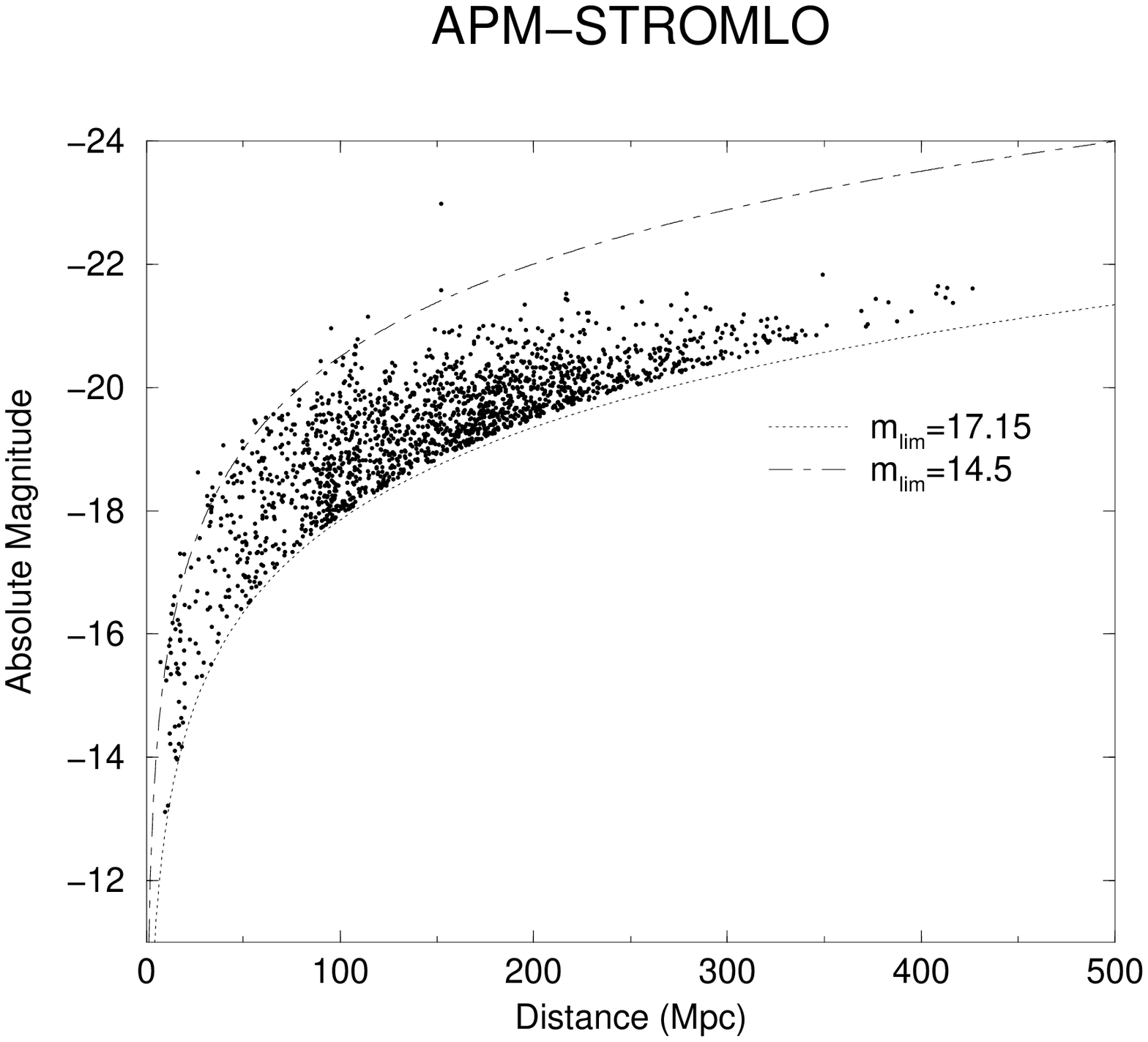}}
\caption{\label{fig21}
Absolute magnitude versus distance diagram for the SARS catalog.
There are shown the two limiting function with apparent magnitude
$m^1_{lim}=14.5$ and $m^2_{lim}=17.15$. The survey is complete up to
$\sim 17-th$ magnitude and does not contain galaxies brighter than $\sim 14.5$.}
\eef

In order to construct VL samples we have adopted two
different procedures. The first is the standard one, i.e. we have 
introduced an upper cutoff in the distance and computed the 
corresponding 
cutoff in absolute magnitude (see Appendix). 
The second consists by putting  two limits in distances
and compute the corresponding two limits in absolute magnitude. 
In such a way we can avoid the selection effects due to the 
fact that in this survey are not included galaxies brighter 
than $14.5$ (this is the same procedure used in the case of LCRS
that is limited by a lower and an upper limit in apparent 
magnitude as SARS).

We have then computed the conditional average density for the  VL samples
with only one cut in absolute magnitude, and we show in Fig.\ref{fig22}
the results.
\bef 
%\vspace{}  
\epsfxsize 8cm
\centerline{\epsfbox{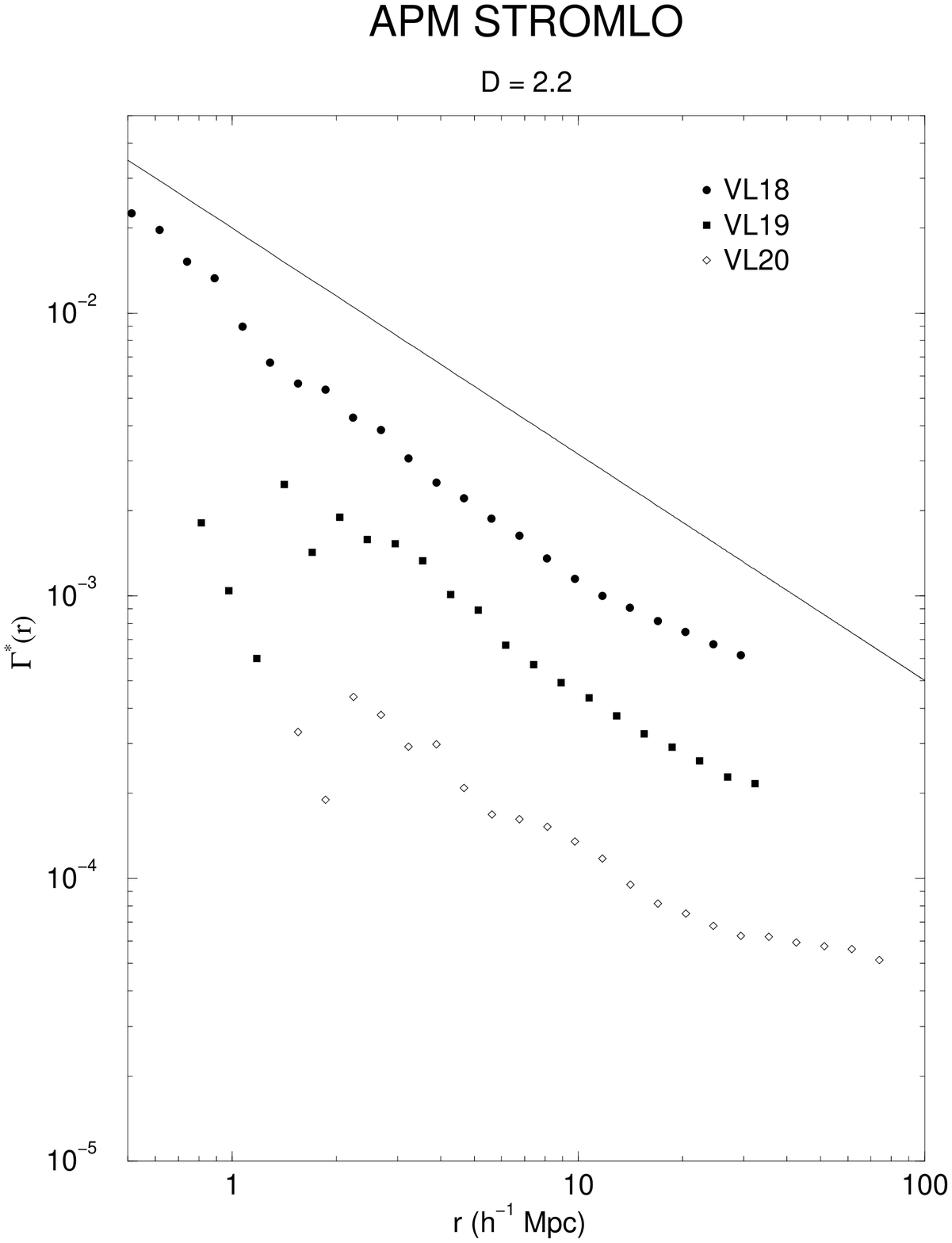}}
\caption{\label{fig22}
The conditional average density computed for some VL sample of
the SARS redshift survey. The reference line has a slope $-\gamma=-0.8$.
The fractal dimension is $D = 2.2 \pm 0.2$, depending on the 
VL sample used.}
\eef
The fractal dimension is $D =2.2 \pm 0.2$ 
up to
$\sim 35 \hmp$.
From the previous figure it follows that the 
conditional density for the deeper VL sample (VL20, where 20 refers to the 
absolute magnitude limit of the VL sample)
has a tendency towards a flattening 
at scales larger than $\gtapprox 30 \hmp$.
Such a flattening in our view is completely spurious and due to the 
sparseness of
 this particular sample. We refer to Sec.\ref{validity} for a 
detailed discussion of this effect.

In order to
 check that the luminosity incompleteness of the sample for
apparent magnitude brighter than $14.5$  affects substantially
the trends found in Fig.\ref{fig22}, we have computed the
conditional average density for the VL samples, with two cuts in distance and 
two in absolute magnitude. These samples  are defined by a lower and an upper
 cut in distance
(for example VL100-200 means that we have cut the VL between the region 
from $100$ to $200 \hmp$ -  see Appendix).
\bef 
%\vspace{} 
\epsfxsize 8cm
\centerline{\epsfbox{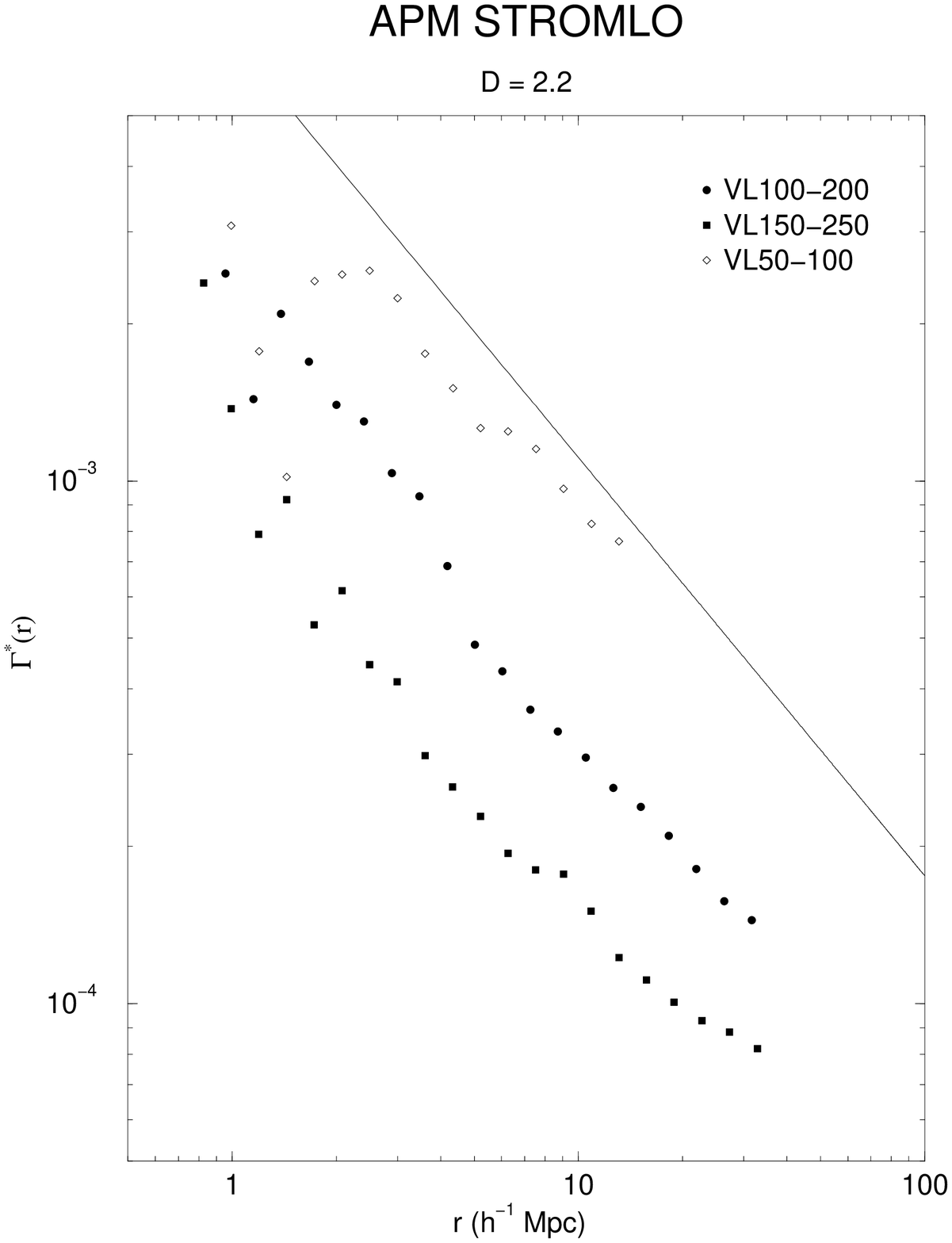}}
\caption{\label{fig23}
The conditional average density computed for some VL sample of
the SARS redshift survey, with two cuts
in distance and absolute magnitude. 
Such a procedure avoids the luminosity incompleteness
of galaxies with apparent magnitude brighter than $14.5$.
The reference line has a slope $-\gamma=-0.8$.
The fractal dimension is $D = 2.2 \pm 0.2$.}
\eef
Even in this case (Fig.\ref{fig23}) the fractal dimension 
turns out to be 
$D = 2.2 \pm 0.2$ up to $R_{eff} \sim 40 \hmp$:
some samples show a tendency towards a flattening.
We can 
show by precise tests that this flattening is spurious 
and it is  due to the 
sparseness and weak statistics of these samples 
(Sec.\ref{validity}).

\subsubsection{LEDA}
\label{gammaleda}

We have studied in a series of papers 
\cite{dmpps96,ledanat,ledabias,ledaps,ledalong,ledasesto}
the statistical properties of there
 LEDA database\footnote{We thank H. Di Nella and G. Paturel 
for their collaboration in the analysis of the LEDA database.}, 
and here we review our main results.
This database  now contains more than 200,000 galaxies with the most
important
astrophysical parameters: name of galaxies, morphological
description,
diameters, axis ratios, magnitudes in different colors, radial
velocities,
21-cm line widths, central velocity dispersions, etc.
In the last 12 years since the inception of LEDA,
more than 75,000 redshifts have been collected, for more than
40,000 galaxies.

 The magnitudes of the galaxies registered in the LEDA database
come from many  different references. It is the
 aim of a database to
collect all new measurements of  galactic observational parameters.
In the paper \cite{pat94} the formulae to reduce twenty
different magnitudes systems to the  photoelectric $b_t$ system of the Third
Reference Catalogue (RC3) are given. This reduction
is done in such a way that the
resulting
magnitudes are completely free of bias or perturbing effects.
Each magnitude is then given with its own error. This error is of course
in the range 0.1 to 1 magnitude depending on
the observations and on the
 comparison
between different
 published values for the same galaxy. The majority
 of the galaxies
have a mean error less than 0.5 magnitude.
The complete list of the references of the
 magnitudes we used
for the purpose
of this paper can be found in \cite{pat95}  (we refer to this paper also
for more
information of the data contained in LEDA).

The main selection effect that must be studied
with great care is the incompleteness of the redshift data
 in the LEDA database, for different directions
 of observation, i.e. a different angular incompleteness.
In  Fig.\ref{fig24}, we show the percentage of galaxies
\bef
 %\vspace{} 
\epsfxsize 10cm
\centerline{\epsfbox{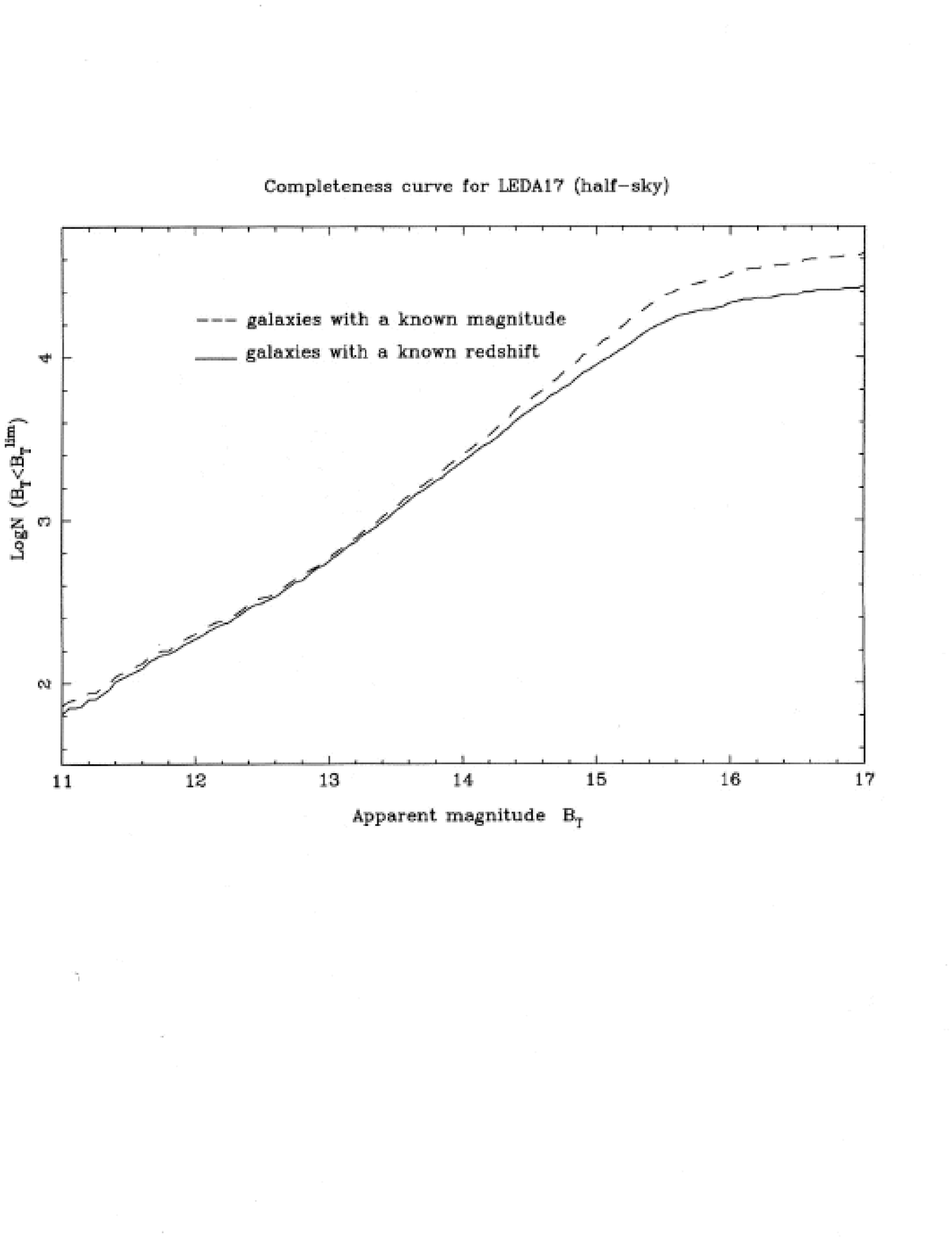}}
\caption{\label{fig24} Fraction between the number of galaxies 
reported in LEDA and the number of galaxies with measured redshift. Up to
$B \sim 14.5$ all the known galaxies are measured. After this
 limit the LEDA database begins to be incomplete.}
\eef
known in angular catalogs compared to the number of galaxies having
a measured redshift.
Up to the limiting magnitude $m_B=14.5$, LEDA has  a
rate of redshift measurements of $90\%$, if we extend the study
by including fainter galaxies up to a limit of $m_B=16$, this
rate falls to $50\%$, and with $m_B$ limit $= 17$, LEDA has only
redshifts the $10\%$ of  galaxies.

To investigate the incompleteness of the LEDA database we have
built three magnitude limited (ML) samples: the first is limited
at $m=14.5$ and it is called LEDA14.5, the second
is limited at $m=16$ (LEDA16), and the third is limited at
$m=17$ (LEDA17). The first ML subsample LEDA14.5
is almost complete, i.e. there are   redshifts for the
$\sim 90 \%$ of the total number of galaxies present in the
correspondent angular
catalog.
The problem that we
  discuss     is whether the incompleteness in
the deeper magnitude limited catalogs 
destroys the statistical relevance of the
samples, or whether it is possible to recover the correct
statistical
properties and to control the effects of such incompleteness.
 In particular we  perform
 several quantitative tests in order to
clarify this crucial point.

In order to avoid the zones of avoidance due to the Milky Way,
we cut the all-sky samples in two  half-sky
catalogs. In this way we  have, for example,
LEDA14.5N that refers to the northern galactic
hemisphere, and
LEDA14.5S that refers to the southern one.
The samples denominated North and
South are respectively selected with the galactic latitude:
$b > +10^{\circ}$ and $b < -10^{\circ}$, as usually done in literature.

 We have then 6 ML samples: LEDA14.5N and LEDA14.5S,
LEDA16N and LEDA16S, LEDA17N and LEDA17S.
We   analyze  separately each of these samples.
Moreover to perform the space distribution analysis, we
extract various VL samples
and we report in the Appendix  their characteristics. In such a way we
have various independent VL samples.

In order to control the effect of the data incompleteness, 
we have firstly
studied the correlation properties of the two VL
samples extracted from 
LEDA14.5.  These samples are nearly complete, and the
incompleteness is no more than $\sim 10 \%$. 
To compare our results
with other known we have computed 
the correlation function in the same
region of the CfA1 redshift survey.
In this case we have a
complete sample because LEDA contains the
CfA1 redshift catalog. We find that the number of points in the
various VL sample of LEDA14.5-CfA1 are 
nearly the same of that of CfA1.
 The conditional density scales as a power law with
 $\gamma=3-D=0.9 \pm 0.2$ ($D = 2.1 \pm 0.2$)
up to the sample limit that is $\sim 20 \hmp$.
The results of our analysis are therefore in perfect
agreement with those of CfA1 (Sec.\ref{gammacfa1}).

The next step is to increase the solid of angle of the LEDA14.5N up
to include all the northern hemisphere. 
We find that the
CF function does not
change neither its slope nor its amplitude,
so that 
 it remains a power law with  the same fractal dimension.
The scaling region extends up to $\sim 30 \hmp$.
Therefore in this way we can control the effect of incompleteness
and conclude that this result confirms
that the small incompleteness
of this sample does not affect the correlation properties of
galaxy distribution.
Moreover, we have done the same analysis for the VL
samples of LEDA14.5S
finding a result in substantial agreement with that of LEDA14.5N.
This implies that the statistical properties of the LEDA14.5
are robust with respect to the small incompleteness
in the galaxy redshifts which characterize this sample.
The fractal dimension in the VL
 sample of LEDA14.5 turns out to be $D = 2.1 \pm 0.2.$

An important  point of the LEDA database analysis is that 
from the law of codimension additivity (Sec.\ref{orthogonal}), 
it follows that if
 we cut points randomly from a fractal distribution, its
fractal dimension does not change. The cutting procedure
can be applied up to eliminate more than $\sim 95   \%$ of the 
structure points without changing the genuine properties of 
the distribution (Sec.\ref{validity}).
From the previous discussion, it follows that if the
incompleteness effects are randomly distributed in space,
they do not  affect the correlation properties of the sample,
unless the percentage of the points present in the sample
is too small.
On the contrary, if there are some systematic effects,
as a poor sampling in certain large scale regions,
the correlation properties can be seriously affected.
This means that if in the incomplete samples we  
find the same
correlation properties of the complete ones, we can be
confident that we are measuring the real properties of galaxy distribution.
Otherwise, if we   detect an eventual cut-off towards
homogenization we have to do a very careful analysis, to understand
whether such a cut-off is spurious, i.e. do to
systematic incompleteness effects, or whether it is an
intrinsic property of the sample.

The amplitude of $\Gamma(r)$ is related to the lower cut-off of
the distribution, i.e.   to the prefactor
of the average density.
To normalize the conditional density in different VL samples
(defined by the absolute magnitude limit $M_{VL}$)
 we divide it by the luminosity factor (we  Sec.\ref{consist} for a detailed 
discussion of such a normalization).
In Fig.\ref{fig25}
\bef 
%\vspace{} 
\epsfxsize 10cm
\centerline{\epsfbox{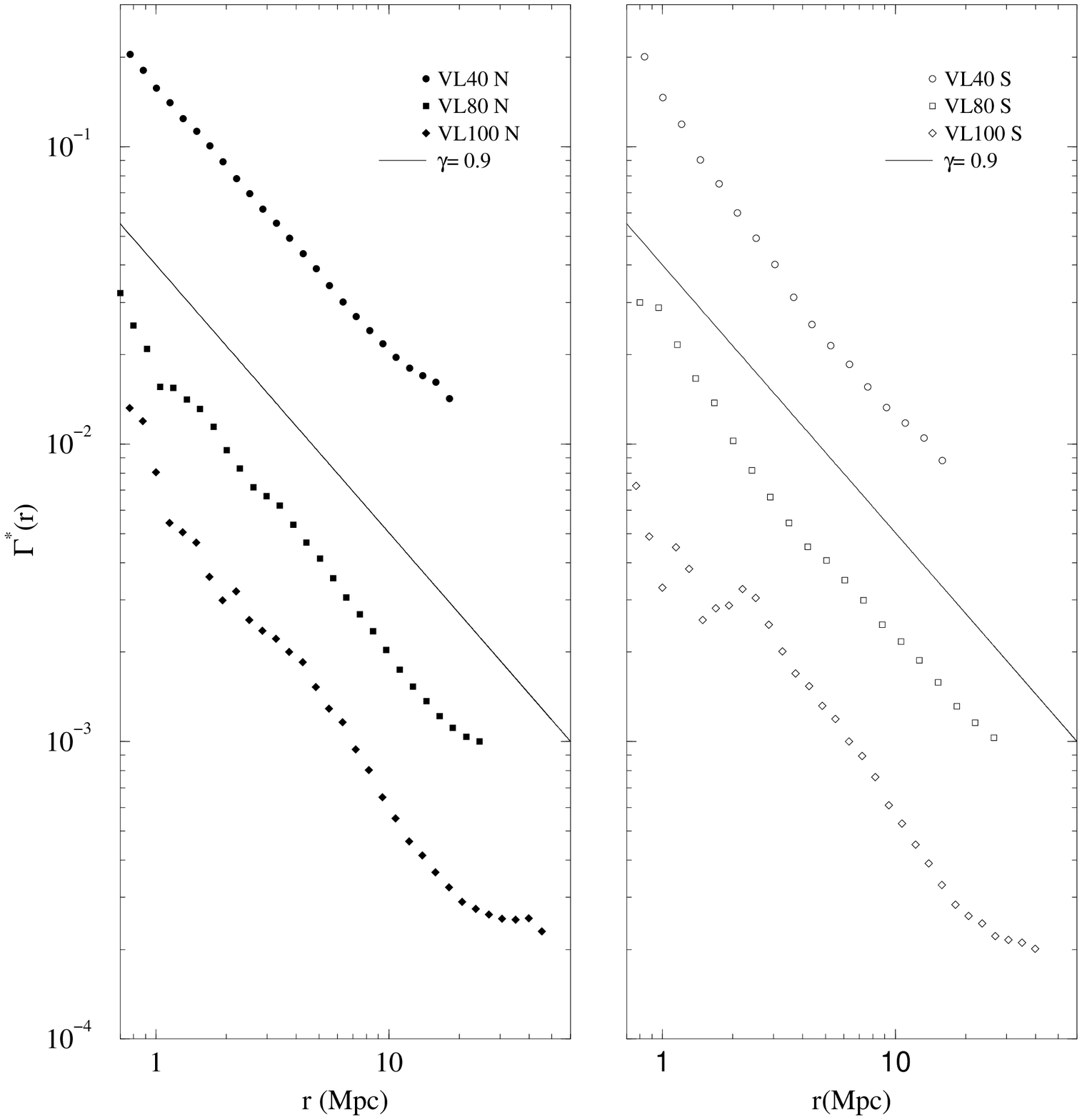}}
\caption{\label{fig25} The conditional average density $\Gamma^*(r)$ 
for various VL samples of the LEDA14.5 sample. $N$ for the VL
 sample of the Northern hemisphere and $S$ for the Southern one.
The reference line has a slope $-\gamma =-0.9$ ($D=2.1$).}
\eef
there are plotted also the different $\Gamma(r)$ for the various
VL samples of LEDA14.5:
we find that there is a nice matching of the
amplitudes and slopes. This implies that we are computing the 
correct
galaxy conditional  density in each subsample, and that such a quantity  
does not depend (or it depends very weakly) on the absolute 
magnitude
of galaxies. Moreover we are confident to be not biased by any 
selection
or incompleteness effect. We have done this kind of analysis for the 
whole north
galactic hemisphere (LEDA14.5N) and for the southern
one (LEDA14.5S): in both cases we obtain the same kind of
behavior for $\Gamma(r)$.

The following step is to analyze the VL samples
of LEDA16  (north and south), and
we have studied the behavior of the correlation function $\Gamma(r)$ in the
various VL samples of LEDA16N and LEDA16S.
We find in both cases that the distribution has long-range correlation
with the same fractal dimension $D = 2.1 \pm 0.2 $, up to $\sim 80-90 \hmp$
(Fig.\ref{fig26}).

We   stress that in some VL samples
of LEDA16 (and also LEDA17)
$\Gamma(r)$ is affected by incompleteness and it slightly
deviates from a power law behavior. However  we find that a power 
law behavior is quite stable in all the VL samples and we   conclude that
is some cases the incompleteness of the database may
affect weakly the behavior of the conditional density.
Such a conclusion is supported also by the power spectrum analysis 
(Sec.\ref{powerspect}).

To compare the amplitude of  $\Gamma(r)$ in this samples we
consider the normalized CF (i.e divided by the luminosity factor).
We find that the amplitude of the normalized CF is
about the $40 \%$ of that found for LEDA14.5. This is because, in this
case, the sampling rate is not $9/10$ as it is for LEDA14.5 but it is
about $3/5$ as we have discussed previously. Hence  
a part this factor, we have that
the CF has the same features of the one  of LEDA14.5.
Even in this case, we conclude that
the incompleteness present in this sample are randomly distributed in the sky,
and the correlation properties are only weakly 
affected by such effects.
\bef %\vspace{} %\vspace{}
\epsfxsize 10cm
\centerline{\epsfbox{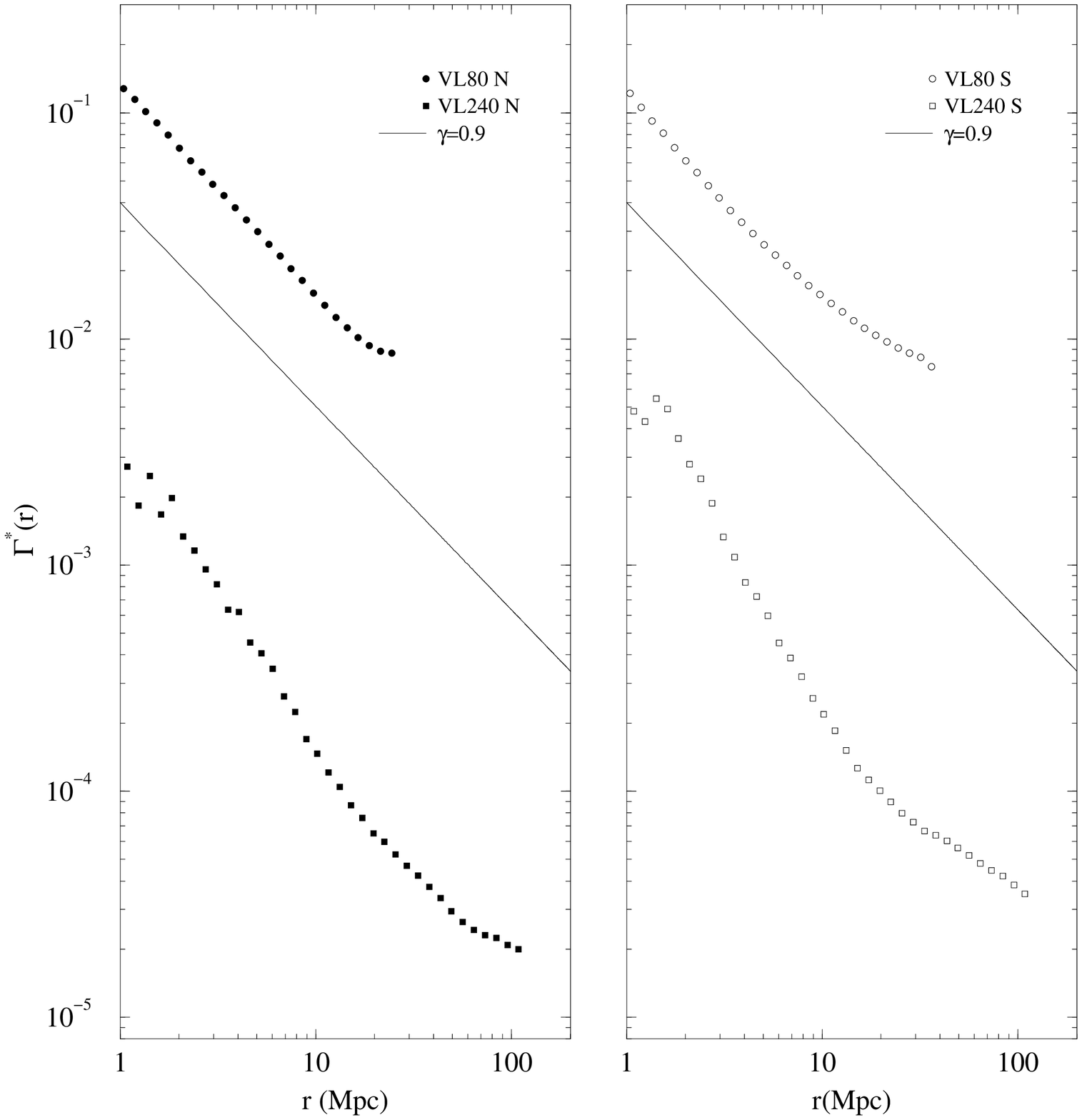}}
\caption{\label{fig26}  The conditional average density $\Gamma^*(r)$
for various VL samples of the LEDA16 sample. $N$ for the VL
 sample of the Northern hemisphere and $S$ for the Southern one.
The reference line has a slope $-\gamma =-0.9$.}
\eef

Finally we consider the VL samples
of LEDA17. For the deepest VL
sample, we find that $\Gamma(r)$
has a power law behavior with
$D = 2.1 \pm 0.2$ 
up to $\sim 150 \hmp$ (Fig.\ref{fig27}).
\bef 
%\vspace{}  
\epsfxsize 10cm
\centerline{\epsfbox{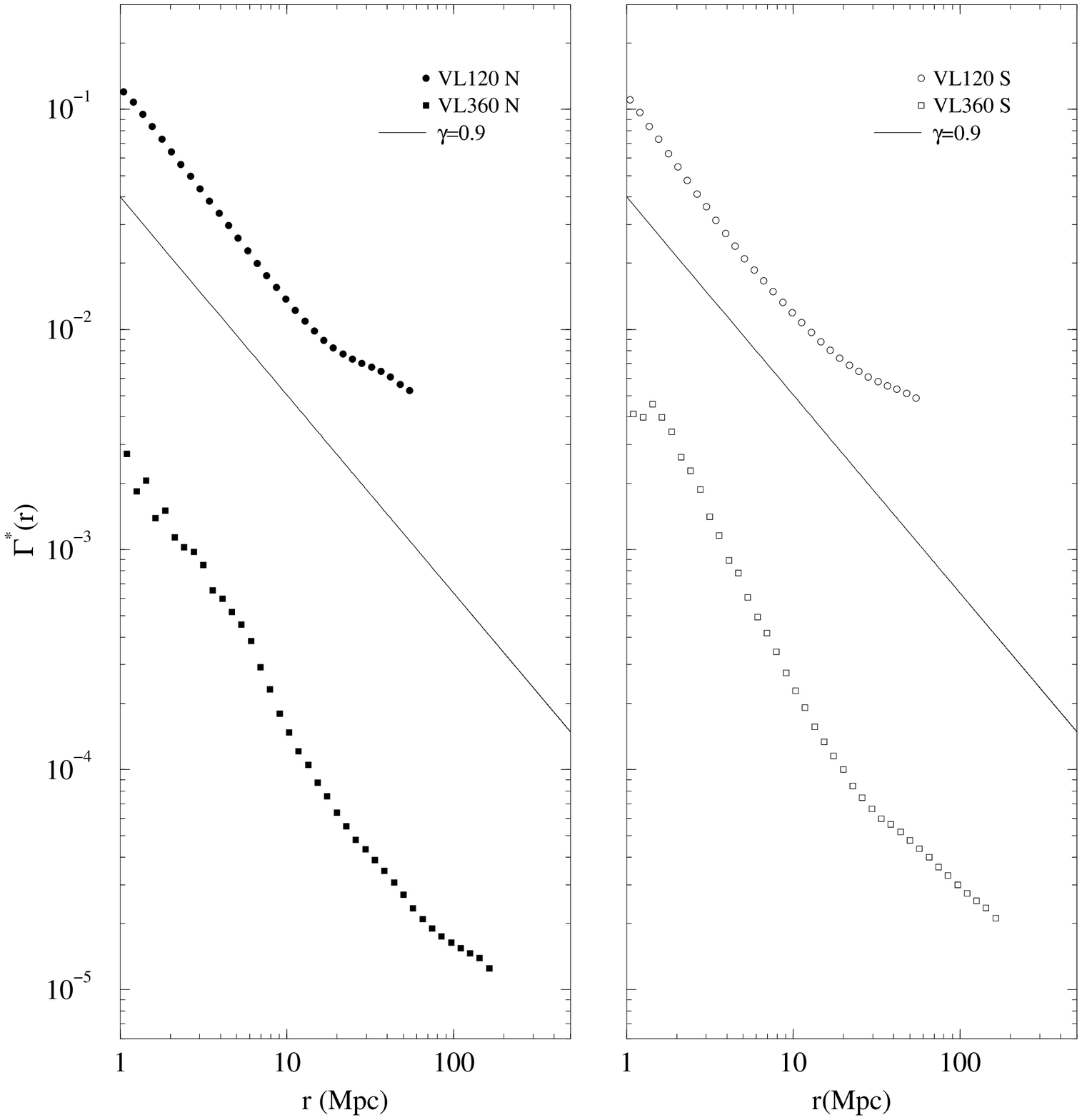}}
\caption {\label{fig27} The conditional average density $\Gamma^*(r)$
for various VL samples of the LEDA17 sample. $N$ for the VL
 sample of the Northern hemisphere and $S$ for the Southern one.
The reference line has a slope $-\gamma =-0.9$.}
\eef
The correlation properties are
substantially the same of the two previous cases. Therefore
we conclude that this analysis show that the power law
correlation cannot be an artifact due to the
incompleteness of the sample.

We have performed another test in order to check the 
effects of the incompleteness.
We have cut a VL sample in two region, for example, one 
up to $ 140 \hmp$ and the other extending from  $140$ to $280 \hmp$. 
We then compute $\Gamma(r)$ in these 
two subsamples, and we find that the
same power law behavior extending on all scales, up to the 
effective depth of the samples (Fig.\ref{fig28}).
\bef %\vspace{} %\vspace{}
\epsfxsize 8cm
\centerline{\epsfbox{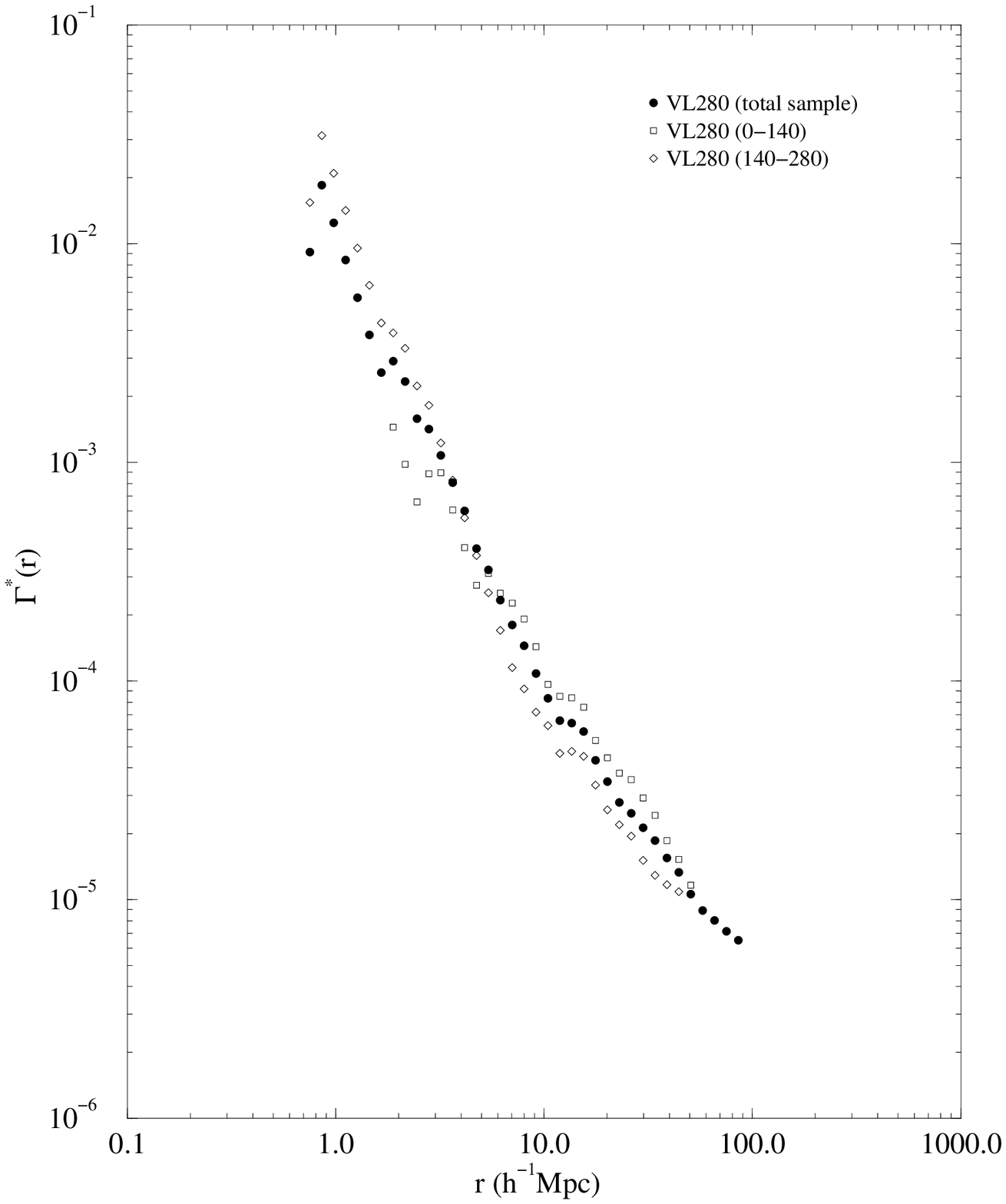}}
\caption {\label{fig28} The conditional average density $\Gamma^*(r)$
for the VL sample VL280 (North) of  LEDA16. 
The filled circles refer to the $\Gamma^*(r)$ of the 
total sample, the squares  for the first half (0-140 $\hmp$)
and the diamonds for the second half (140-280 $\hmp$).}
\eef
This test shows that
the correlation properties of the nearby sample are
the same of the deepest one, and that the signal is stable.
This behavior is exactly what one observes in a fractal 
distribution which shows the same properties 
in the environment of any occupied point.
This is clearly different in the case of 
a radial incompleteness where the nearby points and the 
far away ones have different environments. 
We can conclude again that the incompleteness of the sample 
does not change the genuine behavior of the conditional average density.

We have performed also another test: we have cut one sample of 
LEDA14.5N and of LEDA14.5S at the limiting magnitude $M=-19$.
Then we have measured $\Gamma(r)$: we find the same power law 
behavior in these two 
cases, and also the amplitudes match quite well. Then we have 
repeated the same procedure with LEDA16 and LEDA17, multiplying
the amplitude of $\Gamma(r)$ for $0.5/0.9$ in the first case and
$0.1/0.9$ in the second one. In such a way we have obtained 6 samples,
all with the same magnitude limit $M=-19$: we find that all the amplitudes 
match quite well, confirming again the validity of the catalog.

In order to measure $\xi(r)$, 
we have adopted the same procedure used for the computation of
$\Gamma(r)$. In fact we have firstly used the VL samples
of LEDA14.5-CfA1, to compare the result with
those of CP92. Then we have extended the analysis to the whole
northern and southern hemispheres.
Applying the same steps to LEDA16 and LEDA17, we find
that in the deeper VL sample $r_0$ reached the value of
$45 \pm 5 \hmp$ for a value of $R_{eff}  \approx 150 \hmp$
 scaling as a linear function of the sample
size Fig.\ref{fig29}.
\bef 
%\vspace{}  
\epsfxsize 12cm
\centerline{\epsfbox{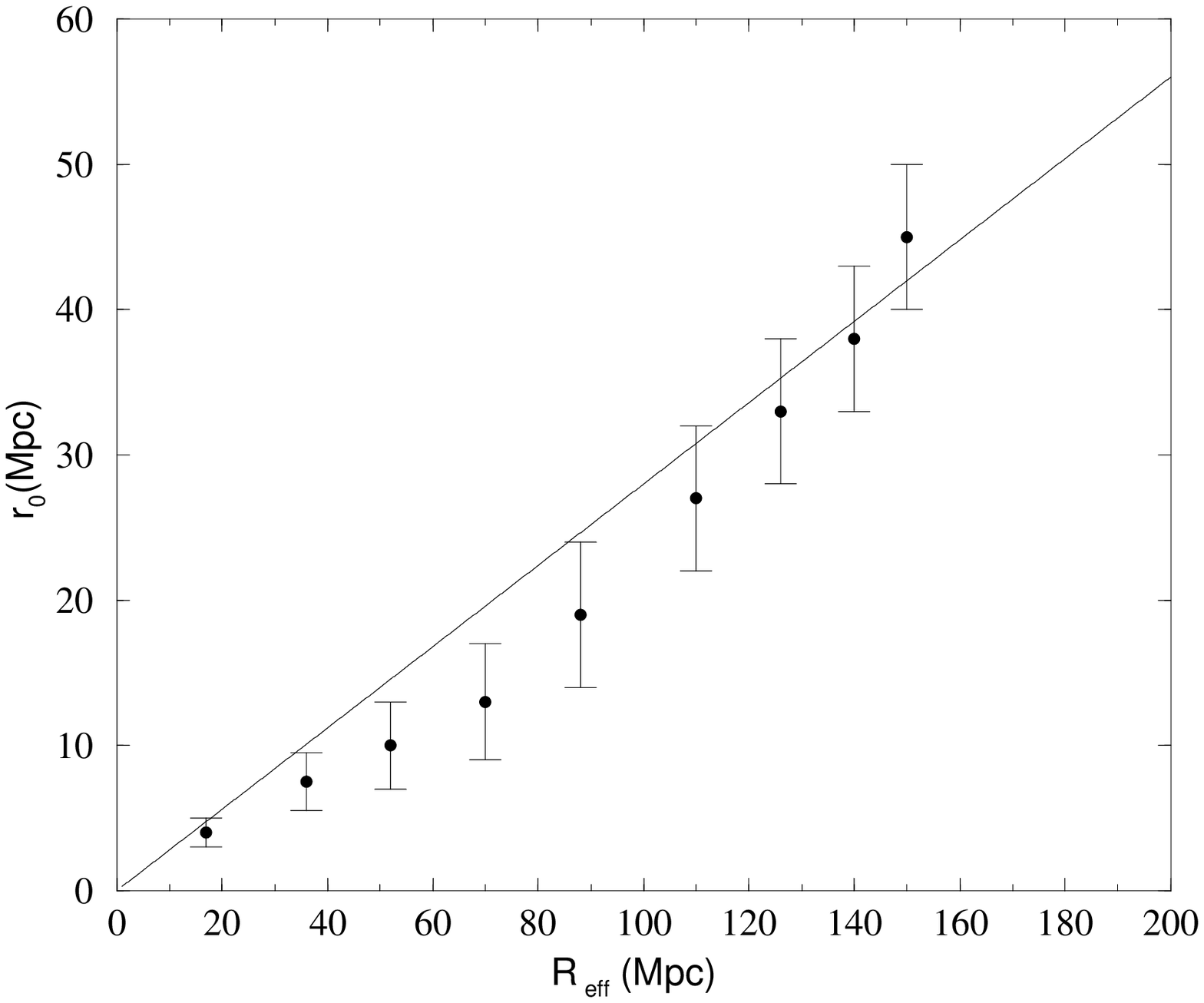}}
\caption{\label{fig29} Behavior of the so-called "correlation 
length" as a function
of the sample size. The reference line has a linear behavior
is in  agreement with the fractal prediction. It is remarkable to note that 
in  the deepest sample 
we find $r_0 \approx 45 \pm 5 \hmp$.}
\eef
This shows  again that the usual $\xi(r)$ is completely misleading
if applied to system characterized by long-range correlations.
The so-called "correlation length" $r_0$ is just an artifact
due to an inconsistent mathematical analysis, and it is only
a fraction of the sample size, so that it has no physical meaning.

\subsubsection{Las Campanas}
\label{gammalcrs}

Las Campanas redshift survey (LCRS) is one of the deeper 
redshift survey available to date. 
This catalog covers a solid angle 
$\Omega = 0.12 \; sr$, and it is made of six $1.5^{\circ} \cdot 80^{\circ}$ 
slices, 3 in the north
galactic hemisphere and 3 
in the south. This survey contains more than
$25000$ redshifts. 
However, there are several selection effects that must be 
considered with great detail. In fact,
  in any slice are mixed measurements
performed with different instruments, which have different 
luminosity selections.
Each slice is made of
 a number of different fields $1.5^{\circ} \cdot 1.5^{\circ}$. 
To each field it is associated a fraction $f$ that gives 
the ratio between the number of galaxies with measured redshift
to the total number of galaxies in the field.
In particular the most of the galaxy measurements are 
performed with  apparent magnitude limit in the range 
$15 \le R \le 17.7$, while 
for some fields 
it has been used the selection $15.0 \le R 
\le 17.3$. Hence
in each slice the fraction of galaxies measured is a  function of 
the angular coordinates and apparent magnitude. 
We refer to \cite{sch96} for a detailed
discussion of the survey construction. 
In Fig.\ref{fig30} we show the absolute magnitude versus distance diagram
for the six slices. It appears the double cut in apparent magnitude.
\bef 
%\vspace{}
\epsfxsize 12cm
\centerline{\epsfbox{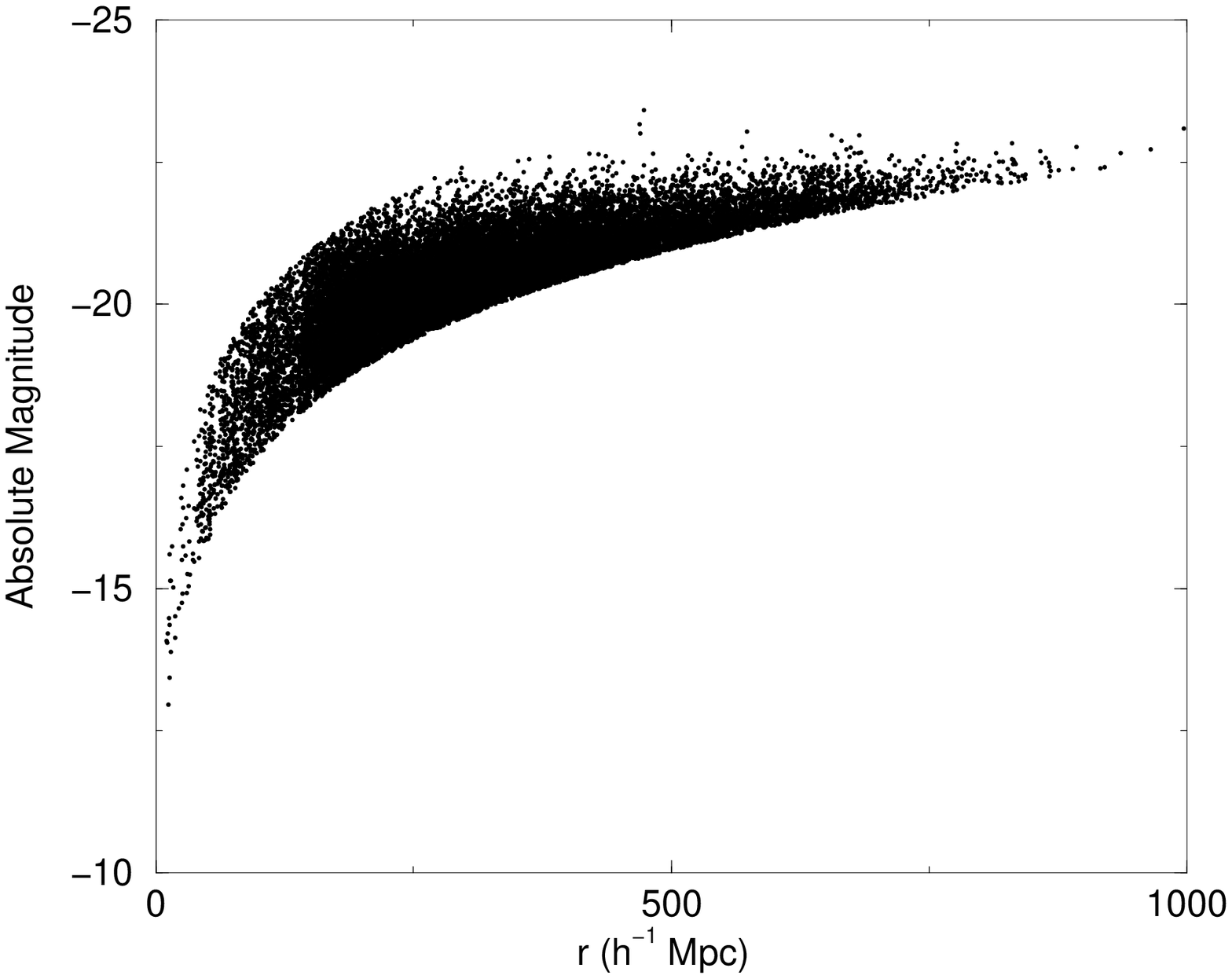}}
\caption{\label{fig30} The absolute magnitude versus distance diagram
for the six slices of LCRS. It is evident the double cut in apparent magnitude.}
\eef
We have computed the conditional average density in each of the six slice
of LCRS.
In order to avoid the various selection effects, and to built an uniform 
sample, we have firstly analyzed the slice at $\delta = -12^{\circ}$ 
in the North galactic cap, that has been obtained with a more 
uniform selection (i.e. it has been done with the 112-fiber spectrograph).
In such a slice the limits on the magnitude are 
$15 \le R \le 17.7$. Then, in order to obtain the same galaxy sampling fraction
among the various 
fields that constitute this slice, we have randomly eliminated
galaxies from the different fields. In such a way we have obtained 
a uniform sample with sampling fraction $f \sim 0.5$.

We have then obtained some VL 
 samples which are listed in the 
Appendix. For the construction of these samples we have an additional problem
because galaxies are selected in a certain range of apparent magnitudes
as in the case of APM.
Hence a VL sample must be
 delimited by two cuts in distance and two in absolute magnitude.
The behavior of the conditional average density is shown in Fig.\ref{fig31}.
\bef 
%\vspace{}
\epsfxsize 10cm
\centerline{\epsfbox{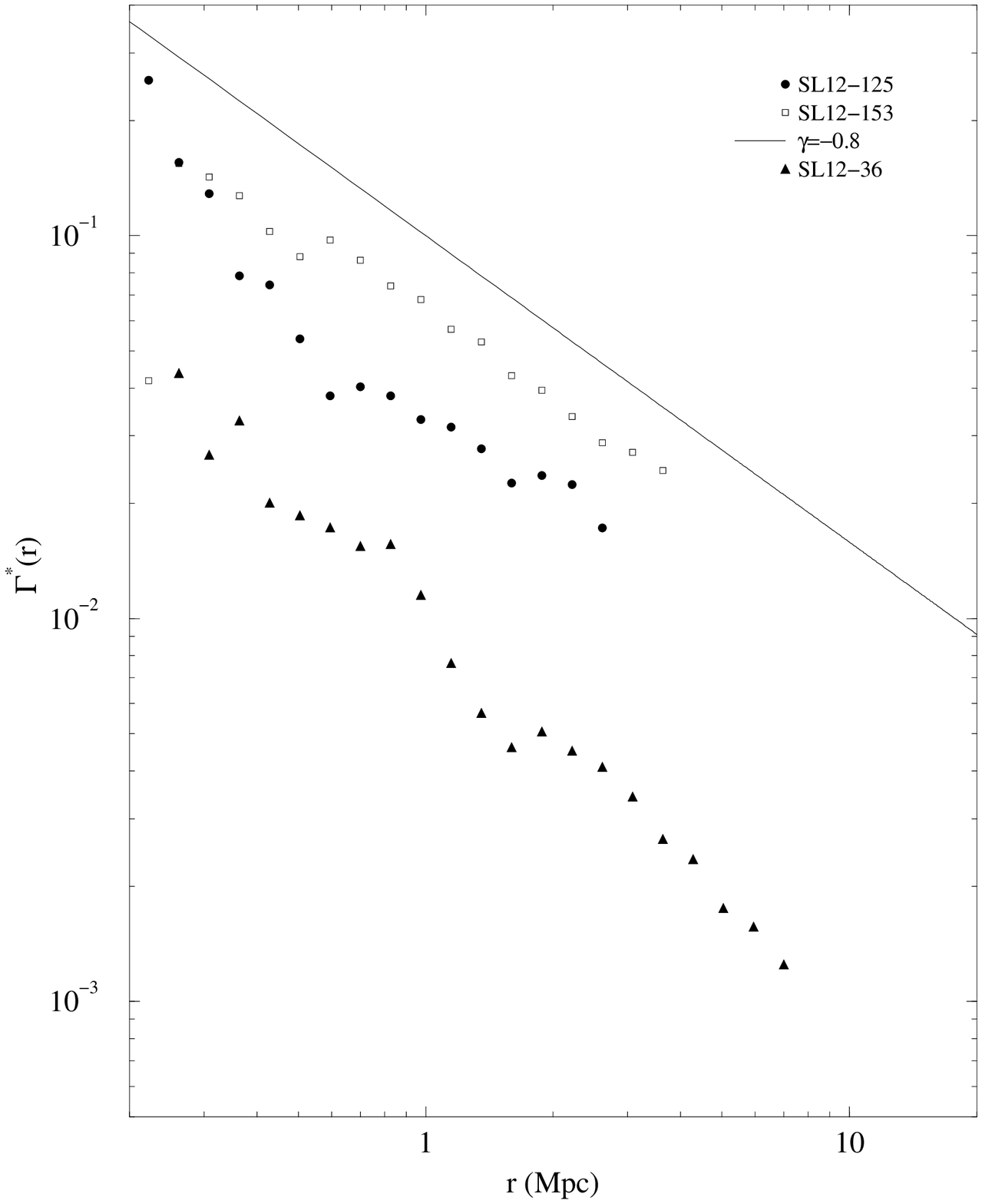}}
\caption{\label{fig31} The conditional average density 
computed in several VL samples of the slice centered at 
$-12^{\circ}$. The reference line has a slope $-\gamma=-0.8$.
}
\eef
A
clear power law behavior is obtained in the range of scales
$0.5 \div 15 \hmp$. The radius of the maximum sphere fully 
contained in the sample volume is in this case $R_{eff} \sim 10 \hmp$.
The fractal dimension turns out to be $D =2.2 \pm 0.2$.
As the radius of the maximum sphere fully included in the sample
volume is $R_{eff} \approx 10 \hmp$, the corresponding 
value for the "correlation length" is $r_0 \approx 4 \hmp$.
We have performed the same analysis for the other five slices
finding similar results, even if the signal is more noisy due 
to the different sampling fraction used in the 
construction of the various
slices.

On the basis of the discussion presented in Sec.\ref{weight}.,
it is clear that
the determination of the correlation function ($\Gamma(r)$ or $\xi(r)$)
at scale $\sim 50 \div 100 \hmp$ or more is {\it completely} affected by 
the use of weighting schemes in treatment of boundary conditions,
and does not contain {\it any}
information about galaxy distribution in this sample \cite{tur96}.

\subsubsection{ESP}
\label{gammaesp}

The ESP\footnote{Our analysis 
of the ESP survey is based on preliminary
data kindly provided by the ESP collaboration. In particular we 
warmly thank the P.I., 
G. Vettolani, for his kindness.}
survey consists of a strip $22^{\circ} \cdot 1.5^{\circ}$ plus a
 near-by area $\:5^{\circ} \cdot 1.5^{\circ}$
(five degrees west of the strip)
in the South Galactic Pole region \cite{vet94,zuc97}.
The limiting magnitude of the survey is $\:b_{J} \le 19.4$ and the
total number of objects is $\:3601$.
 The survey area has been filled with a regular
grid of circular fields with a diameter of
$\: 32$ {\em  arcminutes}:
 this size corresponds to the field of view of the multifiber
spectrograph at the ESO
$\: 3.6 m$ telescope. The volume of the survey is very unusual:
it does not cover a compact solid angle in the sky but,
since it is made by a regular grid of circular fields, 
 it has many
"holes" between a field and the following one.
The total solid angle is $\Omega \approx  0.006 sr$.

We have computed the conditional average density 
in various VL samples of ESP, limited at
the strip $22^{\circ} \cdot 1.5^{\circ}$  (see Appendix).
  The  radius of the maximum sphere  fully enclosed in the 
deepest VL sample 
is $R_{eff} \approx 10 \hmp$, so that it is possible to
compute $\Gamma(r)$ only up to such distance. There is an additional problem
in the case of ESP which requires a detailed investigation. In fact,
as we have already 
stressed, the ESP survey is made of various crucial fields, and
there are "holes" between a field and its neighborhood. 
Unfortunately there is no way to correct for such a selection 
effect. In any case   from Fig.\ref{fig32} it follows that 
\bef 
%\vspace{}
\epsfxsize 8cm
\centerline{\epsfbox{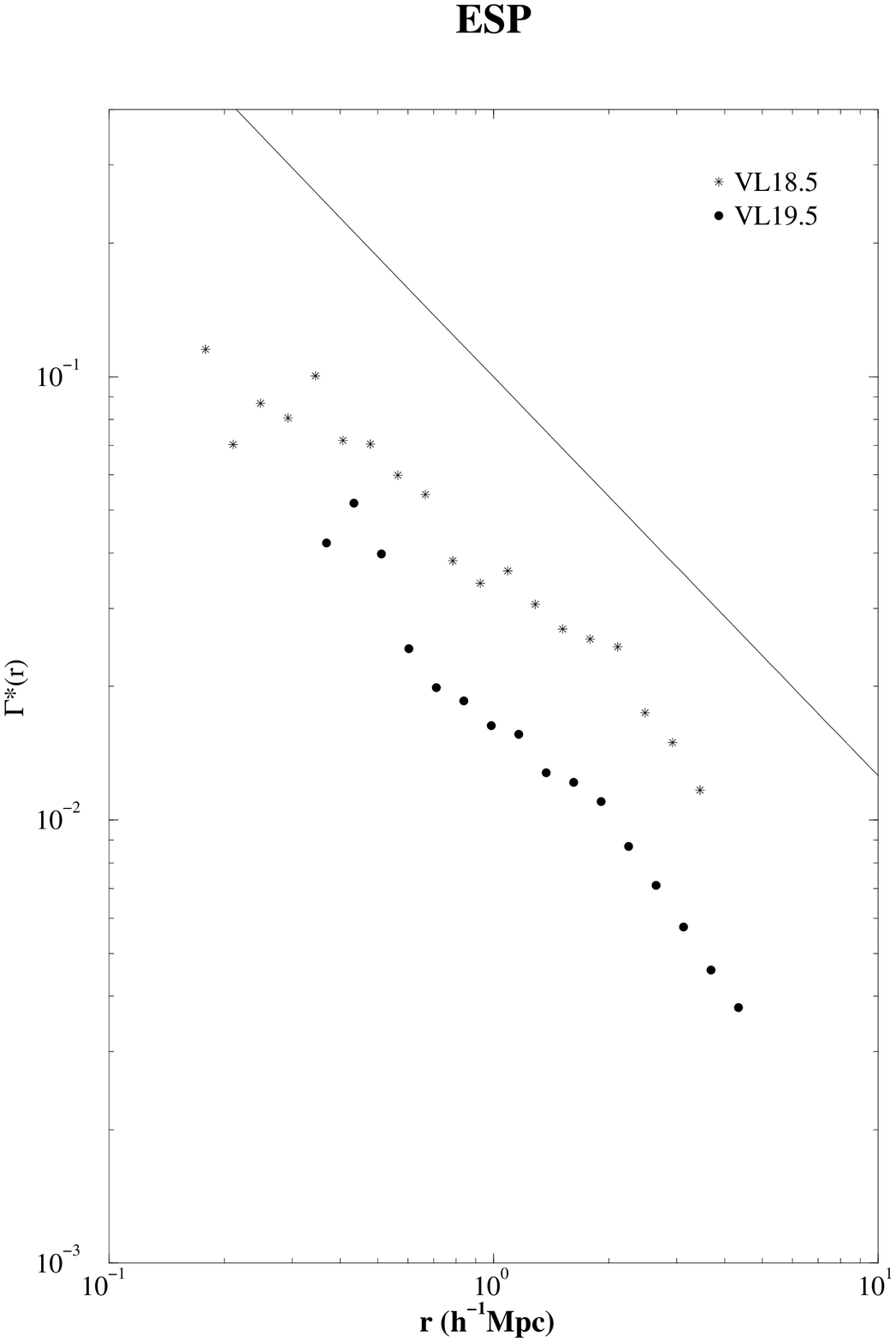}}
\caption{\label{fig32} The conditional average density for various VL
samples of ESP. A clear power law behavior is found up to $\sim 10 \hmp$
with fractal dimension $D =1.8 \pm 0.2 $. The reference line has a slope 
$-\gamma=-1.2$. }
\eef
that a clear power law behavior is found up to $\sim 10 \hmp$         
with $D =1.8 \pm 0.2$. The presence of the holes may
lead to a lower value
 of the fractal dimension in this survey.

\subsubsection{IRAS $2 Jy$ and IRAS $1.2 Jy$ Redshift Surveys}
\label{gammairas}

The IRAS galaxies have been considered to be
the {\it unbiased} tracers
of matter distribution 
\cite{str90,sau91} because the
most infrared active galaxies are the spirals ones which are believed to
be a better representation of the real matter distribution.
The galaxy distribution of  IRAS galaxies
\cite{str92,str96} is dominated by the well-known superclusters:
the Hydra-Centaurus-Pavo-Indus Supercluster
forming together the so-called Great Attractor, and the
Perseus-Pisces Supercluster. There are large voids in front of
Perseus-Pisces Supercluster and off the supergalactic plane.
It is remarkable that  IRAS galaxies {\it do not
fill the voids}, and, on the contrary, they trace the same
structures of the optical
 ones. In particular it has been shown \cite{str92}  very clearly
that the large-scale distribution of IRAS and CfA,
or SSRS, galaxies are very similar, and
belong to the same structures.

The similar space distribution does not correspond
to the same correlation properties of IRAS and
optical samples.  Optical galaxies, in fact,
show long range power-law correlations
up to a certain  distance (see previous sections).
The problem is therefore to understand why   IRAS galaxies, that
are located in the same structures of the optical ones, do not
show the same correlation properties
and, on the contrary, seem to have a very well defined
constant density \cite{str92,fis94} 
with a  value of the so-called "correlation length" of 
$r_0 = 4.5 \hmp$.
In particular, it has been shown that $r_0$ {\it does not} scale with sample
depth \cite{fis94}. This result is quite in contradiction 
with the linear dependence of $r_0$
on  sample depth previously found.

The IRAS 2Jy survey (hereafter I-2)
is a complete sample of galaxies uniformly selected over
most of the sky. We refer the reader to \cite{str90,str92} 
for a detailed discussion of the sample
selection. The sky coverage of the survey is $11.01 sr$.
The selection criteria have been chosen
in relation to the infrared flux, and in
particular there have been selected the galaxies with
apparent flux at $60 \mu m$ ($f_{60}$) greater than
$ 1.936 Jy$. This sample contains $2652$ galaxies.
The IRAS 1.2 Jy survey (hereafter I-12) is a similar catalog, but
it collects all the galaxies with
$f_{60} > 1.2 Jy$, and the number of galaxies is doubled, and contains
 5313 redshifts \cite{str96}.
For the well-calibrated fluxes and the lack of Galactic extinction
in the IRAS wavelengths, these surveys are believed to be
well suited for studies of large scale distribution of galaxies.

We have studied \cite{slgmp96} the space correlation
 properties of the various VL samples (see Appendix)
and in particular we have computed the
conditional average density.
In Fig.\ref{fig33}
there are shown the case of I-2
\bef 
%\vspace{}  
\epsfxsize 8cm
\centerline{\epsfbox{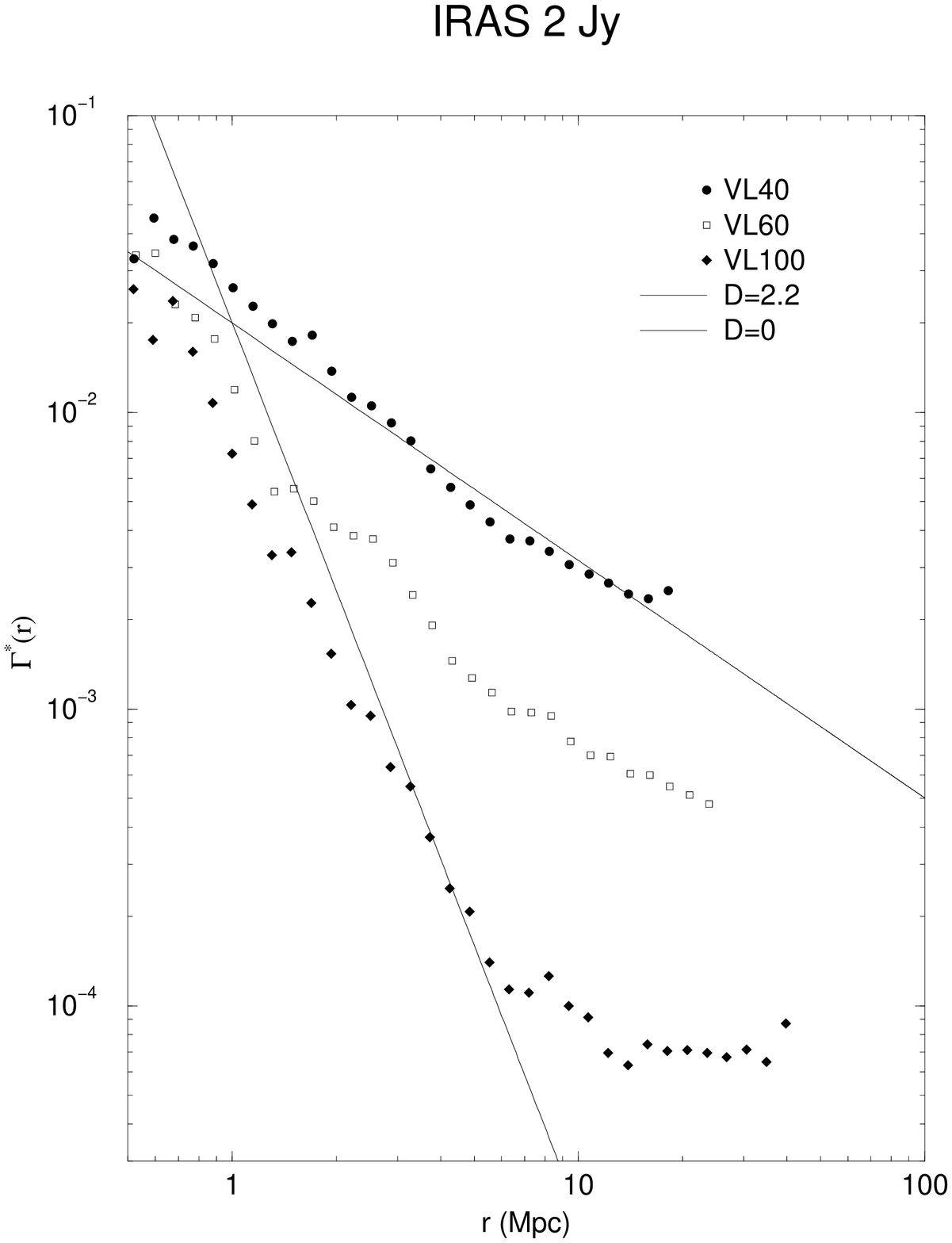}}
\caption{\label{fig33}   The conditional average 
density $\Gamma^*(r)$ computed in various
VL samples of IRAS $2 Jy$ North catalog. The
 reference line has a slope
of $\gamma=3-D=0.8$ (i.e. $D =2.2$). 
The VL at $60 \hmp$ shows power law correlation up to
$25 \hmp$. The VL at $100 \hmp$ shows 
a $1/r^3$ decay at small scales, up to $\sim 6 \hmp$, 
followed by a flat
behavior at larger scales.
}
\eef
respectively, while in Fig.\ref{fig34}
there are the correspondent behaviors for I-12S 
and I-12N.
\bef 
%\vspace{} 
\epsfxsize 8cm
\centerline{\epsfbox{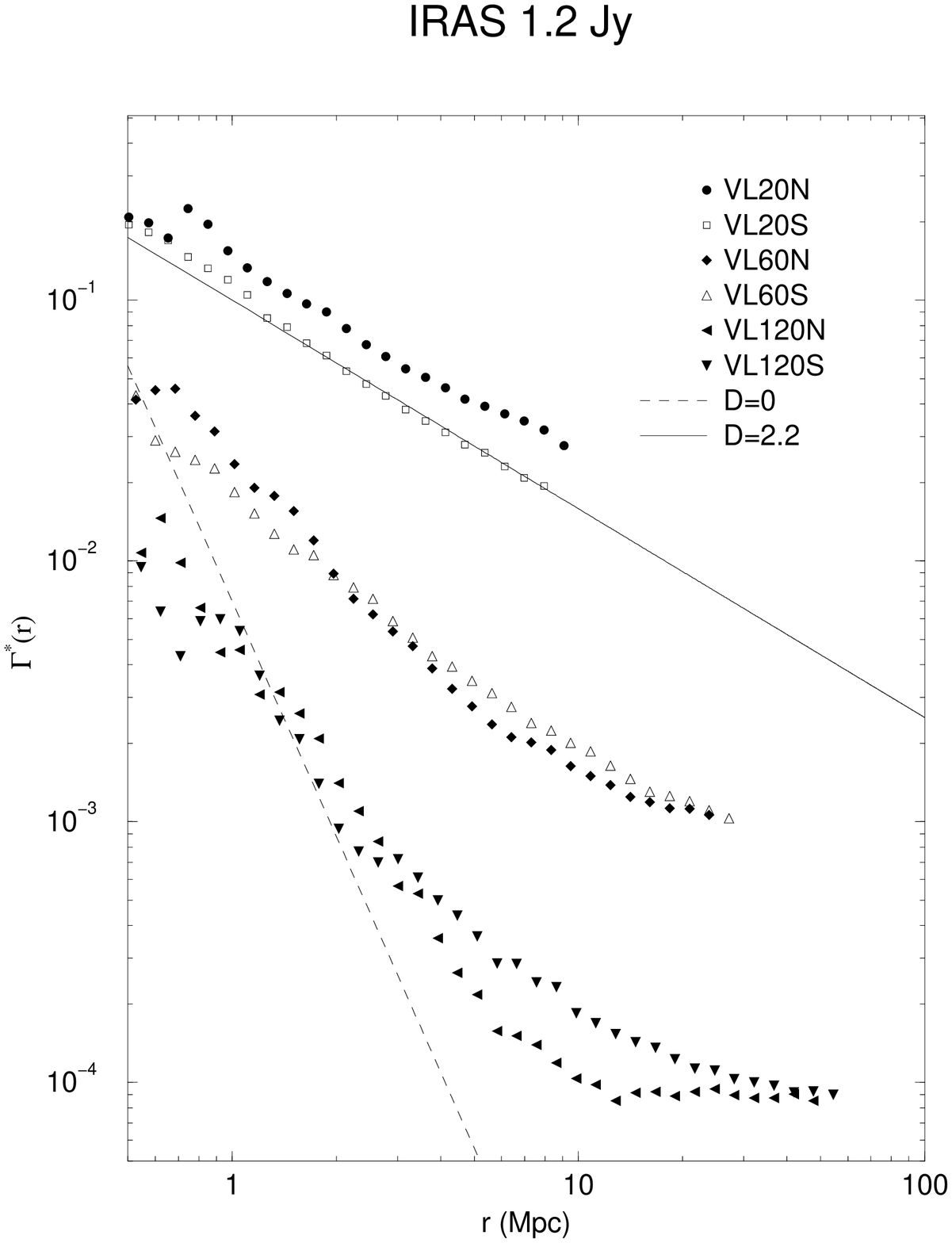}}
\caption{\label{fig34} The same of the previous figure
 but for the VL samples 
of IRAS $1.2 Jy$
North  catalog (N) and of the southern one (S).
The solid line has a slope  corresponding to $D=2.2$.
In this case the sample VL60 shows
power law correlation up to
$\sim 30 \hmp$. The samples at $120 \hmp$ show 
a $1/r^3$ (dashed line) 
decay at small scales, up to $\sim 7 \div 8 \hmp$, 
followed by a flattening.
}
\eef
In the VL at  $60 \hmp$
of I-2N (or I-2S) $\Gamma(r)$ shows a well defined power law
up to $\sim 25 \hmp$ with exponent $\gamma \approx 0.8$,
%that corresponds to a fractal dimension
i.e. $D \approx 2.2$.
 The VL at $100 \hmp$ shows
a $1/r^3$ decay at small scales, up to $\sim 6 \hmp$,
followed by a flat
behavior at larger scales.

 In the VL samples of I-12S and I-12N
  the same power law ($D=2.2. \pm 0.2$) is
shown up to $\sim 30 \hmp$. The deepest samples VL120 (N and S)
show a $1/r^3$ decay at small scales (up to $\sim 7 \hmp$) 
 followed by a flattening.
The point that we  discuss in what follows
is whether such a crossover towards homogeneity
is a real and genuine property of
galaxy distribution or if it is due to some
systematic effect.
The fact that a constant average density seems to be 
reached, corresponds
in terms of $r_0$ to the fact that it does not scale with sample
size, as it has been obtained by \cite{str92,fis94}.
We refer to Sec.\ref{validity} for a detailed discussion 
of the statistical fairness
of the IRAS data: there we   present an explanation of 
the  IRAS correlations in the deeper samples. 

\bef 
%\vspace{}
\epsfxsize 8cm
\centerline{\epsfbox{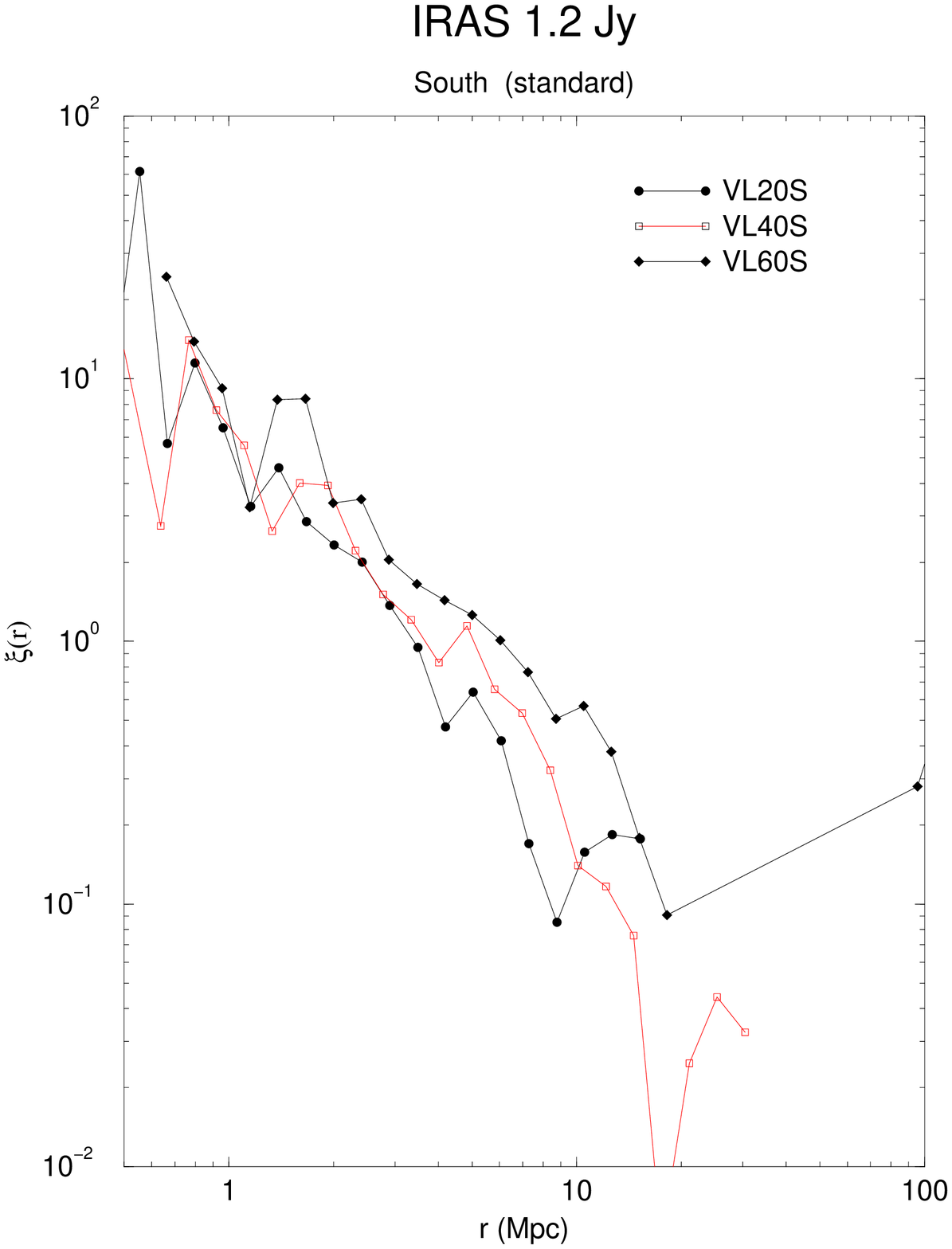}}
\caption{\label{fig35}
The $\xi(r)$ correlation function
computed in the VL samples 
of IRAS $1.2 Jy$ south (S).
The amplitude of $\xi(r)$ shows a dependence
on the sample size, even if the signal is rather noisy.}
\eef
\bef
%\vspace{}
\epsfxsize 8cm
\centerline{\epsfbox{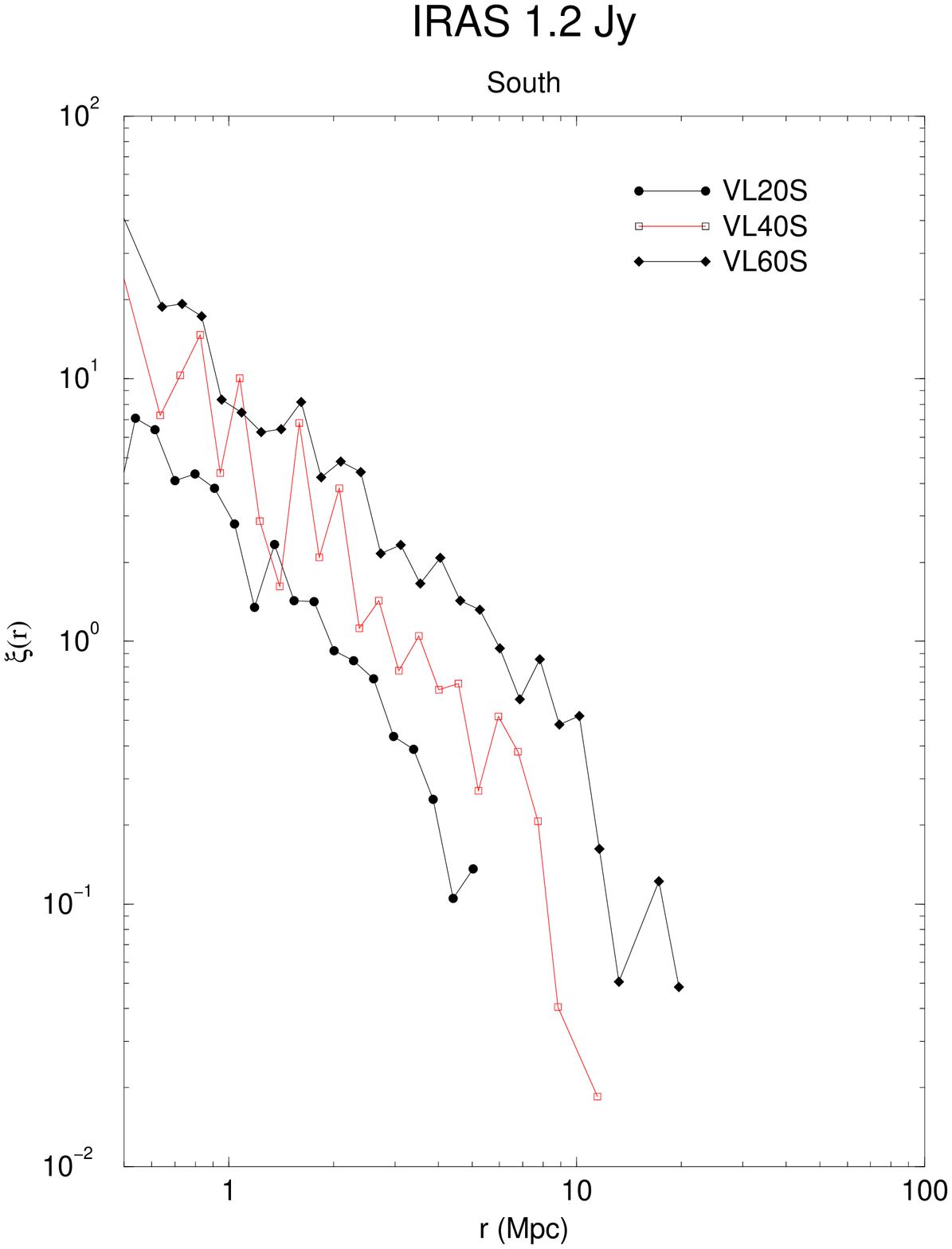}}
\caption{\label{fig36}
The $\xi(r)$ correlation function
computed from the $\Gamma(r)$ function 
($\xi(r)=\Gamma(r)/<n>-1$)
in the VL samples 
of IRAS $1.2 Jy$ south (S).
The amplitude of $\xi(r)$ shows a clear dependence
on the sample size.
}
\eef
In Fig.\ref{fig35} we show the behavior of $\xi(r)$ computed 
with the standard method, while in Fig.\ref{fig36} we show 
the $\xi(r)$ computed in different VL samples from the 
$\Gamma(r)$ function ($\xi(r)= \Gamma(r)/<n> -1$). While in the
letter case   a clear dependence of $r_0$ on the sample size is shown 
($r_0 \approx 8 \hmp$ in the case of VL60). In the former case this dependence,
although still present, is more noisy. This is probably due to the 
treatment of the boundary conditions (Sec.\ref{angdist}.).

\subsection{Analysis of clusters  catalogs}
\label{gammaclu}

With respect to the galaxy catalogs, cluster
surveys offer the possibility to study the large
scale structure of   matter distribution in volumes
much larger, reaching depths
beyond $z\approx 0.2$
 and extending all over the sky.
Using clusters, one can trace matter distribution
with a lower number of objects with respect to
the galaxies,
in the same volume.
For example, in the northern hemisphere we know
$\approx 10^{6}$ galaxies up to $z\approx 0.15$
which correspond to $\approx 500$ rich clusters.
However the problem of cluster catalogs
is their incompleteness, since clusters
are identified
as density enhancements in galaxy {\it angular} surveys and
their distance is usually determined through the
redshift of one or two galaxy members.
It is clear, in fact, that only the measurements of every
clusters (and member galaxies)  allow  us to construct
truly VL samples.
Moreover, as we show in what follows,
there is a strong arbitrariness in the
definition of a {\it cluster}: such an arbitrariness
can be avoided by  using the methods of modern statistical mechanics.

Cluster distribution
shows large scale inhomogeneities and huge voids.
Tully
\cite{tu86,tu92} by investigating
the spatial distribution of  Abell and ACO catalog of 
clusters up to $300 h^{-1} Mpc$,
stressed that there are structures on a scale of $0.1c$ lying in
the plane of the Local Supercluster. Galaxies clump 
in clusters and clusters in superclusters
with extensions comparable to the largest
scales of current samples. An analogous study on
cluster distribution has been performed by
Einasto \etal \cite{ei94}.
Their catalog of superclusters
shows the presence of correlated structures
with extension ranging from few $ Mpc$
to $150  h^{-1} Mpc$.
Very similar results have been obtained by
other  compilations of
Supercluster catalogs \cite{bs83,bb85,pos92,zuc93}.

Natural complementary aspect of presence
of large structures is the presence of
voids. Several authors \cite{bs83,bb85,kf91,ei94}
have
investigated the shape and the dimension of voids
determined by rich clusters and
superclusters. The dimension of a void 
can be  defined
as diameter of empty sphere containing
no clusters.
Clearly also for the
voids there is an arbitrariness, because one can consider only
spherical voids or ellipsoidal voids and so on.
In particular, Einasto \etal \cite{ei94} found that in their
cluster sample (up to $ z\approx 0.1$), the mean radius of
 voids is $50 h^{-1} Mpc$. However, naturally,
voids have elongated
shapes and in one dimension they can exceed this value.
 These authors observe in fact a giant void which can
exceed $300 h^{-1} Mpc$ of length.
A more detailed study
on a single void, is performed by
Lindner \etal \cite{li95}, 
which have investigated the
distribution of galaxies in and around the
closest one, the Northern
Local Void.
The authors found that the dimension of the
voids depends on the objects which have been
used to defined them: voids defined
by clusters are larger than galaxy defined voids, which
are larger than faint galaxy defined voids.
Search of extremely faint dwarf
galaxies in voids has given, up to now,
negative results \cite{li95} 
so that voids {\it are real}
and not due to observational incompleteness.
Moreover one observes an increased 
void size with sample depth, that
may correspond to a
self similarity in the
distribution of voids \cite{elad97}.

As for galaxy distribution, 
the observations of  large scale structures like
superclusters and voids, which extends up to the
limits of the samples investigated, clearly
make questionable  the existence of
the average density.
Usually, the presence of inhomogeneities
on the scale of sample analyzed is interpreted
as incompleteness of the samples themselves.
However, up to now,
 better and more
extensive observations
have confirmed the reality
of these structures,
making the observed agglomeration
sharper and not filling the voids.

For cluster
distributions, the
$\xi(r)$ function is
a power law (Sec.\ref{statmec})
$\xi(r) \approx (r_{0}/r)^{\gamma}
$;
both the exponent and the amplitude $r_{0}$
are uncertain. The exponent $\gamma$
generally appears consistent with the value
$1.8$ observed for galaxy correlation function.
In general high values for $r_{0}$
($20 \div 25 h^{-1} Mpc$)
are obtained
for samples which contain  the richer
Abell/ACO clusters
\cite{bs83,kl83,bah88,hu90,ca92,pos92}
but some authors 
claim that these high values
are overestimates, produced by
systematic biases present
in the Abell/ACO catalog
\cite{su88,su91,ef92,dek89,va94}.
When {\it one corrects} for these biases, a lower
$r_{0} \approx 14 h^{-1} Mpc$ is
obtained.
Lower values of $r_{0}$ ($13 \div 15 h^{-1} Mpc$) 
are obtained also from
the analysis of automated cluster catalogs
(APM, EDCC) and from the
 cluster
catalogs selected from $X-ray$  galaxy survey,
\cite{lu92,gu92,da92,ma90a,ma90b,dab94,ro94,bri94}.
In conclusion, for the samples analyzed so far we find that
$14 h^{-1} Mpc \ltapprox r_{0} \ltapprox 25 h^{-1} Mpc$.

Clusters are then found to be
more aggregate than galaxies,
for which  $r_{0}$ is $\approx
5 h^{-1} Mpc$. 
This {\it  mismatch}
between galaxy and cluster correlations is
another puzzling feature of the usual analysis; clusters,
in fact, are made by galaxies and
many of these are included in the galaxy catalogs
for which the correlation length $ \approx 5 h^{-1}\: Mpc $
was derived.
It is therefore necessary to assume that cluster galaxies
have fundamental differences with respect to the galaxies 
not belonging to a 
cluster. This concept has given rise to the so-called
{\it richness clustering relation}
\cite{bah92,bs83,bah88,bah86}
according to which,
objects with different mass or morphology
would segregate from each other and give rise
to different correlation properties,
i.e. $r_{0i} \approx 0.4<n>^{-1/3}_{i}$
where the index $i$ refers to the system being
considered, and $<n>_{i}$ is its mean spatial density.
This hypothesis has been applied to explain the increase
of $r_{0}$ going from APM and $X-ray$ less rich
clusters to the Abell more rich clusters.
As pointed out from
Coleman \& Pietronero \cite{cp92}
this preposition leads eventually to the paradox that
every object is slightly different from any other and
form, by its own, a morphological class, making totally
meaningless the concept of correlation between galaxies.
Szalay and Schramm 
\cite{sz85}
 have
interpreted the scaling of  $r_{0}$  as signature of
scale invariance properties of the
distribution. This interpretation is
inconsistent, since for
self similar distributions, $r_{0}$
 is proportional to $R_{eff}$,
where $R_{eff}$ is the {\it effective} sample
radius,  and not to
$ <n> ^{-1 /3}$ (Sec.\ref{statmec}).

In the following  we present results of the statistical analysis
on various samples of Abell and ACO clusters \cite{msla97}
\footnote{We thank A. Amici for useful 
collaborations and discussions in the analysis of cluster catalogs.}
The study of the spatial distribution
of clusters requires a complete clusters sample 
 covering a large volume.
There are many available extensive
catalogs of clusters from which one may try
to extract complete subsamples.
Here we present the analyses of  several subsamples
extracted from Abell  and ACO catalogs.

 From Abell catalog we have analyzed
the
Postman sample  
\cite{pos92} that consists of 351 clusters with
the tenth ranked galaxy magnitude ($m_{10}$)$\le 16.5$, and
it  includes
all such clusters which lie north of
$\delta=-27^{\circ} 30'$. The typical
redshift of a cluster with $m_{10}\le 16.5$
is $z\approx 0.09$. To this redshift, incompleteness
in the Abell catalog is considered insignificant
\cite{pos92}.
About half of clusters have
the redshifts based on at least
three independent galaxy spectra;
$25 \% $ of the clusters
have redshift determined from a single spectrum
and $25 \%$ from two galaxy spectra.  
Postman \etal \cite{pos92}
have defined a so called
statistical subsample, which consist of 208
clusters with $z\le 0.08$ and $|b|\ge 30^{\circ}$.
This subsample has been shown to be minimally
affected by selection biases. Then we have considered also the 
 Bachall \& Soneira \cite{bs83} sample
 (BS83 hereafter) that
includes 104 clusters of distance
class $D\le4$ ($z\ltapprox 1$), richness class
$R\ge 1$, located at high
galactic latitude ($|b|\ge 30^{\circ}$)

  From the ACO catalog \cite{ab89} we have
   selected all the clusters with measured redshift
in according to the following constraints:  
$m_{10} \le 16.4, b \le - 20^{ \circ}, \delta \le
-17^{\circ}$.  
 This sample contains 139
clusters up to the limiting distance of $\approx 930 h^{-1} Mpc$.

The completeness of the samples
is usually
estimated from the behavior of the 
  clusters density as function 
redshift.
As shown in  Fig.\ref{fig37} 
\bef 
%\vspace{} 
\epsfxsize 12cm
\centerline{\epsfbox{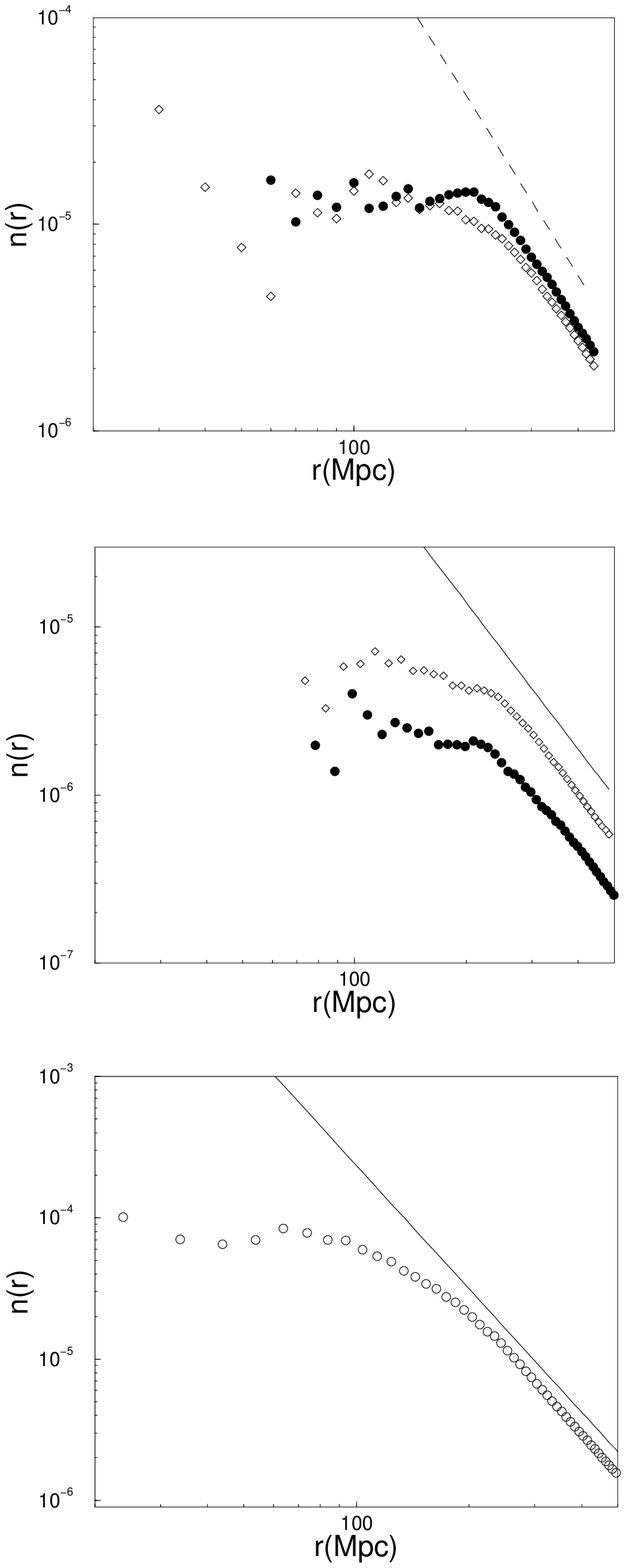}}
\caption{
\label{fig37}  The radial density $n(r)$
from the vertex for
a) Abell catalog Postman sample,
north (diamonds) and south (filled circles)
b) Abell catalog BS83 sample north (diamonds)  and south (filled
circles)
c) ACO catalog sample (empty circles).
the radial density
shows large fluctuations, followed by a $r^{-3}$ decay due to the
incompleteness of the samples.}
\eef
the density, in almost every sample,
presents large fluctuations followed by
a power law decay as  $ \sim r^{-3}$.
At closer distances the density is fluctuating
because of the weak statistics in the samples,  
while at larger  distances the sample is incomplete, i.e.
the number of clusters is almost constant and
the density decrease as $V^{-1}\sim r^{-3}$.

If one compute the radial density in bins, one obtains
larger fluctuations, because this is
a differential
quantity while the integral  density
is much more smooth.
Usually the
fluctuating behavior of the density up to
$z\ltapprox0.1$ is interpreted as
a flat one, i.e. the density of the sample is
considered to be constant and the distribution
homogeneous; consequently the sample
is considered complete and homogeneous
up to this distance and incomplete
beyond.
At this regard, we note that up to the
beginning of the incompleteness
region, all the samples contain few clusters.
The sparsest is the BS83 sample: the
northern galactic part contains 53
clusters up to $230 h^{-1} Mpc$
and the
southern part 24.
The Postman sample
contains 120 clusters up to
$z\approx 0.08$ in the northern part
and 88 in the southern one.
The ACO sample 91 up to the $150 h^{-1} Mpc$.

The study  of cluster spatial   properties
requires an almost complete sample; then it is 
necessary to exclude the
incompleteness region. In the standard approach,
one way to avoid such a problem is to
correct the incompleteness 
by assuming an homogeneous distribution of clusters
up to the sample limits. The observed distribution in the
redshift space
is then weighted with a selection function $p(z)$, that
is the ratio between the observed counts of clusters
in the volume $dV(z)$ at redshift $z$ and those
expected from an homogeneous distribution.
In our analysis, 
{\it we want to avoid any assumption on the distribution itself}
and for this reason
we   limit our analysis up to a depth corresponding
to the beginning of the incompleteness region,
without correcting by means of any selection function.

Another selection effect exists, and it regards
the observed depletion of the surface mean density
of clusters at low galactic latitude (to $|b| \ltapprox 30^{\circ}$).
This is probably due to
obscuration and confusion with high-density regions of stars
of our galaxy \cite{bah88}.
 %(Bachall 1988).
As in the case of redshift incompleteness,
one way to overcome this
incompleteness is to weight the observed distribution with
a latitude selection function
$P(b)$, that is the ratio between
observed surface cluster density at latitude $b$ and the
expected one from an homogeneous distribution.
The normalization of this selection function
is arbitrary,
because it depends on the real density.
Another way is to limit the sample to high galactic latitude region.
We   adopt this standard procedure
(i.e. $|b|\ge 30^{\circ}$ for Abell catalog and $b < -20^{\circ}$
for ACO), that has no assumptions
but  the inconvenient to slightly limit
the sample.
For the Postman sample the limiting 
distance is $\approx 230 h^{-1} Mpc$
corresponding to $z\le 0.08$,
for the BS83 sample we have the same
limiting distance, while for ACO sample
is $\approx 150 h^{-1} Mpc$.
All these limits have been estimated from the
behavior of the radial  density versus the distance.

We have studied the behavior of $\:\Gamma(r)$ and $\:\Gamma^*(r)$
in the samples of Tab.\ref{tabclu}.
The results for the Postman statistical 
subsample is shown in Fig.\ref{fig38}
\bef 
%\vspace 
\epsfxsize 12cm
\centerline{\epsfbox{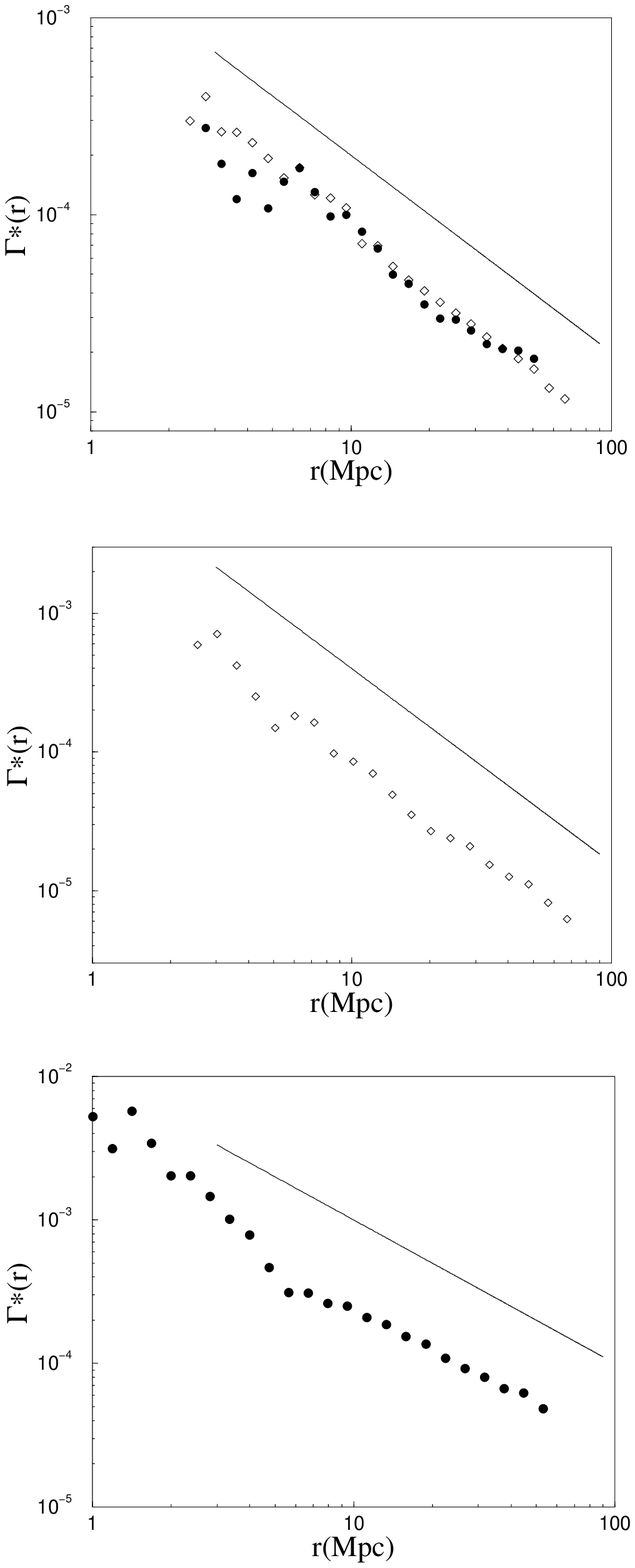}}
\caption{\label{fig38}  
The behavior of  the conditional density $\Gamma(r)$
and average conditional density
$\Gamma^{*}(r)$ for
{\it a)} Abell catalog Postman sample
respectively north (diamonds) and south galactic parts (filled
circles)
{\it b)} Abell catalog BS83 sample north galactic part (diamonds),
{\it c)} ACO catalog sample (filled circles).
The conditional density has a power law behavior  up to the radius
of the largest sphere fully included in the sample.
The reference line has slope $-\gamma = -1.0$.
}
\eef
separately for the southern galactic and northern galactic part.
The southern galactic part has a smaller solid angle
with respect the northern one and hence
a smaller $R_{eff}$.
A well defined power law behavior is detected up
to the sample limit without any tendency towards homogenization.
The codimension is, with good accuracy
$\gamma = 3-D \approx 1.0 \pm 0.2$
so that $\:D \approx 2.0 \pm 0.2$ up
$\approx 70 h^{-1} Mpc$ for
the northern part and up to $\approx 50 h^{-1} Mpc$ for the southern one.
This result agrees well with the 
BS83 sample  for the northern galactic
 part,  where $D\approx 1.7 \pm 0.2$
up to $r\approx 70 h^{-1} Mpc$ (Fig.\ref{fig38}).
We have not reported the analysis
for the BS83 southern galactic part, because
it gives a very noisy result, due to the
very weak statistics of the sample.
The weaker statistics of the BS83 sample
with respect to Postman's one is also the
reason for the small difference
in the estimate of the fractal dimension
$D$ of the two samples.
The exponent 
$\gamma$ has a lower value with respect to the
standard determination, and this is due to the fact
that usually $\gamma$ has been estimated by the $\xi(r)$ analysis 
that has the problems discussed in Sec.\ref{statmec}.

The results from the ACO sample are shown
in the Fig.\ref{fig38}.
The $\Gamma(r)$ is a power law with exponent
$D \sim  2.0 \pm 0.1$ up to $\approx 50 h^{-1} Mpc$.
The fluctuations at $r < 10 h^{-1} Mpc$
are due to the fact that, at these distances,
we are below the minimum 
average distance
between nearest clusters in the sample 
($\approx 15 \hmp$ for Postman and BS83 and 
$\approx 8 \hmp$ for ACO sample), as 
we have discussed in Sec.\ref{angdist}.
Hence we can interpret these fluctuations
as a finite size effect
due to a systematic depletion of points at these distances.
In the correct regime, at distances $>10 h^{-1} Mpc$, where the samples
become statistically
robust, the slope of $\Gamma(r)$ is almost the same for the
Postman  and ACO samples and, with a slight
difference because the poor statistics, for the BS83 sample.

All the samples investigated so far have consistent statistical
properties, i.e. they show a clear
behavior of the correlation function. Note that
long range correlations ({\it fractal})
can be only destroyed by incompleteness,
but not produced by it.
Hence the samples show well defined statistical properties, i.e.
they are {\it statistically fair} samples;
the results of the analysis is that
they are not homogeneous samples, but, on the contrary, fractal.

We have studied the $\:\xi(r)$ in our cluster
 samples. In Fig.\ref{fig39} 
\bef 
%\vspace{} 
\epsfxsize 12cm
\centerline{\epsfbox{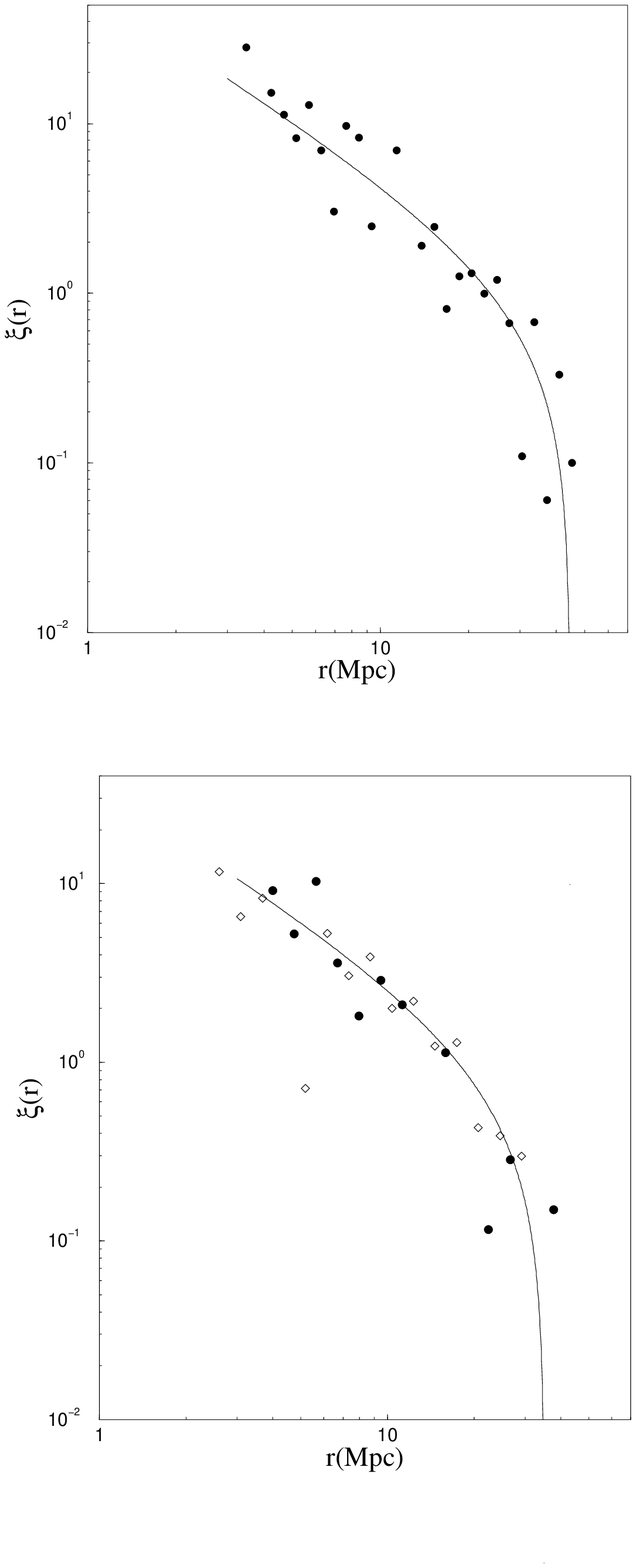}}
\caption{\label{fig39} 
{\it Upper part:} Postman sample (north) (filled circles)
{\it Bottom part:} Postman sample (south) (filled circles) and ACO (diamonds).
The reference line is the functional form of $\xi(r)$ for a fractal
 with $\gamma \approx 1$
}
\eef
we have reported the
$\xi(r)$ for the Postman sample (north and south)
and for the ACO sample.
We have fitted the experimental points
with the functional form given by the fractal prediction 
(Sec.\ref{correlation}).
We found that
$\:\gamma \approx 1$ for the all
analyzed samples. The corresponding $r_{0}$ are
reported in the  Tab.\ref{tabclu}; for the
Postman northern sample  we found
$r_{0}\approx 26 \pm 2$ and for BS83 sample
$r_{0} \approx 27\pm 2$.
These two samples have the same $r_{0}$ because
 they have the same $R_{eff}$.
\begin{table} 
\begin{center}
\centering
\begin{tabular}{|c|c|c|c|c|c|c|}
\hline
     &  &    &         &       &      &        \\
\rm{Sample} & N & $\Omega$ & $<n> $ & $d_{lim}$  & $R_{eff}$&$r_{0}$  \\
&   & $sr$ & $h^{3}Mpc^{-3}$& $(h^{-1}Mpc)$&$(h^{-1}Mpc)$&
$(h^{-1}Mpc)$  \\
\hline
&  &   &   &         &        &      \\
BS83 N & $53$  & $3.1$  & $4.2 \cdot 10^{-6}$
&$230$ &$70$ &$27\pm 2$  \\
Postman N & $120$  & $3.1 $  & $9.5 \cdot 10^{-6}$
& $230$ & $70$ & $26 \pm 2$ \\
Postman S & $88$  & $1.7 $  & $ 1.3 \cdot 10^{-5}$
& $ 230$&$53$& $17\pm 2$   \\
ACO  & $91$    & $2.2$  & $ 3.7 \cdot 10^{-5}$
&$150$&$50$&$17\pm1$ \\
  &  &   &   &         &        &       \\
\hline
\end{tabular}
\caption{\label{tabclu} Features of the various samples analyzed;
column I: the catalog from which the sample has been extracted;
column II: number of clusters in the sample;
column III: solid angle of the sample;
column IV: density of the sample;
column V: sample extension;
column VI: radius of the maximum sphere fully included;
column VII: the correlation length;
column VIII: the mean separation}
\end{center}
 \end{table}
Same conclusions hold for the other
two samples: Postman South and ACO.
Both of them have $R_{eff}$ ($\approx 50 h^{-1} Mpc$) and
the same $r_{0}$ ($\approx 17 h^{-1} Mpc$). 
  In conclusion $r_{0}$ is
simply a fraction of the sample size $R_{eff}$,
without any real physical meaning.

In summary,  the statistical analysis, performed without
any a priori assumptions, shows that Abell samples
have scale invariant properties 
with fractal dimension
$D \approx 2 $ up to $\approx 70 h^{-1} Mpc$, while 
ACO sample up to $\approx 50 h^{-1} Mpc$.
The different limiting distance of the analysis
in the various samples corresponds to the
radius of the greatest sphere fully included in the sample.
No tendency towards homogenization is
detected within the sample limits.
The so-called correlation length $r_{0}$
derived from the $\xi(r)$ analysis,
is simply   proportional to the sample size $R_{eff}$
and then, it is meaningless
in relation with  
the correlation properties of the system.

{\it The mismatch between galaxy and cluster correlation},
i.e. the different correlation length for galaxies and clusters,
is just due to the mathematical inconsistency of the
use of $\xi(r)$ and the correct analysis, in terms
of $\Gamma(r)$ and $\Gamma^{*}(r)$, show that
cluster  correlations are just
the continuation at larger scales of galaxy
correlations.
We conclude that galaxies and clusters are two different
representations of the same self similar structure.
Our conclusion is therefore that 
{\it galaxy clusters extend the correlations of galaxies to
deeper depth}; catalogs of clusters are deeper because clusters are
more luminous than galaxies.
To this end clusters distribution
represents a 
{\it coarse grained}
representation of galaxies, i.e.
it is the same self-similar distribution,
but sampled with a larger scale resolution.
In other words, we consider a cluster of galaxies as
a single object without distinguish the
structure in it.
Hence, we can study the clusters distribution simply
performing
a coarse graining on galaxy distribution.
Usually clusters are instead identified with
some criteria, which are different according to
different observers.
On the contrary, in this way,
we can make the analysis independently
on the definition of {\it cluster} or {\it supercluster}
etc.
The price that one has to pay is however the availability
of a complete galaxy sample.
Same considerations hold, of course, for the
{\it voids distribution}: the void distribution is,
in fact, just
the complement of matter one.

\subsection{Scaling of $r_0$ and luminosity segregation}
\label{lumsegr}

A possible explanation of the shift of $r_0$ with sample size 
is based
on the luminosity segregation phenomenon
\cite{dav88,par94}.
We briefly illustrate this approach.
The fact that the giant galaxies are more clustered than
the dwarf ones,
i.e. that they are located in the peaks of the density field,
has given rise to the proposition
that larger objects may correlate up to larger
length scales and that the amplitude of 
 $\:\xi(r)$ is larger
for giants than for dwarfs one. The deeper VL samples
contain galaxies which are in average
brighter than those in the VL samples with
smaller depths. As the brighter galaxies should have
a larger correlation length the shift of $r_0$ in different samples
can be related, at least partially,
with  the phenomenon of luminosity segregation.

To show that this
is not the case for the
PP survey (for example) and that,
on the contrary,  the shift of
$\:r_0$ is simply due to the
fractal nature of the galaxy distribution,
 we have performed the following test.
We consider a sample of galaxies with
{\it apparent magnitude}
lower $14.5$ (hereafter  PP14.5).
It is evident
that the
VL sample with the same absolute magnitude limit
$M_{lim}$
have different limiting depth $\:d_{lim}$
for the two catalogs according to the formula
\be
\label{ls1}
d_{lim} =  10^{0.2 \cdot (m_{lim} -M_{lim} -25)}
\ee
where $m_{lim}$ is $15.5$ or $14.5$.
Then we construct some VL subsamples for the PP14.5
catalog whose characteristics are
 reported in the Appendix. 
For these VL samples we have done the same analysis
as for the whole catalog PP, and 
the behavior of $r_0$ as a function of $R_{eff}$ 
is again well fitted by the fractal prediction as
for the whole catalog PP.

In Fig.\ref{fig40} we compare 
the values of $r_0$ obtained in VL samples
\bef 
%\vspace{}
\epsfxsize 12cm
\centerline{\epsfbox{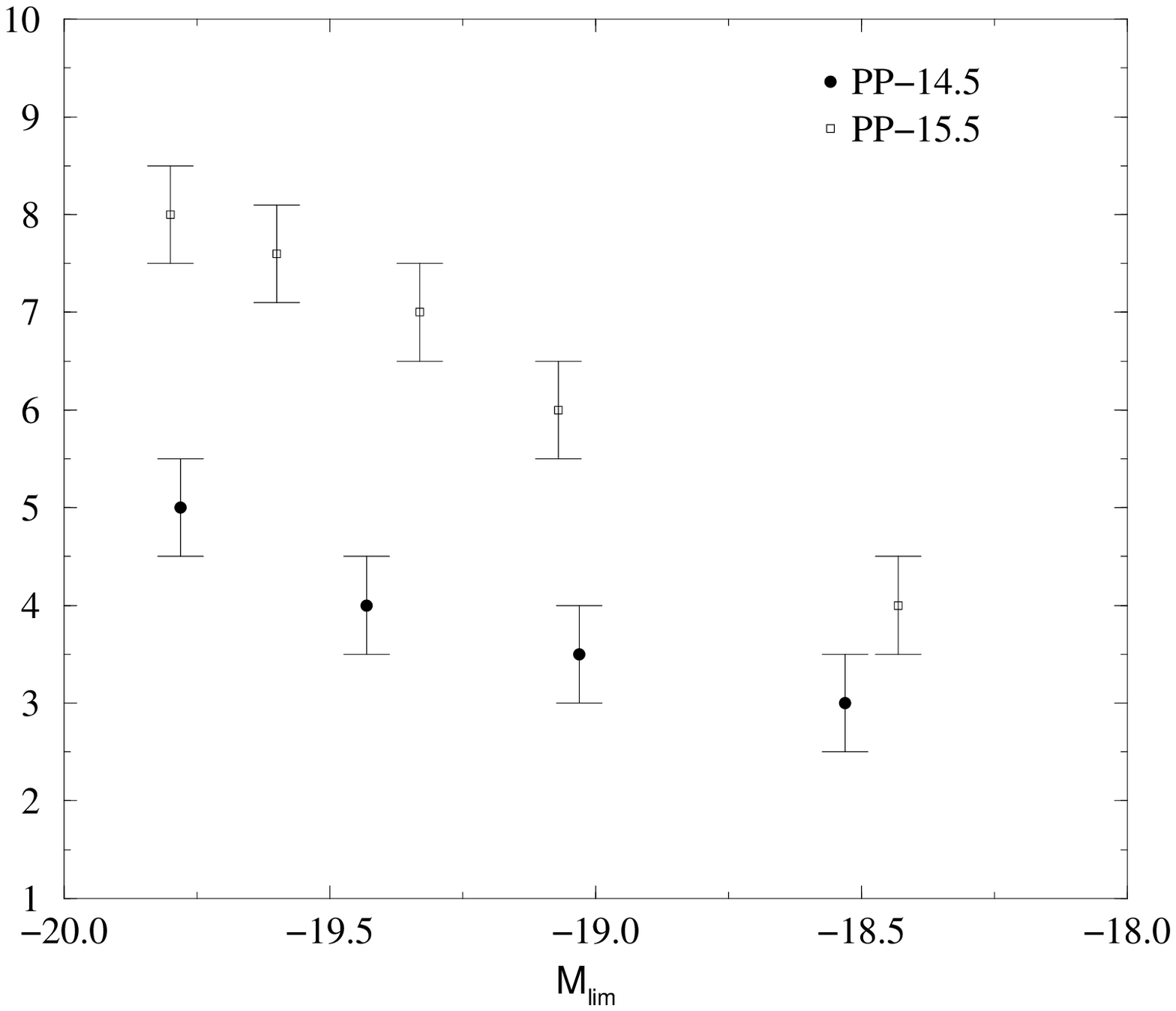}}
\caption{\label{fig40}
 The "Characteristic length scale" $r_0$ ($\xi(r_{0})\equiv 1$)
plotted as a function of the absolute magnitude limit $M_{lim}$
for PP (crosses) and PP14.5 (squares).
To test the luminosity segregation
 hypothesis, one should
find that $r_0$ is the same
for sample with same the $M_{lim}$.
It is clear that there is no agreement between these values.
On the contrary,
the fractal nature of galaxy distribution
explains completely the shift of $r_0$ with sample size.
Hence the hypothesis of luminosity
segregation has no experimental support.
}
\eef
of PP and PP14.5 with the same
absolute magnitude limit $M_{lim}$.
For example the subsample
VL70(b) (we refer with (b) to the 14.5 sample) has the same
absolute
 magnitude limit of the sample VL110 of PP ($M_{lim}=-19.78$).
Hence
these two samples contain galaxies with the same average
absolute magnitude (Appendix). 
If the shift of $r_0$ is due to the luminosity
segregation effect we should not find any difference for $\:r_0$.
 As shown in Fig.\ref{fig40} this is not the case.
In fact, for VL70(b) we find that $r_0 \approx 5 h^{-1}Mpc$
while for VL110 $\:r_0 \approx 8 h^{-1} Mpc$.
On the contrary, if we consider the sample depth dependence of
$\:r_0$ we find that for VL70(b) $R_s \sim 15 h^{-1} Mpc$
that is the same limiting depth of
the subsample VL70 for which
 $\:r_0 \approx 5 h^{-1}Mpc$.

Our conclusion is that luminosity segregation 
cannot be the physical effect
of the shift of $\: r_0$ with sample size
and that, on the contrary, the linear
dependence of $\:r_0$ on depth is naturally described by
the fractal nature of the galaxy distribution.
 The observation
that the giant galaxies are more clustered
than the dwarf ones, i.e. that the massive elliptical
galaxies lie
in the peaks of the density field, is a consequence of the
self-similar behavior of the whole matter distribution
 (Sec.\ref{lumspace}). The increasing
of the correlation length of the $\xi(r)$ has
nothing to do with this effect \cite{cp92}.
{\it  As far as a clear cut-off towards homogeneity 
has not been clearly identified, quantities 
like $r_0$ are meaningless.}

Finally we   stress the conceptual problems of the 
interpretation of the scaling of $r_0$ by the luminosity
 segregation phenomenon.
Suppose we have two kind of galaxies of different masses, one of type $A$
and the other of type $B$. Suppose for simplicity that the mass of 
the galaxies of type $A$ is twice that of the $B$. 
The proposition 
"galaxies of different luminosities (masses) correlate in different 
ways" implies that the gravitational interaction is able to distinguish
between a situation in which there is, in a certain place, a galaxy of type $A$
or two galaxies of type $B$ placed nearby. This is a paradox,
as the gravitational interaction is due the sum of all the masses.
We can go farther by showing
 the inconsistency of the proposition "galaxies
of different luminosities have a different correlation length". Suppose that
the galaxies of type $A$ have a smaller correlation length
than that of the galaxies of type $B$ (Fig.\ref{fig41}).
\bef %\vspace{}
\epsfxsize 8cm
\centerline{\epsfbox{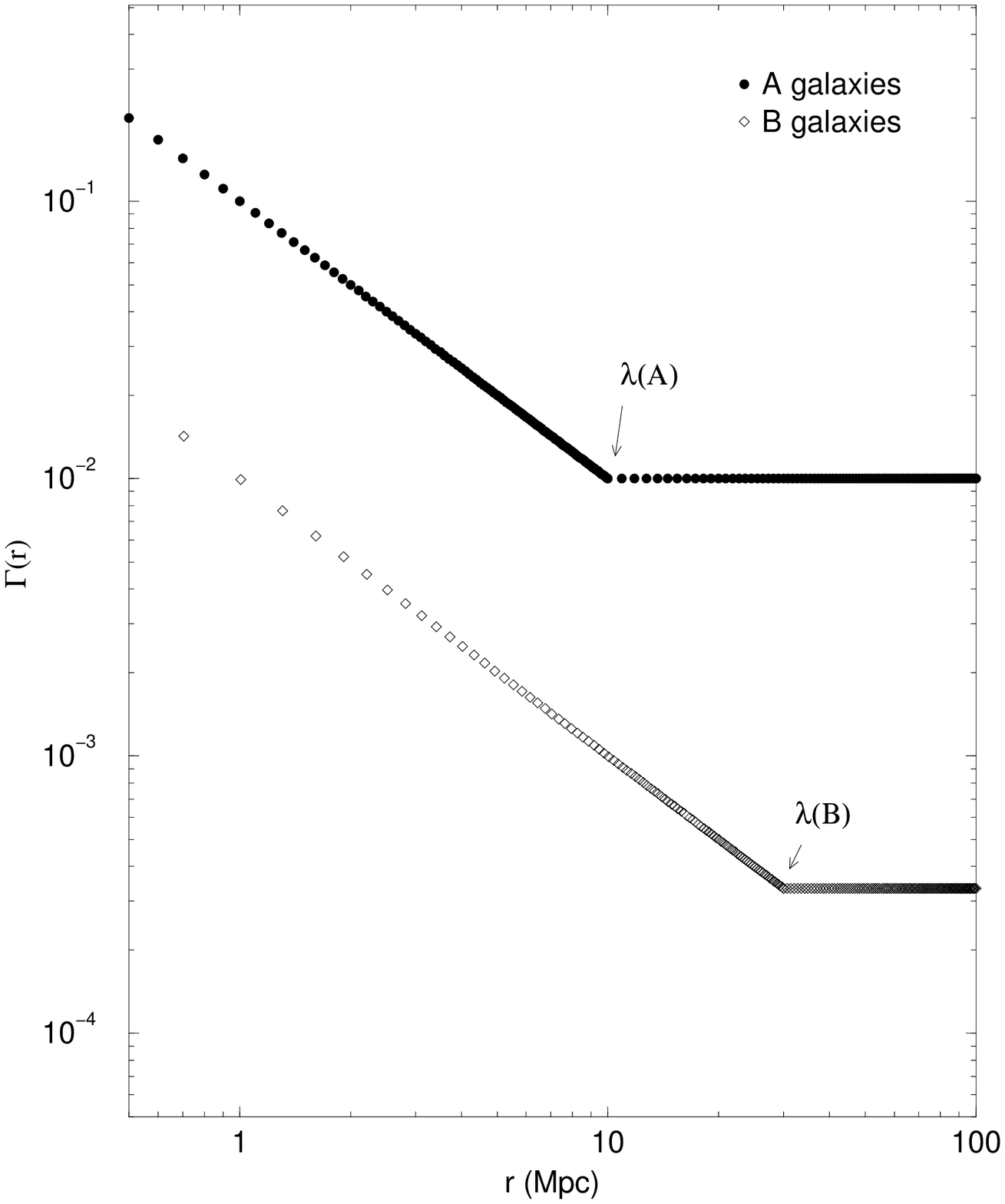}}
\caption{\label{fig41} Luminosity segregation:
in this example we have two kind of objects. Galaxies $A$ have power law
correlation up to $\lambda(A)$ and then they became homogeneous.
Galaxies $B$ have the same behavior, but up to $\lambda(B) > \lambda(A)$.
In this case the voids of $B$ galaxies 
should be full of galaxies $A$.}
\eef
This means that the galaxies of type $B$ are still correlated (in terms of 
the conditional density) when the galaxies of type $A$ are 
homogeneously distributed. This means that the galaxies of type $A$ 
should fill the voids of galaxies of type $B$. This is not the case, as 
the voids
are empty of all types of galaxies, and it seems that the large scale structures
distribution is independent
 on  the galaxy morphological types (Sec.\ref{lumspace}).
We refer to Sec.\ref{powerspect} for a discussion about 
several additional  
tests on luminosity segregation 
we have performed by the power spectrum 
analysis.

\subsection{Tests on the treatment of boundary conditions}
\label{testweight}

In order to test the usual receipt of treatment 
of boundary conditions \cite{dp83}, we have 
computed the conditional density applying the standard 
method. In this case one does not limit the analysis up to the 
radius of the maximum sphere fully contained in the sample volume $R_{eff}$. 
Rather one computes the correlation function up to the maximum 
distance between two galaxies in the sample. In particular the 
correlations are derived from pair counts in redshift space \cite{dav88,gu92}
by using the estimator
\be
\label{w1}
1 + \xi(r) = \frac{DD(r)}{DR(r)} \left( \frac{N_R}{N_D} \right)
\ee
where $DD(r)$ is the number of pair counts 
of galaxies at distance separation $r$, and $DR(r)$ is the number of pairs
consisting of a galaxy and a randomly distributed point, chosen 
within the same catalog window, at distance separation $r$. 
$N_R$ and $N_D$ are the total number of points in the random and galaxy catalogs
respectively. From Eq.\ref{w1} one can easily computes the conditional density
\be
\label{w2}
\Gamma(r) = <n> (1 + \xi(r)) 
\ee
where $<n>$ is the average density in the real sample. 

We have computed Eq.\ref{w1} in the case of an artificial fractal distribution,
generated by the random $\beta$ model algorithm \cite{ben84}. The 
artificial sample has the same geometry of the Perseus-Pisces survey
(a part an arbitrary
 rescaling of distances, we have that $R_{eff} \approx 50 \hmp$), and the 
distance of statistical validity is $R_{eff} \sim 30 \hmp$. We show the
results in Fig.\ref{fig42}.
\bef 
%\vspace{}
\epsfxsize 9cm
\centerline{\epsfbox{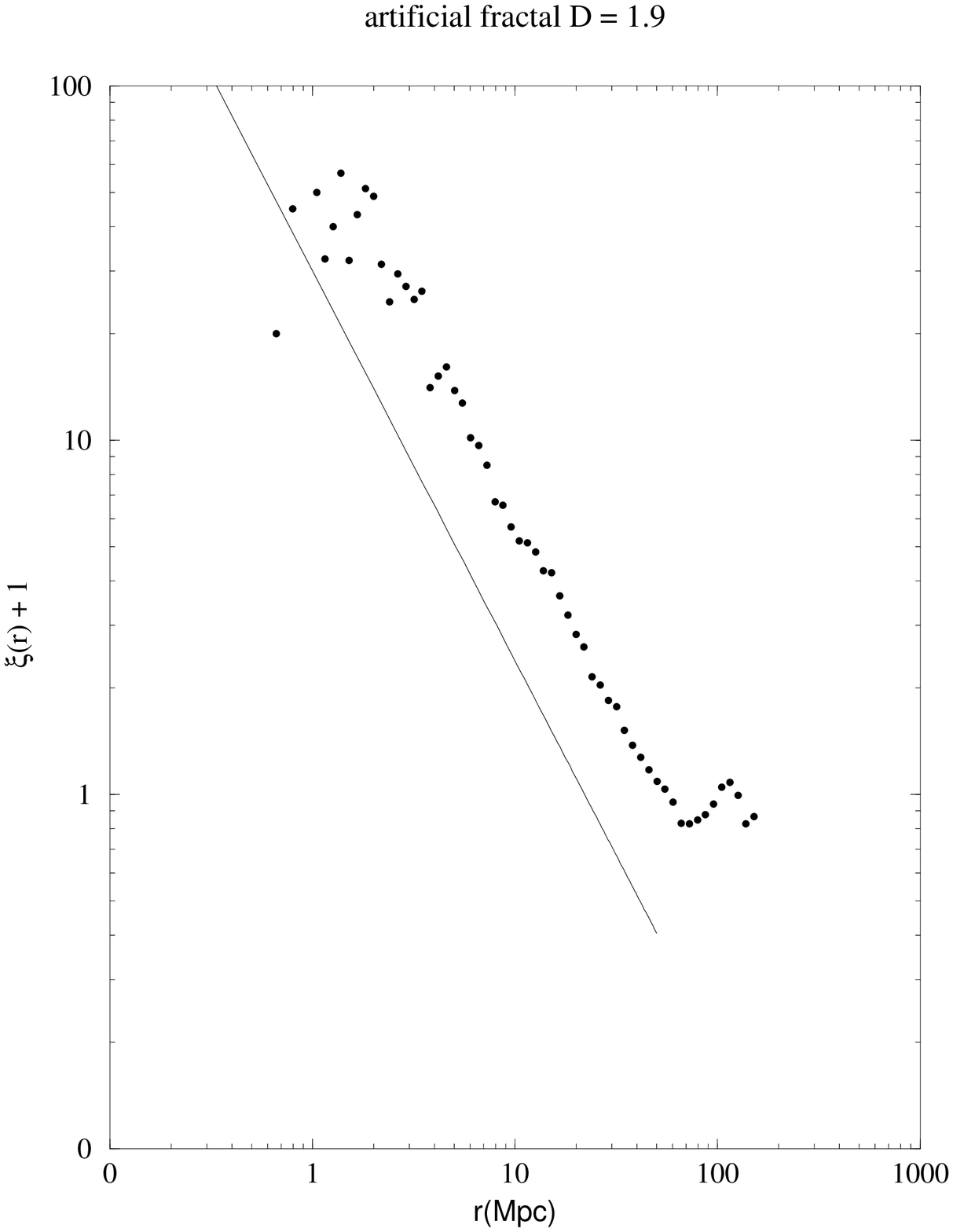}}
\caption{\label{fig42} $\xi(r) + 1$ for an artificial fractal computed 
by the standard method.
 The crossover towards homogenization is
clearly spurious and due to the treatment of boundary conditions.
The reference line has a slope $-\gamma=-1.1$}
\eef
This sample shows well defined power law correlations
up to $R_{eff} \sim 50  \hmp$, with fractal dimension $D \approx 2$. At 
larger scales a clear crossover towards homogenization is shown. 
This is clearly
spurious, because the 
sample is scale invariant by construction  without any characteristic scale.
We have done the same test in the real Perseus-Pisces sample,
finding the same behavior (see Fig.\ref{fig43}, that  
is in agreement with the results of \cite{guz92}).
\bef 
%\vspace{}
\epsfxsize 9cm
\centerline{\epsfbox{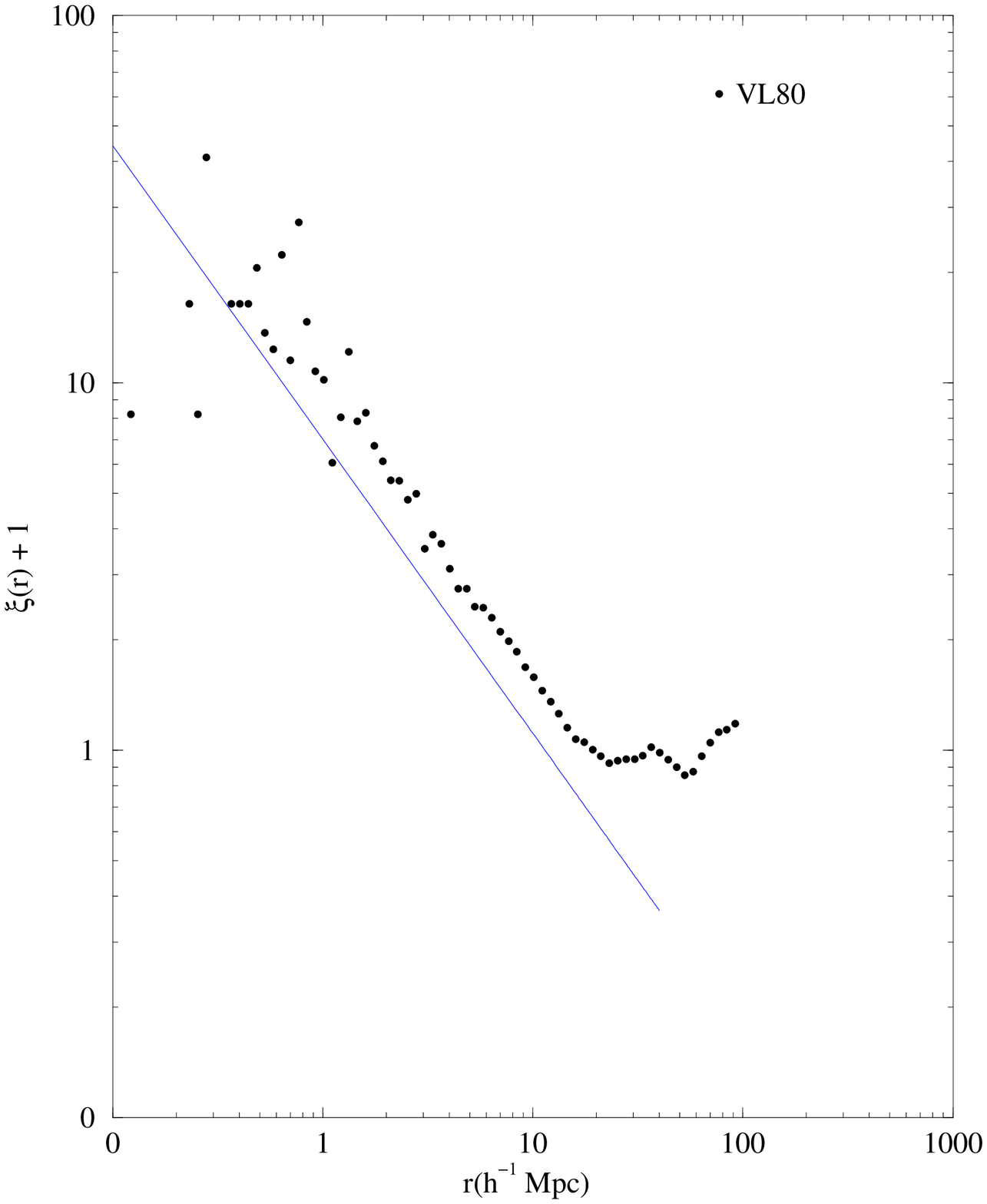}}
\caption{\label{fig43} $\xi(r) + 1$ for the volume limited sample VL80
of the Perseus-Pisces redshift survey, computed 
by the standard method. The crossover towards homogenization is
spurious and due to the inappropriate way of treating
the boundary conditions. The reference line has 
a slope $\gamma=-0.9$.}
\eef

The reason why, by applying the standard procedure we find a
flattening of the conditional density for distances larger than $R_{eff}$,
is the following.  Suppose that we are computing the 
average conditional density $\Gamma^*(r)$ for a sample 
limited by boundaries shown in Fig.\ref{fig44}.
\bef
%\vspace{}
\epsfxsize 8cm
\centerline{\epsfbox{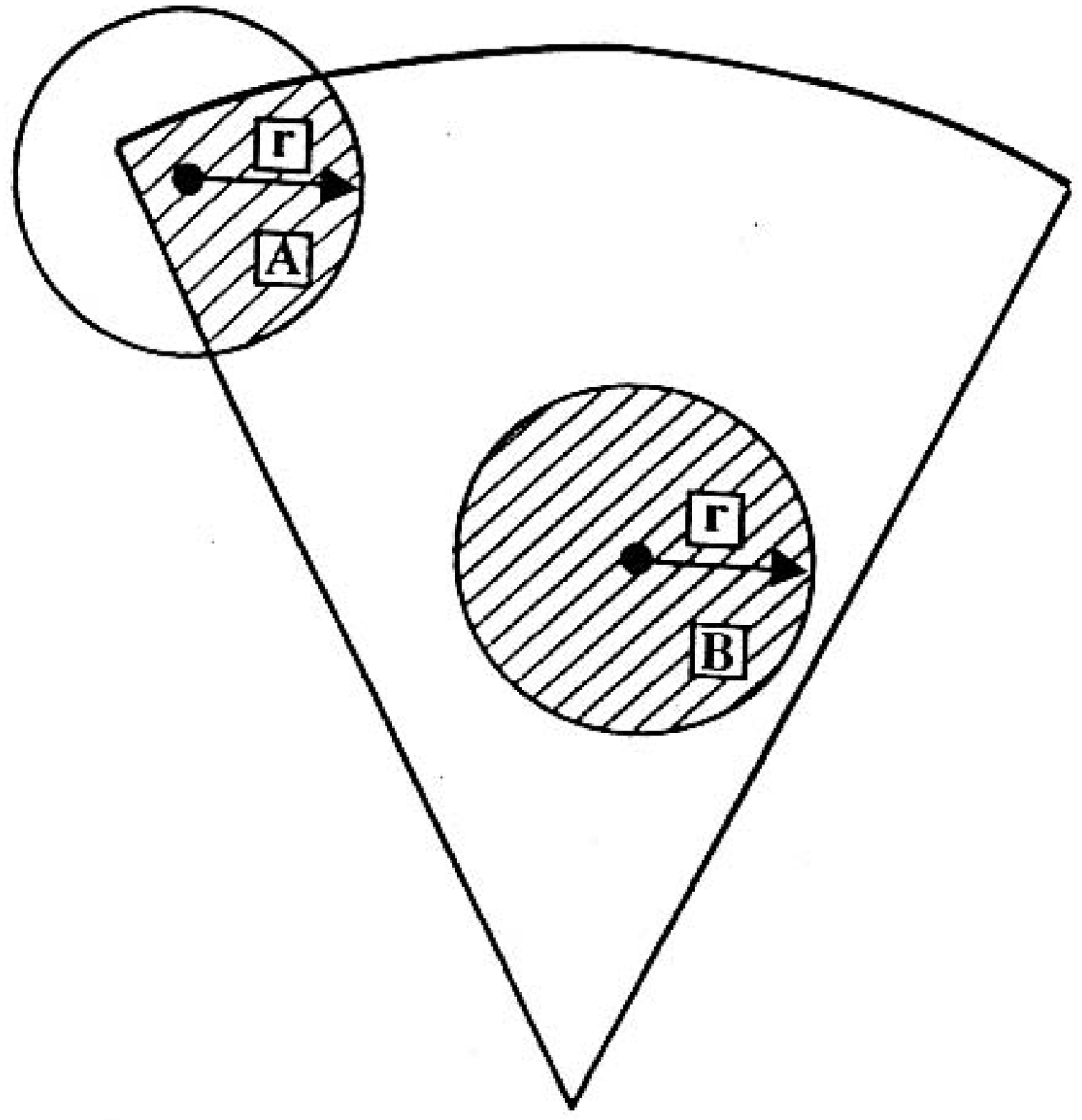}}
\caption{\label{fig44}  The inclusion of the point A in the statistics 
corresponds to an implicit assumption of
homogeneity. It implies in fact that the 
smaller region around A included in the sample, 
and the larger region around B, both contribute
to the average density at distance $r$, each with a weight given by 
its volume (From Coleman \& Pietronero, 1992).}
\eef
For a point B, the sphere centered on B and of radius $r$ is completely
included in the sample, while this is not the
case for the point A. The filled region around A corresponds therefore
to an effective length scale that is actually shorter than $r$ and 
the inclusion of this point, even if properly weighted by its volume, 
in the statistics gives rise to an artificial homogenization of the sample.
If the system has long range correlations, the mixing
of regions of different sizes for the same
nominal distance $r$ cannot be correct \cite{cp92}.

\subsection{Tests on   the conditional density stability versus 
 errors in the apparent magnitude}
\label{errors}

Some authors (e.g. \cite{dav97}) claimed that the long range 
correlations we detected can be due to systhematic selection 
effects in the data or to observational errors.
To check that possible  errors in the apparent magnitude
do not affect seriously the behavior of $\Gamma(r)$
we have performed
 the following tests. We have changed the apparent magnitude
of galaxies in the whole catalog by a random factor $\delta m$
with $\delta m= \pm 0.2, 0.4, 0.6, 0.8$ and $1$. We find that the
number of galaxies in the VL samples change from $5 \%$ up to
$15 \%$ and that the amplitude and the slope of $\Gamma(r)$ are
substantially stable and there are not any significant changes
in their behavior. This is because $\Gamma(r)$ measures a global quantity that
is very robust with respect to these possible errors.

In particular in Fig.\ref{fig45}
we show the behavior of the
conditional density 
in the CfA1 catalog limited at $m_{lim}=14.5$, having 
added a random error up to $\delta =\pm 0.5$ to each galaxy 
in the survey. The correlation properties 
are not substantially affected by such an error.
\bef 
%\vspace{}
\epsfxsize 10cm
\centerline{\epsfbox{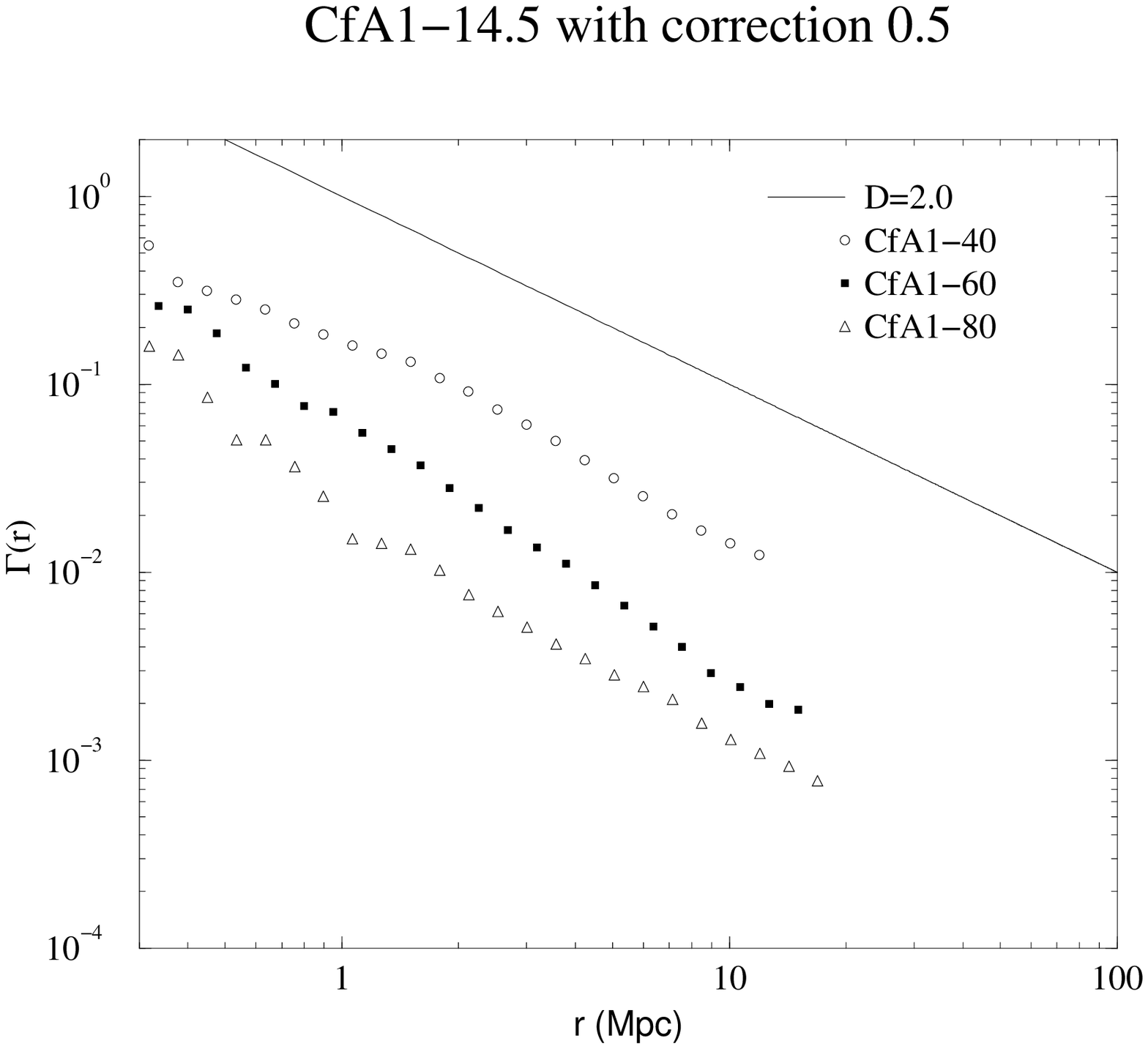}}
\caption{\label{fig45} Behavior 
of the average conditional density 
in the CfA1 catalog limited at $m_{lim}=14.5$, having 
added a random error up to $\delta =\pm 0.5$ to each galaxy 
in the survey. The correlation properties 
are not substantially affected by such an error.}
\eef  
Moreover we have cut the catalog at $m_{lim}=14$ and 
we have added again a random error of $\delta =0.5$ to the apparent 
magnitude. In such a way we have a symmetric situation for the 
galaxies brighter and fainter than $14$. In 
Fig.\ref{fig46} we show the result of such a test.
\bef 
%\vspace{}
\epsfxsize 10cm
\centerline{\epsfbox{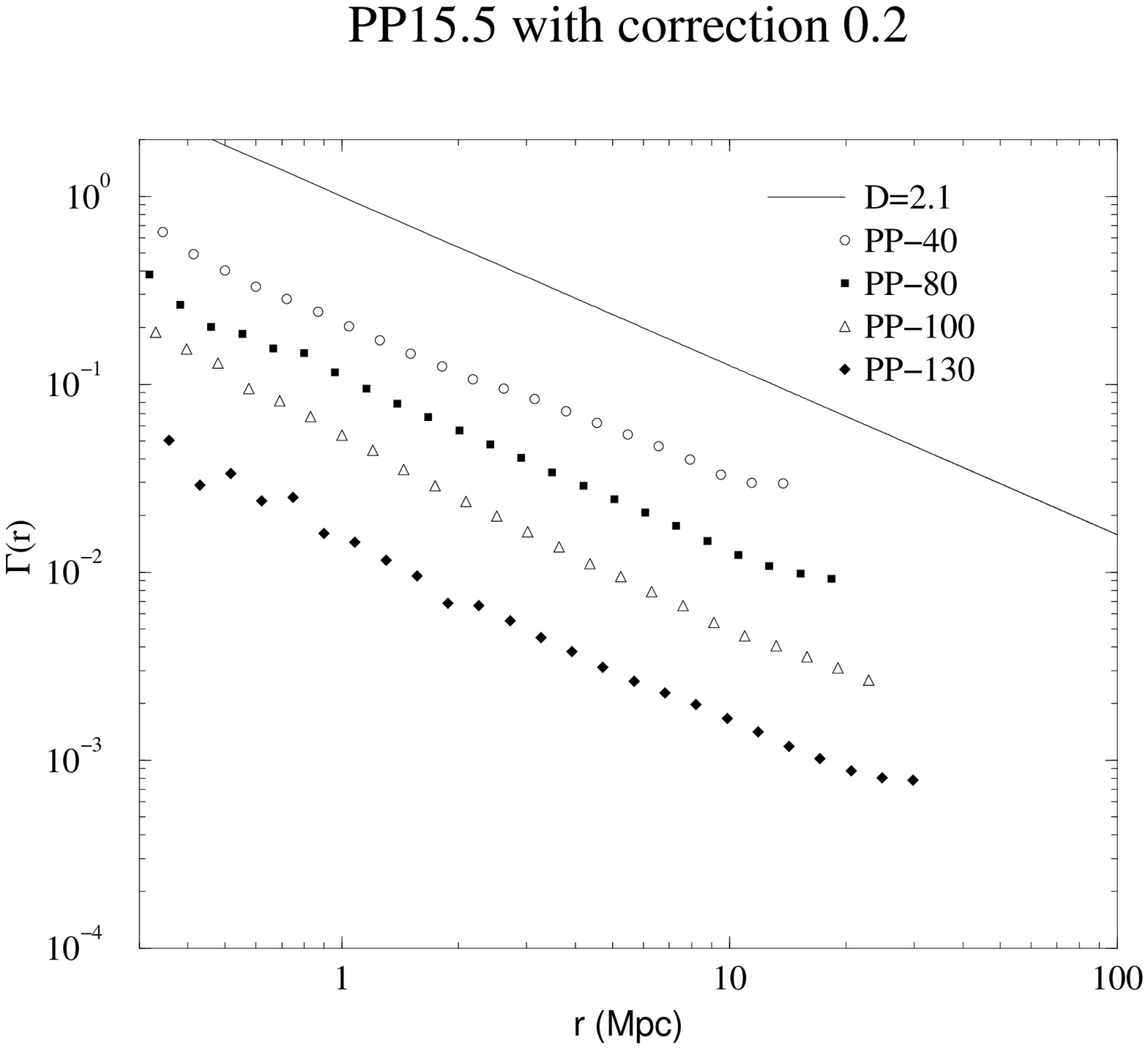}}
\caption{\label{fig46}
Behavior 
of the average conditional density 
in the CfA1 catalog limited at $m_{lim}=14.0$, having 
added a random error up to $\delta =\pm 0.5$ to each galaxy 
in the survey. The correlation properties 
are again stable with respect to these eventual effects
}
\eef  
We have done the 
same tests in the Perseus-Pisces catalog and 
we show
in Fig.\ref{fig47}
and Fig.\ref{fig48}
the results.
\bef 
%\vspace{}
\epsfxsize 10cm
\centerline{\epsfbox{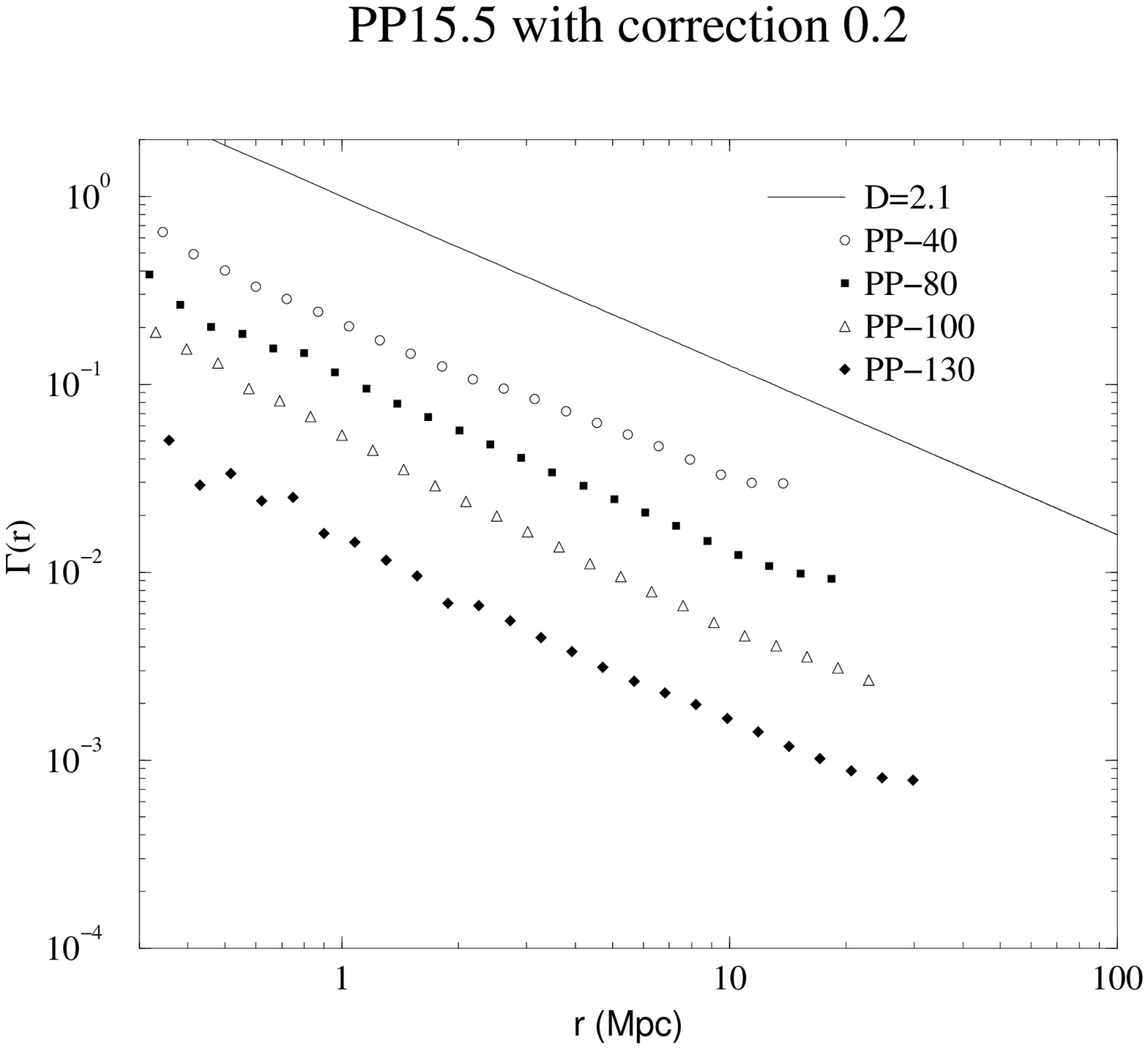}}
\caption{Behavior 
of the average conditional density 
in the Perseus-Pisces  catalog limited at $m_{lim}=15.5$, having 
added a random error up to $\delta =\pm 0.2$ to each galaxy 
in the survey. The correlation properties 
are not substantially affected by such an error 
\label{fig47}
}
\eef  
\bef %\vspace{}
\epsfxsize 10cm
\centerline{\epsfbox{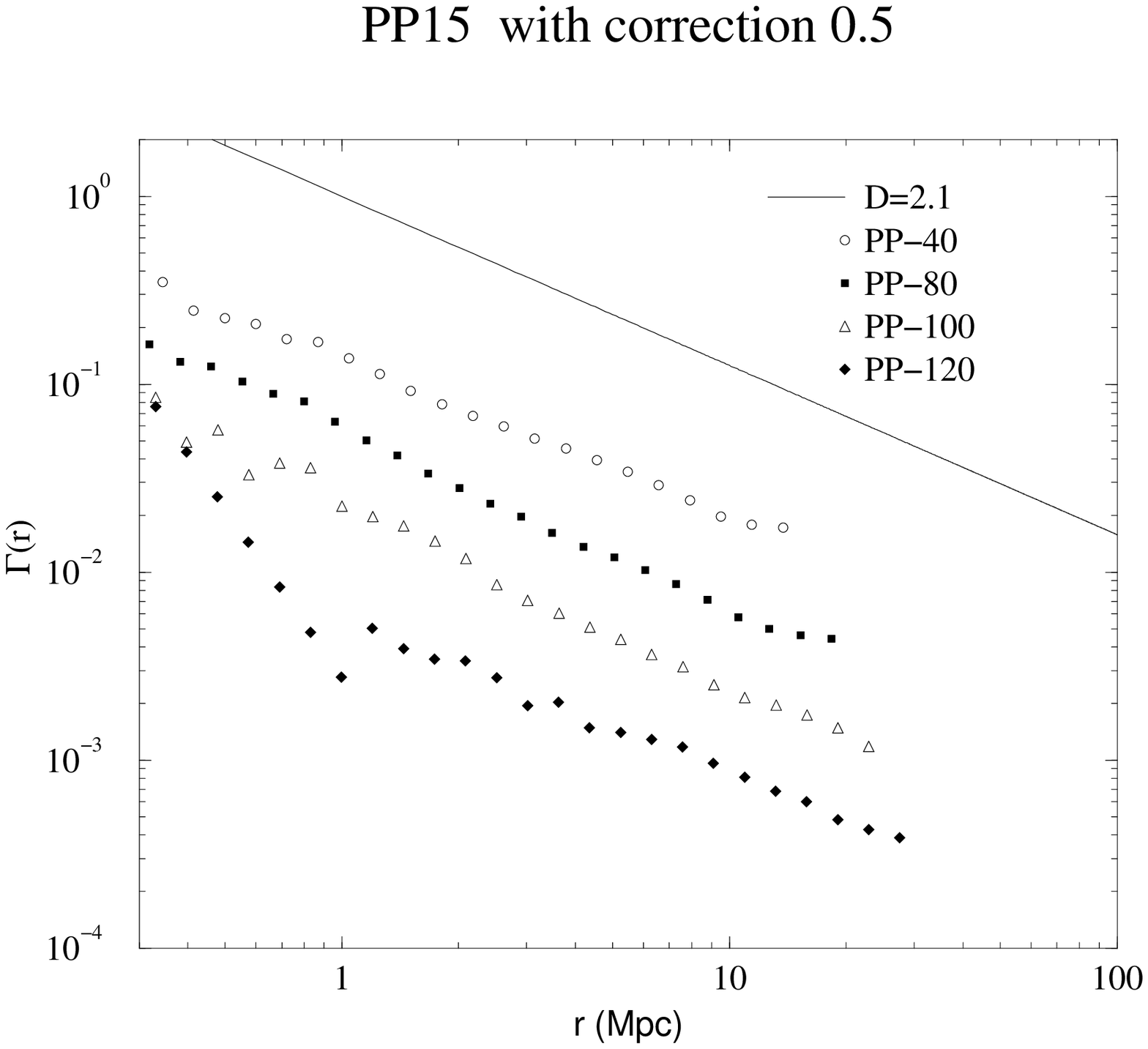}}
\caption{Behavior 
of the average conditional density 
in the Perseus-Pisces catalog limited at $m_{lim}=15.0$, having 
added a random error up to $\delta =\pm 0.5$ to each galaxy 
in the survey. The correlation properties 
are again stable with respect to these eventual effects
\label{fig48}
}
\eef

Moreover to check that the eventual incompleteness
of the catalog for $m \rightarrow m_{lim}=15.5$ do not
affect $\Gamma(r)$ we have done another test. We have changed
again the apparent magnitude of galaxies, but with a probability
density given by $P(m) \sim 10^{\alpha m}$
where $\alpha \sim 0.4 \div 0.6$
is the 
number counts exponent (Sec.\ref{counts}). Even in this case
neither the amplitude nor the slope of $\Gamma(r)$ are changed by
this  eventual incompleteness. These tests confirm that the results
shown in Fig.\ref{fig16} are genuine features of
the galaxy distribution.

We  stress and the end of this section, that a
fractal distribution has a very strong feature: it shows
 power law
correlations up to the sample depth. Such correlations {\it cannot be due}
neither 
by an inhomogeneous sampling of an homogeneous distribution, 
nor by some selection effects that may occur in the observations. 
Namely, suppose that  a certain kind of sampling reduces the 
number of galaxies as a function of distance \cite{pee-pc96}.
 Such an effect in no way can lead to long range correlations,
because when one computes $\Gamma(r)$, one makes an average 
over all the points inside the survey. Of course such a
selection effect may bias the determination of
 the space density determined from
only one point, i.e. the radial 
density as we 
discuss in Sec.\ref{consist}.

%%%%%%%%%%%%%%%%%%%%%%%%%%%%%%%%%%%%%%%%%%%%%%%%%%%%%%%%%%%%%%%%%%%%%%%%%

\section{Power spectrum of galaxy distributions}
\label{powerspect}
 
In a series of papers  \cite{sla96a,sla96b,ledabias,ledaps}
we have studied   the problems of the determination
of the power spectrum in galaxy redshift surveys.
Here we briefly review our main results, and we
present several analyses on real redshift surveys
\footnote{We warmly thank L. Amendola 
for  useful  collaborations and discussions 
 in the analysis of the 
power spectrum properties.}.

First of all  we   stress an important 
conceptual point that makes the determination of the 
power spectrum ambiguous, in the case of fractal structures. 
In fact, if we have a fractal distribution that has 
correlation function that decays 
as a power law $\Gamma(r) \sim r^{D-3}$, its Fourier 
conjugate is a power law too: $\Pi(k) \sim k^{-D}$.
However, there is an additional problem that is usually neglected.
Namely, if we have 
a power spectrum which  decays as power law of the wavenumber, then
it is not needed that the real space distribution
is fractal. The key point is that the existence of 
  correlated structures in real space
implies that the phases 
of the different modes of the power spectrum
are not randomly distributed, but are correlated 
as well. This fact shows the conceptual problem in the determining 
the power spectrum: in fact, in general it is possible to compute the behavior
of the amplitude of the power spectrum at various modes, but the eventual 
phases correlation, at different modes, is very difficult to study. 
All the determinations of the power spectrum in literature refers to the
shape of power spectrum, rather than to the distribution of phases.
Hence the most effective way to study a fractal structure is the
real space correlation analysis.

In the following we stress the technical problems of the 
standard determination of the power spectrum and
we clarify the correct interpretation of
the results. The problems are strictly related to those of
$\xi(r)$, namely that this function is not a power
law for a fractal structure and its amplitude depends on the 
sample size. We introduce the correct quantity that should be
computed in order to perform an analysis without a priori assumptions,
that is the Fourier conjugate of the conditional density.

\subsection{Power spectrum analysis of galaxy distribution}
\label{powergal}

Essentially all the currently
elaborated models of galaxy formation 
(e.g \cite{pee93})
{\it assume large scale homogeneity} and 
predict that the galaxy
power spectrum (hereafter PS), 
which  is {\it the PS  of the density contrast},
decreases both toward small scales and toward large
scales, with a turnaround somewhere in the middle, at a scale $\lambda_f$
that can be taken as separating ``small'' from ``large'' scales. 
Because of the homogeneity assumption, the PS    amplitude
should be independent on the survey scale, any  residual 
variation being attributed to luminosity bias (or to the
fact that the survey scale has not yet reached the homogeneity scale).
However, the crucial clue to this
picture, the firm determination of the 
scale $\lambda_f$, is still missing, although
some surveys do indeed produce a  turnaround
 scale around 100 $\hm$
\cite{be94,fe94}. 
Recently, the CfA2  survey 
analyzed by \cite{par94} 
(hereafter PVGH) 
(and confirmed by SSRS2 \cite{dac94} 
- hereafter  DVGHP), showed a $n=-2$ slope up to $\sim 30 \hm$,
a milder $n\approx -1$ slope  up to 200 $\hm$, and some tentative
indication of flattening on even larger scales. PVGH also find
that deeper subsamples have higher power amplitude,
i.e. that the amplitude scales with the sample depth.

{\it In the following  we argue  that both features, bending and scaling,
are a manifestation of 
the finiteness of the survey volume, and that they
cannot be
interpreted as the convergence to homogeneity, nor to a PS   
flattening.}
The systematic effect of the survey finite size is in fact
to suppress
power at large scale, mimicking a real flattening.
Clearly, this effect occurs whenever galaxies have  not 
a    correlation scale much larger than the survey size, and it has
often been studied in 
the context of standard scenarios \cite{it92,col94}.
 We push this argument further, by showing that
even a fractal distribution of matter, 
which never reaches  homogeneity, shows a sharp flattening
and then a turnaround. Such 
features are  partially corrected, but not quite eliminated,
  when the correction proposed by \cite{pn91}  is applied to the data.
 We show also
 how the amplitude of the
PS    depends on the survey size as long as 
the system shows long-range correlations.

The standard  PS    (SPS) 
measures directly the contributions of different scales to the galaxy
density contrast $\:\delta\rho/\rho$.
It is clear that the density contrast, 
and all the quantities based on it,  is meaningful only when one can define
a constant density, i.e. reliably identify
the sample density with   
the average density of all the Universe.
In other words in  {\it the SPS analysis 
one assumes that the survey volume is large enough
to contain a  homogeneous sample.} 
When this is not true, and we argue that is 
indeed an incorrect assumption
in all the cases investigated so far, a false interpretation of the results may
occur, since both 
the shape and the amplitude of the PS    (or correlation
function) depend on the
survey size.

Let us recall the basic  notation of the PS    analysis.
Following Peebles \cite{pee80}  we imagine that the Universe is periodic
in a volume $\:V_{u}$, with $\:V_{u}$ much 
larger than the (presumed) maximum
correlation length. The survey volume $V\in V_u$
contains $\:N$ galaxies at positions $\:\vec{r_i}$,
and the galaxy density contrast is
\be
\label{eps4}
\delta(\vec{r}) = \frac{n(\vec{r})}{\hat n} -1 
\ee
where it is assumed that exists a 
well defined constant density $\hat n$, obtained
averaging over a sufficiently large scale.
The density function
$\:n(\vec{r})$ 
can be described by a sum of delta functions:
$~n(\vec{r}) = \sum_{i=1}^{N} \delta^{(3)} 
(\vec{r}-\vec{r_{i}})\,.~$
Expanding the density contrast in its Fourier components we have 
\be
\label{eps7}
\delta_{\vec{k}} = \frac{1}{N} \sum_{j \epsilon V} 
e^{i\vec{k}\vec{r_{j}}} - W(\vec{k})\,,
\ee
where
\be
\label{eps7b}
~W(\vec{k}) = \frac{1}{V} \int_V d{\vec{r}} W(\vec{r})
 e^{i\vec{k}\vec{r}}\,~
\ee
is the Fourier transform of the survey window $W(\vec{r})$, 
defined to be unity inside the survey region, and zero outside.
If $\xi(\vec{r})$ is the correlation function of the galaxies
($\xi(\vec{r}) = <n(\vec{r})n(0)>/\hat n^2 -1$),
the true PS $\:P(\vec{k})$ is defined as
the Fourier conjugate of the 
correlation function $\xi(r)$.
Because of isotropy the PS can be simplified to
\be
\label{eps12}
P(k)  =4\pi \int \xi(r)  \frac{\sin(kr)}{kr} r^{2}dr\,.
\ee
The  variance of $\:\delta_{\vec{k}}$ 
is \cite{pee80,pn91,fis93} \be
\label{eps9}
<|\delta_{\vec{k}}|^{2}> = \frac{1}{N} + \frac{1}{V}\tilde P(\vec{k})\,.
\ee
The first term is the usual additional shot noise term
while the second is the true PS convoluted with a 
window function 
which describe the geometry of the sample (PVGH)
\be
\label{eps9b}
\tilde P(\vec{k})= 
{V\over (2\pi)^3} \int
<|\delta_{\vec{k'}}|^{2}>
|W(\vec{k}-\vec{k'})|^2
d^3 \vec{k'}
\,.
\ee
\be
\label{eps9bb}
\tilde P(\vec{k}) = \int d\vec{k'} P(\vec{k'}) F(\vec{k}-\vec{k'}) \,,
\ee
with 
\be
\label{eps9c}
F(\vec{k}-\vec{k'}) = \frac{V}{(2\pi)^3} |W(\vec{k}-\vec{k'})|^2\,.
\ee
We apply now this standard analysis to a fractal distribution.
We recall the  expression 
of the $\:\xi(r)$ in this case is (Sec.\ref{statmec}): 
\be
\label{eps16}
\xi(r) = [(3-\gamma)/3](r/R_{s})^{-\gamma} -1\,,
\ee
 where  $\:\gamma=3-D$. 
 A key point of our discussion is
that  that
on scales larger that $R_s$ the $\xi(r)$ cannot be calculated without
making assumptions on the 
distribution outside the sampling volume.

As we have already mentioned, in a fractal
quantities like 
$\xi(r)$ 
are  scale dependent: in particular both
the amplitude and the shape of $\xi(r)$ depend 
the survey size.
It is clear  that the same kind of 
finite size effects  are also present when computing the SPS, so that 
it is very dangerous to identify real physical features induced
from the SPS analysis without first a firm determination of the
homogeneity scale.
 
The SPS for a fractal distribution 
 model described by
 Eq.\ref{eps16}
inside a sphere of radius $R_s$ is
\be 
\label{eps19}
P(k) =  \int^{R_{s}}_{0} 
   4\pi \frac{\sin(kr)}{kr} \left[ \frac{3-\gamma}{3} 
\left(\frac{r}{R_{s}}\right)^{-\gamma} -1\right] r^{2}dr=
\frac{a_k(R_s) R_{s}^{3-D}}{k^{D}}-\frac{b_k(R_s)}{k^{3}}\,.
 \ee
Notice that the integral has to be evaluated inside $R_s$
because we want to compare $P(k)$  with its {\it estimate}
 in a finite size spherical survey of scale $R_s$. 
 In the  general case, we must deconvolve the
 window contribution 
 from $P(k)$; $R_s$ is then a characteristic window scale. 
Eq.\ref{eps19} shows the two scale-dependent features of the PS. First,
the amplitude of the PS 
depends on the sample depth.
Secondly,
the shape of the PS 
is characterized by two  scaling regimes:
the first one, at high wavenumbers, 
is related to the fractal dimension of the 
distribution in real space, 
while the second one arises only because of 
the finiteness of the sample.
In the case of $\:D=2$  in Eq.\ref{eps19} one has:
\be
\label{eps22}
~a_k(R_s) = \frac{4\pi}{3} (2+\cos(kR_{s}))\,
\ee
and
\be
\label{eps23}
~b_k(R_s) = 4\pi \sin (kR_{s})\,.
\ee
The PS is then a power-law with exponent 
$\:-2$ at high wavenumbers, 
it flattens at low wavenumbers and reaches a maximum at
$k\approx 4.3/R_s$, i.e. at a scale $\lambda \approx 1.45 R_s$.
The scale at which the transition occurs 
is thus related to the sample depth. 
In a real survey, things are complicated by the window function,
so that the flattening (and the turnaround) scale can only be determined
numerically (Fig.\ref{fig49}).
\bef 
%\vspace{}  
\epsfxsize 10cm 
\centerline{\epsfbox{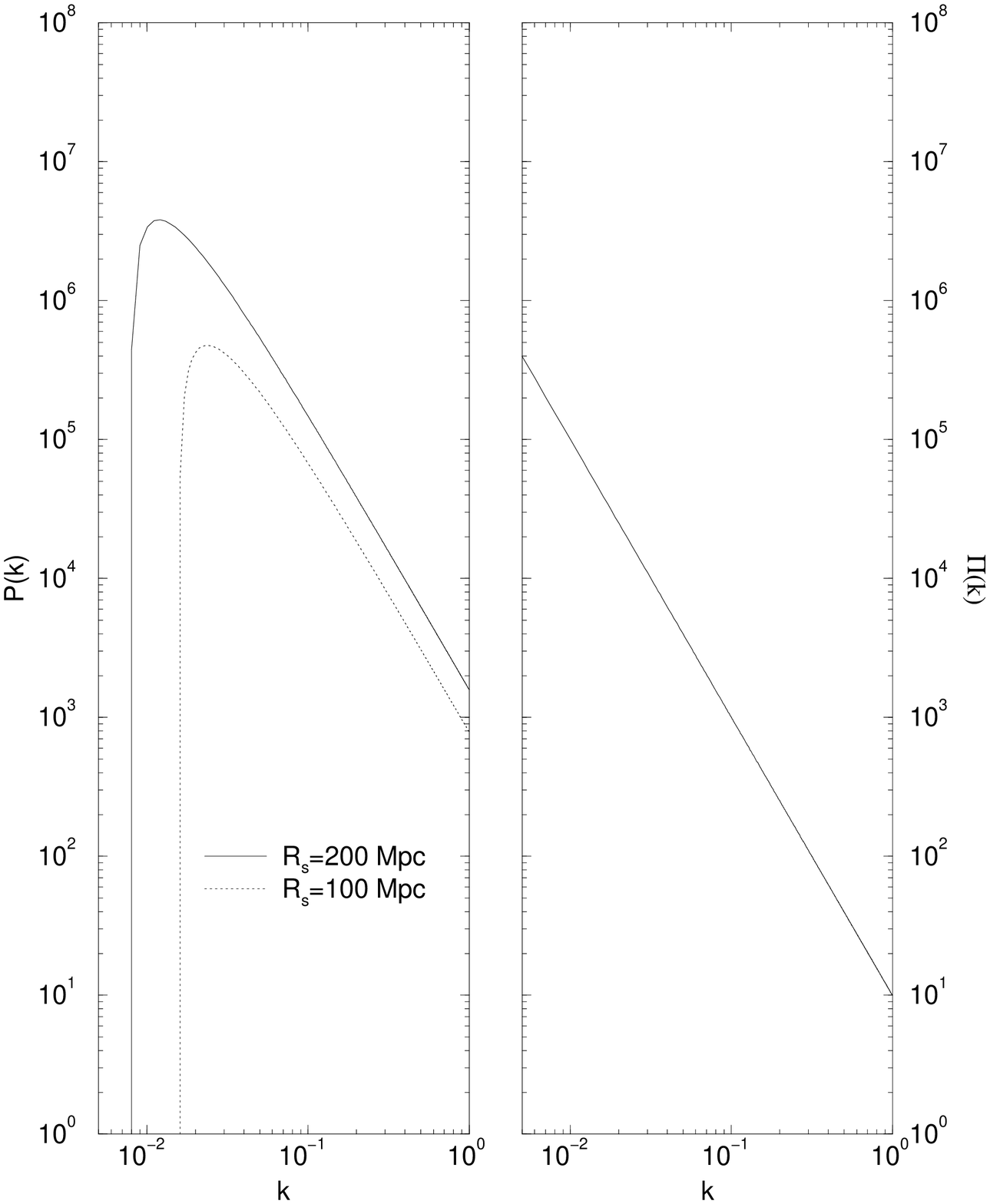}}
\caption{\label{fig49} {\it Left panel:} 
Standard PS    $P(k)$ for an artificial fractal
distribution. It rises as $\lambda^D$
($\lambda=2\pi/k$)
for small scales, it reaches a turnaround at $\lambda\approx 1.5 R_s$
and then it declines to zero as $\lambda\to\infty$.  Moreover 
the sample size dependence of the amplitude of the PS    is shown.
{\it Right panel:} The same of the previous case, 
but with PS    of the density, $\Pi(k)$.}
\eef

In practice, one averages the spectrum over $k$-shells of
thickness $\Delta\le 2\pi/R_s$:
\be
P_a(k)={1\over k^2 \Delta}\int_k^{k+\Delta}
P(k') k'^2 dk'
\ee
For $D=2$ and for $\Delta=2\pi/R_s$ one has
\be\label{fps2}
P_a(k)=(8\pi R_s/3 k^2)[1+f(x)]\approx 8\pi R_s^3/ 3 k^2
\ee
where $f(x)=3[Si(2\pi+x)-Si(x)]/4\pi\ll 1$ for $x>1$
(here $Si(x)$ is the SinIntegral function). Thus, in the case
of a spherical survey, with $D=2$ and  $\Delta=2\pi/R_s$, the
fractal PS goes like the simple power law $k^{-2}$, and
the turnaround feature is removed.
For smaller $\Delta$, the turnaround at small $k$
appears also in the shell-averaged spectrum $P_a(k)$.
We calculated numerically the PS for $D\not =2$, and we found that
the generalization
\be\label{fpsapp}
P_a(k)\approx 8\pi R_s^{3-D}/ 3k^D
\ee
is accurate to better than 10\% for $D$ in the range $(1.7,2.3)$.

However, in practice one has several complications. First, the survey
in general is not spherical. This introduces a coupling with the survey window
which is not easy to model analytically. 
For instance, we found that windows of small
angular opening shift to smaller scales
the PS turnaround. 
This is analogous to what happens with the correlation function
of a fractal: when it is calculated in small angle surveys, the
correlation length $r_0$ decreases. Second, the observations
are in redshift space, rather than in real space. 
The peculiar velocities generally make steeper the PS    slope 
(e.g. \cite{fis93})
 with respect to the real space. Third, in a fractal
the intrinsically high level of fluctuations makes hard
a precise comparison with the theory when the fractal under study
is composed of a relatively small number of points.
In view of these complications, we   compare our results
with the simple fractal-like trend
\be\label{theor}
P(k)=A R_s^{3-D} k^{-D} \; .
\ee

\subsection{Sample-size independent PS   }
\label{powerindep}

As we have already discussed,
the analysis of a distribution on scales at which homogeneity is not reached
must avoid the normalization through the mean density, if the goal is 
to produce
results which are not related to the sample size, and thus misleading.
To this aim,  now we 
consider {\it 
the sample size-independent PS (SIPS) of the density $\:\rho(\vec{r})$,}
a quantity which    gives 
 an unambiguous information 
of the statistical properties of the system.
We first recall the density correlation function (Sec.\ref{statmec})
\be
\label{eps24}
G(\vec{r}) = <\rho(\vec{x}+\vec{r})\rho(\vec{x})> = B r^{-(3-D)}\,,
\ee
where the last equality holds in the case 
of a fractal distribution with dimension {\em D},
and where {\it B} is a constant 
determined by the lower cut-offs of the distribution 
(Sec.\ref{statmec}). 
Defining
the SIPS  as the Fourier conjugate of the correlation function $G(r)$,
\be
\label{eps31}
\Pi(k)  = 4\pi \int G(r)  \frac{\sin(kr)}{kr} r^{2}dr\,,
\ee
one obtains that in a finite spherical volume
\be
\label{eps30}
\Pi(k) \sim B' k^{-D} 
\ee
(where $B'=4\pi B(1-\cos(k R_s))$ if $D=2$)
so that the SIPS is a single power law extending all over the system size,
without  amplitude  scaling 
 with the sample depth (except for $kR_s\ll 1$).
Taking into account the window function, we can write
 the convolution of the SIPS
 with a window function in complete analogy to
Eq.\ref{eps9b}.
In analogy to the procedure above, 
we consider the Fourier transform of the density
\be
\label{eps26}
\rho_{\vec{k}} = \frac{1}{V}\sum_{j\in V} e^{-i\vec{k}\vec{x_j}}\,,
\ee
and its variance
\be
\label{eps27}
<|\rho_{\vec{k}}|^{2}> = \frac{1}{V} \tilde \Pi(\vec{k})+ \frac{1}{N} \,,
\ee
where $\tilde \Pi(\vec{k})$ is the same as in Eq.\ref{eps9b},
with $<|\rho_{\vec{k'}}|^{2}>$ instead of $<|\delta_{\vec{k'}}|^{2}>$.
\be
\label{eps27b}
\tilde \Pi(\vec{k}) 
= \int d\vec{k'} \Pi(\vec{k'}) F(\vec{k}-\vec{k'})\,.
\ee

\subsection{Tests on artificial distributions}
\label{powertest}

In order to  study the 
finite size effects in the determinations of the PS,
we have performed some tests on artificial distributions 
with a priori assigned properties \cite{sla96a}. 
We distribute the sample  in a cubic volume $\:V_{u}$.
We determine $ P(k)$ ($ \Pi(k)$)
defined as the directionally averaged $\: P(\vec{k})$
($\: \Pi(\vec{k})$).
Following PVGH, the estimate of the 
noise-subtracted PS given in Eq. (\ref{eps9b}) 
for a strongly peaked window function is 
\be
\label{eps40}
 P(\vec{k}) = \left( <| \delta_{k}|^2> - \frac{1}{N} \right)
\left( \sum_{\vec{k}} |W_{\vec{k}}|^2 \right) ^{-1} (1-|W_{\vec{k}}|^2)^{-1}\,,
\ee
For the lowest wavenumbers the PS    estimate Eq.\ref{eps40} is not
acceptable, because then the window filter flattens sensibly (see e.g. PVGH).
The factor $\:(1-|W_{\vec{k}}|^2)^{-1}$ has been introduced
by \cite{pn91} %Peacock \& Nicholson (1991)
as an analytical correction to the erroneous identification
of the sample density with the population density. However, the correction 
itself rests on
the assumption that the PS    is flat on very large scales; in
 other words,
it implies the 
flattening, 
which is just the feature we are testing for. PVGH actually 
correct their results by comparing them to the power spectra of
$N$-body simulations; their conclusion is that the PS    correction
is a procedure reliable for wavelengths smaller than $\sim 200\hm$.

We have generated  $D\approx 2$ fractal distributions with the random 
$\:\beta$-model
algorithm \cite{ben84}.
Then we 
have constructed artificial volume limited catalogs with roughly the same
geometry of the CfA2 survey (our model contains no dynamics, so that
we can think of our fractal set 
as lying in redshift space, as the CfA2 galaxies).
We have computed the 
quantity  $\: P(k)$ from Eq.\ref{eps40}
averaging over 50 random  observers (located on one of
the particles) for each  realization. In Fig.\ref{fig50}
\bef 
%\vspace{}  
\epsfxsize 9cm 
\centerline{\epsfbox{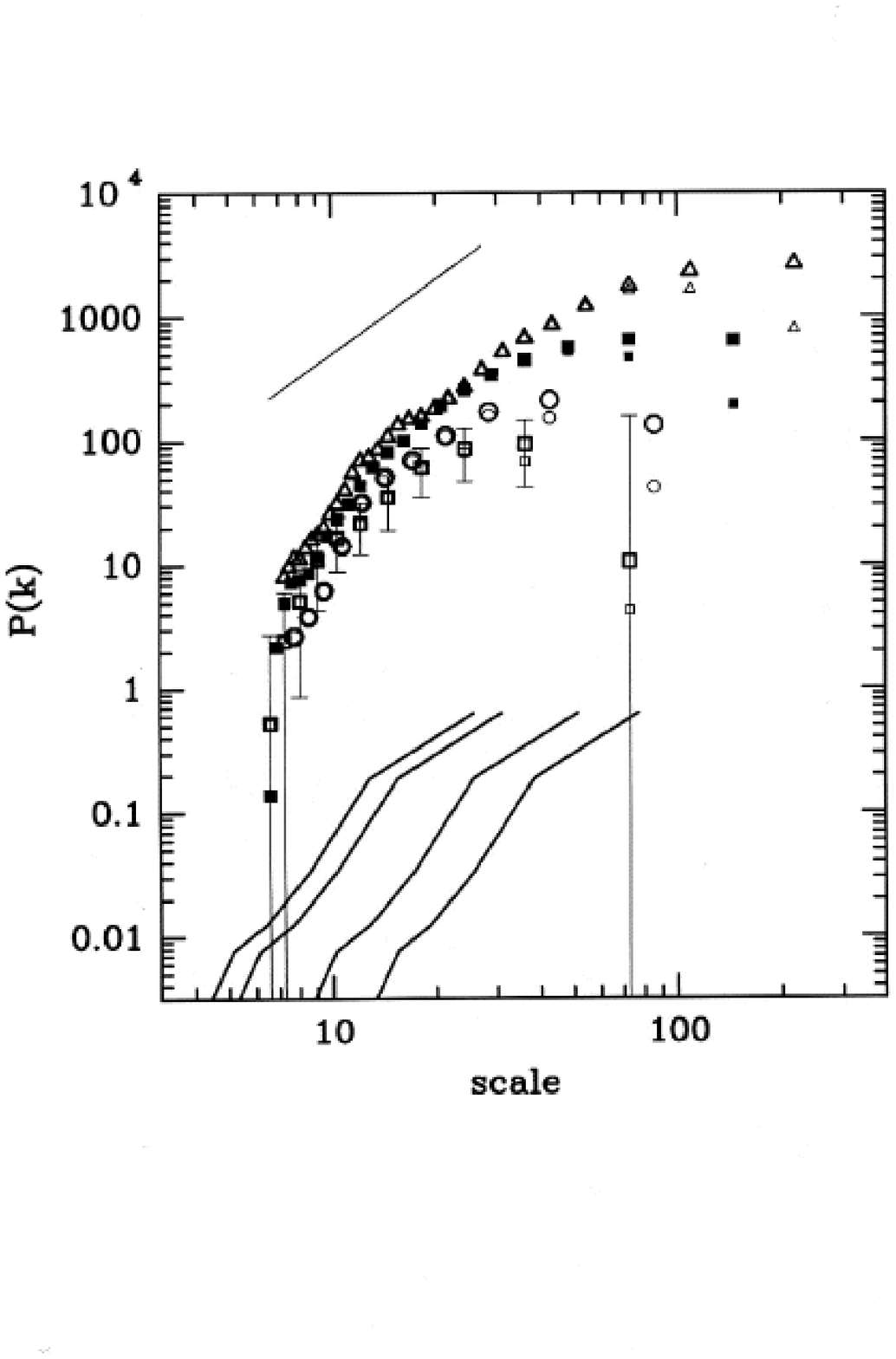}}
\caption{\label{fig50} Top panel: 
Power spectrum for a fractal distribution with dimension 
$\:D=2$ {\it vs.} the scale
$2\pi/k$ without the correction factor (filled squares) and with the
 correction factor (open triangles). 
Here and in what follows the error bars represent the scatter among
the different observers of the same fractal.
The upper set of points 
refer to a subsample of depth  60 cells,
 the lower set to a depth of 30 cells. 
 The dashed lines are least squares linear fit to the data, at small
 scales (slope $D_1$ and amplitude $10^A$) and at
  large scales (slope $D_2$).
At low wavenumbers there is a well defined 
power law behavior with exponent $\:D_1\approx 2$. The flattening at 
high wavenumber is spurious. In the
bottom panel it is shown the window function.
 (From Sylos Labini \& Amendola, 1996).
}
\eef
it is shown the $\: P(k)$ {\it vs.} the scale
$2\pi/k$ with and without the correction factor, 
for some different survey scales $R_s$,
together
with the angle-averaged window PS    $\:|W_k|^2$.
 The slope of the 
PS at high wavenumbers is $\: \approx - D$ in 
agreement with  Eq.\ref{eps19}.
As anticipated, the flattening at low wavenumbers is here completely
 spurious,  i.e. it is
due to the finite volume effects on
the statistical analysis performed. In fact, comparing
the PS at the various sample scales, one finds 
the turnaround of the PS always occurs  near  
the boundary of the sample.
The turnaround scale roughly follows the relation
$\lambda=1.5 R_s$ as predicted previously.
Notice that the  PS starts
flattening 
before the window spectrum flattens; as in PVGH, the change
in slope occurs for $|W_k|^2<0.2$, value that they assumed as preliminary 
condition
for the estimate Eq.\ref{eps40} to be valid.
The amplitude of the PS    scales according to Eq.\ref{eps19}.
In Fig.\ref{fig51} it is shown  
the behavior of $\: \Pi(k)$ 
\bef 
 %\vspace{}
\epsfxsize 10cm
\centerline{\epsfbox{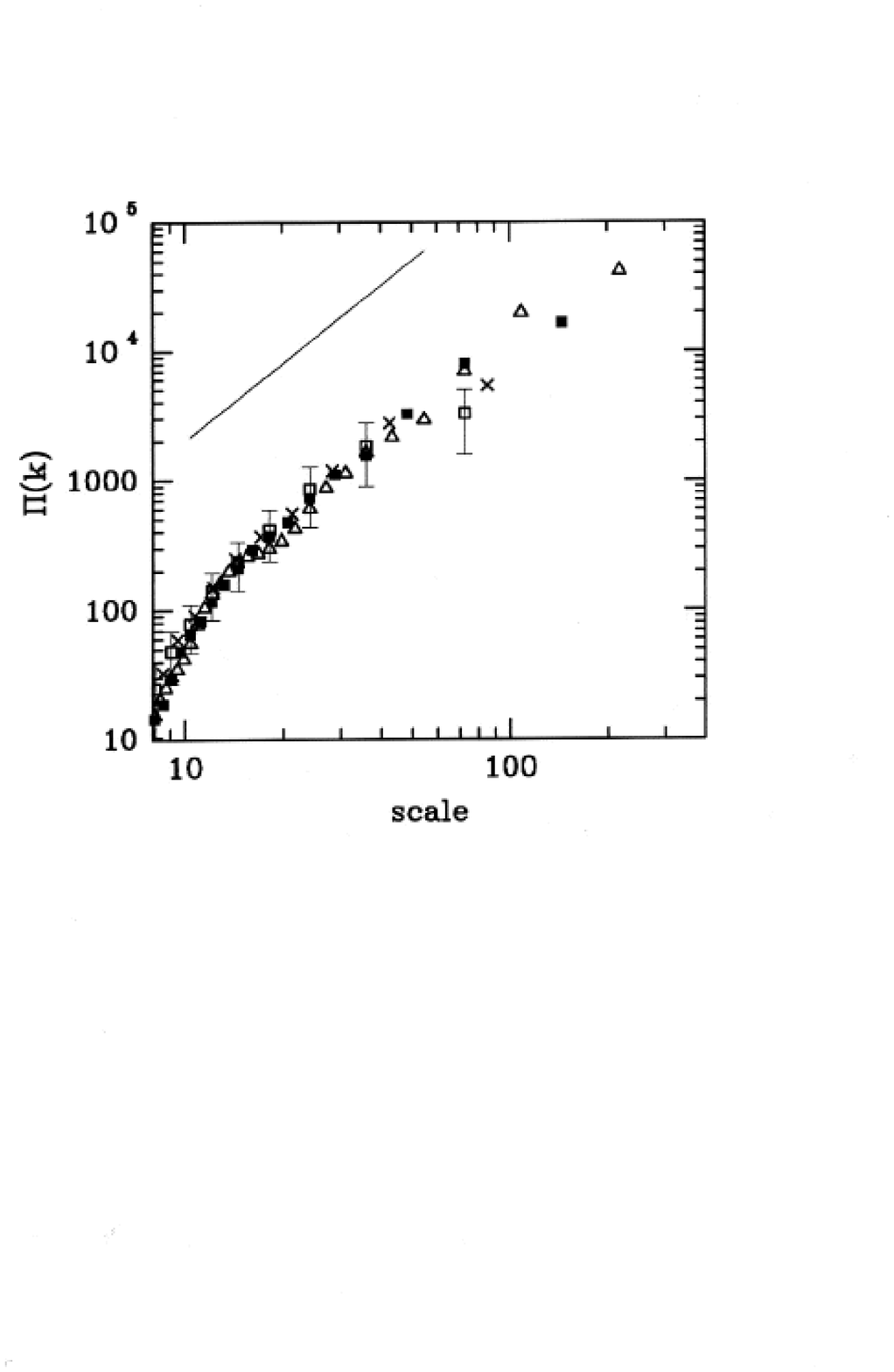}}
\caption{\label{fig51} The scale-independent PS    for the same fractal and 
the same4.2
scales as in Fig.50. 
Now the spectrum is a single power law, up to the 
very few large
scales, at which the flattening of the window PS    make the spectrum
estimator invalid.
 The amplitude of the spectrum is now constant (inside the errors). 
(From Sylos Labini \& Amendola, 1996).}
\eef
computed from Eq.\ref{eps40}
with $<|\rho_k|^2>$ in place of $<|\delta_k|^2>$
and without the Peacock-Nicholson correction:
%(\ref{eps27}): 
as predicted,
its amplitude does not scale with the sample size and it is
characterized by a single power law behavior, up to  very large scales
(where the two terms in the coefficient $A'$
of $\Pi(k)$ are comparable) .

\subsection{CfA2 and SSRS2}
\label{powercfa2}

Let us summarize the results of
PVGH on CfA2, by comparing them with
the analysis of the PS for
a fractal distribution:
i) for $\:k \ge 0.25$ ($\:\lambda \le 25 h^{-1}$ Mpc) the PS
is very close to a power law with slope $\:n=-2.1$. In our view, this is the
behavior at high wavenumbers
connected with the real fractal dimension.
ii) For  $\:0.05\le k \le 0.2$  ($120\hm>\lambda>30 \hm$)  the spectrum
is less steep, with a slope  about $\:-1.1$.
This bending is, in  our view, solely due to the
finite size of the sample.
iii) The  amplitude of the
volume limited subsample CfA2-130
PS is $\:\sim 40\%$ larger
than for CfA2-101.
%; the same trend is found in the $\:\xi(r)$ analysis.
This linear scaling of the amplitude can be
understood again considering that the sample is
fractal with $\:D=2$.
It is important to notice
that this trend is qualitatively
confirmed by the results of Peacock \& Nicholson \cite{pn91}
who find a higher PS amplitude for a deep radio-galaxy survey.
On the contrary the PVGH
explain this fact considering the dependence
of galaxy clustering on luminosity: brighter galaxies
correlate more than fainter ones. They support this
interpretation with the observation that brighter galaxies tend to avoid
underdense regions; they also analyze separately two subsets of
the same volume-limited sample of CfA2, one brighter than the other,
and  find in some cases
a luminosity-amplitude correlation.
However, PVGH do {\it not}
 detect such a luminosity segregation for the two largest
subsamples, CfA2-101 and CfA2-130, to which we are comparing
 here.

In Fig.\ref{fig52}
\bef
 %\vspace{} 
\epsfxsize 10cm 
\centerline{\epsfbox{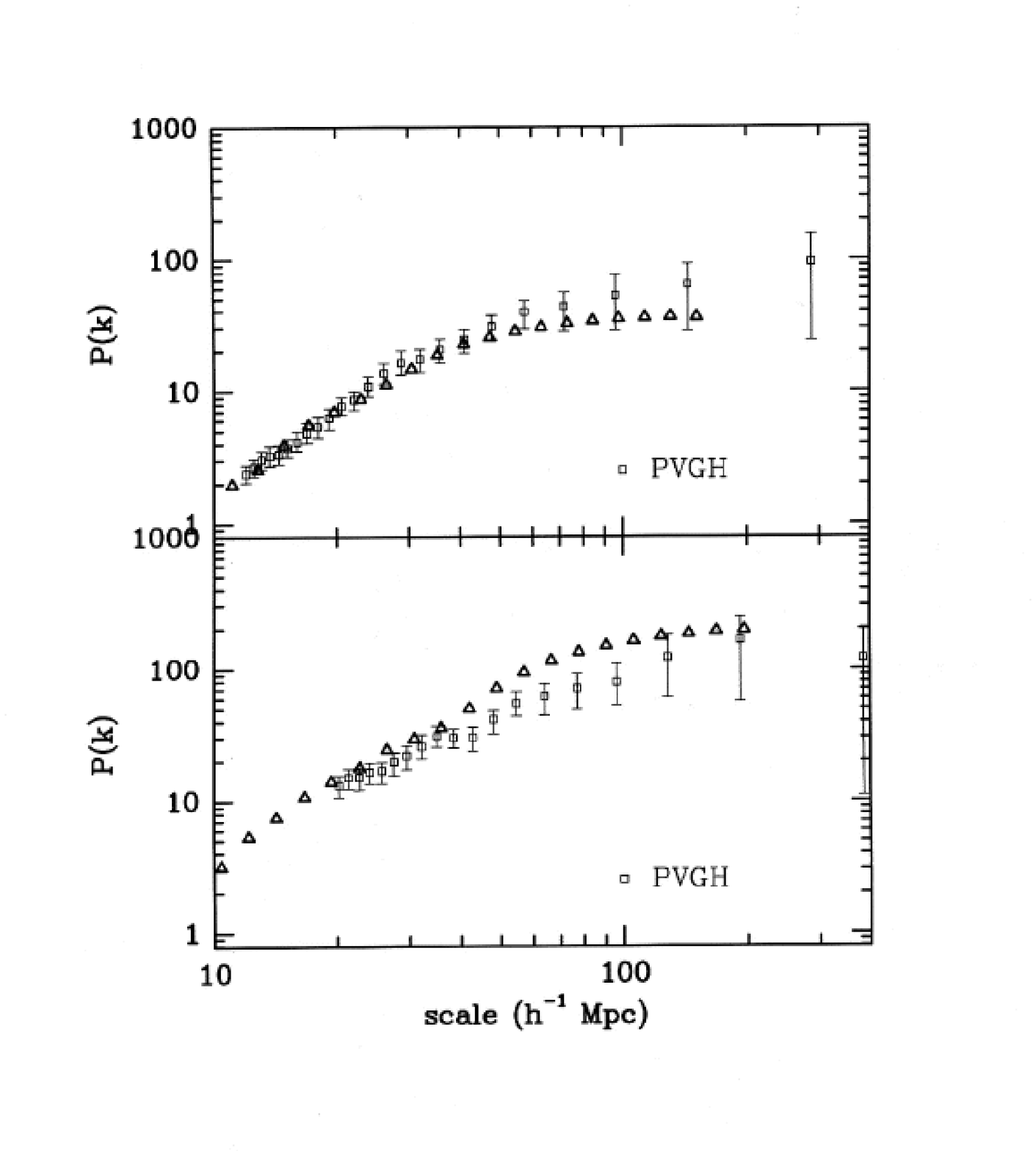}}
\caption{\label{fig52}  
Comparison of the power spectra of fractal distribution (triangles)
with the CfA2 survey (squares). In the top panel 
we plot the PS of the subsample
CfA2-130 (PVGH) along with the PS of our artificial fractal distribution 
(without the error bars for clarity). In the bottom panel, we 
plot CfA2-101 (PVGH) and a subsample of the same fractal as above, with a 
correspondingly scaled depth (from Sylos Labini \& Amendola, 1996).
}
\eef
 we compare directly the PS of our artificial catalogs with
the  PS of the CfA2 subsamples
 obtained generating two volume limited
subsamples at 130 and 101 $\hm$ (PVGH).
The physical scale has been computed matching the CfA2-101 galaxy 
average density.
  Both the shapes and the amplitudes
are  compatible with a fractal distribution:
for CfA2-101 the agreement is excellent; for CfA2-130 the two curves
are compatible inside the errors.

It seems therefore that the amplitude scaling at these scales can be entirely
attributed to the fractal scaling.
Our conclusion is then that the fractal nature of
the galaxy distribution can explain, to the scales
surveyed  in the CfA2 survey, the shift of the amplitude
of the PS.

\subsection{LEDA}
\label{powerleda}

The LEDA database has been described   elsewhere 
(see Sec.\ref{corran} and the Appendix).
Here we only present the subsamples we have 
analyzed\footnote{We thank 
L. Amendola and H. Di Nella 
for their useful  collaboration in the analysis of the 
power spectrum of LEDA.}.
We extract from LEDA 
two magnitude limited samples, one at $m_{lim}=15.5$ and
the other at $m_{lim}=16$. The first cut is to reduce it to
the CfA-North boundaries (R.A.
$8^h<\alpha<17^h$, declination $8.5^0<\delta<44.5^0$,
with a further cut to reduce galactic obscuration, see PVGH),
and is denoted as LEDA-CfA sample. 
This sample contains 4593 galaxies in an angular area of 1.1 sr,
amounting to a completeness level of 70\% (CfA2 North contains
in this region 6478 galaxies, see PVGH). The LEDA-CfA sample is then
cut into several volume-limited subsamples, at depths
60,78,101 and 130 $\hm$, to reproduce the PVGH
results, and at various average magnitudes, as
summarized in Tab.\ref{tabbbleda1}.
The second sample, to be denoted as
LEDA16, contains galaxies with
$m<16$, and is  cut to $b>20^0$ to reduce galactic obscuration.
LEDA16 contains 12801 galaxies
for an angular area of $4.13 sr$, and is estimated to be complete
to 60\%.
\begin{table} 
\begin{center}
\begin{tabular}{|c|c|c|c|}
\hline
SAMPLE      &$M_1$ &  $M_2$	   &  numb.         \\
	 & &	   &  of gal.   \\
\hline
r60m187 & -19 & -18.5 & 330\\
r60m198 &- &-19.28&   301\\
r60m193 &-19.8& -18.9 &360\\
r78m193 &-19.5& -19.1& 492\\
r78m198 &-20.15 &-19.55& 431\\
r78m202 &- &-19.75& 445\\
r100m198 &-19.45 &-19.7 &414\\
r100m202 &-20.5 &-19.98 &428\\
r100m205 &- &-20.13 &431\\
r130m205 &-20.8 &-20.3 &371\\
r130m208 &- &-20.45& 343\\
\hline
\end{tabular}
\caption{\label{tabbbleda1}
The VL subsamples of LEDA-CfA. $M_1$  and  $M_2$ 
denote the brighter and the fainter limits of the 
absolute magnitude inside the sample.}

\end{center}
\end{table}

We extract from LEDA16 several VL samples,
at depths 60, 100, 130 and 150 $\hm$, and with various average magnitudes,
as summarized in Tab.\ref{tabbbleda2}.
 In the choice of the limiting absolute magnitudes
we tried to reach a compromise between maximizing the number of
galaxies, maximizing the range of average magnitudes,
 and also obtaining samples with a comparable number of galaxies within each 
depth-limited survey.
\begin{table} \begin{center}
\label{subsamples1}
\begin{tabular}{|c|c|c|c|}
\hline
SAMPLE      &$M_1$ &  $M_2$	   &  numb.         \\
	 & &	&     of gal.   \\
\hline
 r60m19 
  & - & -18.0  & 3710\\
 r60m185
  & -19.0 & -18.0  & 2045\\
 r60m195
  & -20.4 & -18.95& 1657\\
 r60m20
  & - & -19.55  & 791\\
 r60m202
  & - & -19.78  & 535\\
r100m198                                
  & - & -19.1756  & 4395\\
 r100m193                                
  & -19.44 & -19.1756  & 1057\\
 r100m202                                
  & - & -19.75 & 2134\\
 r100m205                                
  & - & -20.12  & 1044\\
 r100m208                                
  & - & -20.46  & 449\\
 r100m21                                 
  & - & -20.65  & 252\\
 r130m203                                
  & - & -19.8 & 2926\\
r150m206                                
  & - & -20.15  & 1953\\
 r150m202                                
  & -20.29 & -20.1462 & 504\\
 r150m208                                
  & - & -20.5  & 901\\
 r150m21                                 
  & - & -20.65  & 616\\
 r150m212                                
  & - & -20.87  & 331\\
\hline
\end{tabular}
\caption{\label{tabbbleda2}The VL subsamples of LEDA16. 
$M_1$  and  $M_2$ 
denote the brighter and the fainter limits of the absolute 
magnitude inside the sample. }
\end{center} \end{table}

Of course, here a major concern is incompleteness.
There are two major sources of incompleteness in a database
like LEDA. One is purely angular: there can be areas of sky less
sampled, and areas well sampled. This incompleteness can be estimated 
(see below), and can be described by an incompleteness angular factor
$f(\alpha,\delta)$. To our aim it is
important to notice that this source of incompleteness {\it does not
influence the fractal scaling}, because $f$ is independent on scale, 
and can be factored out
of the PS   . The second source occurs when redshifts
inside clusters
are better sampled than redshifts in the field. This incompleteness is
scale dependent, and as such it does change the scale-dependent
amplitude of $P(k)$. However, it is limited to the scale of clusters,
and it is therefore important only on scales less than 10 $\hm$, while we
extend our analysis to much larger scales. Therefore, the two major
sources of incompleteness are likely not to change our results.
So we 
 carried out several quantitative test of incompleteness already 
in \cite{ledaps}. Here,
we test for the incompleteness  effects  in
three ways. We compare the sample LEDA-CfA
with the results from the CfA2, which is complete to 99\%, 
and then we compare the deeper sample LEDA16 with LEDA-CfA.
All these tests consistently show that
the incompleteness does not severely bias our results to the scale
we consider.

The PS estimate is then accurate only up to the scale of the sample.
We  compare our data to the standard homogeneous picture
and to the alternative fractal prediction.
In the homogeneous picture, the  spectrum amplitude, just as the
correlation function, is independent of the sample size $R_s$.
The location of the flattening is also a real
scale\--inde\-pendent feature. As we noticed 
previously in the
fractal picture, on the contrary, the PS scales as $R_s^{3-D}$,
and a turnaround of the spectrum occurs
systematically around $R_s$.

We plot in Fig.\ref{fig53} the four PS of the VL subsamples of LEDA-CfA
\bef 
%\vspace{}
\epsfxsize 12cm 
\centerline{\epsfbox{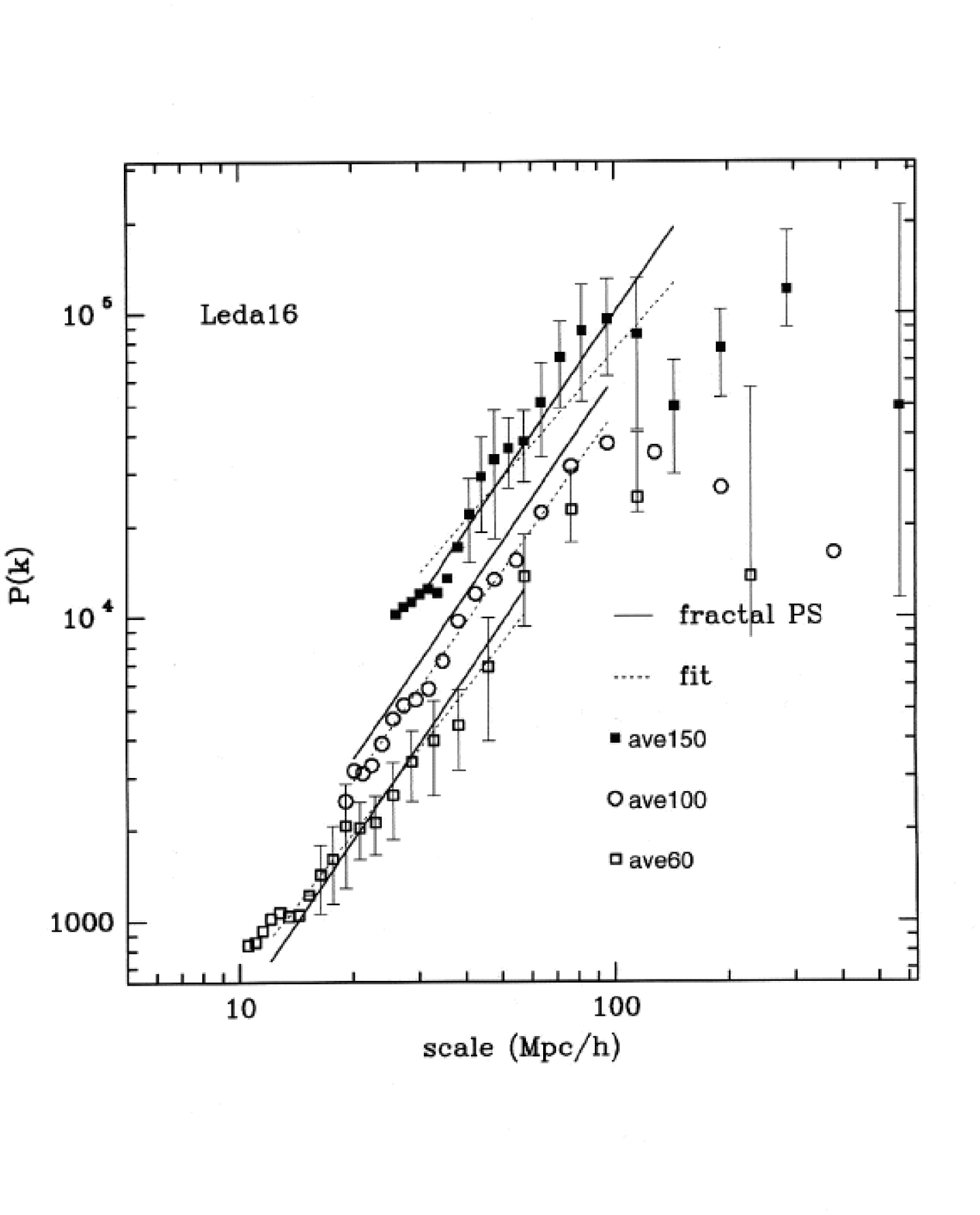}}
\caption{\label{fig53} The four PS of the VL subsamples of LEDA-CfA 
at 60,78, 101 and 130 $\hm$. The flattening, too, moves to
higher and higher scales, as expected in the fractal view}
\eef
at 60,78, 101 and 130 $\hm$, respectively.
(The errors, here and in what follows plots, are the scatter
among fractal simulations with dimension $D=2$, with the same
geometry and the same number of points as the real surveys.)
As reported also in PVGH (see previous section), 
the spectra amplitudes scale 
systematically with depth. The flattening, too, moves to
higher and higher scales, as expected in the fractal view (and puzzling if
interpreted in terms of a constant biasing factor).
 In Fig.\ref{fig54} we compare the 
spectra at
\bef %\vspace{}
\epsfxsize 10cm 
\centerline{\epsfbox{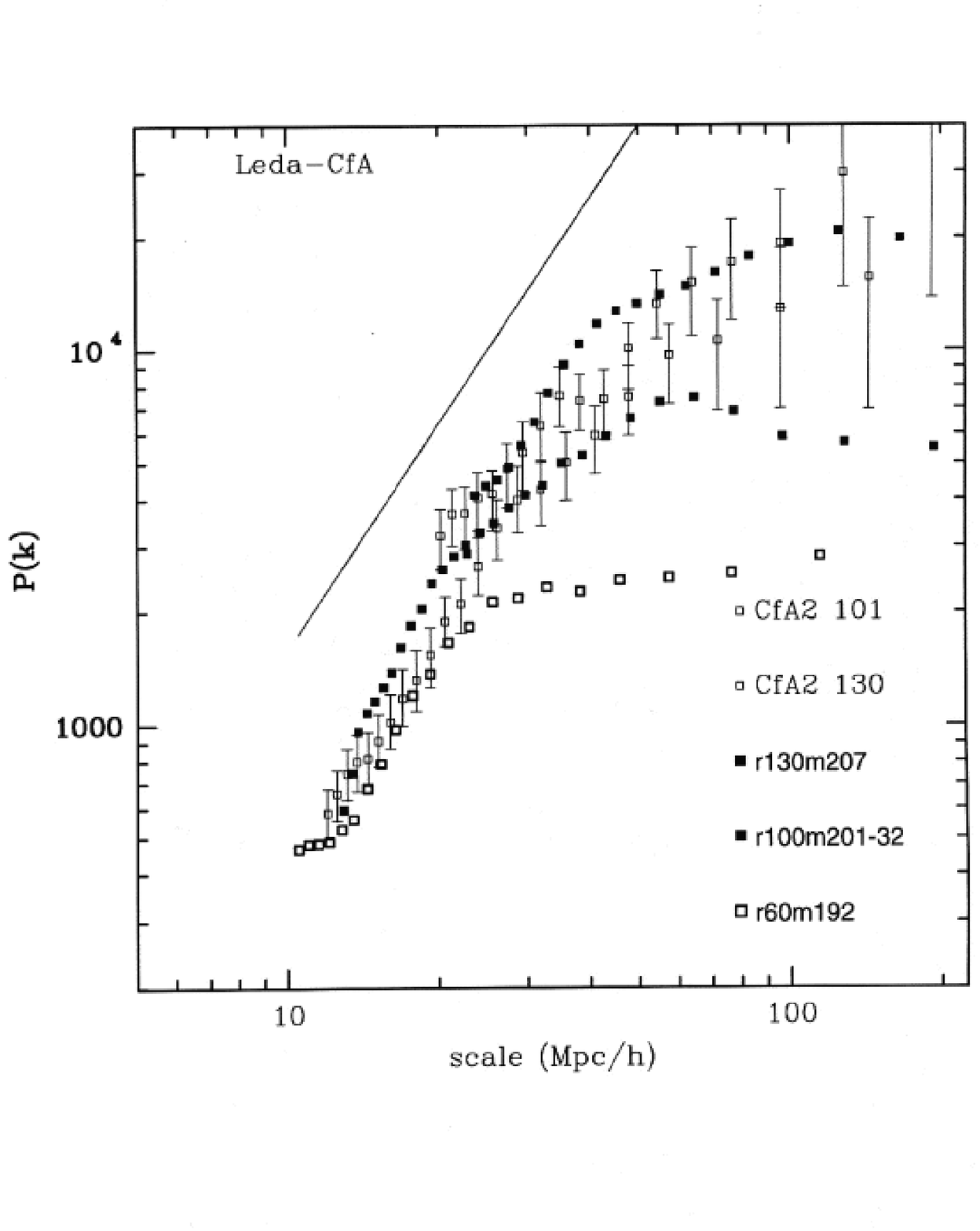}}
\caption{\label{fig54}  We compare the spectra at
101 and 130 $\hm$ with the corresponding CfA spectra (PVGH).  
The agreement is remarkable, given that LEDA is 70\% complete.}
\eef
101 and 130 $\hm$ with the corresponding CfA spectra (PVGH).
The agreement is remarkable, given that LEDA is 70\% complete
at this depth.

However, this scaling effect has been often attributed to luminosity
segregation (see e.g. PVGH \cite{dac94})
% and Da Costa \etal 1994), 
because these volume-limited samples differ in average
magnitude by increments of order $\Delta \hat M\approx 0.5$ .
We calculated then the  PS for all the fixed-depth, varying
magnitude subsamples of LEDA-CfA  
and no evidence of luminosity biasing can be detected.
For as concern LEDA-CfA, then, our conclusion is that the luminosity biasing
does not explain neither the PS scale dependence, nor the 
constant scaling, nor the shift of the flattening scale.
All three features, on the contrary, are naturally expected in
a fractal clustering.

In summary we carried out a detailed investigation of 
the PS    of galaxies in the LEDA database.
The comparison with CfA2 ensures us that the incompleteness
of our samples is not a source of severe distortion, at least down to
about $130 \hm$. We found several interesting
results, which complement the analysis already performed in
\cite{ledaps}.
First, we confirm the scaling of the PS    with $R_s$,
as predicted in a fractal distribution, down to 150 $\hm$, with no
evidence of real flattening.
Second, we find that samples with widely different average luminosities
have very similar spectra. The luminosity bias, where present, is neither
systematic nor constant in scale. A partial exception is detected in
the deepest sample
LEDA16-150 (which is also the less populated, and thus more noisy,  sample).
Third, we find no clear evidence of a trend of $D$ versus absolute
magnitude.

However, we can proceed further.
The LEDA-160  PS shows in fact no sign of convergence to homogeneity.
We may then take into consideration the two deepest subsamples.
If the galaxy distribution is fractal to these scales, the PS should
continue to
scale as $R_s^{3-D}$. We can then scale all the spectra to $360\hm$:
as we show in Fig.\ref{fig55} for the
North samples (the South ones are very similar),
 the fractal scaling gives again the correct
amplitudes (we use here $D=1.8$), although the deepest
samples are rather noisy. Notice that now this joint LEDA-360 PS
spans almost
 five orders of magnitudes! As before, a single power law $P\sim k^{-D}$
fits all the spectrum, over two orders of magnitude in the scale, and with
no sign of flattening. If one considers that we have one single
parameter to model the distribution, namely the fractal dimension $D$, the
alignment of all the spectra along a single line as shown
 in Fig.\ref{fig55} over such a range is indeed quite remarkable.
\bef 
%\vspace{}
\epsfxsize 12cm 
\centerline{\epsfbox{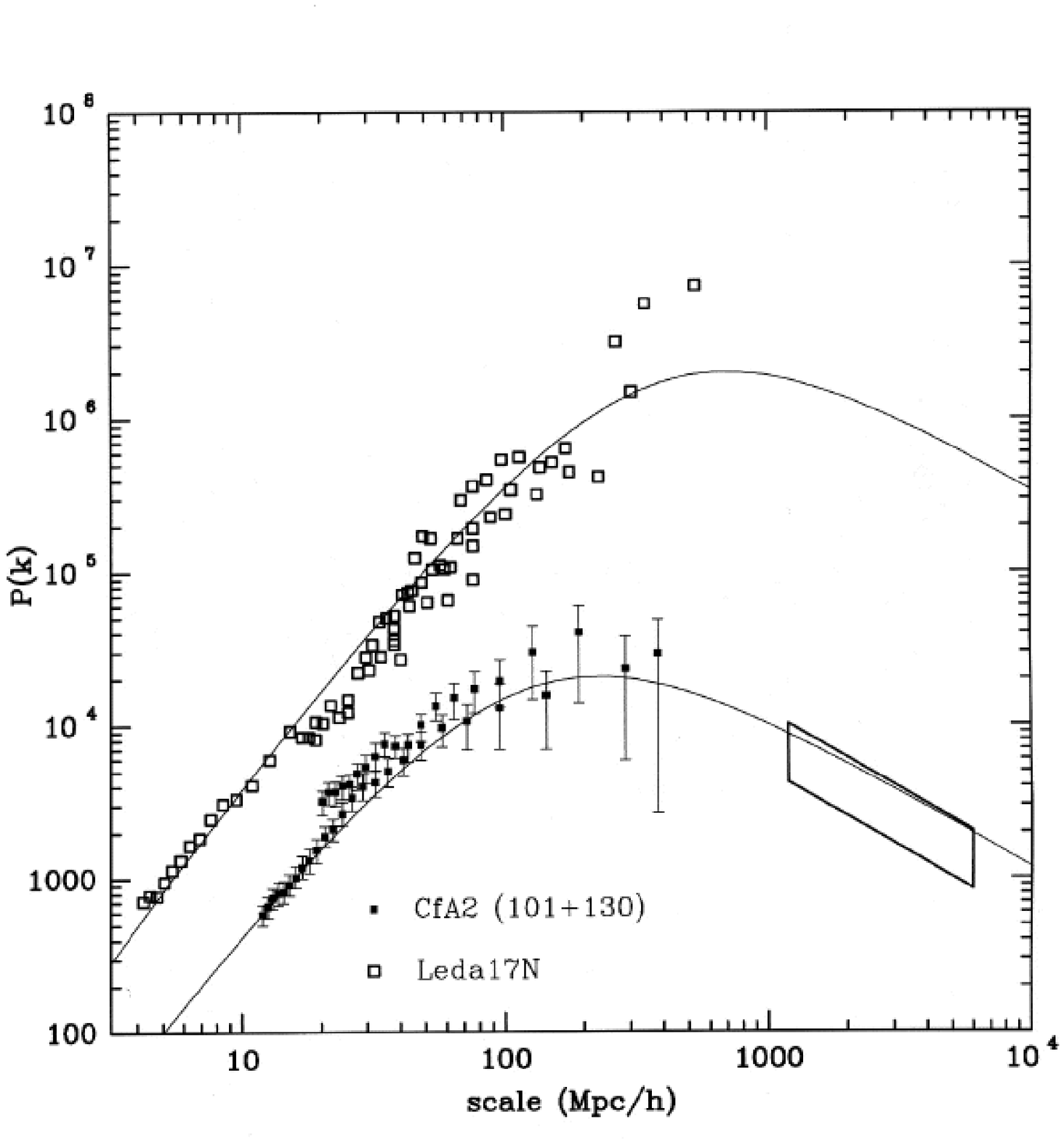}}
\caption{\label{fig55}  All the LEDA spectra to $360\hm$:
the fractal scaling gives again the correct
amplitudes (we use here $D=1.8$), although the deepest
samples are rather noisy. Notice that now this joint LEDA-360 PS
spans almost five orders of magnitudes.}
\eef
In the LEDA-360  PS
the  CDM best fit  gives  $\Gamma\approx .07, A\approx 10^{8.7}$,
as we show in Fig.\ref{fig55}. In terms of $\Omega$ this
result is clearly rather problematic: for $h=0.5$,
we get a value as low as $\Omega=0.14$. And remember
that this is only an upper bound!
Even more problematic the situation is looking at  the bias
factor, defined as the ratio
of the galaxy PS $P_g$ to the matter PS $P_m$, $b^2(\lambda=2\pi/k)
=P_g(k)/P_m(k)$.
To put in agreement our LEDA-360 spectrum with the
COBE data (from \cite{benn94} with $n=1$, $\Omega=1$ and
converted to redshift space) one would need $b(8\hm)\ge 3$,
$b(100\hm)\ge 4.8$, and increasingly  larger values at larger scales.

%%%%%%%%%%%%%%%%%%%%%%%%%%%%%%%%%%%%%%%%%%%%%%%%%%%%%%%%%%%%%%%%%%%%%%%%%%%%

\section{Statistical validity of galaxy  catalogs}
\label{validity}
How many galaxies one needs in order   to characterize correctly
(statistically) 
 the large scale distribution of visible matter ? This fundamental
 question is addressed in this section
 \footnote{We thank A. Gabrielli and 
 A. Amici for   useful  collaborations and discussions 
  in the a characterization of the 
dilution effects on real samples.}, and it  allows us to
 understand some basic properties of the statistical analysis of 
 galaxy surveys. In such a way, we are able to clarify  the
 concept of "fair sample", i.e. a sample which contains a
 statistically meaningful information \cite{slgmp96}.

\subsection{What is a {\it fair sample} ?} 
\label{fairsample}

  We have discussed in the pervious sections the properties of
 fractal structures and in particular we have stressed  the
 intrinsic highly fluctuating nature of such distributions. In this
 perspective, it is important to clarify the concept of a {\it "fair
 sample"}. Often this concept is used as synonymous of a
 homogeneous sample (see for example \cite{dac94})
 so the analysis of  catalogs along the traditional
 lines often leads to the conclusion that we still do not have a
 fair sample and deeper surveys are needed to derive the correct
 correlation properties. A corollary of this point of view is that,
 since we do not have a fair sample, its statistical analysis cannot
 be taken too seriously.

 This point of view is highly misleading because we have seen
 (Sec.\ref{statmec}) that self-similar structures never become homogeneous, so
 any sample containing a self-similar (fractal) structure is 
 automatically be declared "not fair" and therefore impossible to
 analyze. The situation is actually much more interesting, otherwise
 the statistical mechanics of complex systems would not exist.
 Homogeneity is a property and not a condition of statistical
 validity of the sample. A non homogeneous system can have well
 defined statistical properties in terms of scale invariant
 correlations, that may be perfectly well defined. The whole
 study of fractal structures is concerned with this \cite{epv95}.
 Therefore, one should distinguish
 between a "statistical fair sample", which is a sample in which
 there are enough points to derive some statistical properties
 unambiguously and a homogeneous sample, which  is a property that can
 be present or not but which has nothing to do with the statistical
 validity of the sample itself. We have discussed in Sec.\ref{corran}  
 that even the small
 sample like CfA1 is statistically fair up to a distance which can
 be defined unambiguously (i.e $\sim 20 \hmp$).

 In the Sec.\ref{radial} we  define the condition to select a sample large
 enough {\it to manifest the self-averaging properties of a fractal
 distribution}. We now study   a different, but related,
 question. Given a sample with a well defined volume, which is the
 {\it minimum number} of points which it should contain in order to have
 a {\em statistically  fair} sample, even if one computes averages
 over all the points, such as the conditional density or the
 conditional  average density \cite{slgmp96} ?

\subsection{Samples dilution}
\label{dilution}

 Consider a sample which contains a portion of a fractal structure
 with a lower cut-off $B$ (Sec.\ref{statmec}). A real fractal 
structure in general
develops between a lower and/or an upper cut-off, which respectively
 defines the smallest and the largest scale 
between which the self-similarity of the distribution can be found.
(See Sec.\ref{statmec}, and Sec.\ref{radial} 
for a detailed discussion of the role of the
 lower cut-off).
  Given the geometry of the sample (depth $R_s$ and solid angle
 $\Omega$), we investigate what happens if we eliminate {\em
 randomly} more and more points from the sample. The fractal
 dimension and the lower cut-off are not   affected by this
 depletion process, because they are related only to the intrinsic
 properties of the fractal structure. Instead, the
 correlation properties are more and more affected by a statistical
 noise as we cut the points which contribute to the statistics. This
 noise is superimposed to the genuine signal so that $D$ and $B$
 are not changed at all, but the estimate of their values becomes
 noisy.  Obviously, given a finite portion of the original system
 characterized by a lower cut-off $B$, it  exists a {\it maximum}
 value of the number of points which we can eliminate from the
 structure in order to conserve the genuine statistical properties
 of the original distribution.

 At this point we can characterize the statistical information in
 each VL sample more quantitatively. Suppose that
 the sample volume is a portion of a sphere with  solid angle
 $\Omega$ and radius  $R_{VL}$, and that the number of points
 inside  this volume, $N_{VL}$, is 
\be 
\label{sf5} 
N_{VL} = B_{VL}
 \frac{\Omega}{4\pi} R_{VL}^D \; 
\ee 
where $B_{VL}$ takes into
 account the luminosity selection effect (see Sec.\ref{corran}
 and Sec.\ref{radial}). The original
 system, inside the same volume contains 
\be 
\label{sf5b} 
N = B \frac{\Omega}{4\pi} R_{VL}^D 
\ee 
 i.e., there is not any limit in the luminosity distribution.
Hence, the fraction of galaxies present in the sample can be written
 as 
\be 
\label{sf6}
 p = \frac{N_{VL}}{N} = \frac{B_{VL}}{B} \; .
\ee
 In
 this way we can associate to each VL sample a well defined value
 of $p$ (see tables in the appendix).   In a given sample we can
 randomly cut some points in the sample and define $p$ as the
 fraction of points which remains inside (we can clearly have a sample 
 where both these effects are present).

 The crucial point that we examine now,
 is whether the random or luminosity (which we assume to be random in
 in space) cutting of point must stop at a certain limit. Beyond
 this limit one does not have in the sample enough points to
 recover the real statistical properties of the distribution. We 
 call {\em statistical fair sample} a sample which contains a number
 of points for unit volume larger than this limit. The problem is
 how to define this limit, or, in other words, to determine the
 minimal value of $p$ which allows one to recover the genuine
 information, for example, by the two points correlation analysis.
 We can proceed in two independent ways. First  by analyzing
 the correlation properties of the VL samples of real redshift
 surveys and  second  by studying artificial distributions
(see \cite{slgmp96} for a more complete discussion of this problem).

 \subsection{Dilution effects on real redshift surveys} 
\label{dilutionreal}

We have shown in Sec.\ref{corran} that the average 
conditional density of some VL
 samples of various redshift surveys, as IRAS, APM and others,
 show a tendency towards a flattening. 
   The crucial point of the IRAS data analysis, for example,
 is whether these samples contain enough points to allow one to
 detect the real statistical properties of galaxy distribution. To
 this aim,  we compare the statistical properties of the IRAS
 samples with those of well-known optical ones, for
example Perseus-Pisces,
 and we clarify the role of finite size effects in the
 analysis.   First of all it is important to note which is the
 number of galaxies for unit of
 steradian, in the flux limited catalogs  (Tab. \ref{tabiras1}). This gives a
 first idea of the statistical  validity of the sample.
 To this end we note that the two IRAS catalogs, CfA1 and PP
 cover about the same depth but their
 {\it solid angles} are very different. 
 While for CfA1 $\Omega \approx 2$ and for PP $\Omega
 \approx 1$, the IRAS surveys are all-sky, i.e. $\Omega \approx 4
 \pi$.  The number of points for steradian in I-2 is about 
 the $25 \%$ of the CfA1 ones, and for I-12 it is about the $50
 \%$. This consideration gives a first qualitative indication that 
 the IRAS catalogs suffer of {\it a systematic depletion} with
 respect to the optical ones. We can characterize this situation in
 a more quantitative way.  
We can associate to each VL sample a well defined value of $p$ (see
 Tab.\ref{tabiras2}, and the tables in the
 Appendix).  In the VL limited samples of the PP survey we can
 eliminate randomly points up to reach the fraction contained in the
 various IRAS samples, and then we study the behavior of the
 conditional density. We  find  that the correlation function
 has a clear power law behavior up to $30 \hmp$, if the percentage
 remain larger than $1 \div 2 \%$, then it shows a crossover
 towards homogenization. 
 This is
 clearly spurious and due to the fact that we have reached the
 limit of statistical validity, or fairness, of the sample.

In Tab.\ref{tabiras2} we show the characteristics of the VL
 samples of the PP surveys, and the corresponding VL samples of the
 I-12 surveys, in the same region of sky. It appears evident that
 the numbers for the IRAS samples are very small, and correspond to
 the $10 \%$ of those of PP. This systematic {\it dilution effect}
 is the effective cause of the apparent homogenization found in
 IRAS samples. This explains also why  IRAS galaxies are located
 near the brighter optical ones, and do not fill the voids. They
 are not more homogeneously distributed but  they do not belong to
 a {\it statistically fair samples}.   Let us examine  
 the various
 redshift samples.
 \begin{table} \begin{center} 
\begin{tabular}{|c|c|c|c|}
\hline
&      &            &                
 \\ SURVEY       & $\Omega (sr)$    & Number  & Galaxies/ sr    \\
 &      &            &                 \\ 
 \hline
% \hline &      &            &                 \\ 
I-2 Jy     & 4$\pi$ & 2658 & 200 \\ 
%&      &            &                 \\
 I-12 Jy    & 4$\pi$ & 5320 & 450 \\         
%&      &            &                 \\ 
CfA1         & 1.8  & 2000 & 1000 \\
% &      &            &                 \\  
PP           & 1  & 3300 & 3300 \\ 
&      &            &                 \\ 
\hline 
\end{tabular}
 \caption{Number of galaxies in various flux limited surveys 
 \label{tabiras1} } 
\end{center} \end{table}   
\begin{table} \begin{center}
\begin{tabular}{|c|c|c|c|} 
\hline 
&      &            &                 \\ 
Sample           & $R_{VL}$ (Mpc) & Number  & Percentage        \\  
&      &            &                 \\ 
\hline &      &         &                 \\ 
VL40IRAS     & 40 & 27     & 2   \%    \\         
%&      &            &                 \\ 
VL40PP     & 40 & 283      & 15 \%    \\         
  %&      &            &                 \\
 VL60IRAS      & 60 & 113  & 2.6 \%    \\   
     % &      &            &                 \\ 
VL60PP     & 60 & 951   & 22 \\
 %            &      &            &                 \\ 
VL80IRAS  & 80 & 83    & 1.1  \%    \\    
% &      &            &                 \\
 VL80PP     & 80 & 875    & 11 \%    \\   
 &      &            &                 \\  
\hline
 \end{tabular}
 \caption{The VL subsamples of Perseus-Pisces and of I-12 in the
 same sky region \label{tabiras2} }
 \end{center} \end{table}

\subsubsection{Perseus-Pisces}  
\label{dilutionpp}

In the VL samples of the Perseus-Pisces survey we 
eliminate randomly points up to reach the fraction contained in
 the various IRAS samples. We note that the correlation function
 has a clear power law behavior up to $30 \hmp$ if the percentage
 is larger than 
\be 
\label{nn2}
 p \ge 1 \div 2 \% 
\ee 
 then it
 becomes too noisy and it is not possible to recover any power law
 behavior  (Fig.\ref{fig56}).  The quantity $p$ in Eq.\ref{nn2} 
is computed by knowing the value of $B$ in Eq.\ref{sf6}.
This is estimated from the amplitude of the conditional
density and the luminosity selection factor of the sample (see Sec.\ref{corran} and 
Sec.\ref{radial} for a more detailed discussion on the determination of $B$). 
From various VL samples of PP, CfA1 we obtain that $B \approx 15 Mpc^{-D}$.
\bef %\vspace{}
 \epsfxsize 8cm
 \centerline{\epsfbox{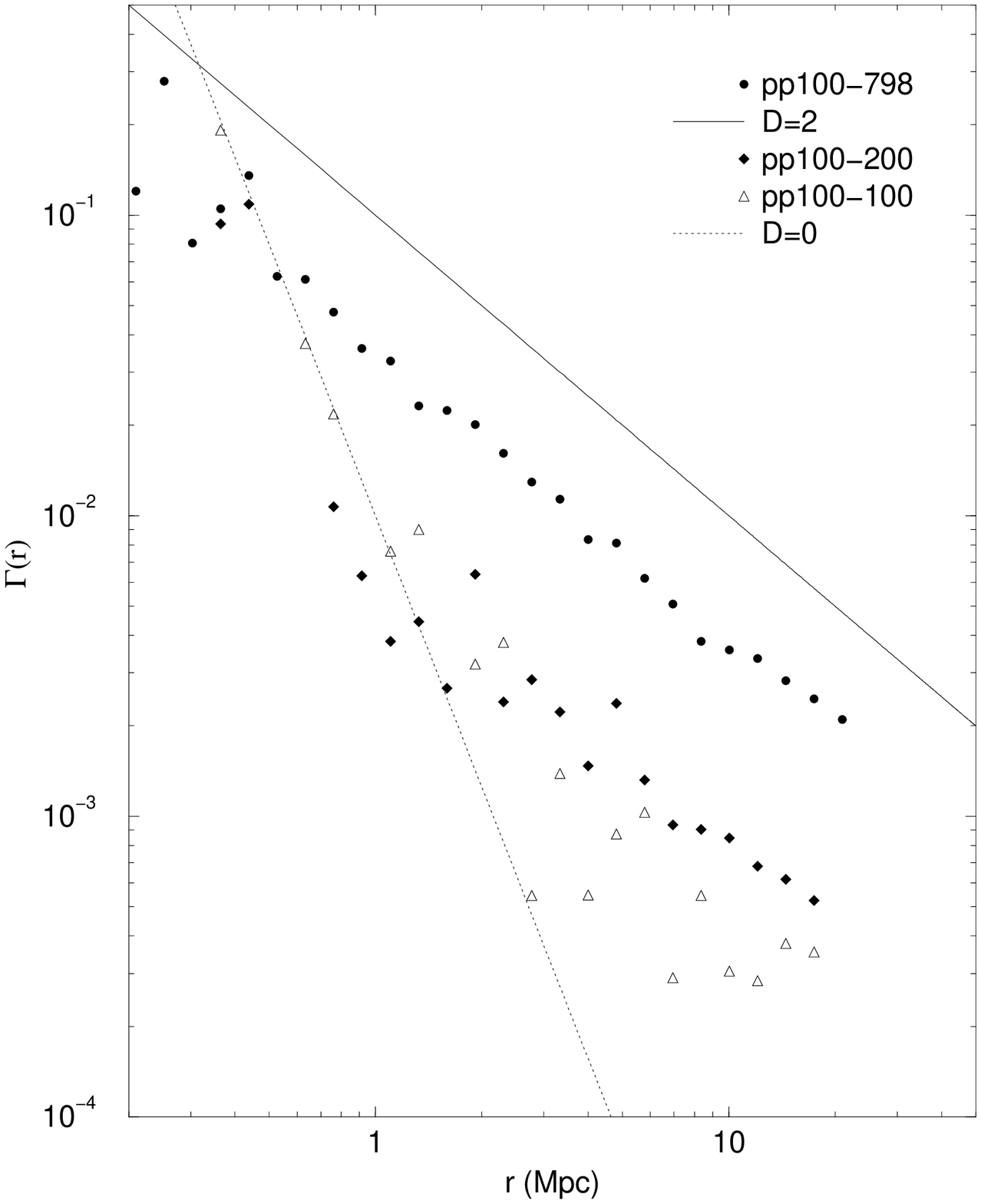}} 
\caption{\label{fig56}   The 
 conditional density $\Gamma(r)$ for VL at 100 $\hmp$ (pp100) of
 Perseus-Pisces. The reference line (solid) has a slope of
 $-\gamma=-0.9$. The whole sample has 798 galaxies and we
 have dilute it up to 100 galaxies. In this case we are well below
 of the condition $p \gtapprox 2 \%$. At small
 scales the behavior  is $1/r^3$ due to the sparseness of the
 sample. At larger distances the signal is too noisy, and does not
 allow one to recover the power law behavior. The sample with 200
 galaxies still exhibits a power law behavior, even if the signal
 is more  noisy.} 
\eef 
The tendency found in Fig.\ref{fig56} is due to the fact that
 we have reached the limit of statistical validity or fairness of
 the sample. At small scale the conditional density
 exhibits a $r^{-3}$ decay that is due to the extreme dilution.
 The apparent homogeneous behavior shown at larger
 scales, in the sample with 100 galaxies,   is related to the fact
 that in a given sample, the mean separation among galaxies
 $\ell$ grows when the number of points decreases ($\ell
 \sim (V/N)^{1/3}$; for a fractal $\ell$ is defined to be 
 the average minimum distance between neighbour galaxies as in Sec.2),
 and when it becomes of the same order of the
 largest void present in original structure in the sample
 volume,  the correlation properties are destroyed. This means that
 the artificial noise introduced by the depletion of points, have
 erased the intrinsic fluctuations of the original system. In this
 situation the system looks like an homogeneous one.

\subsubsection{IRAS}  
\label{dilutioniras}

In Fig.\ref{fig33}  we show the behavior of the conditional
 density for  various in VL of the IRAS 2 Jy survey. 
 The deepest sample VL100, that is the sample with the lower
 number of galaxies,  shows a  $1/r^3$ decay at small scale, due to
 the extreme depletion of galaxies 
at these distances, followed by a almost
 flat behavior:  this is completely spurious. We point out  that the
 other samples  VL40 and VL60 show the correct scaling behavior at
 the same scale at which the $\Gamma(r)$ for VL100 shows a tendency
 towards homogenization.  The same kind of behavior can be found
 for the VL samples of IRAS 1.2 Jy, as shown in Fig.\ref{fig34}. 
 While the samples VL20 and VL60 show a clear power law behavior, 
the sample VL120 (North and South) exhibits a $1/r^3$ decay at
 small distances, and an almost flat behavior at larger ones (i.e.
 for $ r \gtapprox 10 \hmp$).

\subsubsection{SSRS1} 
\label{dilutionssrs1}

In the case of SSRS1 we can find the same kind of behavior for the
 deepest  VL sample (VL120 for which $p \sim 0.45 \%$), as shown in
 Fig.\ref{fig57}. The behavior of the  conditional density has
 a $1/r^3$ decay at small scale, followed by an almost flat
 behavior at larger scales. Even in this case the flattening
 occurs where other VL samples (e.g. VL40, VL80) show a very well
 defined  power law behavior.  
\bef 
%\vspace{}  
\epsfxsize 8cm
\centerline{\epsfbox{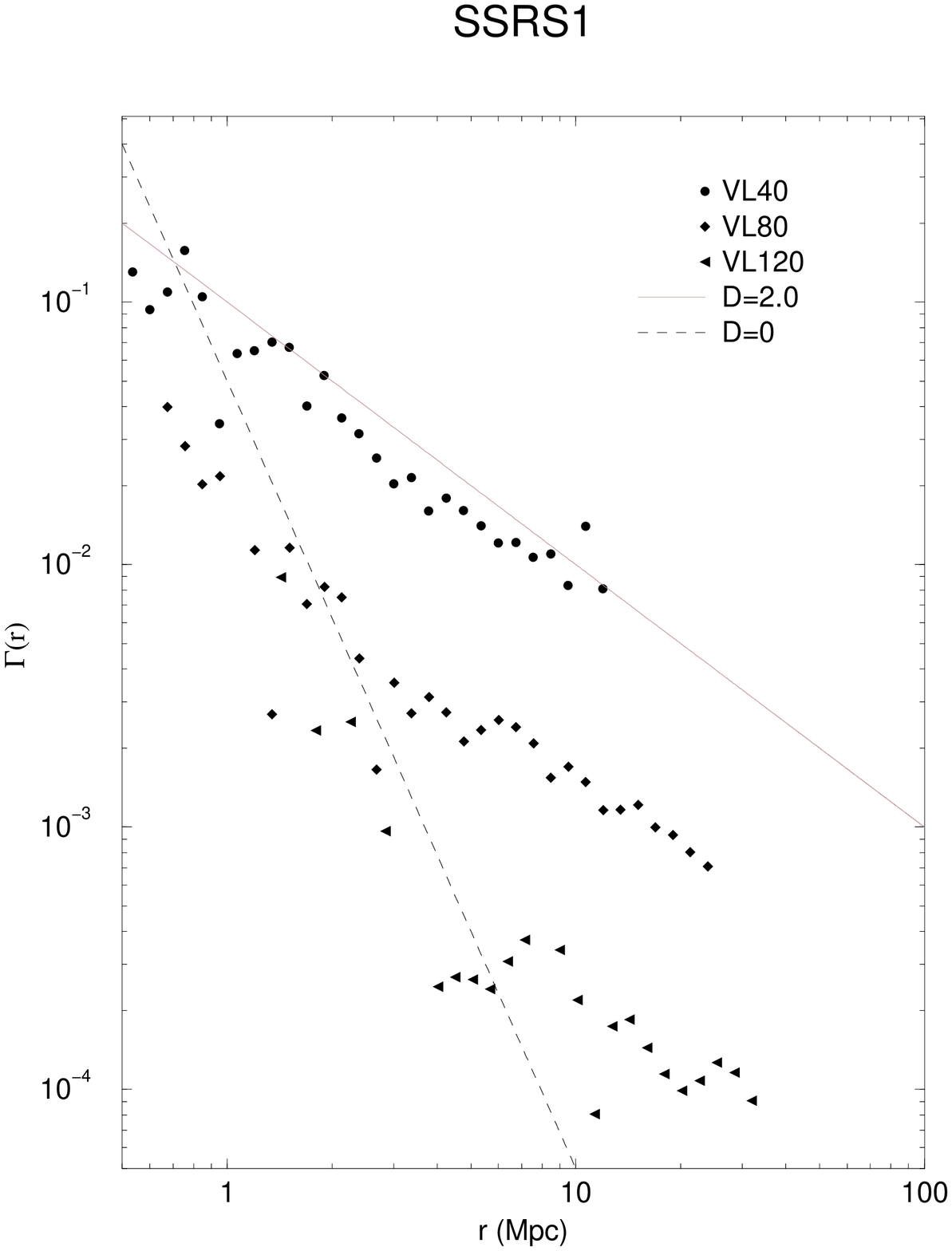}}
\caption{The 
conditional density $\Gamma(r)$ plotted as a function of the length
 scale for various volume limited of  SSRS1: VL40, VL80, VL120. The
 difference in amplitude is simply due to the different luminosity
 selection in the various VL samples, and can be renormalized by
 taking into account the different luminosity factor of these
 subsamples.  A reference slope of $-\gamma=-1$ is indicated by the
 dotted line, which  corresponds to a  fractal dimension of 
 $D\approx 2$. The dashed line has a slope $-3$ (D=0).
 \label{fig57}} 
\eef
 The behavior of $\Gamma(r)$ for the sample VL120, is
 again, spurious and due to the extreme depletion of galaxies.

\subsubsection{CfA1} 
\label{dilutioncfa1}

The case of CfA1 is shown in Fig.\ref{fig58}. 
\bef 
%\vspace{}  
\epsfxsize 8cm 
\centerline{\epsfbox{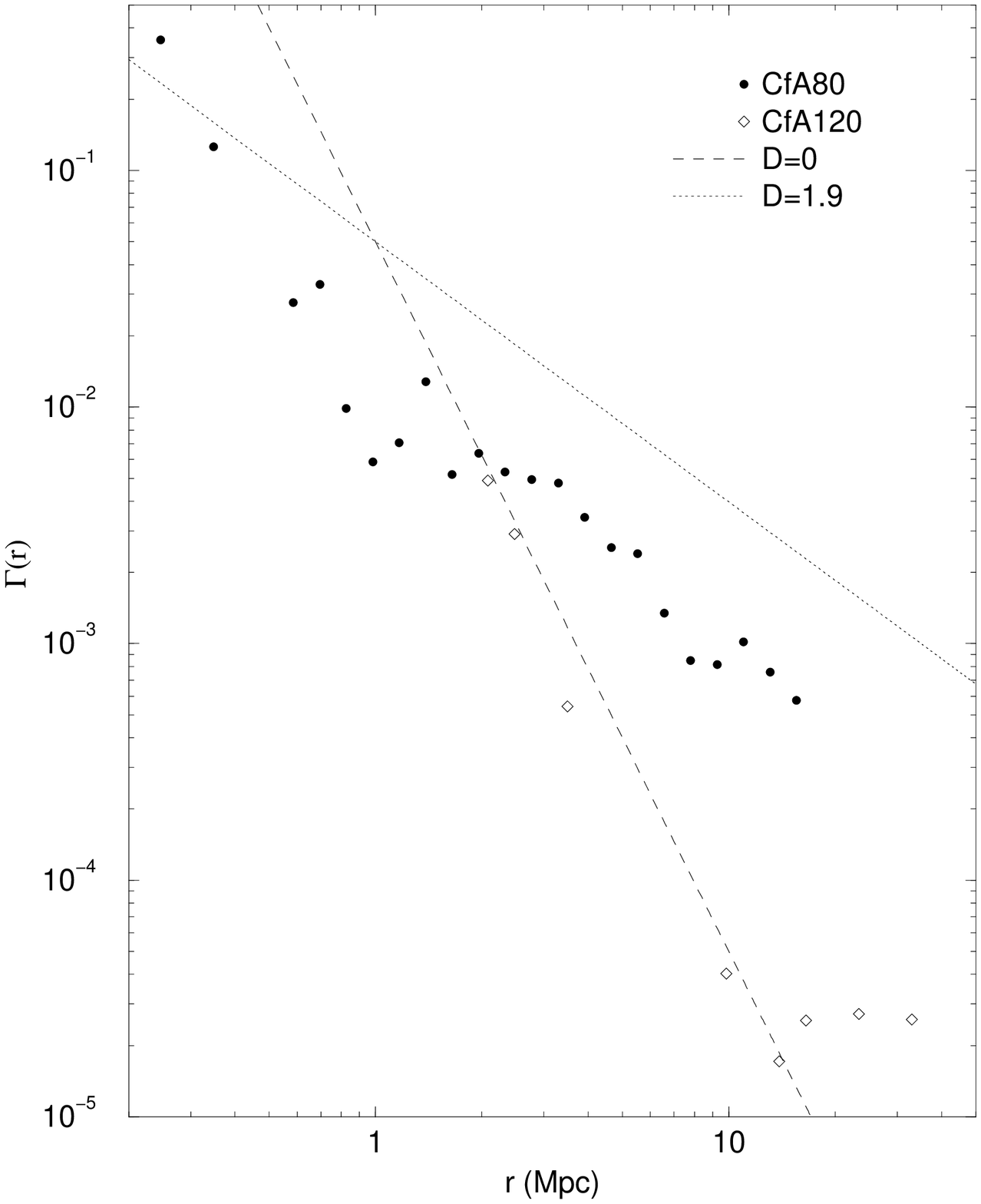}} 
\caption{The conditional density $\Gamma(r)$ plotted
 as a function of the length scale for various volume limited of
 CfA1: VL80 and VL120. A reference slope of $-\gamma=-1.1$ is
 indicated by the solid line, which  corresponds to a  fractal
 dimension of  $D=1.9$ up to $35\:h^{-1} Mpc$. The dashed line has
 a slope $-3$ (D=0). \label{fig58}} 
\eef 
Even in this case the
 deepest and more dilute sample 
 (VL120) the conditional density shows a "pathologic" 
 behavior: a $1/r^3$ decay at small
 distances, followed by a flattening. The conclusion is
 analogous to the previous cases, i.e. this behavior is simply due
 to the fact that  this sample is not a "statistically fair"
 sample because too dilute.

\subsubsection{Stromlo-APM}  
\label{dilutionsars}

Finally we discuss the case of the APM-Stromlo redshift survey.  Even
 in this case only in  the deeper and more dilute samples the
 conditional  density shows a tendency towards a flattening (see
 Fig.\ref{fig21}).  In Fig.\ref{fig59} 
\bef 
%\vspace{}
\epsfxsize 8cm
 \centerline{\epsfbox{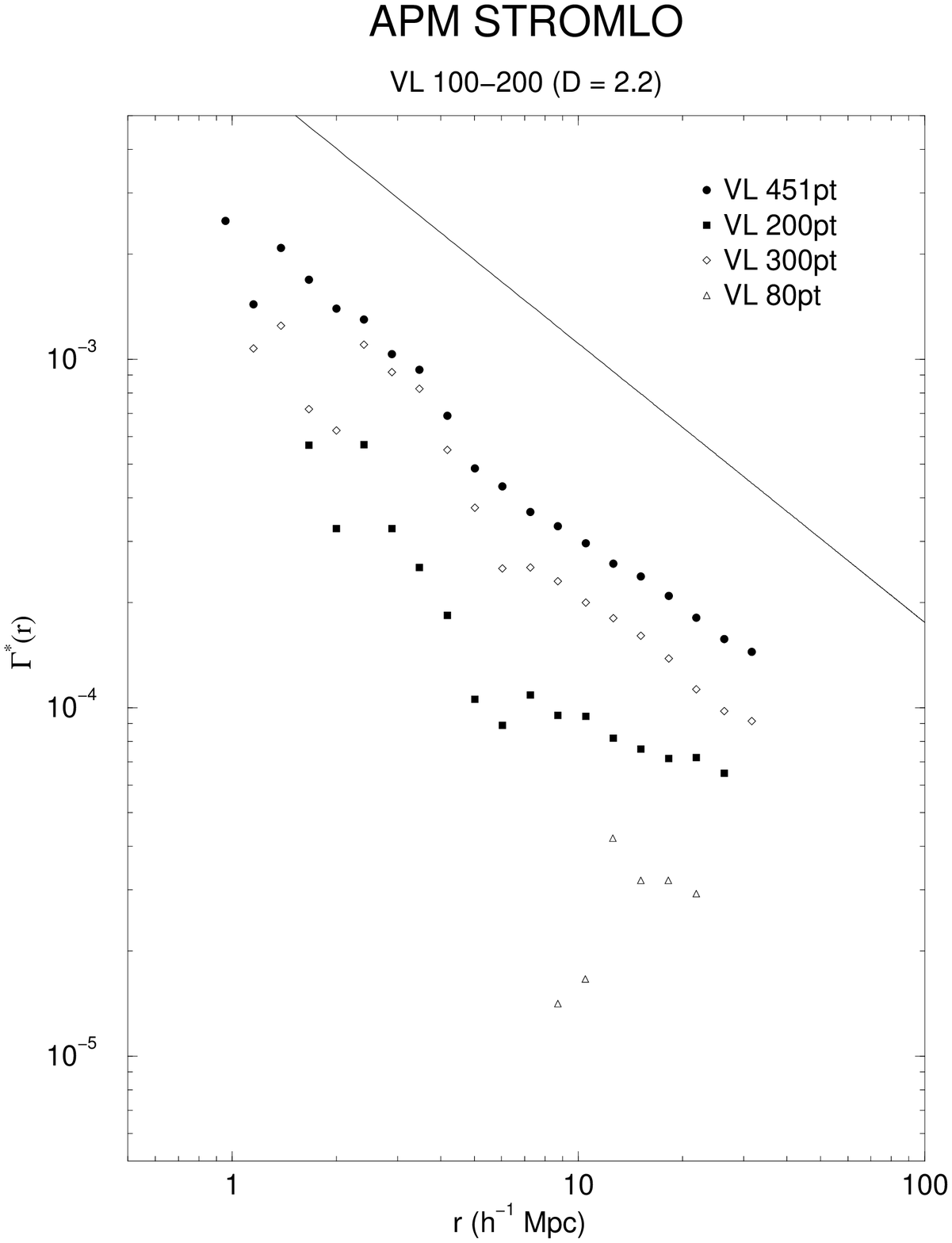}}
 \caption{\label{fig59} The
 average conditional density $\Gamma^*(r)$ the VL12 sample of APM 
 with progressively less galaxies (from 451 to 80 galaxies). The 
 reference line has a slope of $-\gamma=-1$.}
 \eef
 we show the
 behavior of the conditional average density for the VL12 (see
 appendix) in which we progressively eliminate, randomly, galaxies.
 While the whole sample shows very well defined
 fractal correlations,  as the number of galaxies is reduced, the
 signal becomes more and more noisy.  The sample with 200 galaxies
 shows a flattening, while the sample with 80 galaxies contains too
 few galaxies to allow one to reconstruct  the genuine properties
 of the distribution.

 \subsubsection{Artificial fractal distributions: the role of {\it Lacunarity}}
\label{dilutionartific}

 We
 have considered also an artificial catalog with a priori assigned
 properties generated with the random $\beta-$model algorithm
 \cite{ben84}. 
We have found \cite{slgmp96} that also in this case 
 we can recover the right statistical properties only in the limit
 of Eq.\ref{nn2}. Clearly Eq.\ref{nn2} depends on the
 morphological features of the realization of the fractal structure
 and the percentage can weakly fluctuate from a realization to
 another. However, even 
in this case  in the more dilute subsample  the
 conditional density shows a $1/r^3$ decay at small scale, while at
 larger scale 
it reaches an almost flat behavior. This tendency looks like the
 one found in the real catalogs, and it is   due to the fact that
 we have reached the limit of statistical validity of the sample.

The role of lacunarity 
(Sec.\ref{statmec}) is clearly important in the definition
of a statistically fair sample. In fact, if lacunarity of 
the distribution is large enough (Fig.\ref{fig60} upper part),
 this means that the 
system is characterized by having some well defined structures 
and large voids (with respect to the sample size). It is clear that 
in this case one has to eliminate a large amount of points to destroy
these structures and to lead a situation in which the 
remaining points seem to be homogeneously distributed. On the other hand 
if lacunarity is small with respect to the sample size
(Fig.\ref{fig60} bottom part), then the
typical dimension of voids is not so large and the structures are not
well defined as in the 
previous case. In this situation, if one eliminates a certain 
amount of points, it is easy to destroy the fractal long-range correlations.
\bef 
%\vspace{}
\epsfxsize 10cm
\centerline{\epsfbox{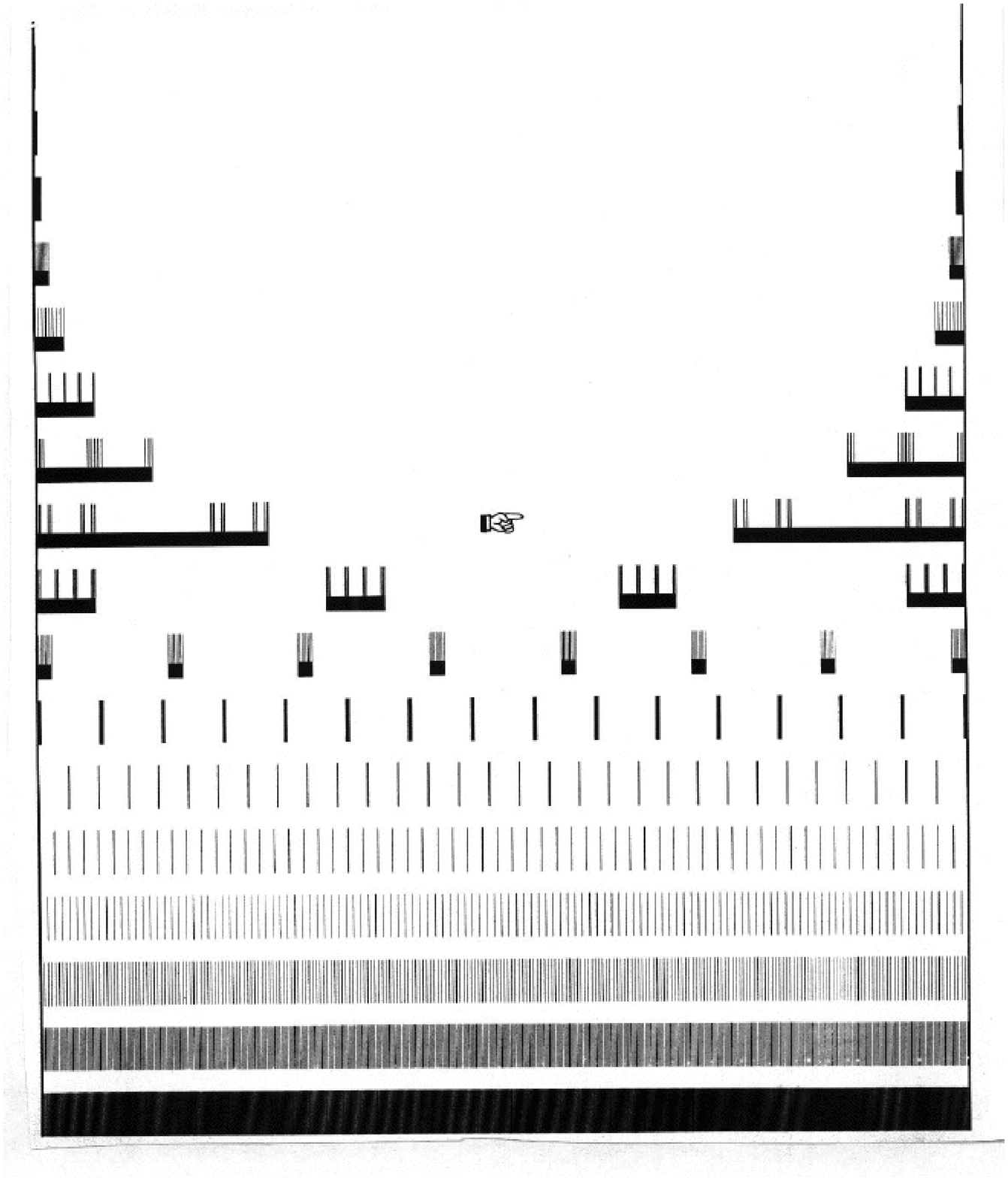}} 
\caption{\label{fig60} Examples of deterministic 
Cantor set with the same fractal 
dimension and different lacunarity. The lacunarity becomes
larger from the bottom to the top. (Courtesy of B. Mandelbrot).} 
\eef 
A  more quantitative study is needed in order to 
characterize properly the role of lacunarity in this respect.

%%%%%%%%%%%%%%%%%%%%%%%%%%%%%%%%%%%%%%%%%%%%%%%%%%%%%%%%%%%%%%%%%%%%%%%%%

\section{The decay of the radial density in wide and in deep surveys}
\label{radial}

In the previous sections we have discussed the methods
that allow one to measure the conditional (average) density
in real galaxy surveys. This statistical quantity is an average
 one, since it is determined by performing an average over all the
 points of the sample. We have discussed   the robustness
 and the limits of such a measurement. We have pointed out 
 that the estimate of the conditional density can be done up to
 a distance $R_{eff}$ which is of the order of the radius of the
 maximum sphere fully contained in the sample volume. This is
 because the conditional density must be computed {\it only 
in spherical shells.}
This condition puts a great limitation to the volume studied,
 especially in the case of 
deep and narrow surveys, for which the maximum depth $R_{s}$ 
can be 
one order of magnitude, or more, than the effective depth $R_{eff}$.

In this section we discuss the measurement of the {\it radial density} 
in VL samples\footnote{We thank A. Gabrielli and A. Amici
for various useful  discussions and for their collaboration 
in the
determination of the radial density properties.} \cite{slgmp96}. 
The determination of such a quantity
allow us to extend the analysis of the space density well
 beyond the depth   $R_{eff}$.
The price to pay is that such a measurement is strongly affected by
 finite size spurious fluctuations, {\it because it is not an average
 quantity. } These finite size effects require a great caution 
 \cite{slgmp96}: the behaviour of statistical quantities (like the 
 radial density and the counts of galaxies as a function of the apparent
 magnitude) that are not averaged out, present new and subtle problems.
  
We briefly also discuss the 
determination of the radial density in magnitude limited (ML) 
 redshift surveys. In particular  we  show 
that such a measurement does not allow one to determine the 
fractal versus homogeneous properties of the sample.
Finally we review all the determinations of the space density
which we have presented in this review (i.e the conditional density
 and the radial density in the different VL samples of the 
various redshift surveys) by  showing
 the consistency of all these measurements. This result is
 particularly important because it shows that all the redshift
 surveys discussed  present consistent properties.
{\it We are able to present  the behavior of the 
galaxy space
 density in the range of distances $0.5\div 1000 \hmp$. This shows
 that all the available redshift samples have the same (in slope
 and amplitude) long-range fractal correlations with dimension $D
 =2 \pm 0.2$, up to limits of
 the available samples.}

\subsection{Finite size effects and the behavior of the radial
 density} 
\label{radialfinite}

  In this section we discuss the general problem of the minimal
 sample size which is able to provide us with a statistically 
 meaningful information. For example, the mass-length relation for
 a   fractal, which defines the fractal dimension, is 
(see Sec.\ref{statmec})
 \be 
\label{efs1} 
D=\lim_{r \rightarrow \infty} 
\frac{\log(N(<r))}{\log(r)} 
\ee  
However this relation is  {\em properly defined only in the
 asymptotic limit}, because only in  this limit  the fluctuations
 of  fractal structures are  self-averaging.  A fractal
 distribution is characterized by large fluctuations at all scales 
 and these fluctuations determine  the statistical properties of
 the structure. If the structure has a lower cut-off, as it is the
 case for any  real fractal, one needs a {\em "very large sample"}
 in order to recover the statistical properties of the distribution
 itself. Indeed, in any real physical problem  we would like to
 recover the asymptotic properties from the knowledge of a  {\em
 finite portion} of a fractal and the problem is that   a single
 finite realization of a random fractal is affected by finite size
 fluctuations due to the  lower cut-off.  

In a   homogeneous distribution we can define, in  average,
 a characteristics volume associated to each particle. This is the
 Voronoi volume \cite{vo08} $v_v$ whose radius $\ell_v$ is of the
 order of the mean particle  separation. It is clear that the
 statistical properties of the system can be defined only in
 volumes much larger than $v_v$. Up to this volume in fact we
 observe essentially nothing. Then one begins to include a few
 (strongly fluctuating) points, and finally, the correct scaling
 behavior is recovered  (Fig.\ref{fig61}). 
\bef %\vspace{} %\vspace{}
\epsfxsize 12cm 
\centerline{\epsfbox{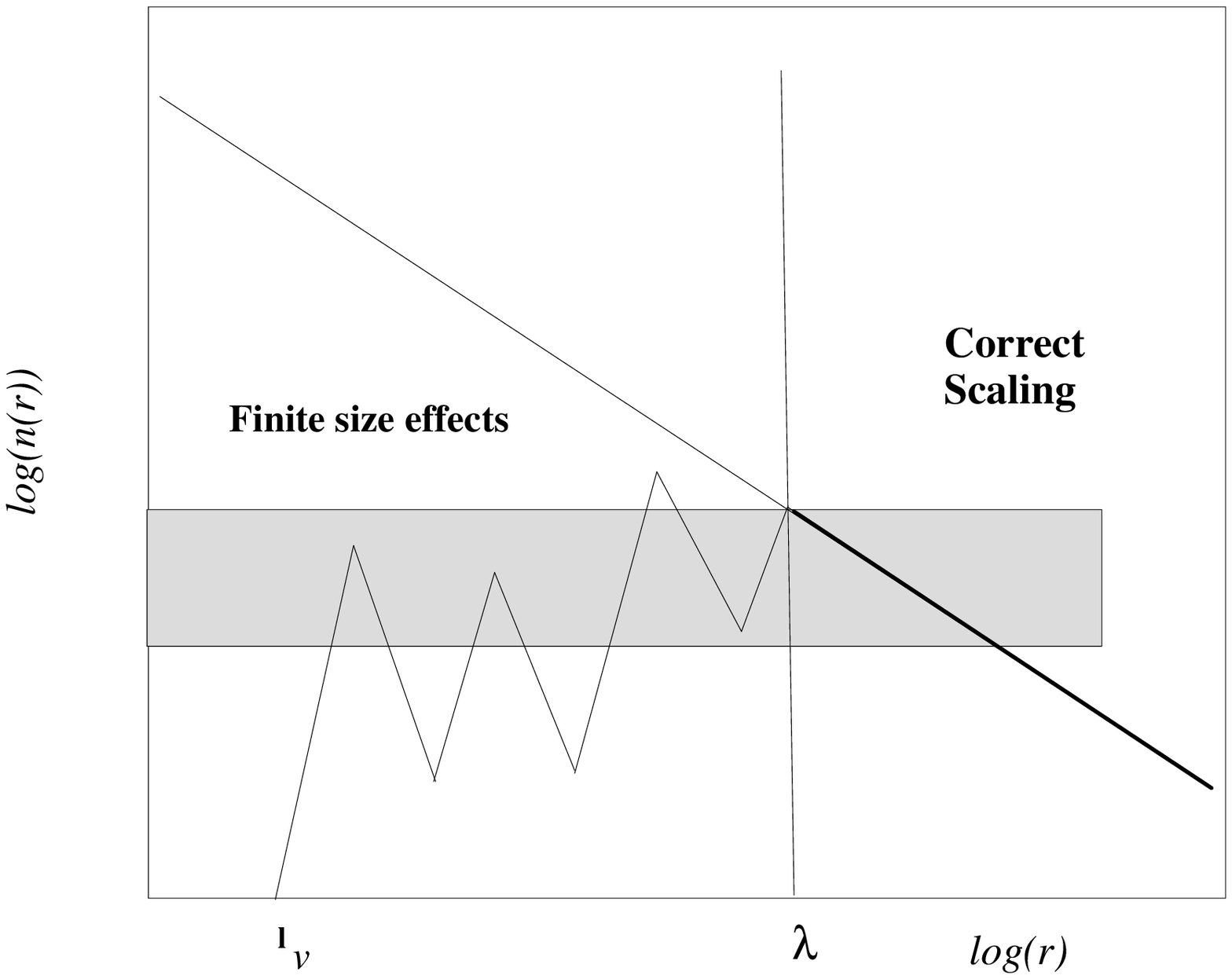}} 
\caption{\label{fig61}  Behavior of the density computed from one
 point, in the case of a fractal
distribution. At small distances
below the average mean separation between neighbor galaxies,
one finds no galaxies. 
Then the number of galaxies starts to grow, but this regime
 is strongly affected by finite size fluctuations. Finally the
 correct scaling region $r \approx \lambda$ is reached. In the
 intermediate region the density can be approximated roughly by a
 constant value.   This leads to an apparent exponent $D \approx
 3$.    This exponent is not real but just due to the size effects.
 } 
 \eef 
For a Poisson sample consisting of $N$ particles inside a volume
 $V$ then the Voronoi volume %(Fig.\ref{vor}) 
is of the order 
\be
 \label{v1} 
v_v \sim \frac{V}{N} 
\ee 
and $\ell_v \approx v_v^{1/3}$. In the case of homogeneous
 distributions, where the fluctuations have  {\em small amplitude}
 with respect to the average density, one  readily recovers the
 statistical properties of the system at small distances, say, $ r
 \gtapprox 5 \ell_v$. 

The case of fractal distribution is more subtle. For a self-similar
 distribution one has, within a certain radius $r_0$, $N_0$
 objects. Following \cite{cp92} we can write the mass-length
 relation between $N(<R)$, the number of points inside a sphere of
 radius $R$, and the distance $R$ of the type (see Sec.\ref{statmec})
\be 
\label{new1} 
N(<R) = B R^D 
\ee 
where the prefactor $B$ is related to the lower cut-offs $N_0$ and
 $r_0$ 
\be 
\label{neww2} 
B=\frac{N_0}{r_0^D} \; .
\ee 
In this case, the prefactor $B$ is defined for spherical  samples.
 If we have a sample consisting in a  portion of a sphere characterized by a solid angle
 $\Omega$, we write Eq.\ref{new1} as 
\be 
\label{new1a} 
N(<R) = B R^D \frac{\Omega}{4\pi} \;.
\ee 

In the case of a finite fractal structure, we have to take into account 
the statistical fluctuations. In a previous 
 paper \cite{slgmp96} 
we have proposed an argument to describe 
the finite size fluctuations based on extension of the 
concept of Voronoi length in the case of  a fractal. Such an argument 
holds for samples in which the scaling region is of the order 
of the finite size effects region.
 Here we present a new and more general argument 
for the description of the finite size small scale fluctuations
(see \cite{mon97b} for a more detailed 
discussion of the finite size fluctuations).

We can identify two basic kinds of fluctuations:
the first ones are intrinsic $f(R)$  and are due to
the highly fluctuating nature of fractal distributions
while the second ones, $P(R)$, are Poissonian fluctuations.
Concerning the first ones, 
one has to consider that 
the mass-length relation is a convolution of fluctuations
which are present at all scales.
For example one   encounters, at any scale,  a 
large scale structure  and 
then a huge void: these fluctuations   affect the power law behavior of 
$N(<R)$. We can quantify these effects as a {\it modulating term} 
around the expected average given by Eq.\ref{new1a}. 
Therefore, in the observations from a single point {\it "i"} 
we   have 
\be
\label{cazzz4} 
[N(<R)]_i = B R^D \frac{\Omega}{4\pi} \cdot f_{\Omega}(R, \delta \Omega) \; .
\ee 
In general it is more useful to focus on the behaviour
of a local quantity (as the number of points in shells) rather
than   an integrated one. However for the purpose of the present
discussion the approximation given by Eq.\ref{cazzz4} is rather good.
This equation  shows that the amplitude of $N(<r)$ is related to 
the amplitude of the intrinsic fluctuations and not only 
to the lower cut-off $B$.
The amplitude  
of the modulating term
is small, compared with the expected value of $N(<R)$
\be 
\label{cazzz1}
\sqrt{|f(R)|^2} < B R^D \frac{\Omega}{4\pi} \;.
\ee
 In general this
fluctuating term  depends on the direction of observation $\Omega$ 
and on  the solid angle of the survey $\delta \Omega$
so that $f(R) = f_{\Omega} (R, \delta \Omega)$.  If we have a 
spherical sample we get
\be
\label{cazzz2}
\frac{1}{4 \pi} \sum_{\Omega} 
f_{\Omega}(R, \delta \Omega) =  f_{4 \pi} (R)  \; .
\ee
In general we expect that $f_{\Omega}(R, \delta \Omega) > f_{4 \pi} (R)$, so 
that larger is the solid angle and smaller is the effect this term.
If we perform the ensemble average of this fluctuating term we 
can smooth out its effect and we have then
\be
\label{cazzz3}
\langle f_{4 \pi} (R) \rangle_i = 1
\ee
 where the average $\langle .... \rangle_i$ refers to all the 
 occupied points in the 
 sample. In such a way
 the conditional density, averaged over all the points
 of the sample,  has a single power law behavior. 
We will discuss the properties of the intrinsic oscillations
in a forthcoming paper \cite{fluc}. Here we stress that 
according to Eq.\ref{cazzz4} the fluctuations in the 
integrated number of points in a fractal, are proportional
to the number of points itself, rather than to the 
root mean square as in a poissonian distribution.
In general \cite{badii84,smith86,sor96,solis97} 
it is possible to characterize these intrinsic fluctuations
as log-periodic oscillations in the power law behavior.
By performing an ensemble average as in Eq.\ref{cazzz3} 
these oscillations can be smoothed out.
However for the purpose of the present paper,
we limit our discussion to the approximation of Eq.\ref{cazzz4}, 
without entering in more details.

 The second $P(R)$
 term is an additive one, and it takes
 into account spurious finite size fluctuations
is simply due to shot noise. In this case we  have that
\be 
\label{cazzz5}
\sqrt{|P(R)|^2} > B R^D \frac{\Omega}{4\pi} \;\;\; \mbox{if}
 \;\;\; R < \lambda \;
\ee
while $P(R) \approx 0$ for $ R \gtapprox \lambda$.  
The ensemble average is, again,  expected to be 
\be
\label{cazzz7}
\langle P(R) \rangle_i = 0 \; .
\ee
This term becomes negligible if the shot noise fluctuations
are small: for example, if
\be
\label{cazzz8}
[N(<R)]_i  > 10 \sqrt{[N(<R)]_i}    \; .
\ee
From this condition and Eq.\ref{cazzz4} we can have a condition on $\lambda$
(neglecting the effect of $f(R)$):
\be
\label{cazzz9}
\langle \lambda \rangle 
\sim \left(10^2 \frac{4 \pi} {B \Omega} \right)^{\frac{1}{D}}
\ee
The {\em minimal statistical length} $\lambda$ is an explicit
 function of the prefactor $B$ and of the solid angle
 of the survey $\Omega$. This scale is a lower limit for the scaling 
region of the distribution: the effect of intrinsic fluctuations,
described by $f(R)$, which are in general non negligible, 
 can modulate the distance at which 
the scaling region is reached. This length depends also, but weakly, 
on the particular
 morphological features of the sample.  
Therefore it is important to stress that Eq.\ref{cazzz9} 
gives an order of magnitude
for $\langle \lambda \rangle $, where we intend the average value over 
all the possible directions of observations. In different 
directions one can have different values for $\langle \lambda \rangle $,
because of the effect of $f(r)$.

 In the case of real galaxy catalogs we have to consider the
 luminosity selection effects. In a VL sample,
 characterized by an absolute magnitude limit $M_{lim}$,  
   the mass-length relation  Eq.\ref{new1a},  can be
 generalized as (see Sec.\ref{corran}) 
\be 
\label{new6} 
N(R,M_{lim}) = B R^D \frac{\Omega}{4 \pi} \psi(M_{lim})
 \ee 
where $\psi(M_{lim})$ is the probability that a galaxy has an
 absolute magnitude brighter than $M_{lim}$ 
\be 
\label{new7} 
0 < \psi(M_{lim})  = \frac{\int_{-\infty}^{M_{lim}} \phi(M) dM}
 {\Psi(\infty)} < 1 
\ee 
where $\phi(M)$ is the Schecther 
luminosity function (normalized to unity as in Sec.\ref{corran}) and
$\Psi(\infty)$ is the normalizing factor 
\be 
\label{new8} 
\Psi(\infty) = \int_{-\infty}^{M_{min}} \phi(M) dM
\ee 
where $M_{min}$ is the fainter absolute magnitude surveyed in the
 catalog (usually $M_{min} \approx -10 \div -11$).

  It is possible to compute the intrinsic prefactor $B$ from the
 knowledge of the conditional density $\Gamma(r)$ (see
 Sec.\ref{corran}) \cite{cp92,slmp96}
\be 
\label{new9} 
\Gamma(r) = \frac{D}{4\pi} B r^{D-3}
 \ee 
computed in the VL samples and normalized for the luminosity factor
 (Eq.\ref{new7}). In the various VL  subsamples of Perseus-Pisces,
 CfA1, and other  redshift surveys (see in what follows)  we find
 that 
\be 
\label{new10} 
B \approx 10 \div 15 (\hmp)^{-D}
 \ee
 depending on the parameters of the Schecther function $M^*$ and
 $\delta$ (see Sec.\ref{corran}).
 From Eq.\ref{cazzz9}, Eq.\ref{new6} and Eq.\ref{new10}  we obtain 
for a typical volume limited sample with $M_{lim} \approx M^*$,
\be
\label{v3} 
\langle \lambda \rangle \approx \frac{(20 \div 60) \hmp}{\Omega^{\frac{1}{D}}} \;.
\ee 
 This is the value of the {\em minimal statistical length} that we
  use in what follows.
In  Tab. \ref{tablambda} we report the value of $\lambda$ for several 
redshift surveys. While in the case of CfA1, SSRS1, 
PP, LEDA and ESP we have checked that there is a 
reasonable agreement with this 
prediction,  the CfA2 and SSRS2 
redshift surveys are not sill published and hence in these
cases we can {\it predict} the value of $\lambda$.

\begin{table} \begin{center} 
\begin{tabular}{|c|c|c|} 
\hline         
&       &     \\ 
Survey & $\Omega (sr)$ & $\lambda (\hmp)$  \\ 
   &    &     \\ 
\hline 
CfA1            & 1.8      & 15 \\  
CfA2 (North)& 1.3      & 20  \\        
SSRS1          & 1.75    & 15  \\    
SSRS2          & 1.13    & 20 \\  
PP                & 1         & 40  \\
LEDA            & 2 $\pi$& 10 \\      
IRAS            & 2 $\pi$ & 15 \\        
ESP             &   0.006  & 300  \\        
\hline
\end{tabular} 
\caption{The {\em minimal statistical length 
\label{tablambda}} $\lambda$ for several redshift surveys }
\end{center} \end{table}

Given the previous discussion, we can now describe in a very simple 
way the behavior of $[N(<r)]_i$, i.e. the mass length relation 
measured from a generic point {\it "i"}. Given a sample 
 with solid angle $\Omega$, we can approximate 
 the effect of the intrinsic and shot noise fluctuations in the following way:
 \be 
 \label{equ22}
(N(<r))_i =   B_1 r^3 \; \; \; \mbox{if} 
\; \;  r \ltapprox \lambda
\ee
i.e. the density is constant up to $\lambda$, while 
\be 
 \label{equ23}
(N(<r))_i  =  B  r^{D} \; \; \; \mbox{if} 
\; \;  r \gtapprox \lambda
\ee
so that by the condition of continuity at $\lambda$ we have 
\be 
 \label{equ24}
B_1= \frac{B}{ \lambda^{3-D}}
\ee
This simple approximation is very useful in the following discussion, 
especially for the number counts.

 \subsection{Density decay from the vertex for galaxy 
 catalogs and pencil beams}    
\label{radialdecay}

To clarify the effects of the spatial inhomogeneities and finite
effects we have studied the behavior of  the galaxy radial
 (number)
density in the VL samples, i.e. the behavior of (using Eq.\ref{new6}
and Eq.\ref{cazzz4})
\be 
\label{g2} 
n_{VL}(r) = \frac{N(<r)}{V(r)} = \frac{3}{4\pi} B r^{D-3} \psi(M_{lim}) 
\cdot f_{\Omega}(r, \delta \Omega)
\ee  
 One expects that, if the distribution is homogeneous the density
 is  constant,  while if it is fractal it decays as power law.

 When one computes the conditional average density (see Sec.\ref{corran}), 
 one
 indeed performs an average over all the points of the survey. In
 particular, as we have already discussed,
 we limit our analysis to a size defined by the radius of the
 maximum sphere fully contained in the sample volume, and  we do not 
 make 
  use  any   treatment of the sample
 boundaries.
 On the contrary Eq.\ref{g2} is computed only from a single  point,
 the origin. This allows us to extend the study of the spatial
 distribution up to very deep scales. 
 The price to pay is that this method is strongly affected by 
 statistical fluctuations and finite size effects.

The effect of the 
finite size spurious fluctuations for a fractal distribution
 is shown Fig.\ref{fig61}:  at small distances one finds  almost no
 galaxies because we are below {\it the average mean separation between 
 neighbor galaxies} $\ell$. Then the number
 of galaxies starts to grow, but this regime is strongly affected
 by finite size fluctuations. Finally the correct scaling region $r
 \approx \lambda$ is reached. This means, for example, that if one
 has a fractal distribution, there is first a raise of the
 density, due to finite size effects and characterized by strong 
 fluctuations, because no galaxies are present before a certain
 characteristic scale. Once one enters in the correct scaling
 regime for a fractal the  density starts to decay as a power law.
 So in this regime of raise and fall with strong fluctuations there
 is a region in which  the density can be approximated roughly
 by a  constant value.  This leads to an apparent exponent $D
 \approx 3$, so that the integrated number grows as   $N(<r) \sim
 r^3$.  This exponent is therefore not a real one 
 but just due to  
 finite size fluctuations.
Of course, depending on the survey orientation in the sky, one can get an
exponent larger or smaller  than $3$, but in general this is 
the more frequent situation (see below).
 Only when a well defined statistical
 scaling regime has been reached, i.e. for $r > \lambda$,  one can
 find the genuine scaling properties of the   structure, otherwise
 the behavior is completely  dominated by spurious finite size
 effects (for seek of clarity in this discussion we do not
 consider the effect of $f(r)$ in Eq.\ref{cazzz4})

The question of the difference between the integration from the
 origin (radial density) 
  and correlation properties  averaged over all points lead
 us to a subtle problem of {\em asymmetric fluctuations} in a
 fractal structure. From our discussion, exemplified by
 Fig.\ref{fig61}, the region between $\ell$ and $\lambda$ 
 corresponds to an underdensity with respect to the real one.
 However we have also showed that for the full correlation averaged
 over all the points, as measured by $\Gamma(r)$,
 the correct scaling is recovered
 at distances appreciably smaller than $\lambda$. This means that
 in some points one should observe an overdensity between $\ell$
 and $\lambda$. However, given the intrinsic asymmetry between
 filled and empty regions in a fractal, only very few points 
 show the overdensity (a fractal structure is asymptotically
 dominated by voids). These few points nevertheless  have, indeed,  an
 important effect on the average values of the correlations. This
 means that, in practice, a typical points shows an underdensity up
 to $\lambda$ as shown in Fig.\ref{fig61}. The full averages instead
 converge at much shorter distances. This discussion shows the
{\it  peculiar and asymmetric nature of finite size fluctuations in
 fractals} as compared to the symmetric Poissonian case.  For
 homogeneous distribution \cite{slgmp96}
the situation is in fact quite different. Below the Voronoi length
 $\ell_v$ there are finite size fluctuations, but for distances $r
 \gtapprox (2\div 4) \ell_v$ the correct scaling regime is readily
 found for the density.
 In this case the finite size effects do not affect too much the
 properties of the system because a Poisson distribution is
 characterized by {\em small amplitude fluctuations}.

\subsubsection{Perseus-Pisces} 
\label{radialpp}
We have computed the $n(r)$ in the various VL samples 
of Perseus-Pisces redshift survey, 
and we show the results   in Fig.\ref{fig62}.
 In the less deeper VL samples (VL70, VL90)   the
 density does not show any smooth behavior because in this case the
 finite size effects dominate the behavior as the distances involved are 
 $r < \lambda $ (Eq.\ref{v3}). At about  the same scales we
  find a very well defined power law behavior by the
 correlation  function analysis (see Sec.\ref{corran}).
 In the deeper VL samples (VL110, VL130)  a smooth
 behavior is reached for distances larger 
than the scaling distance ($\Omega =0.9 \, sr$)
 $r \approx \lambda \sim 50 h^{-1}Mpc$. The fractal dimension is
  $D \approx 2$ as one measures by 
  the correlation 
 function.     
\bef 
 %\vspace{}
\epsfxsize 8cm 
\centerline{\epsfbox{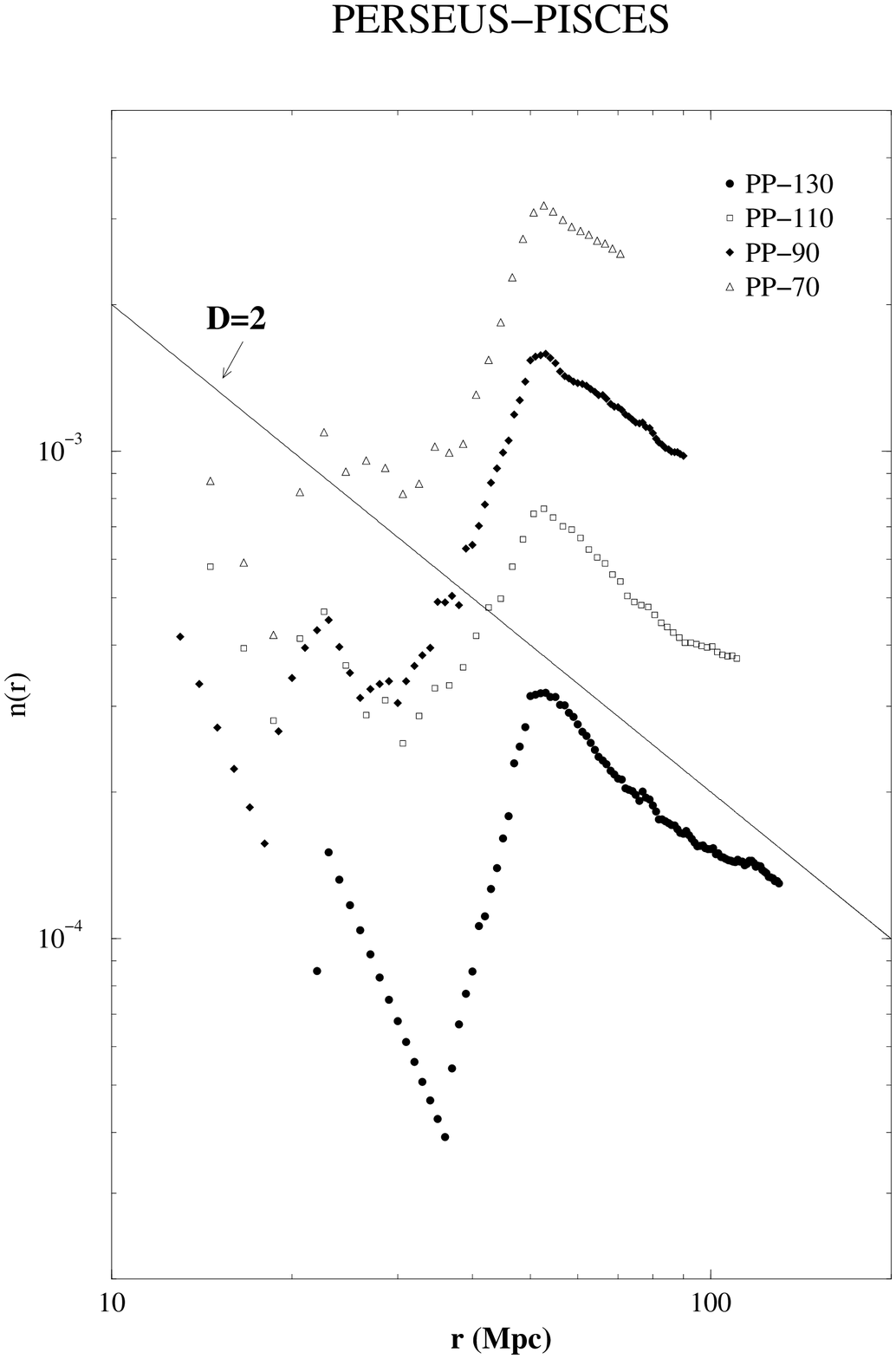}} 
\caption{ The spatial density $n(r)$   computed in the VL sample
 cut at $70, 90, 110, 130  h^{-1}Mpc$ . In the case of VL70 and VL90  the
 density is dominated by large fluctuations and it has not reached
 the scaling regime. In the samples  VL110   and VL130 the density is
 dominated by large fluctuations only at small distances, while at
 larger distances, after the Perseus Pisces chain at $50
 h^{-1}Mpc$, a very  well defined power law behavior is shown, with
 the same exponent of  the correlation function  (i.e. $D=2$)
\label{fig62}}
\eef
 For relatively small volumes it is possible to
 recover the correct scaling behavior for scales of order of
 $\ell$ (instead of $ \sim 10 \ell$) by averaging over several
 samples or, as it happens in real cases, over several points of
 the same sample when this is possible. Indeed, when we compute the
 correlation function we perform an average over all the points of
 the system even if the VL sample is not deep enough to satisfy the
 condition expressed by Eq.\ref{v3}. In this case the lower
 cut-off  introduces a limit in the sample statistics (Sec.\ref{validity}).

\subsubsection{CfA1} 
\label{radialcfa1}
In the case of  CfA1  ($\Omega =1.8$) we obtain that   $\lambda
 \approx 15 \hmp $. We compute the  behavior of the radial
 density and we show the results in Fig.\ref{fig63}: 
\bef 
%\vspace{}
\epsfxsize 8cm 
\centerline{\epsfbox{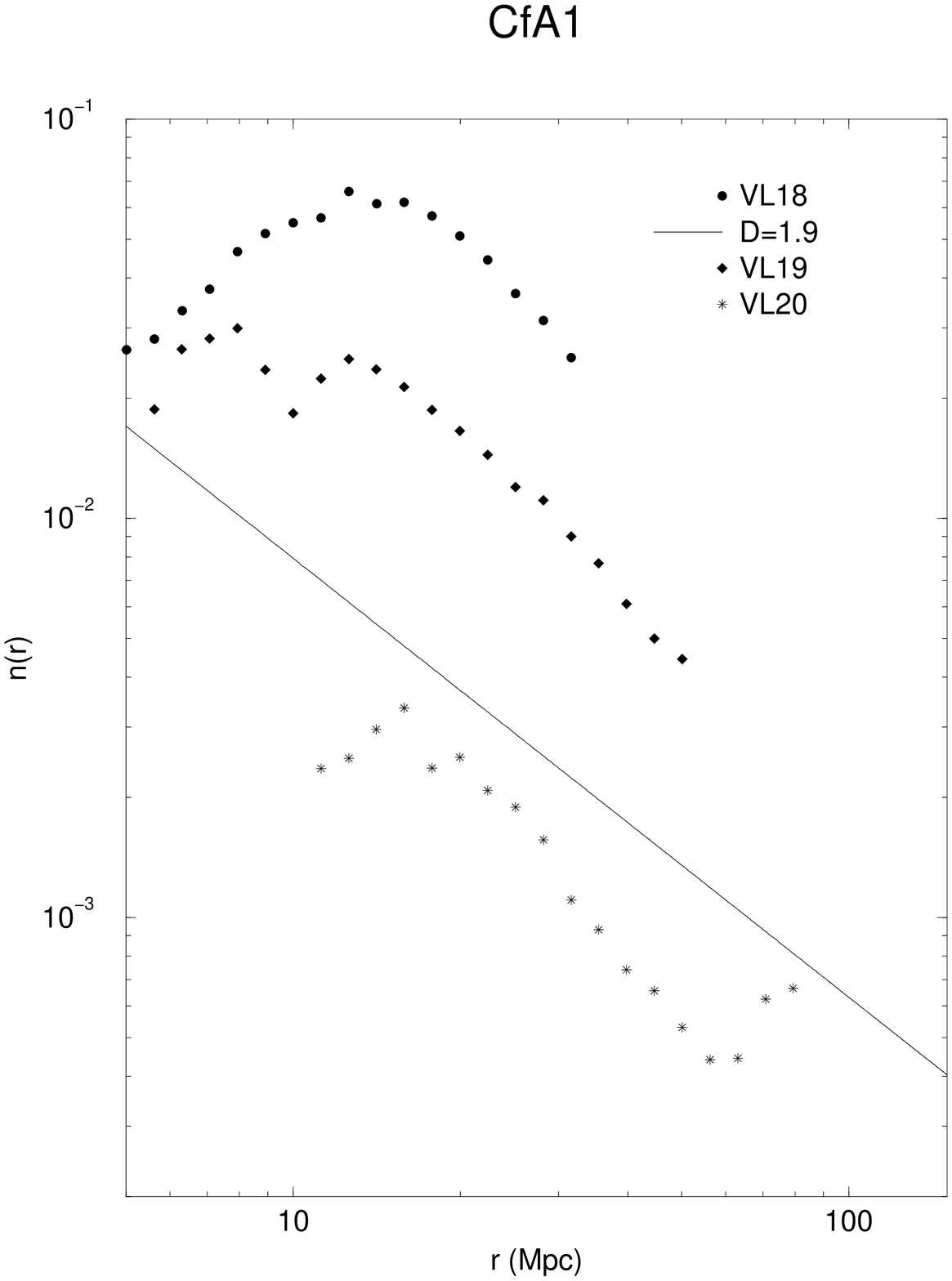}}  
\caption{\label{fig63} The spatial density   $n(r)$     computed
 in various VL sample  of CfA1 (cut in absolute magnitude).
 A very  well defined power law
 behavior with $D \approx 1.9$  is shown for $r \gtapprox 15 \hmp
 \approx \lambda$. } 
\eef 
in the various cases the agreement with Eq.\ref{v3} is quite good.  
The sample VL20 (cut at the absolute magnitude $M=-20$)
shows a raising of the radial density at $\sim 70 \hmp$.
This is a typical fluctuation due to the presence of 
 structures, which characterizes at fractal distribution at any scale, and which is 
not smoothed out because $n(r)$ is not averaged over all the galaxies.
These intrinsic fluctuations are   described by 
the modulating function $f(r)$ in Eq.\ref{cazzz4}.

\subsubsection{SSRS1}  
\label{radialssrs1}
In the case of SSRS1 the solid angle is $\Omega =1.75 \; sr$, and,
as for CfA1,
 we obtain that $\lambda 
\sim 15 \hmp$. We show in Fig.\ref{fig64}
\bef 
%\vspace{}
\epsfxsize 8cm 
\centerline{\epsfbox{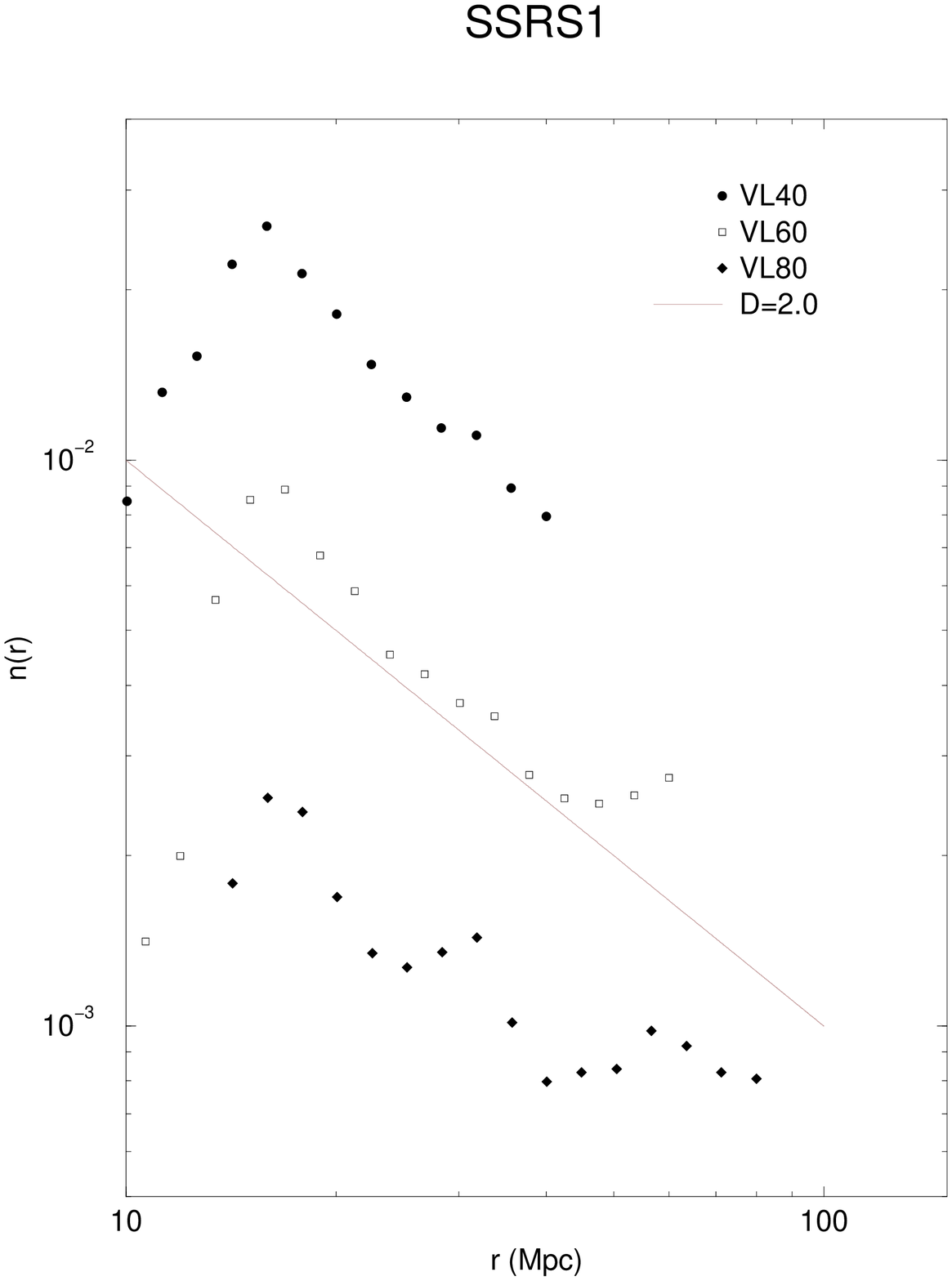}} 
\caption{\label{fig64} The spatial density   $n(r)$     computed
 in some VL samples of SSRS1. A very  well defined power law
 behavior with $D \approx 2$  is shown for $r \gtapprox 15 \hmp
 \approx \lambda$. } 
\eef 
the behavior of $n(r)$ is the VL samples of this catalog.
Even in this case the decay of the radial density shows a $1/r$ 
behaviour (i.e. $D \approx 2$), in agreement with the results obtained by
the conditional density. The sample VL80 is more noisy due to the 
large depletion of points (see Sec.\ref{validity}).

\subsubsection{LEDA}  
\label{radialleda}
LEDA is an all-sky survey, a part the region corresponding to the
galactic plane. The scaling region begins at $\sim 10 \hmp$ and it 
extends up to the limit of the sample, as shown in
Fig.\ref{fig65}.
\bef 
%\vspace{}
\epsfxsize 8cm 
\centerline{\epsfbox{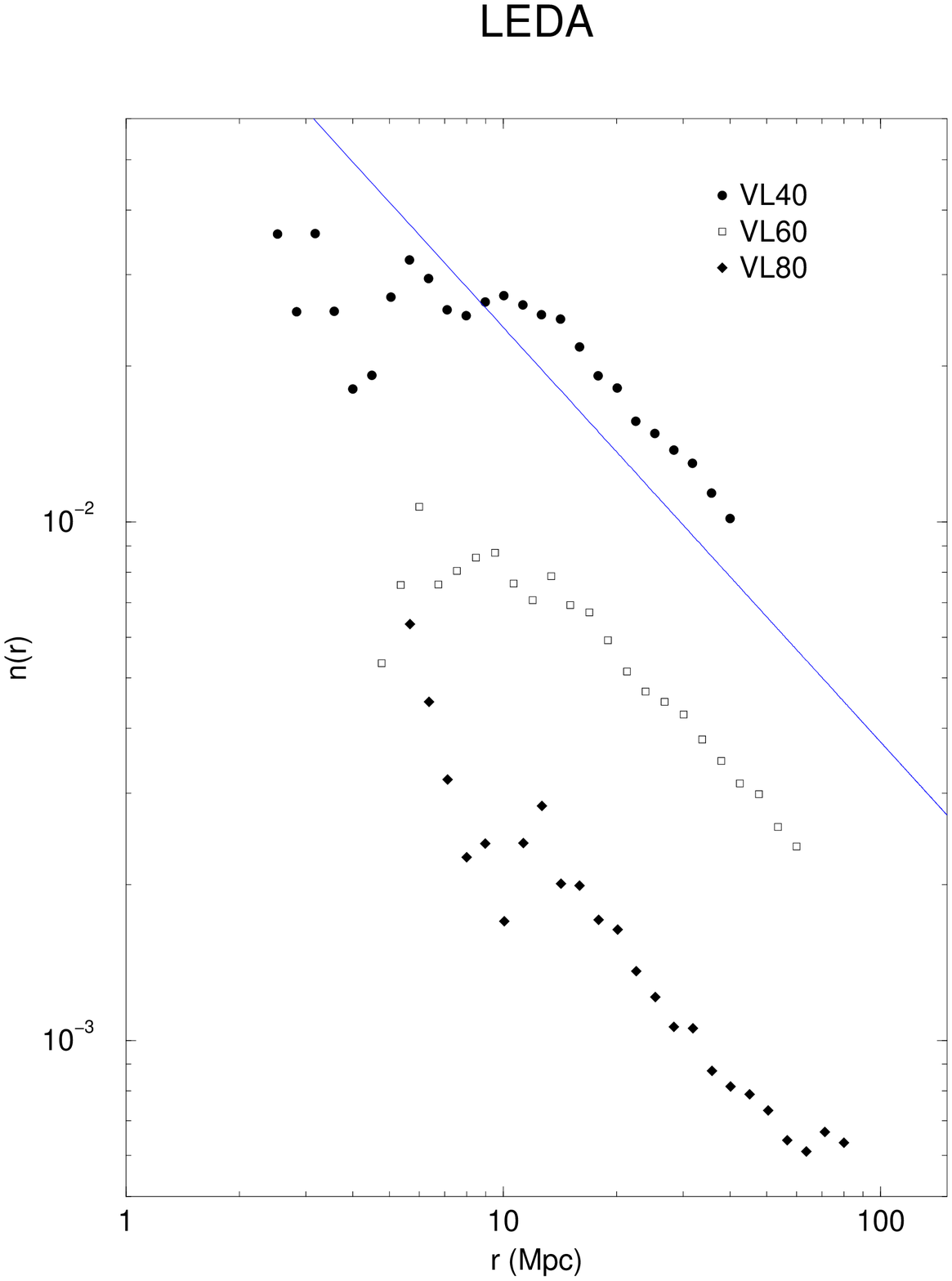}}  
\caption{\label{fig65} 
The spatial density   $n(r)$     computed in some 
VL samples of LEDA145. A very  well defined power law behavior with
$D \approx 2$  is shown for $r \gtapprox 10  \hmp \approx \lambda$.
} 
\eef 
The scaling behavior is reached for distances larger
$r \gtapprox 15 \hmp$, and the fractal dimension is $D \approx 2$ even in
this case. The sample VL80 shows a fluctuation near the boundary,
and has a $1/r^3$ decay up to $\sim 10 \hmp$ due to 
the depletion of points in this sample.

\subsubsection{Las Campanas Redshift Survey}
\label{radiallcrs}
In the case of LCRS it is not possible to compute the radial
density. This is because the survey is limited by a double cut in
apparent magnitude, a lower and an upper one (see Sec.\ref{corran}). Hence it
is possible to have VL samples in a certain range of distances, 
the lower limit not being the origin. Therefore, in this case it is 
possible to study the derivative of $n(r)$
\be
\label{lcrsnr1}
\frac{dn(r)}{dr} \sim r^{-\gamma -1} \; .
\ee
Clearly this is not an integrated quantity, as $n(r)$, and
 hence it is much noisy than the radial density.
We
 show in Fig.\ref{nrlcrs} the results of such determinations.
\bef 
%\vspace{}
\epsfxsize 8cm 
\centerline{\epsfbox{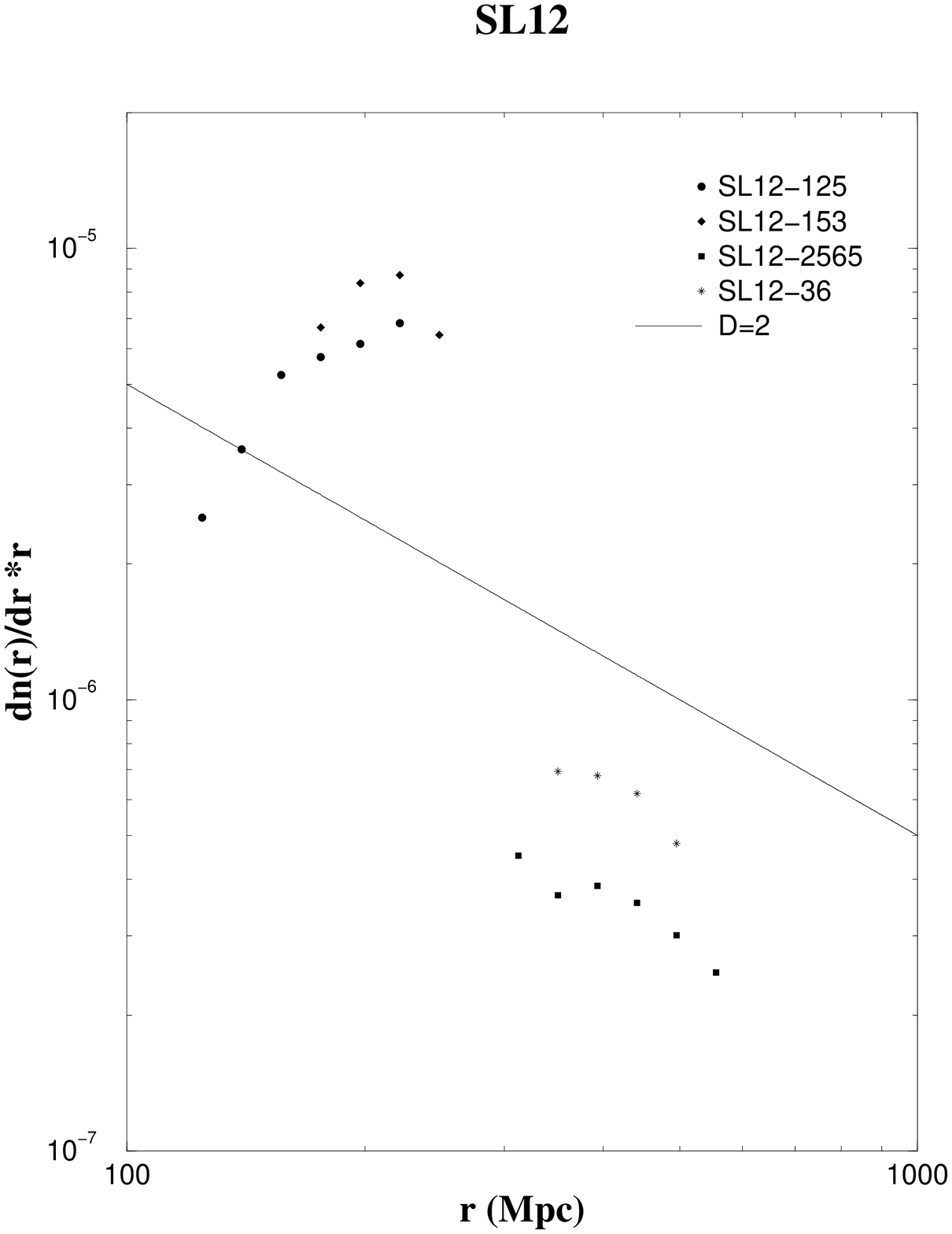}} 
\caption{\label{nrlcrs} The spatial density  
 $dn(r)/dr \cdot r$     computed
 in some  VL samples of LCRS.
The solid line has a slope $-\gamma = -1$.}
\eef 
Unfortunately it is not possible to obtain a VL sample that
covers a large extension in depth. In this case
it is not possible to detect any specific value for $\lambda$. 
  There are large fluctuations in the space distribution (Fig.\ref{nrlcrs})
and the samples SL12-2565 and SL12-36 show a power law behavior in the 
range $\sim 200 \div 600 \hmp$ with $D \approx 2$. 
This behavior is not so very well defined 
 because 
$dn/dr$ is neither an integrated quantity (as $n(r)$), nor it is averaged
as $\Gamma(r)$, and  the extension in depth is  small in log scale.
The samples SL12-125 and SL12-153 show a highly fluctuating behavior for
$dn/dr$. This is due to the fact that 
the scaling is not reached at all in these samples.
We   stress that an homogeneous distribution at these    scales 
should 
exhibits a flat behavior {\it without large fluctuations}.

\subsubsection{Stromlo-APM Redshift Survey} 
\label{radialsars}

The same problem of LCRS is present also in the case of Stromlo-APM
Redshift Survey, for what concerns the construction of the VL
 samples. Even in this case we have to estimate the derivative of
radial density, rather than the radial density itself. The
 behavior of $dn(r)/dr \cdot r$ for various VL samples 
of SARS is shown in Fig.\ref{fig67}.
\bef 
%\vspace{}
\epsfxsize 8cm 
\centerline{\epsfbox{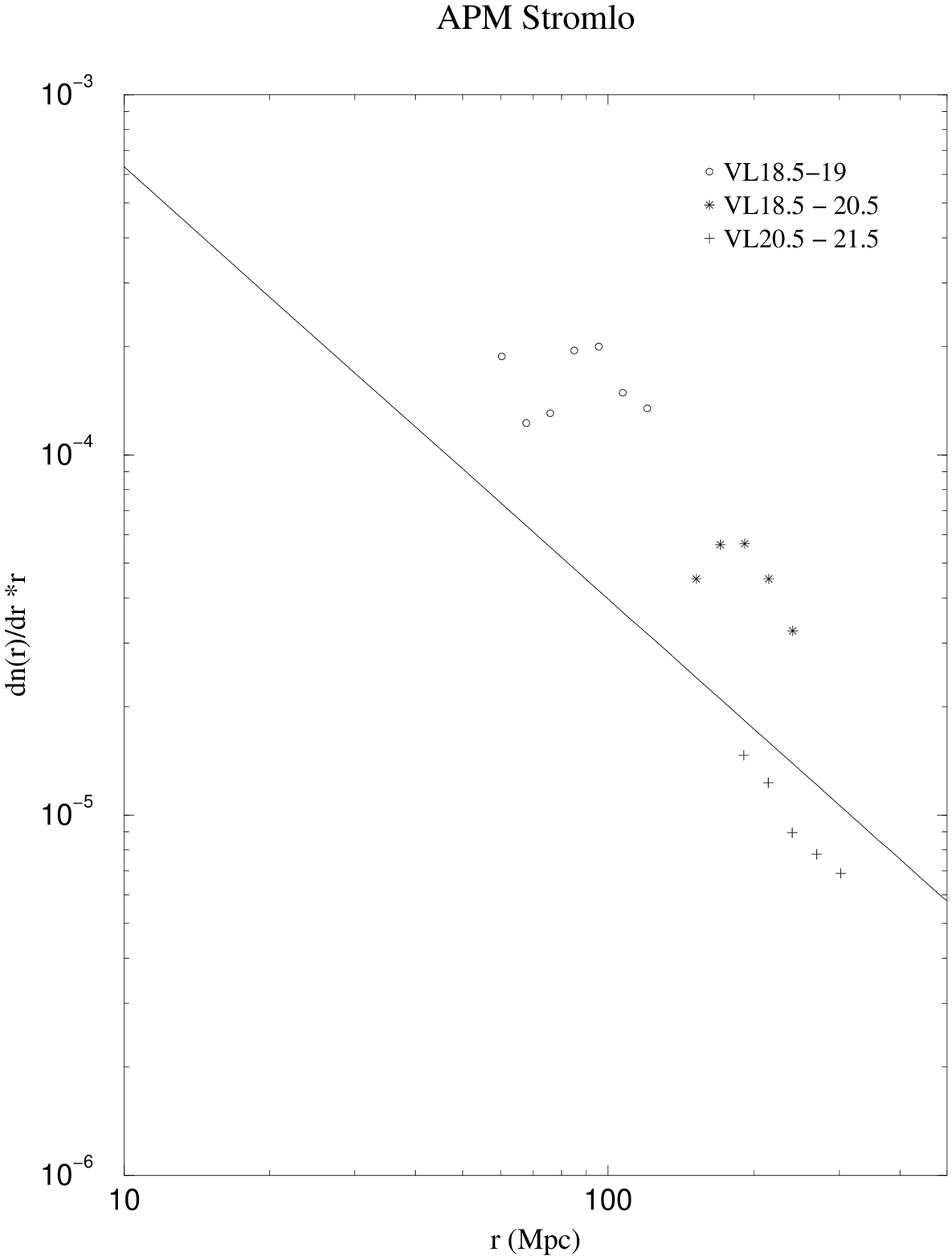}}  
\caption{\label{fig67} The spatial density   $dn(r)/dr \cdot r$     
computed in
 a VL samples of APM. The solid line has
a slope $-\gamma = -1$.}
\eef 
In this case there are the same problems 
discussed for the case of LCRS: a very fluctuating behavior 
and a small range of distance where it is possible to study
the power law decay of the radial density. These results are again 
{\it  compatible} with a fractal distribution and not with an 
homogeneous one.

\subsubsection{ESP}  
\label{radialesp}

As the redshifts involved are quite large a particular care is
 devoted to the construction of the VL subsamples for 
the case of ESP. We have used
 various {\it distance-redshift} ($d(z)$) and {\it
 magnitude-redshift} ($m(z)$) relations  \cite{bslmp94}. 
\bigskip

   {\it 1) Friedmann-Robertson-Walker relations}  

If we assume a
 particular cosmological model as the FRW, we use the following
 comoving distance-redshift relation 
\be d(z,q_0)=\frac{c}{H_0}\frac{zq_0+(q_0-1)(\sqrt{2q_0z+1}-1)}{q_0^2
(z+1)} \; \; h^{-1}Mpc 
\ee 
and we have used the following values of the deceleration parameter
 $q_0=0.03, 0.5$ and $1$. The data are selected in the blue-green
 and even if the redshift of galaxies are moderate ($z\le0.2$)
 K-corrections are needed to compute the absolute magnitude of
 galaxies \cite{vet94}. The corresponding {\it magnitude-redshift}
 relation is
\be
\label{eqmlesp}
 m-M=25+5\log_{10}\left(d(z,q_0)(1+z)\right) + a(T)z
 \ee
 where we have used the functional forms of the K-correction
 $a(T)z$ as a function of redshift from \cite{sha84}. $a(T)$
 depends on the morphological type $T$ and goes from $a \sim 2$
 for the {\it Scd } galaxies to $a \sim 3.7$ for the {\it E/S0}
 galaxies.
 It is not possible to apply the K-correction to each
 morphological type because over the $\:17^{th}$ magnitude it is
 not possible to recognize the Hubble type from visual inspection.
 To overcame this  problem (following \cite{vet94}) we have adopted
 a statistical approach: we have assumed various percentage of late 
 and early-type galaxies. The
 observed percentage in nearer samples is $\:70\%$ of late and
 $\:30\%$ of early type of galaxy. We have performed a number of
 tests by varying these percentages to show that the final result
 depends weakly on the adopted values. We note that varying these
 percentages change the number of points in the VL 
 subsamples as the absolute magnitude can change of about $1$ in
 absolute value.

The behavior of the {\em number-distance relation} in the standard
 FRW model depends strongly on the value of the deceleration
 parameter $\:q_{0}$ for high redshift $\:z>0.5$, while for $\:z
 <0.2$ the relativistic corrections are very small for all
 reasonable $\:q_{0}$.

All the sample show a 
highly fluctuating behavior for the space density (Fig.\ref{fig68}).
\bef 
%\vspace{}
\epsfxsize 8cm 
\centerline{\epsfbox{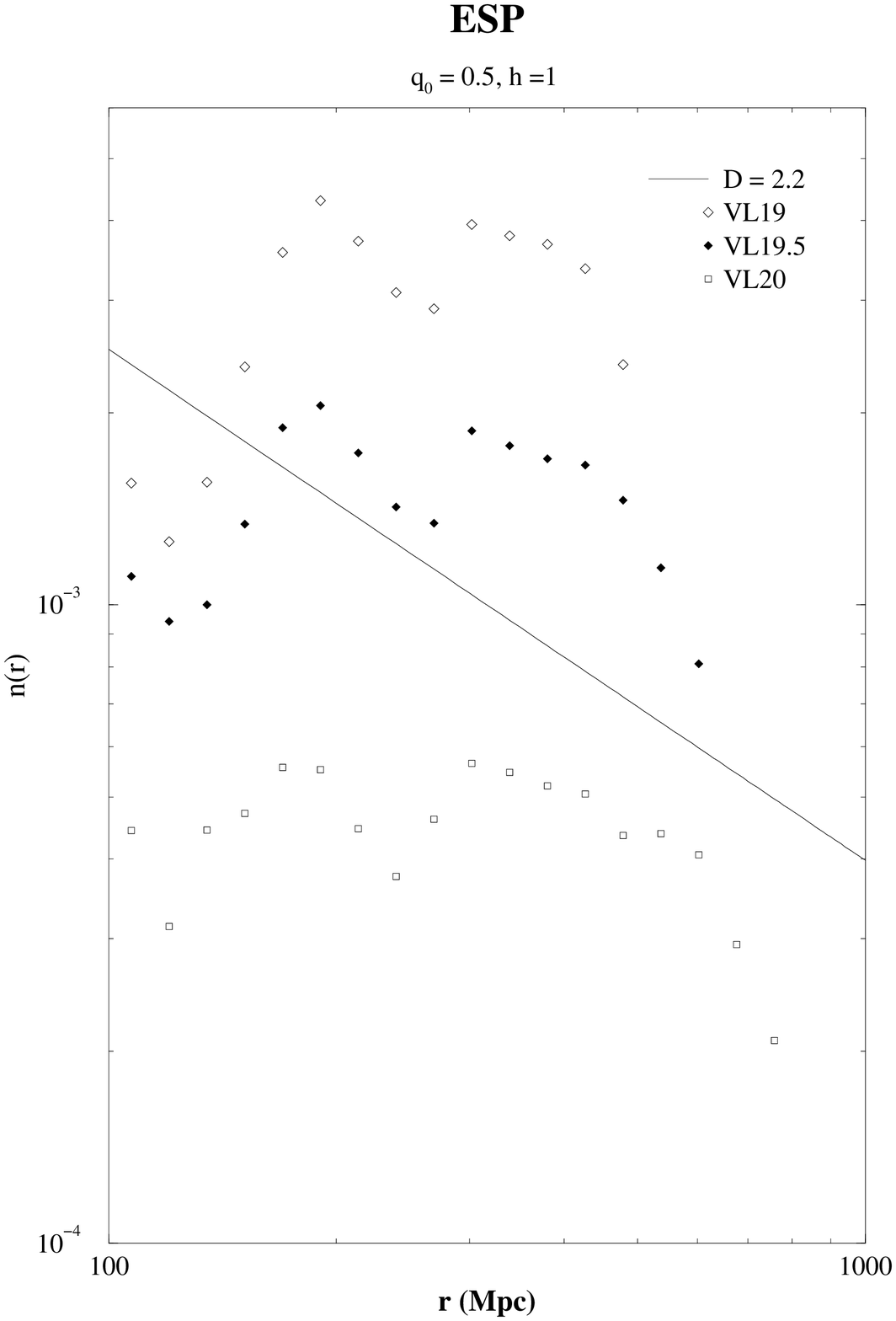}} 
\caption{\label{fig68} The spatial density 
  $n(r)$     computed
 in a VL samples of ESP, using $q_0=0.5$.
The power law behavior with $D \approx 2$ is defined
for $r \gtapprox 300 \hmp$ in all the samples.} 
\eef
At $\lambda \sim 300 \hmp$ the scaling region is
reached. Again, as in the previous case, we   stress
that an homogeneous distribution (say at $ 50 \hmp$)
would show a very smooth and flat behavior for 
$n(r)$ at these distance scales. This is clearly not the case.
\bigskip

 {\it 2.) The Euclidean relations} 

It is very, interesting, in our opinion to study also the case of
 a purely Euclidean space. In this case we can write the $d(z)$
 relation simply as: 
\be 
d=\frac{c}{H_0}z
 \ee and the $m(z)$ relation (without K-corrections): 
\be 
m-M=5\log_{10} \left( d(z)  \right) +  25 \; .
\ee    
In this case the volume grows as $\sim z^3 \sim d^3$.     
  The  behavior of the radial density in the Euclidean case is 
 reported in Fig.\ref{fig69}.
\bef 
%\vspace{}
 \epsfxsize 8cm 
\centerline{\epsfbox{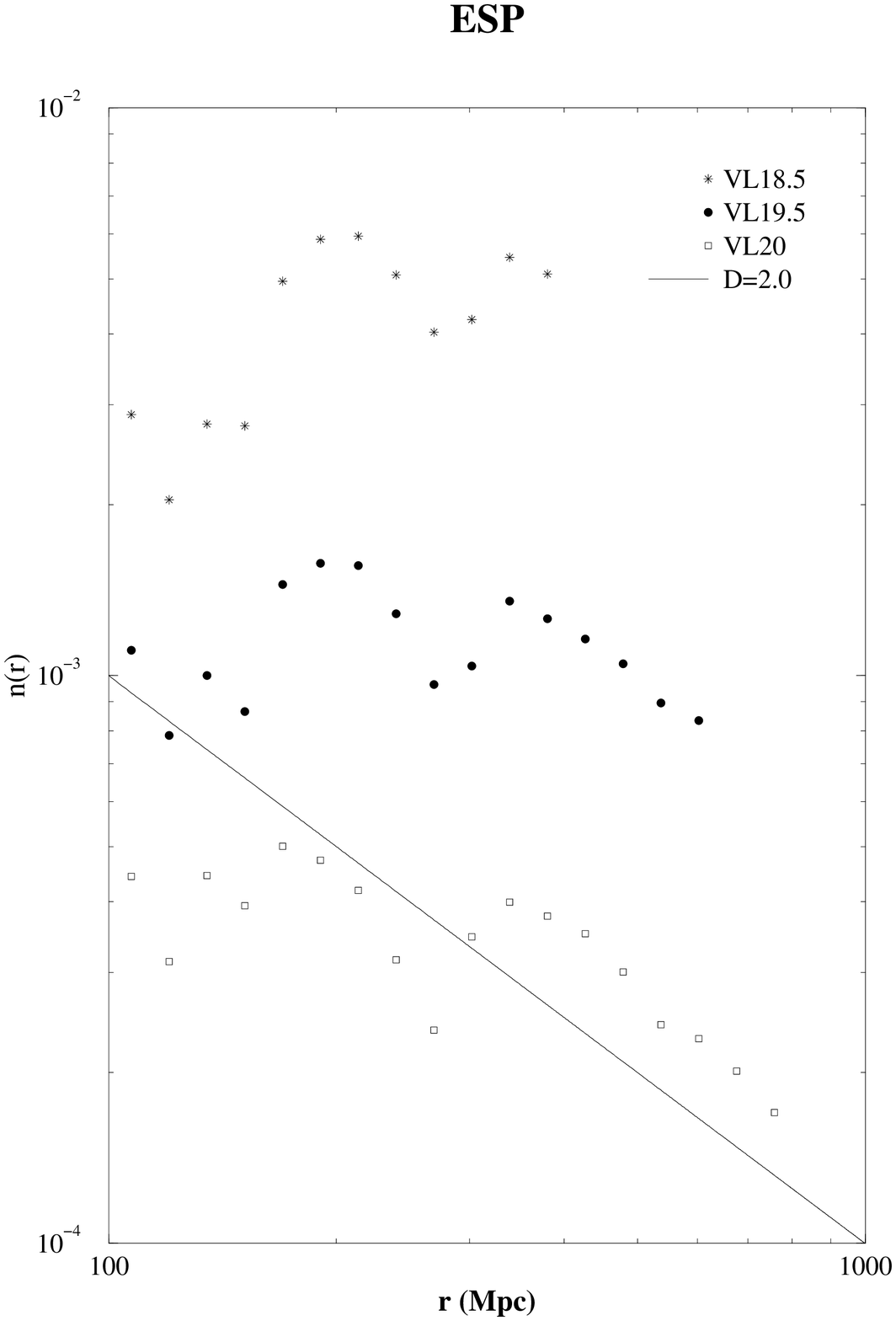}} 
\caption{\label{fig69} The spatial density   $n(r)$     computed
 in a VL samples of ESP. A the power law behavior
 with $D \approx 2$  is shown for $r \gtapprox 300 \hmp \approx
 \lambda$, even though with strong fluctuations}
\eef 
 From  this figure it
 is possible to see  the effect of the finite size
 fluctuations: the scaling region begins at 
$\lambda \sim 300 \div 320 h^ {-1}Mpc$ for all the VL subsamples. For
 smaller distances the statistical fluctuations dominate the behavior
 of $n(r)$. The different normalization in the various VL
 subsamples is due to the different absolute magnitude limit
 $M_{lim}$ which defines each VL subsample. To normalize the
 behavior of $n(r)$ we divide it for a luminosity factor (see
 Sec.\ref{consist}). It seems that in this case the power law behavior 
for $r \gtapprox \lambda \approx 320 \hmp$
is better defined than in the previous case.
The fractal dimension turns out to be $D \approx 2.2$.

 The {\it K-correction} can change the absolute magnitude of a unit. This
 correction is a systematic effect for each morphological type. As
 we put these correction in a random manner we are mixing a
 systematic effect with a random correction. In this way we can
 only{\it  check the stability of the results} but we cannot hope
 to obtain a better fit (see Fig.\ref{fig70}). 
Hence we conclude that
 the fractal dimension is $\:D \approx 2$ is a
 substantially stable result.
\bef 
%\vspace{}
\epsfxsize 8cm 
\centerline{\epsfbox{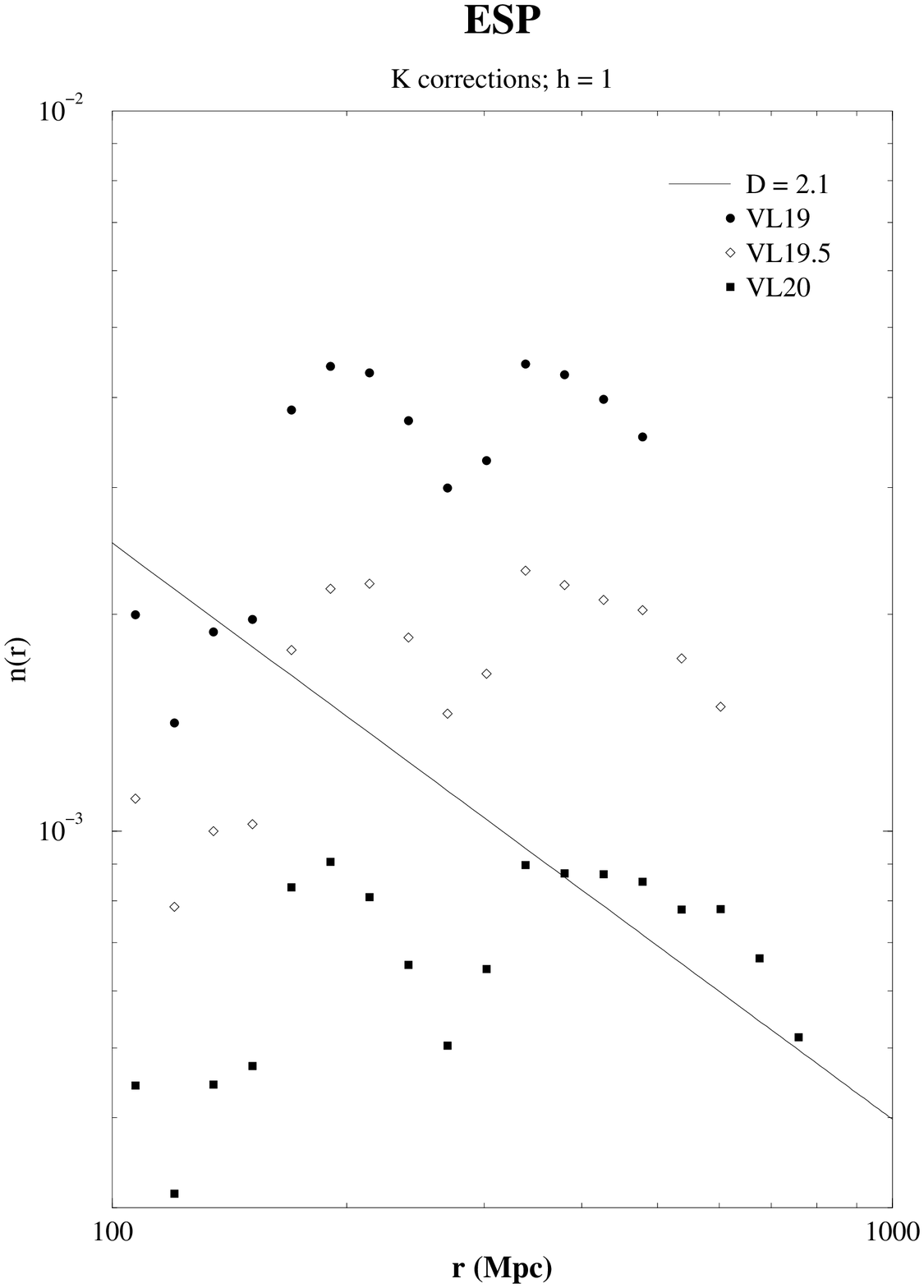}} 
\caption{\label{fig70} The spatial density   $n(r)$     computed
 in a VL samples of ESP applying the random K-corrections
as explained in the text.
 A very  well defined power law behavior with $D \approx 2$  is
 shown for $r \gtapprox 320 \hmp \approx \lambda$.
}
\eef

\subsubsection{IRAS}
\label{radialiras}

As for the case of LEDA, also IRAS is an all sky survey. However
the IRAS samples are limited by the weak statistics which 
characterizes these surveys, as discussed in Sec.\ref{corran} 
and Sec.\ref{validity}. 
We show in Fig.\ref{fig71} 
\bef 
%\vspace{}
\epsfxsize 8cm 
\centerline{\epsfbox{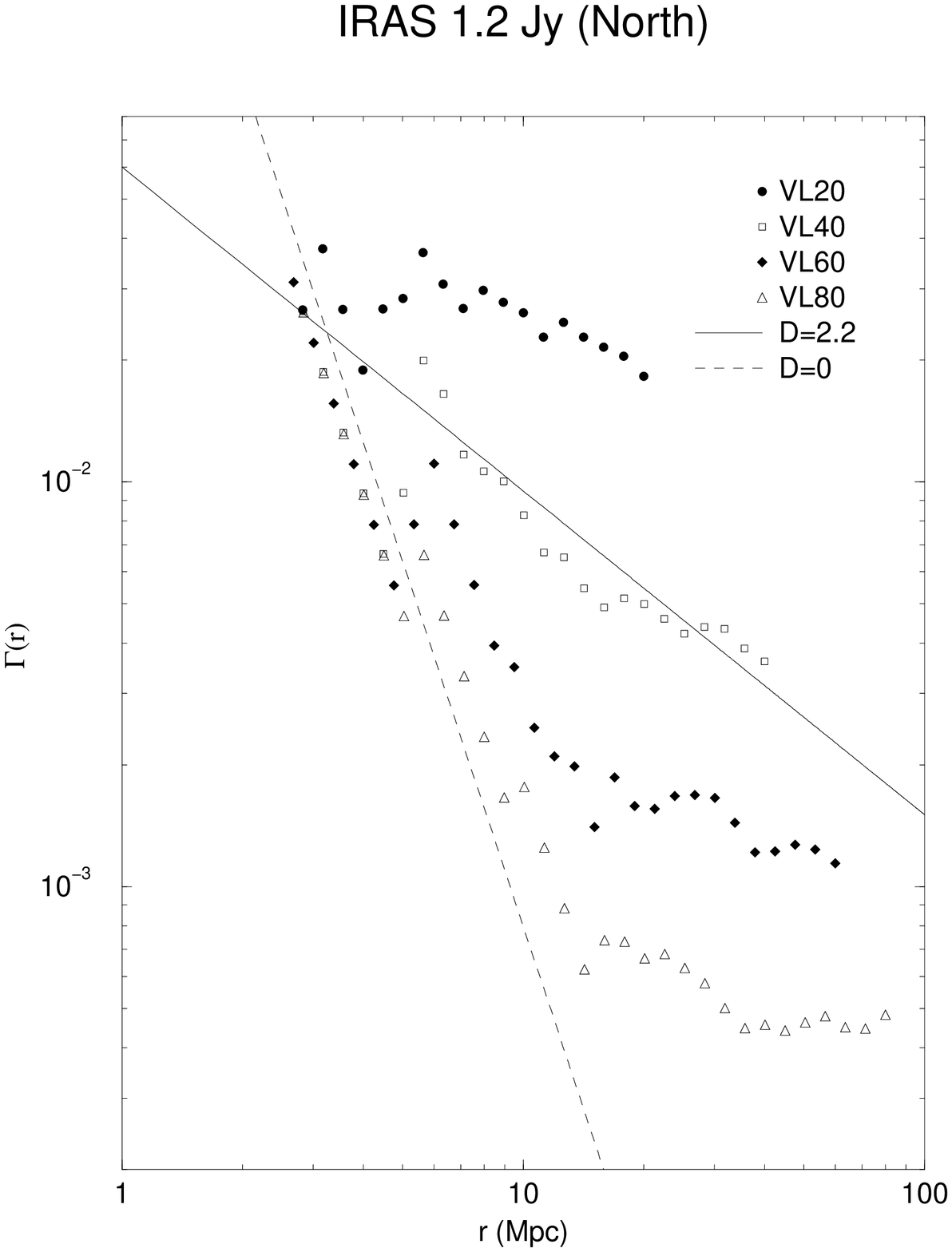}} 
\caption{\label{fig71} The spatial density   $n(r)$     computed
 in a VL samples of IRAS 1.2 Jy. The power law behavior with $D \approx 2$
is shown for distances $\gtapprox 10\hmp$.
}
\eef 
the results of the determination of the radial density
in various VL samples of the survey IRAS $1.2 Jy$ (North).
A power law behavior is defined for $r \gtapprox 10 \hmp$, as in the 
case of LEDA, with $D \approx 2$, up to $\sim 60 \hmp$. The deepest VL sample 
(VL80) shows an $1/r^3$ decay at small scale and a flat behavior 
at larger scales. This is again interpreted as due to the 
extreme dilution of galaxies in this sample (see Sec.\ref{validity}).

\subsection{Radial density in magnitude limited surveys} 
\label{radialml}

It is usually measured in literature the number of galaxies
 $N_{ml}(r)$ as a function of distance in a magnitude limited
 surveys. Some authors (e.g. Davis \cite{dav97})
 claims that from the behavior
 of    $N_{ml}(r)$ it is possible to distinguish between a fractal
 and an homogeneous distribution. Contrary to this result, we 
 show that such a measurement depends only on the
 luminosity selection function of the survey, and very weakly from
 the value of the fractal dimension. 

Assuming, as in Sec.\ref{corran}, the independence of the  space density on
the luminosity one, we can compute the number of galaxies in the bin of
distances $\Delta R$ in a magnitude limited survey ($m_{lim} $ is the
 apparent magnitude limit) as
\be
\label{ml1}
N_{ml}(R) = \int_R^{R+\Delta R}  \rho(r) d^3 r 
 \int_{-\infty}^{M(r)} \phi(M) dM
\ee
(we neglect relativistic corrections), 
where 
\be
\label{ml2}
M(r) = m_{lim} -5 \log(r) -25
\ee
From Eq.\ref{ml1} it follows that  $N_{ml}(r)$is a function of 
four parameters: the fractal dimension $D$ and the prefactor $B$ 
and the two parameters
 of the luminosity function $\delta$ and $M^*$. 
In Fig.\ref{fig72}
 we show the behavior of  $N_{ml}(r)$ with different values of the
 three parameters, in a survey with fixed $m_{lim}$.
The major effect is due to 
 the luminosity selection function rather than to the space
 distribution. This is the reason why one has to use VL samples
 rather than ML ones, in order to study the behavior of the space
 density.
\bef 
%\vspace{}
 \epsfxsize 8cm
\centerline{\epsfbox{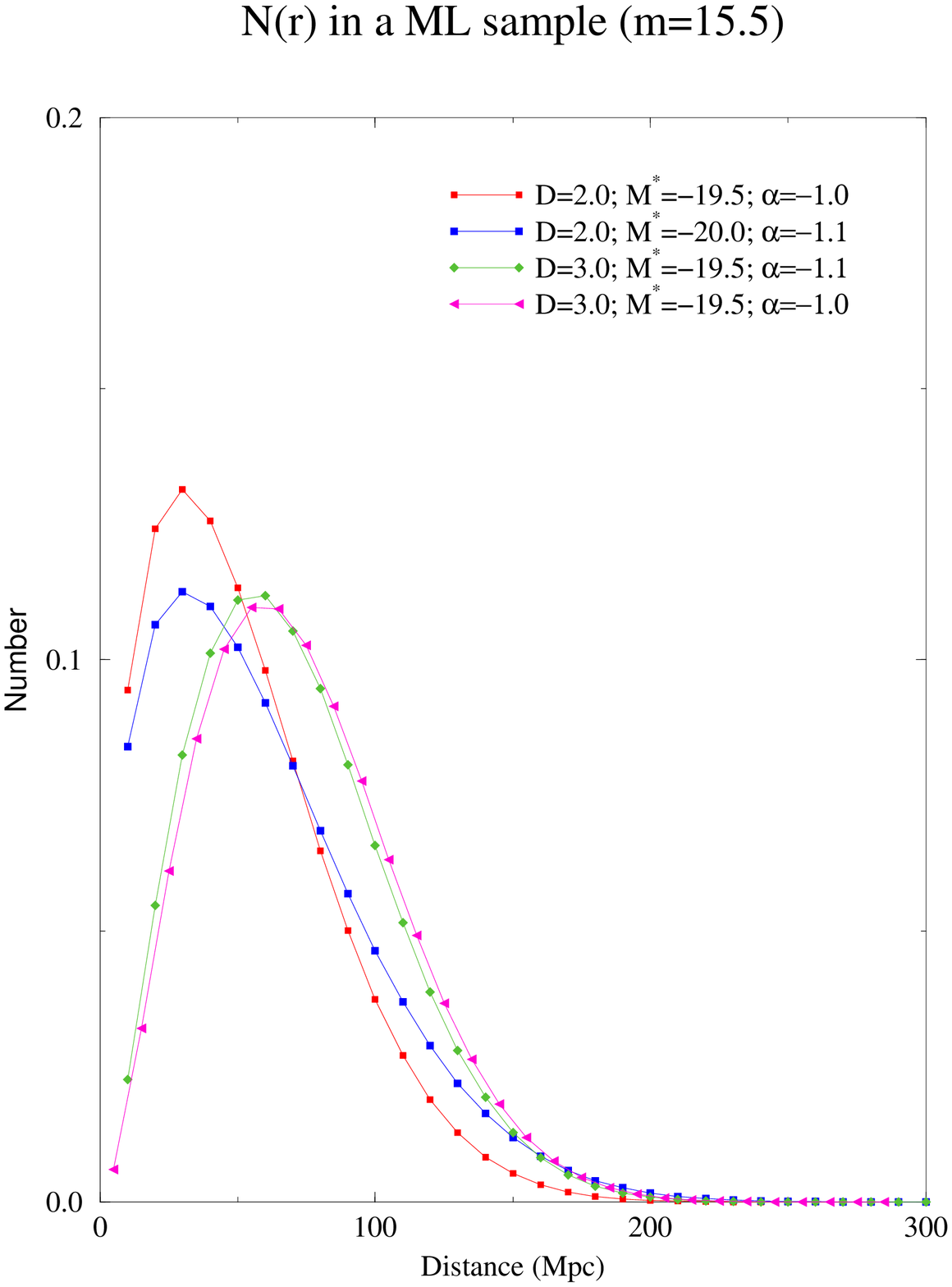}}  
\caption{\label{fig72} 
The spatial density   $N_{ml}(r)$ (per bin of $10 \hmp$)
 in a magnitude limited survey ($m_{lim}=15.5$) for
 various values of the parameters $D$,
 $\delta$ and $M^*$. It is
 clear that the behavior of  $N_{ml}(r)$ is mostly due 
on  the luminosity
 selection function of the survey, and hence $\delta$ and $M^*$,
 rather than on the fractal dimension $D$. This is the reason why
 one has to use VL samples in order to study the behavior of the
 space density. } 
\eef 
All the determinations of the space density (or of the amplitude of 
luminosity function) in magnitude limited samples are seriously affected 
by this bias (see for example \cite{zuc97}).

\subsection{Pencil beams and very deep surveys}
\label{radialpencil}

Deep "pencil beams" surveys 
cover in general very narrow angular size ($ \sim 1^{\circ}$)
and extend to very deep depths ($ z\gtapprox 0.2$). 
These narrow shots through deep space provide
a confirmation of strong inhomogeneities in the galaxy distribution.

One of the most discussed results obtained from pencil-beams surveys 
has been the claimed detection of a typical 
scale in the distribution of galaxy structures, corresponding to a
characteristic separation of $128 \hmp$ \cite{bro90}. 
However in the last five years several other surveys, in different
regions of the sky, do not find {\it any} evidence for such a periodicity.
In particular, Bellanger \& De Lapparent \cite{bd95}, by analyzing 
a sample of 353 galaxies in the redshift interval 
$0.1 \ltapprox z \ltapprox 0.5$, concluded that 
these new data contain {\it no evidence for a periodic signal} on a scale 
of $128 \hmp$. Moreover they argue that the low sampling rate of 
\cite{bro90} is insufficient for mapping the detailed large-scale structure
and it is the real origin of the apparent periodicity.

On the other hand, Willmer \etal \cite{wil94} detect four of the five nearest peaks 
of the galaxies detected by \cite{bro90}, because their survey 
is contiguous to that of \cite{bro90}, in the sky region near the north 
Galactic pole. Moreover Ettori \etal \cite{et96} 
in a survey oriented in three small regions around the South Galactic Pole
do not find any statistically periodic signal distinguishable 
from noise.  Finally 
Cohen \etal \cite{coh96} by analyzing a sample of 140 objects up to $z \sim 0.8$, 
find that there is no evidence for periodicity 
in the peak redshifts.

In a pencil-beam survey one can study 
the behavior of the {\it linear density}
along a tiny but very long cylinder. 
The observed galaxy distribution corresponds therefore to the 
intersection of the full three dimensional galaxy distribution with one
dimensional cylinder \cite{cp92}. In this case, from the law of
codimension additivity \cite{man82,cp92} (see Sec.\ref{statmec}), one
obtains that the fractal dimension of the intersection is given by
\be
\label{pb1}
D_I = D + d_{pc} -d \approx D+1-3 \approx 0
\ee
where $D$ is the galaxy distribution fractal dimension, 
embedded in a $d=3$ Euclidean space, and $d_{pc} =1$ is the 
dimension of the pencil beam survey. 
This means that the set of points visible in a 
randomly oriented cylinder has dimension $D_I \approx 0$.
In such a situation the power law behavior is no longer present and the 
data should show a chaotic, featureless nature strongly dependent
on the beam orientation. If the galaxy distribution   
becomes instead homogeneous above 
some length, shorter than the pencil beam depth, 
one has the regular situation $D_I = 1$ 
and a well defined density
must be observed. 

We  stress that in {\it any} of the available
pencil beams surveys, one can detect   tendency towards 
an homogeneous distribution. Rather, all these surveys show a
very fluctuating signal, characterized by the presence of galaxy structures. 
Some authors \cite{sch96} claim to detect the {\it end greatness} (i.e.
that the galaxy structures in the deep pencil beams are not so different
from those seen in nearby sample - as the Great Wall), or that
\cite{bd95} the dimension of voids does not scale with sample size, by the 
visual inspection of these surveys. However one should consider 
in these morphological analyses, that one is just looking at a convolution of the 
survey geometry, which in general are characterized by very narrow solid angles, 
 and large scale structures, and that in such a situation,
a part from very favorable cases, one may detect portions of 
galaxy structures (or voids).

Finally we note that if the periodicity would be present, 
the amplitude of the different peaks is very different from each other, and 
an eventual transition to homogeneity in a periodic lattice, should be, for
example, ten times the lattice parameter, i.e. $\gtapprox 1000 \hmp$ !

\subsection{Consistency of the various  catalogs and summary of the 
 galaxy correlation properties: fractal behavior from $0.5\hmp$ to
 $1000\hmp$} 
\label{consist}

In this section and in Sec.\ref{corran} we have discussed two different
 determinations of the galaxy space density, i.e the radial density
 and the conditional density respectively. Now we  
show the consistency of these measurements in each of the presented
 galaxy samples, and then we show the consistency of the
 correlation properties in the various  galaxy catalogs. 
In particular, the new analysis shows that all the available data 
are consistent with each other and show fractal correlations with
 dimension $D = 2.0 \pm 0.2$ up to the deepest scale probed up to
 now by the available redshift surveys, 
i.e. $ \sim 1000 \hmp$. The distribution of 
luminous matter
 in the Universe is therefore fractal and not homogeneous. The
 evidence for this is very strong up to $\sim 150 \hmp$, due to
 the statistical robustness of the data, and progressively weaker
 (statistically) at larger distances due to the limited data
 available.

\subsubsection{Normalization of the density in different VL samples} 
\label{normden}

In Sec.\ref{corran} we have already 
defined the luminosity factor which  is associated
to each VL sample. Both the conditional density and the radial
 density measure the space density of galaxies. As long as the
 space and the luminosity density can be considered independent,
 the normalization of $\Gamma(r)$ and $n(r)$ in different VL
 samples can be simply done by dividing their amplitudes for the
 corresponding luminosity factors. 
Of course such a normalization is parametric, because it depends on
 the two parameters of the luminosity function $\delta$ and $M^*$
(see Sec.\ref{corran} and Sec.\ref{lumspace}). 
For a reasonable choice of these two
 parameters we find that the amplitude of the conditional and radial
 density matches quite well in different VL samples. In particular:
\begin{itemize}

\item (1) For CfA1, PP, ESP, LEDA and APM the parameters of the
 luminosity function are $\delta= -1.1$ and $M^* = -19.5$. All these
 surveys have been selected in the B-band, and hence the luminosity
 function is the same. 

\item (2) For SSRS1 the selection criteria have been chosen in the
 apparent diameter $d$, and hence we use the linear-diameter    
 function,
 rather than the luminosity function for the normalization of the
 amplitude in different VL samples. We use the diameter function:
\be
\label{diamfunc}
N(D)dD=N_0\cdot exp(-\delta_d \cdot D)dD
\ee
where $D$ is the absolute diameter (in Kpc), 
$\delta_d=0.109$ and $N_0$ is a constant \cite{pat95}.
The shape of this linear-diameter function is similar to 
the Schecther one for magnitudes.

\item (3) Galaxies in LCRS have been measured in the $r$ band, so
 that we have used the luminosity function given by 
\cite{sch96}. i.e. a Schecther like function with parameters
$\delta= -0.9$ and $M^* = -20.03$.

\item (4) The two IRAS surveys have been measured in the near
 infrared. In this case we have normalized the amplitude of the
 density using the $IR$ luminosity function  \cite{sau90}.
\end{itemize}

We find that in all the cases (1)-(4) the density amplitude  
 in the different VL samples match quite well 
(see Figs.\ref{fig73}-\ref{fig74}). 
\bef 
%\vspace{}
\epsfxsize 12cm 
\centerline{\epsfbox{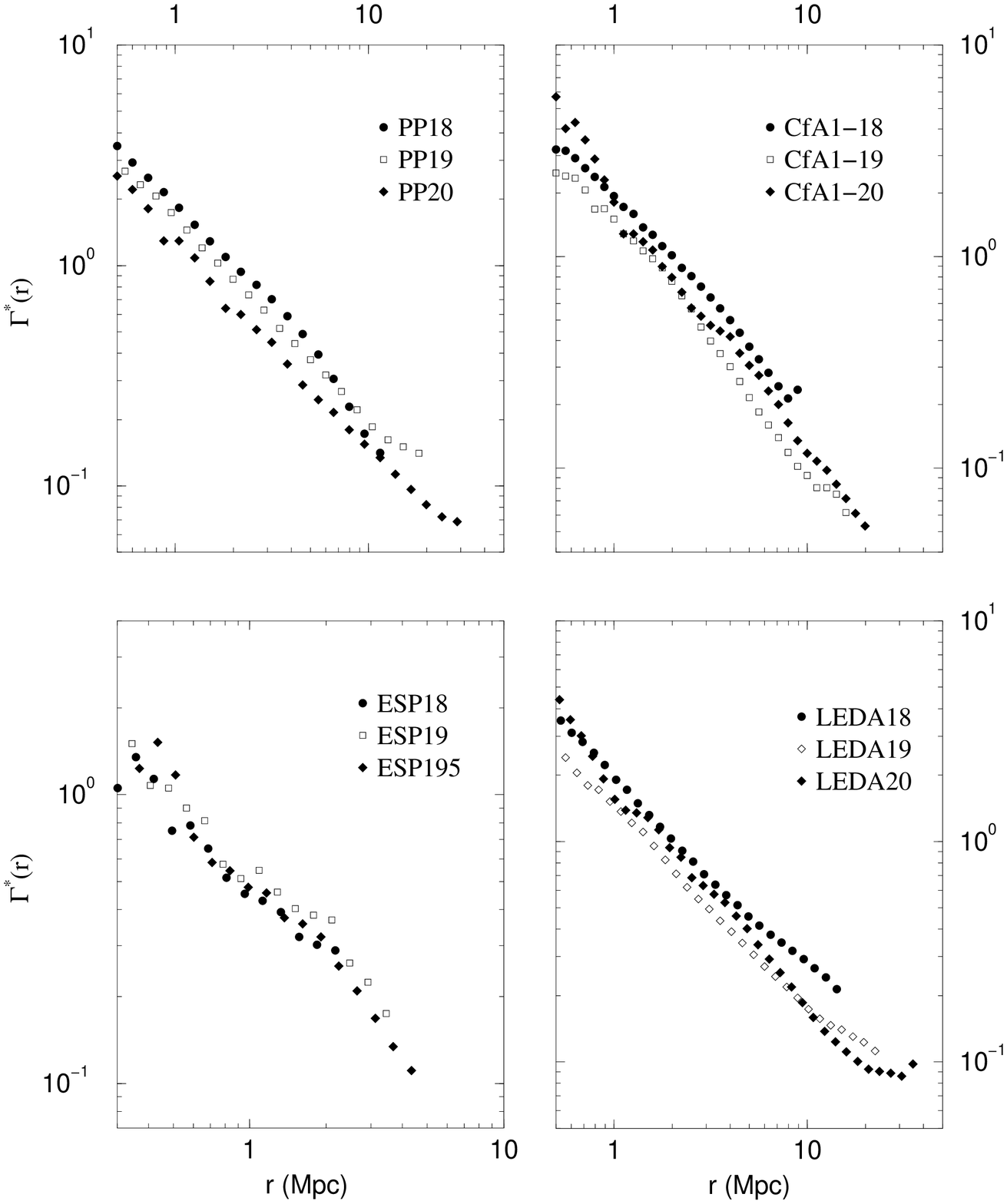}}  
\caption{\label{fig73} 
The spatial conditional average density 
$\Gamma^*(r)$ computed in some 
VL samples of Perseus-Pisces, CfA1, ESP and LEDA
and normalized to the corresponding
luminosity factor. } 
\eef 
\bef 
%\vspace{}
 \epsfxsize 12cm 
\centerline{\epsfbox{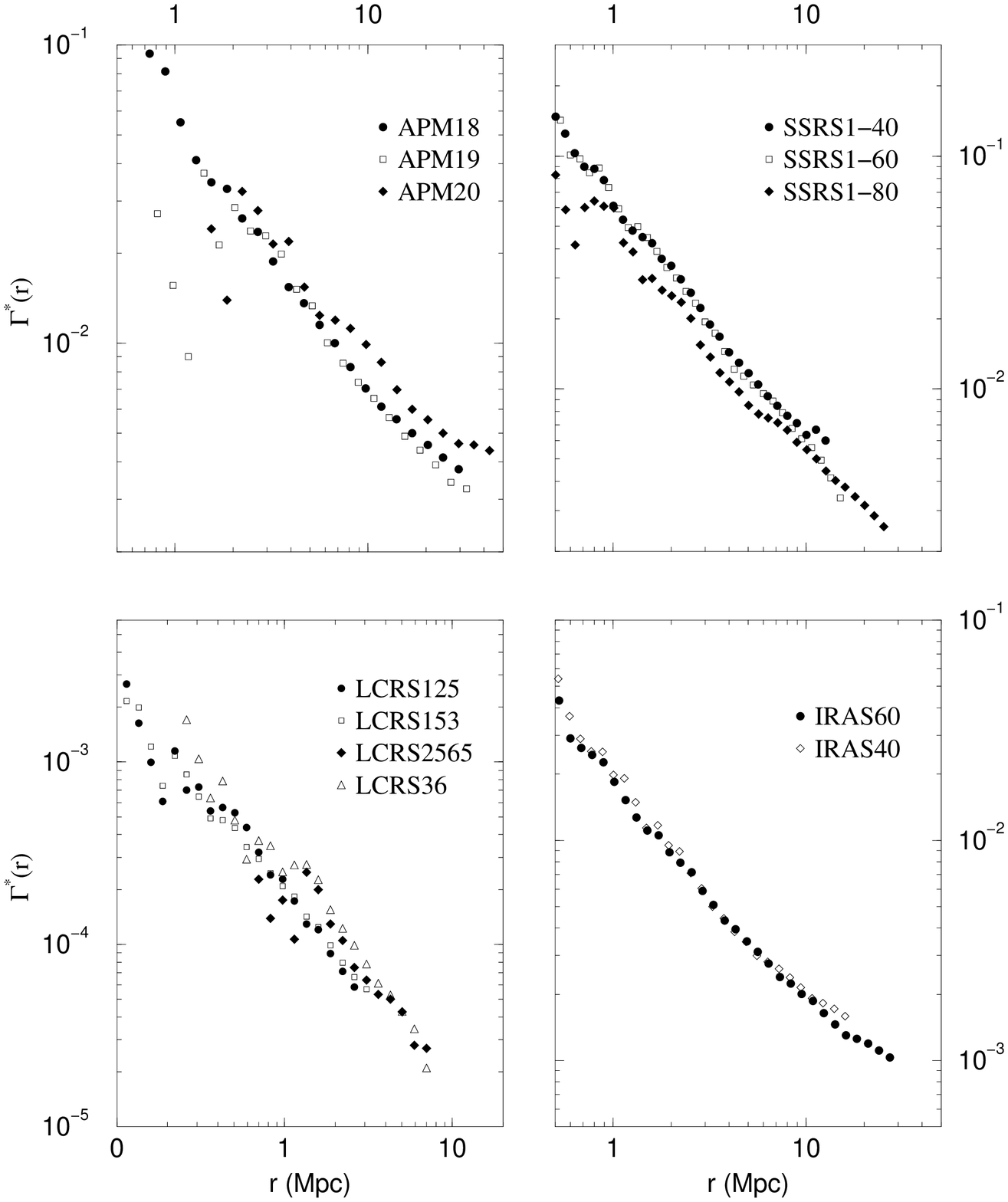}}  
\caption{\label{fig74} 
The spatial conditional average density $\Gamma^*(r)$ 
computed in some 
VL samples of APM, LCRS, SSRS1, IRAS $1.2 Jy$ 
and normalized to the 
luminosity factor. } 
\eef

\subsubsection{Normalization of the average 
conditional density to the radial
 density} 
\label{normcon}
\bef %\vspace{}
\epsfxsize 8cm 
\centerline{\epsfbox{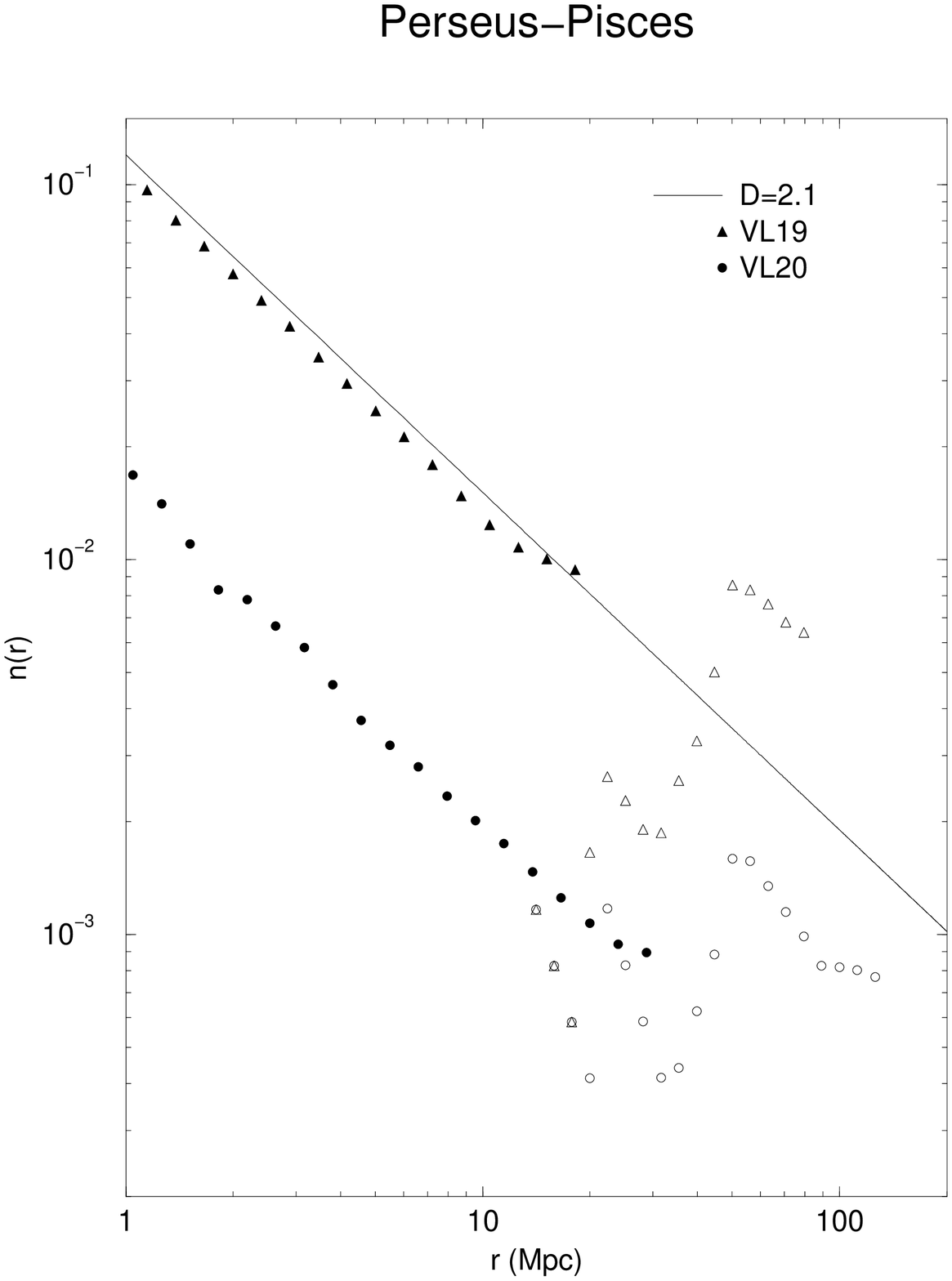}}  
\caption{\label{fig75} 
The conditional average density $\Gamma^*(r)$ (filled points)  and the radial density 
 $n(r)$ (open points)  computed in a VL sample of Perseus-Pisces.} 
\eef 
\bef 
%\vspace{}
\epsfxsize 8cm 
\centerline{\epsfbox{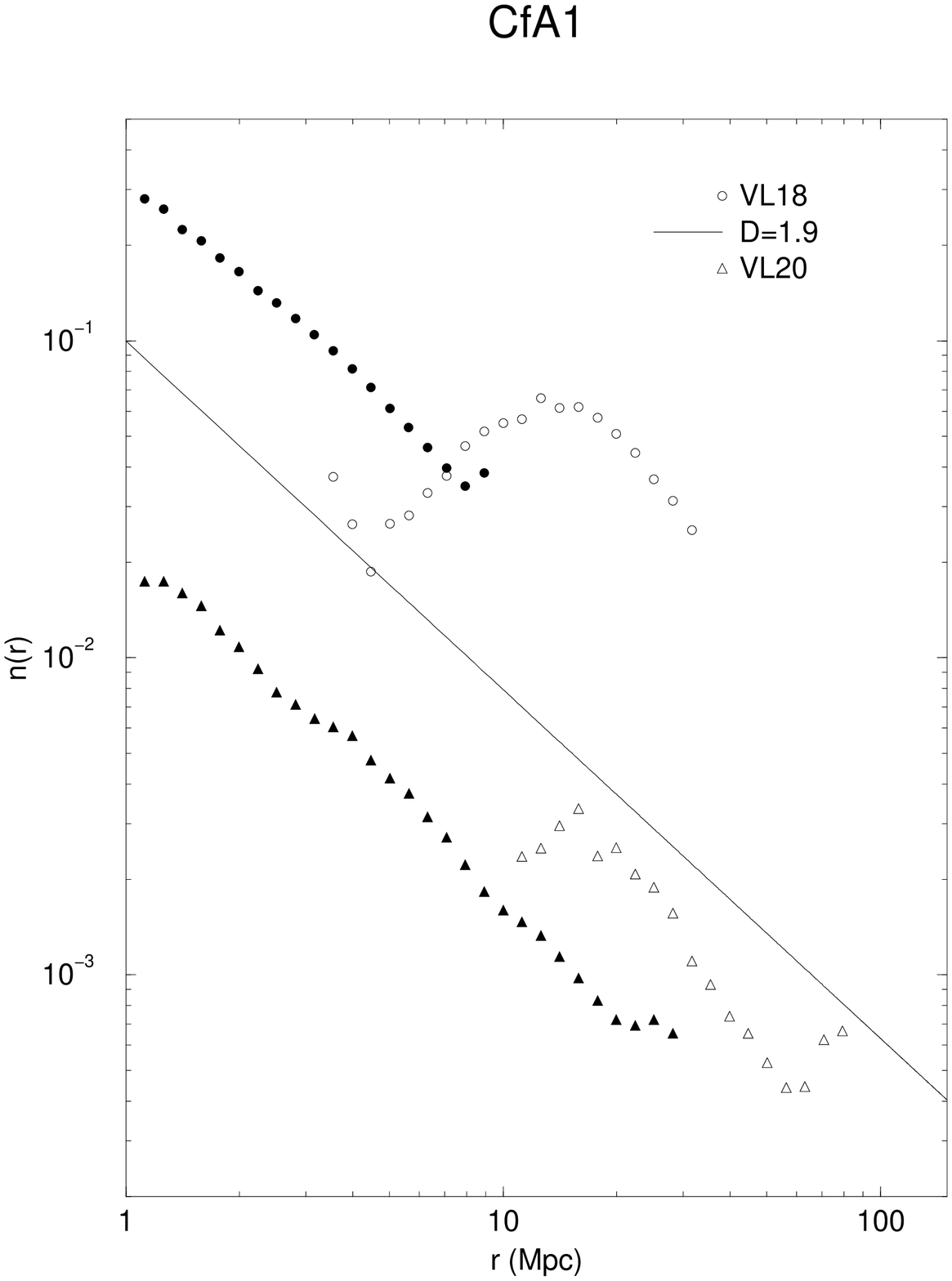}}  
\caption{\label{fig76}
The conditional average density $\Gamma^*(r)$ 
(filled points) and the radial density 
 $n(r)$ (open points)
 computed in a VL sample of CfA1. } 
\eef 
The normalization of the conditional density to the radial density
 should take into account only a trivial $(3/D)$
 factor. 
However for example in the case of PP and 
CfA1 there is a 
certain different (about a factor $2 \div 5$) between 
the amplitude of the conditional density and of the radial one, even 
if one has properly normalized the amplitude of these functions, as 
shown in Fig.\ref{fig75} 
and  Fig.\ref{fig76}.
As we have already mentioned in (see Eq.\ref{cazzz4}), 
this difference can be understood considering the fact that
$N(<R)$ is modulated by a function $f(R)$ which takes into account 
the presence of not smoothed out fluctuations at all scales.
Usually, galaxy surveys are pointed towards some great large scale
 structures. If one computes the conditional density in spherical
 shells,  one is properly making the average, taking into
 account the structures and the voids present in the sample. If, on
 the other hand, one measures the radial density from one point only, one
 is biased by the fact that the survey, by construction, does not
 contain voids and structures of the same size. The dimension of the
 structures is limited by the boundaries of the sample. Hence, if the
 survey contains a large scale structure cannot contain also a
 corresponding large scale void. This asymmetry is the real effect
 which shifts the amplitude of the radial density of a small factor
 beyond that of the conditional density. However we  
 stress that only the conditional density measures the correct
 amplitude, while the radial density is biased by the modulating 
 effect of $f(R)$, and hence by the geometry of the survey.
\bef 
%\vspace{}
 \epsfxsize 8cm 
\centerline{\epsfbox{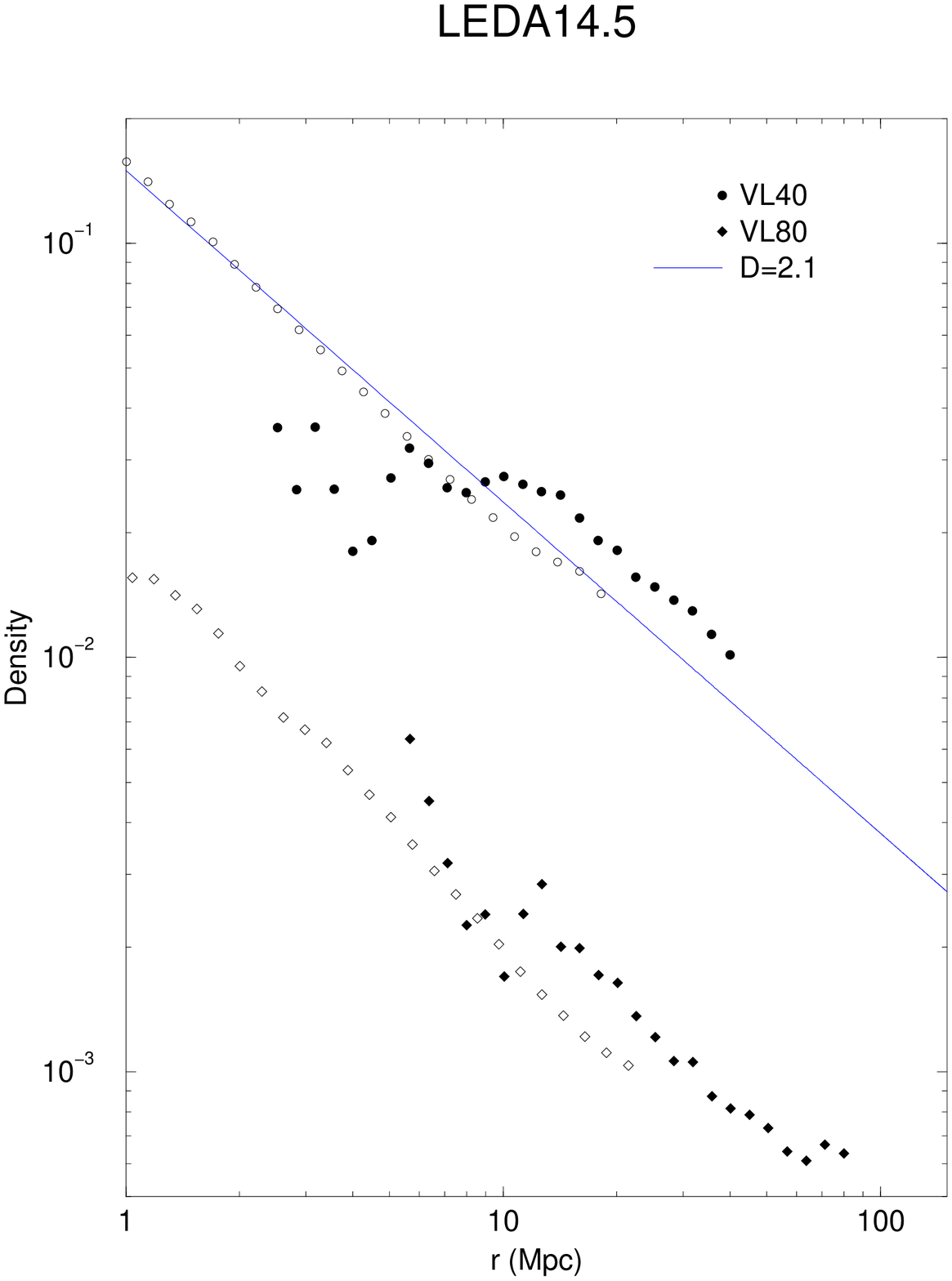}}  
\caption{\label{fig77} 
The conditional density   $\Gamma(r)$  and the radial density 
 $n(r)$   computed in a VL sample of LEDA14.5. The amplitude in this
 case match quite well because LEDA is an all-sky survey.  } 
\eef 
\bef 
%\vspace{}
\epsfxsize 8cm 
\centerline{\epsfbox{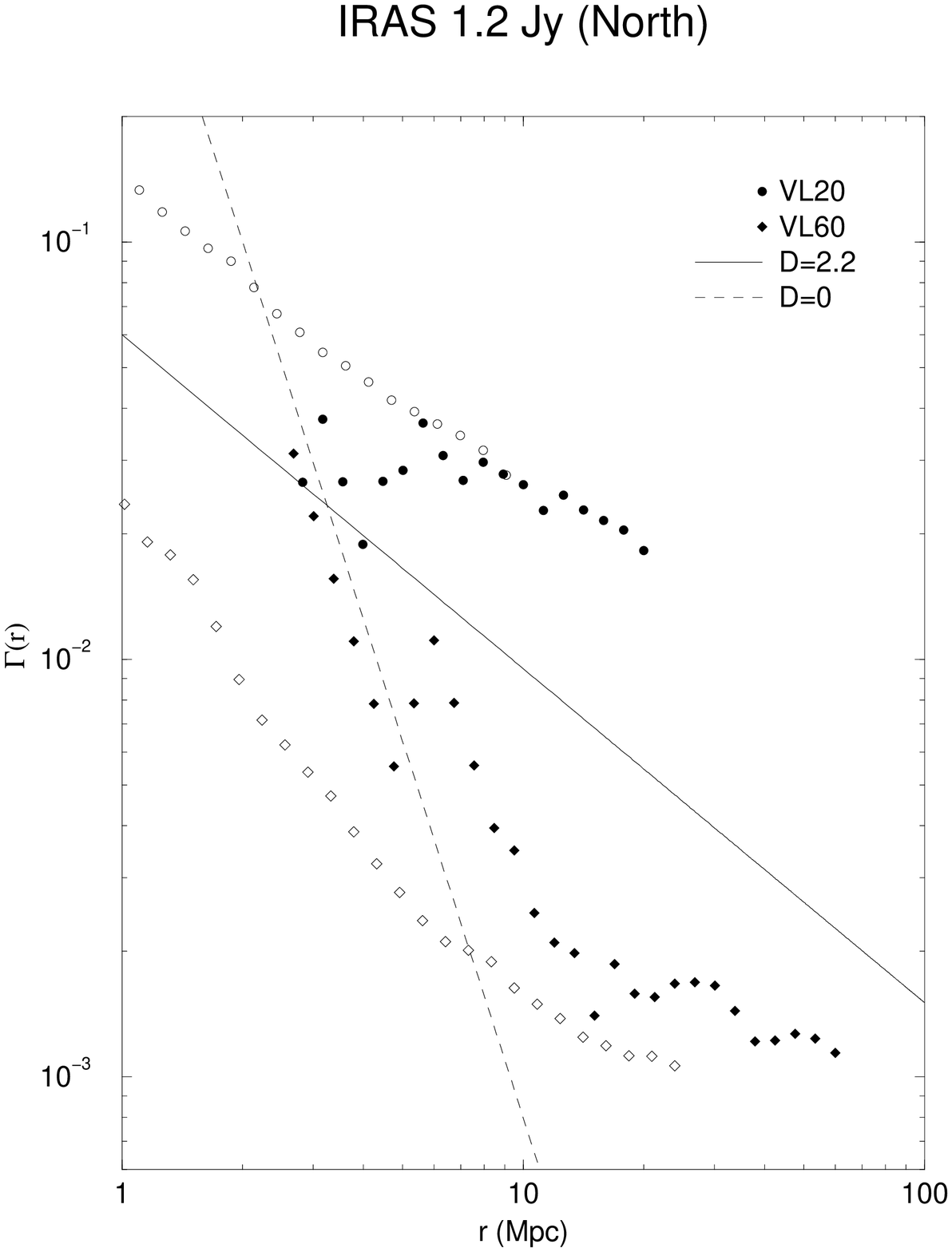}}  
\caption{\label{fig78} 
The conditional average density $\Gamma^*(r)$  and the radial density 
$n(r)$ computed in a VL sample of IRAS 1.2 Jy. The amplitude in this
case match quite well because IRAS is an all-sky survey.  } 
\eef 
 If the surveys are all-sky ones, then the amplitudes
 of the conditional and the radial density should match better,
because in this case by measuring $n(r)$ one properly
averages voids and structures. In other terms, as we have
previously discussed, this means that  $f_{\Omega}(R, \delta \Omega) > f_{4 \pi}$. 
This can be seen, for example, in the case of LEDA (Fig.\ref{fig77}), 
or IRAS (Fig.\ref{fig78}).

\subsubsection{Normalization of the density in different surveys} 
\label{normtot}
\bef 
%\vspace{}
 \epsfxsize 8cm 
\centerline{\epsfbox{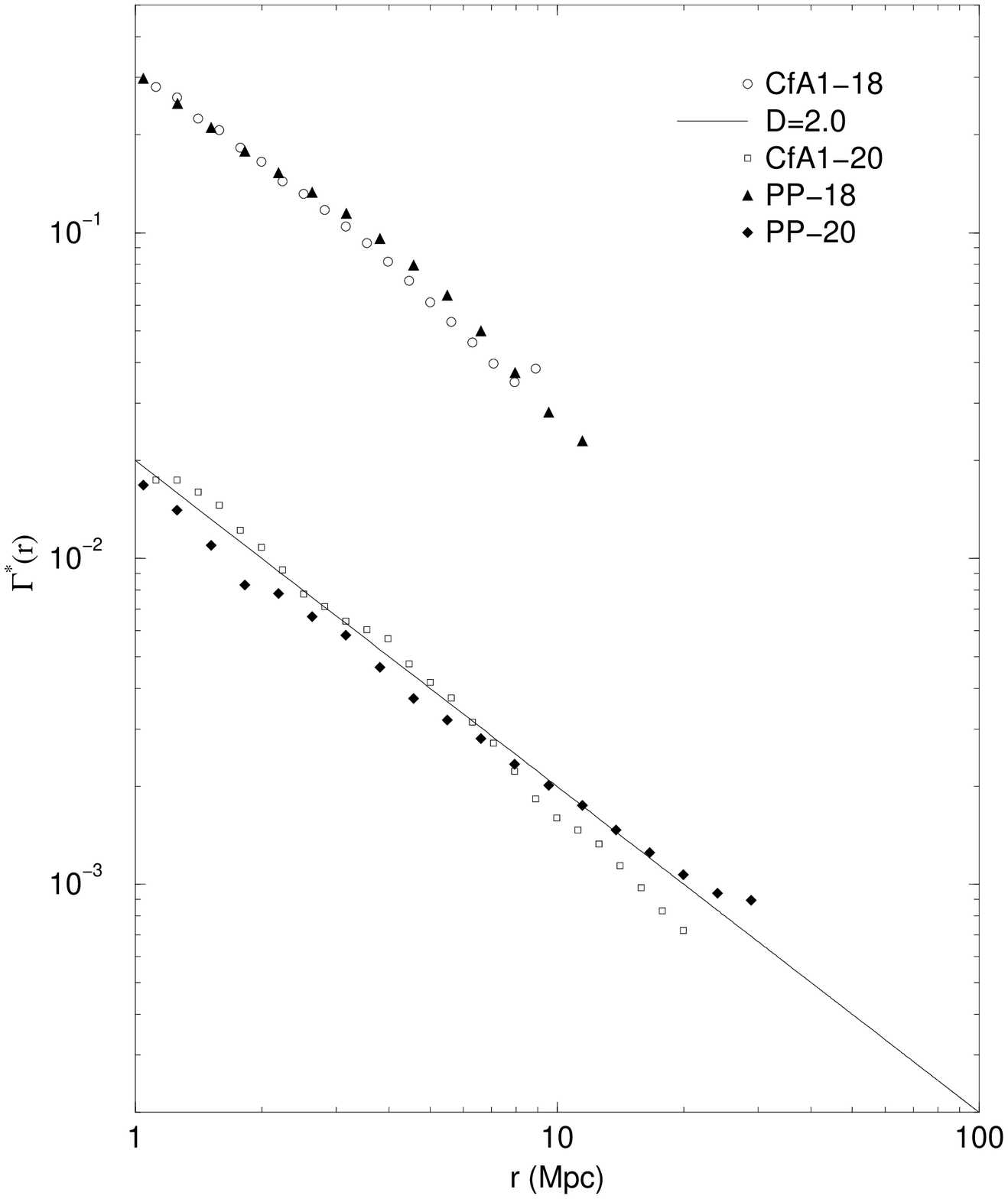}}  
\caption{\label{fig79} 
The spatial density   $\Gamma(r)$     computed in some 
VL samples of CfA1 and Perseus-Pisces cut at the same absolute
magnitude. The solid line corresponds to a fractal dimension $D=2$} 
\eef 
The normalization of $\Gamma(r)$ (or $n(r)$) in the surveys
CfA1 and Perseus-Pisces 
  can be done without any
 additional factor. If one compares VL with the same cut in
 absolute magnitude no normalization is needed (see
 Fig.\ref{fig79}).

For the case of Stromlo-APM Redshift Survey one has to take into
 account the random sampling, and hence one has to multiply the
 amplitude of the density for $20$ (see Sec.\ref{corran}). The same 
factor must be taken into account in the
 case of LEDA, even if the factor varies according to the sample
 considered (see Sec.\ref{corran} for a detailed discussion of the LEDA
 criteria).

The case of ESP is more complex: because of the presence of 
the "holes" in the survey one obtains a lower value of 
amplitude of the correlation function 
(see Sec.\ref{corran}). For this
reason the comparison is not possible as in the previous cases.
As we are dealing with a fractal distribution, we cannot estimate the missing points
through a simple factor, as one may do in the case of an homogeneous distribution.

\bef %\vspace{}
 \epsfxsize 14cm 
\centerline{\epsfbox{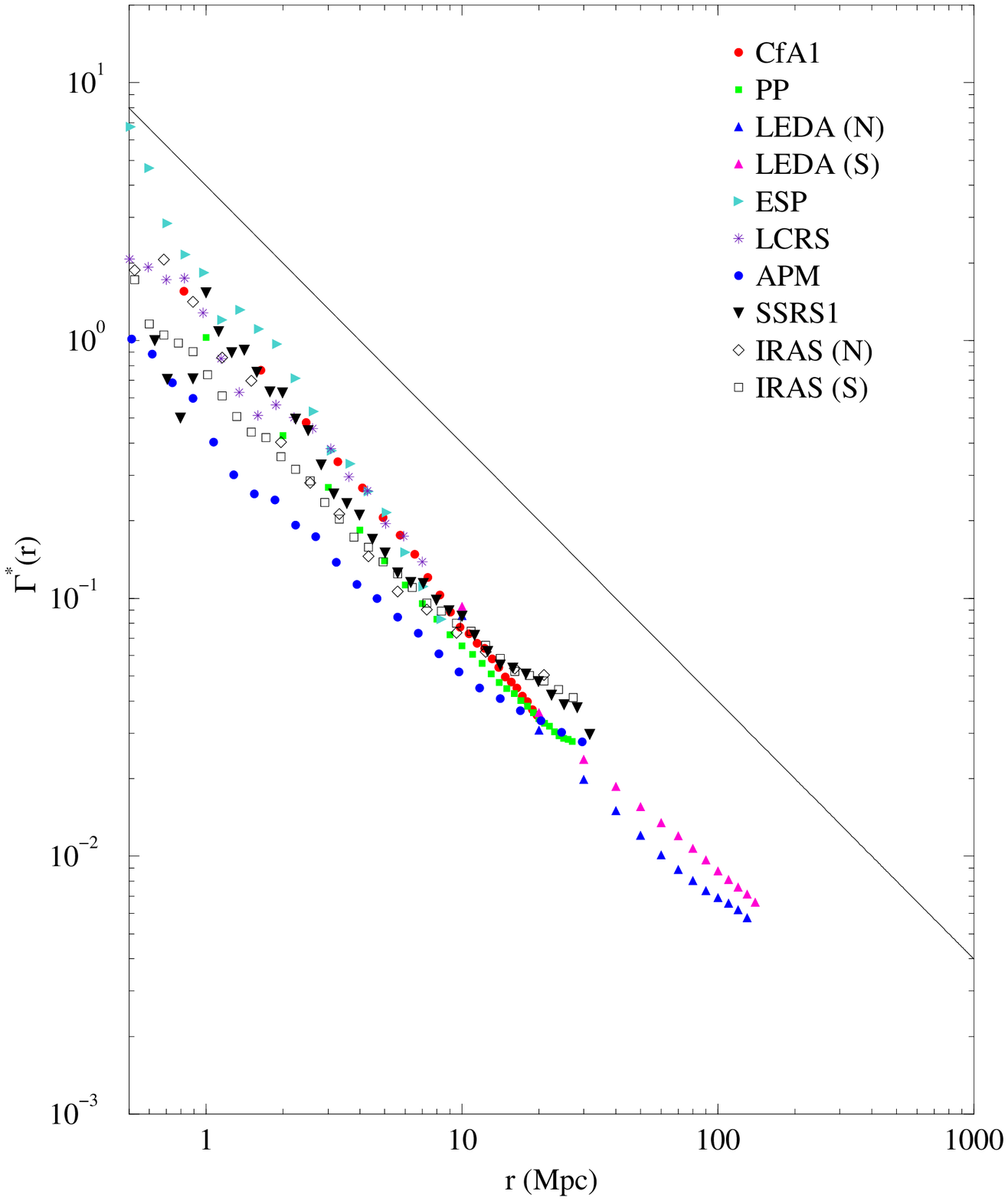}}  
\caption{\label{fig80} 
The spatial density   $\Gamma(r)$     computed in some 
VL samples of CfA1, PP, LEDA, APM, ESP, LCRS, SSRS1, IRAS 
and ESP
and normalized to the corresponding factor, as explained in the
text. } 
\eef 
In the case of SSRS1, LCRS and IRAS 
 the normalization can be done considering the different luminosity
 selection criteria adopted in each case. For the case of LCRS one
 has to take into account also the random sampling adopted in this
 case. We show in Fig.\ref{fig80} the normalization of all the
 surveys.
In Fig.\ref{fig81}
\bef 
%\vspace{}
 \epsfxsize 9cm 
\centerline{\epsfbox{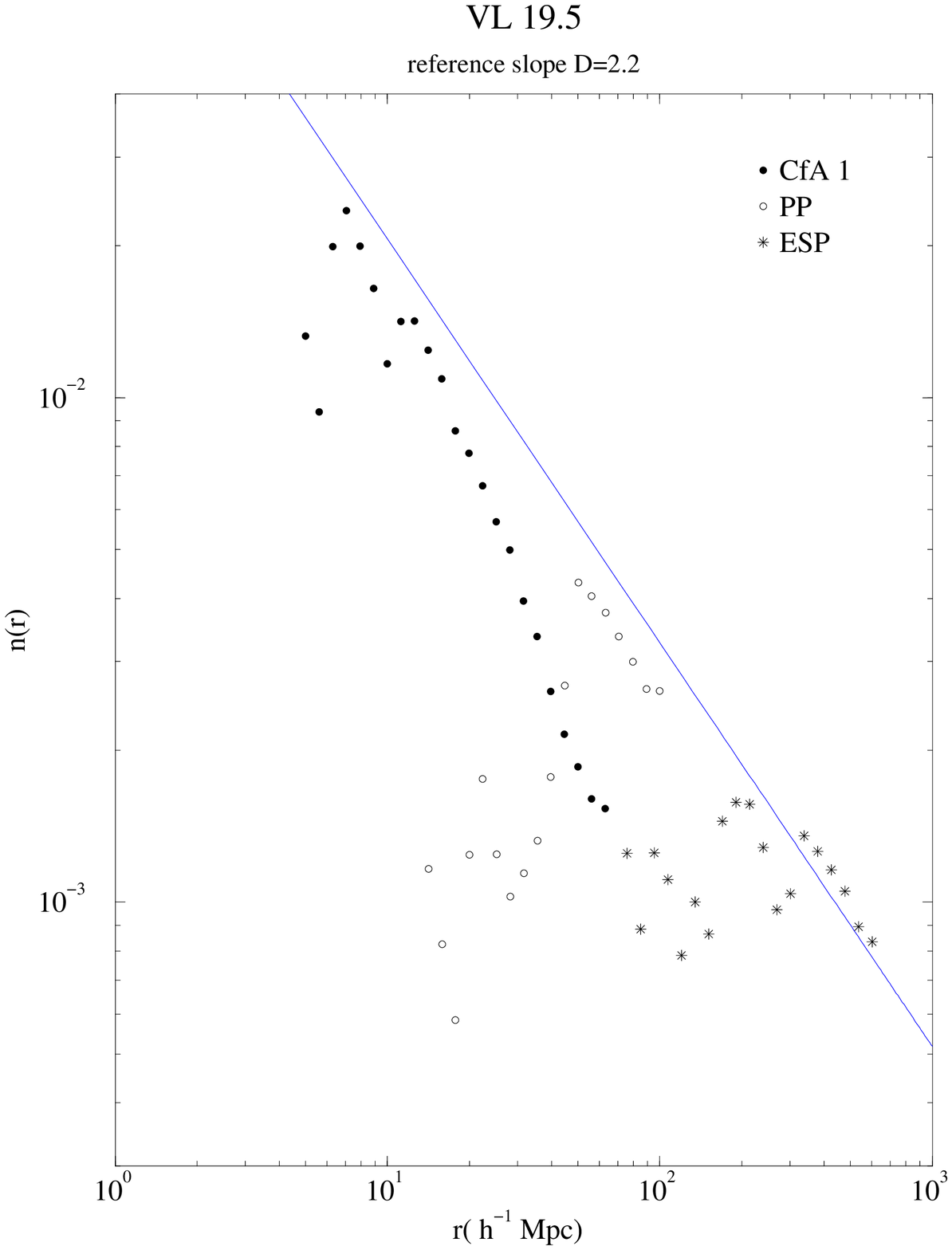}}  
\caption{\label{fig81} 
The spatial density   $n(r)$     computed in some 
VL samples with the same cut in absolute magnitude of CfA1, PP 
and ESP.} 
\eef 
we show the determinations of the radial density in VL with the same
cut in absolute magnitude, in 
different surveys: PP, CfA1, and ESP.
In these cases the match of the amplitudes and exponents
is quite good. 
\bef 
%\vspace{}
 \epsfxsize 14cm 
\centerline{\epsfbox{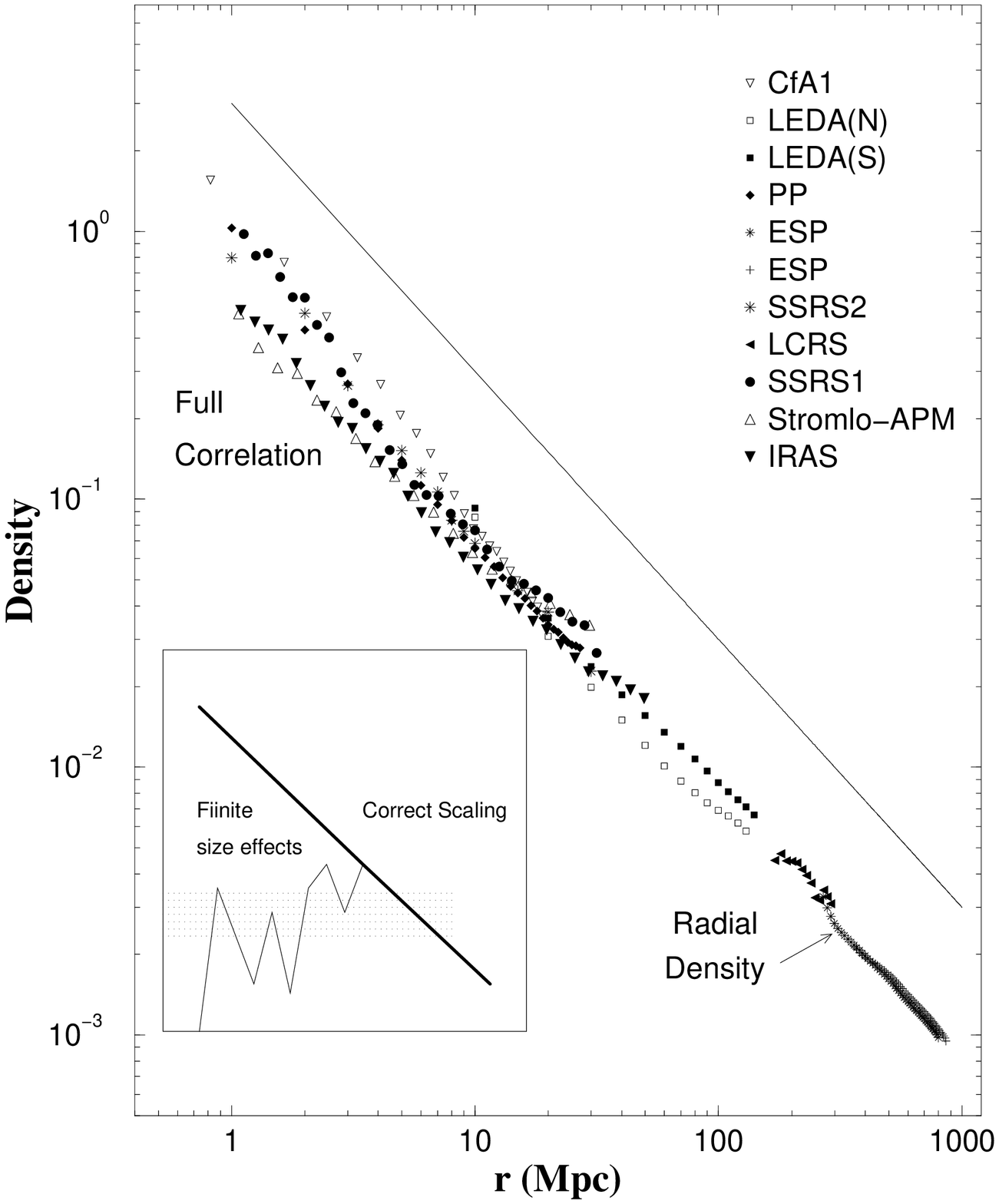}}  
\caption{\label{fig82} Full correlation analysis for the various
 available redshift surveys in the range of distance $0.5 \div 1000
 \hmp$. A reference line with slope $-1$ is also shown, which 
corresponds to fractal dimension $D = 2$. 
 } 
\eef 
\bef 
%\vspace{}
\epsfxsize 13cm 
\centerline{\epsfbox{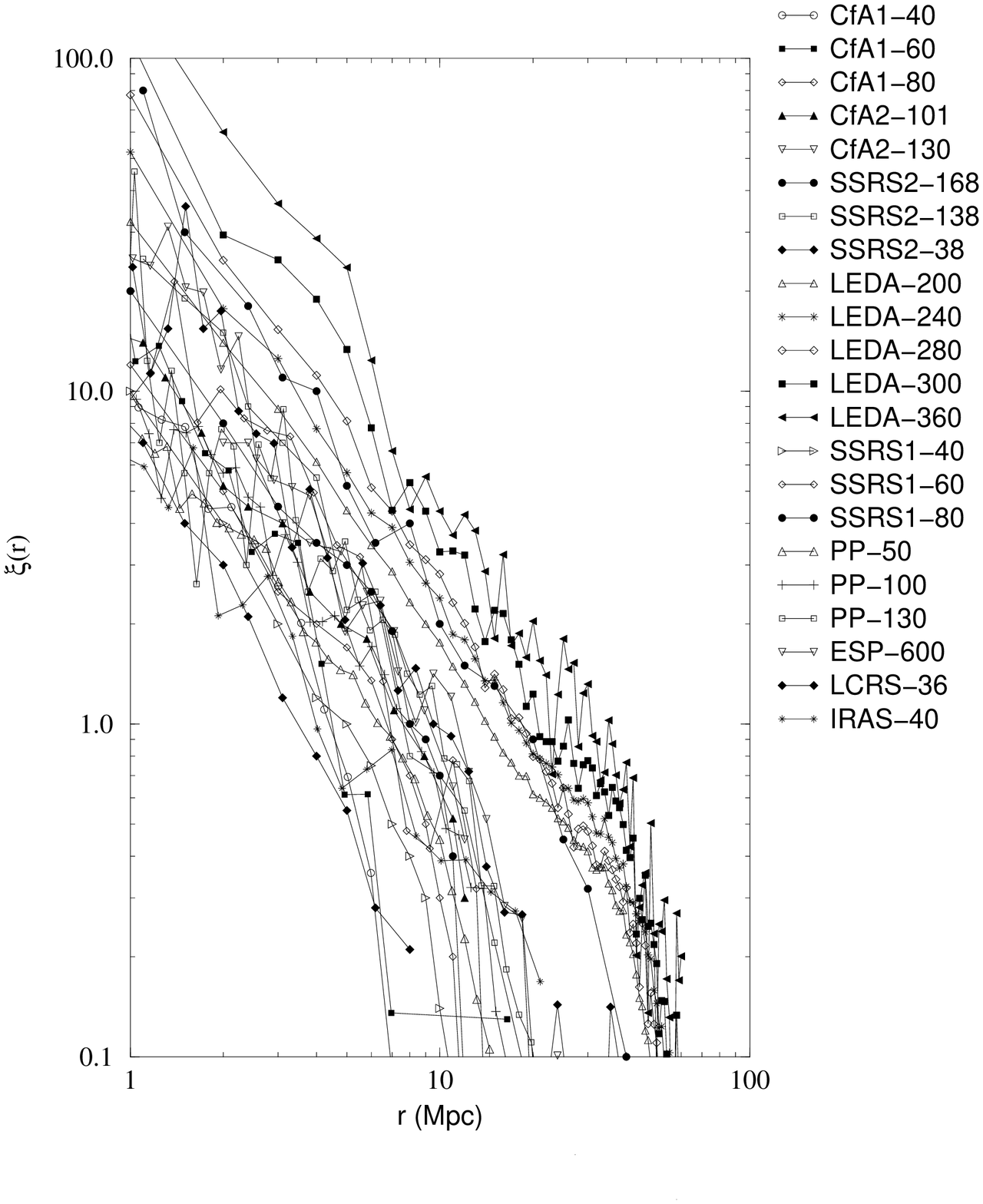}}  
\caption{\label{fig83}
Traditional analyses based on the function $\xi(r)$
of the same galaxy catalogs of the previous figure.
 The usual 
analysis is based on the a priori untested assumptions of 
analyticity and homogeneity. These properties
are not present in the real galaxy distribution and 
the results appear therefore rather confusing. 
This lead to the impression that galaxy catalogs are not good
enough and to a variety of theoretical problems like the 
galaxy-cluster mismatch, luminosity segregation, linear and 
non-linear evolution, etc. This situation changes completely and 
becomes quite clear if one adopts the more 
general conceptual framework that is at the basis 
the previous figure}
\eef 
\bef %\vspace{}
 \epsfxsize14cm 
\centerline{\epsfbox{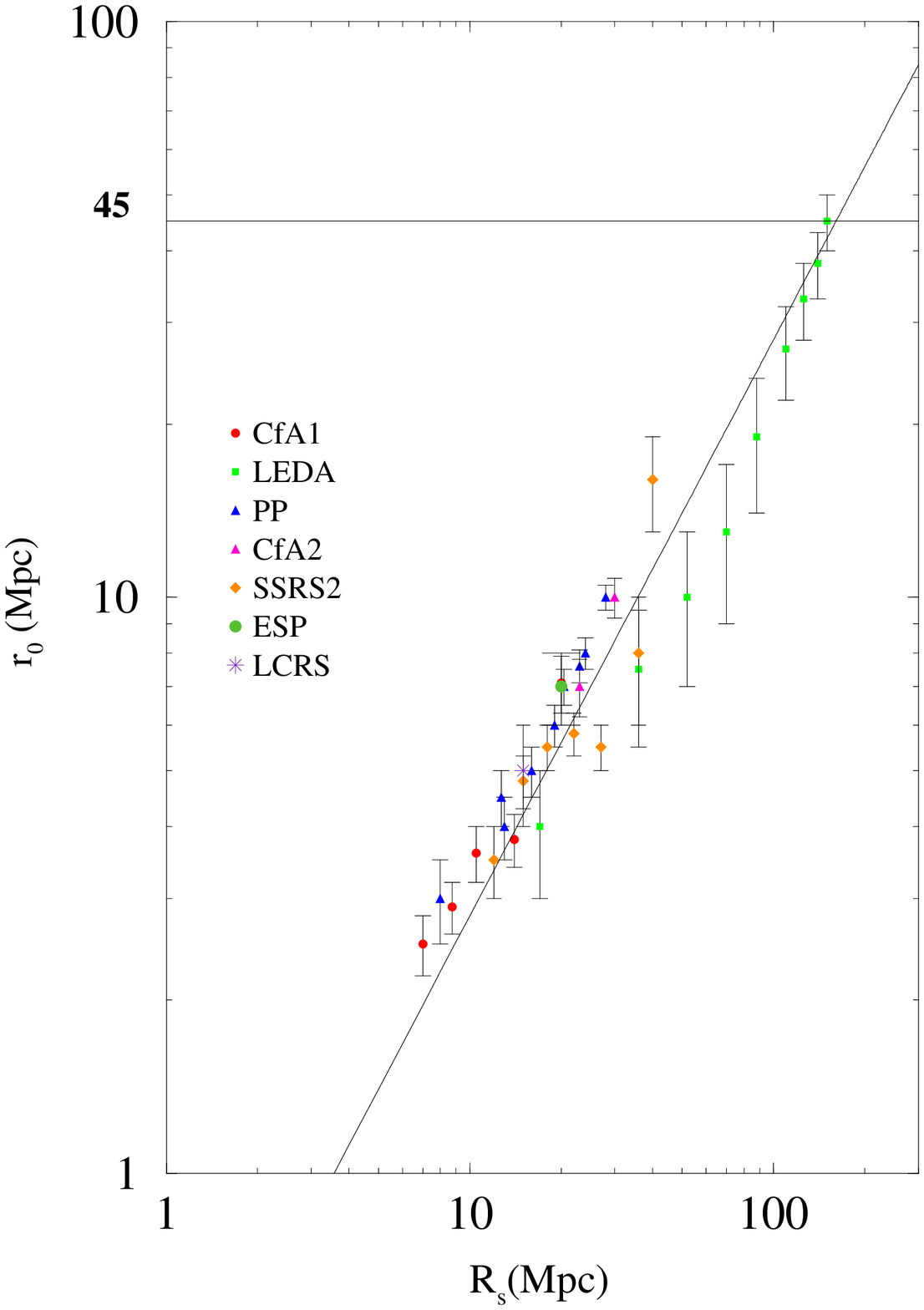}}  
\caption{\label{fig84}
Determinations of the so called 
"correlation-length" $r_0$ in the available redshift surveys.
The agreement among the different surveys is quite good.
A linear dependence of $r_0$ on the sample size $R_{eff}$
is shown. The observed is in agreement with the fractal prediction.}
\eef 

Finally in Fig.\ref{fig82}
we show the final diagram with all the available surveys. It is
 remarkable to stress that the amplitudes and the slopes of the
 different surveys match quite well. From this figure we conclude
 that galaxy correlations show very well defined fractal properties
 in the entire range $0.5 \div 1000 \hmp$ with dimension $D = 2 \pm
 0.2$. Moreover all the surveys are in agreement with each other.

 It is interesting to compare the analysis of Fig.\ref{fig82} with 
the usual one, made by the function $\xi(r)$, for the same 
galaxy catalogs. This is reported in Fig.\ref{fig83}.

 From this point of view, the various data 
the various data appear to be in strong disagreement with 
each other. This is due to the fact that the usual analysis
looks at the data from the prospective of analyticity and large
scale homogeneity (within each sample). These properties have never
been tested and they are not present in the real galaxy
distribution so the result is rather confusing (Fig.\ref{fig83}).
Once the same data are analyzed with a broader perspective the
situation becomes clear (Fig.\ref{fig82}) and the data of 
different catalogs result in agreement with each other. It is 
important to remark that analyses like those of Fig.\ref{fig83}
have had a profound influence in the field in various ways: 
first the different  catalogs appear in conflict with each other.
This has generated the concept of {\it not fair samples} and a 
strong mutual criticism about the validity of the data 
between different authors. In the other cases the
discrepancy observed in Fig.\ref{fig83} have been 
considered real physical problems for which various technical
approaches have been proposed. These problems are, for example, 
the galaxy-cluster mismatch, luminosity segregation, 
the richness-clustering relation and 
 the linear non-linear evolution of the perturbations 
corresponding to the {\it "small"} or  {\it "large"}
amplitudes of fluctuations. All this problematic
situation {\it is not real} and it arises only from a 
statistical analysis based on 
inappropriate and too restrictive
 assumptions that do not find 
 any correspondence in the 
physical reality. It is also
important to note that,
even if the galaxy distribution 
would eventually became 
homogeneous at larger scales, the use of the above statistical
concepts  is anyhow inappropriate for the range of scales 
in which the system shows fractal correlations as those 
shown in Fig.\ref{fig82}.

In Fig.\ref{fig84} we report the various determinations 
of the so called "correlation;  length" $r_0$ as a function of the sample size
in the available redshift surveys. Also in this case  $r_0$ is 
just a linear fraction of the sample depth, as expected in the fractal case.
{\it It is worth to notice that all the available 
catalogs give a consistent results for such a measurement.}

%%%%%%%%%%%%%%%%%%%%%%%%%%%%%%%%%%%%%%%%%%%%%%%%%%%%%%%%%%%%%%%%%%%%%%%%%%

\section{Number counts and angular correlations}
\label{counts}

  The most complete information about galaxy distribution comes
 from the full three dimensional samples, while  the angular
 catalogs have a poorer qualitative information, even if usually
 they contain  a much larger number of galaxies. However, one of
 the most important tasks in observational astrophysics,  is the
 determination of the  $\log N-\log S$ relation for different kind
 of objects: galaxies in the  various spectral  band (from
 ultraviolet to infrared), radio-galaxies, Quasars, X-ray sources
 and $\gamma$-ray bursts. This relation gives the number $N(S)$
 (integral or differential)  of objects, for unit  solid angle,
 with {\it apparent flux} (larger than a certain limit) $S$. The
 determination of such a quantity avoids the measurements of the
 distance, which is always a very complex task. However we show in
 the following that the behavior of the $\log N-\log S$ is
 strongly biased by some statistical finite size effects 
due to small scale fluctuations\footnote{We thank A. Gabrielli
for his useful  collaboration in the analysis of the 
number counts.} \cite{slgmp96}.

The counts of galaxies as  a function of the apparent magnitude are
determined from the Earth only, and hence it is not possible
to make an average over different observers. As we have 
already discussed in the previous section, this kind of measurement 
is affected by intrinsic fluctuations that are not smoothed out at any scale.
Moreover at small scale there are finite size effects which may
seriously perturb the behavior of the observed counts. Following the 
simple argument about the behaviour of the radial density 
we have presented in the previous section, we consider here the 
problem of the galaxy-number counts. For a more detailed 
discussion on this subject we refer to a forthcoming paper
\cite{fluc}.

 We present in this section a
 new interpretation of this basic relation at the light of the 
 highly inhomogeneity nature of galaxy distribution, and we show
 its compatibility with the behavior of  counts of  galaxies in
 different frequency bands, radiogalaxies, Quasars and X-ray
 sources. Moreover, we consider also the case of  $\gamma-$ ray
 burst  counts behavior showing that it presents the same feature
 of the galaxy counts. {\it Our  conclusion is that the counts of all
 these different kind of objects are compatible with a fractal
 distribution of visible matter up  to the deepest observable
 scale.}

Once we have clarified the correct interpretation of the number counts, 
we have all the elements to give the correct reinterpretation 
also to the angular catalogs. These catalogs are qualitatively inferior 
to the three dimensional ones because 
they only correspond to the angular projection
and do not contain the third coordinate, even if they usually
contain more galaxies.
We show that  angular projections  
mix different length scales and this gives an artificial randomization of
the galaxies. 
This implies that the angular projection of a fractal 
is really homogeneous 
relatively large angles. Clearly this is an artificial 
effect and from a smooth angular projection one cannot
 deduce whether the 
real distribution is also smooth. We study this problem both by
analyzing real
galaxy catalogs,
 and we point out several other subtle
 effects which enter in the angular analysis.

\subsection{Galaxy number counts data} 
\label{countsdata}

  Historically \cite{hu26,pee93} the
 oldest type of data about galaxy distribution is given by the
 relation between the number of observed galaxies $N(>S)$  and
 their apparent brightness $S$. It is easy to show that,
under very general assumptions one gets 
(see Sec.\ref{countsbasic})
 \begin{equation} 
\label{eq1} 
N(>S)  \sim  S^{-\frac{D}{2}}
 \end{equation} 
where $D$  is the fractal dimension of the galaxy
 distribution. Usually this relation is written in terms of the
 apparent magnitude $m$ ($S \sim 10^{-0.4 m}$ - 
note that bright galaxies
 correspond to small $m$). In terms of $m$, Eq.\ref{eq1} becomes
\be 
\label{nn1} 
\log N(<m)   \sim \alpha m 
\ee 
 with $\alpha =
 D/5$ \cite{bslmp94,pee93}. Note that $\alpha$ is the
 coefficient of the exponential behavior of Eq.\ref{nn1} and we
 call it "exponent" even though  it  should not be confused
 with the exponents of power law behaviors. In Fig.\ref{fig85} 
\bef 
%\vspace{}
\epsfxsize14cm
\centerline{\epsfbox{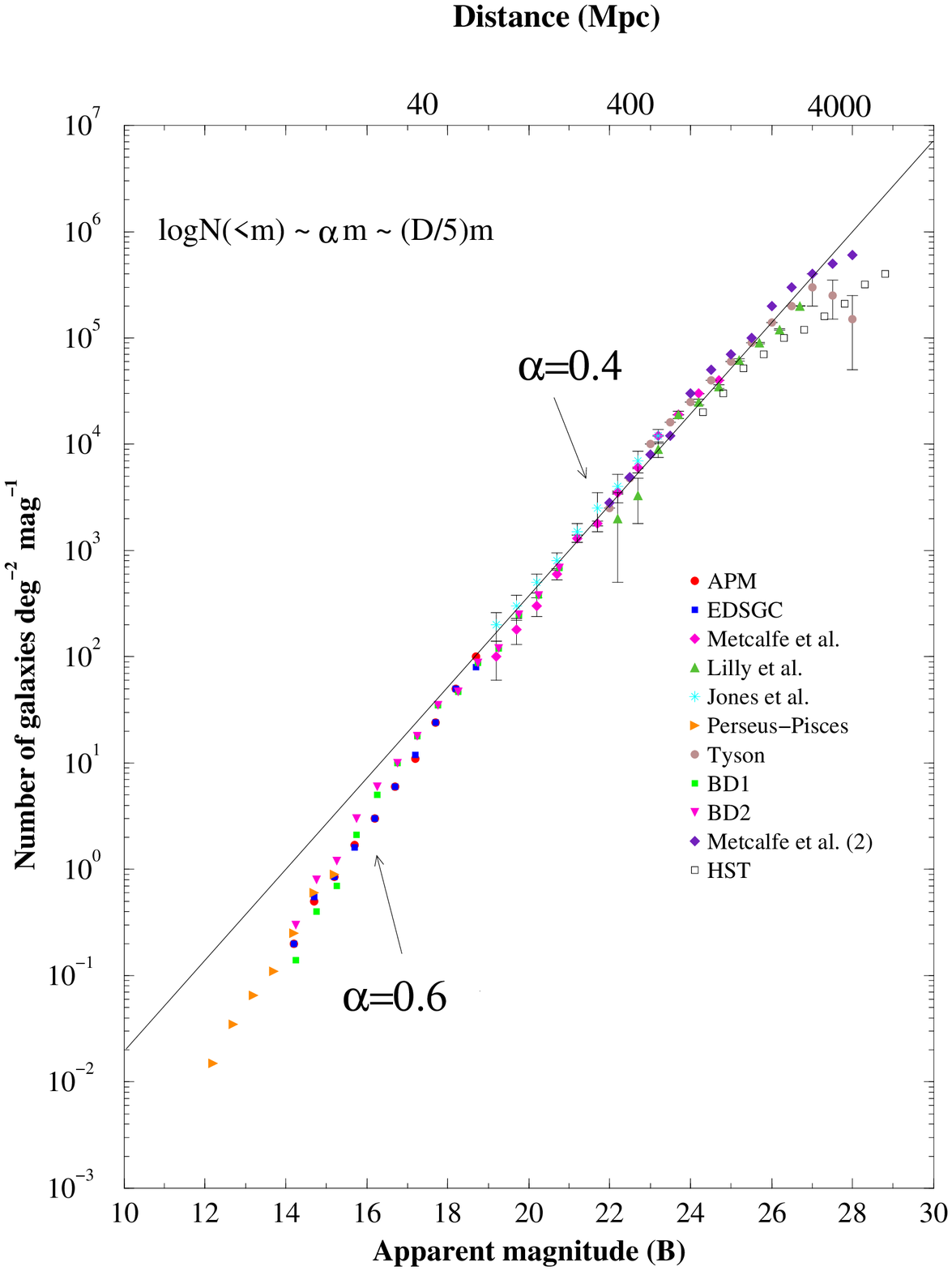}}
\caption{ \label{fig85} The galaxy number
 counts in the $B$-band, from several surveys. 
In the range $12 \ltapprox  m \ltapprox
 19$ the counts show an exponent $\alpha \approx 0.6 \pm 0.1$,
 while in the
 range $19 \ltapprox m \ltapprox 28$ the exponent is $\alpha 
 \approx 0.4$. The solid line is computed from the determination
of the amplitude of the conditional density at small scale, 
the fractal dimension  $D=2.2$, and from the knowledge
of the luminosity function.
The distance is computed for a galaxy with $M=-16$ 
and we have used $H_0=75 km sec^ {-1} Mpc^{-1}$.}  
\eef
 we have collected all the recent observations
 of $N(<m)$ versus $m$ in the
 $B$-spectral-band $m_B$ 
 \cite{ms91,mg94,ty88,lc91,jf91,dr94,cg90,co91,gc93,co94,gl94,mb86}.
 At bright and intermediate magnitudes
 ($\:12 \ltapprox m_B \ltapprox 18$), corresponding to small
 redshift ($\:z<0.2$),  one obtains $\:\alpha \approx 0.6$, while
 from $m_B \sim 19$ up to $m_B \sim 28$ the counts are well fitted
 by a smaller exponent with $\alpha \approx  0.4$.  The usual
 interpretation \cite{pee93,sha84,yo93,be92,yt88,yp95}  is that
 $\alpha \approx  0.6$ corresponds to $D \approx 3$ consistent with
 homogeneity, while at large scales galaxy evolution and space time
 expansion effects are invoked to  explain the lower value $\alpha 
 \approx 0.4$. On the basis of the previous discussion of the VL
 samples,   this interpretation is untenable. In fact,
 we know for sure that, at least  up to 
$\sim 150  \hmp $ there
 are fractal correlations, as we have discussed 
in the previous sections,
 so one would eventually expect the opposite
 behavior. Namely  small value of $\alpha \approx 0.4$ (consistent
 with $D \approx 2$) at small scales followed by a crossover to an
 eventual homogeneous distribution at large scales ($\alpha 
 \approx 0.6$ and $D  \approx 3$). 

 The GNC in the (red) $R$-band
 shows an exponent $\alpha\approx  0.37-0.41$ in the range
 $20<R<23$ \cite{ty88,jf91,ms91}. 
 Moreover Gardner {\em et
 al.} \cite{gc93}  have studied the GNC in the (infrared) {\em
 K}-band in the range $\:12 \ltapprox K \ltapprox 23$, and they
 show that the slope of the counts changes at $K \approx 17$ from
 $\:0.67$ to $\:0.26$ (see also \cite{co94,gl94,mb86,so94}).
 Djorgovski {\em et al.} \cite{dj95} found that the slope of the
 GNC is little bit higher than \cite{gc93}, i.e.  $\alpha =0.315
 \pm 0.02$ between $K=20$ and $24$ mag.

The situation is therefore
 quite similar in the different spectral bands. The puzzling
 behavior of the GNC represents an important apparent 
contradiction we
 find in the data analysis.   We argue here  that this  apparently
 contradictory experimental situation can be fully understood on 
 the light of the small scale effects in the space distribution of
 galaxies (Sec.\ref{radial}). For example a fractal distribution
 is non analytic in
 every occupied point: it is not possible to  define a meaningful
 average density because  we are dealing with intrinsic
 fluctuations which grow with  as the scale of the system itself.
 This situation is qualitatively different from an homogeneous 
 picture, in which a well defined density exists, 
and the
 fluctuations represent only  small amplitude perturbations. The
 nature of the fluctuations in these two cases is 
completely
 different, and for fractals the fluctuations themselves define all
 the statistical  properties of the distribution.
 This concept has 
 dramatic consequences in the following discussion as well as in
 the  determination of various observable quantities, such as the
 amplitude of the two point  angular  
correlation function.   It is
 worth to notice that the
  small scale effects are usually neglected in
 the study of fractal structures because one can generate large
 enough (artificial) 
structures to avoid these problems. In Astrophysics the
 data are instead intrinsically 
 limited and, as we have already mentioned in the 
previous two sections, 
 a detailed analysis of finite size effects is very important. We
 discuss     the problems of finite size effects in the
 determination of the asymptotic properties of fractal
 distributions,  considering explicitly the problems induced by the
 lower cut-off (Sec.\ref{radial}).

\subsection{Galaxy counts: basic relations} 
\label{countsbasic}

We briefly introduce some basic relations which are 
 useful later. We can start by computing
the expected GNC in
 the simplest case of a magnitude limited (ML) sample. A ML sample
 is obtained by measuring all the galaxies with apparent magnitude
 brighter than a certain limit $m_{lim}$. In this case we have (for
 $m < m_{lim}$) 
\be 
\label{q3} 
N(<m) = 
B \Phi(\infty) 10^{\frac{D}{5}m} 
\ee where 
\be 
\label{q3a} 
\Phi(\infty) =
 \int_{-\infty}^{\infty} \phi(M) 10^{-\frac{D}{5}(M+25)} dM \;.
\ee  
We
 consider now the case of a volume limited (VL) sample. A  VL 
 sample consists of  every galaxy in the volume which is more luminous
 than a certain limit, so that in such a  
sample there is no
 incompleteness for an observational luminosity selection effect
 \cite{dp83,cp92}. Such a sample is 
defined by a certain
 maximum distance $R$ and an absolute magnitude limit given by: 
\be
 \label{q3b}
 M_{lim}=m_{lim}-5\log_{10}R -25 - A(z)
\ee
 ($m_{lim}$ is the
 survey apparent magnitude limit). By performing the calculations
 for the number-magnitude relation,   we obtain 
\be
\label{q4b}
 N(<m)= A(m) \cdot 10^{\frac{D}{5}m} + C(m) 
\ee 
where $A(m)$ is 
\be
 \label{q4c} 
A(m) =B \int_{M(m)}^{M_{lim}}\phi(M)
 10^{-\frac{D}{5}(M+25)}dM
\ee 
and $M(m)$ is given by $M(m) = m - 5 \log(R) -25$,
  and it is a function of $m$.
 The second term is 
\be 
\label{q5e}
C(m)= BR^{D}\int_{-\infty}^{M(m)} \phi(M) dM \;.
\ee 
This term, as $A(m)$, 
depends on the VL sample considered. We assume a luminosity
 function with a Schecther shape (see Sec.\ref{corran}). 
 For $M(m) \gtapprox M^{*}$ we have that $C(m)
 \approx 0$, and $A(m)$ is nearly constant with $m$. This happens
 in particular for the deeper VL samples for which $M_{lim}\sim
 M^*$. For the less deeper VL ($M_{lim} > M^*$) samples these terms
 can be considered as a deviation from a power law behavior only
 for $m \rightarrow m_{lim}$. 
 If one has $\phi(M) = \delta(M-M_0)$ then it is simple to show that
 $\log(N(<m)) \sim (D/5)m$ also in each  VL sample.

\subsection{Galaxy counts in redshift surveys}
\label{countsred}

We have studied   the GNC in 
the Perseus-Pisces
 redshift  survey \cite{hg88} in order 
 to clarify the role of spatial inhomogeneities and finite
 size effects. In the previous sections 
 we have analyzed the spatial
 properties of galaxy distribution in this sample by measuring
the conditional (average) density and the radial density.
Let us briefly summarize our main results.

 When one
 computes the conditional average density, one indeed
 performs an average over all the points of the survey, as we have
 discussed     in Sec.\ref{corran}. 
 On the contrary the radial density is computed only from
 a single  point, the origin (Sec.\ref{radial}). 
This allows us to extend the study of
 the spatial distribution up to very deep scales: 
the price to pay is that this method
 is strongly affected by  statistical fluctuations and finite size
 effects. Analogously, when one computes $N(<m)$,  one does not
 perform an average but  just counts the points from 
the origin. As
 in the case of the radial density
 $n(r)$ also $N(<m)$ is strongly affected by
 statistical fluctuations  due to finite size effects.
 as well as intrinsic oscillations that are not
 smoothed out.
  We are now
 able to 
 clarify how the behavior of $N(<m)$, and in particular its
 exponent,  are influenced  by these effects.

We show in
 Fig.\ref{fig86}, 
\bef 
%\vspace{}
\epsfxsize 10cm
\centerline{\epsfbox{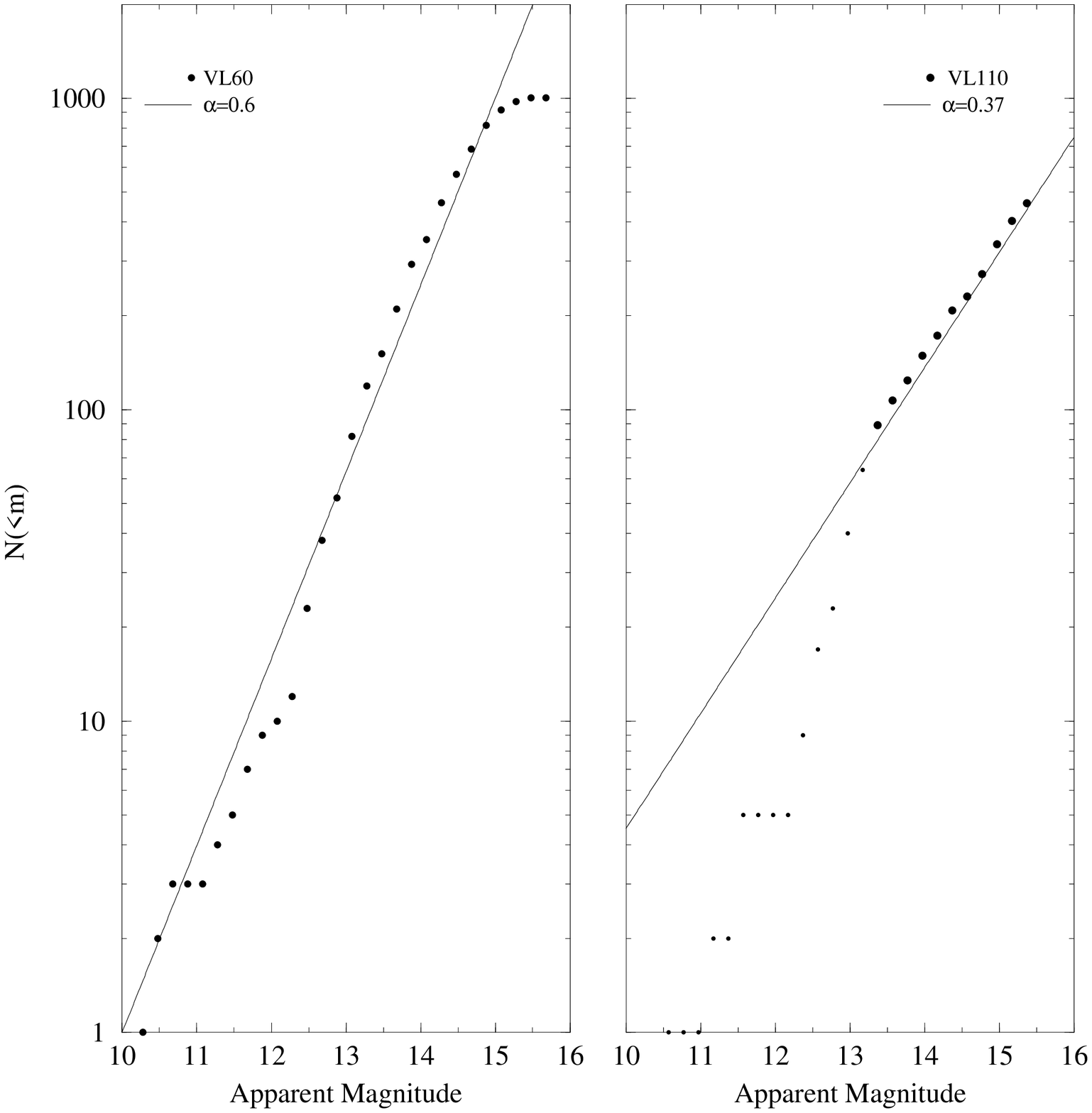}}
\caption{\label{fig86}
$(a)$
 The Number counts $N(<m)$ for the VL sample VL60. The 
 slope is $\alpha \approx 0.6$. In this case occurs a flattening
 for $m \rightarrow m_{lim}$. 
$(b)$ The Number
 counts $N(<m)$ for the VL sample VL110. The slope is
 $\alpha \approx 0.4$, a part from the initial fast growth 
due to weak statistics. This
 behavior corresponds to a well defined define power law behavior of
 the density with exponent $D \approx  5 \alpha \approx 2$.
 } 
\eef 
and 
Fig.\ref{fig87} 
\bef 
%\vspace{}
\epsfxsize 10cm
\centerline{\epsfbox{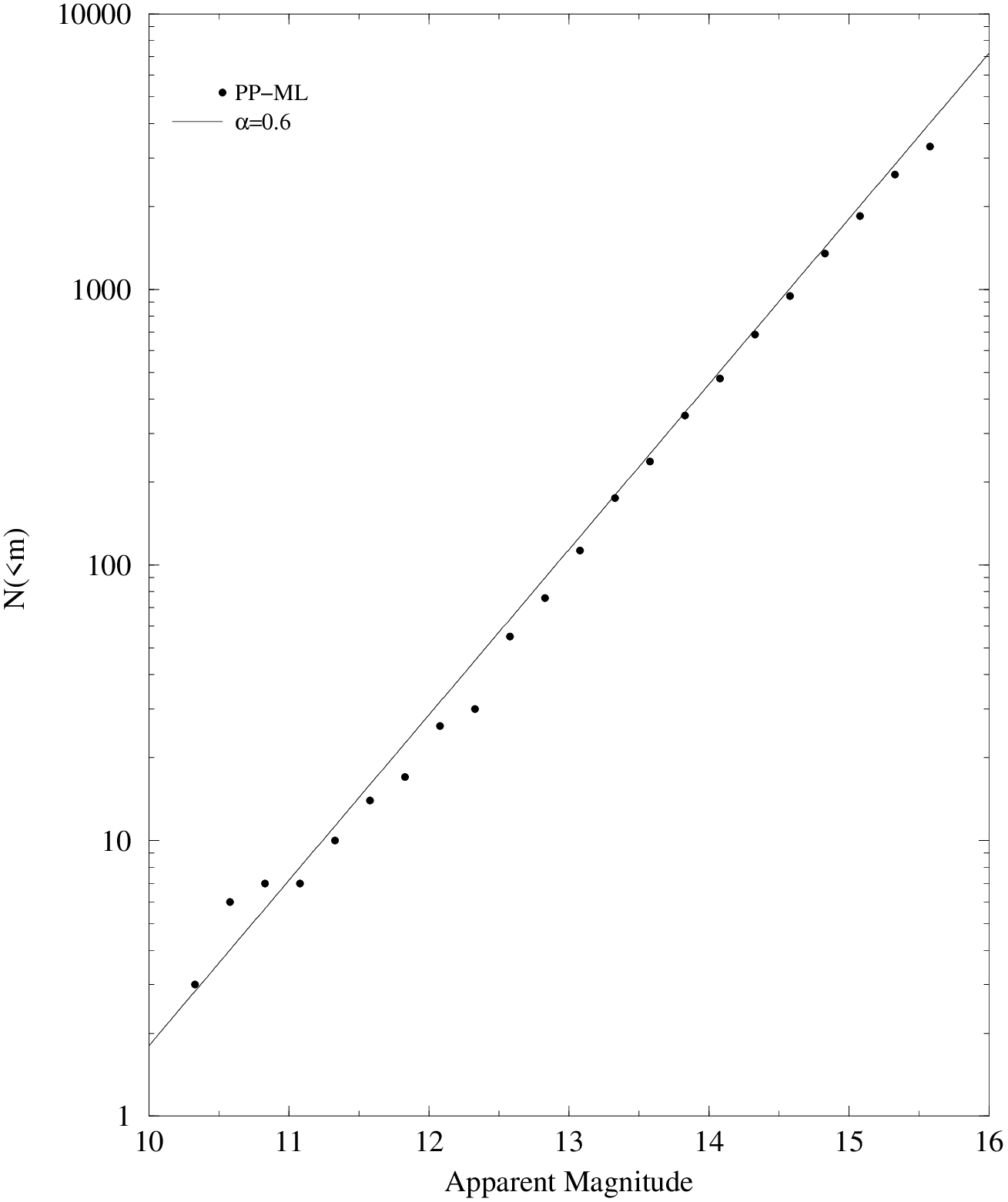}}
 \caption{\label{fig87}
 The Number counts $N(<m)$ for the whole magnitude
 limit sample. The slope is $\alpha \approx 0.6$ and it is clearly
 associated only to fluctuations in
 the spatial distribution rather
 than to a real homogeneity in space. } 
\eef 
the behavior of
 $N(<m)$ respectively for the various VL samples  
 and for the whole
 magnitude
 limit sample. In  VL60 there
 are very strong inhomogeneities in the behavior of $n(r)$ (see
 Fig.\ref{fig62}) and
 these are associated with a slope $\alpha \approx 0.6$ for the
 GNC. (The flattening for $m \rightarrow m_{lim}$ is just to a
 luminosity selection effect  that is  explained in Sec.\ref{countsbasic}). 
 For VL110 the behavior of the density is much more
 regular and smooth, so that it shows  indeed a clear power law
 behavior. Correspondingly the behavior of $N(<m)$ is well fitted
 by  $\alpha \approx 0.4$. Finally the whole magnitude limit sample
 is again described by an exponent $\alpha \approx 0.6$. 

 We have
 now enough elements to describe the behavior of the  GNC. The
 first point is that the exponent of the GNC is strongly related to
 the space distribution. Indeed, what has never been taken into
 account   before is  the role of finite size effects \cite{slgmp96}.
 The behavior of the GNC is determined by 
 a convolution of the space density
 and the luminosity function, and the space density enters
 in the GNC as an integrated quantity.  The problem is to consider
 the correct space density in the interpretation of   data analysis. 
 In fact, if the density has
 a very fluctuating behavior  
in a certain region of length scale, 
 as in  the case shown in Fig.\ref{fig62}, its integral  over this
 range of length scales is almost equivalent to a flat one. This
 can be seen also in Fig.\ref{fig61}:  at small 
distances one finds 
 almost no galaxies because one is below the mean  minimum separation
 between neighbor galaxies. Then
 the number of galaxies starts to grow, but this regime is strongly
 affected by finite size fluctuations. Finally the correct scaling
 region $r \approx \lambda$ is reached. This
 means for example that
 if one has a fractal distribution, there is first a raise of
 the density, due to finite size effects and characterized by
 strong  fluctuations, because no galaxies are present before a
 certain characteristic scale. Once one enters in the correct
 scaling regime for a fractal the  density becomes to decay as
  a power law. So in this regime of raise and fall with strong
 fluctuations there is  a region in which  the density can be
 approximated roughly by a  constant value.  This leads to an
 apparent exponent $D \approx 3$, so that the integrated number
 grows as   $N(<r) \sim r^3$ and  it is associated in terms of GNC,
 to $\alpha  \approx 0.6$ (Fig.\ref{fig88}).
\bef 
%\vspace{}
\epsfxsize 12cm
\centerline{\epsfbox{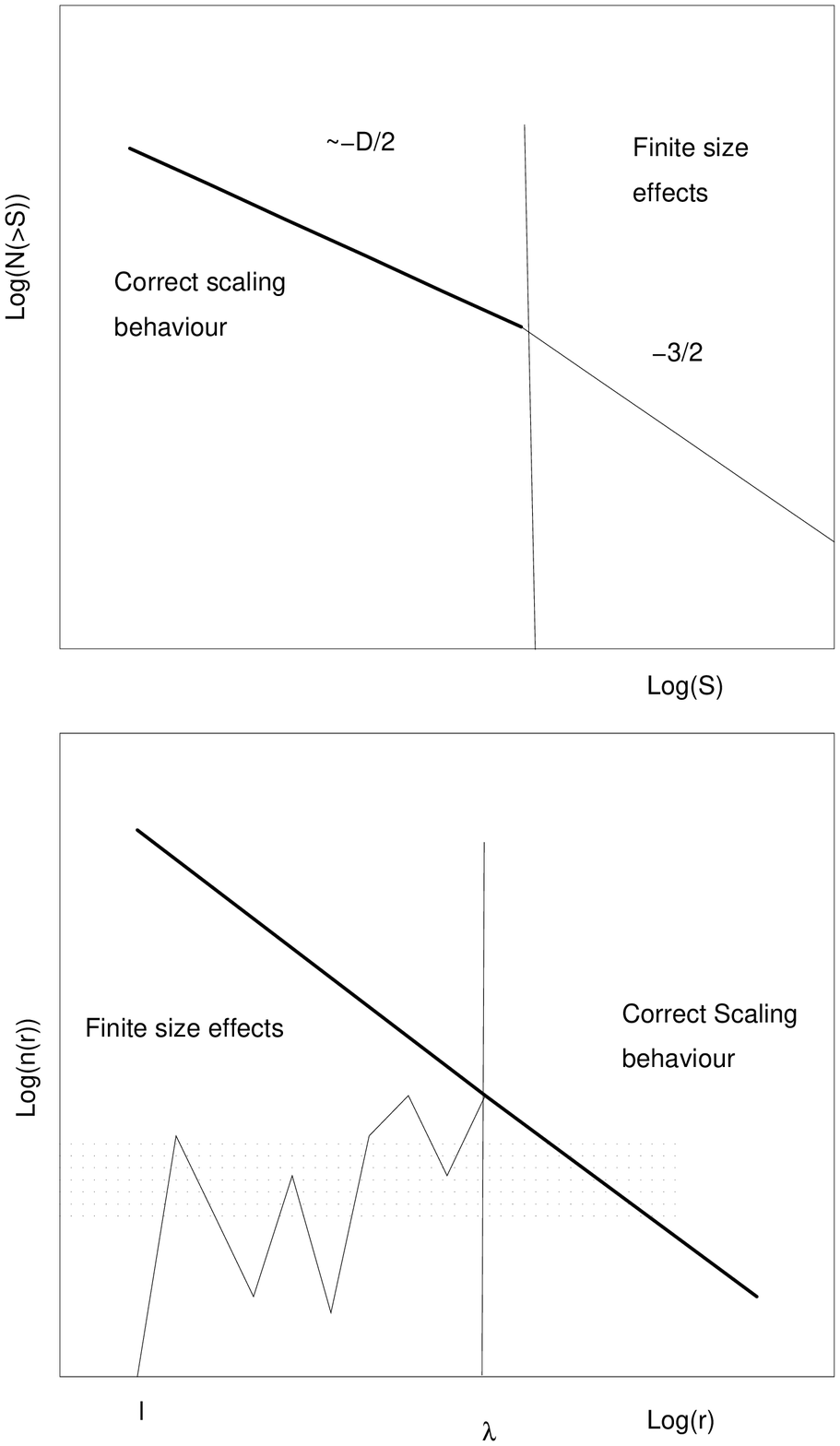}}
 \caption{\label{fig88}
 The number counts $N(<m)$ in a magnitude
 limit sample, together with the behavior of 
the space density.
At small scale the density is characterized by
having strong fluctuations which lead to a 
slope $\alpha \approx 0.6$. This is clearly
 associated only to fluctuations in the 
spatial distribution rather
 than to a real homogeneity in space. At larger scales (faint end)
the correct scaling behavior is recovered and $\alpha = D/5$. } 
\eef 
This exponent is therefore not a real one but
 just due to finite size fluctuations. Only when a well defined
 statistical scaling regime has been reached, i.e. for $r >
 \lambda$,  one can find the genuine scaling properties of the  
 structure, otherwise the behavior is completely  dominated by
 spurious finite size effects.  
 In the VL samples where $n(r)$ scales
 with the asymptotic properties (Fig.\ref{fig86}) the GNC grows also
 with the right exponent ($\alpha=D/5$).

 If we now consider instead
 the behavior of the GNC in the whole magnitude limit 
 (hereafter ML) survey, we
 find that the exponent is $\alpha  \approx 0.6$ (Fig.\ref{fig87}). 
 This behavior can be understood by 
 considering that at small
 distances, well inside the distance $\lambda$ defined by
 Eq.\ref{v3}, the number of galaxies present in  the sample is
 large because there are  galaxies of all magnitudes. Hence the
 majority of galaxies correspond to small distances ($r < \lambda$)
 and the distribution has not reached the scaling regime  in
 which the  statistical self-averaging   properties of the system
 are present. For this reason in the ML sample the finite size
 fluctuations dominate completely the behavior of the GNC.
 Therefore this behavior  in the ML sample is associated with
 spurious finite size effects rather than to real homogeneity.
 We discuss in a more quantitative way 
 the behavior in ML surveys later.

\subsubsection{Test on finite size effects: the averege $N(<m)$}
\label{countstest}

To prove that the behavior found in Fig.\ref{fig87}, i.e. that the
 exponent $\alpha \approx 0.6$ is connected to large fluctuations
 due to finite size effects in the space distribution and not to 
 real homogeneity, we have done the 
following test. We have adopted
 the same procedure used for the
 computation of the correlation
 function (see Sec.\ref{corran}), i.e. we make an average
 for $N(<m)$ from all the points of the sample rather than counting
 it from the origin only.  

To this aim we have considered a VL
 sample with $N$ galaxies and we have built $N$ independent
 flux-limited surveys in the following way.
 We consider each galaxy
 in the sample as the observer, and for each observer we have
 computed the apparent magnitudes of all the other galaxies. To
 avoid any selection effect we consider only the galaxies  which
 lie inside a well defined volume around the observer. This volume
 is defined by the maximum sphere fully contained in the sample
 volume with the observer as a center.  

Moreover we have another
 selection effect due to the fact that our VL sample has been built
 from a ML survey done with respect to the origin. To avoid this
 incompleteness we have assigned to each galaxy a constant
 magnitude $M$. In fact, our aim is to show that 
the inhomogeneity
 in the space distribution plays the fundamental role that
 determines the shape of the $N(<m)$ relation, and the functional
 form of the luminosity function enters in Eq.\ref{q4c} only as an
 overall normalizing factor.  

Once we have computed $N_i(<m)$ from
 all the points $i=1,..,N$ we then compute the average. We show in
Fig.\ref{fig89} 
\bef %\vspace{} 
\epsfxsize 10cm
\epsfysize 10cm
\centerline{\epsfbox{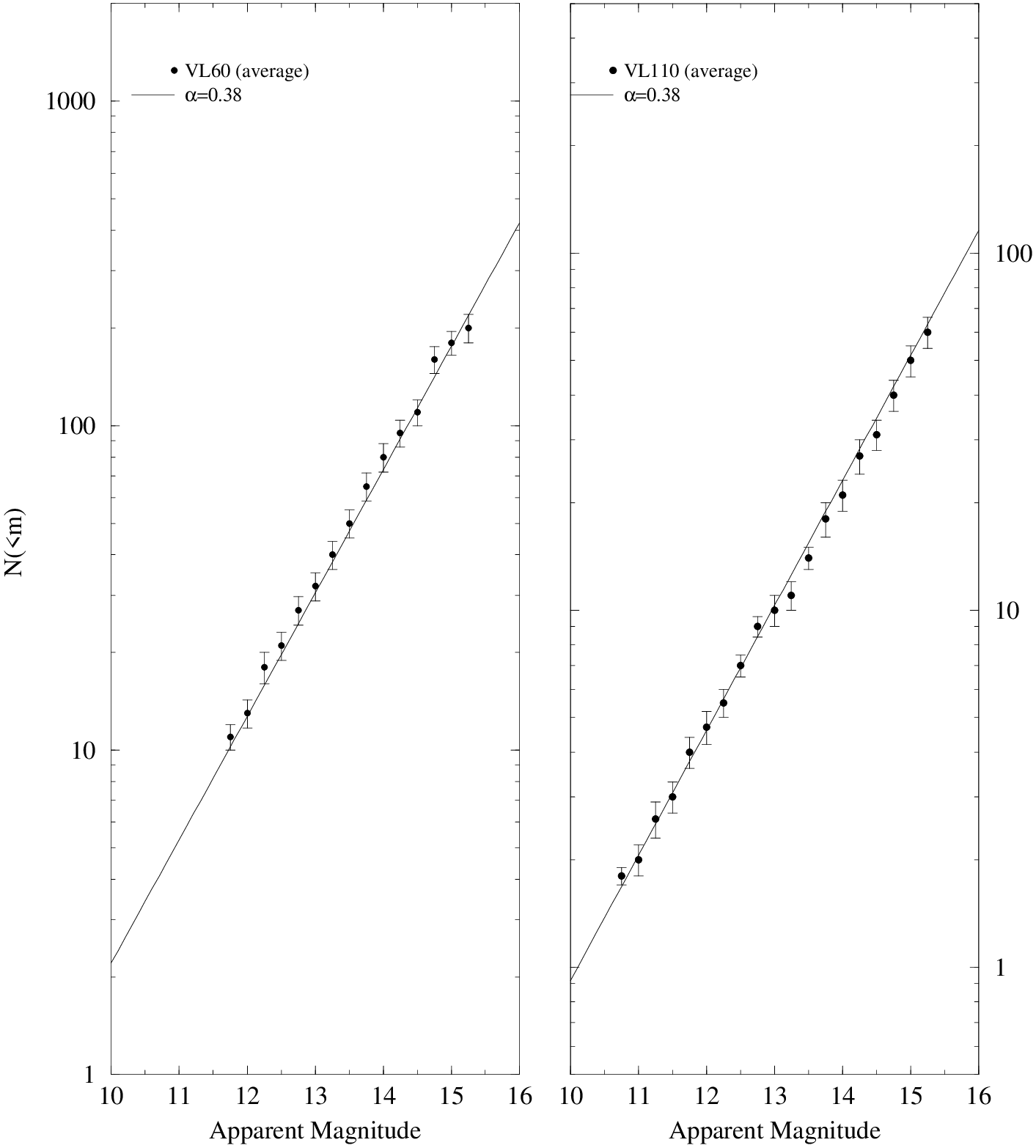}} 
\caption{\label{fig89}  
{\it Left panel}  The average $N(<m)$ in the VL sample VL60. 
 The squares crosses
 refer to  the average $N(<m)$ computed assigning to all the
 galaxies the same absolute magnitude $M_{0}=M^*$. The reference
 line has a slope   $\alpha=0.4$.  {\it Right panel} The average
 $N(<m)$ in the VL sample VL110. The squares crosses refer to  the
 average $N(<m)$ computed assigning to all the galaxies the same
 absolute magnitude $M_{0}=M^*$. The reference line has a slope 
 $\alpha=0.4$ } 
\eef
 the results for VL60 and VL110: a very well
 defined exponent $\alpha=D/5\approx 0.4$ is found in both cases.
 This is in fully agreement with the average space density (the
 conditional average density $\Gamma(r)$) that shows $D \approx 2$
 in these VL samples.  

We have also performed various tests on
 artificial distributions with a  priori assigned properties. Using
 the random $\beta-$model algorithm  and
  we have performed  the analysis by assigning  to
 each point of the system the same absolute magnitude. 
The results
 are in complete agreement with the previous findings: if we do not
 perform any average  the exponent of the
 GNC is strongly affected
 by the presence of fluctuations due to finite size effects and we
 obtain $\alpha \approx 0.6$, while if we compute  the GNC by
 making an average over all the points of the structure we find
 again that the relation $\alpha = D/5$ holds in a very well
 approximation (see \cite{slgmp96} for more details).
  On the contrary in the
 homogeneous case one does need to
 perform any average to recover the correct scaling properties of
 $N(<m)$, because the system reaches very soon 
($r \gtapprox \ell$) the correct scaling properties.

\subsubsection{Behavior of   galaxy counts in magnitude limited samples}
\label{countsml}

  We are now able to
 clarify the problem of ML catalogs. Suppose to have a certain
 survey characterized by a solid angle $\Omega$  and we ask the
 following question: up to which apparent magnitude limit 
 $m_{lim}$ we have to push  our observations to obtain that the
 majority of the galaxies lie in the statistically significant 
 region ($r \gtapprox  \lambda$) defined by Eq.\ref{v3}.  Beyond
 this value of $m_{lim}$ we should
 recover the genuine properties
 of the sample because, as we have
 enough statistics, the finite
 size effects self-average. From the previous condition for
 each solid angle $\Omega$ we can find an apparent magnitude limit
 $m_{lim}$ 
 so that finally
 we are able to obtain $m_{lim}=m_{lim}(\Omega)$ 
in the following
 way.

In order to clarify the situation, 
we can now compute the expected value of the counts if 
we use the approximation for the behavior of the mass-length relation 
given by Eqs.\ref{equ22}-\ref{equ24}. Suppose, for seek of clarity,
also that 
$\phi(M)=\delta(M-M_o)$, with $M_o=-19$. 
We define
\be\label{eq41}
\lambda= 10^{0.2 (m_{\lambda} - M -25)}
\ee
where $\lambda$ is given by Eq.\ref{v3}.
Then the differential counts are given by
\be
\label{eq42}
\left(\frac{dN}{dm }\right)_i  =  \frac{\log_e 10} {5} 3 B_1 \cdot 
   10^{ \frac{3}{5} m} \cdot 10^{ - \frac{3}{5}  (M_o+25)}
\; \; \; \mbox{if} 
\; \;  m \ltapprox m_{\lambda}
\ee
and 
\be\label{eq43}
\left(\frac{dN}{dm }\right)_i =  \frac{\log_e 10}{5} BD 
  \cdot 10^{\frac{D}{5} m} \cdot 10^{ - \frac{D}{5}  (M_o+25)}
\; \; \; \mbox{if} 
\; \;  m \gtapprox m_{\lambda}
\ee
If $M_o=-19$ and $\lambda \sim \frac{30}{\Omega^{1/D}} (\hmp) $ we have 
\be
\label{eq44}
m_{lim} = m(\Omega) \approx 14 - \frac{5}{D}\log \Omega
\ee

In order to give an estimate of such an effect 
if $\Phi(M)$ has a Schechter like shape, we can 
impose the condition that, in a ML sample, the peak of 
the selection function, which occurs at distance $r_{peak}$,
satisfies the condition
\be
\label{cayy1}
r_{peak} > \lambda
\ee
where $\lambda$ in the minimal statistical 
length defined by Eq.\ref{v3}.
The peak of the survey 
selection function occurs for $M^* \approx -19$ 
(Sec.\ref{radial}) and then we have 
$r_{peak} \approx 10^{\frac{m_{lim}-6}{5}}$.
From the previous relation and from Eq.\ref{cayy1} and Eq.\ref{v3}
we easily recover Eq.\ref{eq44}.

 The magnitude $m(\Omega)$
separates the $0.6$ behavior, strongly dominated by
intrinsic and shot noise fluctuations, from the asymptotic $0.4$
behavior. Of course this is a crude approximation
in view of the fact that $\lambda$ has not a well defined value,
but it depends on the direction of observation and not only on the solid
angle of the survey. However the previous equations gives a 
reasonable description of real data.

We show in Fig.\ref{fig90} the condition given by Eq.\ref{eq44}.
\bef 
%\vspace{}
\epsfxsize 10cm
\centerline{\epsfbox{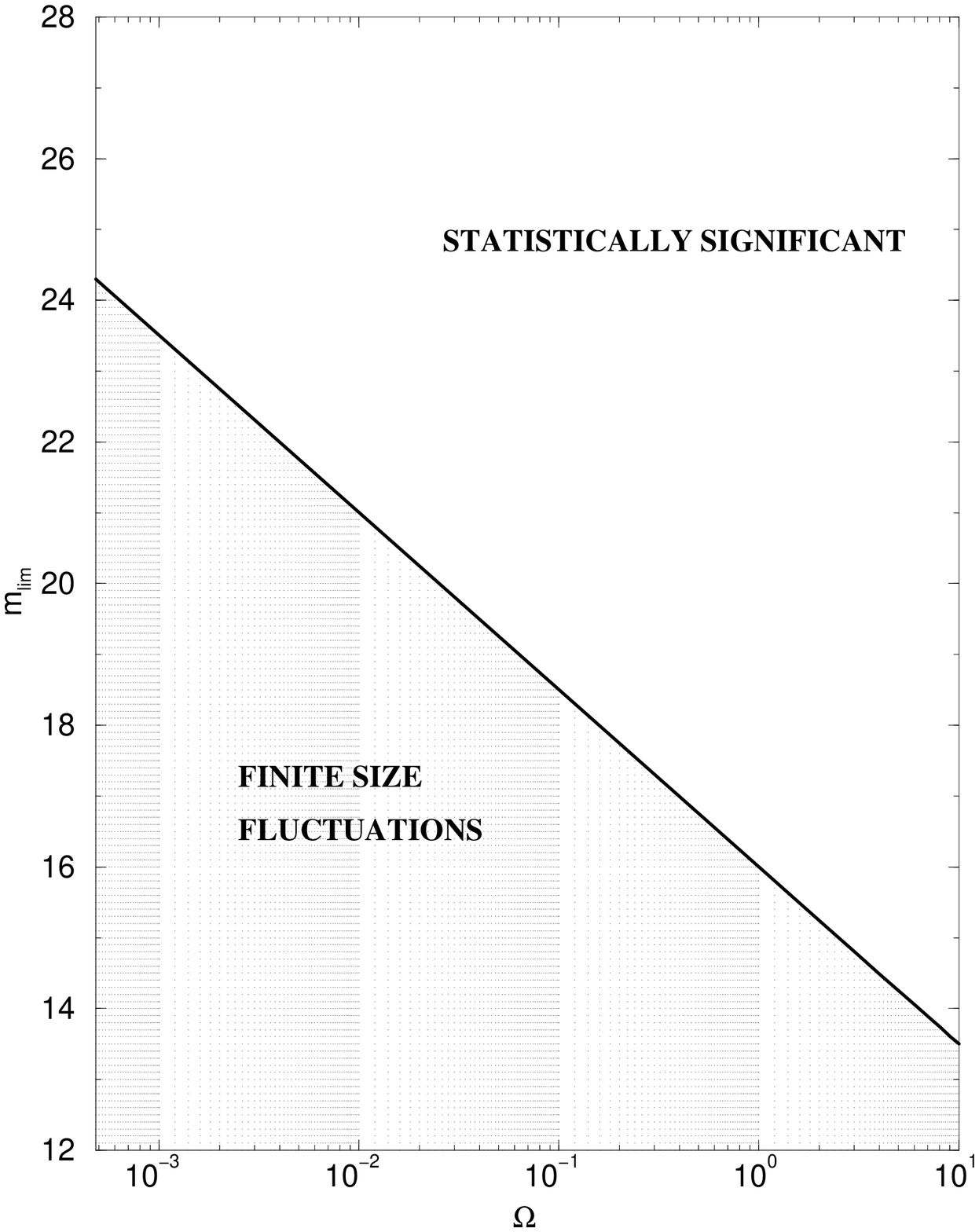}}
 \caption{\label{fig90} If a survey
 defined by  the apparent magnitude limit $m_{lim}$ and the solid
 angle $\Omega$ lie in the  statistically significant  region it is
 possible to obtain the self-averaging properties of the
 distribution also with the integral from the vertex. 
Otherwise one needs a redshift survey which contains the three
 dimensional information, and then one can perform average.
 Only in this way it is possible to smooth out the finite size
 effects. } 
\eef 
From the previous figure it follows that for $m > 19$
 the statistically significant region  is reached for almost {\em
 any} reasonable value of the survey solid angle. This implies that
 in deep surveys, if we have enough statistics, we   readily 
 find the right behavior ($\alpha =D/5$) while it does not happens
 in a self-averaging way for the nearby samples. Hence the exponent
 $\alpha \approx 0.4$ found in the deep surveys ($m>19$) is a {\em
 genuine feature of  galaxy distribution}, and corresponds to real
 correlation properties.  
In the nearby surveys $m < 17$ we do not
 find the scaling region in the ML sample for almost {\em any}
reasonable
 value of the solid angle. Correspondingly the value of the
 exponent is subject to the finite size effects, and to recover the
 real statistical properties of the distribution one has to perform
 an average.  

From  the previous discussion it appears now clear
 why a change of slope is found at $m \sim 19$: this is just a
 reflection of the lower cut-off of the fractal structure and in
 the surveys with $m_{lim} > 19$ the self-averaging properties of
 the distribution cancel out the finite size effects. This result
 depend very weakly on the fractal dimension $D$ and on the
 parameters of the luminosity function $\delta$ and $M^*$ used. Our
 conclusion is therefore that the exponent $\alpha \approx 0.4$ for
 $m > 19$ is a genuine feature of the galaxy distribution and it is
 related to a fractal dimension $D \approx 2$, which is found for $m
 < 19$ in redshift surveys only by 
{\em performing averages}.  We note
 that this result is based on the 
assumption that the Schecther
 luminosity function  holds also at high redshift, or, at
 least   to $m \sim 20$. This result
 is confirmed by the analysis
 of Vettolani \etal \cite{vet94} who found that the luminosity
 function up to $z \sim 0.2$ is in
 excellent agreement with that
 found in local surveys \cite{dac94}.

 Finally a comment on the {\it amplitude} of counts.
In Fig.\ref{fig85} the solid line represents the behavior of
Eq.\ref{q3}. The prefactor $B$ has been determined as is 
Sec.\ref{radial}, while the fractal dimension is $D=2.2$. The
parameter of the luminosity function are $\delta = -1.1$ and $M^*=-19.5$
as usual. The agreement at faint
 magnitudes ($ m \gtapprox 19$) is quite good. At bright
magnitudes one usually 
underestimates the number of galaxies because of
finite size effects, even in some cases the number counts
can be larger than the predicted value. This is 
related to the asymmetry of the fluctuations in a 
fractal structure, as explained     in  Sec.\ref{radial}.
For example we report in Fig.\ref{fig91} various 
determinations of the GNC in different regions of the sky
(from \cite{b96}).
\bef 
%\vspace{}
\epsfxsize 10cm
\centerline{\epsfbox{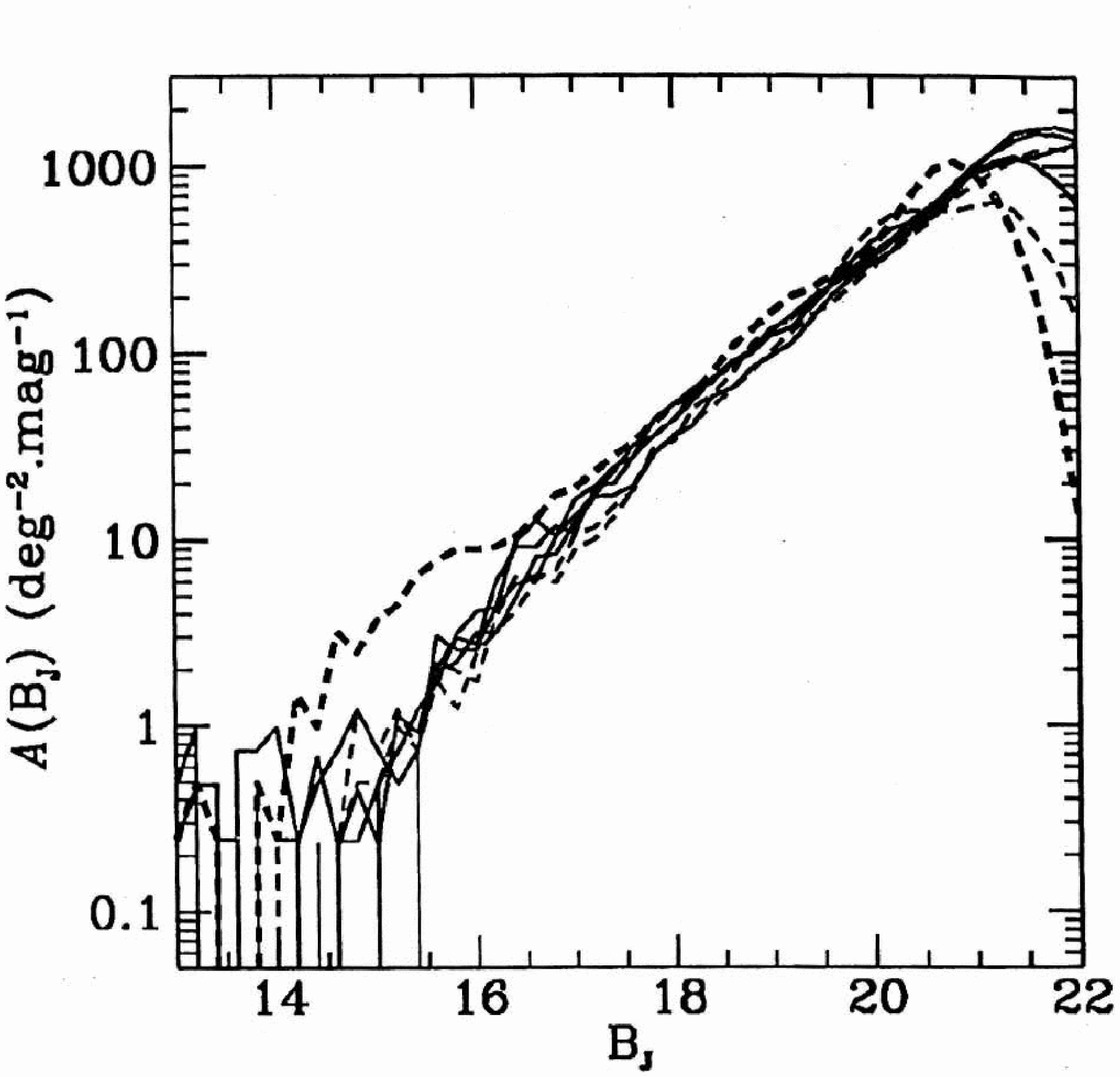}}
 \caption{\label{fig91} Determinations of the GNC in
various sky regions (from Bertin \& Dennefeld, 1996).
 The bright end 
wildly fluctuates around the predicted value, while
at fainter magnitudes the behavior is quite smooth and
is the same for the different fields.} 
\eef
The fact that at faint magnitudes the behavior is quite regular, is due 
to the smoothing of spatial fluctuations for the luminosity effect. 
Namely at a 
given apparent magnitude there are contribution from galaxies located at very
different distance, as the luminosity function of galaxies is spread 
over several decades of luminosities 
(see Sec.\ref{angred}). On the other hand at the bright end there 
are contributions only from   nearby galaxies,
 and is such a way the space 
finite size fluctuations are not smoothed out.

In the data shown in Fig.\ref{fig85}  K-corrections have been not 
 applied. 
Such corrections
 must be present because of the Hubble
 distance-redshift relation. Namely
the spectrum of a certain galaxy at 
redshift $z$ is shifted towards
 red  of a certain 
amount, according to 
the Hubble law. There are several galaxies (E/S0) with steep
spectra and hence for
 these one detects a lower value of the intensity of the "true"
one because of the 
shift of the maximum of the spectrum. However there are 
several other galaxies
 (Sdm, Scd) with flat spectra. In some case the 
 shift due to the Hubble law 
may produce an higher intensity 
while in other a decrease of the apparent flux. 
 The K-corrections 
are model dependent corrections: 
the sign can be in both directions,
 i.e. towards an increasing or  a decreasing
of the absolute magnitude \cite{kkcor}. Moreover the 
effect of such corrections is in general  
not so important for the distortion of
 the number counts relation
\cite{yo93}.

\subsubsection{Galaxy Counts in the various spectral bands}
\label{countsspec}

 Finally we report in Tab.\ref{tabande}
\begin{table} \begin{center}
 \begin{tabular}{|c|c|c|c|c|c|c|} 
\hline 
&       &        &      &            &     &            
\\  Photometric Band         & Driver  & Metcalfe  & Tyson & Lilly &     Djorgovski  & Jones\\ 
&       &       &       &            &        &         \\ 
\hline 
%&       &       &      &            &            &     \\ 
U      &  -      &      -&      -&      -&      -&0.5\\ 
%&       &       &      &            &                 &\\ 
B     & 0.43 & 0.49  & 0.45 & 0.38 & -&-\\  
%&       &       &      &            &           &      \\ 
V     & 0.39 & -  & - & - & -&-\\  
%&       &       &      &         &   &                 \\
R     & 0.37 & 0.37  & 0.39 & - &-& -\\ 
% &       &       &      &            & &                 \\
I     & 0.34 & -  & 0.34 & 0.32 & -&-\\
  %&       &       &      &            &    &             \\ 
K     & - & -  & - & - & 0.32 &-\\ 
\hline
\end{tabular} 
\caption{\label{tabande} The exponents of galaxy counts in different spectral bands 
derived by different authors.} 
\end{center} \end{table}
 the exponents of the galaxy counts in different frequency bands,
 at faint magnitudes. The {\it faint end exponent}
 is lower than
 $0.6$ in all the case, and it is in the range $0.3 \div 0.5$, so
 that $D$ is in the range $1.5 \div 2.5$. These differences  can 
be probably related to the oscillations that may affect
the behavior of the radial density as computed from a single point.
In particular, as we have already discussed in Sec.\ref{radialfinite},
in the determination of  the mass-length relation from one point only 
(the same argument can be extended to the case of galaxy counts),
there are intrinsic fluctuations that are not smoothed
out by any averaging procedure. It is important to stress again
that these fluctuations are proportional to the number of points,
rather than to its root mean square, as in a Poisson distribution.
The effect of the intrinsic fluctuations may strongly depend
on the geometry of the survey, and this is why we can have different
determinations of the number counts exponent, as well as 
fluctuations in the amplitude \cite{fluc}. Only 
a three dimensional analysis, in which it is possible
to compute average quantities, allows one to clarify the situation.

\subsection{Counts of X-ray sources, Radio galaxies, 
Quasars and $\gamma$-ray  burst sources}   
\label{countsobj}

 In observational astrophysics there are a lot of data which are
 only angular ones, as the measurements of distances is general a
 very complex task. We briefly discuss here the distribution of
 radiogalaxies, Quasars and the $\gamma$-ray burst distribution.
 Our conclusion is that all these data are compatible with 
a fractal structure with $D \approx 1.6 \div 2.0$ similar to that
 of galaxies.  

\subsubsection{Radio galaxies}
\label{countsradio}

The majority of  catalog radio sources are
 extragalactic and   some
 of the strongest are at cosmological distances.
 One of the most important information on radio galaxies
 distribution has been obtained from the sources counts as a
 function of the apparent flux \cite{con84}. 
Extensive surveys of sources have been made at 
various frequencies in the range  $  0.4
 \div 5 \; GHz$. In Fig.\ref{fig92} 
\bef 
%\vspace{} 
\epsfxsize 12cm
\centerline{\epsfbox{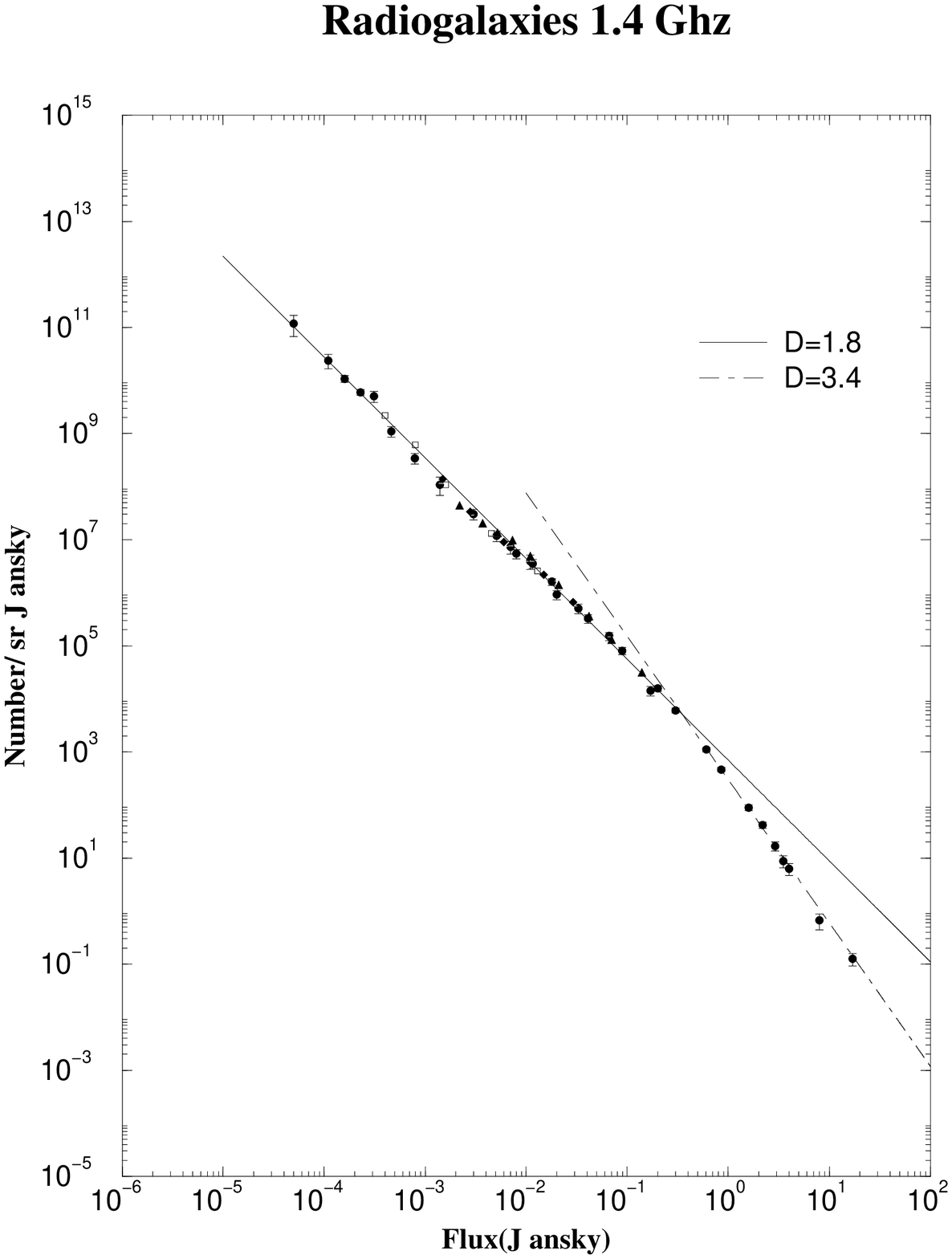}}
 \caption{\label{fig92} Normalized differential sources counts at
 $\nu = 1.4 Ghz$. {\it Abscissa} log flux density (Jy). {\it
 Ordinate}  log differential number of sources $n(S)$. The solid
 line represents the behavior of a fractal  structure with
 $D=1.8$. The amplitude has been computed from the knowledge
of the luminosity function (see text). 
The agreement is excellent, except in the bright fluxes
 region due to the presence 
of finite size fluctuations. (From Condon, 1984). } 
\eef 
we show a compendium of sources counts at $\nu=1.4
 Ghz$ \cite{con84}. The differential counts are plotted against the
 apparent flux. In the bright flux region there is
 a deviation from a power law function, while for four decades the
 agreement with a fractal distribution with  $  D \approx 1.8$ is
 excellent. Such a behavior has been usually explained in
 literature as an effect of sources/space-time
 evolution. Here we propose that
 the radio galaxies are fractally distributed, as galaxies, with
 almost the same fractal dimension. 
The deviation at bright fluxes
 in this picture is explained as a spurious effect due to  the
 small scale fluctuations, as in the
 case of galaxy counts. In the
 other frequency bands the situation is nearly the same
 \cite{con84}. The simple picture of a fractal distribution of
 radio sources is therefore fully compatible with the experimental
 situation. 

%  {\bf AMPLITUDE}  

It is simple to show that the differential number of 
radio galaxies for unit flux (in Jansky) and 
unit steradian, in an Euclidean space,  is given by:
\be
\label{eqradio1}
n(S)=\frac{1}{(3\cdot 10^9)^2} 
\left(\frac{1}{4\pi}\right)^{\frac{D+2}{2}}
\frac{DB_R}{2} S^{-\frac{D+2}{2}} \int \phi(L) dL
\ee
where we have taken, as usual, 
$n(r) = (B_RD/4\pi) r^{D-3}$ and $S=L/(4 \pi r^2)$
($L$ is the intrinsic luminosity).
From the knowledge of the parameters of the luminosity function (see
 \cite{con84}) we have that a good approximation of the data
reported in Fig.\ref{fig92} is given by $D=1.8$ and $B_R=0.1 Mpc^{-D}$.
Such a low value for $B_R$ implies that the 
density of radio galaxies is about 100 lower than that of 
optical galaxies. 
When it will be possible to have  a complete redshift 
sample of Radio galaxies, i.e. a well defined volume limited
sample, one will be able to measure directly $B_R$ and $D$ from
the knowledge of the conditional density and hence to compare those
values with the ones we have measured from the  number flux 
relation.

\subsubsection{Quasars}
 \label{countsquas}

In the case of Quasars the
 situation is almost the same. In fact we find that at bright
 magnitudes ($ 14.75 < B < 18.75$) the exponent of the counts is 
$\alpha \approx 0.88$, while at faint magnitude it is $\alpha
 \approx 0.3$ \cite{ht90}. 
A break in the integral Quasars counts 
has been found by \cite{jon96}
in the very deep ROSAT survey. This is the 
deepest optically identified 
X-ray survey yet performed. The break 
in the Quasar $\log N-\log S$ relation
occurs at 
$\sim 2  \cdot 10^{-14} \; erg \; cm^{-2} \; s^{-1}$ 
($0.3 \div 3.5$ keV), and the slope 
changes from $D=3$ to $D=2.07$.
The break in these counts occurs at 
approximately the same flux as the break in the counts
of X-ray sources (see Sec.\ref{countsxray}).

Even in this case we can interpret such
 a behavior as due to the fractal distribution in space with  
$  D \approx 1.5 \div 2$.   
This is a crude indication of the fractal behavior of
Quasar space distribution: the data are not statistically robust as in the
case of radio galaxies, and further studies are needed on the 
three dimensional distribution.

\subsubsection{X-ray sources} 
\label{countsxray}

The counts of X-ray sources have been measured recently by the Rosat 
Satellite \cite{rosat}. The data are not spread over a
large interval in apparent flux (about four orders of magnitude), but
there 
is a clear evidence of a change of slope at $\sim 2.5 \cdot 10^{-14}
erg\; cm^{-2} \; s^{-1}$, well beyond the flux limit of the survey.
Even in this case the effect of finite size fluctuations
must be present and we can interpret this change of slope 
as an evidence of fractal distribution of the X-ray sources with 
dimension $D \approx 1.8$, as in the case of optical galaxies. 

\subsubsection{$\gamma$ ray bursts}
\label{countsgray}  

Finally we comment the
 distribution of $\gamma$-ray burst (GRB). This is a long-standing
 problem in Astrophysics and after 20 years  of intense studies and
 observations it is still mysterious: the origin of GRBs, in our
 galaxy or from the Cosmos, is a matter 
of debate \cite{fis95,bri96,har95}. We argue here that from the present
 angular and intensity data it is
 possible to show that the space
 distribution of $\gamma$-ray
 bursts sources is fully {\it compatible}
 with a fractal structure with $D \approx 1.7$. This result
 clarifies the statistical 
analysis of the available data  and
 points out a fundamental aspect of the $\gamma$-ray bursts sources
 distribution.   

From the angular catalogs recent available
 \cite{mee95}  it is possible to study three
 statistical quantities. The first one
 is the number versus
 apparent intensity distribution which  shows a deviation from the
 Euclidean behavior at low fluxes \cite{mee95,mee92}. The
 second is the $V/V_{max}$ test \cite{sch88} which  again provides an
 evidence that  the spatial distribution of sources is not
 homogeneous \cite{mee92,mee95}. 
 Finally the angular
 distribution is isotropic within the statistical limits
 \cite{bri93,mee95}  and there is not any evidence for an
 angular correlation or a clustering of bursts towards the galactic
 center or along the galactic plane of bursts \cite{mee95,now94,qua93,mee95b}.
 These results together
 indicate that the bursts sources are distributed isotropic but not
 homogeneously \cite{mee95}. We argue here that these three
 evidences are fully compatible with a fractal distribution of
 sources with $D \approx 1.7$.  

In the standard interpretation of cosmological origin of
$\gamma$ ray bursts, the sources must 
be at very high redshift. In such
a way, due to the modification of the Euclidean Geometry in the Friedmann
models, one should obtain a deviation 
from the purely $-3/2$ behavior
for the counts. However the problem is that the sources must be of
very high intrinsic energy, if they 
are at so large distances.
This seems to be the most important 
problem for a cosmological
origin of such sources. We show that the sources can be at 
low redshift  (say $z \ltapprox 1$), giving the new
interpretation of the counts relation. 

The first observational evidence is the
 number of burst as a function of the apparent flux $N(>S)$. At
 bright apparent flux, which are associated to small distance of the
 sources, one sees an exponent $-3/2$, which  seems to  be in
 agreement with the homogeneous  case. 
This is just a spurious
 effect which arises form the fact this quantity is computed without
 performing any average.  At faint apparent fluxed, one is
 integrating the density in the correct 
scaling regime, and in this
 region the genuine statistical properties of the system can be
 detected. From the $N(>S)$ relation in the limit of low $S$ we can
 estimate a fractal dimension $D \approx 1.7 \pm 0.1$.   An
 equivalent test on the homogeneity versus fractal properties is
 given by the $V/V_{max}$ distribution \cite{sch88}, where 
\be
 \frac{V}{V_{max}} = \left( \frac{C}{C_{lim}}
 \right)^{-\frac{3}{2}} \; . 
\ee 
In this ratio $V$ is the volume
 contained in a sphere extending to the location  bursts and
 $V_{max}$ is the volume of the sphere extending to the maximum 
 distance at which the same burst would be detectable by the
 instrument, whose limiting flux is $C_{lim}$. 
It is simple to show
 that if the spatial distribution of
 sources is described by a fractal
 structure, then we have 
\be 
<\frac{V}{V_{max}}> =
 \frac{D}{D+3} 
\ee 
Even in this case the data show \cite{mee95}
 that  $< \frac{V}{V_{max}}> = 0.33 \pm 0.01$ which  in terms of
 fractal dimension means $D \approx 1.5 \pm 0.1$. This value is
 somewhat smaller than the 
fractal dimension estimated with the
 $N(>S)$. This difference is probably due to the fact that this
 test is integrated while the $N(>S)$ is a differential one.  

 Let
 us came to the third evidence, i.e. a substantially  isotropic
 angular distribution and a lack of any correlation at small
 angles. 
 The projection of
 a fractal distribution on the unit sphere conserves the
 correlations properties (see Sec.\ref{angcorr} ) only in 
the small angles approximation,
 while at large angular scale the long range correlation are
 destroyed by projection effects. 
The angular correlation function
 $\omega(\theta)$ has a power law behavior in the small angles
 approximation ($\theta
 < 10^{\circ}$).  
In the available sample \cite{mee95} the number of points is
 $N=1122$ distributed over the whole sky. This means that at
 angular distance smaller than
 $\sim 20^{\circ}$ one does not find any other object in average,
 and therefore it is not possible to study the angular correlation 
function at such low angular separation.

  These results indicate
 together that the $\gamma$-ray burst sources are fractally
 distributed in space with $D \approx 1.7$. This is value is very
 similar to that the fractal dimension of the galaxy distribution
 is space  which  is $D \approx 2$ up to 
some hundreds Megaparsec.
 This "coincidence" can be seen as an indication that the
 $\gamma$-ray bursts sources are associate to a population of
 objects distributed as the visible galaxies. These
objects should have low redshift, and distance up to some
hundreds Megaparsec, depending on the intrinsic luminosity of 
the sources. We  stress that recently 
a correlation between the $\gamma$ ray bursts 
at the 3B catalog and the 
Abell clusters has been found \cite{kp96}.
 This evidence points
towards a cosmological origin of bursts, and can be in agreement with 
our interpretation of the counts distribution. In fact, the
sample of bursts considered are located within $\sim 600 \hmp$
\cite{kp96} where the modification of the Euclidean Geometry are 
negligible. 

We stress that a
 larger sample of bursts  may allow one a better determination of
 the fractal dimension and, if the number of objects for steradian
 will became larger, it may be possible to study also the angular
 correlations in the small angle approximation.

\subsubsection{Compendium of counts}
\label{countscomp}

In Tab.\ref{tabbande} we summarize the behavior of the number counts
for various astrophysical objects. The small scale exponent corresponds to
the bright end of 
the number counts relation, while the large scale exponent to the 
faint end. The "small scale exponent" {\it is not} a real 
exponent and wildly fluctuates for the different objects. On the 
contrary  the large scale
exponent {\it is} a real
 exponent and its value is in the range $1.8 \ltapprox D 
\ltapprox 2.2$ for almost all the cases. The counts of all these 
different objects is therefore compatible with a fractal distribution in space.

This implies a completely new 
interpretation of the counts. At small scale 
the exponent $D \approx 3$ (i.e $\alpha \approx 0.6$ 
 in the case of magnitude counts)
is due to finite size effects and not to
 a real homogeneity: this has be shown
to be the case with very specific tests.
 On the other hand, at larger scale
 the value $D \approx 2$ 
(i.e. $\alpha \approx 0.4$) corresponds to 
the correct correlation
properties of the samples, that we can find 
by the complete correlation 
analysis in the three dimensional space. 
This implies that galaxy evolution,
 modifications of the 
Euclidean Geometry and the K-corrections 
{\it are not} very relevant 
in the range of 
the present data. In addition the fact 
that the exponent $\alpha \approx 0.4$ 
holds up to magnitude $27 \div 28$ for galaxies seems to indicate that 
the fractal may continue up to distances $\sim 4000 \hmp$. Quite
 a remarkable fact
if one considers 
that the Hubble radius of the Universe is supposed to be  $  4000 \hmp$.
Moreover such a 
behavior can be found for galaxies in the different photometric bands, 
as well as for other astrophysical objects. 

\begin{table} \begin{center} 
\begin{tabular}{|c|c|c|c|c|}
 \hline
 &       &            &&   \\ 
Objects     & Small  scale  behavior &  & Large scale behavior &\\
 &       &            &&   \\
 \hline 
%\hline 
%&       &            &&   \\
 U-band$^{1}$ & $ ? \ltapprox U \ltapprox 18$  & {\bf $D \approx
 ?$}        &  $18 \ltapprox U \ltapprox 24$  &{\bf $D \approx 2.5$}           \\ 
%&       &            &&   \\
B-band$^{2}$ & $12 \ltapprox B \ltapprox 18$ & {\bf $D \approx 3$}      & $18 \ltapprox B \ltapprox 28$ 
& {\bf $D \approx 2$}              \\ 
%&       &            &&   \\ 
V-band$^{3}$ &    $ %? \ltapprox V
 \ltapprox 20$&  {\bf $D \approx ?$}    & $22 \ltapprox V \ltapprox 25$
 & {\bf $D \approx 1.95$}               \\ 
%&       &            &&   \\ 
R-band$^{4}$&     $15 \ltapprox R \ltapprox 18$  &{\bf $D \approx
 2.8$}   &$20 \ltapprox R \ltapprox 26$  &{\bf $D \approx 1.85$}
                \\ 
%&       &            &&   \\ 
I-band$^{5}$ &   $? \ltapprox I \ltapprox 19$ & {\bf $D \approx
 ?$}      &  $19 \ltapprox I \ltapprox 25$  & {\bf $D \approx 1.7$}
              \\ 
%&       &            &&   \\ 
K-band$^{6}$ &  $12 \ltapprox K \ltapprox 17$ &  {\bf $D \approx
 3.35$}     &  $20 \ltapprox K \ltapprox 24$  & {\bf $D \approx
 1.6$}             \\
 %&       &            &&   \\ 
%\hline
&       &            &&   \\ 
Quasars$^{7}$ & $14.75 \ltapprox B \ltapprox 18.75$ &  {\bf $D
 \approx 4.4$}      & $19 \ltapprox B \ltapprox  23$ & {\bf $D
 \approx 1.5$}              \\ 
%&       &            &&   \\
 Radio$^{8}$ $\nu=1.4 Ghz$ &  $1 \ltapprox S \ltapprox 10$ &{\bf $D
 \approx 3.44 $}      & $10^{-5} \ltapprox S \ltapprox 1$ & {\bf $D
 \approx 1.8$}         \\ 
%&       &            &&   \\ 
Radio$^{8}$ $\nu=0.61 Ghz$ &$1 \ltapprox S \ltapprox 10$ &{\bf $D
 \approx 3.8$}        &  $10^{-3} \ltapprox S \ltapprox 1$ & {\bf
 $D \approx 1.6$}              \\
 %&       &            &&   \\
 Radio$^{8}$ $\nu=0.408 Ghz$ &$1
 \ltapprox S \ltapprox 10$ &{\bf $D \approx 3.7$}        & 
 $10^{-3} \ltapprox S \ltapprox 1$ & {\bf $D \approx 1.5$}   
           \\
           %&       &            &&   \\ 
Radio$^{8}$ $\nu=5.0 Ghz$ &$1 \ltapprox S \ltapprox 10$ &{\bf $D
 \approx 3.4$}        &  $10^{-4} \ltapprox S \ltapprox 1$ & {\bf
 $D \approx 1.8$}              \\
 %&       &            &&   \\
X-ray sources$^{9}$ & $5 \cdot 10^{-13} \ltapprox S \ltapprox
 10^{-12}$ & {\bf $D \approx 3.4$} &  $10^{-16} \ltapprox S
 \ltapprox 5\cdot 10^{-13}$ & {\bf $D \approx 1.8$}              \\
   %&       &            &&   \\ 
%\hline
 %&       &            &&   \\ 
$\gamma-ray$ bursts $^{10}$ & $10 \ltapprox S \ltapprox 100$  &
 {\bf $D \approx 3$}       &  $10^{-1} \ltapprox S \ltapprox 10$  &
 {\bf $D \approx 1.7$}              \\ 
 &       &            &&   \\ 
\hline
\end{tabular}
 \caption{ \label{tabbande} The exponents of counts for different
 kinds of astrophysical objects (see text). (Ref.1-6: see Table 8;
   Ref.7:
 Hartwick  \& Schade,  1990; Ref.8: Condon,1984; Ref.9:  
 Hasinger \etal, 1993; Ref.10: Meegan \etal 1995)  }
 \end{center} \end{table}

\subsection{Discussion of the angular 
correlations and their reinterpretation}  
\label{angcorr}

We have now enough elements to give the correct 
reinterpretation
also of the angular catalogs. These catalogs are
quantitatively inferior to the 3-d ones because they 
correspond to the angular projection and 
do not contain the third coordinate. However,
the fact that they contains more galaxies with 
respect to the 3-d
catalog has led some authors to assign an excessive importance 
to these catalogs and, again, they are supposed to show 
the much sought homogeneity. 
Actually the interpretation of
the angular catalogs is quite
 delicate and ambiguous for a variety 
of reasons which are usually 
neglected.
 
\subsubsection{Angular properties of three dimensional samples}
\label{angred}
In Fig.\ref{fig93}
\bef 
%\vspace{}
\epsfxsize 12cm 
\centerline{\epsfbox{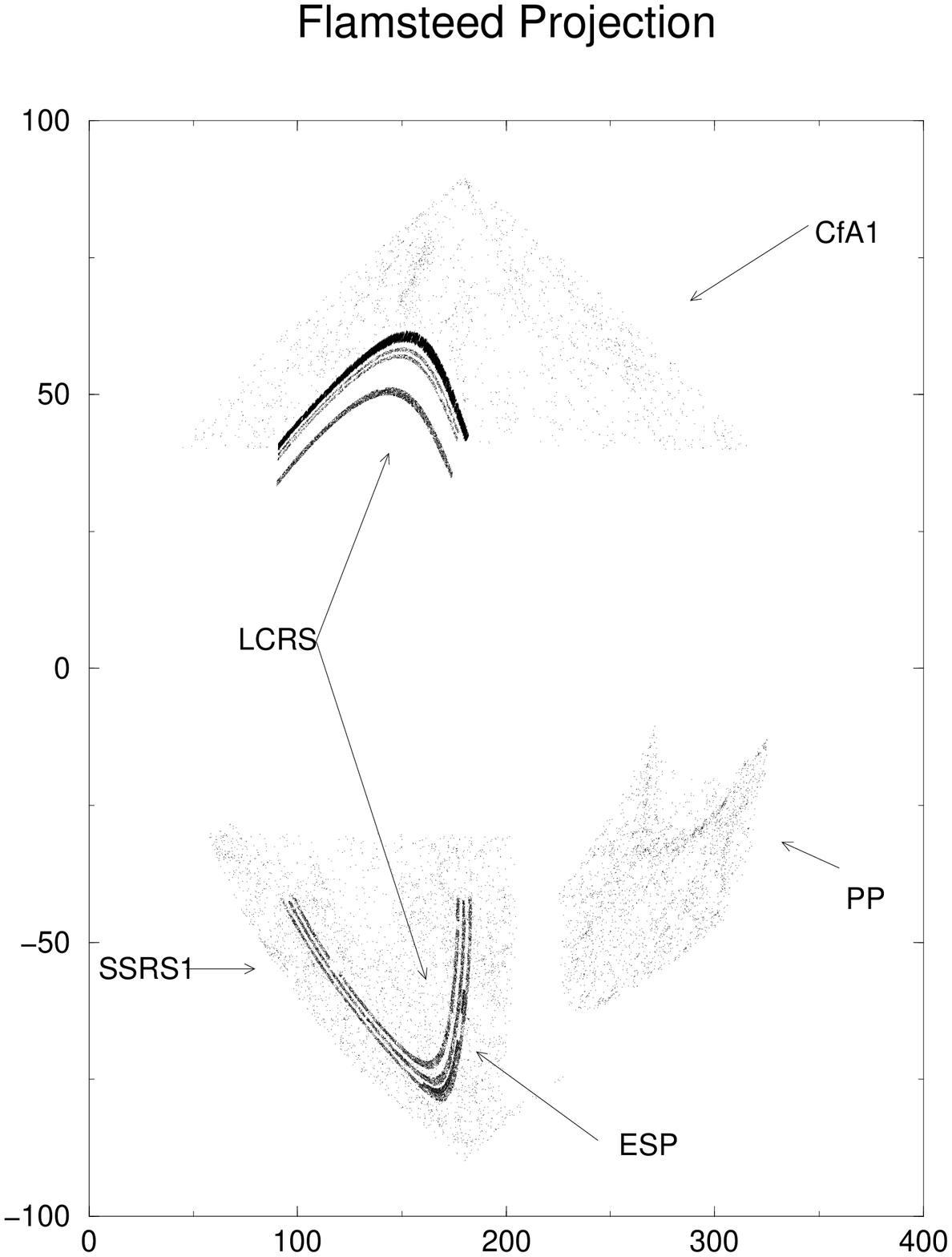}}  
\caption{\label{fig93} Angular distribution
(Flamsteed projection) of the various redshift surveys presented in this review.}
\eef
we show the angular projection of the various redshift samples
we have discussed.  
No large voids are present in the angular distribution, 
which appears quite uniform. The problem we address 
in what follows 
concerns the origin of the angular homogeneity, which appears in contradiction
with the irregular three dimensional space properties.

The standard method used to analyze angular catalogs,
is based on the assumption that galaxies are correlated only at small
distances. In such a way  the effect of the large spatial inhomogeneities
is not considered at all. Under this  assumption, that is
not
supported by any experimental evidence, it is possible to derive the
Limber equation \cite{lim70,lim71}. In practice,
the angular analysis is performed by computing the two 
point correlation function
\be
\label{eqa1}
\omega(\theta) = \frac{\langle n(\theta_0)n(\theta_0+\theta) \rangle}
{\langle n \rangle} -1
\ee
where $\langle n \rangle$ is the average density in the survey.
This function is the analogous of the $\xi(r)$ for the 3-d analysis.
The results of such an analysis are quite similar to the 
three dimensional ones \cite{pee80,pee93}. 
In particular, it has been obtained that, 
in the limit of small angles,
\be
\label{eqa2}
\omega(\theta) \sim \theta^{-\gamma+1}
\ee
with $\gamma \approx 1.7$ (i.e. $\gamma_a=\gamma-1=0.7$).
 It is possible to show \cite{pee80} that,
in the Limber approximation Eq.\ref{eqa2}, the angular correlation
function 
 corresponds to
$\xi(r) \sim r^{-\gamma}$ for its three dimensional counterpart
(in the case $\gamma > 1$).

We now study the case of {\it a self-similar angular distribution}
so that, if such properties are present in real catalogs, we are 
able to recognize them correctly.
Of course, if the distribution is homogenous, we are able to
reproduce the same results obtained by the 
$\omega(\theta)$ analysis.
Hereafter we consider the case of small angles ($\theta \ltapprox 1$), that
is quite good the  for the catalogs investigated so far.  
The theorem  for orthogonal projections discussed in Sec.2 can be extended 
to the case of angular projections in the limit of 
small angles ($\theta < 1$). Therefore in the following in Eq.\ref{epro1}
we have $D'=D_a$ 

In this case the number of points up to an angle $\theta$ scales as 
\be
\label{eqa3}
N(\theta)= B_a \theta^{D_a} 
\ee
where $D_a$ is the fractal dimension corresponding to the 
angular projection and $B_a$ is the lower cut-off of the 
distribution. Eq.\ref{eqa3} holds from every occupied point, and 
in the case of an homogenous distribution we have $D_a=2$.
Following Coleman \& Pietronero \cite{cp92}
we define the conditional average density as
\be
\label{eqa4}
\Gamma(\theta)= \frac{1}{S(\theta)} \frac{dN (\theta)} {d\theta} = 
\frac{BD_a}{2\pi} \theta^{-\gamma_a}
\ee
where $S(\theta)$ is the differential solid angle element 
and $\gamma_a=2-D_a$
is the angular correlation exponent (angular 
codimension). The last equality 
holds in the limit $\theta < 1$. From the very definition of 
$\Gamma(\theta)$ we have that
\be
\label{eqa5}
\omega(\theta)=\frac{\Gamma(\theta)}{\langle n \rangle} -1 \; .
\ee
A first important consequence of eq.\ref{eqa5} is that 
if $\Gamma(\theta)$ has a power law behavior, while $\omega(\theta)$ is
a power law minus one. This corresponds to a break in the log-log plot
for angular scales such that $\omega(\theta) \ltapprox 1$. We show in 
Fig.\ref{fig93bis} the behaviour of such a quantity. 
\bef 
%\vspace{}
\epsfxsize 12cm 
\centerline{\epsfbox{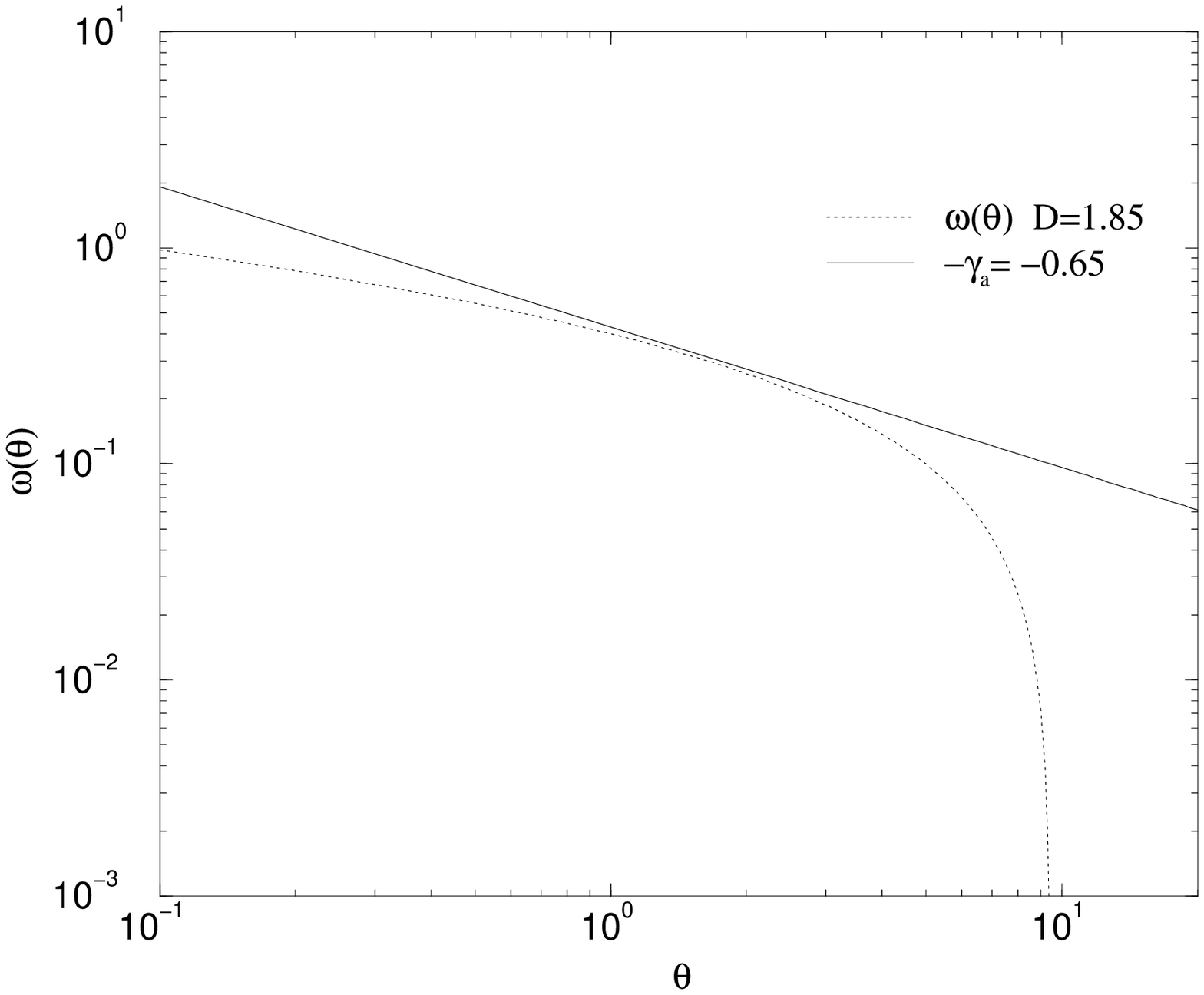}}  
\caption{\label{fig93bis}  In this figure we show the behaviour of
$\omega(\theta)$ (dotted line)
 is the case of a fractal structure with $D_a=1.85$
($\gamma_a=0.15)$.
It can be seen that the exponent obtained by fitting this function 
with a power law behavior (solid line) is higher than the real one 
($\gamma =-0.65$). Also the break in the power law behaviour is 
completely artificial.  }
\eef

The codimension found by fitting $\omega(\theta)$
with a power law function is higher than the real one. This is an important 
effect that has never been considered before. As we show below, this is 
actually
the case for real catalogs.  The second important point is that the break
of $\omega(\theta)$ in the log-log plot is clearly artificial and does not
correspond to any characteristic scale of the original distribution.
The problem is that in the case of a scale-invariant distribution the 
average density in Eq.\ref{eqa1} is not well defined, as it depends 
on the sample size \cite{cp92}.

 We have studied the angular 
conditional density
in the various magnitude limited samples.
The results are shown in 
Fig.\ref{fig94} Fig.\ref{fig95} 
Fig.\ref{fig96} and Fig.\ref{fig97}.
\bef 
%\vspace{}
\epsfxsize 8cm 
\centerline{\epsfbox{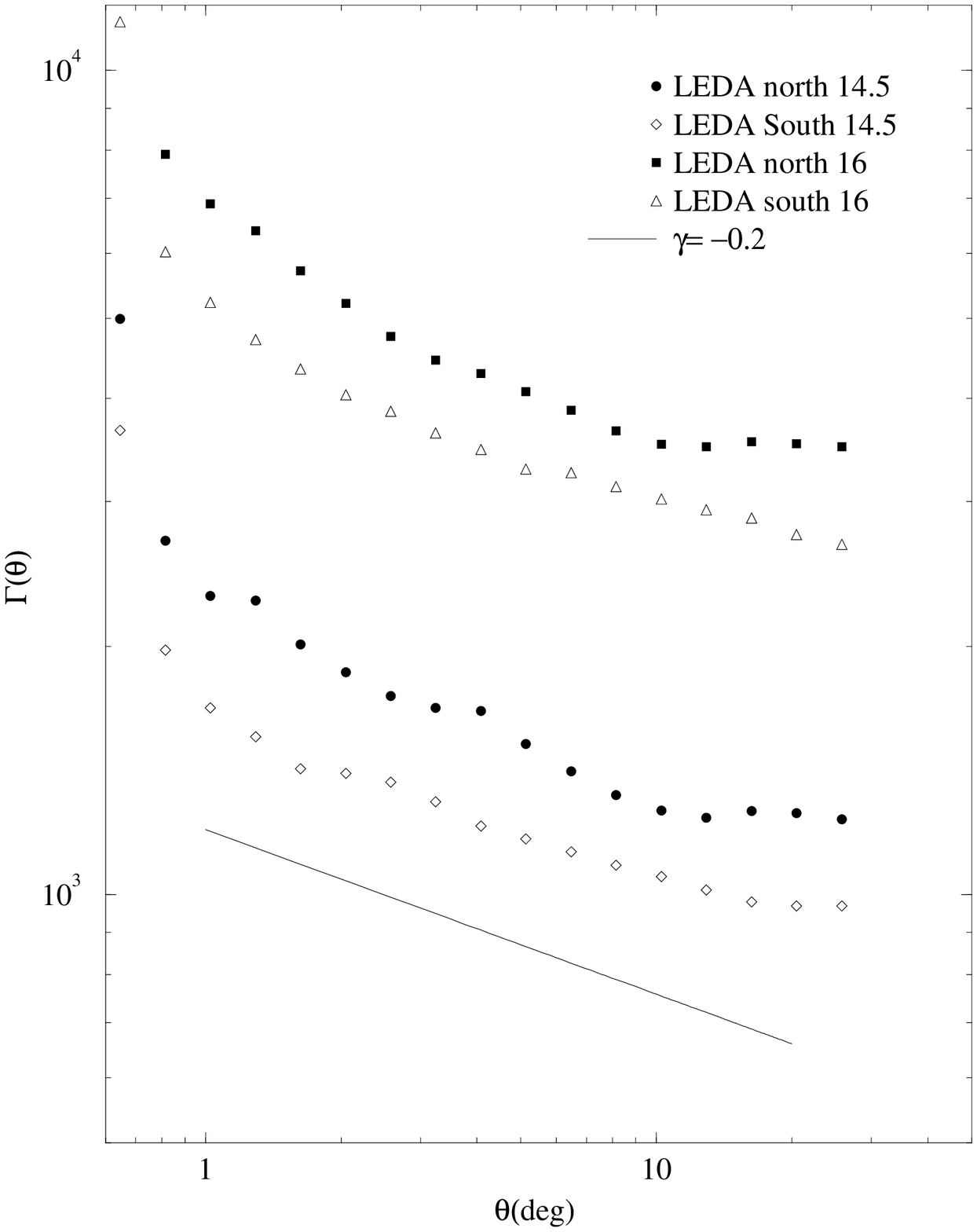}}  
\caption{\label{fig94} Angular correlation function $\Gamma(\theta)$
 for the magnitude limited 
samples LEDA14.5 and LEDA16
 (North and South). The reference line has a slope $-\gamma_a=-0.2$ 
which corresponds to a fractal dimension $D=1.8$ in the 3-d space. 
The flattening at large angular separations 
is due to projection effects and  does not 
correspond to any real feature of the 3-d distribution.}
\eef
In the case of $D < 2$ the exponent 
of the angular correlation function is expected to 
 be $-\gamma_a=D-2$
\cite{pee93}, i.e.
 \be
\Gamma(\theta) \sim  \theta^{-\gamma_a}
\ee
\bef 
%\vspace{}
\epsfxsize 8cm 
\centerline{\epsfbox{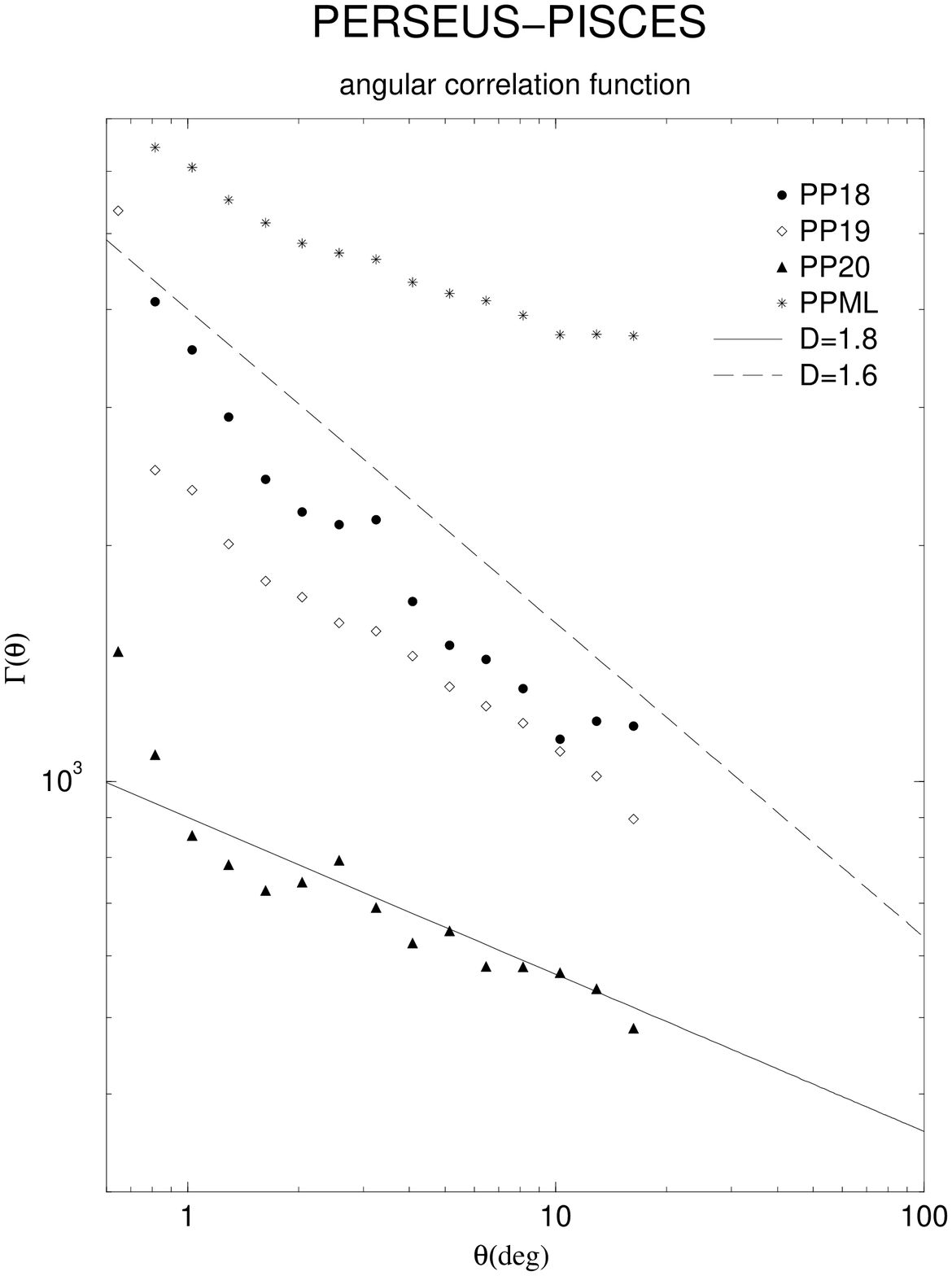}}  
\caption{\label{fig95} Angular correlation function $\Gamma(\theta)$ for the various VL  
samples (PP18, PP19, and PP20) and for the whole ML catalog (PPML)  of Perseus-Pisces. The 
solid line has a slope $-\gamma_a=-0.2$, while the dotted line has a slope $-\gamma_a=-0.4$.
 }
\eef
 
\bef
 %\vspace{}
\epsfxsize 8cm 
\centerline{\epsfbox{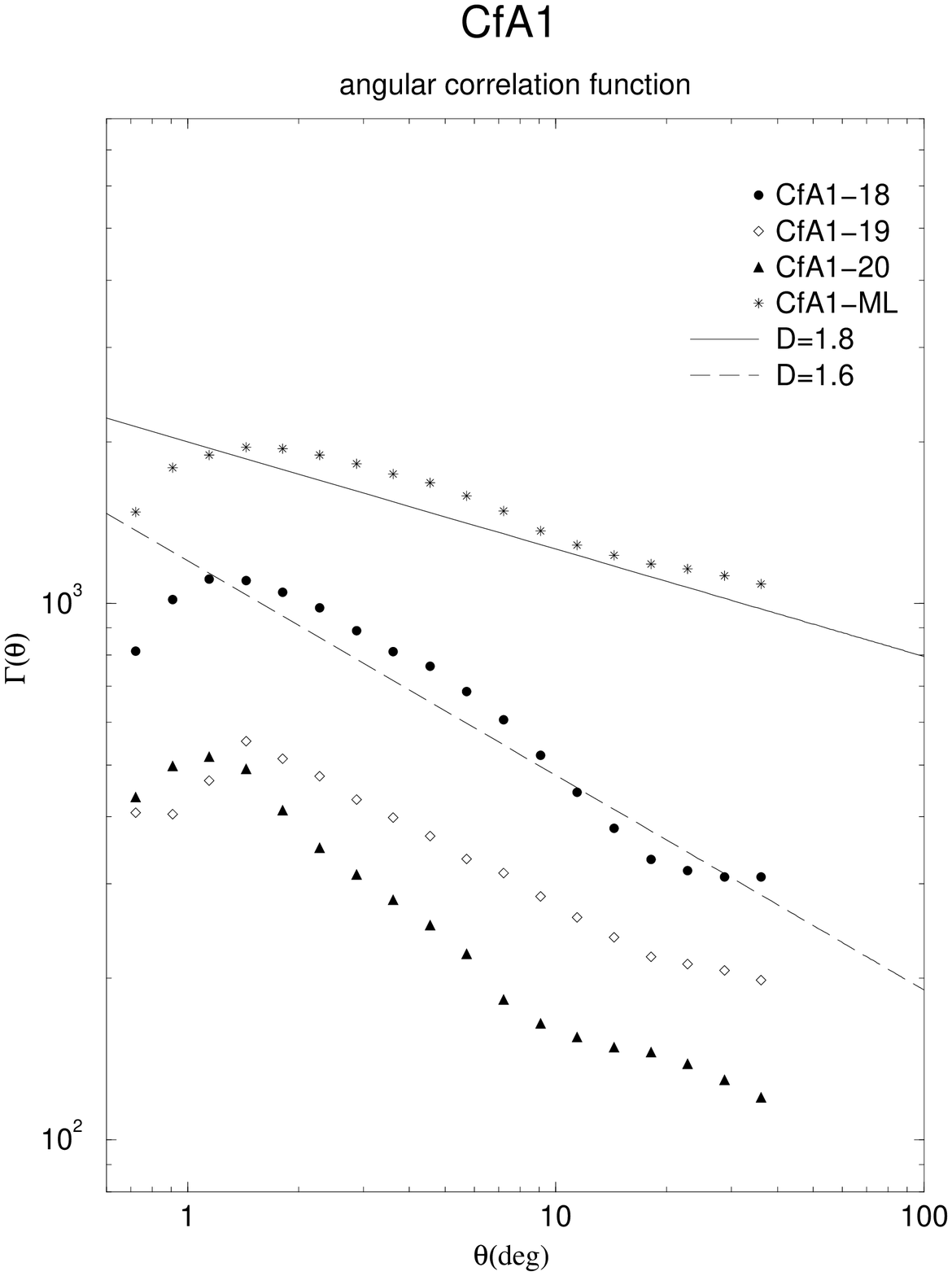}}
\caption{\label{fig96}
Angular correlation function $\Gamma(\theta)$ for 
the various VL   samples (CfA1-18, CfA1-19, and 
CfA1-20) and for the whole ML catalog (CfA1-ML)  of 
CfA1. The solid line has a slope $-\gamma_a=-0.2$, 
while the dotted line has a slope $-\gamma_a=-0.4$. }
\eef
\bef 
%\vspace{}
\epsfxsize 8cm 
\centerline{\epsfbox{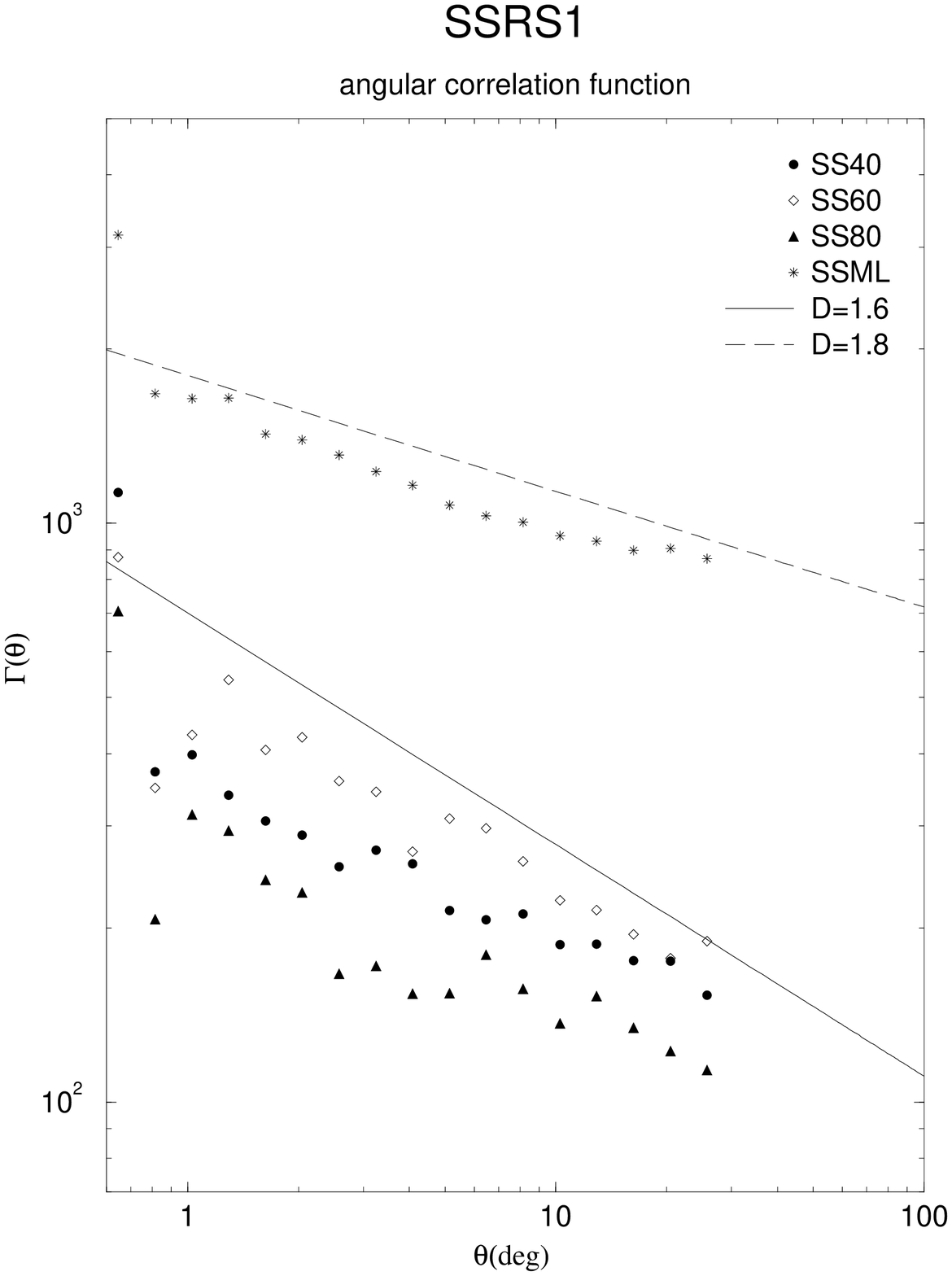}}  
\caption{\label{fig97} Angular correlation function $\Gamma(\theta)$ for the various VL   
samples (SS40, SS60 and SS80) and for the whole 
ML catalog (SSML)  of SSRS1. The solid line has
a slope $-\gamma_a=-0.4$, while the dotted line has a slope $-\gamma_a=-0.2$. }
\eef

According to the theorem discussed in Sec.\ref{orthogonal}
the orthogonal projection of a fractal structure with dimension $D$
in the $d=3$ Euclidean space, has dimension $D-2$, i.e. 
it should be compact for $D \ge 2$. The angular projection is  
different from the orthogonal one for the following reason. 
In the projection of at 3-d distribution on a planar infinitesimal  area $dA$,
we have contribution, up to the distance $R$,
from a volume which growths as $dV= dA \cdot R$.
On the other hand, in the projection on a infinitesimal solid angle $ d \Omega$ 
we have contribution, up to  $R$,  from a 3-d volume $dV = d\Omega\cdot R^3$. 
While in the first case the number of points
which contribute to the projection 
growths as $N_{ort} \sim R^{D-2}$, 
in the second case $N_{ang} \sim R^D$.  The qualitatively difference 
behavior of the number of points in these two case, is one of the reason 
of the uniformity of angular distribution.

Although  we have measured that the fractal dimension
in the three dimensional Euclidean space is  $D=2$, the 
angular correlation exponent is $-\gamma_a \approx -0.2$.
This can be due to the fact that the result $-\gamma_a=D-2$
is valid in the limit of  an asymptotic
 fractal structure, i.e.
if the volume of the three dimensional 
sample goes to infinity.
The finite size effect, due to the
 fact that the observed three 
dimensional volume is limited, can lead 
to a higher 
absolute value for the angular correlation exponent. 
The reason is essentially the following. 
We can approximate a limited 
fractal distribution in 3-d, as the intersection
of the asymptotic 3-d  structure 
with a plane, with a certain fixed thickness, 
which has dimension $d=2$.
If the thickness of the plane is small enough, 
the resulting distribution has dimension 
$D_i=D+2-3 \approx D-1 \approx 1$ in the 3-d space.
The angular  projection of  a 
structure with $D_i \approx 1$ has still 
dimension $1$.
Hence, in this case, the angular correlation exponent is 
$-\gamma_a \approx -1$ rather than $\gamma_a =0 $
which corresponds to the limit of having a 3-d sample extended enough.

In order to test that this is the case we have cut various
catalogs at
progressively smaller distances, finding that the 
angular correlation exponent is
 higher (in absolute value) 
as the three dimensional sample is limited in distance
(see Fig.\ref{fig98} and Fig.\ref{fig99}). 
\bef %\vspace{}
\epsfxsize 8cm 
\centerline{\epsfbox{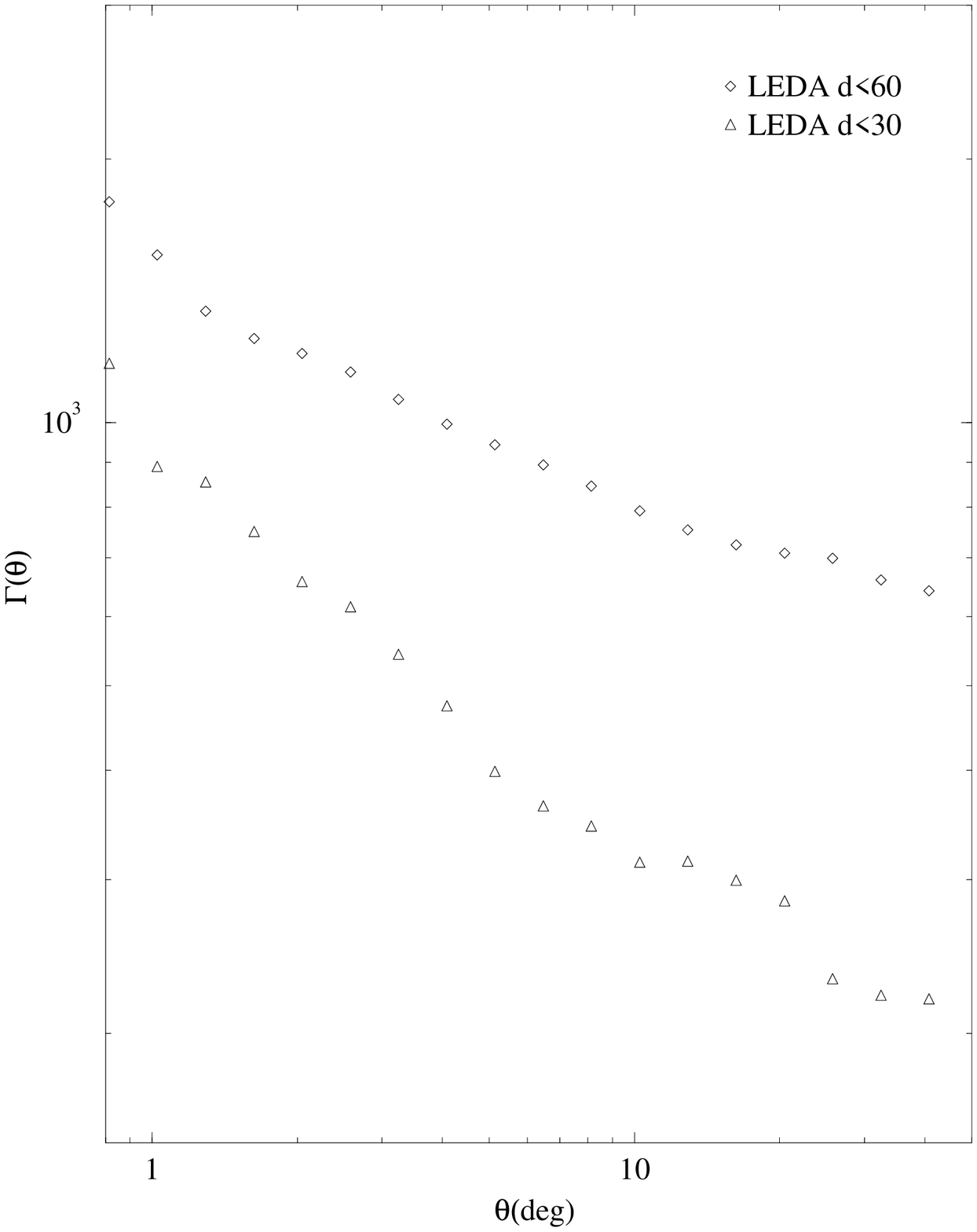}}  
\caption{\label{fig98}  Angular correlation 
function $\Gamma(\theta)$ for the ML catalog 
LEDA-14.5, cut at $r<60 \hmp$ and $r< 30 \hmp$. It is clearly
shown the difference in the correlation exponent.}
\eef
\bef
 %\vspace{}
\epsfxsize 8cm 
\centerline{\epsfbox{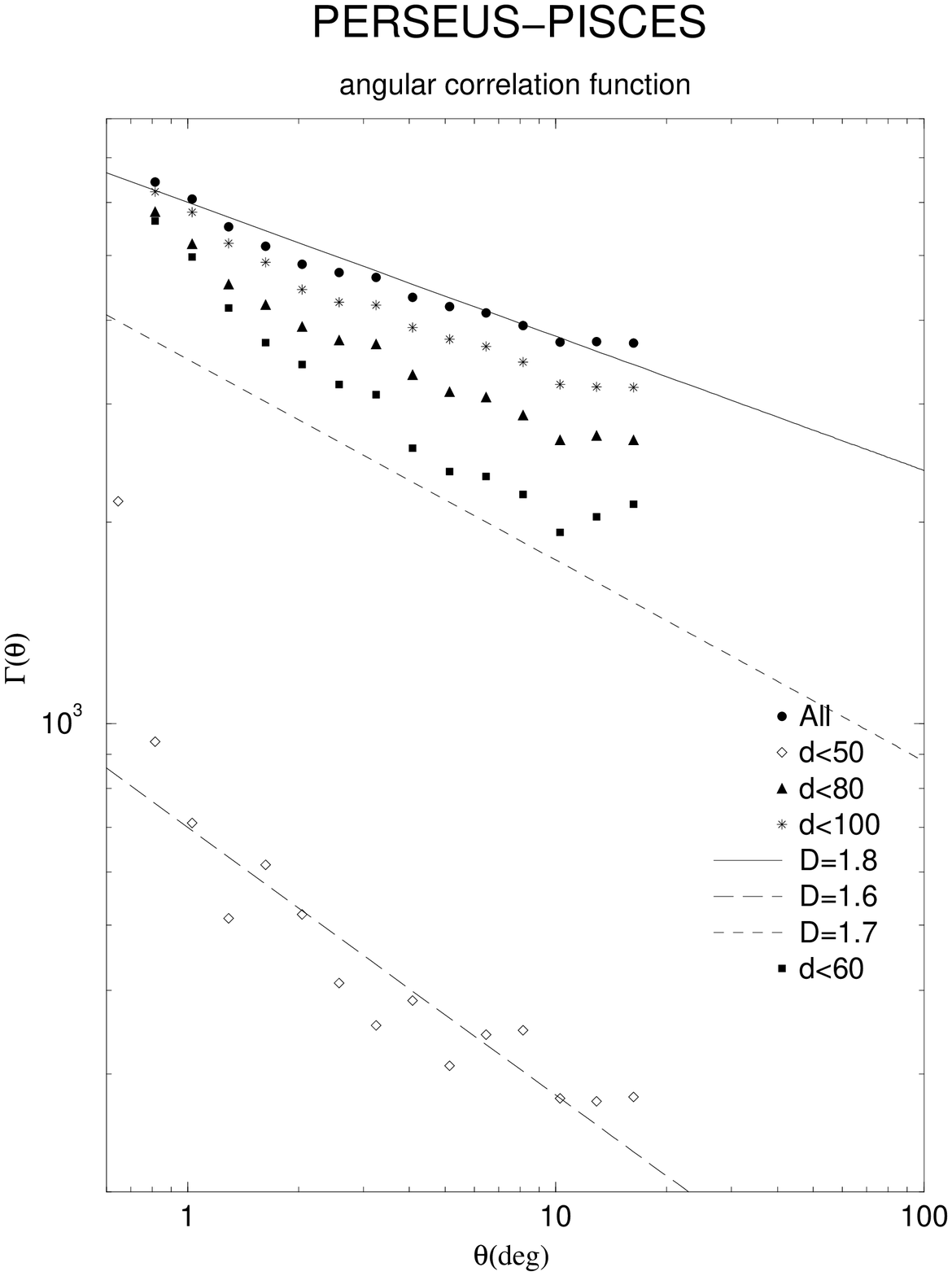}}  
\caption{\label{fig99}  Angular correlation 
function $\Gamma(\theta)$ for ML Perseus-Pisces catalog 
cut at difference distances.  It is clearly
shown the difference in the correlation exponent. The solid 
line has a slope $-\gamma_a=-0.2$, while the
dotted line has a slope $-\gamma_a=-0.4$.}
\eef
In other words when the thickness of the 3-d
 sample is some few $\hmp$ ($\sim 20 \div 40$),
the angular correlation function shows a trend 
$-\gamma_a \rightarrow -1$.
Otherwise when the thickness is $\sim 100 \hmp$ 
or more, we get $\gamma_a \rightarrow 0$. 
As the galaxies which contribute to the angular
 projection are located in a large 
volume of space, due the mix of different length 
scales, the projection appears 
quite uniform (i.e. $\gamma_a \sim 0$). 
As shown in Fig.\ref{fig94} the angular 
correlation function
presents a flattening at large angles. 
Such a behavior  is  completely spurious
as the three dimensional sample is 
a piece of fractal up to its 
limiting distance. This is due to the fact that 
the angular catalogs are qualitatively inferior 
to the three dimensional ones because 
they do not contain the third coordinate, even if they usually
contain more galaxies.
Angular projections, which are magnitude limited,  
mix different length scales and this gives an artificial randomization of
the galaxies. 
This implies that the angular projection of a fractal is really homogeneous at 
relatively large angles. Clearly this is an artificial 
effect and from a smooth angular 
projection one cannot deduce whether the 
real distribution is also smooth. 
For a detailed discussion of the scaling of the angular correlation amplitude, as
well as other effects of angular projection, we refer the reader
to \cite{dp91,cp92}.

\subsubsection{The large angles problem: 
angular projection of fractal structures}  
\label{anglarge}

Dogterom \& Pietronero \cite{dp91} (see also \cite{cp92}) studied
     the surprising  and subtle  properties of the angular
 projection of a fractal distribution. They find that the angular
 projection produces an {\it artificial crossover towards homogenization
 with respect to the angular density}. 
This crossover  is artificial
 (just due to the projection)  as  it
 does not correspond to any
 physical features of the three dimensional distribution. Moreover
 they showed that there is an explicit dependence of the angular
 two point correlation function $\omega(\theta)$ on $\theta_M$ the
 sample angle: this effect has never been taken into account in the
 discussion of real angular catalogs. These arguments show that it
 is very dangerous to make any definite conclusion just  from the
 knowledge of the angular distribution. By the way 
this is the reason why one has to measure redshifts, 
which, of course,  is not an easy task.

 However, there is a point
 of the discussion which remains still open. In fact, some authors
 \cite{pee93,mad90} claim that one of the most important
 facts that disprove the existence of  fractal correlations at
 large scales, is the scaling of the amplitude of the two point
 angular correlation function (ACF) with sample depth, in the small
 angles approximation. We can now clarify this puzzling situation.

Before considering the problem of the angular correlation function,
let us consider the angular projection of an artificial fractal. Some authors
(e.g. Peebles \cite{pee93}) pointed out that a 
fractal with dimension $D$
significantly less than three {\it cannot}
 approximat  the observed 
isotropic angular distributions of deep
 samples. In particular Peebles 
\cite{pee-pc96}
showed that a fractal with dimension 
$D \approx 2$ has  large-scale angular 
fluctuations which are not compatible with
 the observed angular maps. We   stress that there are various
problems, which are usually neglected (e.g. \cite{pee-pc96})
in constructing an artificial distribution
with the properties of the real one:
\begin{itemize}

\item The first point is that in generating an artificial 
fractal structure a very important   role is
  played by {\it lacunarity}. As we have seen in 
  Sec.\ref{statmec} and Sec.\ref{validity}.,
 even if the fractal dimension
is fixed, one can have very different angular distributions
depending on the value of the lacunarity. 
In fact, if lacunarity is large,
 and the sample is characterized by having voids 
 of the order of the sample size, it is clear 
that the angular distribution is 
highly inhomogeneous. On the other
 hand, if lacunarity is small 
(with respect to the sample size)
one can obtain more uniform angular 
projections (see Fig.\ref{fig7}).
A low value of the 
lacunarity should therefore be used for
 the reproducing galaxy distribution, because the
real galaxy distribution has indeed a 
low value of the lacunarity:
in the available samples, the dimension 
voids is  smaller than the 
survey volume \cite{elad97}. (See  \cite{durrer} 
for more details).

\item
The second important point which should be considered is that the real angular
distributions are {\it magnitude limited ones}, i.e. contains all the galaxies
with apparent magnitudes brighter than a certain limit $m_{lim}$. 
This implies a mixing of
length scales due to the fact that galaxies have a very 
spread luminosity function, and 
their absolute luminosity can change of more than a factor ten. 
For example 
suppose that  $m_{lim}=14$, then one gets contributions from galaxies 
in the range of distances $\sim 1.5 \div 50 \hmp$
 (see Fig.\ref{fig100}).
\bef 
%\vspace{} 
\epsfxsize 12cm
\centerline{\epsfbox{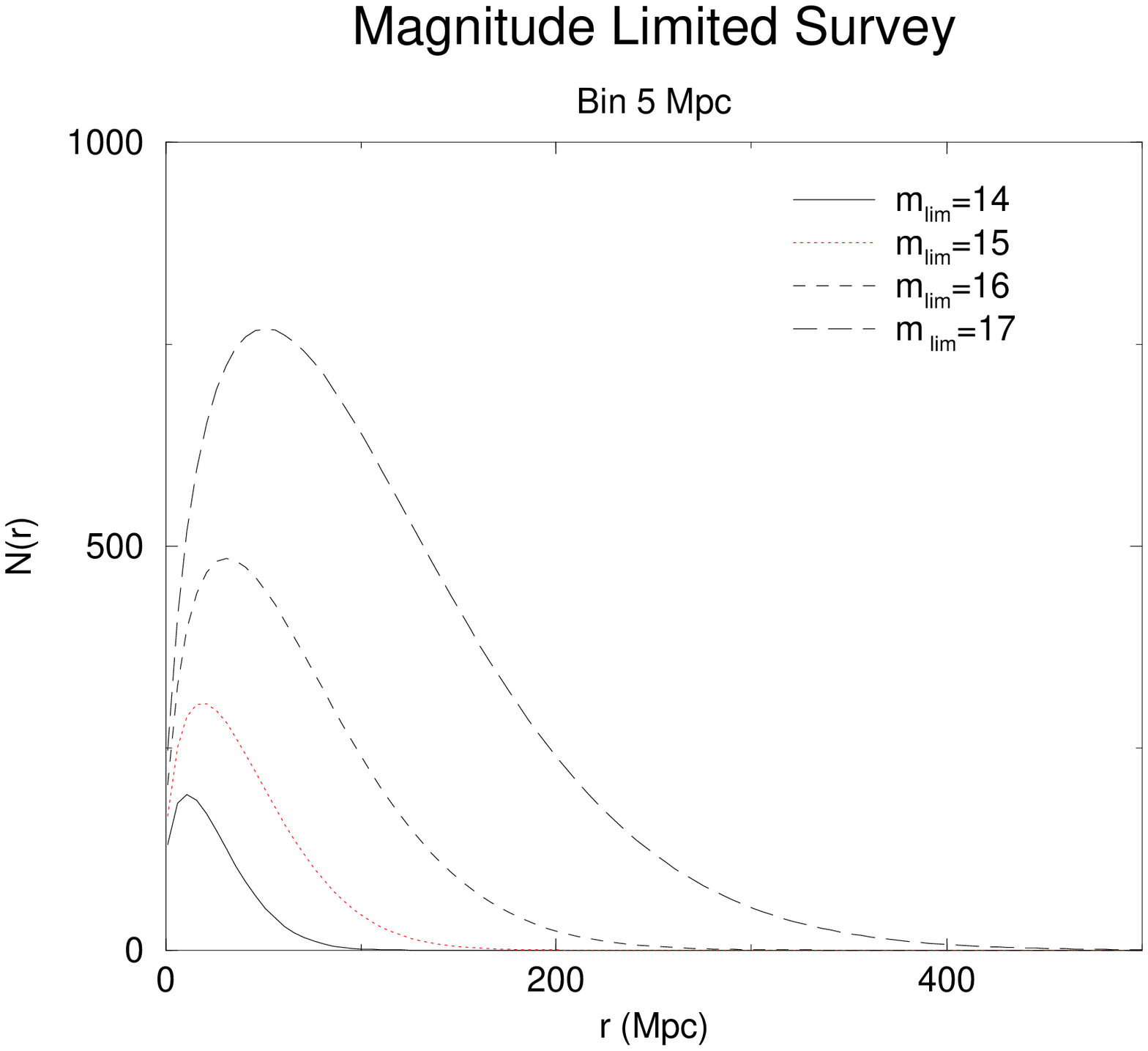}}
 \caption{\label{fig100}
Number of galaxies (per bin of $5 \hmp$) 
for magnitude limited samples
with different limit $m_{lim}$. As the apparent magnitude limit increases
the number of galaxies at large distance, 
which contributes to the 
magnitude limit angular sample, growths. 
This is the origin 
of the 
mixing of luminosity and space
 distributions which leads to an apparent homogeneous
behavior for the angular distribution
} 
\eef 
However if, for example,   $m_{lim} = 17$ then the range of distances 
from which one has important contributions to the angular 
distribution, rapidly grows. 
This  implies that
 galaxies in angular catalogs correspond
to extremely different distances. Correspondingly 
the projection is a 
complex convolution of various 
luminosities and distances. 
This is another important 
reason why if one looks an angular 
map limited at galaxies
with apparent magnitude
 brighter than $14$ one sees large scale fluctuations. 
On the other hand a catalog 
(as the angular APM catalog) limited at $m_{lim} > 18$,  
is quite  smoother, because the mixing of length scales is large. 

\item An important point, recently pointed out by Durrer \etal \cite{durrer}, 
is the following. The angular lacunarity of a 3-d fractal set 
with $D \approx 2$ can be very small, 
even if there are large voids in the space distribution. Moreover the angular
fluctuations depend on whether the galaxies are projected with 
apparent or fixes size, being more uniform in the latter case.

\end{itemize}

\subsubsection{Rescaling of the amplitude of 
 $\omega(\theta)$} 
\label{angsmall}

 Assuming
 that the fractal correlation are only present at small scales,
 i.e. that $\xi(r) = (r_0/r)^{\gamma}$, it is possible to show that,
if $\gamma > 1$, 
 in the small angle approximation ($ \theta << 1$) one has for the
 homogeneous case that \cite{pee93} 
\be 
\label{acf2}  
\omega(\theta)
  \sim  \theta^{1-\gamma} (r_0/D_*)^{\gamma} 
\ee 
where the depth
 $D_*$ is 
\be 
\label{acf3}
 D_*= \left( \frac{L_*}{4 \pi f}
 \right)^{1/2}.
 \ee
 $L_*$ is the cut-off of the Schecther
 luminosity function and $f$ is the limiting flux density of the
 survey. In the case of a fractal distribution with $D <2$ it is
 easy to show that instead of Eq.\ref{acf2} we have \cite{pee93} 
\be
 \label{acf4} 
\omega(\theta)  \sim \theta^{1-\gamma}
 \ee
 so that
 the difference between the homogeneous and the fractal case is
 that in the first case the amplitude of the ACF depends on the
 sample depth $D_*$ while in the second case it is constant.
 Peebles \cite{pee93} claims that, since in the real angular
 catalogs one observes the scaling of the amplitude \cite{mad90,gp77}, 
 this provides an evidence against the fractal behavior. We
 show now that this conclusion is not correct because it does not
 take into account the finite size effects in real galaxy surveys,
 as it is the case of the GNC. 

In fact, the amplitude of the ACF is
 strongly related to the behavior of the angular number density, i.e. to 
 $N(<m)$.  Eq. \ref{acf4} is obtained under the assumption that the  
 density for  a fractal scales as $r^{-\gamma}$: this is true for
 the {\em average conditional density}  in the case of an ideal
 fractal distribution, if the correct scaling regime  is reached.
 As previously discussed, 
 the conditional density computed from a single point is instead 
 strongly affected by finite size effects up to the characteristic
 {\em minimal statistical length} $\lambda$. 

 In order to
 illustrate this point we present the analysis of the ACF for the
 Perseus-Pisces redshift survey \cite{hg88}.  In Fig.\ref{fig101} 
\bef
 %\vspace{}
\epsfxsize 10cm
\centerline{\epsfbox{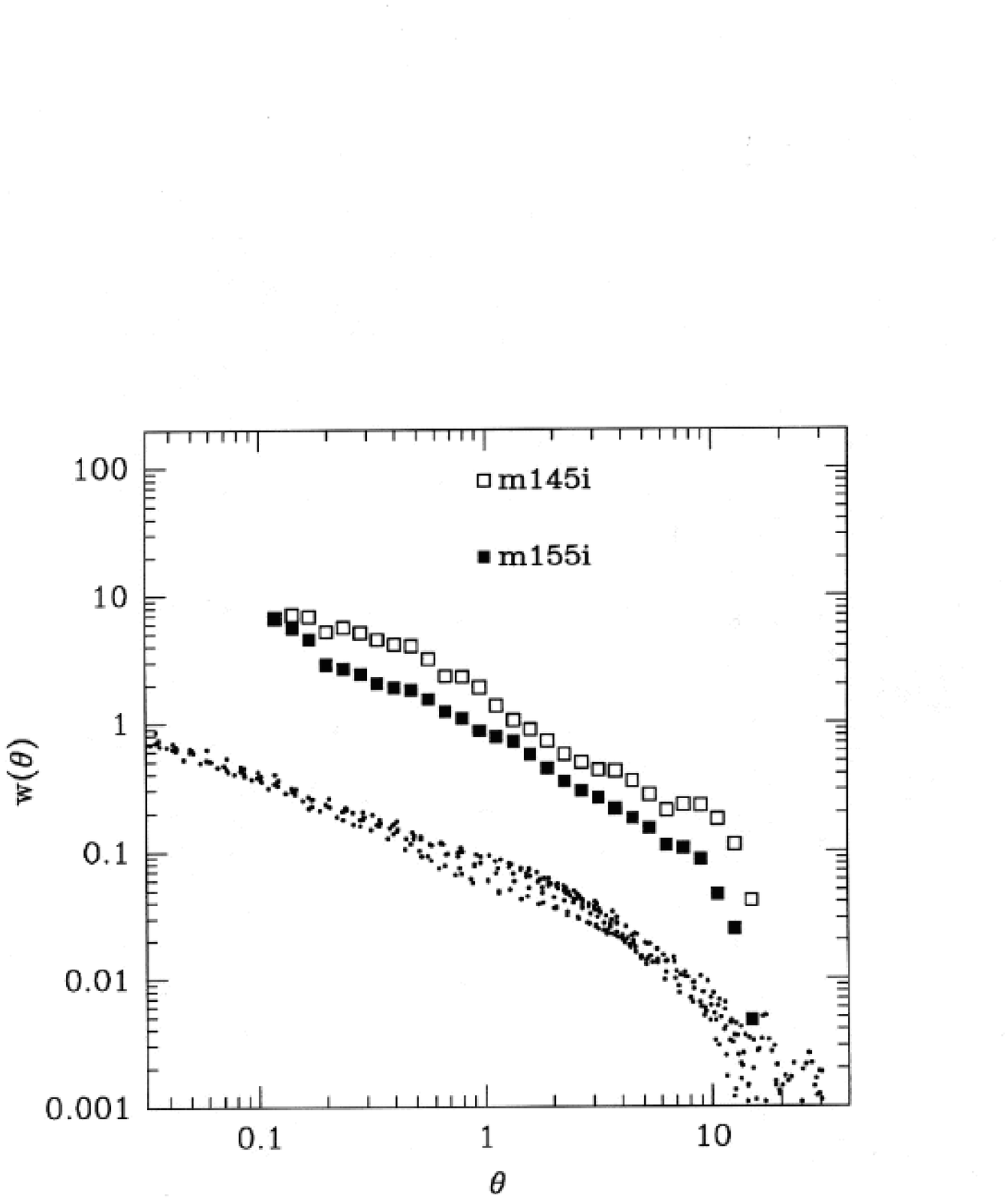}}
 \caption{ \label{fig101}  The angular
 correlation function for Peruses Pisces limited at $m_{lim} 14.,5$
 ($m145i$) and $15.5$ ($m155i$). The points refers to different
 samples of the APM catalog scaled to the ACF of the Lick survey
 (Maddox \etal, 1990). The scaling with depth of the amplitude of
 the angular correlation function in this case is due to a finite
 size effects and it is not a proof of homogeneity in space, as we
 know from the space analysis that this sample has fractal behavior
 up to its deeper depth. }
 \eef 
 we show the behavior of the ACF
 $\omega (\theta)$ for the whole ML survey. There
 is a clear scaling of the amplitude with the apparent magnitude
 limit of the survey  ($m_{lim}=14.5, 15.5$ respectively). We know
 from the space analysis that the galaxy
 distribution is fractal in
 this sample and therefore this trend 
 is not a consequence of
 homogeneity, but only of the finite size effects  which are
 especially large for the counting from the vertex.

 In summary the
 apparent homogeneity inferred
 from the behavior of the angular correlation amplitude as a function of
 depth,  has the same origin as the exponent $\alpha \approx 0.6$ of the galaxy
 counts at bright magnitudes (small scales). Both arise from finite
 size effects (for a more 
 detailed study on the ACF we refer the
 reader to \cite{ams95}) 
 and do not correspond to the real
 statistical properties.

 \section{Luminosity and space distributions}
 \label{lumspace}
 In a well-known review on the galaxy luminosity  function (LF)
 Binggeli \etal  \cite{bin88} state that {\em "as the distribution
 of galaxies  is
 known to be inhomogeneous on all scales up to a least $100 h^{-1}
 Mpc$, a rich cluster  of galaxies is like a Matterhorn in a grand
 Alpine landscape of  mountain ridges and valleys of length up to
 100 Km"}. The aim of this section is to consider this point of
 view in the light  of the concept of multifractality of the mass
 distribution. We show how the main observational aspects of galaxy
 luminosity and space distributions are strongly related and can be
 naturally linked and explained  as a multifractal (MF)
 distribution. The concept of MF is appropriate to discuss physical
 systems with local  properties of self-similarity, in which the
 scaling properties are  defined by a continuous distribution of
 exponents. Roughly speaking one can visualize this property as
 having different scaling properties for different regions of the
 system (see \cite{cp92,slp96} for a more detailed discussion).

The fundamental point we discuss in 
 this section is that {\it the whole matter
 distribution, i.e.  weighing each point by its mass, is 
 self-similar}.  This situation  requires the generalization of the
 simple fractal scaling  to a MF distribution in which a  
 continuous set of exponents is necessary to  describe the spatial
 scaling  of peaks of different weight  (mass or luminosity). In
 this respect the mass and space  distributions become naturally
 entangled with each other.

\subsection{Galaxy space and luminosity distributions} 
\label{lumspacegal}
  We briefly review  here the main features of the luminosity and
 space distributions of   galaxies  together with the various 
 morphological properties which can  be naturally  embedded in a MF
 scheme.  
\smallskip

1. The  main statistical tool to study the spatial distribution of
 galaxies is  the two point correlation function (CF) and we refer
to Sec.\ref{corran} for a detailed discussion of its determinations.
\smallskip   

 2. One of the main characteristics of galaxy  surveys is that one
 finds that  groups of galaxies comprise at least $70\%$ of all
 galaxies not being part of clusters, and {\em truly isolated
 galaxies} are very rare. Tully \cite{tul88}
 by analyzing the {\em Nearby Atlas of Galaxies}  \cite{tf87} finds
 that essentially all galaxies can be grouped into clouds and that 
 roughly $70\%$ of these can be assigned to groups.
 \smallskip

  3. Various studies \cite{ede89,bin90,fb94} of the
 spatial distribution of {\em dwarf galaxies}  show that these 
 galaxies fall into the structures defined by the luminous ones and
 that there is no evidence of segregation of bright and faint
 galaxies on large scale: dwarf galaxies are not more uniformly
 distributed than giants and  the dwarfs, as the giants, belong to
 clouds, groups or clusters. There is evidence that  the dwarfs
 fall well into the large scale patterns suggested by the giants
 consisting of filaments, walls and arcs. In particular, there is
 no evidence for them to fill voids \cite{thu87,bot88}. 
 Disney \& Phillips \cite{dp87} pointed out that galaxies of very
 {\em low surface brightness} (LSB) are entirely missed  for an
 observational selection effect, and that what one can see is the
 {\em "tip of the iceberg"}. 
 Moreover Bothun \etal  \cite{bot88} 
 concluded that the patterns of Large Scale
 Structures appears to be mostly independent on galaxy surface
 brightness. Binggeli \etal \cite{bin88}  stressed that  most of
 the galaxies which are entirely missed because of their low surface
 brightness seem  to be also of low total brightness, so that this
 observational bias has the effect of a lower cut-off in
 brightness.
  \smallskip

 4. An observation that is particularly  important from a
 theoretical point of view  is  the behavior of the {\em
 giant-to-dwarf ratio}  as a function of the local density: we ask
 whether   dwarf galaxies exist in low density regions,  where
 giants are rare, or if they are only found as satellites of giants
 so that the  giants-to-dwarfs ratio does not depend  on
 environmental density. There is a clear  experimental indication
 that the dwarf-to-giant ratio  depends on  the local density
 \cite{bin88,bin90,fb94}.
 \smallskip 

5. Einasto and Einasto \cite{ee85}   found that  the {\em brightest
 galaxies} in groups and clusters are brighter  than in the field
 by up to 1 magnitude: the brightest galaxies lie preferentially in
 dense environments.  In particular Dressler \cite{dre84} pointed
 out that the  most luminous  elliptical galaxies  
 usually reside in the clusters  cores, at local density maxima,  
 and are not present in low density fields,  so that these objects 
 seem to be the product of dense environments. 
\smallskip   

6. The fact that  giant  galaxies are "more clustered" than the
 dwarfs  has been interpreted as corresponding to a larger value of
 the amplitude  of the correlation function for the giants than for
 dwarfs: this is the so-called {\it "luminosity segregation"}
 phenomenon  (see Sec.\ref{radial}).
On the contrary we show here  that the segregation of giant
 galaxies in clusters arises as a consequence of self-similarity of
 matter  distribution, and that in this case the only relevant
 parameter is the {\it exponent}  of the correlation function,
 while the amplitude is a spurious quantity that has no direct
 physical meaning and depends explicitly on the  sample size.  
\smallskip 

7.  There is evidence that galaxies in  different environments  
 are morphologically  different and may followed   different
 evolutionary paths.  There is  in particular a  predominance of
 early type of galaxies in rich clusters: high density regions are
 dominated by E  and S0 galaxies which themselves are hard to find
 in the field. Numerous studies have analyzed the variation in the
 population fraction and its possible relationship with cluster
 morphology \cite{oem74}; the  {\em morphological segregation} has
 been studied   systematically by  Dressler \cite{dre80}, who
 examined the variations in the relative  fractions of E, S0, and
 spiral  galaxies as a function of the {\em local  density} and
 hence quantified the so-called  {\em morphology-density relation}.
 He discovered that the local density  of galaxies governs the
 mixture of Hubble types in any local environment of a cluster,
 independently of cluster global parameters like richness or size. 
 The correlation of the morphological mix with  local density is
 continuous and monotonic.

This behavior has been shown to extend continuously over 6 orders
 of  magnitude in space density from rich clusters to low density
 groups \cite{desou82,pg84}.   The main features of the  
 {\em morphological segregation}  is the decrease in spiral
 population with increasing local density and the increase with
 density of the fraction of S0  and elliptical galaxies.    Several
 authors \cite{oem74,dre80,hg88,iov93}
 found that the relative abundance of elliptical, lenticular and
 spiral galaxies in clusters and their  peripheries is a function
 of the local density:  $80\%$ of   field galaxies are spirals and 
  $15\%$ of   galaxies   in rich clusters show spiral structure. 
 The morphology-density relation in rich clusters  is continuous
 over six orders of magnitude in space density and, 
 correspondingly, the galaxy density is a continuous parameter:
 the consequence is that the separation between  the luminosity
 function and the space density  is seriously questionable .  The
 morphology-density relation is found to hold also for dwarf
 galaxies \cite{fb94}.
 \smallskip  

8. The characteristics of morphology segregation can
 also be described by a comparison of the  {\em angular correlation
 function} for representative  samples of different morphology.
 Davis {\it et al.} \cite{dav76}  found that elliptical-elliptical 
 angular correlation function can be described by a power  law with
 a slope significantly steeper than the one of  the  corresponding
 spiral sample. Moreover the slope of the  angular correlation
 function which characterizes the S0-S0 clustering is  intermediate
 to other classes.   Giovanelli \etal \cite{gio86},  by analyzing the
 {\em Perseus-Pisces} redshift survey  found that the slopes of
 $w(\theta) \sim A \theta^{\beta}$  are significantly steeper for
 early type of galaxies: for early galaxies $\beta = -0.90$, while
 for early spirals $\beta = -0.65$  and for late spirals $\beta =
 -0.37$.

In the following we  show that  all these 
morphological evidences are  
   naturally explained within the  context of a  multifractal
 description, providing in the process  a quantitative mathematical
 description of the phenomena.

 \subsection{Standard analysis of the Luminosity Function}  
\label{lumspacelf}

The differential luminosity function gives the
 probability of finding a galaxy with luminosity  in the range
 $[L,L+dL]$ in the unit volume ($Mpc^{-3}$). In    literature (see
\cite{bin88} 
 for a review) one finds several methods
 to determine the LF for field galaxies and cluster galaxies.
 Special emphasis is devoted to the systematic differences in   the
 LF for the various Hubble types. Here we are interested in the 
 determination of the {\em general} LF defined as the sum over all
 Hubble types  for field galaxies.  

  Let $\nu(L,\vec{r})$ denote
 the number of galaxies  lying in volume $dV$ at $\vec{r}$ that
 have intrinsic luminosity  between $L$ and $L+dL$. The main
 assumption generally used  \cite{bin88}  is that galaxy
 luminosities  are not correlated with spatial location.  
  Under such an hypothesis one can write  
\be 
\label{mfe2}
 \nu(L,\vec{r}) dL dV = \phi(L) D(\vec{r}) dL dV 
\ee 
where
 $\phi(L)$ gives the fraction of galaxies per unit luminosity
 having intrinsic luminosity  in the interval ($L$,$L+dL$), and
 $D(\vec{r})$ gives the number of galaxies of all luminosities  per
 unit volume  at $\vec{r}$ and it is related to the 
amplitude of $\phi(L)$.

The so-called {\em classical method}
 to determine the LF, is based also 
on the hypothesis that the galaxy
 distribution in the  samples under analysis has reached 
 homogeneity so that the  average density $n_0$ of galaxies in
 space is constant and well defined. This method  is highly
 sensitive to  the spatial inhomogeneities in the distributions of
 galaxies which should distort the shape of the LF.  For this reason
 many authors in the past \cite{fel77}  excluded a region of solid
 angle containing strong "inhomogeneities" in galaxy distribution
 as the Virgo cluster. 

   Given the highly irregular character of 
 galaxy distribution  in all the recent redshift surveys  
(see Sec.\ref{corran}) the assumption of constant density 
 and homogeneous distribution  is questionable and,   in fact, the
 amplitude of the LF, which is the average galaxy  number density,
 is a strongly fluctuating and not {\em well defined} quantity in
 the available samples.  For this reason all new methods to
 determine the LF aim at a  separation between the shape and the
 amplitude. In particular the so-called {\em
 inhomogeneity-independent}  methods have been developed with the
 intent to  determine only the shape of the LF.  

The basic idea is
 to consider the ratio of galaxies having intrinsic luminosity
 between $L$ and $L+dL$ to the total number of galaxies brighter
 than $L$. If Eq.\ref{mfe2} holds then
 \be 
\label{e3}
 \frac{\nu(L,\vec{r})dLdV}{\int_{L}^{\infty}\nu(L',\vec{r})dL'dV}
 = 
\smallskip
 \frac{\phi(L)D(\vec{r})dLdV}{\int_{L}^{\infty}\phi(L')D(\vec{r})dL'dV} = 
\smallskip 
\frac{\phi(L)dL}{\Phi(L)} \sim d\log\Phi(L) \; .
 \ee 
By
 differentiating the integrated LF $\Phi(L)$ one  obtains the
 differential LF $\phi(L)$.   This technique allows recovery of 
 the shape for the LF undisturbed  by  space inhomogeneities.  
 Usually the LF is assumed to be described by an analytical
 function. The most popular is the one proposed by Schecther \cite{sch76}
 \be
 \phi(L)d(L/L^{*}) = \phi^{*}(L/L^{*})^{\alpha}
 exp(-L/L^{*}) d(L/L^{*}) 
\ee
 where $L^{*}$ is the cut-off,
 $\phi^{*}$ is the normalization constant  (amplitude) and 
 $\alpha$ is the exponent. By Eq.\ref{mfe2} 
we have separated the space and the 
luminosity distributions: the amplitude of the luminosity function
in now related to the behavior of the space density, while the 
constant $\phi^*$ is given by the normalization condition
\be
\int_{-\infty}^{\infty} \phi(L) dL =1 \; .
\ee
The determination of the amplitude is then directly related to the 
space density.

The LF has been measured by several
 authors in different  redshift surveys 
\cite{del89,mar94,lov92,dac94,vet94}
and the agreement between the various
 determinations in very different volumes is excellent. The best
 fit parameters  are $\alpha =-1.13$ and $M^*_{bj} =-18.70$.

 We discuss this point later.     
 We  
 show that  not only the homogeneity assumption is  inappropriate
 for the determination of the LF,  but also that the assumption in
 Eq.\ref{mfe2}  is not satisfied by the actual distribution of visible
 matter.  As the available samples show structures  as large as the
 survey depth we    see  that  {\em not only the amplitude of 
$\nu(L,\vec{r})$
  but also the cut-off $L^*$ of the LF are dependent on the
 sample depth}.  

Our essential points is  the following. 
The
 galaxy luminosities  are strongly correlated with their 
 positions in space. This  clear observational fact  can be studied
 quantitatively with the MF formalism. In particular  in such a
 scheme one can  determine analytically the  shape  and the
 amplitude of the LF, and unify the various observational issues in  
 quantitative mathematical scheme.

\subsection{Galaxy luminosity distribution in space: Multifractality}  
\label{lumspacemf}
We now briefly introduce
 the multifractal picture is
 a refinement and generalization of the fractal properties 
\cite{gp84,pv87,ben84,cp92,bslmp94,slmp96}
which arises naturally
in the case of self-similar distributions. If one does not
 consider the mass one has a simple set  given by the galaxy
 positions (that we call the {\it support}  of the measure
 distribution). Multifractality instead becomes interesting and a
 physically  relevant property when one includes the galaxy masses
 and consider the entire matter distribution \cite{pie87,cp92}.
%(Pietronero, 1987; CP92). 
In this case the measure distribution  is defined by
 assigning  to each galaxy a weight which is  proportional to its
 mass. The question  of the self-similarity versus homogeneity of
 this set can be  exhaustively discussed in terms of the single
 correlation exponent which corresponds to the fractal dimension  of
 the support of the measure distribution. 
Several
 authors 
\cite{mj90} instead considered the eventual
 multifractality  of the support itself. However the physical
 implication of  such an analysis is not clear,  and it does not
 add much to the  question above.    In the more complex case of MF
 distributions the scaling properties can be different for
 different regions of the system and one has to introduce a
 continuous  set of exponents to characterize the  system  (the
 multifractal spectrum). The discussion     presented in the
 previous sections  was meant to distinguish between homogeneity and
 scale invariant properties;  it is appropriate also in the case of
 a multifractal. In the latter case the correlation functions we
 have considered would correspond to a single exponent  of a
 multifractal spectrum of exponents, but the issue of homogeneity
 versus scale invariance (fractal or multifractal) remains exactly
 the same. 

 Suppose that the total volume of the sample consists of
 a {\em 3-}dimensional cube of size $L$. The  density distribution
 of visible matter is described by 
 \be 
\label{mf1}
 \rho(\vec{r})= \sum_{i=1}^{N} m_{i} \delta(\vec{r}- \vec{r_{i}}) 
\ee 
where
 $m_{i}$ is the mass of the $i$-th galaxy and $N$ is the number of
 points in the sample and $\delta(\vec{r})$  is the Dirac delta
 function. We assume that this distribution  corresponds to a
 measure defined on the set of points which have the correlation
 properties described by $\Gamma(r) \sim r^{D-3}$ (Sec.\ref{corran}).
 It is possible to define the
 dimensionless  normalized density function
 \be 
 \label{mf2}
 \mu(\vec{r}) = \sum^{N}_{i=1} \mu_{i} \delta(\vec{r}-\vec{r}_{i})
 \ee
 with $\mu_{i} = m_{i}/M_{T}$ and $M_{T} = \sum^{N}_{i=1}
 m_{i}$, the total mass in the sample. We divide this volume into boxes of linear
 size $l$. We label each  box by the index $i$ and construct for
 each box the function 
\be 
\label{mf3} \mu_{i}(\epsilon)  =
 \int_{{\it i-th box}} \mu(r)dr 
\ee
 where $\epsilon = l/L$ and $0
 < \mu_i<1$. The definition of the box-counting fractal dimension
 is 
\be 
\label{mf4}
 \lim_{\epsilon \rightarrow 0} \mu_{i}(\epsilon)
 \sim \epsilon^{\alpha (\vec{x})} 
\ee 
where $\alpha (\vec{x})$ is
 constant and equal to $D$  in all the occupied boxes in the case
 of a simple fractal. 
 %In Eq.\ref{mf4} the  exponent $\alpha (\vec{x})$ % (a sort of local 
 %fractal dimension)
  This
 exponent  fluctuates widely with the position $\vec{x}$ in the 
 case of MF.  In general we   find  several boxes with a measure
 which scales with the same exponent $\alpha$. These boxes form a
 fractal subset  with dimension $f$ which  depends on the  exponent
 $\alpha$, i.e. $f=f(\alpha)$. Hence the number of boxes which have
 a measure $\mu$ that scales with exponent in the range [$\alpha ,
 \alpha + d\alpha$] varies with $\epsilon$ as 
\be 
 \label{mf5}
 N(\alpha, \epsilon)d\alpha \sim \epsilon^{- f(\alpha )} d\alpha.
 \ee  
The function $f(\alpha)$ is usually \cite{pv87} %(Paladin \& Vulpiani 1987) 
a single humped function with the maximum at $max_{\alpha}
 f(\alpha) = D$,  where $D$ is the dimension of the support. In the
 case of a single fractal, the function $f(\alpha)$ is reduced to
 a single point: $f(\alpha) = \alpha = D$.

  In order to analyze 
 the mass distribution of galaxies, obviously one needs to know the
 density distribution  $\rho(\vec{r})$. The mass of each galaxy may
 be related to its total luminosity in a simple way 
\be 
M= k(i) L^{\beta} 
\ee 
where $k$ is the mass to light ratio and depends on
 the galaxy  morphological type $i$. In relation with the MF
 properties,   $k$ plays a little role because the important
 quantity  is the range of galaxy mass, which can be as large as a
 factor $10^6$ or more. Therefore a modification of $k$  produces
 small effects on a logarithm scale. The exponent $\beta$ is  more
 important,  and here we assume \cite{fg79} %(Faber and Gallagher 1979) 
 that
 $\beta \approx 1$. However a different value of $\beta$   should
 not change the MF nature of the mass distribution, if it is
 present in the sample, but only the parameters of the spectrum.  
   From a practical point of view one does not determine directly
 the spectrum of exponents $[f(\alpha), \alpha]$;  it is more
 convenient to compute   its Legendre transformation $[\tau(q),q]$
 given by 
\be 
\left\{ \begin{array}{l} \label{mf12} \tau(q)=q\cdot
 \alpha(q)-f(q)\\ \frac{d\tau(q)}{dq}=\alpha(q) \end{array} \right. \; . 
 \ee 
In the case of a simple fractal one has  $\alpha=f(\alpha)=D$.
 In terms of the Legendre transformation this corresponds to  
\be
 \tau(q) = D(q)(q-1)
 \ee 
i.e. the behavior of $\tau(q)$ versus $q$ is a
 straight line with coefficient  given by the fractal dimension. 
 The analyses  carried out on CfA1 \cite{cp92}
 and Perseus-Pisces \cite{slmp96} 
 redshift surveys provide 
unambiguous evidence for a MF behavior as 
shown by the  non linear    behavior of $\tau(q)$ 
in Fig.\ref{fig102}.
\bef
 %\vspace{}
\epsfxsize 8cm 
\centerline{\epsfbox{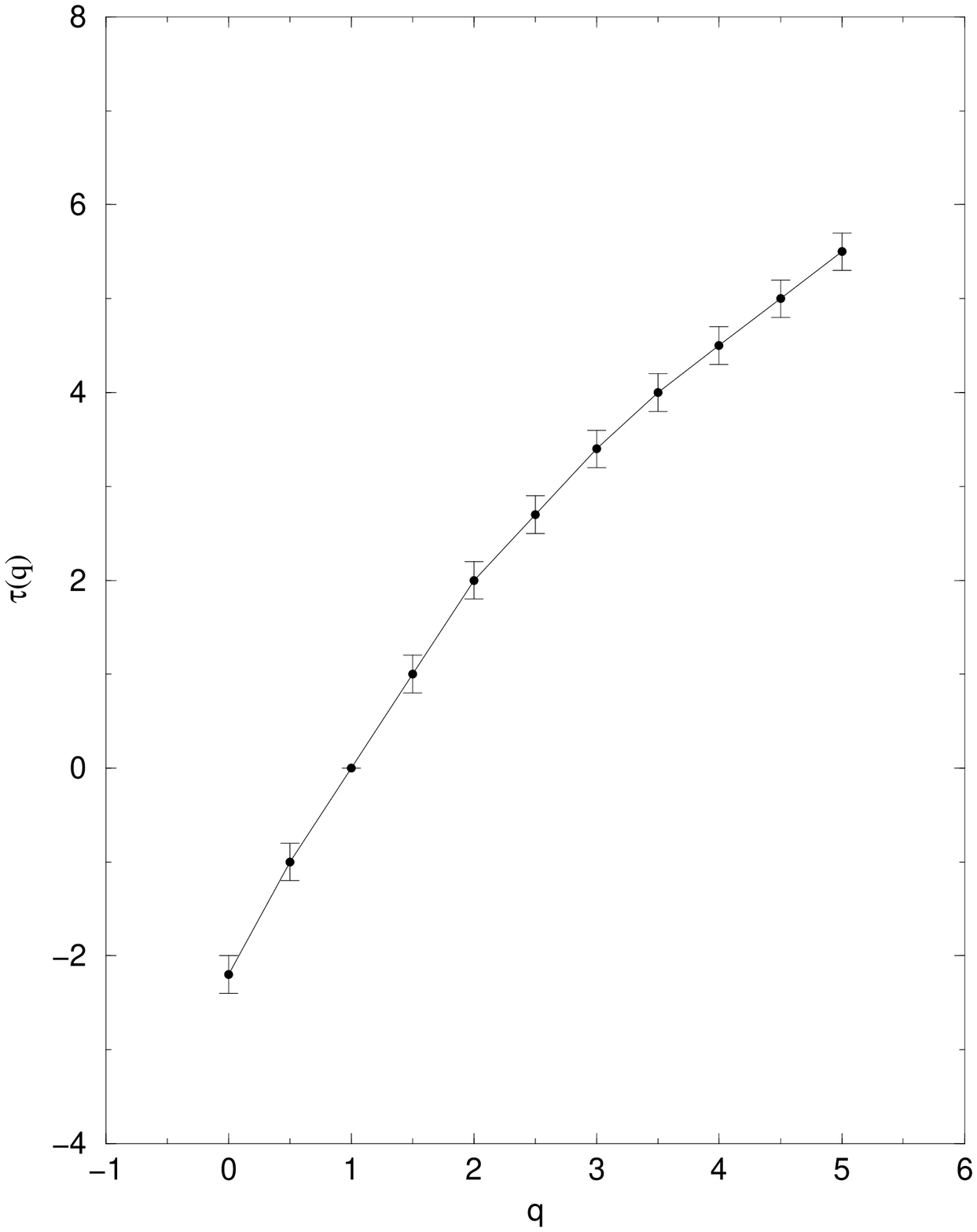}}
\caption{\label{fig102} The scaling 
exponents $\tau(q)$ as a function of the moment $q$
for the Perseus-Pisces Redshift survey (form Sylos Labini \etal (1996)).
The multifractal behavior is shown by the change of slope. For negative
 momenta, the
data are erratic because they are dominated by the smallest galaxies not
present in the sample.}
\eef 

We have
 performed a test to check the MF nature of the sample,  by randomizing
 the absolute magnitudes of the galaxies. By doing this one
 destroys the correlations between the spatial locations and
 magnitudes of galaxies. As shown in Fig.\ref{fig103}
\bef %\vspace{} 
\epsfxsize 8cm 
\centerline{\epsfbox{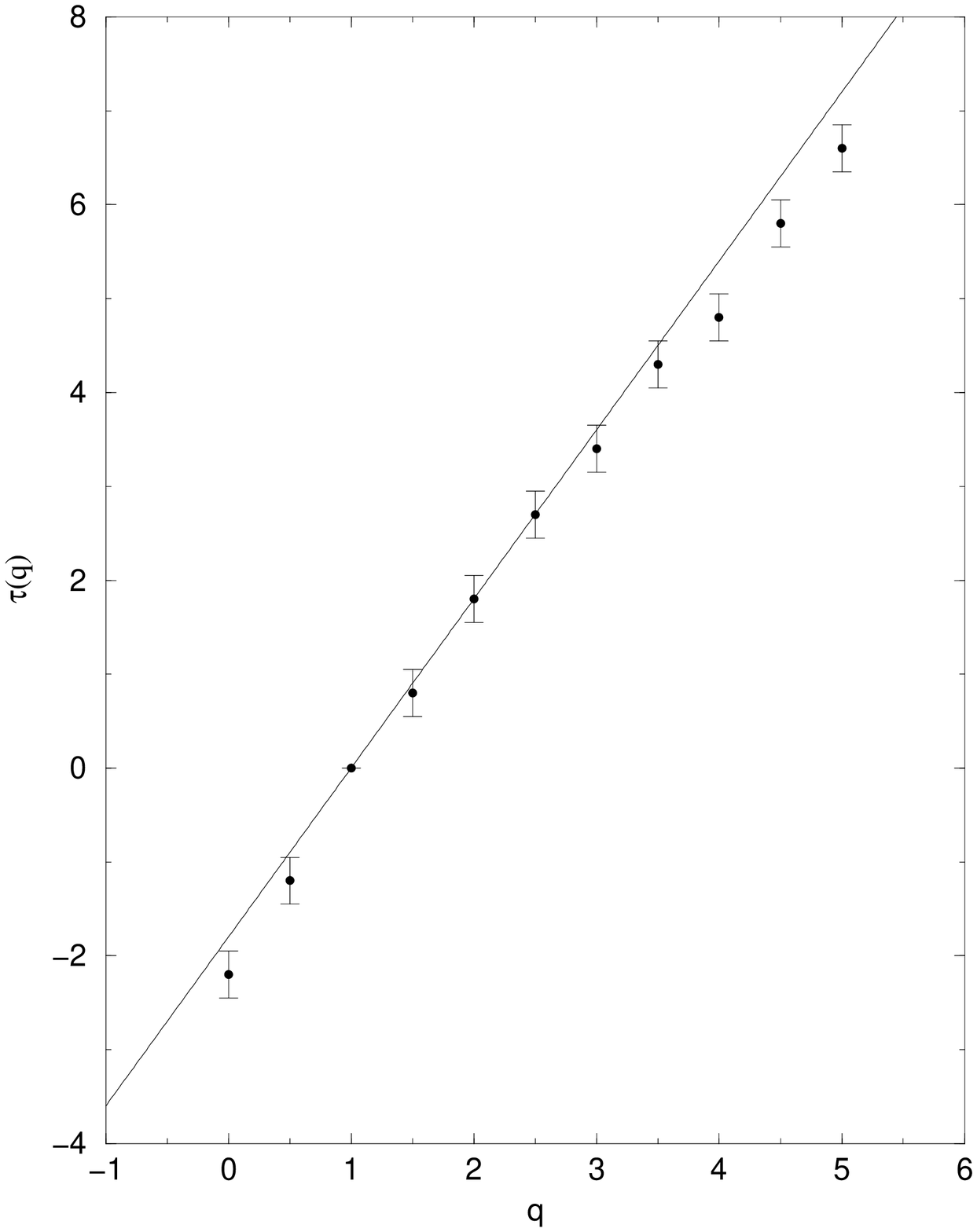}}
\caption{\label{fig103} The
 scaling exponent $\tau(q)$ as a function of the moment $q$: in
 this case we use the same sample previously analyzed, but we have first
 randomized the absolute magnitude. By doing this we destroy the
 correlation between galaxy space locations and luminosities
 (masses). The reference line has slope $2$ in agreement with the
 simple fractal properties. This result supports the real nature of
 the multifractal behavior. Moreover with this formalism it is
 possible to study the correlation between masses and space
 locations.}
\eef 
 in this case the
 dependence of $\:\tau(q)$ versus $\:q$ is linear, as for a simple
 fractal, with slope $\approx 2$. 

 The MF implies a strong correlation between
 spatial and mass distribution so that the number of objects with
 mass $\:M$ in the point $\:\vec{r}$  per unit volume
 $\:\nu(M,\vec{r})$, is a function of space and mass and {\it it is
 not separable in a space density multiplied by a mass (or
 luminosity) function} \cite{bin88}.
 This
 means that  we {\it cannot} express the number of galaxies
 $\:\nu(M,x,y,z)$  lying in volume $\:dV$ at {\it (x,y,z)} with
 mass between {\it M} and {\it M + dM} as in Eq.\ref{mfe2}.
  Moreover  we
 cannot define a well defined average density, independent on
 sample depth as for the simple fractal case.  It can be shown (see
Sec.\ref{lumspacelf2})
 \cite{slp96} that the mass
 function of a MF distribution, in a well defined volume, has
 indeed
 a Press-Schecther behavior whose exponent $\:\delta$ can be
 related to the properties of  $\:f(\alpha)$. Moreover the fractal
 dimension of the support is $\:D(0)=f(\alpha_{s})=3-\gamma$. Hence
 by the knowledge of the whole $\:f(\alpha)$ spectrum one obtains
 information on the correlations in space as well as on the mass
 function.  The phenomenon of morphological segregation naturally
 arises from the MF nature of galaxy distribution, due to the
 self-similar behavior of the mass distribution.

\subsection{The luminosity  function and its relation to the 
space distribution}  
\label{lumspacelf2}
We have seen
 in the previous section the basic formulae which describe  the
 scaling properties of a multifractal measure (MF). Suppose now we
 have a MF sample  in a well defined volume $V$,  and we want to
 study the behavior of the number  of boxes with measure in the
 range  $\mu$ to $\mu+d\mu$, having fixed the  partitioning of the
 measure  with boxes of size $\epsilon$. By  changing variables and
 using Eq.\ref{mf4}, then the measure distribution (Eq.\ref{mf5})
 becomes
\be 
\label{f3} 
N_{\epsilon}(\mu) d\mu \sim
 \epsilon^{-f(\alpha(\mu))} \frac{1}{|\log(\epsilon)|}
 \frac{d\mu}{\mu}. 
\ee
 From this equation it follows  that  the
 distribution of the  measure, at fixed resolution $\epsilon$, does
 not scale as a power law in  $\mu$, because the exponent
 $f(\alpha(\mu))$  is a complex function of $\mu$. The
 self-similarity of the distribution is recovered  by looking at
 the measure distribution as a function of the scale  $\epsilon$. 

  Suppose we fix the box dimension    at the scale  $\epsilon$:
 for example, we can suppose that  this can be the galactic scale,
 or the cluster  scale. The function $N_{\epsilon}(\mu)$ is 
 bell-shaped and convex  with a maximum corresponding  to the point
 at which 
\be  
 \label{f4} 
\frac{\partial
 N_{\epsilon}(\alpha)}{\partial\mu} =  \frac{\partial
 N_{\epsilon}(\alpha)} {\partial\alpha}\frac{\partial
 \alpha}{\partial\mu}  = -(f'(\alpha)+1) N_{\epsilon}(\alpha)
 \frac{1}{\mu}  = 0
 \ee 
this condition corresponds 
\be 
\label{f5} 
\left( 
\frac{\partial f(\alpha) }{\partial\alpha}\right)_{\alpha_{c}} = -1.
 \ee
 The
 maximum of $N_{\epsilon}(\mu)$ fixes the most probable value of
 $\mu$.  Well beyond this maximum the function can be well fitted 
 by a power law.  In practice this is the only observable part of
 the measure distribution in the case of galaxies because the
 higher values of $\alpha$ correspond to the smallest galaxies that
 are not present in the sample \cite{cp92}. For still higher values of
 $\mu$ the function  shows an exponential-like  decay. The tail is
 fixed by the point at which the derivative Eq.\ref{f5} has a
 maximum.  This happens for $\alpha = \alpha_{min}$, namely at the
 value corresponding to the box which contains the maximum measure
 (i.e. the strongest singularity) 
\be 
\label{f6} 
\mu^{*}  \sim
 \epsilon^{\alpha_{min}} \;. 
\ee
 In order to compute the exponent
 characterizing the leading power law behavior  we study the
 derivative of $\log(N_{\epsilon}(\mu))$ with  respect to
 $\log(\mu)$. By performing the logarithmic derivative of
 Eq.\ref{f3} we obtain 
\be 
\label{f7}
 \frac{\partial \log
 (N_{\epsilon}(\mu))}{\partial\log(\mu)} = - \left( \frac{ \partial
 f(\alpha) } {\partial \alpha} +1\right) \; .
 \ee  
We can try to fit
 Eq.\ref{f3} with a power law function of $\mu$, plus an
 exponential tail. From Eq.\ref{f7}  we can define an effective
 exponent $\delta$, which depends  explicitly on $\mu$. This
 implies that the power  law approximation can be considered as a
 local fit  
\be  
\label{f8} \delta =- \left(\frac{\partial
 f(\alpha)}{\partial \alpha} +1\right). 
\ee 
This leads to $\delta
 = 0$ for $\alpha=\alpha_{c}$ and $\delta=$-1 for $\alpha_{0}$ such
 that $f(\alpha_{0})=D(0)$.  Locally we can expand $f'(\alpha)$ in
 power series of $\mu$, so that the measure distribution (MD)  of
 Eq.\ref{f3} in a certain range of $\mu$, is well fitted by a power
 law function  with a cut-off
 \be 
\label{f9}
 N(\mu) \sim
 \mu^{\delta} e^{-\frac{\mu}{\mu^{*}}}.
 \ee 
The exponent $\delta$
 depends on the shape of the derivative of $f(\alpha)$, as well as
 on the value of $\alpha$ around  which one develops $f(\alpha)$.
 We shall see,  with the help of computer simulations  which the
 value of $\delta$ is usually  in the range $[-2,-1]$ for a wide
 range of  $f(\alpha)$ spectra \cite{slp96}.

 \subsection{Numerical simulations and 
reinterpretation of luminosity segregation
in terms of multifractal} 
\label{lumspacnumsim}

In order to examine a more general case  
%As a next step towards reality
  we now  consider a    random multiplicative
 measure  generated by a random  fragmentation process using  $m$
 normalized generators $\chi^{(\gamma)}$ and $\gamma =1,..,m$  in
 the $d$-dimensional Euclidean space 
(Fig.\ref{fig104}). The analytical shape
 of the measure distribution can be computed  from the knowledge of
 $f(\alpha)$, determined by the Legendre transformation of
 $\tau(q)$ (computed from the  $\chi^{(\gamma)}$). It is
 interesting to see how a MF distribution  naturally leads to the
 various morphological properties which we have discussed in
 Sec.\ref{lumspacegal}.
 \bef 
 %\vspace{}
\epsfxsize 14cm 
\centerline{\epsfbox{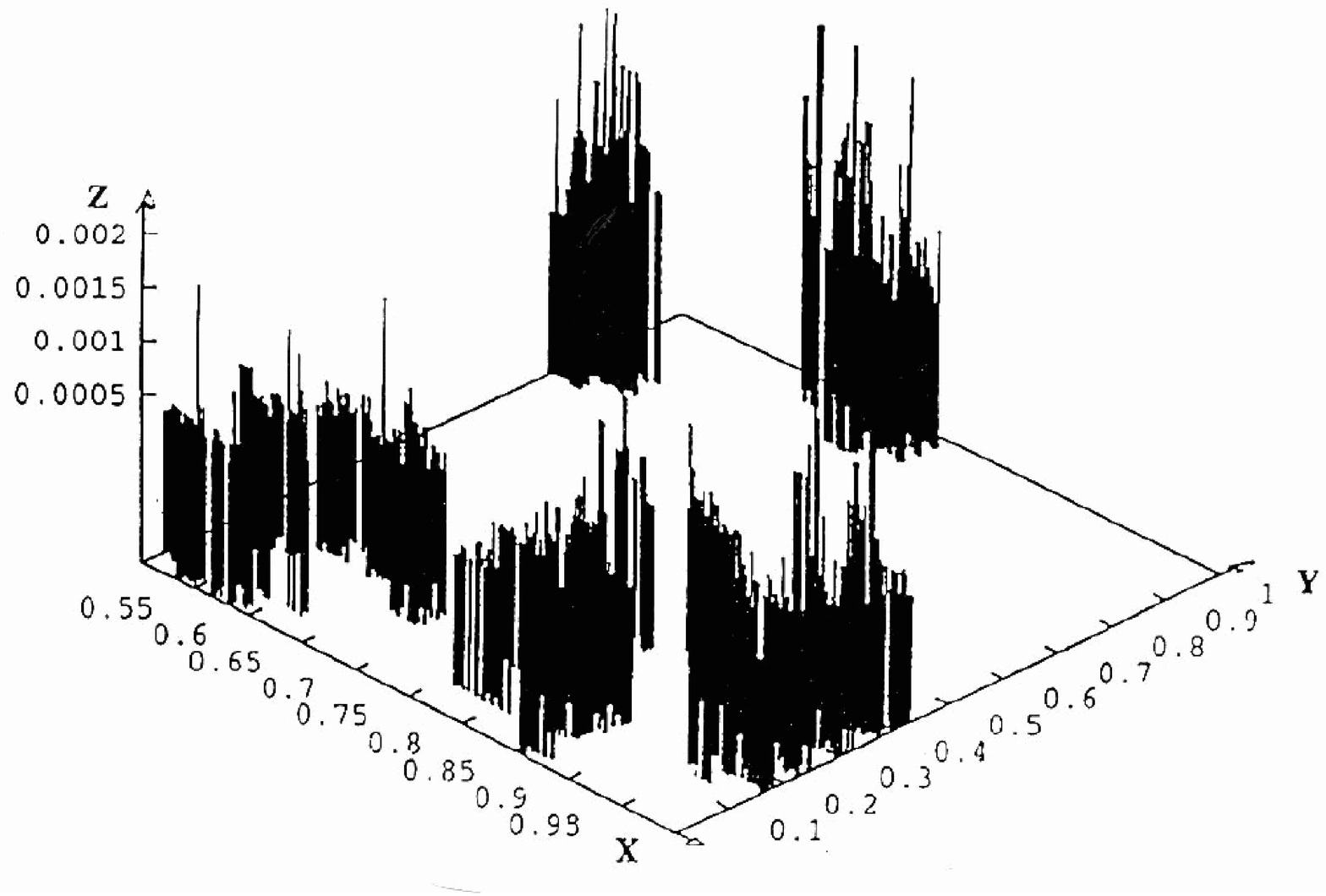}}
\caption{\label{fig104} A random multifractal obtained by
the $random-\beta-model$ algorithm in the two dimensional
Euclidean space. }
\eef
\bigskip  

   {\it (1) The two point number-number correlation
 function} 
\smallskip 

 We have seen that the characteristics of a
 MF distribution is  the $f(\alpha)$ spectrum of exponents. The
 relation of the properties of this distribution with the usual
 correlations analysis  is straightforward. The exponent that 
 describes the power law behavior of the  number density
 corresponds to the support of the MF distribution and  therefore
 it is related to the maximum of $f(\alpha)$  (see 
Sec.\ref{fairsample}).
 \bigskip   

{\it (2) Luminosity function} 
\smallskip  

The function $N(\mu)$ (Eq.\ref{f3}) for the random 
$\beta$ model of Fig.\ref{fig104} can be fitted by a power
law with $\delta =-1$ and with an exponential
cut-off \cite{slp96}.    
 Consider a  portion of
 a MF distribution in  a spherical volume of radius $R$.  The
 number of singularities of type $\alpha$  in the range $[\alpha,
 \alpha+ d \alpha]$, is given by  Eq.\ref{mf5}. The corresponding
 average space density scales therefore as  $<\nu(\alpha,R)> \sim
 R^{(f(\alpha)-3)}$. This relation implies  that the space density
 is a complex function  of $\alpha$ and $R$.  Although not strictly
 valid,  a separation between a space and a    luminosity
 distribution is  useful in the analysis of real catalogs. We
 approximate such a separation as follows. Integrating 
 Eq.\ref{mf5}  in $d\alpha$ we obtain that  the  total number of 
 boxes of size $\epsilon$, divided for the volume, is
\be  
 n(R)
 =  \frac{N(\epsilon(R))}{V} = \frac{3}{4\pi} B R^{D(0)-3} 
\ee
 where the last equality follows from the mass-length relation.

The probability
 that a certain box (at a given scale $\epsilon$) has the measure
 in the range ($\mu$, $\mu+d\mu$)  is determined by  Eq.\ref{f3}
 (or Eq.\ref{f7}). Hence   the average number of boxes for unit
 measure  and unit volume can be written as  
\be 
\label{new2}
 <\nu(R,\mu)> = \frac{3}{4\pi} B R^{D(0)-3} \mu^{\delta} 
 e^{-\frac{\mu}{\mu^*}} 
\ee
 This equation can be read as the {\it
 average} probability of having a galaxy of a certain luminosity
 and in a certain volume, in a MF Universe.   

The amplitude of
 $\nu(R,\mu)$ is related  to the lower cut-offs  of the
 distribution and    of the sample size, and therefore has
 no special physical meaning. We stress in what follows (see {\it
 (5)}) that the shape of the LF is not completely independent  on
 the size of the sample volume. This implies that the approximation
 of Eq.\ref{new2}  does not strictly hold, because it does not 
 consider the correlation between  space and luminosity
 distribution, which  for MF  we know to be an important feature 
 (see {\it (6)}). Nevertheless we stress that Eq.\ref{new2} can be
 used in practice  in the analysis of real redshift surveys, with
 great  accuracy. In fact, the result  of Eq.\ref{new2} has been
 obtained under the approximation of Eq.\ref{f9}, while in the more
 general case the cut-off $\mu^*$ depends explicitly on the sample
 size.  However such a behavior is very weak in the available
 samples.  

 In summary, in  order to study a MF distribution the
 three dimensional volume must be well defined:  only in such a
 kind of volume one may define the scaling properties of the MF.
 This implies that  an understanding of the distortions of the LF
 shape requires that   volume limited samples should be used,
 rather than  magnitude limited ones. Several authors, by analyzing
 the mean density in redshift surveys (i.e. the amplitude of the
 LF), concluded that  the samples are not large enough to be {\it
 fair} because the fluctuations are too large  \cite{dac94} \cite{mar94}.
 Our conclusion is that in
 magnitude limited  samples it is still possible to use the
 inhomogeneity-independent  technique to determine, if the galaxy
 distribution is MF, the  shape of the LF, but its is certainly not
 possible to recover the  amplitude of the LF, which is related to
 the space distribution  via the average density. To this end
 one should consider  a volume limited sample and normalize for the
 global luminosity selection which is related to such a sample
 (see Sec.\ref{corran}).
  \bigskip

{\it (3) Morphological and 
 luminosities properties: a new interpretation of
 "luminosity segregation"}
\smallskip  

Massive  galaxies are mostly
 found in rich clusters while field  galaxies are  usually spirals
 or gas rich dwarfs (see Sec.\ref{statmec}).  These observational properties 
 are consistent with multifractality, i.e. with the self-similar
 behavior of the whole  matter distribution (Fig.\ref{fig104}).
 The largest peaks are located in the largest clusters.  For
 the self-similarity each point of the structure belongs to a
 cluster or to a group of galaxies,  because a certain portion of
 the  fractal distribution is always made of smaller and smaller
 structures. Moreover the observations that the dwarf  and
 low-surface-brightness galaxies do not fill the voids,  is
 consistent with the fact  that the galaxy distribution continues 
 to be fractal  even for the lowest peaks of the MF. In this
 picture the giant-to-dwarf ratio depends on the environmental
 density. In fact, the dwarf galaxies can belong  to the rich
 clusters where the giants lie, but they can also be in small
 groups. The morphological-segregation can be seen  as the 
 self-similar character  of the matter distribution.
 Multifractality is a  description that can be useful for the
 statistical characterization  of the system, but, of course, it
 cannot explain     various evidences  which require a more
 appropriate morphological analysis.  

  From our result we can
 conclude that the fractal nature of galaxy distribution in the
 available samples, can account for the scaling of $r_0$ with
 sample depth. In this sense  the luminosity segregation, intended
 as a different clustering  properties of brighter and fainter
 galaxies in terms of the  amplitude of the standard correlation
 function (see Sec.\ref{corran} and Sec.\ref{powerspect})
has no experimental support. The linear dependence of $r_0$
 on the sample size can be completely  explained by the fractal
 nature of galaxy distribution.
 The correct perspective to describe the different
 clustering of brighter and fainter galaxies  is  the MF picture,
 for which we have given ample evidences,  and  which implies that
 massive  are mostly into large clusters  as observed. The
 quantitative characterization of such a phenomenon is therefore in
 terms of the exponent of the correlation function  rather than its
 amplitude. In particular the brighter galaxies should have a
 greater correlation exponent than the fainter ones (see {\it
 (4)}).
 \bigskip  

{\it (4) Multifractal spectrum and Multiscaling}
 \smallskip  

Previously we have introduced the MF spectrum
 $f(\alpha)$  and now we clarify its basic properties.
 Multifractality implies that if we select only the largest peaks
 in the measure distribution, the set defined by these peaks may
 have different fractal dimension  than the set defined by the
 entire distribution. One can define a cut-off in the measure and
 consider only those singularities which are above it. If the
 distribution is MF the fractal dimension decreases as the cut-off
 increases. We note that, strictly speaking, the presence of the
 cut-off can lead (for a certain well defined value of the  cut-off
 itself) to the so-called {\it multiscaling} behavior of the MF
 measure \cite{jan91}.
In fact, the presence of a
 lower cut-off  in the calculation of the generalized correlation
 function affects the single-scaling regime of $\chi(\epsilon,q)$
 for a well determined value of the cut-off $\alpha_{cut-off}$ such
 that $\alpha_{cut-off} < \alpha_c$,  and this function exhibits a
 slowing varying exponent proportional to the  logarithm of the
 scale $\epsilon$. However some authors \cite{mar95}
%(Martinez {\it et al.}, 1995) 
misinterpret the multiscaling of a MF distribution as the
 variation of the fractal dimension with the  density of the
 sample.   The fractal dimension $D$ of the support corresponds to
 the  peak of the $f(\alpha)$-spectrum  and raising the cut-off
 implies a drift of $\alpha$ towards $\alpha_{min}$ so that
 $f(\alpha)<D$. This behavior can be connected with the
 different correlation exponent found by the angular correlation
 function for the elliptical, lenticular and spiral  galaxies.

In particular the observational evidence is that the
 correlation exponent is higher for elliptical than for spiral
 galaxies: this trend is compatible with a  lower fractal dimension
 for the more massive galaxies than for  the smaller ones, in
 agreement with a  MF behavior. \bigskip   

 {\it (5) Scaling of the maximum mass} 
\smallskip 

 As shown in \cite{cp92} an important
 feature of the $f(\alpha)$  spectrum is represented by the value
 at $\alpha_{min}$:  $f(\alpha_{min})$. This exponent corresponds
 to the scaling of the  maximum singularity
 \be  
 \label{sc1}
 \mu_{max}(\varepsilon) \sim \varepsilon^{\alpha_{min}} 
\ee 
where
 $\mu_{max}(\varepsilon)$ id the maximum measure among all the
 boxes corresponding to a  gridding of size $\varepsilon$. The
 corresponding maximum density $\rho_{max}$ is therefore
 $\rho_{max} \sim \mu_{max}/\varepsilon^{3}$ where $\varepsilon
 \sim 1/R_s$ and $R_s$ is the total size of the system. Under the
 physical assumption that the maximum  mass $M_{max}$ is related 
to the maximum density one can conclude that:
 \be 
\label{sm2}
 M_{max} \sim R_s^{3-\alpha_{min}} 
\ee
 i.e. that the maximum galaxy
 mass we can observe in a certain sample is related to the size of
 the sample itself. To study  such an effect the sample size should
 be varied over a large range  of length scales. In practice the
 depth of the VL samples  which can be extracted from the available
 redshift surveys does not allow one to detect the scaling implied
 by  Eq.\ref{sm2} so that the inhomogeneity-independent   method in
  magnitude limited samples  remains the most suitable to study
the shape of 
 the LF.  
\bigskip  

{\it (6) Correlation between space and luminosity distribution} 
\smallskip   

 In \cite{slmp96}  we have performed a test to check the 
MF nature of the observed Perseus-Pisces catalog. 
We have randomized  the absolute magnitudes of the 
galaxies, i.e. we have fixed the galaxy position and we
 have assigned to each galaxy an absolute  magnitude 
chosen at random among all the other galaxies.  By doing 
this one destroys the correlations between the spatial
locations and magnitudes of galaxies.   As shown in
 Fig.\ref{fig103},
 in this case the behavior of $\:\tau(q)$ versus $\:q$ is linear,
 as for a simple
 fractal, with slope $\approx 2$. Instead in Fig.\ref{fig104} 
we show that the original distribution as a curved shape for $\tau(q)$. 
This result shows that the locations of galaxies  are intrinsically 
correlated with their luminosities, i.e. there exists   a luminosity-position
 correlation.  The MF framework provides a mathematical tool 
to study  such a correlation.   

\subsection{Theoretical implications} 
\label{lumspacgaltheo}

 In this section we have shown that it is possible to 
  frame  the main properties of the galaxy space and 
luminosity  distribution  in a unified scheme, by  using
 the concept of multifractality (MF). In fact, the
 continuous set of exponents $[\alpha,f(\alpha)]$ 
which describes a MF distribution can characterize 
completely  the galaxy distribution when one considers
  the mass (or luminosity) of galaxies in the analysis. In 
this way many observational evidences are linked together
 and arise naturally from the self-similar properties of the 
distribution.   Considering a MF  distribution, the usual 
power-law space correlation properties correspond just
 to  a single exponent of the $f(\alpha)$ spectrum: such 
an exponent simply describes the space distribution of 
the support of the MF measure.  Furthermore the shape
 of the luminosity function (LF), i.e. the probability of 
finding a  galaxy of a certain luminosity per unit volume, 
is related to the $f(\alpha)$ spectrum of exponents of the
 MF. We have shown that, under MF conditions, the LF
 is well  approximated by a power law function with an
 exponential tail. Such a function corresponds to the 
Schecther LF observed in real galaxy catalogs. In
 this case the  shape of the LF is almost independent
 on the sample size. Indeed we have shown that a 
weak dependence on sample size   is still present 
because the cut-off of the Schecther function  
for a MF distribution turns out to be related 
to the sample  depth: $L^*$ increases with
 sample depth. In practice as this quantity
 is a strongly fluctuating one, in order to
 study its dependence on the sample size
 one should have a  very large sample 
and should vary the depth over a  
large range of length scales. Given
 this situation  a sample size independent  shape of the LF can be
 well  defined using the inhomogeneity-independent method in 
magnitude limited  samples. Indeed such a technique has been 
introduced to take into account the highly irregular  nature  of
 the large scale galaxy distribution. For example  a fractal 
distribution is non-analytic in each point and it is not possible 
to define a meaningful average density. This is because the
 intrinsic fluctuations which characterize  such a distribution 
can be large as the sample itself, and the  extent of the 
largest structures is limited only by the  boundaries of 
the available catalogs.   Moreover if the distribution is
  MF, the amplitude of the  LF depends on the sample
 size as a power law function. To determine the amplitude 
of the LF, as well as the  average density, one should have
 a well defined volume limited sample,  extracted from  
a three dimensional survey (Sec.\ref{corran}).

These results have important consequences from a 
theoretical point of view. In fact, when one deals 
with self-similar structures the relevant  physical 
phenomenon which leads to the scale-invariant 
structures is characterized by the {\it exponent} 
and {\it not the amplitude} of the  physical 
quantities which characterizes such distributions. 

Indeed, the only relevant and meaningful quantity is the  exponent 
of the power law correlation function (or of the space density), 
while the amplitude of the correlation  function, or of the space 
density and of the LF, is just  related to the sample size and to 
the lower cut-offs of the distribution.  The geometric 
self-similarity has deep implications for the  non-analyticity 
of these structures. In fact, analyticity or regularity would 
imply that at some small scale  the profile becomes smooth 
and one can define  a unique tangent. Clearly this is impossible 
in a self-similar structure because at any small scale a new 
structure appears and the  distribution is never smooth.
 Self-similar structures are therefore intrinsically irregular 
at all scales and correspondingly one  has to change the 
theoretical framework into one which is  capable of 
dealing with non-analytical fluctuations. This means
 going from differential equations to something like 
the  Renormalization Group to study the exponents.
 For example the so-called "Biased theory of galaxy
 formation" \cite{kai84}
is implemented considering 
the evolution of  density fluctuations within an analytic 
Gaussian framework,  while the non-analyticity of fractal 
fluctuations  implies a breakdown of the central limit 
theorem which is the  cornerstone of Gaussian processes 
\cite{pie87,cp92,epv95,bslmp94}.

 In this scheme {\it the space correlations and the 
luminosity function are then two aspects of the 
same phenomenon, the MF distribution of visible matter}. 
The more complete and direct way to study such a  
distribution, and hence at the same time the space  
and the luminosity properties, is represented by the 
 computation of the MF spectrum of exponents. 
This is the natural objective  of theoretical investigation 
in order  to explain the formation and the distribution 
of galactic structures. In fact, from a theoretical point
 of view one would like  to identify the dynamical 
processes which can lead to such a  MF distribution. 

As a preliminary step in this direction we have  
developed a simple stochastic model 
\cite{slp95a,slp95b}
in order to study which are the fundamental physical effects which  lead to
 such a MF structure in an aggregation process. This is a very complex
 problem but to study it correctly one has to use the appropriate
 concepts and statistical tools.  If a crossover towards homogeneity 
would eventually be detected,  this would not change the above 
discussion but simply introduce  a crossover into it. The  
(multi)fractal nature of the observed structures would in 
any case,  require a change of theoretical perspective.

\section{Conclusions and theoretical implications}
\label{conclusion}
Cosmology, as a part of Physics, is an experimental science and, for
 this reason, all the assumptions and the relations of the various models
 must have a solid empirical basis. Since  twenty years, there has been
a very fast development of the experimental techniques in this field.
For example in 1980 only one thousand redshifts of galaxies were 
known, while now
there are available about 100 thousand and in five year the number
 of redshifts will be more than  one million. The possibility of studying
 the large scale 
structure of the Universe  in 
 various electromagnetic bands (from the radio and 
microwaves to the 
$\gamma$ and X rays) with satellite instruments, improved
 dramatically our knowledge of the Universe. However, despite
 this great observational effort, the most popular theoretical models,
 and in particular the Hot Big Bang theory and related
 galaxy formation models, encounter strong difficulties and
new observations requires  new "ad hoc" explanations and, 
often, the introduction of new free parameters.

From an experimental point of view, there are four main facts in cosmology
which are \cite{bslmp94}: 
i) the space distribution of galaxies 
and clusters, 
ii) the cosmic microwave 
background radiation (CMBR), 
iii) the linearity of the redshift-distance relation, usually 
known as the Hubble law,
iv) the abundance of light elements.

The theoretical problem is to explain these four 
independent experimental
 evidences through an  unique and self-consistent picture.
The current idea of standard Cosmology \cite{wei72,pee93} 
is  that in the observable Large
Scale Structure distribution,
isotropy and homogeneity do not apply to the Universe
   , but only to a "smeared-out " Universe
averaged over regions of order
$\:\lambda_{0}$.
One of the main problems
of observational cosmology is, therefore,
the  identification of $\:\lambda_{0}$,
The standard model {\it assumes} the homogeneity of matter
 distribution and deduces, 
from the solution of 
Einstein's field equations, the linearity of Hubble law.
Moreover, in 
the Hot Big Bang scenario, the CMBR is explained as 
the relict of a
primordial epoch, when the matter and radiation were
in thermodynamical equilibrium.
{\it The lack of an observational evidence in favor of 
homogeneity is therefore a crucial point for the standard
Big Bang scenario. } We briefly review in what follows the main 
theoretical implications of our results.

\subsection{Cosmological Principle}

The main cosmological theories are based on the assumption
of the homogeneity of matter distribution.
The reason is essentially the following.
The basic hypothesis of a post-Copernican Cosmological theory is that
{\em all the points} of the Universe have to be essentially equivalent:
this hypothesis is
required in order to avoid any privileged {\em observer}.
This assumption has been implemented
by Einstein in the so-called
 Cosmological Principles (CP): {\em all the positions}
in the Universe have to be essentially equivalent,
so that the Universe is homogeneous.
This situation implies also the condition of spherical
symmetry about every point, so that the Universe is also Isotropic.
There is a hidden assumption in the formulation of the CP with regard to
the hypothesis
that all the points are equivalent. Namely,
the condition that all the occupied points
are statistically equivalent with
respect to their environment corresponds to the property of
Local Isotropy. It is generally believed that the
Universe cannot be isotropic about every
point without being also homogeneous \cite{wei72}.
Actually, Local Isotropy does not necessarily implies homogeneity;
in fact a topology theorem 
states that homogeneity
is implied by the condition of local isotropy
together with {\em the assumption
of the analyticity or regularity} for the distribution of matter.
A fractal distribution, being locally isotropic around each point of
the distribution, is perfectly compatible with a weaker 
version of the CP \cite{man82,cp92,sl94}.
In addition this also implies that the tests
of dipole moment saturation are only tests of local
isotropy, but not homogeneity. We have shown
in fact that the dipole moment tends to saturate
  also in a fractal structure, depending on the morphological
features of the distribution,  while the monopole
moment grows as the power law of the sample size
\cite{sl94}. This is an important property of stochastic fractal
structures which plays an important role 
in the interpretation of the large scale peculiar
velocities.

 \subsection{ The Hubble - de Vaucouleurs Paradox}
\label{hdvpar}

The basic assumption of the Big Bang is 
the hypothesis that the universe is spatially homogeneous and isotropic \cite{wei72}.
Homogeneity of matter distribution plays the central role in 
the standard expanding universe model, 
because homogeneity implies that the recession velocity is 
proportional to distance \cite{wei72,pee93}.
This means that linear velocity-distance law 
$v_{exp}=Hl$ 
is valid at such distance scales where matter distribution can be 
considered on average as uniform.

   For a long time it was believed the deflections from homogeneity occur 
only inside "inhomogeneity cells" with size (about 5-10 Mpc)
corresponding to a few average distances between galaxies.
 Hence the homogeneity of the universe can be understood in the same sense 
as that of a fluid or gas, i.e. as a "smeared-out" universe averaged 
over cells of inhomogeneity.  Expansion of such a  cosmological fluid
 naturally obeys the linear velocity-distance relation.

From our results (see Sec.\ref{radial} for a summary)
 the size of the inhomogeneity cell has increased upwards to 
at least 
 150 $\hmp$ or more. Thus in the context of the standard model,
one would expect strong deflections from 
the linearity of the redshift-distance relation deep
 inside inhomogeneities \cite{hw72,fang91} 
while observations show an almost strictly linear 
Hubble law starting immediately beyond the 
Local Galaxy Group \cite{san95}. We have  explored the apparent 
contradiction between the linearity of the Hubble law and the 
inhomogeneity of galaxy distribution and discuss 
its implications for Cosmology in various papers \cite{bslmp94,bpslt96}
and here we briefly review our main results.\footnote{We warmly 
thank Y. Baryshev and P. Teerikorpi
for very useful discussions and collaborations 
in the study of the relation between the Hubble law 
and the inhomogeneous matter distribution.}

\subsubsection{Standard Model and the linear velocity-distance law}
\label{standmod}

   According to the Standard Model the universe is homogeneous, isotropic
 and expanding \cite{wei72,pee93}.
Homogeneity of matter distribution is the 
heart of the standard Cosmology because it allows one to introduce the space 
of uniform curvature in the form of the Robertson-Walker (RW) line element.
 This line element 
leads immediately to the strict linear 
velocity-distance law \cite{har93,pee91}. 
Indeed, the proper distance $l$, at cosmic 
time $t$, of a comoving body with fixed coordinate 
from a comoving observer is $l = R(t)r$. 
So the expansion velocity $v_{exp} = dl/dt$
defined as the rate of change of the proper distance $l$ is
\be
\label{ehd2}
v_{exp} = H l = c \cdot  l/l_{H} 
\ee
where $H = \dot{R}/R$ is the Hubble constant, $l_H = c/H$
is the Hubble 
distance.  In this way, the linear velocity-distance relation of 
Eq.\ref{ehd2} 
is a  rigorous consequence of spatial homogeneity
and valid for all distances $l$.  
In particular, for $l > l_H$, the expansion velocity $v_{exp} > c$.  
Such an expanding and curved space is consistent with 
special relativity locally and general relativity globally \cite{har93}.

In the expanding space the wavelength of an emitted photon is 
progressively stretched, so that the observed redshift $z$ is 
given by Lemaitre's redshift law
\be
\label{ehd3}
z = R(t_{obs})/R(t_{em})- 1
\ee
which is a consequence of the radial null-geodesic of
the RW-line element. For $z \ll 1$ (Eq.\ref{ehd3}) yields 
$z \approx dR/R \approx H_0 dt \approx l/l_H$, 
and from Eq.\ref{ehd2} one gets the approximate velocity-redshift 
relation which is valid for small redshifts
\be
\label{ehd4}
v_{exp} \approx cz 
\ee
It is important to emphasized that the 
expansion velocity-redshift relation 
differs from the 
relativistic Doppler
 effect.  So, the space expansion redshift
mechanism in the standard model is quite distinct from the usual 
Doppler mechanism.  We stress this, because homogeneity
alone does not tell which   is the cause of the redshift.

\subsubsection{Observed Hubble Law}
\label{hubbleobs}

In the famous paper \cite{hu29} 
 Hubble found a roughly linear
 relation between spectral line displacement 
$z =(\lambda_{obs}-\lambda_{em})/\lambda_{em}$, expressed according 
to the spectroscopic tradition as radial velocity $cz$, 
and distance $d$, which can be written as:
\be
\label{ehd5}
cz = H_0 d 
\ee
Hubble's 'radial velocity' was a technical term 
which did not refer to any interpretation of redshift, though he 
had in mind especially the de Sitter effect \cite{smith79}. 
Space expansion, Doppler mechanism, and de Sitter effect yield
at first order Eq.\ref{ehd5}.  Hence 
in case of motion
\be
\label{ehd6}
V \approx H_0 d
\ee
where $V$ is either space expansion velocity $v_{exp}$ 
or usual velocity, and 
$d$ is the distance measured by astronomer.  It has been 
natural to interpret 
this law as reflecting Eq.\ref{ehd2} and Eq.\ref{ehd4} 
 in the context of the standard 
Cosmology, and to regard the coefficient of 
proportionality $H_0$ in Eq.\ref{ehd5} 
as the present value of the theoretical Hubble constant $H$ 
from Eq.\ref{ehd2}.

Since its discovery, the validity of the Hubble law has been 
confirmed within an ever increasing distance 
interval where local and more distant distance indicators may be
tied together.  Recently, several new Cepheid distances have been measured 
to local galaxies, thanks to the HST programs \cite{free96,tamm96}.
Along with 
previous Earth-based Cepheid distances, 
methods like Supernovae Ia and 
Tully-Fisher have been better calibrated than
before \cite{tamm96}, 
and confirm the linearity with good accuracy up to 
$z \approx  0.1$ \cite{san95}.  Brightest cluster galaxies
trace the Hubble law even deeper, up to $z \approx 1$, and
radio galaxies have provided such evidence at still larger
redshifts \cite{san95}.

It is well known that there are small deviations $\delta V$ from
the Hubble law, connected with local mass concentrations such as
the Virgo Cluster, and, possibly the Great Attractor.

However, these perturbations are still only of the order
$\delta V/V_H \sim 0.1$, while in  the general 
field the Hubble law has been suggested to be quite smooth,
with $\delta V$ around $50 km/s$ \cite{san95}.

There have been proponents of global non-linearity of the 
Hubble law \cite{devac1,segal76}. 
However, it has been convincingly shown that such
 deviations from linearity are easily produced by the Malmquist bias in 
magnitude limited samples \cite{tee75,son79,san95} and when 
one by-passes this problem using suitable methods, the linearity is recovered.
The second main advancement is concerned with measurement of the value
of the Hubble constant $H_0$. Since Hubble's $559 km/s/Mpc$, 
$H_0$ has decreased and according to the most
recent studies  \cite{tamm96}
 utilizing the increased set of Cepheid-calibrators, 
seems to be stabilized at about $55 km/s/Mpc$.

\subsubsection{The Paradox}
\label{paradox}

   The Hubble and de Vaucouleurs laws describe very different 
aspects of 
the Universe, but both have in common universality and 
observer independence.  
This makes them fundamental cosmological laws.  
Hence, it is important to compare the range of distance scales where
they exist.  In Fig.\ref{fig105} we display these laws together.
\bef 
 %\vspace{}
\epsfxsize11cm 
\centerline{\epsfbox{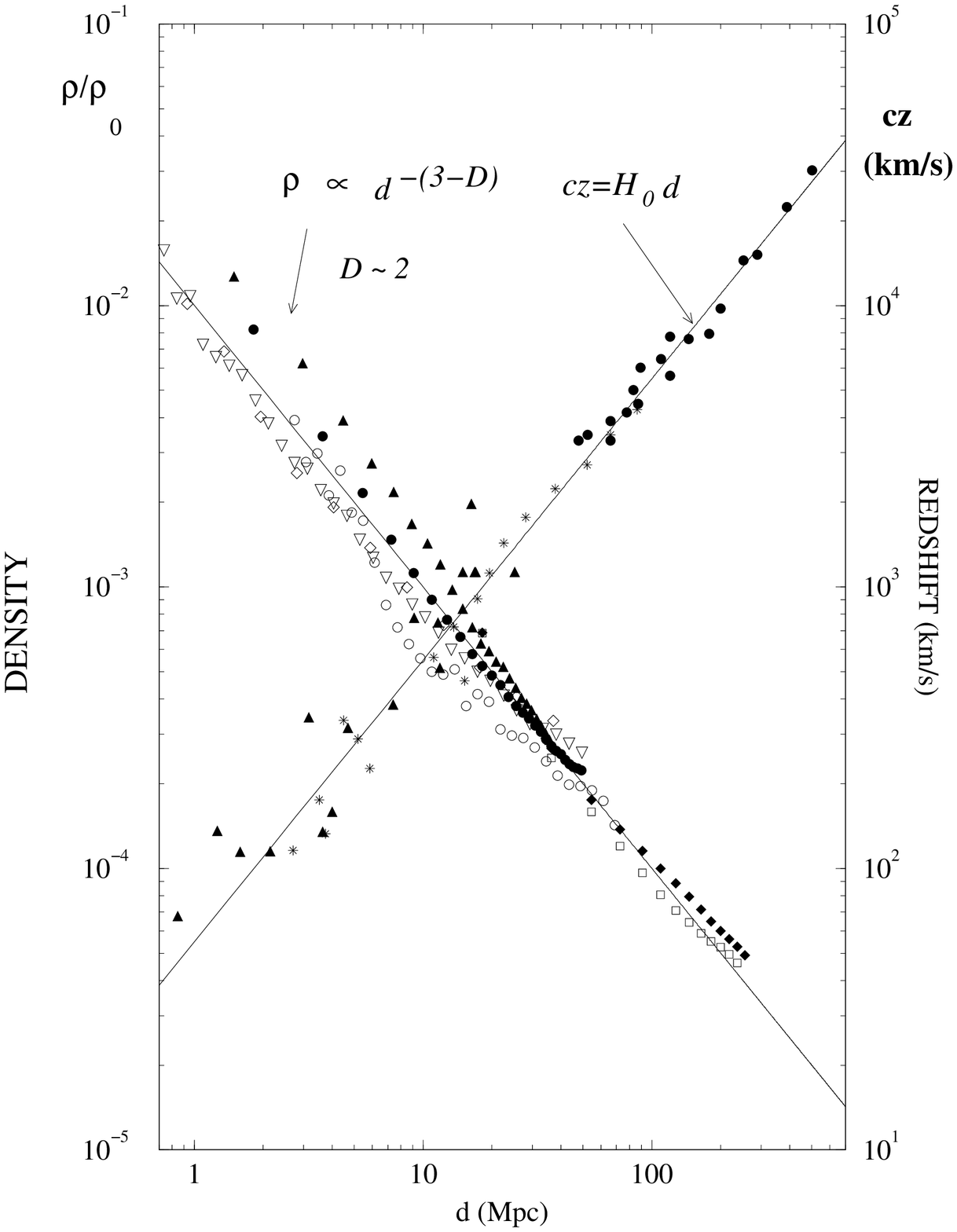}}
\caption{\label{fig105} Hubble 
redshift-distance and de Vaucouleurs density-distance 
laws in the distance scales from $1$ to 
$500 Mpc$
 ($H_0=55 km/s/Mpc$). 
 The Hubble law (increasing from left to right) is constructed
from: galaxies with Cepheid-distances for $cz > 0$ (triangles), 
galaxies with Tully-Fisher
(B-magnitude) distances (stars), galaxies with SNIa-distances for $cz > 3000
km/s$ (filled circles).
 TF-distance points are generally averages of a few tens of galaxies
from the "unbiased plateau" of the method of normalized distances.
Redshift $cz$ is reduced to the Local Group center and contains
the small correction due to the Virgo infall velocity field.  The solid
line  corresponds to the Hubble law with  $H_o = 55 km/s/Mpc$.
The de Vaucouleurs law (decreasing from left to right) in the normalized
form is constructed from 
the computation of the conditional average density.
The dotted line  corresponds to the de Vaucouleurs law with
 correlation exponent $\gamma = 1$, i.e. $D=2$.
}
 \eef

  The puzzling conclusion from Fig.\ref{fig105} is that the strictly linear 
redshift-distance relation is observed deep inside the 
fractal structure.  This empirical fact presents a profound
challenge to the standard model where the 
homogeneity was the basic explanation of the Hubble law, and "the 
connection between homogeneity and Hubble's law was the first success of 
the expanding world model" \cite{pee91}. 
This also reminds of the natural reaction 
 "In fact, 
we would not expect any neat relation of
proportionality between velocity and distance 
[for such close galaxies] \cite{wei77}. 
However, contrary to the expectations, modern data show a good linear 
 Hubble law even for nearby galaxies.

\subsubsection{Implications}
\label{implications}

   We 
discuss shortly the cosmological significance of
 the HdeV diagram, using a minimum of assumptions.  Especially, 
we illustrate its importance by discussing two opposite
possibilities, depending on which is more fundamental law in 
Fig.\ref{fig105}.  There may be a homogeneously
distributed dark matter, producing the linear Hubble law, while
the fractal structure is just confined to the less important
luminous matter.  Or, the Universe may be dominated, in terms of mass, 
by the fractal structure, observed for the visible matter.

\subsubsection*{Alternative I: Homogeneous dark matter}

The observed scale-invariant density-distance law is related to the 
luminous matter, i.e. the stars of the galaxies. Because 
the fractal distribution is inhomogeneous at all scales, this 
implies within the standard 
model that there must  be something
other than visible galaxies which is uniformly distributed in the 
universe.  
This could well be a homogeneous substratum of dark matter.  
Of course, this is not the first time  when a dark mass component
is proposed from other considerations, however
we emphasize the simple argument following from Fig.\ref{fig105}.

Assume that the Universe contains a large amount of its mass in the 
form of some unknown dark matter which has always filled the space 
uniformly and carried at rest in the comoving system 
the primordial seed density 
fluctuations around which later were
to form the luminous galaxies.  Then only at scales where the 
homogeneous dark mass density presently dominates over the 
luminous, inhomogeneous 
mass, the linear Hubble law can be observed.  In other words, only
in scales where the matter distribution looks homogeneous, one can speak 
about the Hubble law.
This simple argument is quite strong: a dark
matter component coupled to galaxies is not sufficient.  One needs
a truly homogeneous substratum, both at small scales in view of the good 
Hubble law, and at very large scale where one already
starts approaching the Hubble length scale.

The argument about the required dominance of the homogeneous
substratum can be made more quantitative within the 
linear perturbation theory which predicts that in the scale where 
the Hubble velocity is $V_H$ and density perturbation is $\delta \rho$,
the peculiar velocity $\delta V$ has attained a value
\be
\label{pek1}
\delta V/V_H = \Omega^{0.6} \delta \rho / \rho_0
\ee
where $\rho_0$ is the mean cosmological density. E.g., for 
$\Omega =1$, it is clear that small $\delta V / V_H$ 
(good Hubble law) needs small $\delta \rho / \rho_0$.
Because at small scales the fluctuations of the inhomogeneous 
galaxy distribution are comparable to the average density 
$<\rho_{gal}>$, the uniform dark mater substance must have density 
high enough  and fluctuations small enough in 
order to reduce $\delta \rho / \rho_0$ to its small value
suggested by $\delta V/V_H$. This means in practice 
that the dark matter density must not be less than the galaxy density in 
the scale concerned.

With $\Omega <1$ the mean density is less than the critical one,
which more than compensates the influence of the factor $\Omega^{0.6}$
in Eq.\ref{pek1} ($\delta \rho$ coming from galaxy 
distribution remains the same), and the above argument is still valid.

How smooth is this dark substratum and what is its density,
 or at what minimum scales it 
may be regarded as homogeneous and dominant?

Because the Hubble law (in Fig.\ref{fig105}) for the field galaxies 
around the Local Group (LG) starts around $1 \div  2 Mpc$,
this gives the scale where the dark substratum is already 
close to homogeneous and
dominant.  Also the detailed analysis of the velocity field of
 the LG galaxies in \cite{sandage87} showed the existence
of the expansion down to $\sim 1.5 Mpc$ from the LG.  Additional,
quite strong evidence for such a small size of the scale where the Hubble
law still exists, is given by the very small internal dispersion
 ($\sigma<60$km/s) inferred from the Hubble diagram
 in the local velocity field \cite{san95}.
The scale should be less than $\sigma/H_{o}$,
as one may see by considering the line of sight
traversing cells where the Hubble law collapses.

 In order to estimate the required minimal density of
 the uniform dark matter, we use two estimates of 
the mass tied to LG galaxies.  From the rotation curves the
masses of the Galaxy and M31 yield \cite{zheng91} 
the sum $\approx 6.3 \cdot 10^{11} M_{\bigodot}$.
 If one smooths out this mass in the volume $4 \pi /3 (1.5Mpc)^3$ 
one finds an average density $\rho(r) \approx 0.4 \cdot 10^{-29} g cm^{-3}$.
This value should be exceeded by the density of the uniform dark matter.
As the critical density for the standard model 
is $\rho_{crit} = 1.9 \cdot 10^{-29}h^{2} g cm^{-3}$, this means that the
density parameter for dark matter $\Omega_{dark} > 0.6$ for $h = 0.55$. 
The second mass estimate comes from dynamical studies of the LG yielding
\cite{sandage87,valt93}  the combined mass $\approx 3 \cdot 10^{12} M_{\bigodot}$.
  Similarly, this gives $\Omega_{dark} > 3$.

  The first lower limit $\Omega_{dark} > 0.6$ is interestingly close to
the critical value $\Omega = 1$.  The dynamical limit $\Omega_{dark} > 3$
seems to contradict the available data for the classical cosmological tests
\cite{san95}.

\subsubsection*{ Alternative II: Matter is tied to galaxies}

\subsubsection*{(a) Spherical symmetric inhomogeneities}

Instead of looking at the HdeV paradox with the aid of the 
dark substratum, one may choose to
 investigate the implications of inhomogeneous 
cosmologies.  Spherical symmetrical inhomogeneity has 
been used \cite{hw72,fang91} to represent isotropic fractal, 
with the conclusion that there is a strong 
deflection from linear Hubble law
 inside a fractal inhomogeneity cell.  The Lemaitre-Tolman-Bondi
 (LTB) solution of Einstein equations was 
utilized \cite{rib93} for modeling a relativistic fractal Cosmology, and it 
was concluded that it is difficult to get
a linear redshift-distance law.

We   stress that, in order to be
solvable, these models introduce one or more privileged points that
are the center of the LTB metric. This breaks the symmetry of the problem
and one models the fractal structure with an analytical density behavior.
In these models the Hubble law is just an accident: this 
probably occurs because the
isotropy of the distribution is broken rather than because a fractal
is really incompatible with linearity of Hubble law. In fact, one
may conjecture that the Hubble law is a consequence of isotropy rather than of
homogeneity. Clearly this situation (non analytical matter distribution) has
never been studied and it requires very complex models to be analyzed.

   To save the linearity of the redshift-distance relation it 
was suggested \cite{mof92} that 
our Galaxy is situated in a large void. 
However, this would put us in a privileged position, while a typical 
place of a galaxy, belonging to the fractal structure,
 is within a fractal cluster \cite{sl94}.  
So, in the light of 
the LTB model the Hubble law is an accidental or luckily observed 
phenomenon of the universe.

\subsubsection*{(b) Fractal universe and cosmological gravitational redshift}

One should not ignore the alternative 
possibility that the de Vaucouleurs law
could be fundamental for Cosmology \cite{devac1}.
Similarly as the Cosmological Principle of homogeneity 
predicts Hubble law in the standard model, 
it is tempting to ask whether the more general fractal distribution can 
predict Hubble law in some other context.

   Even though such a Cosmology has not been yet developed, 
one can safely say that a Hubble law due to expansion is not expected,
because of the global non-uniformity.  If the linear Hubble
 law is, however, observed, its origin must be dominated by
 something other than expansion.

   The only known mechanism, other than Doppler effect and space expansion,
to produce redshift is gravitational effect.  Bondi \cite{bon47} 
demonstrated that cosmological redshift is due to two causes:
relative motion and gravitation effect.  The spectral shift depends
 on the relative motion of source and
 observer, and also on the global matter distribution.
  In fact, it was proposed \cite{haw62} 
that within a static homogeneous Einstein model there 
is the gravitational redshift $z$ which is 
proportional to the square of the distance.

  It should be emphasized that the gravitational 
part of the cosmological spectral shift is redshift  \cite{bon47,haw62} 
and not blueshift, as it has been supposed \cite{zn84}. 
This choice follows from a
 causality argument: the emission of 
a photon precedes its absorption by observer's light detector. 
It is also a strict consequence 
of the general Mattig's relation 
between proper distance and redshift \cite{bslmp94}.

   It has been pointed out \cite{bar81} that while in the homogeneous case
the gravitational redshift increases as the square of distance, 
 inside a hierarchical structure with
correlation exponent $\gamma = 1$, 
the redshift 
is linearly proportional to distance, producing the Hubble $(z - d)$
 relation.  Indeed, if $\rho(d) \sim 1/d$, 
then mass $M \sim d^2$, 
and gravitational potential 
$\phi \sim M/d \sim d$. 
So the gravitational redshift $z \sim \phi/c^2 \sim d$.
   In this case the value of the Hubble constant is 
determined by 
\be
\label{ehd8}
H_{grav} = 2 \pi G \rho_0 d_0/c
\ee
where $\rho_0$ and $d_0$ are the density and radius of the lower 
cut-off in the fractal structure, $G$ 
is the gravitational constant.  
If the product $\rho_0 d_0 = 1/2\pi g cm^{-2}$ 
is valid in the fractal universe, then $H_{grav} =  G/c = 69 km/s/Mpc$.  
It is an interesting fact 
that for an average galaxy with 
$\rho_0 = 5 \cdot 10^{-24} gcm^{-3}$ and $d_0=10 kpc$ 
this condition is fulfilled.
  Naturally, this explanation of the paradox is based on a non-standard
approach to the cosmological redshift \cite{bslmp94}.

\subsection{The Cosmic Microwave Background Radiation}
\label{cmbr}

The best known background radiation is the
Cosmic Microwave Background Radiation 
(CMBR) \cite{pw65}. After Relict and COBE experiments, 
the CMBR ani\-so\-tro\-pi\-es and spectrum are well
known \cite{mat90,mat94,gus90,fix94}. The perfect thermal Planck spectrum
 of the CMBR and its very small
anisotropies 
are among the most important 
observational cosmological facts.

Few groups have detected CMBR anisotropies at various angular scales,
while many other have obtained upper limits only. The amplitude
of these fluctuations, after having taken into account the 
various spurious contributions,
is of the order of some $\:\Delta T/T \sim 10^{-5}$. However
 it is very hard to decide if the signals observed by Relict 1
 \cite{stru92}, %(Strukov et al., 1992),
 by COBE \cite{smoo92} and other groups,
are CMBR anisotropies or a mixture of CMBR and spurious signals. 
These results pose some trouble for the  Standard  Model. In fact, 
in the standard scenario of galaxy formation, small amplitude primordial 
perturbations to the energy density are amplified by gravitational instability
as the Universe expands and they are predicted to 
leave a detectable imprint on the CMBR. Such low
amplitude measured fluctuations are not 
compatible with the standard baryonic matter 
and hence in  many theories
of galaxy formation one must introduce some 
exotic kind 
of non baryonic matter, that 
has a weaker interaction with photons 
than the baryonic one.
Such non baryonic dark matter
should 
contribute to the density  of the universe, with more 
than $95 \%$ of the total density.
 Moreover, on large angular scales (few degrees),
the dominant mechanism by which 
density fluctuations induce anisotropy in the 
CMBR is the Sachs-Wolf effect  \cite{sw67}.
 This is a gravitational effect 
due to the presence of matter between the CMBR and the observer. 
For this reason, the existence
of large-scale structures can represent  a 
serious problem due to their 
incompatibility (in the standard scenario)
 with such low amplitude
temperature fluctuations \cite{bslmp94}. 

An important point in this discussion is represented by the isotropic nature
of fractal structures and in particular it is important to notice that 
the presence of highly isotropic distribution of 
objects as optical galaxies, radio-galaxies, and so on is {\it compatible} with 
an inhomogeneous and scale invariant distribution in the three
dimensional space. However a complete discussion of the CMBR goes 
beyond the scope of this review.

\subsection{Galaxy bulk flows}
\label{bulk}

An important kind of large-scale measurement of 
motions is the so-called "bulk" flow which is the estimate
of the net velocity $\:<V>$ of galaxies within a large volume.
There are evidences \cite{cou93}
%(Courteau et al., 1993) 
of large-scale coherent flows, and 
  a coherent motion on very large 
volume ($\:\sim 150 h^{-1}Mpc$) has been found, 
that implies an increase of $\:<V> $
on scales much larger than that of the GA.
If the CMBR dipole is due to Doppler effect then the 
Local Group (LG) has a velocity of $\:620 km sec^{-1}$ 
with respect to the
rest frame represented by the CMBR. The origin of this motion can be 
due to some anisotropic mass fluctuation whose amplitude is expected 
(in an homogeneous picture) to decrease with increasing scale. Hence the bulk 
flows of galaxies contained in very large volume should be at rest with 
respect to the CMBR. The results of Lauer \& Postman \cite{lp94}
show, on the contrary, that the LG motion relative to
 an Abell clusters sample
is inconsistent with the velocity of the LG inferred form the CMBR dipole,
and imply that the CMBR dipole anisotropy
 is generated by a very large 
scale mass concentration 
beyond $\:100 h^{-1} Mpc$.
Moreover Mathewson \etal \cite{mat96}
found that there is not backside infall
into the GA and that there is evidence of a bulk flow of $\:600 km sec^{-1}$
in the direction of the GA on  a 
scale at least $\:60 h^{-1} Mpc$. 
Recently  Mathewson \& Ford \cite{mf94}
found that the flow is not uniform over the GA region and that it seems
to be associated with a denser region which participates in the flow too.

From the data now available it 
emerges that the {\it full extent of the galaxies flows
is still uncertain} and not detected and the origin of
these large amplitude and coherent length peculiar motions
is very unclear in the standard scenario. In fact 
it is very hard to reconcile these results
with an homogeneous picture,
in which the bulk flows have to be small on large scales
as the mass fluctuations have to be of small amplitude on the 
large scale.
On the contrary, in a fractal distribution,
which is intrinsically inhomogeneous,  there are fluctuations
in the distribution of mass at all scales; in this case 
large-scale coherent flows are limited only by the 
the property  of  local isotropy that 
characterized a fractal structure and implies that the dipole 
(the net gravitational force) 
saturates beyond a certain scale \cite{sl94}.

\subsection{Discussion}
\label{concltheo}

In summary our main points are:

\begin{itemize}

\item
 The highly  irregular galaxy distributions with large structures and 
voids strongly point to a new statistical approach in which the 
existence of a well defined average density is not assumed a priori and 
the possibility of non analytical properties should be addressed 
specifically.

\item
 The new approach for the study
 of galaxy correlations in all the available  catalogs 
shows that their properties are actually compatible with each other 
and they are statistically valid samples. The severe discrepancies 
between different  catalogs which have led various authors to consider 
these  catalogs as {\it not fair}, were due to the inappropriate methods of 
analysis.

\item

 The correct two point correlation analysis shows well defined fractal 
correlations up to the present observational limits, from 1 to 
$1000\hmp$ with fractal dimension $D \simeq 2$.
Of course the statistical quality and 
solidity of the results is stronger up to 
$100 \div 200 \hmp$ and 
weaker for larger 
scales due to the limited data. It is remarkable, 
however, that at these larger scales one observes exactly the continuation
of the correlation properties at the small and intermediate scales.

\item
 These new methods have been extended also to the analysis of the 
number counts and the angular  catalogs, which are shown to be fully 
compatible with the direct space correlation analysis. The new analysis of 
the number counts suggests that fractal correlations may extend also to 
scales larger than $1000\hmp.$

\item
The inclusion of the galaxy luminosity (mass) leads to a distribution 
which is shown to have well defined multifractal properties. This leads 
to a new, important relation between the luminosity function and that 
galaxy correlations in space.

\item
{\it New perspective on old arguments.}
On the light of these results we can now take 
a standard reference volume in the field (i.e. \cite{pee93})
and consider 
 the usual arguments invoked for homogeneity from a new point
of view. 
These arguments are: {\bf (a)} number counts: we have seen in Sec.7 that 
the small scale exponent of number counts is certainly not related to 
homogeneity but to small scale fluctuations. The real exponent of the number 
counts is instead the lower one (i.e. $\alpha \approx 0.4$) that indeed 
corresponds to the three dimensional 
(i.e fractal with $D \simeq 2$) 
correlation properties. {\bf (b) }
$\delta N/ N$ is
small for various observations. This point is exactly the same
as the fact $r_0$ is a spurious length. In the absence of 
a reference average one cannot talk about "large" or "small" 
amplitude of fluctuations. In addition, for any distribution, even
 a fractal one $\delta N/N$ is always small for sizes comparable
to the total sample because the average is computed from the sample itself.
{\bf (c)} Angular correlations. We have seen in Sec.7
 that angular correlations
are ambiguous in two respects: 
first the angular projection of a fractal is really
uniform at large angles due to projection effects, 
second the angular data are strongly affected by the finite size 
fluctuations which provide an additional artificial homogenization, as in the 
case of the number counts. The inclusion of these effects reconciles quite
 naturally the angular catalogs with the fractal properties in the three
dimensional ones. 
{\bf (d)} X-ray background. The argument that $\delta N/N$ becomes very small
for the X-ray background combines the two problems discusses before:
angular projections and reference average. This angular uniformity 
is analogues, for example, to the Lick angular sample, 
and certainly is
not a proof of real homogeneity. 

Finally one should note that all these arguments are {\it indirect}
and always require an interpretation based on some assumptions.
The most {\it direct} evidence for the properties of galaxy distribution 
arises from the correct correlation analysis of the 3-d volume limited samples
that has been the central point of our work.

\end{itemize}

\subsection{Theoretical Implications}
\smallskip

From the theoretical point of view the fact that 
we have a situation characterized by {\it self-similar structures} 
 implies that we should not use concept
 like $\xi(r)$, $r_0$, $\delta N/N$ and certain properties of
the power spectrum, because they are not suitable to represent 
the real properties of the observed structures. 
To this end also the N-body simulations
should be considered from a new perspective.
One cannot talk about "small" or "large" amplitudes 
for a self-similar structure because of the lack of a reference value like the 
average density.
The Physics should shift from {\it "amplitudes"} towards {\it "exponent" }
and the methods of modern statistical Physics should be adopted.
This requires the development of constructive interactions between two fields.
\bigskip

{\it Possible Crossover.}
We cannot exclude of course, that visible matter 
may really become homogeneous at some large scale not 
yet observed. Even if this would 
happen the best way to identify the
eventual crossover is by 
using the methods we have described and  not the usual ones.
From a theoretical point of view  
the range of fractal fluctuations, extending at least over
three decades ($1 \div 1000 \hmp$), should  anyhow be
addressed with the new theoretical concepts.
Then one should study the (eventual) crossover to homogeneity 
as an additional problem.
For the moment, however, no tendency to 
such a crossover is detectable from the experimental 
data and it may be reasonable to consider also more radical
 theoretical frameworks in which homogenization may 
simply not exist at any scale, at least for  
luminous matter.
\bigskip

{\it Dark Matter.} 
 All our discussion  refers to luminous matter. 
It would be nice if the new picture for the visible universe 
could reduce, to some extent, 
the importance of dark matter in the theoretical framework.
At the moment however this is not clear. We have two possible 
situations: {\it (i)} if dark matter is essentially 
associated to luminous matter, then the use of FRW metric is not
justified anymore.  This does not necessarily imply that there is 
no expansion 
or no Big Bang. 
It implies, however, that  these phenomena should be
described by more complex models. {\it (ii)}  If dark matter
 is homogeneous and luminous matter is fractal then, 
at large scale, dark matter  dominate the gravity 
field and the FRW  metric is again valid. 
The visible matter however remains self-similar and non analytical 
and it still requires the new theoretical methods 
mentioned before.

\subsection{Predictions for future surveys}
\label{predictions}

\begin{table} \begin{center}
\caption{\label{tabpre}
The volume limited samples of various 
 catalogs (not still published and analyzed) are characterized by the following 
parameters:
- $R_{VL} (\hmp)$ is the depth of the VL sample considered
with absolute magnitude limit $M_{VL}$
- $\Omega$ is the solid angle
- $R_{eff} (\hmp)$ is the radius of the largest sphere 
that can be contained in the  catalog volume. 
This gives the limit of statistical validity of the sample.
- $r_0(\hmp)$  is the length at which $\xi(r) \equiv 1$.
(distance are expressed in $\hmp$).
}
\begin{tabular}{|c|c|c|c|c|c|}
\hline
     &      &          &    &              &        \\
\rm{Sample} & $\Omega$ ($sr$) & $R_{VL} $ & $M_{VL} $ & $R_{eff} $& $r_0$ \\
     &       &    &    &    &  		         \\
\hline
CfA2  & 1.83  &  101 & -19.5   & 22   & 7    \\
%      &       &      &         &      &      \\
CfA2  & 1.83  &  160 & -20.5   & 36  &  12   \\
 %     &       &      &         &      &      \\
SLOAN & $\pi$ &  400 & -19     & 185  &  60  \\
  %    &       &      &         &      &      \\
SLOAN & $\pi$ & 600  & -20   &  275 &  90 \\
    %  &       &      &         &      &      \\
2dF (South)   & 0.28  &  550 & -19   & 50   &   15  \\
   % &       &      &         &      &      \\
2dF (South)   & 0.28  & 870  & -20   & 100  &  30   \\
      &       &      &       &      &      \\
\hline
\end{tabular}
\end{center} \end{table}

According to the standard interpretation, the length 
$r_0 \simeq 5 \hmp$ characterizes the physical properties of 
galaxy distributions. Therefore deeper samples like CfA2 and SLOAN
should simply reduce the error bar, which is 
now about considered 
to be $10 \%$. A possible variation of $r_0$ with absolute magnitude,
due to a luminosity bias,  is 
considered plausible but it has 
never been quantified. This should be checked by
varying independently absolute magnitude and depth of the volume limited samples.
However, from this interpretation, the value of $r_0 = 5 \hmp$, 
corresponding to a volume limited of CfA1 with $M=-19.5$, should not change 
when considering in CfA2 and SLOAN volume limited samples with the 
same solid angle $\Omega$ and the same absolute magnitude limit ($M=-19.5$).

In our interpretation, instead,  $r_0$ is  spurious, and it scales 
linearly with the radius $R_s$ of the largest sphere fully contained 
in the volume limited samples. Therefore we predict for the
volume limited sample of CfA2 with $M=-19.5$ (with a solid angle of 
$\Omega=1.1 \; sr$ \cite{par94}) $r_0 \approx 7 \hmp$ (if, in the final 
version of the survey the solid angle is $\Omega =1.8$, 
the value of $R_s$
 increases accordingly, and the value of $r_0$ is shifted up 
to $\sim 9 \hmp$).
Note however that for the deepest 
volume limited CfA2 sample ($M \ltapprox -20$) we 
predict instead $r_{0} \approx 15 \div 20 \hmp$.
For the volume limited sample of the 
full SLOAN with $M=-19.5$
 ($\Omega = \pi$), 
our prediction is
that $r_0 \approx 65 \hmp$. It is clear that 
however, the first SLOAN slice   gives smaller values 
because the solid angle is be small.
In Tab.\ref{tabpre} we report the predictions for $r_0$ in the next future 
surveys.

\section*{Acknowledgements}
Various parts of this work have been done in collaboration with
L. Amendola, A. Amici, Yu.V. Baryshev, P. Coleman, 
H. Di Nella, A. Gabrielli, B. Mandelbrot, P. Teerikorpi
and we  thank them very warmly.
We also thank for useful discussions, comments, criticisms 
and  suggestions 
 D.J. Amit, T. Buchert, A. Cappi, L. Da Costa,  C. Di Castro,
M. Davis, R. Durrer, J-P. Eckmann, R. Giovanelli, M. Haynes, 
L. Guzzo, E. Lerner, F. Melchiorri, M. Munoz,  
 G. Parisi, G. Paturel, J.P.E. Peebles, C. Perola, D. Pfenniger,  
 M. Ribeiro, G. Salvini,  G. Setti, 
 N. Turok, F. Vernizzi, P. Vettolani, H. Wagner,  and G. Zamorani. 
This work has been partially supported by the Italian Space Agency (ASI).
F.S.L is particularly grateful to R. Durrer and J.-P. Eckmann for their kind 
hospitality.

\newpage

\section*{Appendixes}
\label{appen}
In the following we report the details of the various 
samples we have analyzed in this paper. 

\subsection*{CfA1}
The sample contains 1845  galaxies with 
$m_{ph} \le 14.5$  in the northern hemisphere in a
region of $1.83 \; sr$ \cite{huc83}. We have extracted 
same VL sample, and we report in Tab.\ref{tab1cfa1} 
the characteristics. The noticeably high density area 
in the center of the catalog
is the nearby Virgo cluster.

\subsection*{Perseus-Pisces}

The
{\it Perseus-Pisces}
redshift survey  collects the positions and
the redshifts for the galaxies in the region
$22^{h}< \alpha <4^{h}$ and $\:0^{\circ}<\delta <
45^{\circ}$ \cite{hg88}. %(Giovanelli, \& Haynes 1988).
The survey consists mainly of highly accurate
{\it 21 cm} H I line redshifts.
The radio data are complemented with optical observations
 of early-type galaxies carried out at the 2.4 m telescope
of the MacGraw-Hill Observatory, plus a number of redshifts
provided by J.Huchra and other smaller sources in
the public domain.
The catalog used comprises those redshifts
obtained before 1991 December, for a total of 5183 galaxies.
Among them, 3854 have Zwicky magnitudes of 15.7 or brighter.
From this sample, we have excluded the data in the region more affected
by extinction; hence
we have considered the data only in the ranges
$[22^{h}< \alpha <3^{h} 10^{m}]$ and $[0^{\circ}<\delta <
42^{\circ}30']$, with $m \leq 15.5$.

We have studied galaxies with
corrected velocity in the range \mbox{$0-13,000$} \mbox{$Km \dot sec^{-1}$}.
The apparent magnitudes of galaxies have been corrected for
the extinction, using the absorption maps produced by
Burstein \& Heiles \cite{gio86}.
%(Giovanelli et al., 1986).
The final
 sample,
with apparent magnitude less than $15.5$,
contains $N=3301$ galaxies (that we call hereafter
PP 15.5).
With these data, we have produced some VL subsamples
whose characteristic are reported in  
Tab.\ref{tab1pp} and Tab.\ref{tab2pp}

\begin{table} \begin{center}
\caption{\label{tab1cfa1} VL samples of CfA1. 
Apparent magnitude limit =14.5
Total number of galaxies =1843 (cut in absolute magnitude)}
\begin{tabular}{|c|c|c|c|}
\hline
     &      &         &            \\
{\rm Sample} & $d_{lim} (h^{-1}Mpc)$
& $M_{lim}$ & $N$ \\
     &      &         &            \\
\hline
%     &      &         &              \\
VL18 &  $31.3$ & $-18.00$ & $480$  \\
%     &      &         &            \\
VL185 &  $39.8$ & $-18.50$ & $424$  \\
 %    &      &         &            \\
VL19 &  $50.1$ & $-19.00$ & $335$  \\
 %    &      &         &          \\
VL195&  $63.1$ & $-19.50$ & $231$  \\
 %    &      &         &             \\
VL20&  $79.4$ & $-20.00$ & $200$ \\
  %   &      &         &            \\
VL205&  $100.0$ & $-20.50$ & $98$ \\
     &      &         &            \\
\hline
\end{tabular}
\end{center} \end{table}

\begin{table} \begin{center}
\caption{\label{tab1pp} Perseus-Pisces: Apparent magnitude limit =15.5
Total number of galaxies =3301 }
\begin{tabular}{|c|c|c|c|c|c|}
\hline
     &      &         &       &      &        \\
{\rm Sample} & $d_{lim} (h^{-1}Mpc)$
& $M_{lim}$ & $N$ & $R_{s} (h^{-1}Mpc)$ & $r_{0} (h^{-1}Mpc)$\\
     &      &         &       &      &        \\
\hline
%     &      &         &       &      &        \\
VL40 &  $40$ & $-17.53$ & $291$ & $8$ & $3\pm 0.5$\\
 %    &      &         &       &      &        \\
VL50 &  $50$ & $-18.03$ & $871$ & $12$ & $4.5\pm 0.5$\\
  %   &      &         &       &      &        \\
VL60 &  $60$ & $-18.43$ & $990$ & $13$ & $4\pm 0.5$ \\
     %&      &         &       &      &        \\
VL70 &  $70$ & $-18.77$ & $975$ & $16$ & $5\pm 0.5$\\
    % &      &         &       &      &        \\
VL80 &  $80$ & $-19.07$ & $894$ & $19$ & $6\pm 0.5$\\
    % &      &         &       &      &        \\
VL90 &  $90$ & $-19.33$ & $780$ & $20.4$ & $7\pm 0.5$\\
    % &      &         &       &      &        \\
VL100 & $100$ & $-19.57$ & $688$ & $23$ & $7.6\pm 0.5$\\
    % &      &         &       &      &        \\
VL110 & $110$ & $-19.79$ & $451$ & $24$ & $8\pm 0.5$ \\
    % &      &         &       &      &        \\
VL120 & $120$ & $-19.98$ & $291$ & $28$ & $10\pm 0.5$\\
     &      &         &       &      &        \\
\hline
\end{tabular}
\end{center} \end{table}

\begin{table} \begin{center}
\caption{\label{tab2pp}  Perseus-Pisces: Apparent magnitude limit =15.5
Total number of galaxies =3301
 (cut in absolute magnitude)}
\begin{tabular}{|c|c|c|c|}
\hline
     &      &         &            \\
{\rm Sample} & $d_{lim} (h^{-1}Mpc)$
& $M_{lim}$ & $N$ \\
     &      &       &              \\
\hline
  %   &&      &                     \\
VL18 &  $50.1$ & $-18.00$ & $895$  \\
    % & &     &                   \\
VL185 &  $63.1$ & $-18.50$ & $1050$  \\
    % &  &    &                     \\
VL19 &  $79.4$ & $-19.00$ & $975$  \\
    % &   &   &                  \\
VL195&  $100.0$ & $-19.50$ & $798$  \\
    % &    &  &                      \\
VL20&  $125.9$ & $-20.00$ & $468$ \\
     &     & &                     \\
\hline
\end{tabular}
\end{center} \end{table}
Then we construct some VL subsamples for the PP14.5
catalog (apparent magnitude limit $14.5$) 
whose characteristics are reported in Tab.\ref{tab3pp}.
\begin{table} \begin{center}
\caption{ Perseus-Pisces: Apparent magnitude limit =$14.5$
Total number of galaxies =$865$ \label{tab3pp}}
\begin{tabular}{|c|c|c|c|c|c|}
\hline
     &      &         &       &      &     \\
\rm{Sample} & $d_{lim} (h^{-1}Mpc)$ & $M_{lim}$ & $N$
& $R_s  (h^{-1}Mpc)$ & $r_{0}  (h^{-1}Mpc)$\\
     &      &      &        &           &     \\
\hline
  %   &      &         &       &      &           \\
VL40(b) & $40$ & $-18.53$ & $114$ & $8.0 $ & $3.0 \pm 0.5$ \\
  %   &      &         &       &      &           \\
VL50(b) & $50$ & $-19.03$ & $318$ & $9.0 $ & $3.5\pm 0.5$ \\
   %  &      &         &       &      &           \\
VL60(b) & $60$ & $-19.43$ & $278$ & $12.0 $ & $4.5\pm 0.5$ \\
   %  &      &         &       &      &           \\
VL70(b) & $70$ & $-19.78$ & $205$ & $15.0$ & $5.0\pm 0.5$ \\
   %  &      &         &       &      &           \\
\hline
\end{tabular}
\end{center} \end{table}

\subsection*{LEDA}

We refer to Sec.\ref{gammaleda} for a detailed discussion of the 
properties of this database, and for a list of reference 
where it is possible to find further informations.
Here we report several tables with the characteristics of various 
VL samples extracted from this database. 

In Tab.\ref{tableda1} we show the characteristics 
of the VL samples extracted form the database limited at
the apparent magnitude $14.5$ (LEDA14.5N), in 
the same sky region of 
CfA1.
\begin{table} \begin{center}
\caption{The VL subsamples of LEDA14.5N-CfA1 ($\Omega =1.8$)
\label{tableda1} }
\begin{tabular}{|c|c|c|c|c|}
%\begin{tabular}{lllll}
\hline
 SAMPLE      &   $d_{lim} (h^{-1}Mpc)$ & $M_{lim}$ & $N$&  p       \\
     &      &         &       &               \\
\hline
     &      &         &       &               \\
VL40      & 40 & -18.54     & 445    & 8   \%    \\
   %  &      &         &       &              \\
VL60      & 60 & -19.43     & 320     & 2   \%    \\
   %  &      &         &       &        \\
VL80      & 80 & -20.07     & 226     & 1.1 \%    \\
    &      &         &       &                 \\
\hline
\end{tabular}
\end{center} \end{table}

In Tab.\ref{tableda2} we show the properties of the VL samples
(limited in distance)
extracted from the database limited at $14.5$ (LEDA14.5N 
for the northern hemisphere, and LEDA14.5S for the southern one).
The solid angle is $\sim 5 \; sr$. 
\begin{table} \begin{center}
\caption{The VL subsamples of LEDA14.5N and LEDA14.5S
(North N=4703, South N=4163)  \label{tableda2}}
\begin{tabular}{|c|c|c|c|c|}
%\begin{tabular}{lllll}
\hline
     &      &         &       &               \\
SAMPLE      &   $d_{lim} (h^{-1}Mpc)$ & $M_{lim}$ & $N$&  p       \\
%SAMPLE      & VL & Limiting   &  nb. of & \%        \\
%North-South & Mpc&  abs. mag. &  gal.   & percent.  \\
\hline
     &      &         &       &               \\
VL40-N      & 40 & -18.54     & 1096    & 8   \%    \\
  %   &      &         &       &               \\
VL40-S      & 40 & -18.54     & 683     & 6   \%    \\
   %  &      &         &       &               \\
VL60-N      & 60 & -19.43     & 838     & 2   \%    \\
   %  &      &         &       &               \\
VL60-S      & 60 & -19.43     & 1088    & 2.3 \%    \\
   %  &      &         &       &               \\
VL80-N      & 80 & -20.07     & 542     & 1.1 \%    \\
   %  &      &         &       &               \\
VL80-S      & 80 & -20.07     & 571     & 1.1 \%    \\
     &      &         &       &               \\
\hline
\end{tabular}
\end{center} \end{table}
In Tab.\ref{tableda3} 
we show the properties of the VL samples
(limited in absolute magnitude)
extracted from the database limited at $14.5$.
\begin{table} \begin{center}
\caption{\label{tableda3} The VL subsamples of LEDA14.5N and LEDA14.5S
(North N=4703, South N=4163)  (cut in absolute magnitude)}
\begin{tabular}{|c|c|c|c|}
\hline
     &      &         &            \\
{\rm Sample} & $d_{lim} (h^{-1}Mpc)$
& $M_{lim}$ & $N$ \\
     &      &         &            \\
\hline
 %    &      &         &              \\
VL18S &  $31.3$ & $-18.00$ & $578$  \\
  %   &      &         &            \\
VL18N &  $31.3$ & $-18.00$ & $1138$  \\
   %  &      &         &            \\
VL185S &  $39.3$ & $-18.50$ & $712$  \\
   %  &      &         &            \\
VL185N &  $39.3$ & $-18.50$ & $1120$  \\
  %   &      &         &            \\
VL19S &  $49.3$ & $-19.00$ & $1034$  \\
  %   &      &         &          \\
VL19N &  $49.3$ & $-19.00$ & $1095$  \\
  %   &      &         &          \\
VL195S&  $61.8$ & $-19.50$ & $1160$  \\
  %   &      &         &             \\
VL195N&  $61.8$ & $-19.50$ & $844$  \\
  %   &      &         &             \\
VL20S&  $77.4$ & $-20.00$ & $675$ \\
 %    &      &         &            \\
VL20N&  $77.4$ & $-20.00$ & $634$ \\
   %  &      &         &            \\
VL205S&  $96.9$ & $-20.50$ & $371$ \\
  %   &      &         &            \\
VL205N&  $96.9$ & $-20.50$ & $282$ \\
     &      &         &            \\
\hline
\end{tabular}
\end{center} \end{table}

Tab.\ref{tableda4} is the same 
of Tab.\ref{tableda2} for the database
limited at the apparent magnitude $16$ (LEDA16N and LEDA16S)
\begin{table} \begin{center}
\caption{
The VL subsamples of LEDA16N and LEDA16S \label{tableda4}}
\begin{tabular}{|c|c|c|c|c|}
%\begin{tabular}{lllll}
\hline
     &      &         &       &               \\
SAMPLE      &   $d_{lim} (h^{-1}Mpc)$ & $M_{lim}$ & $N$&  p       \\
%SAMPLE      & VL & Limiting   &  nb. of & \%    \\
%North-South & Mpc&  abs. mag. &  gal.   & perct.\\
\hline
     &      &         &       &               \\
VL40-N      & 40 & -17.04     & 2845    & 23 \% \\
  %   &      &         &       &               \\
VL40-S      & 40 & -17.04     & 1598    & 13 \% \\
  %   &      &         &       &               \\
VL80-N      & 80 & -18.57     & 4550    & 9  \% \\
  %   &      &         &       &               \\
VL80-S      & 80 & -18.57     & 4264    & 8  \% \\
%     &      &         &       &               \\
VL120-N     & 120& -19.48     & 3565    & 3  \% \\
%     &      &         &       &               \\
VL120-S     & 120& -19.48     & 3458    & 3  \% \\
%     &      &         &       &               \\
VL160-N     & 160& -20.13     & 1669    & 1  \% \\
 %    &      &         &       &               \\
VL160-S     & 160& -20.13     & 1898    & 1  \% \\
     &      &         &       &               \\
\hline
\end{tabular}
\end{center} \end{table}
Tab.\ref{tableda5} is the same 
of Tab.\ref{tableda3} for the database
limited at the apparent magnitude $16$ (LEDA16N and LEDA16S)
\begin{table} \begin{center}
\caption{\label{tableda5} The VL subsamples of LEDA16N and LEDA16S
(North N=13362, South N=11794)  (cut in absolute magnitude)}
\begin{tabular}{|c|c|c|c|}
\hline
     &      &         &            \\
{\rm Sample} & $d_{lim} (h^{-1}Mpc)$
& $M_{lim}$ & $N$ \\
     &      &         &            \\
\hline
 %    &      &         &              \\
VL18S &  $61.8$ & $-18.00$ & $4402$  \\
 %    &      &         &            \\
VL18N &  $61.8$ & $-18.00$ & $4011$  \\
%     &      &         &            \\
VL185S &  $77.4$ & $-18.50$ & $4632$  \\
%     &      &         &            \\
VL185N &  $77.4$ & $-18.50$ & $4827$  \\
%     &      &         &            \\
VL19S &  $96.9$ & $-19.00$ & $4481$  \\
%     &      &         &          \\
VL19N &  $96.9$ & $-19.00$ & $4678$  \\
%     &      &         &          \\
VL195S&  $121.0$ & $-19.50$ & $3724$  \\
%     &      &         &             \\
VL195N&  $121.0$ & $-19.50$ & $3761$  \\
%     &      &         &             \\
VL20S&  $150.9$ & $-20.00$ & $2463$ \\
  %   &      &         &            \\
VL20N&  $150.9$ & $-20.00$ & $2143$ \\
  %   &      &         &            \\
VL205S&  $187.8$ & $-20.50$ & $1305$ \\
  %   &      &         &            \\
VL205N&  $187.8$ & $-20.50$ & $901$ \\
     &      &         &            \\
\hline
\end{tabular}
\end{center} \end{table}
 Tab.\ref{tableda6} is the same 
of Tab.\ref{tableda2} for the database
limited at the apparent magnitude $17$ (LEDA17N and LEDA17S)
\begin{table} \begin{center}
\caption{
The VL subsamples of LEDA17N and LEDA17S \label{tableda6}}
\begin{tabular}{|c|c|c|c|c|}
\hline
     &      &         &       &               \\
SAMPLE      &   $d_{lim} (h^{-1}Mpc)$ & $M_{lim}$ & $N$&  p       \\
\hline
     &      &         &       &               \\
VL40-N      & 40 & -16.03     & 3766    &  31 \%   \\
  %   &      &         &       &               \\
VL40-S      & 40 & -16.03     & 1975    &  17 \%   \\
  %   &      &         &       &               \\
VL120-N     & 120& -18.48     & 7388    &  7\%     \\
   %  &      &         &       &               \\
VL120-S     & 120& -18.48     & 6913    &  6 \%    \\
   %  &      &         &       &               \\
VL160-N     & 160& -19.13     & 5919    &  3\%     \\
   %  &      &         &       &               \\
VL160-S     & 160& -19.13     & 5731    &  3\%     \\
   %  &      &         &       &               \\
VL200-N     & 200& -19.65     & 3914    &  1.3 \%  \\
   %  &      &         &       &               \\
VL200-S     & 200& -19.65     & 4118    &  1.4 \%  \\
     &      &         &       &               \\
\hline
\end{tabular}
\end{center} \end{table}
Tab.\ref{tableda7} is the same 
of Tab.\ref{tableda3} for the database
limited at the apparent magnitude $17$
(LEDA17N and LEDA17S)
\begin{table} \begin{center}
\caption{\label{tableda7} The VL subsamples of LEDA17N and LEDA17S
(North N=15407, South N=14101  (cut in absolute magnitude)}
\begin{tabular}{|c|c|c|c|}
\hline
     &      &         &            \\
{\rm Sample} & $d_{lim} (h^{-1}Mpc)$
& $M_{lim}$ & $N$ \\
     &      &         &            \\
\hline
  %   &      &         &              \\
VL18S &  $96.8$ & $-18.00$ & $7518 $ \\
   %  &      &         &            \\
VL18N &  $96.8$ & $-18.00$ & $8022$  \\
   %  &      &         &            \\
VL185S &  $121.0$ & $-18.50$ & $7695$  \\
   %  &      &         &            \\
VL185N &  $121.0$ & $-18.50$ & $8083$  \\
  %   &      &         &            \\
VL19S &  $150.9$ & $-19.00$ & $6940$  \\
   %  &      &         &          \\
VL19N &  $150.9$ & $-19.00$ & $6954$  \\
  %   &      &         &          \\
VL195S&  $187.8$ & $-19.50$ & $5356$  \\
  %   &      &         &             \\
VL195N&  $187.8$ & $-19.50$ & $5002$  \\
   %  &      &         &             \\
VL20S&  $233.1$ & $-20.00$ & $3390$ \\
   %  &      &         &            \\
VL20N&  $233.1$ & $-20.00$ & $2767$ \\
  %   &      &         &            \\
VL205S&  $288.5$ & $-20.50$ & $1637$ \\
   %  &      &         &            \\
VL205N&  $288.5$ & $-20.50$ & $1187$ \\
     &      &         &            \\
\hline
\end{tabular}
\end{center} \end{table}

\subsection*{IRAS}

The IRAS 2Jy survey (hereafter I-2)
is a complete sample of galaxies uniformly selected over
most of the sky. We refer the reader to Strauss \etal, 1990,
Strauss \etal, 1992, for a detailed discussion of the sample
selection. The sky coverage of the survey is $11.01 sr$.
The selection criteria have been chosen
in relation to the infrared flux, and in
particular there have been selected the galaxies with
apparent flux at $60 \mu m$ ($f_{60}$) greater than
$ 1.936 Jy$. This sample contains $2652$ galaxies.
The IRAS 1.2 Jy survey (hereafter I-12) is a similar catalog, but
it collects all the galaxies with
$f_{60} > 1.2 Jy$, and the number of galaxies is doubled, and contains
 5313 redshifts \cite{str96}.
% (Strauss \etal, 1995).
For the well-calibrated fluxes and the lack of Galactic extinction
in the IRAS wavelengths, these surveys are believed to be
well suited for studies of large scale distribution of galaxies.

The redshift are converted to the Local Group reference frame
%\cite{yah77}
%(Yahil \etal, 1977) 
and no further corrections have been made
for peculiar motions.
In order to eliminate the possible selection effect due to the residual
galactic extinction, we have separated the catalogs in two
half sky samples.
In this way we have for example,
I-2N (I-12N) which refers to the northern hemisphere, and
I-2S (I-12S) which refers to the southern one.
% The sample I-2N
%contains XXX galaxies, and the sample I-2S YYY galaxies. 
The sample I-12N contains $N=2851$ galaxies and the sample I-12S
contains $N=2480$ galaxies.

We have then built the volume limited samples (VL)
from these flux limited surveys
(Table \ref{tableiras1},
 Table \ref{tableiras2}).
                 
\begin{table} \begin{center}
\begin{tabular}{|c|c|c|c|} %{llll}
\hline
&      &            &                 \\
Sample (North-South)      & $R_{VL}$ (Mpc) & Number  & Percentage        \\
&      &           &                 \\
 \hline
% &      &            &                 \\
VL40-N      & 40 & 189     & 1.5   \%    \\
 %              &      &            &                 \\
VL40-S      & 40 & 111     & 0.9   \%    \\
    %           &      &            &                 \\

VL60-N      & 60 & 166     & 0.6  \%    \\
%               &      &            &                 \\

VL60-S      & 60 & 163     & 0.6  \%    \\
   %            &      &            &                 \\

VL80-N      & 80 & 145     & 0.3  \%    \\
      %         &      &            &                 \\

VL80-S      & 80 & 117     & 0.24 \%    \\
         %      &      &            &                 \\

VL100-N     & 100 & 155     & 0.2  \%    \\
            %   &      &            &                 \\

VL100-S     & 100 & 123     & 0.2 \%    \\
&      &            &                 \\
\hline
\end{tabular}
\caption{The VL subsamples of I-2N and I-2S \label{tableiras1}}
\end{center} \end{table}

\begin{table} \begin{center}
\begin{tabular}{|c|c|c|c|c|} %{lllll}
\hline
&      &            &        &         \\
Sample (North-South)      & $R_{VL} (\hmp)$ & Number  & Percentage   &
 $\ell $    \\
 &      &            &                 &\\
\hline
%&      &            &          &       \\
VL20-N      & 20 & 243     & 7.8  \% & 4   \\
%      &  &      &            &                 \\
VL20-S      & 20 & 160     & 5.2 \%  &5  \\
   %    & &      &            &                 \\
VL40-N      & 40 & 384     & 2.4 \% & 70    \\
      %  &&      &            &                 \\
VL40-S      & 40 & 266    &2.2 \% & 8   \\
 %       &&      &            &                 \\
VL60-N      & 60 & 411   & 1.5  \% & 10    \\
    %    & &     &            &                 \\
VL60-S      & 60 & 471    & 1.7 \%   & 10 \\
       % &  &    &            &                 \\
VL80-N     & 80 & 412    & 0.8  \% & 18    \\
%&      &    &        &                 \\
VL80-S     & 80 & 378    & 0.7 \%  & 14   \\
%&      &     &       &                 \\
VL100-N      & 100 & 352   & 0.4  \% & 18    \\
   %     & &     &            &                 \\
VL100-S      & 100 & 360    & 0.5 \%   & 18 \\
      %  &  &    &            &                 \\
VL120-N     & 120 & 283    & 0.3  \% & 23    \\
%&      &    &        &                 \\
VL120-S     & 120 & 378    & 0.3 \%  & 22   \\
&      &     &       &                 \\
\hline
\end{tabular}
\caption{The VL subsamples of I-12N and I-12S \label{tableiras2}}
\end{center} \end{table}

\subsection*{Las Campanas Redshift Survey}

The Las Campanas Redshift Survey contains over 26000 galaxy spectra, with
an average redshift $z=0.1$, over 700 square degrees of the sky.
The survey is composed by 6 slices $1.5^{\circ}$ in declination by $80^{\circ}$
in right ascension.
In the North galactic hemisphere the 3 slices are centered at 
$\delta = -3^{\circ}\;, -6^{\circ}\;, -12^{\circ}$, while 
in the South hemisphere at $\delta = -39^{\circ}\; , -42^{\circ}\;, -15^{\circ}$ .
The $20 \%$ of the data was obtained
using a 50-object fiber system, while the remaining $80 \%$ of the data was 
taken with 
a 112-object system. 
The survey galaxies were selected in the $r-$band.
For the 50-fiber 
data the isophotal magnitude limits are $16 < m <17.3$, while 
for the 112-fiber data the isophotal magnitude limits are $15 < m < 17.7$
\cite{sch96}.

When there are more objects that meet the photometric criteria
than fibers available in a given $1.5^{\circ} \cdot 1.5^{\circ}$ field
(which composes each slice) the targets galaxies are chosen random from within 
the photometric boundaries. In such a way each it is possible to associate 
to each field a {\it galaxy sampling fraction}. The more uniform slice
is the one centered at $\delta = -12^{\circ}$, and we examine 
this slice in what follows, by extracting some VL samples. 

As the survey is limited by an upper and a lower magnitude,
in order to construct VL samples, one has to put to limits in distance
and absolute magnitude. We show The characteristics of the VL samples 
for the  $\delta = -12^{\circ}$
slice in Tab.\ref{tablelcrs1}. For example the sample SL12-2565 is limited 
in the distance range $250 \div 650$.

\begin{table} \begin{center}
\begin{tabular}{|c|c|c|c|} %{llll}
\hline
        &                         &                  &            \\
Sample  & $\Delta R_{VL} (\hmp)$  & $\Delta M_{lim}$ & N       \\
        &                         &                  &               \\
\hline
  %      %&                         &                  &      \\
SL12-2565   & $250 \div 650$         & $-21.36 \div -22.04$  & 507   \\
    %     &                         &                  &         \\
SL12-153 & $150 \div 300$ & $-19.68 \div -20.93$  & 1132    \\
       %  &                         &                  &       \\
SL12-125& $120 \div 500$ & $-19.29 \div -20.05$  & 565  \\
         %&                         &                  &              \\
SL12-36 & $300 \div 600$ & $-21.19 \div -22.44$  & 651     \\
      &                         &                  &      \\ 
\hline
\end{tabular}
\caption{The VL subsamples of the LCRS (slice SL12) 
catalog with a double cut in distance and absolute magnitude \label{tablelcrs1}.
Total number of galaxies 5189.}
\end{center} \end{table}

\subsection*{ESP}
This galaxy survey consists of a strip of 
$22^{\circ} \cdot 1^{\circ}$ plus a nearby area of
$5^{\circ} \cdot 1^{\circ}$ five degrees west of the strip, in the South
Galactic Pole region. The right ascension limits are 
$22^{h} 30^m$ and $ 01^{h} 20^m$ at mean declination $\delta = - 40^{\circ}$.
This area has been filed with a regular grid of circular fields
with a diameter of $32 \; arcminutes$ \cite{vet94} (this size corresponds
to the field of view of the multifiber spectrograph used).
The limiting magnitude of the survey is $b_j \le 19.4$. The  catalog has
a $95 \%$ completeness at $b_j \le 20.5$, a photometric 
accuracy of $\sim 0.03$ magnitudes 
and an estimated stellar contamination  $\le 10 \%$.
The total number of object is of the order of $4000$. 
The characteristics of the VL samples for the  larger 
strip are shown in Tab.\ref{tabesp1}, while 
 in Tab.\ref{tabesp2} there are shown the numbers for the whole 
 catalog.

\begin{table} \begin{center}
\caption{Apparent magnitude limit = 19.4
Total number of galaxies =3175
\label{tabesp1}Euclidean without K-corrections.
}
\begin{tabular}{|c|c|c|c|}
\hline
     &      &         &            \\
{\rm Sample} & $d_{lim} (h^{-1}Mpc)$
& $M_{lim}$ & $N$ \\
     &      &       &              \\
\hline
    % &&      &                     \\
ES300&  $300$ & $-17.98$ & $641$  \\
     %& &     &                   \\
ES400 &  $400$ & $-18.61$ & $1006$  \\
  %   &  &    &                     \\
ES500&  $500$ & $-19.09$ & $871$  \\
     %&   &   &                  \\
ES600&  $600$ & $-19.49$ & $671$  \\
   %  &    &  &                      \\
ES700&  $700$ & $-19.82$ & $413$ \\
     %&     & &                     \\
ES800&  $800$ & $-20.11$ & $204$ \\
     &     & &                     \\
\hline
\end{tabular}
\end{center} \end{table}

\begin{table} \begin{center}
\caption{\label{tab1esp} Apparent magnitude limit =19.4
Total number of galaxies =2583, first region, 
% (in the sky region $22^{\circle}
%\cdot 1^{\circle}$)
Euclidean without K-corrections. 
(cut in absolute magnitude) \label{tabesp2}}
\begin{tabular}{|c|c|c|c|}
\hline
     &      &         &            \\
{\rm Sample} & $d_{lim} (h^{-1}Mpc)$
& $M_{lim}$ & $N$ \\
     &      &       &              \\
\hline
    % &&      &                     \\
VL18 &  $302$ & $-18.00$ & $551$  \\
     %& &     &                   \\
VL185 &  $380.2$ & $-18.50$ & $862$  \\
     %&  &    &                     \\
VL19 &  $478.6$ & $-19.00$ & $759$  \\
     %&   &   &                  \\
VL195&  $602.5$ & $-19.50$ & $508$  \\
     %&    &  &                      \\
VL20&  $758.6$ & $-20.00$ & $201$ \\
     %&     & &                     \\
 VL205&  $955$ & $-20.50$ & $32$ \\
     &     & &                     \\
\hline
\end{tabular}
\end{center} \end{table}

\subsection*{Stromlo-APM Redshift Survey}
In this survey it has been selected galaxies randomly at the rate 1 in 20 
from the complete APM 
magnitude limited catalog \cite{lov92}. The catalog 
covers 4300 square degrees of the southern sky
defined by $21^{h} \ltapprox \alpha \ltapprox 5^h$ and 
$ -72^{\circ}.5 \ltapprox \delta \ltapprox -17^{\circ}.5$.
The survey is essentially complete to magnitude limit
of $b_j =17.15$ and consists of 1769 galaxies.
In Tab.\ref{tableapm1} we show the characteristics of the VL
samples limited in absolute magnitude. An important selection effects exists,
because all the galaxies with apparent magnitudes brighter than $\sim 14.5$ are 
not included in the survey. In order to construct VL samples 
which are not biased by this effect, we have put two 
limits in distance, as for the case of LCRS (Tab.\ref{tableapm2}, 
where WL12 means the VL sample limited by the distance range 
 $100\div 200  \hmp$).

\begin{table} \begin{center}
\begin{tabular}{|c|c|c|c|c|c|} %{llllll}
\hline
       &                  &            &        &              &          \\
Sample & $R_{VL} (\hmp)$  & $M_{lim}$  & N      & $p$   & $\ell (\hmp)$ \\
\hline
       %&		  & 	       &         &            &          \\
VL18   &              107 & -18.0      & 325     &$2 \%$      & 11       \\
      % &		  & 	       &         &            &          \\  
VL185  &              135 & -18.5      & 406     & $1.7  \% $ & 13       \\
%       &		  & 	       &         &            &          \\   
VL19   &              170 & -19.0      & 506     & $1.3 \%  $  & 16       \\
  %     &		  & 	       &         &            &          \\  
VL195  &              214 &-19.5       &575      & 1.0  \%    & 19   \\
    %   &		  & 	       &         &            &          \\   
VL20   &              269 & -20.0      & 403     & 0.4 \%     & 27   \\
      % &		  & 	       &         &            &          \\  
VL205  & 339              &-20.5       & 176     & 0.1  \%    & 44   \\
  %     &		  & 	       &         &            &          \\     
VL21   & 427              & -21.0      & 339     & 0.02 \%    &   88   \\
       &		  & 	       &         &            &          \\
\hline
\end{tabular}
\caption{The VL subsamples of the APM catalog \label{tableapm1}}
\end{center} \end{table}

\begin{table} \begin{center}
\begin{tabular}{|c|c|c|c|c|c|} %{llllll}
\hline
        &                         &                  &     &       &                 \\
Sample  & $\Delta R_{VL} (\hmp)$  & $\Delta M_{lim}$ & N   & $p$   & $\ell (\hmp)$    \\
        &                         &                  &     &        &                 \\
\hline
%        &                         &                  &     &        &                 \\
WL12    & $100 \div 200$         & $-19.35 \div -20.5$  & 451   & 1.0 \% & 19  \\
   %      &                         &                  &     &        &                 \\
 WL13    & $100 \div 300$ & $-20.5 \div -20.23$  & 136   & 1.0 \% & 44 \\
      %   &                         &                  &     &        &                 \\
 WL515    & $50 \div 150$ & $-18.7 \div -19.0$  & 104   & 1.0 \% & 23  \\
         %&                         &                  &     &        &                 \\
 WL1525   & $150 \div 250$ & $-19.81 \div -21.4$  & 367   & 1.0 \% & 24  \\
%         &                         &                  &     &        &                 \\
 WL23     & $200 \div 300$ & $-20.2 \div -22$  & 170  & 1.0 \% &36  \\
   %      &                         &                  &     &        &                 \\
 WL2535     & $250 \div 350$ & $-20.6\div -22.5$  & 66   & 1.0 \% & 60  \\
      %   &                         &                  &     &        &                 \\
 WL51     & $50 \div 100$ & $-17.85 \div -19.0$  & 170   & 1.0 \% & 13  \\
         %&                         &                  &     &        &                 \\
 WL115     & $100 \div 150$ & $-18.7 \div -20.5$  & 303   & 1.0 \% & 15  \\
  %       &                         &                  &     &        &                 \\
 WL153     & $150 \div 300$ & $-20.2 \div -21.38$  & 230   & 1.0 \% & 35  \\
         &                         &                  &     &        &                 \\ 
\hline
\end{tabular}
\caption{The VL subsamples of the APM catalog 
with a double cut in distance and absolute magnitude
 \label{tableapm2}}
\end{center} \end{table}

\subsection*{SSRS1}
The SSRS1 is the counterpart of the CfA1 catalog in the southern
hemisphere. The selection criterion is such that this survey contains all
the galaxies with $D(0) \le 0.1$, where $D(0)$ is a "face-on" diameter
defined by \cite{dac91}
\be
\label{ssrs1_1}
\log D(0) = \log D_1 -0.235 A(T) \log (D_1/D_2)
\ee
and $D_1$ is the major diameter, while $D_2$ is the minor one,
$T$ is the morphological type and $A(T) = 0.894$ if $ T \ge 0$
or $A(T)=0.950$ for $T < 0$. The northern declination boundary of
the catalog is $\delta = -17^{\circ}.4$, and to avoid bias due to the absorption
in the Galactic plane it has been selected galaxies below the galactic
latitude $ b = -30^{\circ}.0$. The sample contains
1773 objects distributed over $1.75 \, sr$. In Tab.\ref{tabssrs11}
there are shown the characteristics of the VL samples.

\begin{table} \begin{center}
\begin{tabular}{|c|c|c|c|c|c|} %{lll lll}
\hline
&                &           &   &       &                 \\
Sample  & $R_{VL}$ (Mpc) & $D_{lim}$ & N & $p$   & $\ell (\hmp)$    \\
        &                &           &   &       &                 \\
\hline
   %     &                &           &   &       &                 \\
        &                &           &   &       &                 \\
 VL60   & 60             & 21.9      &346&5.5 \% & 7    \\
%        &                &           &   &       &                 \\
 VL80   & 80             & 29.2      &241&2.1 \% & 11   \\
   %     &                &           &   &       &                 \\
 VL100      & 100 & 36.6 &195   & 1.1  \% & 14   \\
      %   &                &           &   &       &                 \\
 VL120     & 120 & 44 & 134    & 0.5 \%  & 21   \\
       &                &           &   &       &                 \\
\hline
\end{tabular}
\caption{The VL subsamples of the SSRS1 catalog 
\label{tabssrs11}}
\end{center} \end{table}


\begin{thebibliography}{99}

\bibitem{pie87} Pietronero L., Physica A, 144, (1987) 257

\bibitem{cp92}
 Coleman, P.H. \& Pietronero, L., Phys.Rep. 231, (1992) 311  

\bibitem {pee93} Peebles P. J. E., (1993) 
Principles of physical cosmology, Princeton Univ. Press

\bibitem{wei72} Weinberg S.E. (1972) 
Gravitation and Cosmology, Wiley \& Sons, New York

\bibitem{char1} Charlier C.V.L., Arkiv. for mat. Astron. Physik 4, 
(1908) 1

\bibitem{char2} Charlier C.V.L., 
Arkiv. for mat. Astron. Physik 16, (1922)1

\bibitem{devac1} De Vaucouleurs, G., Science, 167, (1970) 1203 
 
\bibitem{devac2} De Vaucouleurs, G., Publ. A.S.P. 83, (1971)113

\bibitem{bslmp94} Baryshev, Y., Sylos Labini, F., 
Montuori, M., Pietronero, L. Vistas in Astron. 38, (1994)  419

\bibitem{klein} Klain O., Nature 211
(1966) 1338


\bibitem{lerner} Lerner E., IEEE Trans on
Plasma Sci. PS-14 (1986) 609 


\bibitem{wil81} Wilson K.G.,  Phys. Rep. 12, (1974) 75

\bibitem{amit86} Amit D., (1978) "Filed Theory, the Renormalization 
Group and Critical Phenomena" (Mc Graw-Hill, New York)

\bibitem{man82} Mandelbrot B., (1982)
 The Fractal Geometry of Nature,
Freeman, New York

\bibitem{epv95} Erzan A., Pietronero L., Vespignani A.,  
Rev. Mod. Phys.  67, (1995) 554 

\bibitem {sl94} Sylos Labini F., 
Astrophys. J., 433, (1994) 464


\bibitem{huc83} Huchra, J.,  Davis, M., Latham, D., Tonry, J.
 Astrophys. J. .Suppl., 52, (1983) 89.

\bibitem{dp83} Davis, M., Peebles, P. J. E.  
Astrophys. J., 267, (1983) 465

\bibitem{dav97} Davis, M.,  in the Proc of the 
Conference "Critical Dialogues in Cosmology" N. Turok Ed. (1997)
World Scientific

\bibitem{pmsl97}  Pietronero, L., Montuori, M. and Sylos Labini, F., 
in the Proc of the 
Conference "Critical Dialogues in Cosmology" N. Turok Ed. (1997)
World Scientific

\bibitem{cps88}
 Coleman, P.H.  Pietronero, L.,\& Sanders, R.H., 
Astron. Astrophys. Lett. 245, (1988), 1

\bibitem{bs83} Bachall N.A., \& Soneira R.M., Astron. Astrophys.
270, (1983), 20


 


%%%%%%%%%%%%%%%%%%  sec3.tex%%%%%%%%%%%%%%%%%%%%%%%

\bibitem {slgmp96} Sylos Labini F., Gabrielli A., Montuori M., 
Pietronero L.,  
Physica A,   226, (1996) 195


\bibitem{dmpps96} Di Nella, H., Montuori, M., Paturel, G.,
 Pietronero, L., Sylos Labini, F.  Astron. Astrophys. Lett., 308 (1996)
 L33

\bibitem{pee80}  Peebles, P.J.E.  
Large Scale Structure of the Universe , Princeton Univ. Press (1980)

\bibitem{sw95} Strauss M.A., and Willick J.A., (1995),
 Phys. Rep. 261, 271


\bibitem{lim70} Limber D.N. 1953 Astrophys.J., 117 (1953), 655

\bibitem{lim71} Limber D.N. 1953 Astrophys.J., 119 (1953), 655

\bibitem{mad90} Maddox {\em et al.}, 
Mon.Not.R.Astr.Soc.  242, (1990) 43

\bibitem{ben84}  Benzi, R., Paladin, G.,
Parisi, G., Vulpiani, A. J.Phys. A., 17, (1984) 3251.

\bibitem{sie89} Siebesma A.P., Ph.D. Thesis, (1989) Groningen

\bibitem{man96} Mandelbrot B., 
Astrophys Lett \& Comm (1997), in print

\bibitem{fal90} 
Falconer K, "Fractal geometry", J. Wiley \& Sons (1990)

\bibitem{mon97b} Montuori M. \& Sylos Labini F., (1997) preprint

\bibitem{gp84} Grassberger, P. and Procaccia I, 1984 Physica D 13, 34

\bibitem {del88}  De Lapparent, V., Geller, M. J.,
Huchra, J. P. Astrophys. J. 332, (1988) 44

\bibitem{dac94} Da Costa, L. N. {\em et al.} 
 Astrophys. J. 424, (1994) L1

\bibitem{par94} Park, C., Vogeley, M.S., Geller, M., Huchra, J. 
 Astrophys. J., 431, (1994) 569

\bibitem{dac88} Da Costa, L. N. {\em et al.} 
  Astrophys. J., 327, (1988) 544

\bibitem{hg88} 
Haynes, M., Giovanelli, R., (1988) 
in "Large-scale motion in the Universe", 
Eds. Rubin, V.C., Coyne, G., Princeton University Press, Princeton

\bibitem{sch96} Schectman S. {\em et al.}, 
Astrophys. J., (1996) 470,172
 
\bibitem{str96} Fisher K.,   {\em et al.}, 
Astrophys.J. Suppl., (1996) 100, 69

\bibitem{bro90}  Broadhurst, T. J., \etal 
 Nature, 343, (1990) 726

\bibitem{vet94} Vettolani, G., {\em et al.} (1994)
 Proc. of Scloss Rindberg 
workshop  Studying the Universe with Clusters
of Galaxies

\bibitem{zuc97} Zucca E., {\em et al.} (1997) 
Astron. \& Astrophys., in print


\bibitem{pee89} Peebles P.J.E., Physica D, 38, (1989) 273

\bibitem{slp96} Sylos Labini F. \& Pietronero L., 
Astrophys.J.,  (1996), 469, 28

%%%%%%%%%%%%%%  section 4  %%%%%%%%%%%%%%%%%%%

\bibitem{sch76} Schecther, P., 
Astrophys.J. 203, (1976) 297

\bibitem{dav88} Davis, M. {\em et al.}  
 Astrophys.J.Lett., 333, (1988) L9

\bibitem{vog92} Vogeley M.S., Park C., Geller m.J. \& Huchra J.P.
Astrophys. J., 391, (1992), L5


\bibitem {slmp96} Sylos Labini F., Montuori M., 
Pietronero L.,  Physica A,  230 (1996)  336 


\bibitem{guz92} Guzzo L., {\em et al}  
Astrophys. J., 382, (1992) L5

\bibitem{dac91} Da Costa N., Pellegrini P., Davis M., Miksin A.,
Sargent W.L and Tonry J.L. 
Astrophys. J. Suppl., 75 (1991), 935

\bibitem{mau92} Maurogordato S., Schaeffer R. and Da Costa 
L.N., Astrophys.J. 390, (1992), 17

\bibitem{ben96} Benoist C., Maurogordato S., Da Costa L.N.,
Cappi A., and Schaeffer R., Astrophys. J., (1996), 472, 452

\bibitem{lov92} Loveday J., Peterson B.A., Efstathiou G. and Maddox
 S.J., Astrophys.J., 390 (1992), 338

\bibitem{lov96} Loveday J., Peterson B.A, Maddox S.J. and Efstathiou G. 
Astrophys.J. (1996), in press 


\bibitem{ledanat} Di Nella H., Sylos Labini F., Montuori M.,
Paturel G. and Pietronero L., (1996) preprint

\bibitem{ledabias}  Amendola L., Di Nella H., 
Montuori M., Sylos Labini F., (1997) Fractals, in print

\bibitem{ledaps}  Amendola L., Di Nella H., 
Montuori M., Sylos Labini F., (1997) {\it preprint}
 
\bibitem{ledalong}  Montuori M., Sylos Labini F., 
Di Nella H., Amendola L., (1997) {\it preprint}
 
\bibitem{ledasesto} Di Nella, H.   Sylos Labini, F.  Proc. of the workshop 
"Observational cosmology" G. Giuricin, F. Mardiriossan, and M. 
Mezzetti eds.

\bibitem{pat94}  Di Nella H., and Paturel G., 
Comptes redus de l'Academie des Sciences Paris,
serie II, t319, (1994) 57

\bibitem{pat95} Paturel g., Vauglin I., Garnier R., Marthinet M.C., 
Petit C., Di Nella H., Bottinelli L., Gouguenheim L., Durand N., 
in "Databases and On-Line data in astronomy" Eds. Egert D. and
 Albrecht M., (1995) Knluwer Academic Publisher

\bibitem{tur96} Tucker D.L. \etal, Mon.Not.R.Astr.Soc.   (1997) in print

\bibitem{str90} Strauss, M.A.  {\em et al.}, 
Astrophys.J.  361, (1990) 49

\bibitem{sau91} Saunders W. \etal Nature 349, (1991) 32

\bibitem{str92} Strauss M.A., {\em et al.}, 
Astrophys.J.Suppl. 83, (1992) 29

\bibitem{fis94} Fisher K., {\em et al} 
 Mon.Not.R.Astr.Soc. 266, (1994) 50

\bibitem{tu86} Tully B. R.,  , Astrophys.J
  303 , (1986) 25

\bibitem{tu92} Tully, B. R., Scaramella, R., Vettolani G.
\& Zamorani G., Astrophys.J  388, (1992) 9

\bibitem {ei94} Einasto J. et al.,  
  Mon.Not.R.Astr.Soc.  269, (1994) 301

\bibitem{bb85}   Batusky D. J. \& Burns, J. O., 
Astrophys.J  299, (1985) 5 

\bibitem{pos92}  Postman M., Huchra J. P., \& Geller M. J., 
 Astrophys.J  384, (1992) 404

\bibitem{zuc93} Zucca E., Scaramella, R., Vettolani G.,
Zamorani G.,   Astrophys.J  407, (1992) 470

\bibitem{kf91} Kauffmann G. \&, Fairall A. P.,
 Mon.Not.R.Astr.Soc.  248, (1991) 313

\bibitem{li95} Lindner U., Einasto J., Einasto, M., 
Wolfram F., Klaus F., Tago E.,  
Astron. Astrophys.,  301, (1995) 329 

\bibitem{elad97} H. El-Ad \& T. Piran (1997)
preprint (astro-ph/9702135)

\bibitem{kl83} Klypin, A. A. \& Kopylov, A.I., 
Soviet Astr. Lett.,  9, (1983) 41

\bibitem {bah88} Bachall N. A.,  
Astron. Astrophys.  Ann. Rev.,   26, (1988) 63

\bibitem {hu90}Huchra J. P., Henry J. P., Postman M., 
Geller M.J.,  
Astrophys.J. 365, (1990), 66

\bibitem {ca92} Cappi A. \& Maurogordato S., 
Astron.Astrophys.  259, (1992) 423

\bibitem {su88} Sutherland W. J., 
 Mon.Not.R.Astr.Soc. , 234, (1988) 159

\bibitem {su91} Sutherland W. J \& Efstathiou, G.,
 Mon.Not.R.Astr.Soc, 248, (1991) 159

\bibitem {ef92}Efstathiou G., Dalton G. B., 
Sutherland W. J., Maddox S. J.,  
Mon.Not.R.Astr.Soc., 257, (1992) 125

\bibitem {dek89} Dekel A., Blumenthal, G. R.,
Primack,  J. R., Olivier S., 
Astrophys.J.,  338, (1989) L5

\bibitem {va94} Van Harlem M. P., (1996),
preprint 

\bibitem {lu92} Lumsden S. L., Nichol R. C.,
Collins C.A., Guzzo L., 
Mon.Not.R.Astr.Soc., 258, (1992) 1

\bibitem {gu92} Guzzo L., Collins C.A., Nichol R.C., 
Lumsden S. L., 
Astrophys.J., 393, (1992) L5

\bibitem {da92} Dalton G. B.,  Efstathiou G.,  Maddox S. J
Sutherland W. J., Astrophys.J.,  390, (1992)  L1

\bibitem {ma90a}  Maddox S. J Efstathiou G., Sutherland W. J.,
 Mon.Not.R.Astr.Soc.,  243, (1990) 629

\bibitem {ma90b}  Maddox S. J Efstathiou G., Sutherland W. J.,
Mon.Not.R.Astr.Soc., 246, (1990) 433

\bibitem {dal94} Dalton G. B., Efstathiou G., Maddox S. J
Sutherland W. J., Mon.Not.R.Astr.Soc., 269, (1994) 151

\bibitem {dab94} Dalton G. B., Efstathiou G., Sutherland W. J., Maddox S. J., Davis M.,
Mon.Not.R.Astr.Soc.,  271, (1994) L47

\bibitem {ro94} Romer A.K. et al.,  
Nature, 372, (1994) 75

\bibitem{bah92} Bachall N. \& West M.J.
Astrophys.J. 392, (1992) 419

\bibitem {bri94} Briel, U. G. \&  Henry, J. P., 
Astron.Astrophys., 278, (1994) 379

\bibitem {bah86} Bachall N. A. \& Burgett W. S., 1986
Astrophys.J., 300, (1986) L35

\bibitem {sz85} Szalay A. S. \& Schramm D. N.,
Nature,   314, (1985) 718 

\bibitem {msla97} Montuori M., Sylos Labini F. and Amici A.,
Physica A (1997), in print

\bibitem {ab89} Abell G. O., Corwin H.G. Jr., Olowin R. P.,
 Astrophys.J.,  70, (1989) 1

\bibitem{pee-pc96} Peebles P.J.E., Private Communication (1996)

%%%%%%%%%%%%%%%% SEC5.TEX %%%%%%%%%%%%%%%%%%%%%%%%%%%%%%%%%%%%%%%%%%

\bibitem{sla96a} Sylos Labini  F. \& Amendola L. 
Astrophys.J. 468 (1996) L1  

\bibitem{sla96b} Sylos Labini  F. \& Amendola  L.
Astrophys.Lett and Comm., (1997) in print

\bibitem{be94} Baugh C.M. and Efstathiou G. 
Mon.Not.R.Astr.Soc., 267 (1994) 323

\bibitem{fe94} Feldman H. Kaiser N. \& Peacock J.
 \etal (1994) Ap.J. 426, 23  

\bibitem{it92} Itho M., Suginohara T. \& Suto Y., 
PASJ 44, (1992), 481

\bibitem{col94} Colombi S., Bouchet F.R. and Schaeffer R.
Astron.Astrophys., 281 (1994), 301

\bibitem{pn91} Peacock, J.A., Nicholson, D.
 Mon.Not.R.Astr.Soc, 235 (1991) 307 

\bibitem{fis93} Fisher et al. 1993,
Astrophys.J.,  402, (1993) 42

\bibitem{benn94} 
Bennett C.L. \etal Astrophys.J., 436 (1994), 423

%%%%%%%%%%%%%%%%%%%% sec6.tex %%%%%%%%%%%%%%%%%%%%%%%%%%%%%%%%


\bibitem{vo08} Voronoi, J. Reine Angew. Math. 134, (1908) 198


\bibitem{fluc} Sylos Labini F., \etal (1997) in preparation 

\bibitem{badii84} Badii R., and Politi A. 
Physics Letters A (1984) 104, 303

\bibitem{smith86} Smith L.A., Fournier J.-D.  and Spiegel E.A., 
Physics Letters A (1986) 114, 465

\bibitem{sor96} Sornette D., Johansen A., Arneodo A., Muzy J.F. and 
Saleur H., Phys. Rev. Letters (1996) 76, 251

\bibitem{solis97} Solis F.J. and Tao L., 
(1997) preprint (cond-mat/9703051)

\bibitem{sha84} Shanks T., Stevenson, P.R.F., Fong, R.,
MacGillivray, H.T.,  Mon.Not.R.Astr.Soc.,  206, (1984) 767

\bibitem{bd95}Bellanger C. \& De Lapparent V.
Astrophys.J.,  455, (1995) L1

\bibitem{wil94} Willmer C.N.A. \etal 
Astrophys.J.,  437, (1994) 560

\bibitem{et96} Ettori S., Guzzo L. and Tarenghi M. 
 Mon.Not.R.Astr.Soc. (1996), in press

\bibitem{coh96} Cohen J.G. \etal, (1996) in prints (astro-ph/9608121)

\bibitem{sau90} Saunders W., Rowan Robinson M., Lawrence A., 
Efstathiou G., Kaiser N., Ellis R.S. and Frenk C.S.
 Mon.Not.R.Astr.Soc.  242, (1990) 318

%%%%%%%%%%%%%%%%	SEC7.TEX 	%%%%%%%%%%%%%%%%%%

\bibitem{hu26} Hubble, E.  Astrophys.J. 
64, (1926) 321

\bibitem{ms91}  Metcalfe, N., Shanks, T., Fong, R., Jones L.R., 
 Mon.Not.R.Astr.Soc.  249, (1991) 498

\bibitem{mg94} McGaugh, S.,  Nature, 367, (1994) 538.

\bibitem{ty88} Tyson, J.A., Astron.J. , 96, (1988) 1

\bibitem{lc91} Lilly, S. J., Cowie, 
L. L. and Gardner, J. P., Astrophys.J. . 369, (1991) 79

\bibitem{jf91} Jones L. R., Fong, R., Shanks, T., Ellis, R. S., 
\& Peterson, B. A.,   Mon.Not.R.Astr.Soc.  , 249, (1991) 481

\bibitem{dr94} Driver S.P., Phillipps S., Davies J.I., Morgan I. and 
Disney M.J. Mon.Not.R.Astr.Soc.   266, (1994) 155

\bibitem{yp95} Yoshii, \& Peterson, Astrophys.J,. 444, (1995)15 

\bibitem{yt88}  Yoshii, Yu.,
Takahara, F. Astrophys. J. 326, (1988)1

\bibitem{cg90} Cowie L., Gardner, J. P., Lilly S. J., McLean, I. Astrophys.J.
 360, (1990) L1

\bibitem{co91}  Cowie, (1991) in Observational test of Cosmological 
Inflation, ed. T. Shanks, A. J. Bandy, R. S. Ellis, C. S. Frenk, \&
A. W. Wolfendale (Dordrecht: Kluwer), 257

\bibitem{yo93} Yoshii, Astrophys.J, 403, (1993) 552

\bibitem{be92}  Broadhurst, T. J., Ellis, R., S. \& Glazebrook, K. 
Nature, 355, (1992) 55

\bibitem{gc93} Gardner, J.P., Cowie, L., Wainscoat, R. J.,  
Astrophys.J., 415, (1993) L9

\bibitem{co94} Cowie L. L., {\em et al.},  Astrophys.J.,  434, (1994) 114

\bibitem{gl94} Glazebrook K., {\em et al.},   Mon.Not.R.Astr.Soc.   266,
(1994) 65

\bibitem{mb86} Mobasher B., {\em et al.},   Mon.Not.R.Astr.Soc.   223, 
(1986) 11

\bibitem{so94} Soifer, B.T.,  {\em et al.},   Astrophys.J., 420, (1995)L1

\bibitem{dj95} Djorgovski, B. T., {\em et al.}, Astrophys.J., 438, (1995) 
L133

\bibitem{b96} Bertin E. and Dennefeld M., 
Astron.Astrophys., 317, (1996) 413


\bibitem{kkcor} Pence W. Astrophys.J., 203, (1976) 39

\bibitem{con84} Condon J. J. Astrophys.J., 
287, (1984) 461

\bibitem{ht90} Hartwick F. D. A. \& Schade D.   Ann.Rev Astron. Astrophys., 28, 
(1990) 437

\bibitem{rosat} Hasinger G. \etal 
Astron.Astrophys., 275 (1993), 1

\bibitem{fis95} Fishman G. \& Meegan C. 
Ann.Rev Astron. Astrophys., 33, (1995) 415

\bibitem{bri96} Brigss M.S.   Astrophys \& Space Sci. (1996) in press

\bibitem{har95} Hartmann D.H.   Ann.Rev Astron. Astrophys., 33, (1995) 
225

\bibitem{jon96} Jones L.R., \etal Mon.Not.R.Astr.Soc., (1996) in press
(astro-ph/9610124)

\bibitem{mee95} Meegan C. \etal   Astrophys.J.  Suppl. (1995) in print

\bibitem{mee92} Meegan C. \etal Nature 355, (1992) 143

\bibitem{sch88} Schmidt M., Higdon J.C. \& Hunter G.,  Astrophys.J., 329, (1988) 
L85

\bibitem{bri93} Briggs M.S., Astrophys.J., 407, (1993) 126

\bibitem{now94} Nowak M.A.   Mon.Not.R.Astr.Soc.   266, (1994) L45

\bibitem{qua93} Quashnock J. \& Lamb D.   Mon.Not.R.Astr.Soc.   265, 
(1994) L59

\bibitem{mee95b} Meegan C. \etal   Astrophys.J., 446,(1995)  L15

\bibitem{kp96} T. Kolatt, T. Piran  Astrophys.J. (1996) in print  

\bibitem{dp91} Dogterom, M., Pietronero,  L., Phys. A 171, (1991) 239

\bibitem{gp77} Groth, E.J. \& Peebles, P.J.E. Astrophys.J. . 217, (1977) 
385

\bibitem{durrer} Durrer R., Eckmann J-P., Sylos Labini F., 
Montuori M., and Pietronero L., submitted to Phys. Rev. Lett. (1997)

\bibitem{ams95} F. Amendola, L. Montuori, M., Sylos Labini, F. 
(1996),  in preparation 

%%%%%%%%%%%%% SEC9.TEX %%%%%%%%%%%%%%%5


\bibitem{bin88}  Binggeli, B., Sandage, A.,
Tammann, G. A.  
The luminosity function of galaxies.
 Astron. Astrophys. Ann. Rev.  26, (1988) 509

\bibitem{tul88} Tully R.B. (1988) in in "Large scale motion in the
Universe", ed. Rubin V.C and Coyne G., Princeton

\bibitem{tf87}
 Tully R.B. \& Fisher J. R., (1987) 
"Nearby Galaxies Atlas"
Cambridge: Cambridge University Press

\bibitem{ede89}
 Eder J. A. {\it et al.},  Astrophys.J,. 340, (1989) 
29

\bibitem{bin90} 
 Binggeli B., Tarenghi M., Tammann G.A.   Astron.Astrophys. 
 228,(1990) 42

\bibitem{fb94}
 Ferguson H.C. \& Binggeli B.,   Ann. Rev. Astron.Astrophys. 
6, (1994) 67 

\bibitem{thu87} Thuan, T. X., Gott III, J. R., Schneider, S. E. 
Astrophys. J. Lett 315,(1987) L93-7.

\bibitem{bot88} 
Bothun G. D. {\it et al.}, Astrophys.J.  308, (1988) 
510

\bibitem{dp87}
 Disney M. \& Phillips S., Nature 329, (1987) 203

\bibitem{ee85}
 Einasto M. \& Einasto J., Mon.Not.R.Astr.Soc. , 226, (1987) 543

\bibitem{dre84}
 Dressler A., Ann.Rev. Astron.Astrophys.   313, (1984) 42

\bibitem{oem74} Olmer A. Astrophys.J. . 194, (1974) 1

\bibitem{dre80}Dressler A., Astrophys.J., 236, (1980) 351

\bibitem{desou82} Da Souza {\it et al.}, Astrophys.J., 263, (1982) 557

\bibitem{pg84} Postman M. \& Geller M., Astrophys.J., 281, (1984) 95

\bibitem{iov93} Iovino A. \etal Mon.Not.R.Astr.Soc.  265 (1993), 21 

\bibitem{dav76} Davis M. \& Geller M., Astrophys.J. . 208, (1976) 13

%\bibitem{gio88} Giovanelli R. \& Haynes M.P. Astron.J. 97, (1988) 633

\bibitem{gio86} Giovanelli R., Haynes M.P.
Chincarini G.  Astrophys.J. 300, (1986) 77

\bibitem{fel77} Felten J., Astron.J.  82, (1977) 861

\bibitem{del89}  De Lapparent V., Geller M. \& Huchra J., Astrophys.J.,
 343, (1989) 1
 
\bibitem{mar94} Marzke R.O., Huchra J. \& Geller M. 
Astrophys.J., 428, (1994) 43

\bibitem{pv87} Paladin, G., Vulpiani, A.   Phys. Rep. 156, (1987) 147

\bibitem{mj90}  Martinez, V. T., 
Jones, B. J. T Mon. Not. R. astr. Soc., 242, .(1990) 517

\bibitem{fg79} Feber S.M. \& Gallagher J.S.
Astron.Astrophys. Ann.Rev. 17, (1979) 135

\bibitem{jan91} Jensen M.H., Paladin G., Vulpiani A. Phys.Rev.Lett. 67,
 (1991) 208

\bibitem{mar95}
Martinez {\it et al.}, Science 269, (1995) 1245

\bibitem{kai84} Kaiser N., Astrophys.J. . 284, (1984) L9

\bibitem{pt86} Pietronero L. \& Tosatti E., Eds (1986)
Fractal in Physics 
North-Holland, Amsterdam

\bibitem{slp95a}
 Sylos Labini F. \& Pietronero L. (1995) in  "Birth of the Universe
and fundamental physics" F. Occhionero ed., p.317, Springer Verlag.

\bibitem{slp95b}
 Sylos Labini F. \& Pietronero L. 
Astrophys.Lett \& Comm., (1995)   36 , 49
%%%%%%%%%%%%%%%%%%%%%%%% SEC11.TEX %%%%%%%%%%%%%%%%%%%%%%%%%%%%%%%%%%%5

\bibitem{san95} Sandage, A. in The Deep Universe (eds. Binggeli, B., Buser, R.) 
(Springer, 1995)

\bibitem{bpslt96} Baryshev, Y.,  Sylos Labini, F., Montuori M.,  
Pietronero L., and Teerikorpi P.,   (1997) preprint


\bibitem{har93}  Harrison E. Astrophys.J., 403, (1993) 28

\bibitem{pee91} Peebles P.J.E. \etal. Nature, 352, (1991) 769 

\bibitem{hu29} Hubble E., Nat. Astron. Sci. Proc. 15 (1929), 168

\bibitem{smith79}  Smith, R.W. J.Hist.Astron. 10, (1979) 133 


\bibitem{segal76} Segal, I. Mathematical Cosmology and Extragalactic Astronomy 
(Academic Press, New York 1976)
 
\bibitem{tee75} Teerikorpi, P. Astron. Astrophys. 45, 91975) 117 

\bibitem{son79} Soneira, R. M. Astrophys.J. (1979) L63 

\bibitem{wei77} 
Weinberg, S., The First Three Minutes, p.26 (Basic Books, New York 1977)


\bibitem{sandage87} Sandage, A., Astrophys.J. 317, (1987) 557 

\bibitem{zheng91} Zheng, J-Q., Valtonen, M.J., Byrd, G.G., Astron.Astrophys.
 247, (1991) 20 

\bibitem{valt93} Valtonen \etal., Astron.J. 105, (1993) 886 

\bibitem{hw72} Haggerty M.J., Wertz J.R. Mon.Not.R.Astr.Soc., 155, (1972) 495 

\bibitem{fang91} Fang L.L., \etal.Astron.Astrophys., 243, (1991) 283

\bibitem{rib93} Ribeiro M.B. Astrophys.J., 415, (1993) 469

\bibitem{mof92} Moffat, J.W., Tatarski, D.C., Phys. Rev. D, 45, 
(1992) 3512


\bibitem{bon47} 
Bondi H. Mon.Not.R.Astr.Soc., 107 (1947) 410 


\bibitem{haw62} Hawkins, G.S. Nature, 194, (1962) 563

\bibitem{zn84} 
Zeldovich Yu.B., Novikov I.D. Relativistic Astrophysics Vol.2
(The University of Chicago Press)(1984), p. 97

\bibitem{bar81} Baryshev Yu.V. Izvestiya SAO, 14, (1981) 24 
%%%%%%%%%%%%%%%%%%%%%%%%%%%%%%%%%%%%%%




%%%%%%%%%%%%%%%%%%%%%%%%%%%%%%%%%%%%%%%


 \bibitem{free96} Freedman W.L., in the Proc of the 
Conference "Critical Dialogues in Cosmology" N. Turok Ed. (1997)
World Scientific

\bibitem{tamm96} Tammann G.A., in the Proc of the 
Conference "Critical Dialogues in Cosmology" N. Turok Ed. (1997)
World Scientific




\bibitem{pw65} Penzias A.A. \& Wilson R.W.
Astrophys.J. 142, (1965) 419 

\bibitem{mat90}  Mather, J. C. \etal 
Astrophys.J.Lett, 354, (1990)
L37

\bibitem{mat94}  Mather, J. C. \etal 
Astrophys.J., 420, (1994)
439

\bibitem{gus90} Gush H.P., Hlpern M., Wishhow E.H.
Phys Rev. Lett., 65 (1990), 537

\bibitem{fix94}  Fixsen D.J., \etal Astrophys. J. 420 (1994),
445


\bibitem{stru92} Strukov \etal, Sov. astron. Lett. 18 (1992), 153

\bibitem{smoo92} Smooth G. \etal, Astrophys J. 396 (1992), L1

\bibitem{sw67} Sachs R.K. \& Wolf A.M. Astrophys.J. 147, (1967) 73



\bibitem{cou93}  Courteau, S., Faber, S.M., Dressler, A., Willick, J.A.
 Astrophys.J.Lett. ,412, (1993) L51 

\bibitem{lp94} Laurer T.R. \& Postman M., 
Astrophys. J., 425 (1994), 418

\bibitem{mat96} Mathewson, D.S.,  Ford V.L., Buchhorn M.
Astrophys.J.Lett., 389, (1992) L5

 \bibitem{mf94} Mathewson, D.S., Ford, V.L.
Astrophys.J.Lett., 434, (1994) L39



%%%%%%%%%%%%%%%%%%%%%%%%%%%%%%%%%%%%%%
\end{thebibliography}
\end{document}